\documentclass[11pt,ngerman,english]{scrreprt}
\usepackage[T1]{fontenc}
\usepackage[latin9]{inputenc}
\usepackage{color}
\usepackage{babel}
\usepackage{verbatim}
\usepackage{pdfpages}
\usepackage{units}
\usepackage{cite}
\usepackage{scrhack}
\usepackage{amsmath}
\usepackage{amssymb}
\usepackage{graphicx}
\usepackage{setspace}
\usepackage{bm}
\usepackage{esint}
\usepackage[unicode=true,pdfusetitle,
 bookmarks=true,bookmarksnumbered=false,bookmarksopen=false,
 breaklinks=false,pdfborder={0 0 1},backref=false,colorlinks=false]
 {hyperref}
\usepackage{geometry}
\makeatletter
\newcommand{\be}{\begin{equation}}
\newcommand{\ee}{\end{equation}}
\newcommand{\bea}{\begin{eqnarray}}
\newcommand{\eea}{\end{eqnarray}}
\newcommand{\ba}[1]{\begin{array}{#1}}
\newcommand{\ea}{\end{array}}

\newcommand\Tr{\mathrm{Tr}}

\newcommand{\non}{\nonumber\\}

\global\long\def\p{\partial}
\global\long\def\L{{\cal L}}
\global\long\def\ve{\varepsilon}
\global\long\def\vf{\varphi}
\global\long\def\cl{\circlearrowleft}
\global\long\def\eh{\hat{\mathbf{e}}}

\global\long\def\bt{\bar{\theta}}
\global\long\def\gt{\tilde{g}_{8}}
\global\long\def\at{\tilde{A}}
\global\long\def\t{\theta}

\usepackage[normalem]{ulem}  

\numberwithin{equation}{section}
\numberwithin{figure}{section}
\numberwithin{table}{section}

\newcommand{\eqhat}{\mathrel{\hat{=}}}
\usepackage{listings}

\@ifundefined{showcaptionsetup}{}{%
 \PassOptionsToPackage{caption=false}{subfig}}
\usepackage{subfig}
\makeatother
\usepackage{listings}
\definecolor{codegreen}{rgb}{0,0.6,0}
\definecolor{codegray}{rgb}{0.5,0.5,0.5}
\definecolor{codepurple}{rgb}{0.58,0,0.82}
\definecolor{backcolour}{rgb}{0.95,0.95,0.92}

\lstdefinestyle{mystyle}{
backgroundcolor=\color{backcolour}, 
keywordstyle={\color{blue}},
commentstyle={\color{magenta}\itshape},
emphstyle={\color{red}},
numberstyle=\tiny\color{codegray},
breaklines=true,
keepspaces=true,
basicstyle={\ttfamily},
tabsize=2,
numbers=left,                    
numbersep=5pt,
stringstyle={\color{codegreen}},
identifierstyle={\color{cyan}}}

\addto\captionsenglish{}
\addto\captionsngerman{}

\begin{document}
\selectlanguage{ngerman}%
\begin{titlepage}
\vspace{-1cm}
  \begin{center}
	\begin{figure}[ht]
	\begin{center}
	\includegraphics[scale=0.3]{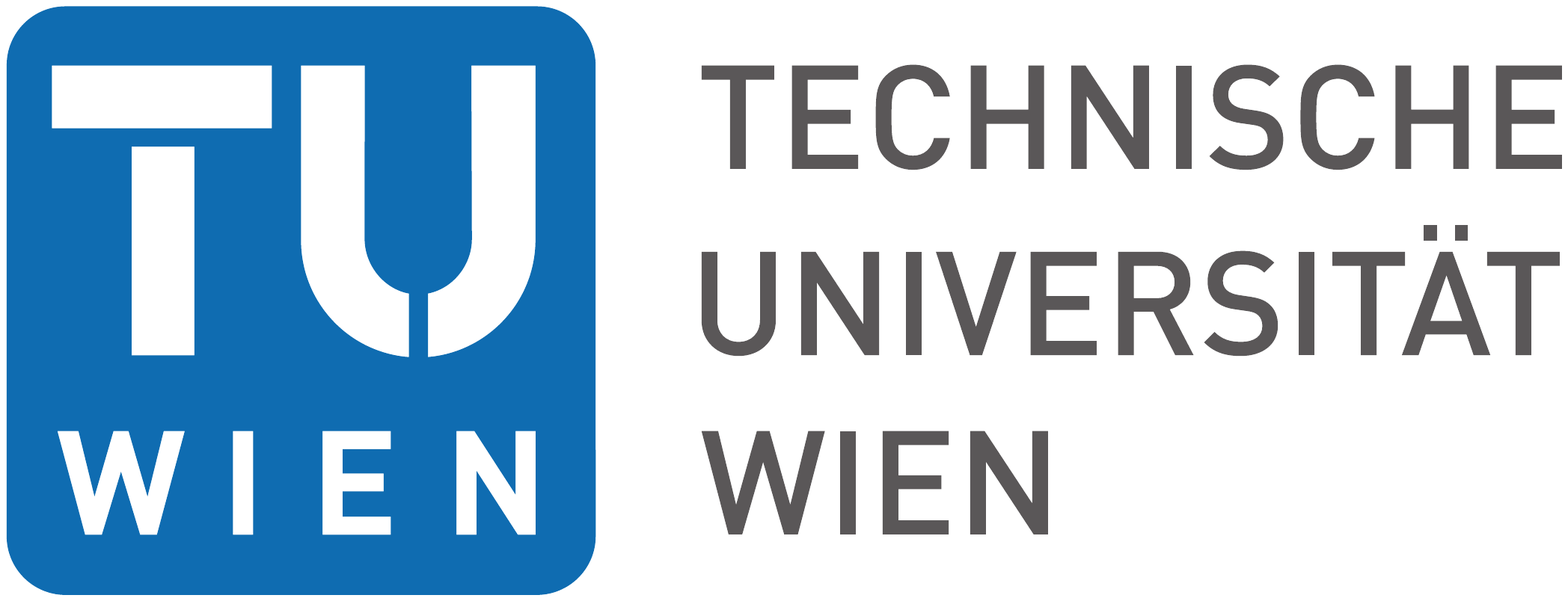}\\[1.0cm]
	\end{center}
	\end{figure}	
\vspace{-2cm}
	{\LARGE DISSERTATION\\[1.0cm]}

    {\LARGE\textbf{Multicomponent Superfluids and Superconductors in Dense Nuclear and Quark Matter}\\[1.0cm]}
   
	\begin{center}    
      {\normalsize ausgeführt zum Zwecke der Erlangung des akademischen Grades eines Doktors der Naturwissenschaften unter der Leitung von}\\[1.5cm]
    \end{center}

	\begin{center}    
      {\normalsize 	Priv.Doz.\ Dr.phil.nat.\ Andreas Schmitt\\
		Mathematical Sciences and STAG Research Centre,\\
		University of Southampton, UK\\[1.5cm]}
    \end{center} 
	\begin{center}    
      {\normalsize eingereicht an der Technischen Universit{\"a}t Wien,\\
      Fakult{\"a}t f{\"u}r Physik} \\[1.5cm]     
     
    \end{center} 
	\begin{center}    
      {\normalsize von\\
        Dipl.-Ing.\ Alexander Haber, BSc\\
        Matrikelnummer 00826860 \\
		Forsthausgasse 10/A2/22\\
		1200 Wien}\\[2.5cm]
    \end{center} 

	\noindent\begin{tabular}{ll}
	\makebox[2cm]{Wien, am 20.08.2018}\hspace*{4cm} & \makebox[4cm]{\hrulefill}
	 \end{tabular}
    
  \end{center} 
\end{titlepage}

\selectlanguage{english}%
\thispagestyle{empty}\newpage{}

\null\vfill%
\noindent\parbox[t]{1\columnwidth}{%
\raggedright{\large\itshape "In the beginning the Universe was created.\\
This has made a lot of people very angry\\ and been widely regarded as a bad move."
\par\bigskip} \Large{Douglas Adams, The Restaurant at the End of the Universe}\par
}

\vfill\vfill \clearpage

\newpage{}

\section*{Abstract}
\addcontentsline{toc}{section}{Abstract}
Matter at intermediate baryon densities and low temperatures is notoriously hard to tackle theoretically. Whereas lattice methods cannot cover more than rather small densities, perturbative methods are only applicable at much higher densities. The regime of intermediate chemical potential at low temperatures in the QCD phase diagram is therefore out of reach of first-principle methods. Whereas earth bound laboratories, like particle accelerators as the LHC at CERN, are capable of studying the high temperature/low density part of the phase diagram, we have to rely on stellar objects to investigate dense nuclear and quark matter at low temperatures. Compact stars can serve as a unique laboratory for this regime. 
In compact stars, nuclear as well as quark matter possibly form complicated interacting multicomponent systems of superfluids and superconductors. Due to their fast rotation and high magnetic field, phenomena like hydrodynamic instabilities and the formation of vortices/flux tubes become of phenomenological interest. In this thesis, I try to investigate these multicomponent system in a consistent multi-fluid treatment. By starting from a field-theoretical, bosonic model, the phase structure of a two-fluid system, e.g.~ consisting of superfluid neutrons and superconducting protons, is explored. The two fields are coupled via a density and a derivative coupling, which leads to the entrainment effect. Consequently, hydrodynamic instabilities, which might serve as trigger for pulsar glitches, are calculated for two-fluid systems. Technically, this amounts to calculating the sound modes of the system, which is identical to the slope of the Goldstone dispersion relation in the case of two superfluids. An analysis shows that the dynamical two-stream instability can only occur beyond Landau's critical velocity, i.e., in an already energetically unstable regime. A qualitative difference is found for the case of two normal fluids, where certain transverse modes suffer a two-stream instability in an energetically
stable regime if there is entrainment between the fluids. 

By incorporating a gauge field and taking into account the charge of one scalar field, the influence of a superfluid on the magnetic phase structure of a superconductor is studied. This amounts to calculating various critical magnetic fields from a temperature-dependent Ginzburg-Landau potential derived from the field-theoretical description of the two-component system. A calculation of the critical magnetic fields for the transition to an array of magnetic flux tubes, based on an approximation for the interaction
between the flux tubes shows that the transition region between type-I and type-II superconductivity
changes qualitatively due to the presence of the superfluid: the phase transitions at the upper and lower critical fields in the type-II regime become first order, opening the possibility of clustered flux tube phases. Since all calculations are performed in a fully relativistic setup,
the presented results are very general and not only of potential relevance for (super)fluids in neutron stars but, in the
non-relativistic limit, in the laboratory.

Color-superconducting quark matter can effectively be described as a multicomponent\linebreak (color-)superconductor as well.  
Color-flavor locked (CFL) quark matter expels color-magnetic fields due to the Meissner effect.
One of these fields carries an admixture of the ordinary abelian magnetic field and therefore flux
tubes may form if CFL matter is exposed to a magnetic field. As in the nuclear system, a Ginzburg-Landau approach for three massless quark flavors is employed. Based on the weak-coupling
expressions for the Ginzburg-Landau parameters, the regime where CFL is a
type-II color superconductor is identified and the radial profiles of different color-magnetic flux tubes are computed.
Among the configurations without baryon circulation, a new solution that is energetically
preferred over the flux tubes previously discussed in the literature in the parameter regime relevant
for compact stars, is found. Within the same setup, a new defect in the 2SC phase, namely
magnetic domain walls, is described. These emerge naturally from the previously studied flux tubes if a more
general ansatz for the order parameter is used. Color-magnetic defects in the interior of compact
stars allow for sustained deformations of the star, potentially strong enough to produce detectable
gravitational waves.

\section*{Preface}

\addcontentsline{toc}{section}{Preface}
The results in this thesis have been largely obtained in collaboration with my supervisor, Andreas Schmitt\footnote{Dr.~Andreas Schmitt, Mathematical Sciences and STAG Research Centre, University of Southampton, UK, a.schmitt@soton.ac.uk}. The investigation of hydrodynamic instabilities was additionally carried out in collaboration with Stephan Stetina\footnote{Dr.~Stephan Stetina, TU Wien, Vienna, Austria, stephan.stetina@tuwien.ac.at}.%

\vspace*{1cm}

The research parts of this thesis are based on the following publications:
\begin{itemize}
\item A.~Haber, A.~Schmitt, S.~Stetina, Phys.Rev.~\textbf{D93} (2016) no.2, 025011, Ref.~\cite{Haber:2015exa} 
\item A.~Haber, A.~Schmitt, EPJ Web Conf.~\textbf{137} (2017) 09003, Ref.~\cite{Haber:2016ljn} 
\item A.~Haber, A.~Schmitt, Phys.Rev.~\textbf{D95} (2017) no.11, 116016, Ref.~\cite{Haber:2017kth} 
\item A.~Haber, A.~Schmitt,  J.Phys.~\textbf{G45} (2018) no.6, 065001, Ref.~\cite{Haber:2017oqb}

\end{itemize} 
\newpage{}
\section*{Acknowledgments}
\addcontentsline{toc}{section}{Acknowledgments}
I want to use this page to thank all the people who supported me all over the years and made this thesis possible.

First and foremost, I own deep gratitude to my supervisor Andreas Schmitt, who sparked my fascination for the physics of compact stars with his great lecture years ago and supported me throughout my entire time at the Institute for Theoretical Physics, even after his move to Southampton. Whether it concerned physics, calculations, coding, writing, preparing talks, networking, traveling  to conferences and the transition to the post-doc level, Andreas was always incredibly helpful, patient and kept my fascination for the topic alive. Thank you very much, it was fun to work with you for (nearly) the last five years.

I also want to thank Nils Andersson for his hospitality and financial support which enabled me to visit University of Southampton three times for a total duration of more than half a year, and his insight in several scientific discussions.

Furthermore, I would like to thank my office colleagues in Vienna, David M{\"u}ller, who accompanied me since the very beginning at the TU Wien ten years ago in 2008, Frederic Brünner, Christian Ecker and Alex Soloviev, for countless discussions on math, physics, coding and in general for many fun hours in an enjoyable office atmosphere. Also thank you to all the other members of the Institute for Theoretical Physics, especially the "bridge gang", who made my experience at the university a very pleasant one.
I am also very thankful to my office colleagues in Southampton, who took me in from the very beginning and made my stay in the UK a fun episode of my studies.

Additional thanks go to Stephan Stetina and Florian Preis for passing on their wisdom and experience as my predecessors.

Thanks for valuable comments to Mark Alford, Ian Jones, Armen Sedrakian and Andreas Windisch.

Finally I want to thank my family, who supported me morally and financially during all my studies. A lot of love to my wife, who supported and encouraged me when times were difficult and stressful, and kept up with my countless stays abroad, my long working hours and my bad mood when sometimes things were not running so smoothly. 

This thesis was financially supported by the Austrian Science Fund (FWF) under project no. W1252, and from the NewCompStar and PHAROS networks, COST Actions MP1304 and  CA16214.

\newpage{}
\section*{Units and Conventions}

\addcontentsline{toc}{section}{Units and Conventions}In this thesis
natural Heaviside-Lorentz units are used. This means that we set $\hbar=c=k_{B}=1$,
and choose electron volts as the unit of energy. As a consequence,
lengths are given by inverse energies, $[l]=\text{eV}^{-1}$. Sometimes,
for comparison with existing literature, femtometers are used as an
unit of length. One femtometer, sometimes called one Fermi, is $1\,\mbox{fm}=10^{-15}\,\mbox{m}$
and corresponds to $1\,\mbox{fm}=1\frac{\hbar c}{\mbox{MeV}}\thickapprox197.327\,\mbox{MeV}^{-1}$.
Temperatures are given in eV as well, where for a rough comparison
$295$K (room temperature) corresponds to $295\,\text{K}\approx\frac{1}{40}\,\text{eV}$.
In order to be able to compare the strength of the occurring magnetic
fields to the masses, $[qB]=\mbox{GeV}^{2}$ with the electric charge $q$, and $[m]=\mbox{GeV}$
is used. In natural units, the electric charge is dimensionless. This
is not true if one wants to compare the field strengths to the astrophysical
literature, where Gaussian units are common. For the proton, the charge
is given by the elementary charge, which can be calculated from the
fine structure constant, $\alpha=\frac{e^{2}}{4\pi}\thickapprox\frac{1}{137}$,
leading to $e\thickapprox0.30$ in our system of units. Using this,
one obtains the conversion for the magnetic field strength, $qB=0.1\,\mbox{GeV}^{2}$
is then equivalent to $B=1.7\times10^{19}\,\mbox{G}$ (Gauss). The
Gauss as a unit for magnetic field strengths is very common in astrophysics
and can be related to the SI unit Tesla via $1\,\text{G}=0.1$
mT. \footnote{The difference between Gauss (G) and Heaviside-Lorentz (HL) units for the gauge fields is $A_\mathrm{ G}^\mu=\sqrt{4\pi}A_\mathrm{ HL}^\mu$, and for the charges $q_\mathrm{ G} = q_\mathrm{ HL}/\sqrt{4\pi}$ (such that the product $qA^\mu$ is the same in both units). This implies for the elementary charge $e_\mathrm{ HL}=\sqrt{4\pi\alpha}\simeq 0.3$, and $e_\mathrm{ G}=\sqrt{\alpha} \simeq 0.085$.}

For the metric, the mostly-minus convention of particle physics is used,\be 
g^{\mu\nu}=\mbox{diag}(1,-1,-1,-1)\, .
\ee
Three-vectors will be denoted by bold letters, e.g.~$\mathbf{a}$,
where their norm will be denoted by the corresponding normal letter,
i.e.~$\left|\mathbf{a}\right|=a$. Whenever this notation might be ambiguous, especially for Greek letters, the notation $\vec{a}$ will be used instead. Four-vectors will either be written
with a greek index, e.g.~$k^{\mu}$ where the index is lowered and
raised with the Minkowski metric $g_{\mu\nu}$, or by capital letters,
e.g.~$K$. The Minkowski product will be abbreviated by a dot, $K^{\mu}Q_{\mu}=K^{\mu}Q^{\nu}g_{\mu\nu}:=K\cdot Q$.

Throughout this thesis, most of the thermodynamical quantities are
expressed as densities, since the thermodynamic limit is applied.
For readability, we will simply write "free energy" instead of
"free energy density" if it is clear in the given context.

\newpage{}

\tableofcontents{}

\newpage{}

\pagenumbering{arabic} \setcounter{page}{1}

\part[Introduction]{Introduction\\[2cm]\protect\includegraphics[scale=0.5]{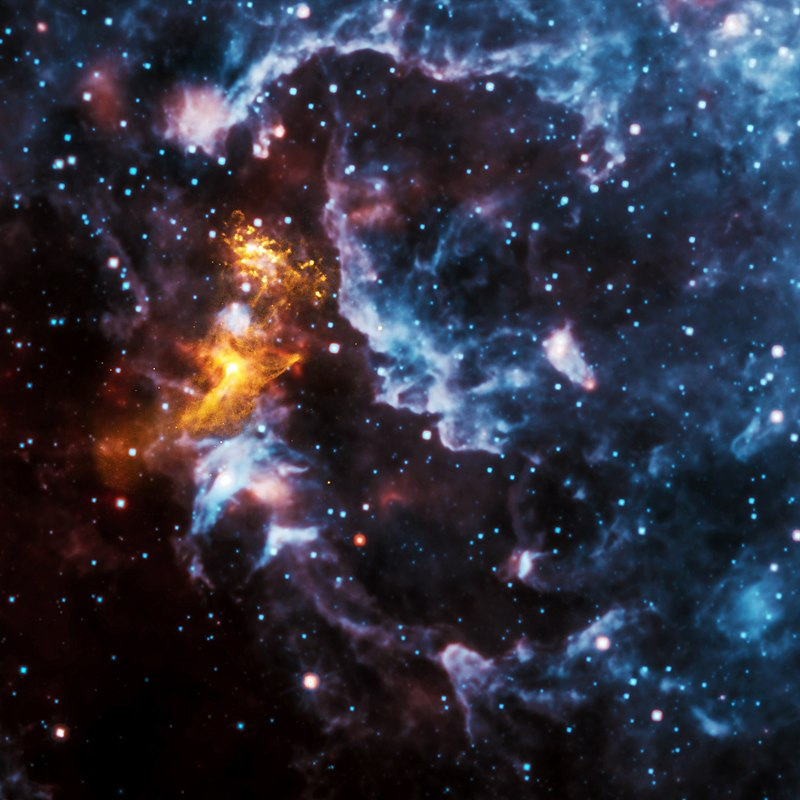}\\[-0.2cm]\tiny{Pulsar PSR B1509-58 makes gas glow and illuminates rest of the nebula. Source: NASA/CXC/SAO (X-Ray); NASA/JPL-Caltech (Infrared)}}

\chapter{The QCD Phase Diagram}
\label{chap:QCDphasedia}

\begin{figure}[t]
\centering{}\subfloat[Phase diagram of water \cite{chem_book}]{\begin{centering}
\includegraphics[width=0.45\textwidth]{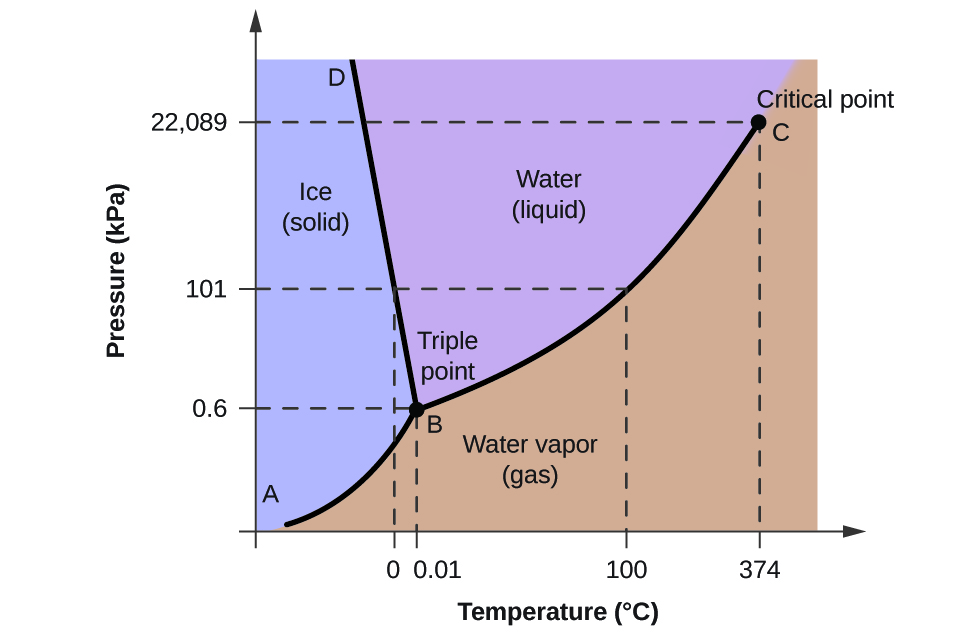}
\par\end{centering}
}\hfill{}\subfloat[QCD phase diagram for three-flavor quark matter.]{\begin{centering}
\includegraphics[width=0.45\textwidth]{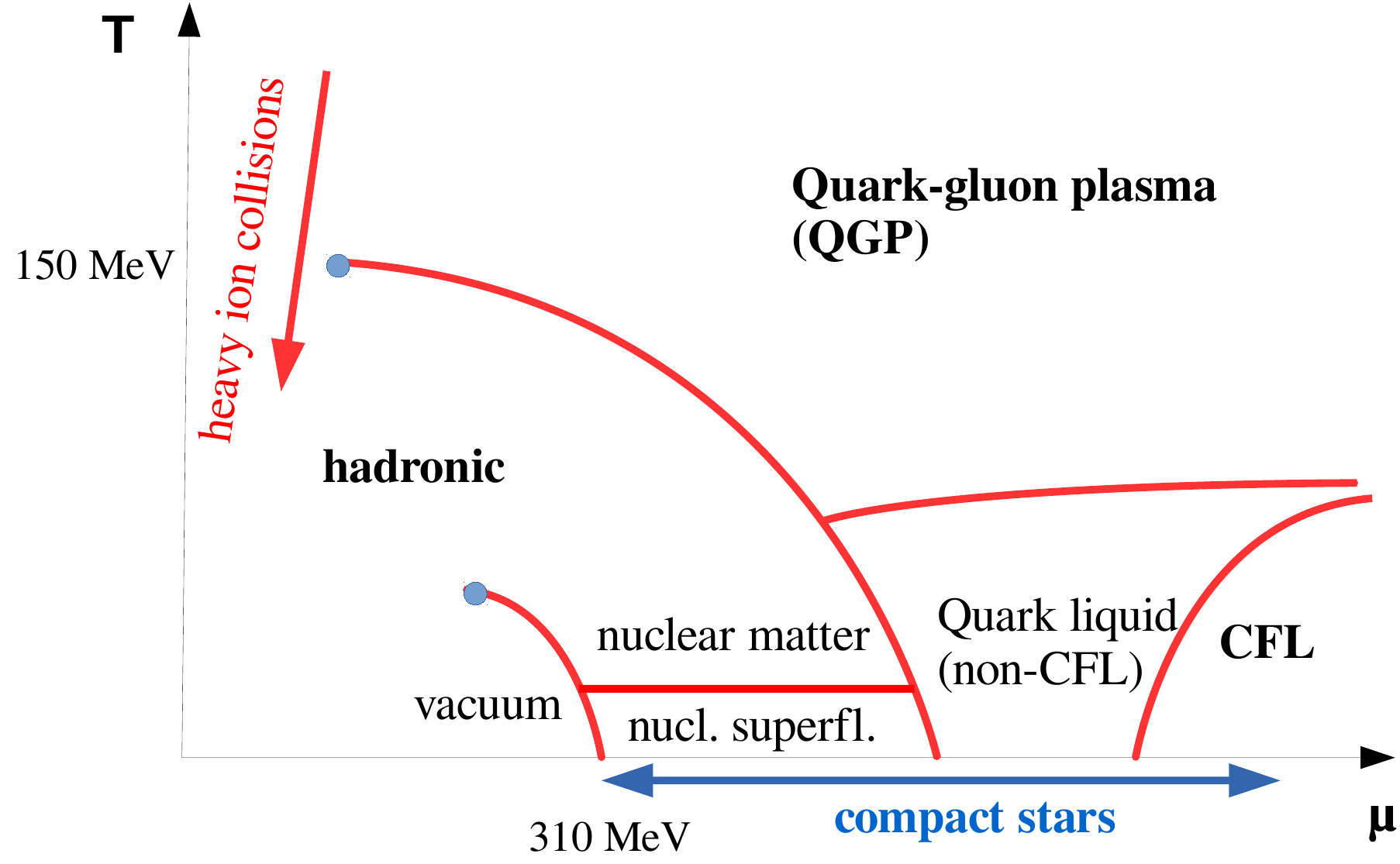}
\par\end{centering}
}\caption{\label{fig:QCDphaseDia}On the left hand side a standard phase diagram
of water, with the gas (brown), solid/ice (blue) and liquid (purple) phase as function of temperature and pressure. On the right hand side the highly speculative QCD phase diagram in the plane of temperature and quark chemical potential $\mu$.}
\end{figure}

In order to describe the world around us, physicists have been using
the concept of "elements" and their phases for millennia. In many
areas of science, phase diagrams are used to draw a map of the thermodynamically
distinct states a single material can be found in at different external
parameters, like temperature $T$ and pressure $P$. In the simplest
cases, a phase diagram of a single substance shows, as a function of two parameters,
the various states, like liquid, gaseous and solid. These phases are often separated by phase transition lines between them, which can end in critical points.
A very common example is the phase diagram of water, shown in Fig.~\ref{fig:QCDphaseDia}.
In principle, these diagrams can contain additional parameters and
become multi-dimensional or contain a plethora of different information.

\begin{figure}[t]
\begin{centering}
\includegraphics[width=8cm]{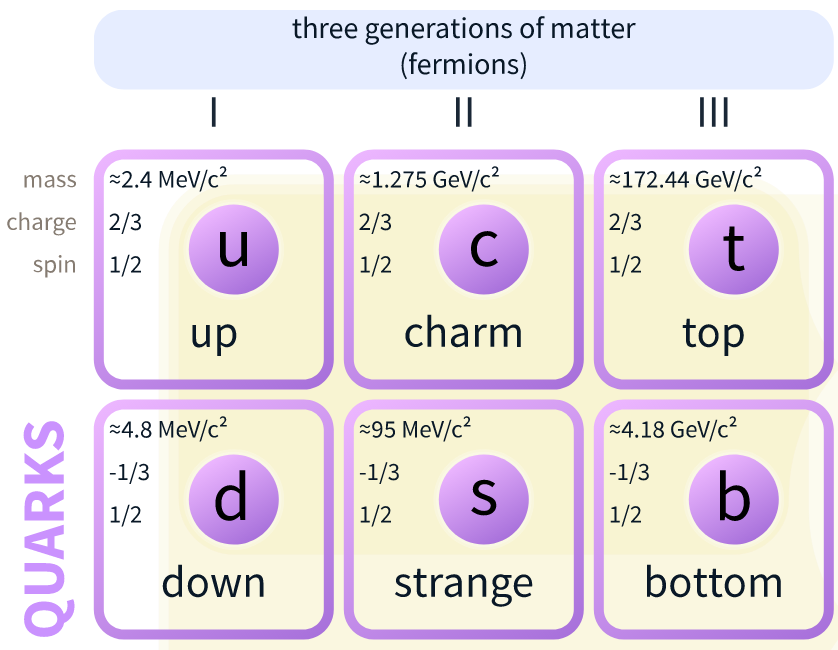}
\par\end{centering}
\caption{\label{fig:quarks}The three quark families and their properties.
Image credit: \cite{quarks_pic}}

\end{figure}

Due to the advances of particle physics in the last century, more
and more fundamental theories of matter aim to describe the entirety
of what the world around us is built of in a single theory. The standard
model of particle physics is the most complete and successful attempt
in history of mankind to reach this goal. The standard model describes
successfully and with a never before achieved accuracy three out of
the four fundamental interactions which govern our universe: the electromagnetic
interaction, which is responsible for next to all chemical and every-day processes,
the weak interaction, responsible for nuclear fusion and fission,
and the strong interaction. The latter one is the dominating force
at the level of neutrons and protons, holding together atomic nuclei
and controlling the behavior of quarks and gluons, the fundamental constituents of hadrons 
in the standard model. Hadron is therefore a name for every particle which consists of quarks, in contrast to leptons like the electron which are considered to be fundamental. Quarks come in three different "\textit{colors}"
and six "\textit{flavors}". Besides their electromagnetic charge, which can be either positive or negative,
quarks possess another kind of charge associated with the strong interaction that comes in three instead
of one opposing variant like the electric charge.
In analogue to the concept of additive color, we call these color-charges
red, green and blue. A bound object of either three quarks with one
of each color (called baryon), or two quarks, one with color and the
other with the corresponding anti-color (called mesons), are said
to be color neutral. All particles we observe in nature are such color
neutral objects, a phenomenon which is called confinement. In addition,
there are six different flavors of quarks, grouped into three families.
Ordered by their increasing mass, the quarks are called up, down, strange, charm,
bottom and top quark. A graphical summary of their properties can
be found in Fig.~\ref{fig:quarks} and in the booklet of the particle
data group (pdg), Ref.~\cite{pdg2017}. We observe a significant
jump in the quark masses from the strange to the charm. Thus the up,
down and strange quarks are often referred to as light quarks, whereas the charm, bottom and
top quarks are called heavy quarks. For the purpose of this thesis, the latter
three are too heavy to occur in the described systems and are therefore
neglected from now on. All further statements refer to three flavor
quark matter of up, down and strange quarks. 

The theory of the strong interaction, which describes quarks and their
interactions, is called quantum chromodynamics (QCD) and based on
the QCD Lagrangian:
\begin{equation}
\L=\bar{\psi}_{QCD}\left(i\gamma^{\mu}D_{\mu}-m\right)\psi_{QCD}-\frac{1}{4}F_{\mu\nu}^{a}F_{a}^{\mu\nu}\,,
\end{equation}
where the QCD spinor $\psi_{QCD}$ is a multi-dimensional object in
color-, flavor- and Dirac space and $F_{\mu\nu}^{a}$ is the gluon
field strength tensor. Consequences of this formula will be discussed in more detail
throughout this thesis. Here, it rather serves a pedagogical purpose:
although the theory can be written down in an incredibly compact way,
it is far from trivial and to a big extent not possible at all to
use this expression directly to compute properties of our world from
first principles. However, putting together our knowledge from experiments,
theory and observations, it is possible to draw a sketch of the most
fundamental phase diagram of the world: the \textbf{QCD phase diagram} does
not only show the phases of a specific material, it summarizes our
understanding of all (infinite) nuclear and quark matter as a function of temperature and chemical potential, a quantity which can be roughly translated to
density or pressure. Here, we focus on isospin symmetric matter, which means that the number of up and down quarks is equal, but in principle the QCD phase diagram can be extended and include isospin asymmetric matter. In the following I am going to describe briefly
the features we believe to be hidden in the QCD phase diagram and
how we can learn something about them. For an extensive review about
the challenges and perspectives of QCD see Ref.~\cite{Brambilla:2014jmp},
for a review on the QCD phase diagram itself see Ref.~\cite{Hands:2001ve}.
\begin{itemize}
\item \textit{Low densities.} If we start in the origin at vanishing temperature
and density, we are in the vacuum, where no matter exists at all.
Due to the critical end point of the liquid-gas transition, which is indicated by the shorter curved
line with the blue endpoint and separates the vacuum from the hadronic
phase at vanishing temperature, the vacuum is continuously connected
with the hadronic phase. This means, even at a too small chemical potential (i.e.\ lower than indicated by the solid line), 
we will find a small baryon density at finite temperature. Going up
the temperature axis, we will reach the cross-over to the deconfined phase
at approximately $T_{\mathrm{QCD}}=150$ MeV. The region of small densities
and high temperatures is accessible in two ways: numerically,
lattice QCD provides a first-principles approach to QCD, which has provided us with tremendous
insight to QCD at low densities. Due to the sign problem, a computation
at high densities is currently out of reach. The second approach to
this part of the phase diagram is experimental: heavy ion colliders,
like currently the large hadron collider LHC at CERN in Geneva, the
relativistic heavy ion collider RHIC at Brookhaven National Laboratory
(BNL) in the US, or in the future the PANDA experiment at the Facility
for Antiproton and Ion Research (FAIR) in Darmstadt, Germany and NICA
in Dubna, Russia \cite{NICA}, explore various density regimes in
the phase diagram. Especially the last two upcoming experiments aim
for higher densities and lower temperatures in order to find the critical
endpoint of QCD; the blue point at the end of the dominating curved
line separating the hadronic from the deconfined phase.
\item \textit{Baryon onset of nuclear matter}. If we follow the $T=0$ axis
to the right, we hit a first-order phase transition from the vacuum
to the hadronic phase, an infinite, symmetric phase of homogeneous nuclear matter.
This transition can be found  at a baryon chemical potential\footnote{In the diagram, the quark chemical potential is used which is one third of the baryon chemical potential.} $\mu_{B}=922.7$ MeV,
which results from the mass of the nucleon $m_{N}=939$ MeV reduced by the binding
energy of homogeneous nuclear matter, $E_{B}=-16.3$ MeV. The shape
of the transition can be influenced by other parameters, like magnetic
fields, see for instance Refs.~\cite{Haber:2014ula,Haber:2014zba}. This is practically
also more or less the point in the diagram where we actually live:
the matter we and most of our surrounding world is made of, is sitting
in a mixed phase of nuclear matter clumped into more or less spherical
nuclei immersed in vacuum (and some electrons). Consequently, every
earth-bound low energy nuclear experiment (and in principle every
"ordinary" experiment on earth) directly and indirectly tests
this region of the phase diagram.
\item \textit{Deconfinement and chiral transition}. If one increases temperature
and density, one reaches the deconfinement transition, which separates
the confined hadronic phase, where quarks are bound to hadrons, and
the deconfined phase. There, temperature and densities are so high,
that the individual quarks cannot be assorted to specific bound states
any more. We are now dealing with a soup of quarks and gluons, the so
called quark-gluon plasma. This phase can be reached in heavy
ion collisions. At asymptotic densities and temperatures, QCD is asymptotically
free: with rising chemical potential and temperature, quarks interact arbitrarily weak with each other \cite{Gross:1973id}.
This phase can be described by perturbative methods. In addition,
there is the restoration of chiral symmetry. Fundamental particles,
like quarks, have an additional property which we call chirality or
handedness, from the ancient Greek word for hand. This basically means
that quarks come in two copies, a so called left-handed and a right-handed
version. In contrast to the weak interaction, the strong interaction
treats massless left- and right-handed particles equally, a phenomenon
we call chiral symmetry. However, the ground state of QCD does not
respect this symmetry, which is related to the concept of spontaneous
symmetry breaking (SSB) and the formation of a chiral condensate consisting
of  quark antiquark pairs. This mechanism is mainly responsible for
the mass of the nucleons. Whether the chiral phase transition, the
transition where the chiral condensate is destroyed and chiral symmetry
is restored, is for low densities identical to the deconfinement transition is an open
question. Additionally, chiral symmetry is broken explicitly by finite
quark masses and is consequently at most considered an approximate symmetry as long as the chemical potential is still large compared to the masses.
\item \textit{Low temperatures}. Following the $T=0$ axis to high densities,
we stumble across a new phenomenon: \textbf{superfluidity} and \textbf{superconductivity}.
Right after the baryon onset, we expect different superconducting
and superfluid nuclear phases, where neutrons and protons form Cooper
pairs and condense. At higher densities, we reach color-superconducting
phases: quarks form Cooper pairs in a great variety of different ways
and condense to form a color superconductor. At very high densities,
we reach the ground state of dense and cold matter, the color-flavor
locking (CFL) phase, a color superconductor with maximal high residual
symmetry \cite{Alford:1997zt,Alford:2007xm}. These phases are going
to be investigated in more detail in the last part of this thesis. Theoretically,
at intermediate densities it is very hard to describe these regions. We have to rely on effective
descriptions like relativistic mean field models or the Nambu\textendash Jona-Lasinio
model \cite{NJL} for instance, which treats the quark-quark interaction
contact-like. "Experimentally", or rather observationally, there
is another intriguing option: \textbf{compact stars}. Compact stars are incredibly
dense stellar objects, whose behavior has to be explained on the microphysical
level and can therefore be used as a unique laboratory for dense
QCD at low densities.
\end{itemize}

\chapter{Compact Stars}

\begin{figure}[t]

\begin{centering}
\includegraphics[scale=0.5]{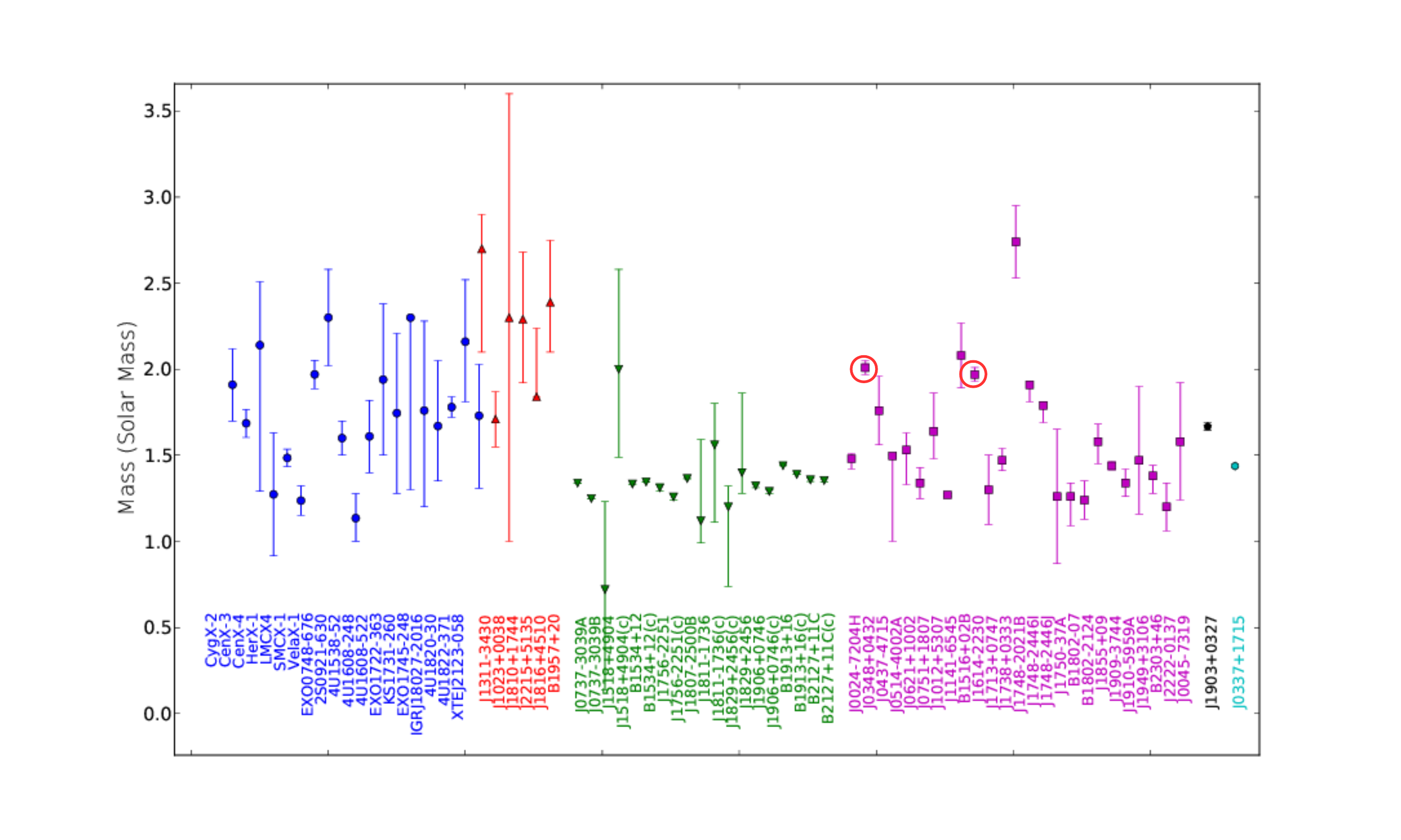}
\par\end{centering}
\caption{\label{fig:ns_masses}Table of neutron star mass measurements at a
$68\%$ confidence level, originally published in Ref.~\cite{Watts:2014tja}.
The color coding indicates different kinds of sources. Due to the
large systematic errors in some measurements, the current accepted
record holders (circled in red), measured using Shapiro
delay, are PSR J1614-2230 with $M=1.97\pm0.04\,M_{\odot}$ (see Ref.~\cite{Demorest:2010bx})
and PSR J0348+0432 with a mass of $M=2.01\pm0.03\,M_{\odot}$(see
Ref.~\cite{Antoniadis:2013pzd}).}
\end{figure}

Compact stars are the
collapsed cores of large stars, born in a supernova explosion. Due to their neutron dominated composition, they are often called neutron stars in the literature. Neutron
stars are the densest objects known to mankind after black holes,
in which matter is crushed beyond the limits of a description within particles of
the standard model. They have typical radii of $R\approx12$ km and
masses between one and a few solar masses, $M\approx1-3\,M_{\odot}$
with a typical value of $M\approx1.4\,M_{\odot}$, where the mass
of the sun is given by $M_{\odot}=1.989\cdot10^{33}$ g. For a list
of currently known neutron star masses see Fig.~\ref{fig:ns_masses},
general literature on compact stars from a particle physics point
of view can be found in Refs.~\cite{Glendenning:1997wn,Reddy:2004jg,Haensel:2007yy,Schmitt:2010pn},
which are mainly used as sources in this section. 
 Using the latter numbers, we can estimate the average density of
a neutron star to reach around 
\begin{equation}
\rho\simeq7\cdot10^{14}\,\text{g\,cm}^{-3}\approx2.8\rho_{0}\,,
\end{equation}
where $\rho_{0}=2.5\cdot10^{14}\text{g\,cm}^{-3}$ represents the
density of heavy nuclei, the nuclear ground state density. Although
this might not sound impressive at first sight, one has to consider
that the matter we are made of is mostly "empty" considering the
mass distribution. A sugar cube of neutron star matter has approximately
the same mass as the estimated total mass of the entire earth's population. 

The term neutron star originates in the traditional picture of the composition
of these stellar objects: neutron rich nuclear matter. However, even
in this simplified view a certain amount (up to $10\,\%$) of the
baryon density is contributed by protons. Due to the requirement of
neutrality (the electromagnetic force would push away any significant
amount of charge, see Chap.~$2$ note $1$ of Ref.~\cite{Schmitt:2010pn}),
the same number density of particles with negative elementary charge
$e$ has to be present. Moreover, the core of a compact star might
contain a significant amount of non-nuclear matter: hyperons (baryons
with strange quark content), meson condensates or color-superconducting
quark matter. Consequently, the term "compact star" should be preferred
in order to include these exotic objects. On top of that, the possibility
of a strange quark star has been discussed extensively in the literature,
for a review see Ref.~\cite{Weber:2004kj}. 

\begin{figure}[t]
\centering{}\includegraphics[scale=0.4]{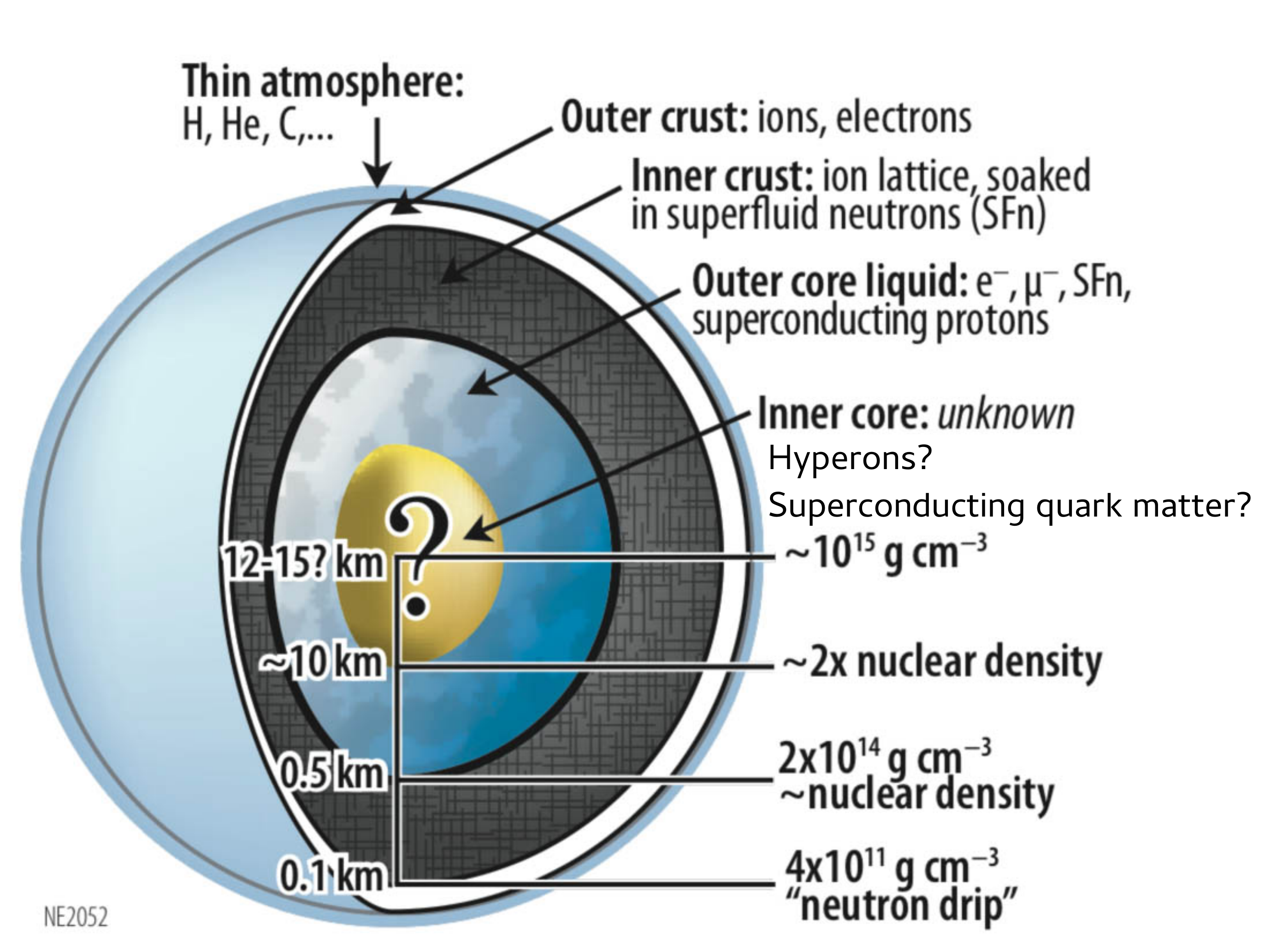}\caption{Possible composition of a compact star. Superfluidity most likely
starts to occur in the inner crust in the form of superfluid neutrons.
In the outer core, protons are believed to form a superconductor,
rendering the liquid a complicated interacting multi-fluid system.
The composition of the inner core is fairly unknown, where hyperons
and color-superconducting quarks are prime candidates. Picture taken
from Ref.~\cite{NICER}.}
\end{figure}

Compact stars have some very interesting properties which help us
to observe them in the first place.
\begin{itemize}
\item Due to the conservation of angular momentum in the collapse, they
are \textbf{rotating}, some of them with an astonishing frequency.
The current record holder is PSR J1748-244ad\footnote{The label can be read as follows: PSR indicates that the object is
a "Pulsating Source of Radio emission" (either neutron star or
white dwarf), the letter J the use of the J2000 coordinate system
which is followed by the right ascension and, after the minus sign
$-$ which indicates that the object can be seen in the southern hemisphere
($+$ for northern), the declination. The extra letters are added
if two stars are too close to each other.} at a rotation period of only $1.39$ ms, which means that its equator rotates
approximately with $24\%$ of the speed of light \cite{Hessels1901}.
The upper limit of rotation, where the neutron star would tear itself
apart, is given by the Kepler frequency. According to Refs.~\cite{1999A&A...344..151H,Haensel:2009wa},
this critical frequency is above $\nu\gtrsim1$ kHz, which is much
higher than the observed maximum value of $\nu\approx716$ Hz. The
reason for this gap is not fully understood and an open question in
neutron star research. It is worth mentioning that the Kepler frequency
only applies to rigidly spinning stars. By allowing for differential
rotation, higher frequencies might be possible. However, this is not
a stable configuration since viscous effects will always redistribute
the angular momentum until uniform rotation is achieved. 
\item Conservation of magnetic flux leads to huge \textbf{magnetic fields}
in compact stars. Surface fields are found to be around $B\approx10^{12}$
G, but even larger fields have been observed. Stars with fields of
the order of $B\approx10^{15}$ G are called magnetars \cite{Duncan:1992hi,Rea2011},
and it is conceivable that the magnetic field reaches even higher
values in the core of the star with a theoretical stability limit
of \textbf{$B\approx10^{18}-10^{20}$ }G in the core \cite{Lai1991}.
The combination of strong magnetic fields and rotation leads to the
lighthouse effect - the defining phenomenon of a pulsar. The emitted
electromagnetic radiation is geared towards the magnetic axis, which
does not align necessarily with the axis of rotation. The beam of
radiation therefore spins around the star's rotation axis, hitting
our telescopes on earth periodically like the light of a lighthouse
- therefore the term pulsar: pulsating star. Most of the compact stars
we know are observed as pulsars. Additionally, magnetic fields lead
to a stress on the crust and deform the star if they are strong enough,
leading to the emission of gravitational waves \cite{Glampedakis:2017nqy}.
These deformations are called "magnetic mountains", which can reach
heights of a few centimeters. 
\item In a particle physics language, neutron stars are \textbf{cold}. Compared
to room temperature of course they are quite the opposite, with temperatures
reaching up to $T\approx10^{11}$ K after the supernova, which corresponds
to $T\approx10$ MeV. Mainly driven by neutrino emission, they cool
down to the keV range within days. Compared to the other scales which
we have to consider, the quark chemical potential (several hundred
MeV) and the characteristic scale of QCD $\Lambda_{QCD}\approx150$
MeV, this can be considered small, $T\ll\mu_{q}$. This allows us
to work in the zero temperature approximation in most of this thesis,
$T=0$. Observing the temperature of a star for a longer time can
be used to learn something about the microscopical transport properties
of dense matter \cite{Potekhin:2015qsa}.
\end{itemize}
All of these properties can be deducted from the electromagnetic radiation
we are receiving from these stars. On August $17^{th}$ 2017 however,
a new window to the sky opened: in an event named $GW170817$, the
gravitational wave detectors LIGO and VIRGO observed the first direct
signal of a binary neutron star inspiral \cite{PhysRevLett.119.161101}.
It is worth mentioning that the inspiral of the Hulse-Taylor binary
system served as an indirect proof for gravitational waves long before
the first LIGO event $GW150914$ in 2015. The orbit of this binary
system is decaying at a rate absolutely consistent with the expected
energy loss due to gravitational wave emission predicted by general
relativity (GR) \cite{Weisberg:2010zz}. Gravitational wave astronomy
will play a major role in the future investigation of compact stars,
but currently the sample size is still too small. Nevertheless, the
observation already impacted other fields, like cosmology and astronomy,
and some constraints on the behavior of matter inside the star can
be inferred \cite{Bauswein:2011tp,Takami:2014zpa,Agathos:2015uaa}.
In the following, I will explain some of the intriguing phenomena
we observe and how we can relate them to the underlying microscopic
properties of dense matter. Although most of them rely on the observation
of electromagnetic waves, possible sources of gravitational waves
will be mentioned. 

\section{Observations and Microscopic Physics}

In this section I am going to explain some of the common features
of compact stars and how we can use them to learn something about
fundamental physics. Although I will focus on phenomena relevant to
the research part of this thesis, I will outline other selected topics
for pedagogical purposes. Additionally, I will emphasize measurements
with strong relation to superfluidity respectively superconductivity
and hints for the existence of quark matter in compact stars.

\subsection{Mass-Radius Curves and the Equation of State}

\begin{figure}[t]

\begin{centering}
\includegraphics[width=8cm]{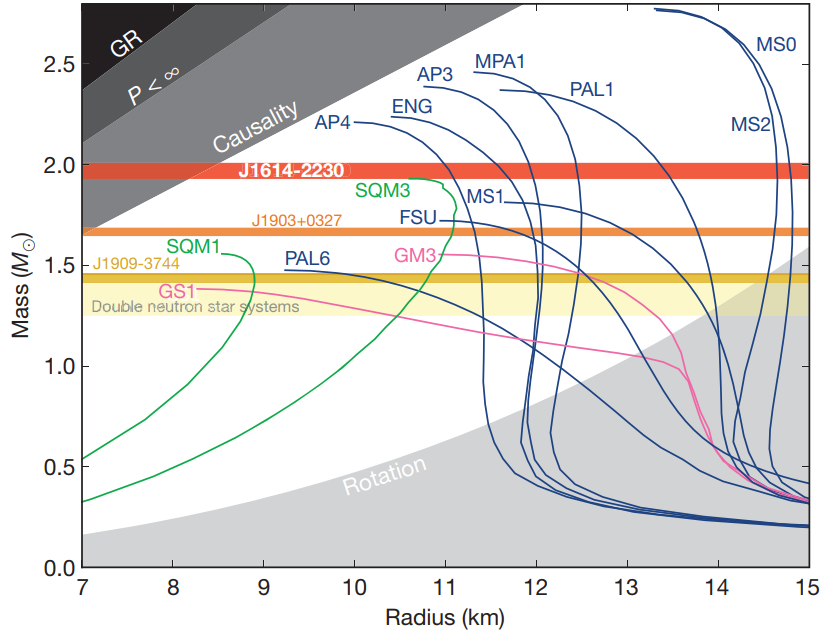}
\par\end{centering}
\caption{$M-R$ curve taken from Ref.~\cite{Demorest:2010bx}. The blue lines
correspond to nucleonic equations of state, pink denotes nucleonic
plus some exotic matter (hyperons, kaon condensates), and green represents strange
quark matter. The shaded areas are excluded by theoretical constraints
from GR, requirements on the pressure and spin
and the speed of sound, which has to be smaller than the speed of
light (causality). Every curve which does not intersect with the red
band, the mass of one of the heaviest known pulsars PSR J1614-2230,
is effectively ruled out.\label{fig:MRcurve}}
\end{figure}

During the collapse of the core after the supernova, gravity compresses
the matter inside the star until we reach an equilibrium configuration:
the matter inside the star counteracts the gravitational pull
with its internal pressure. If the initial mass is too big, the collapse
cannot be stopped and we end up with a black hole. For a standard
compact star, the gravitational energy is more than $10\,\%$ of the
star's mass, $E_{grav}\simeq0.12\,M_{star}$, which means that GR effects
have to be taken into account. The differential equation
for a non-rotating compact star in hydrostatic equilibrium is called
the Tolman-Oppenheimer-Volkov (TOV) equation \cite{Oppenheimer1939,tolman_TOV}:
\begin{equation}
\frac{dP(r)}{dr}=-\frac{G\varepsilon(r)m(r)}{r^{2}}\left[1+\frac{P(r)}{\varepsilon(r)}\right]\left[1+\frac{4\pi r^{3}P(r)}{m(r)}\right]\left[1-\frac{2Gm(r)}{r}\right]^{-1}\,,
\end{equation}
with Newton's constant $G$, and where the differential mass $m(r)$,
the pressure $P(r)$ and the energy density $\varepsilon(r)$ are
given as a function of the radius of the star. Additionally, the equation
for the encapsulated mass as a function of the stellar radius reads
\begin{equation}
\frac{dm(r)}{dr}=4\pi r^{2}\varepsilon(r)\,.
\end{equation}
At this stage we do not have enough information to solve this set of coupled equations; we have to relate the pressure and the energy density.
This additional input is provided by the equation of state (EOS),
which encodes the microscopic  properties of the matter inside the
star in the form $P(\varepsilon)$. In order to solve the TOV equation,
we have to specify the boundary conditions. By denoting the actual
radius of the star by $r=R$ we can specify the boundary
conditions at the edge of the star:
\begin{align}
m(R) & =M\,,\\
P(R) & =0\,.\label{eq:P(R)=00003D0}
\end{align}
By integrating the mass function up to the surface of the star we
obtain its total mass $M$. On the surface, the internal pressure
has to vanish, otherwise the star would blow itself apart. Eq.~(\ref{eq:P(R)=00003D0})
actually defines the surface of the star and allows us to determine
its radius. In the center of the star, the integrated mass $m(r)$
is zero, whereas the pressure takes a finite, unknown value $P_{0}$.
\begin{align}
m(0) & =0\,,\\
P(0) & =P_{0}\,.
\end{align}
By varying the central pressure $P_{0}$, we can compute a curve in
the $M-R$ plane, the so-called mass-radius curve. For a given central
pressure and an EOS, the solution of the TOV equations will provide
us with a point in the diagram, varying $P_{0}$ and keeping the EOS
fixed leads to a continuous curve. Choosing different equations of
state for various nuclear matter models, quark or hyperonic matter,
leads to a huge variety of curves in the $M-R$ diagram. All of these
curves possess a stable maximum of mass at a certain radius beyond
which the pressure of matter cannot counteract the gravitational
pull anymore and the star collapses. An example of such a diagram
can be seen in Fig.~\ref{fig:MRcurve}. 
The varying slope of the majority of the presented $M-R$ curves shows
some interesting behavior of nuclear matter: adding ordinary matter
on earth to a sphere will in most cases increase the radius of the
sphere. This behavior can be seen in the very steep part of blue curves,
which slightly bend to the right. However, for the biggest part nuclear
matter seems to be rather compressible: throwing matter onto a compact
stars will lead in many cases to a decrease of its size.

In principle, measuring the exact radius and mass of a few stars should
allow us to determine the equation of state. However, although several
masses of neutron stars are well known, measuring the extension of
an object of approximately $20$ km in diameter million kilometers
away from Earth turns out to be rather complicated. As shown at the
beginning of this chapter, the mass can be measured rather accurately
in some cases. The measurements of PSR J1614-2230 with $M=1.97\pm0.04\,M_{\odot}$\cite{Demorest:2010bx}
and PSR J0348+0432 with a mass of $M=2.01\pm0.03\,M_{\odot}$\cite{Antoniadis:2013pzd}
are considered especially robust because of the little model building
involved. Both of these stars are in a binary system with a white
dwarf. Whereas the computation of the lighter mass from Shapiro delay relies solely on
the correctness of general relativity, the second
observation depends on the modeling of white dwarf stellar atmospheres
and white dwarf cooling, which are fairly well known. Since the behavior
of dense matter should be unique, every EOS curve in Fig.~\ref{fig:MRcurve}
which does not intersect with the horizontal red line, representing
the mass of PSR J1614-2230, is effectively ruled out. Especially the
pink lines including hyperonic matter evidently seem to fail this
requirement. Intuitively, this can be understood as follows: every
new appearing (fermionic) particle species will soften the EOS, i.e.\ will only allow for lighter stars. This is a direct consequence of
Pauli's exclusion principle. Filling the Fermi sphere with the same
species of particles does not only get energetically more and more
costly, it will also increase the Fermi momentum and thus the pressure
in the star. Any new species opens a new Fermi sphere that can be
filled from below instead of adding new particles to an already existing
one. Hence, as soon as the chemical potential becomes high enough
in the star to produce a new species, we expect the Fermi sphere of
the new particle to be populated, leading to a softening of the EOS.
The fact that we expect hyperons in a compact star (the chemical potential
is most probably high enough to produce them) combined with the non-ability
to explain the two solar mass stars within hyperonic models, is called
the \textbf{hyperon puzzle}. Although new models including for instance
three-body interactions between the hyperons achieve sufficiently
high masses, the necessary tuning of model parameters is considered
questionable, leaving the hyperon puzzle a fascinating unsolved question
of neutron star physics. For an extensive review including further
references on this topic see Ref.~\cite{Chatterjee:2015pua}. 

The discussion above shows an easy example how we are able to learn
a lot about microscopical physics and dense matter by restricting
the equation of state by an astronomical observation like a mass measurement.
The existence of neutron stars with masses larger than one solar mass
also disproves a very common misconception in popular science descriptions
of neutron stars: the stability of the star solely based on the neutron degeneracy
pressure. Since neutrons are fermions, they will form a degeneracy
pressure based on Pauli's exclusion principle. Very often, including
the Wikipedia article on neutron stars \cite{wiki:xxx}, it is stated
that the collapse of the star is prevented by this pressure. Already
in the original paper of Oppenheimer and Volkoff Ref.~\cite{Oppenheimer1939}
in 1939 it was shown that a non-interacting Fermi gas, where the only
source of repulsion is the degeneracy pressure, cannot sustain masses
larger than $M\lesssim0.7\,M_{\odot}$. We can therefore conclude
that the neutron-neutron interaction has a repulsive component which stabilizes the star.

Nevertheless, a mass measurement alone does not provide us with sufficient
information, it only puts constraints on the equation of state. Earthbound
experiments can only provide limited insight as well. Dense matter
is largely isospin asymmetric, meaning that there are much more neutrons
than protons. As mentioned at the very beginning of this chapter,
neutrons make up at least $90\,\%$ of the nucleonic matter. This difference
can be expressed with help of the isospin asymmetry parameter or neutron
excess parameter $\delta=\left(n_{N}-n_{P}\right)/n\,,$ where $n_{N}$
denotes the neutron, $n_{P}$ the proton and $n$ the total density.
The nuclear symmetry energy, which can be expanded in $\delta$, influences
the equation of state significantly. Unfortunately, the largest somewhat
stable nuclei on earth barely reach a neutron-proton fraction of $2$,
making the measurement of the symmetry energy very difficult. For
a review on the nuclear symmetry energy and its influence on various
aspects of particle physics see Ref.~\cite{Baldo:2016jhp}. 

For all these reasons, a stellar radius measurement is highly desirable.
Most methods are based on the behavior of hot spots. Hot spots can
be created for instance via accretion: if matter is accreted from
a binary companion, most of the (charged) matter will follow the magnetic
field lines of the star and heat the crust of the star in the vicinity
of the magnetic poles. These hot spots slightly alter the photon flux
we are receiving on Earth and change the wave form. The deduction
of the radius from these observations requires GR corrections (due
to the effect of gravitational lensing we can observe parts of the
hot spot which are geometrically hidden behind the star), modeling
of the neutron star surface and good statistics. However, a lot of
progress has been made in the last decade. Especially the installation
of the NICER (Neutron star Interior Composition ExploRer) instrument
on the international space station ISS in June 2017 will provide a
lot of insight \cite{NICER}. NICER enables rotation-resolved spectroscopy
in the soft X-ray band and is expected to measure up to $5$ different
neutron star radii within a precision of $5-10\,\%$, which corresponds
to an angular resolution in the range of nanoarcseconds. For additional
information on how the measurement of neutron star radii and masses
can help us to determine the EOS see for instance Refs.~\cite{Steiner:2010fz,Ozel:2015fia},
a review on radius determination using X-ray timing is presented in
Ref.~\cite{RevModPhys.88.021001}.

\subsection{Effects of Superfluidity and Superconductivity}

One of the most striking phenomena, which is most likely connected
to the existence of a superfluid in the inner crust of compact stars,
is \textbf{pulsar glitches}. Because the electromagnetic radiation
we are receiving from pulsars is driven by rotation, measuring their
rotation frequencies $\Omega=2\pi/P$ with the period $P$ can be
done very accurately. Some millisecond pulsars spin in an incredible
constant fashion, their frequency is more stable than the best atomic
clocks on Earth \cite{1997A&A...326..924M}. Nevertheless, pulsars
generally tend to spin down due to the loss of rotational energy that
is radiated away. Surprisingly, from time to time we observe sudden
jumps $\Delta\Omega$ in the frequency, followed by a slow partial
relaxation to a regular spin down rate. The relative size of these
jumps vary over many orders of magnitudes, ranging from $\Delta\Omega/\Omega\approx10^{-5}$
to $\Delta\Omega/\Omega\approx10^{-12}$. From all pulsars we observe,
approximately $10\,\%$ show this behavior and we have observed seven
pulsars with more than ten events \cite{Haskell:2015jra}. 
\begin{figure}[t]
\begin{centering}
\includegraphics[width=8cm]{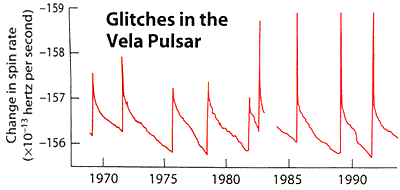}
\par\end{centering}
\caption{Glitches in the Vela pulsar over a time span of more than twenty years. Modified version from Ref.~\cite{lyne_graham-smith_2012} taken from Ref.~\cite{vela_glitch}. }

\end{figure}

The leading and commonly accepted model for pulsar glitches has been
presented the first time in Ref.~\cite{Anderson:1975} and presents
one of the strongest hints for the existence of a nuclear superfluid
in compact stars. As will be explained in much more detail in the
following chapters, superfluids exhibit some striking phenomena. Most
interestingly in this context is the existence of superfluid vortices.
A superfluid in a rotating bucket will not follow the rotation of
the bucket, instead it will store angular momentum in the form of
microscopic superfluid vortices. These vortices cannot move freely
when the superfluid is embedded for instance in a solid lattice, as
is the case of the neutron superfluid in the ion lattice of a neutron
star crust. Instead, they will be pinned to the ion lattice. Whether they are actually pinned between or to the ions of
the lattice is an open question \cite{Wlazlowski:2016yoe}. When the star, or to be more precise the crust, is spinning down, the number of superfluid vortices has to diminish in principle, since the stored angular momentum in
vortices is proportional to the external rotation frequency. Additionally,
vortices are protected from decaying by topological properties.
Due to the pinning force, vortices cannot leave the system and a
lag between the superfluid and the crust builds up. In other words,
the amount of angular momentum stored in the superfluid is bigger
than the rotation of the crust would induce in equilibrium. At some
point, this lag becomes critical, a significant amount of vortices
unpin and transfer angular momentum from the nucleonic superfluid
to the crust, which we observe as pulsar glitch. Despite the model's
ability to explain the observed data qualitatively, the underlying
mechanism for the collective unpinning is fairly unknown. Several
triggers are discussed in the literature, including crust quakes,
vortex avalanches and hydrodynamic instabilities. For a review on
glitches and trigger mechanisms see Ref.~\cite{Haskell:2015jra}.
Hydrodynamic instabilities like the two-stream instability are an
interesting trigger mechanism first discussed in Ref.~\cite{Andersson:2002zd}
and are highly relevant for this thesis. I will discuss the possible
existence of two-stream instabilities in relativistic multi-fluid
systems extensively in the research section of this thesis which is
based on publication \cite{Haber:2015exa}. Although a model for pulsar
glitches solely based on star quakes has been studied in the literature,
its shortcomings in explaining the large and frequent glitches in
some pulsars and the slow post-glitch relaxation time lead to the
commonly accepted theory described above \cite{Haskell:2015jra}.

Another observable strongly correlated to superconductivity and superfluidity
is \textbf{cooling}. Cooling in compact stars is predominantly driven
by neutrino emission, since photons are very ineffective except in the very late
stages of the star's existence. The mean free path of neutrinos in
dense matter is of the same order or even longer than the radius of the star, meaning
that these weakly interacting particles can carry away energy more
or less freely. The most efficient process for cooling is called direct
Urca process, which emits neutrino and anti-neutrinos due to the (inverse)
beta decay: a neutron decays into a proton, an electron and an anti-neutrino or a proton captures an electron and turns into a neutron
plus an electron neutrino,
\begin{equation}
n\to p+e+\bar{\nu}_{e}\,,\qquad p+e\to n+\nu_{e}\,.
\end{equation}
Because of momentum conservation, this process is strongly suppressed
in nuclear matter at lower densities since it is sensitive to the
proton fraction. By the presence of a spectator nucleon, which balances
out the momentum conservation equation, the processes can occur in
its modified form, the modified Urca. However, the formation of Cooper
pairs in a nuclear superfluid leads to the existence of an energy
gap $\Delta$ in the quasi-particle spectrum. The neutrino emissivity
$\varepsilon_{\nu}$ will be exponentially suppressed for temperatures
far below the critical temperature $T_{c}$, $\varepsilon_{\nu}\propto\exp\left(-\Delta/T\right)$.
Although it appears that superfluidity therefore reduces the cooling
rate, an new formation channel for neutrinos opens:
the constant breaking and formation of neutron (or proton) Cooper
pairs, which results in the emission of neutrino-antineutrino pairs. This process is called pair
breaking and formation process or short PBF. Superfluidity can therefore
also make positive contributions to neutron star cooling.

The combination of a superconductor and magnetic fields can lead to
the formation of \textbf{magnetic flux tubes}. The tension between the flux
tubes can lead to the formation of magnetic mountains too. Furthermore,
superfluid vortices can scatter on flux tubes, which is important
for the movement of vortices in models for pulsar glitches. Electrons,
which are important for transport phenomena due to their low mass,
can scatter off vortices as well. This effect is called mutual friction \cite{Alpar:1984zz}.

\section{Quark Matter in Compact Stars}
\label{sec:QMinCS}
Since the 1960s it has been known that nuclear matter undergoes phase
transitions to qualitatively quite different states at densities a
few times the nuclear saturation density \cite{Ivanenko:1969gs,Ivanenko:1969bs}. Compared to
terrestrial standards, these phases are "\textit{exotic}". One of the most intriguing exotic phase
is three flavor deconfined quark matter. As conjectured in the QCD phase diagram
in Chap.~\ref{chap:QCDphasedia}, nuclear matter at low
temperatures may undergo a first-order phase transition
to deconfined (superconducting) quark matter. In a strongly simplified
picture, the density of nucleons, which are made up of quarks, becomes
so high that they "overlap" and we cannot assign the quarks to
single nucleons anymore. Quark matter stars basically come in two
very distinct fashions: strange quark stars and hybrid stars. 

\textbf{Strange quark stars} are a possible consequence of the strange
quark matter hypothesis \cite{Bodmer:1971we,Witten:1984rs,Farhi:1984qu}.
This hypothesis basically states that nuclear matter is only a metastable
state, with a liquid of deconfined strange quark matter being the
absolutely stable ground state at zero pressure. Between the hadronic phase and the strange quark matter phase there has to exist a huge energy barrier. It originates in the ineffectiveness of the weak interaction: a conversion from
an ordinary nucleus to a stranglet (a "clump" of strange quark
matter) requires a nearly simultaneous conversion of a huge number
of up and down quarks. Since flavor is conserved in QCD (the number
of each quark flavor stays constant), this transition can only be
mediated via the weak interaction. If only a small amount of quarks
is converted there is no stable nucleation because hyperons decay
quickly in vacuum. If for any reason a stable stranglet hits a neutron
star, it can serve as a nucleation core for the successive conversion
of the neutron star into a strange quark star. In a binary star merger,
stars are partially ripped apart, which should lead to a certain amount
of ejected stranglets. It has been argued that if the hypothesis were
true every star should be a strange quark star by now, which means
a single observed neutron star disfavors the hypothesis, although
more recent publications question the amount of ejected strangelets
in these events \cite{Bauswein:2008gx}. An interesting property of
strange quark stars is their self boundedness: since strange quark matter
is assumed to be stable at $P=0$ they are not necessarily held together
by gravity. A strange star would consist almost entirely of strange
quark matter with a strangelet crust\cite{Alford:2008ge,Jaikumar:2005ne}
or a thin nuclear crust. For a more detailed discussion of the strange
quark matter hypothesis within the MIT bag model \cite{Chodos:1974pn,Chodos:1974je}
see Chap.~2.2.1 of Ref.~\cite{Schmitt:2010pn}, a review on strange
quark matter in compact stars is presented in Ref.~\cite{Weber:2004kj}. 

The second kind of quark stars are called \textbf{hybrid stars}. Hybrid stars consist of a quark matter
core, which is normally supposed to be color-superconducting, and
a nuclear matter outer core and crust. The exact form of the interface
is not known and depends on the surface tension between the nuclear
and the quark phase. Since theoretical predictions vary strongly, it is unknown whether a sharp interface or mixed phases form \cite{Alford:2001zr,Palhares:2010be,Pinto:2012aq}.
Unfortunately, quark matter cores are hard to observe directly, since
a lot of the phenomena we observe are related to the physics of the
outer layers of the star. Depending on the exact form of the phase
transition and the interface, a strong first-order phase transition
from nuclear to quark matter (and maybe even between different color
superconducting phases \cite{Alford:2017qgh}) can manifest itself directly
in the $M-R$ curves in the form of disconnected branches \cite{Alford:2015gna,Alford:2015dpa}.
In the mass-radius curve, we could possibly observe (depending on
some undetermined parameters of the quark matter equation of state)
two stars with the same gravitational mass but very distinct radii.
In a simplified picture, the star undergoes a phase transition to
a hybrid star because it accretes matter (so its baryonic mass actually
increases), but is compressed because of the strong interaction in
the quark core. The additional negative binding energy of the quark
phase leads to a smaller gravitational mass that we observe. Other
bulk observables or properties which could provide information on the nature of
the core include temperature and bulk and shear viscosity. 

\begin{figure}[t]
\begin{centering}
\includegraphics[width=0.45\textwidth]{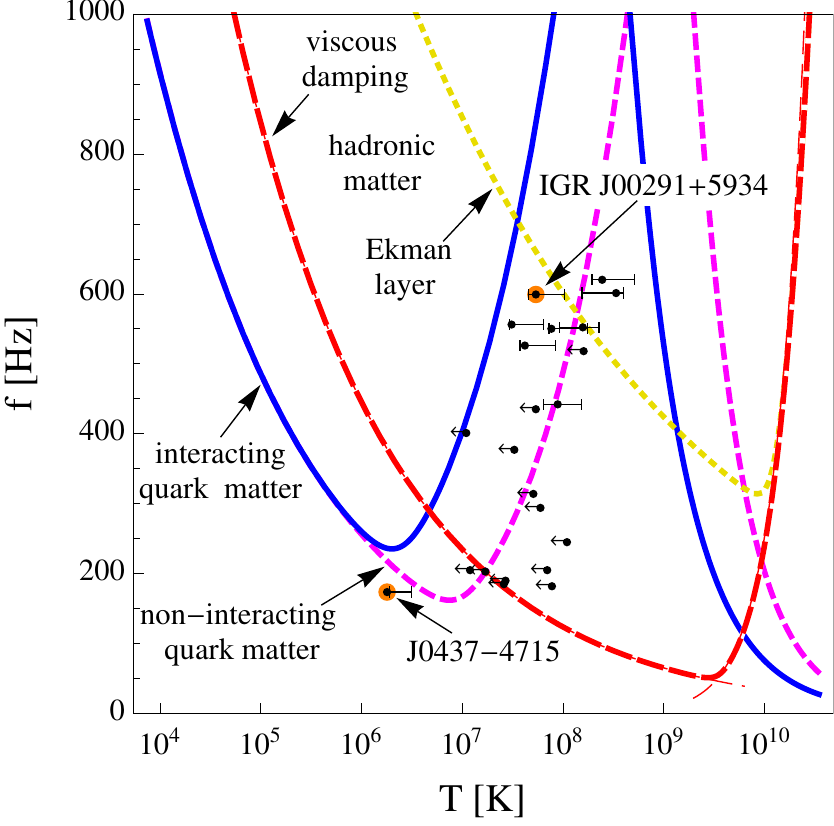}
\par\end{centering}
\caption{\label{fig:rmode:inst}R-mode instability window in the plane of temperature
and spin frequency of the star taken from Ref.~\cite{Alford:2013pma}.
The red, dashed line shows the standard instability window, the pink
short-dashed lines non-interacting and the solid blue lines interacting
quark matter. The black dots, representing actual stellar observations,
are all found outside of the instability window if interacting quark
matter is assumed in the core.}
\end{figure}

The latter two are crucial microscopic ingredients to the solution
of the \textbf{r-mode puzzle}. R-modes (short for Rossby) are non-radial
pulsation modes which are unstable with respect to the emission of
gravitational waves and have first been discussed in Refs.~\cite{Andersson:1997xt,Andersson:1998qs},
for reviews see Refs.~\cite{Andersson:2000mf,Haskell:2015iia}. Even
at an arbitrarily small rotation frequency, the star can find a lower
energy and angular momentum configuration by amplifying the mode. The instability is thereby driven by the emission of
gravitational waves but can only operate if the growth of the mode
is not damped by viscosity. Due to the temperature dependence of bulk
and shear viscosity, one can compute an r-mode instability window
bounded by bulk viscosity on one side and by shear viscosity on the
other, in the plane of spin frequency and temperature. Within this
window, a star is thought to be unstable and should rapidly spin down
and cool until it leaves the unstable zone. However, calculations of the
instability window using generic equations of state combined with
observations show numerous stars within the unstable region. Besides
various different solutions including the coupling of the r-mode to its superfluid counterpart \cite{Gusakov:2013jwa}, a first calculation
of the instability window provided by interacting quark matter seems
promising, see Ref.~\cite{Alford:2013pma} and Fig.~\ref{fig:rmode:inst}.
Also the damping of the modes due to a phase lag at the nuclear
matter - quark matter interface has been discussed \cite{Alford:2014jha}.

As mentioned earlier, quark matter in compact stars is most likely
a \textbf{color superconductor}. Color superconductivity and superfluidity can have similar effects like ordinary superfluidity and superconductivity.
For instance, it has been shown that color-magnetic flux tubes can
lead to "color-magnetic mountains", which could emit gravitational
waves detectable with future telescopes \cite{Glampedakis:2012qp}.
Also the Urca process exists in quark matter and is affected by color
superconductivity, leading to modified cooling curves. A lot of the
effects of color superconductivity on observables depend on the details
of the superconducting phase. Before I am going to explain the details
of color superconductivity in the last part of this thesis, I am going to discuss superfluidity and
superconductivity first in general and then in some detail in nuclear
matter in the following chapters and parts.

\chapter{Superfluidity and Superconductivity}
\label{chap:SF_SC_gen}
\begin{figure}[t]
\begin{centering}
\includegraphics[width=0.45\textwidth]{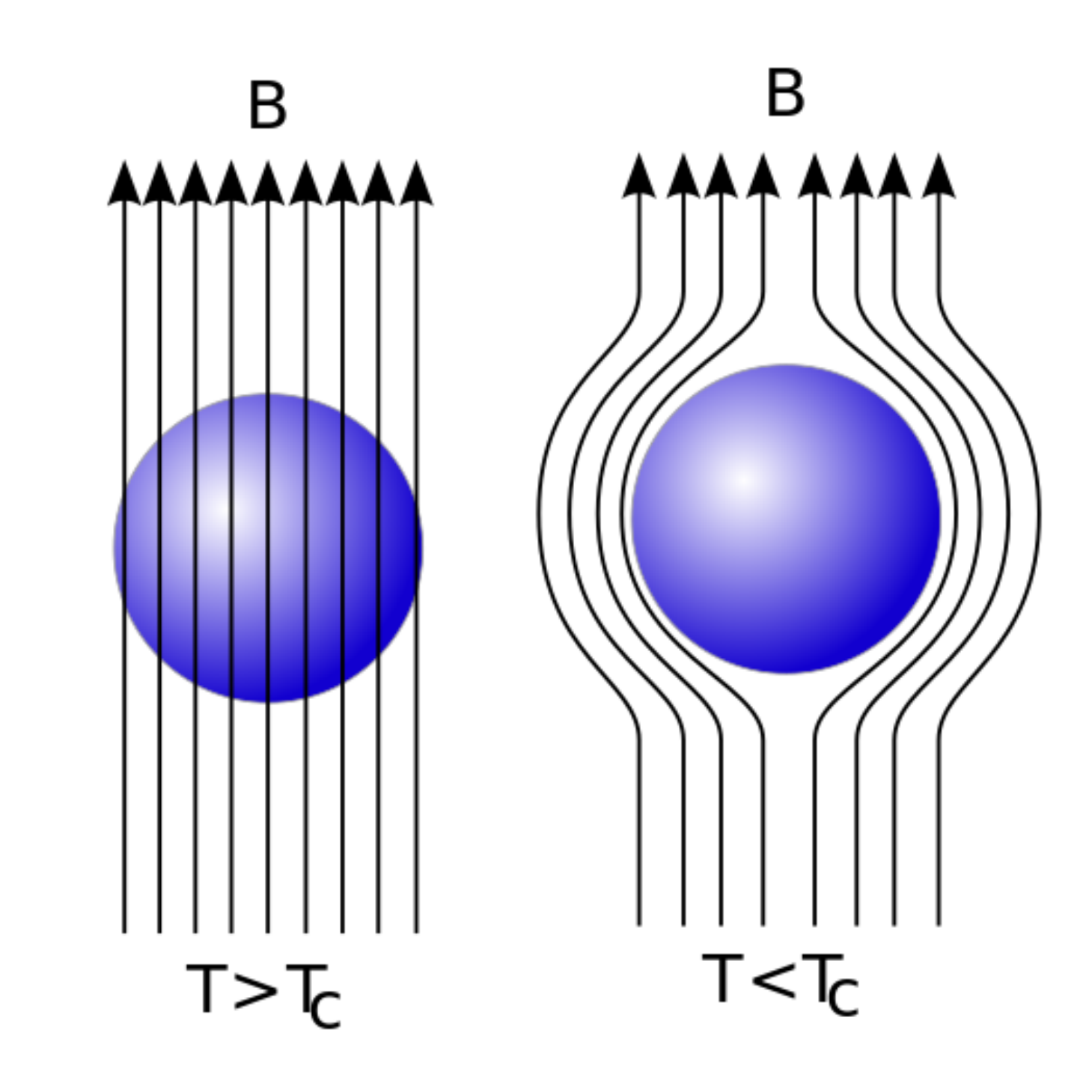}
\par\end{centering}
\caption{Graphical presentation of the Meissner effect from Ref.~\cite{meissner_pic}. A superconducting sample
is placed in a magnetic field. Below the critical temperature, magnetic
fields are expelled completely, above the field can penetrate the
material according to its magnetic properties. The same effect can
be seen if the sample is put into the magnetic field already at
$T<T_{c}$.}
\label{fig:meissner_schematic}
\end{figure}

Superfluidity and superconductivity are pure quantum effects, thus
low temperatures are required to observe these peculiar phenomena.
It is therefore no surprise that the discovery of superconductivity
was preceded by the liquefaction of helium at $4.2$ K in 1908 by
Heike Kamerlingh Onnes. Only three years later in 1911, Kamerlingh
Onnes discovered that the resistance of mercury drops at a critical
temperature $T_{c}$ of $4.20$ K from $0.1\,\text{\ensuremath{\Omega}}$
below the sensitivity of his experiment at approximately $10^{-6}\,\text{\ensuremath{\Omega}}.$
The critical temperature is a material dependent constant.Perfect conductivity is one of the defining properties of superconductivity.
It can best be studied in persistent current experiments: an induced
current in a superconducting ring has been observed to flow without
measurable decrease for a year and a lower bound of $10^{5}$ years
for the characteristic decay time has been inferred. Another defining
property, discovered in 1933 by Meissner and Ochsenfeld \cite{Meissner1933}, is perfect
diamagnetism: on the one hand, magnetic fields are excluded from
entering a superconductor. This could be easily explained by perfect
conductivity and Lenz's law, which states that a time varying magnetic
field will produce a current in a conductor in such a way that the
resulting field opposes the change that produced it. On the other
hand, magnetic fields in an originally normal sample are getting expelled
once the material is cooled below the critical temperature (in field
cooling). This effect is called \textbf{Meissner effect} and implies
the existence of a critical magnetic $H_{c}$ field at which superconductivity
is destroyed. In this simplified explanation the existence of type
II superconductors has been completely neglected, but I will explain
the behavior of superconductors in magnetic fields in much more detail
later in this chapter, since it is a crucial ingredient of this thesis.
In Fig.~\ref{fig:meissner_schematic}, we see a graphical representation of this effect.

Surprisingly, it took another 16 years until superfluidity in helium-$4$
was discovered in 1927 by Wolfke and Keesom who observed a jump in
the specific heat at $2.17$ K \cite{wolfke}. Kapitza and independently Allen and
Misener then performed groundbreaking experiments establishing the
remarkable nature of superfluid helium in 1938 \cite{1938Natur.141...74K,1938Natur.141...75A}. Two of the key-experiments led to the invention of the two-fluid model, which was able to solve the viscosity paradox. Immersing a torsion pendulum in superfluid helium showed finite damping times, whereas the capillary experiment suggested vanishing viscosity for a superfluid. In such an experiment, two containers are connected by a very thin tube, called a "superleak". Above the critical temperature, helium is not able to flow trough the capillary because of viscous effects. By cooling it down, it is possible to observe the fluid to flow from one container to the other. In the two fluid picture, this paradox is solved as follows: superfluid helium consists of a superfluid component and a normal fluid component, which originates in the excitations of the fluid. At zero temperature, the normal fluid component therefore vanishes. However, at finite temperatures below the critical value, the normal fluid component leads to a damping of the pendulum, but in the capillary experiment there is always the superfluid component which flows into the second container. 
Kapitza also was the first to use the phrase superfluidity, anticipating a 
theoretical connection of superfluidity to superconductivity. This was rather remarkable since a theoretical description of fermionic
superconductivity was only provided in the 1950s in works of Bardeen,
Cooper and Schrieffer who shared the Nobel prize in 1972 \cite{BCS}. We are going
to see that a superconductor can be largely seen as an electrically
charged superfluid. Therefore, I will use the terms superfluidity
and superconductivity synonymously from now on, whenever a distinction
is not necessary. For both phenomena, we have to distinguish between
the fermionic and the bosonic case. Bosons can condense into a Bose-Einstein
condensate (BEC), which is a coherent macroscopic quantum state. In a BEC, the majority of the particles occupy the same energy level, the ground state.
Coherence means that the bosons in the BEC can be described by a single-particle
wave function with a single phase instead of a complicated many-body
quantum wave function. We are going to exploit this feature in order
to derive a phenomenological theory for bosonic superfluidity and
superconductivity called  Ginzburg-Landau (GL) theory, which has been
originally developed by Ginzburg and Landau in the early 1950s \cite{GLtheory}.
Identical fermions cannot occupy the same quantum state due to Pauli's exclusion
principle, they obey Fermi-Dirac statistics instead of Bose-Einstein
statistics like bosons. Instead, they occupy higher and higher energy
levels, consecutively filling the Fermi sphere. As a consequence,
fermions have to undergo an "intermediate" step before condensation:
the formation of Cooper pairs. The Fermi surface is unstable towards
the formation of these pairs, as long as there is an arbitrarily small attractive
interaction between the particles. In electronic superconductors,
this attractive interaction is mediated by the ion lattice, i.e.~by
electron-phonon interaction. In nuclear and quark Cooper pairing,
the strong interaction between the particles provides an attractive
channel directly. Cooper pairs can, to some extend, be seen as bosons
and form a condensate, enabling fermionic superfluidity.%
 The condensate is the main ingredient to dissipationless
transport of charge, it allows the superfluid to flow through the
capillary and the superconductor to have vanishing resistance. In
the following chapters, I am going to present in some more detail
the bosonic description of superfluids and the behavior of superconductors
in magnetic fields. 
This entire chapter, including the introduction above and the more
technical discussion below, largely follows the textbook of Tinkham,
Ref.~\cite{tinkham2004introduction}, and the Springer Lecture Notes
in Physics of Schmitt, see Ref.~\cite{superbook}.

\section{Landau's Critical Velocity\label{sec:Landau's-Critical-Velocity}}

As described above, the formation of a coherent condensate allows
us to transport charge, for instance mass in a superfluid, without
friction. For now we are going to restrict ourselves to the superfluid
case. As we are going to see in the next part, the condensation is
connected to the breaking of a global symmetry. Following Goldstone's
theorem, the system therefore possesses a gapless mode which can be
excited by arbitrarily low temperatures. This mode is called a Goldstone
mode. In principle, a gapless mode is very easy to excite, for instance it will be populated for even arbitrarily low temperatures, and in principle could lead to dissipation. However, Landau was able to find a general condition for these excitations
to still enable dissipationless transport. Consider a superfluid in
a capillary moving with velocity $\mathbf{v}_{s}$. The energy of
an excitation in the rest frame of the fluid (where $\mathbf{v}_{s}=0$
but the capillary moves with $-\mathbf{v}_{s}$) is given by $\varepsilon_{p}>0$
with momentum $\mathbf{p}$. The entire energy of the system in the
lab frame is given by the kinetic energy of the fluid $E_{kin}$ plus
the energy of the excitation transformed into the lab frame. We are
going to use the non-relativistic Galilean transformation here to do
so instead of the more general, relativistic Lorentz transformation,
\begin{equation}
E=E_{kin}+\varepsilon_{p}+\mathbf{p}\cdot\mathbf{v}_{s}\,.
\end{equation}
If there is dissipation, the fluid loses energy which happens whenever
\begin{equation}
\varepsilon_{p}+\mathbf{p}\cdot\mathbf{v}_{s}<0\,,
\end{equation}
which can only be negative if at least its minimum is negative, i.e.~when
$\mathbf{p}$ and $\mathbf{v}_{s}$ are exactly anti-aligned, $\varepsilon_{p}-pv_{s}<0$.
We can use this criterion for calculating the maximal critical velocity
below which the superfluid supports dissipationless transport:
\begin{equation}
v_{c}=\min_{p}\frac{\varepsilon_{p}}{p}\,,
\end{equation}
which can be rewritten as the solution of the equation
\begin{equation}
\frac{\p\left(\varepsilon_{p}/p\right)}{\p p}=0\Rightarrow\frac{\p\varepsilon_{p}}{\p p}=\frac{\varepsilon_{p}}{p}\,.
\end{equation}
This equation can be interpreted geometrically, which will be interesting
for the discussion of instabilities in multi fluid systems later on. Imagine
plotting the quasi-particle excitations in the $\varepsilon_{p}-p$
plane. We now draw a straight line from the origin along the $p-$axis
and start rotating it upwards. If we can do so by a finite angle before
intersecting with the dispersion relation there is a finite critical
velocity. Consequently, a purely quadratic dispersion relation, like
$\varepsilon_{p}=p^{2}/(2m)$ for a non-relativistic free particle does
not support superfluidity. Goldstone modes in general have a linear
dispersion relation in the origin, therefore they do not destroy superfluidity.
If there are no further contributions for higher momenta, the critical velocity is then directly given by the slope of the Goldstone
mode at $p=0$.
\begin{figure}[t]
\begin{centering}
\subfloat[Quadratic dispersion relation]{\begin{centering}
\includegraphics[width=0.45\textwidth]{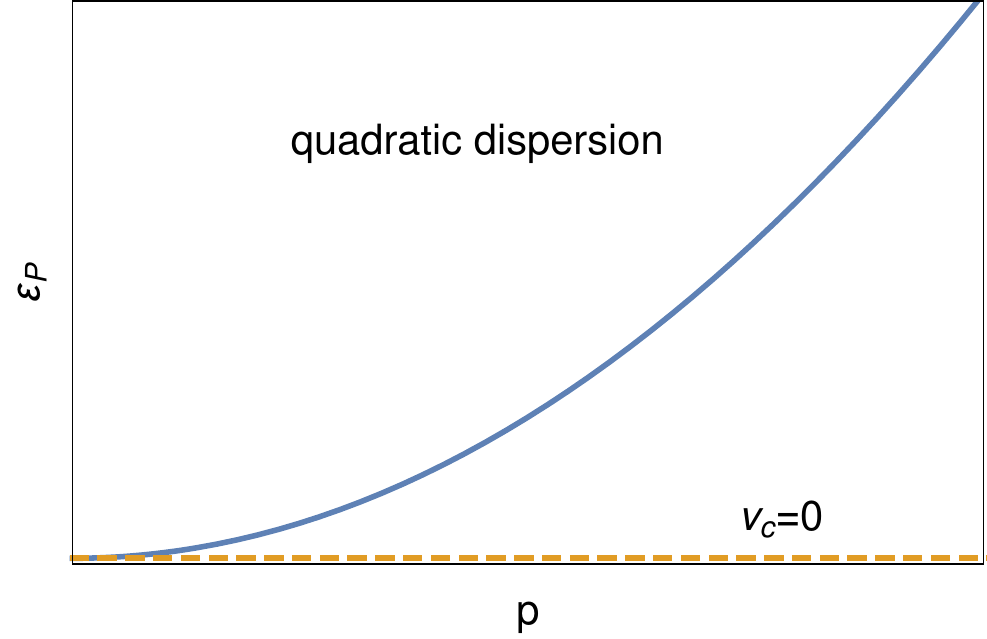}
\par\end{centering}
}\hfill{}\subfloat[Schematic dispersion relation of $^{4}\text{He}$]{\begin{centering}
\includegraphics[width=0.45\textwidth]{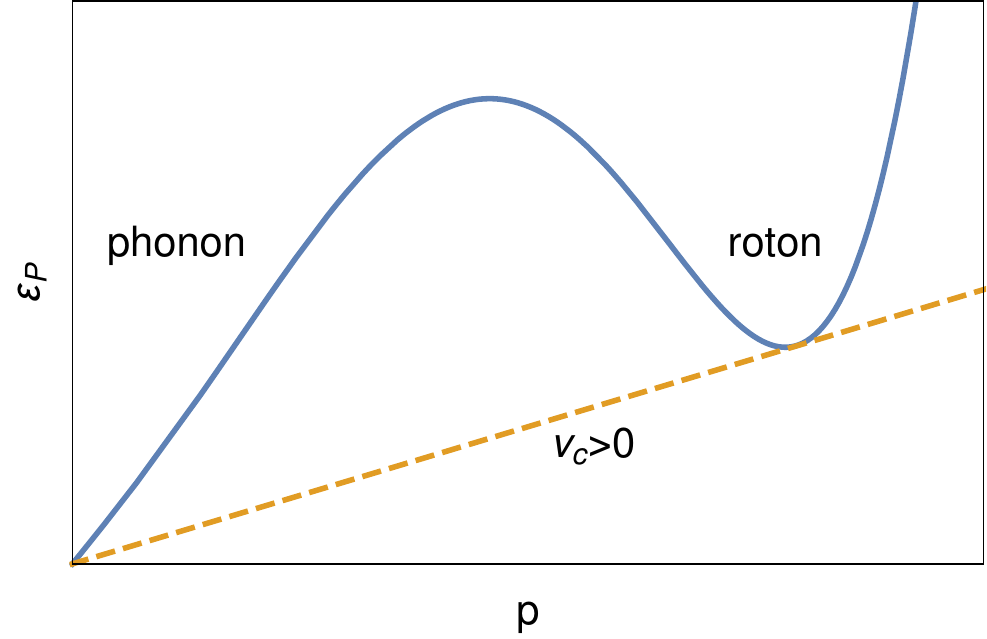}
\par\end{centering}
}
\par\end{centering}
\caption{Landau's critical velocity for a purely quadratic dispersion relation on the left and a schematic
dispersion relation of superfluid $^{4}\text{He}$ on the right. In the right panel
one can observe the phonon, i.e.~the Goldstone mode, and the roton
contribution. Landau's critical velocity for a quadratic dispersion relation is zero, whereas the dispersion relation
of helium-$4$ is linear in the origin and thus allows for a finite critical velocity.}

\end{figure}
Note that the existence of a finite critical velocity is only a necessary
but not a sufficient condition for superfluidity. The dispersion relation
of a free relativistic particle is gapped by its mass, $\varepsilon=\sqrt{p^{2}+m^{2}}$,
but is certainly no superfluid. Only if there is a condensate and
a finite critical velocity the system supports superfluidity.

\part[Technical Background and Two-Component Model]{Technical Background and Two-Component Model\\[2cm]\protect\includegraphics[scale=0.8]{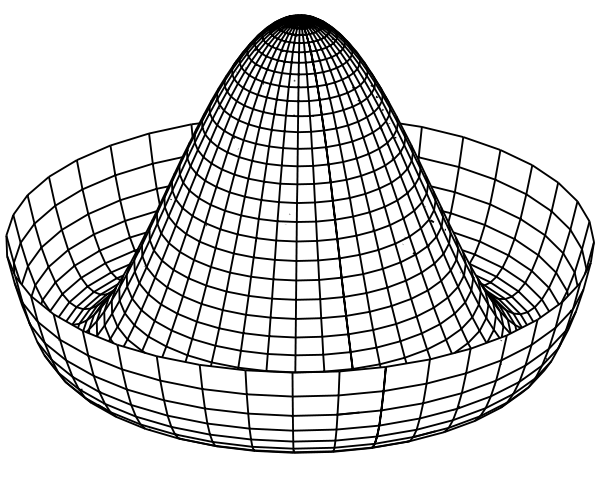}\\[-0.4cm]\tiny{Mexican hat potential of $\phi^4$-model}}

In this part of the thesis, the basics of Ginzburg-Landau theory are introduced. Afterwards, I will start with connecting the well
known machinery of quantum field theory (QFT) to the models of superfluidity
and superconductivity. For an introduction into
quantum field theory see Refs.~\cite{Srednicki:2007qs,zee_qft-nut}.
Most of this part relies on the foundations of thermal (quantum) field
theory, the combination of statistical physics and quantum field
theory. An extensive introduction can be found in the textbook of
Kapusta and Gale, see Ref.~\cite{kapusta_tft}. 

After doing so, I will introduce the main model of this thesis, which
describes two interacting superfluids. We are going to gauge
one of the fields later on in order to describe a mixed system of a superfluid and a superconductor.
The first chapters of this discussion follow the textbooks of 
Schmitt Refs.~\cite{Schmitt:2010pn,superbook} and Srednicki
Ref.~\cite{Srednicki:2007qs}. Large parts of this model have been
developed in the publications Refs.~\cite{Alford:2013ota,Alford:2013koa}
and been discussed in more details in the phd thesis of Stetina,
see Ref.~\cite{Stetina:2015exa}. Furthermore, a slightly condensed
description of the calculations presented in this part concerning
the two-fluid model can be found in our publications on which this
thesis is based on, see Refs.~\cite{Haber:2015exa,Haber:2016ljn,Haber:2017kth,Haber:2017oqb}. 
\chapter{Ginzburg-Landau Theory}
\label{sec:Ginzburg-Landau-Theory}
Ginzburg-Landau theory was originally proposed by the famous
Russian physicists Vitaly Ginzburg and Lev Landau in Ref.~\cite{GLtheory}
as phenomenological theory that describes superconductivity. Later,
it has been shown that it can be obtained from the microscopic BCS
theory assuming, among others, temperatures close to the critical
temperature, i.e.~a small gap \cite{BCSfromGL}. It is based on
Landau's theory for second-order phase transitions.
Landau formulated his theory in terms of the free energy of the system,
which he postulated to be analytic and to obey the symmetries
of the underlying Hamiltonian. The free energy is then written down as an expansion
of the order parameter and its gradients. This expansion assumes that
the order parameter is small, hence it cannot be applied to first
order phase transitions because the order parameter can jump to an
arbitrary value at the transition. Thus, Landau's theory works best for second
order phase transitions and the free energy is technically only correct
close to the transition. Nevertheless, qualitative results can be
obtained for a larger parameter range. A general form for the free
energy density can be written down in the form
\begin{equation}
\frac{F_{GL}}{V}=\alpha\left|\vf\right|^{2}+\frac{\beta}{2}\left|\vf\right|^{4}+\frac{1}{2m}\left|-i\mathbf{\nabla}\vf\right|^{2}\,,
\end{equation}
with the order parameter $\vf$, an effective mass $m$, and where $V$ is the total volume of
the system which is sent to infinity in the thermodynamic limit. Here,
we already used symmetry arguments to rule out odd terms in $\vf$.
The coefficients $\alpha$ and $\beta$ generally dependent on thermodynamic
variables like temperature and chemical potential. Note that the overall
unit of $\nicefrac{F}{V}$ is $\left[\nicefrac{F}{V}\right]=m^{4}$,
which leads to $\left[\vf\right]=m^{\nicefrac{3}{2}}$, $\left[\alpha\right]=m$
and $\left[\beta\right]=m^{-2}$ in the non-relativistic setup used
in this part of the thesis. In order to describe superconductivity,
Ginzburg and Landau identified the order parameter in the general
free energy with the superconducting wave function. The square of
the wave function accordingly describes the density of the superconducting
charge carriers $n_{s}$, 
\begin{equation}
n_{s}=\left|\vf\right|^{2}\,.
\end{equation}
Taking into account the effect of the magnetic field on the charged
particles, we minimally couple the gauge field $\mathbf{A}$ to the
gradient and add the energy of the magnetic field, yielding
\begin{equation}
\frac{F_{GL}}{V}=\frac{F_{N}}{V}+\alpha\left|\vf\right|^{2}+\frac{\beta}{2}\left|\vf\right|^{4}+\frac{1}{2m}\left|\left(-i\mathbf{\nabla}-q_{s}\mathbf{A}\right)\vf\right|^{2}+\frac{\mathbf{B}^{2}}{8\pi}\,.\label{eq:F_GL_single}
\end{equation}
In an electron superconductor, $m$ and $q_{s}$ correspond to
the mass and the charge of the Cooper pair, thus $m=2m_{e}$ and
$q_{s}=-2e$ with the electron mass $m_{e}$ and the elementary charge
$e$. The prefactor of the magnetic field contribution depends on
the chosen units, in SI units it is proportional to $\mu_{0}$, the
magnetic permeability of the vacuum. The free energy of the normal
phase at zero magnetic field is given by $F_{N}$, and the magnetic
field and the vector potential are connected via the curl operator,
\begin{equation}
\mathbf{B}=\mathbf{\nabla}\times\mathbf{A}\,.
\end{equation}
By varying the GL free energy with respect to $\vf$ and the gauge
potential we obtain the Ginzburg-Landau equations, which are the Euler-Lagrange
equations of motion (EOM) of the theory:
\begin{align}
\alpha\vf+\beta\left|\vf\right|^{2}\vf+\frac{1}{2m}\left(-i\mathbf{\nabla}-q_{s}\mathbf{A}\right)^{2}\vf & =0\,,\label{eq:GLeom1}\\
\nabla\times\mathbf{B} & =4\pi\mathbf{j}_{s}\,,\label{eq:GLeom2}
\end{align}
 with the supercurrent 
\begin{equation}
\mathbf{j}_{s}=\frac{q_{s}}{m_{s}}\text{Re}\left\{ \vf^{*}\left(-i\mathbf{\nabla}-q_{s}\mathbf{A}\right)\vf\right\} \,.\label{eq:GL_superj}
\end{equation}
The first equation takes the form of a non-linear Schrödinger equation.

\section{Characteristic Length Scales}

We can investigate the theory by computing simple solutions of the GL
equations and study the emerging characteristic length scales. Without
magnetic field, the solution can be assumed to be homogeneous, yielding
\begin{equation}
\left|\vf_{0}\right|^{2}=-\frac{\alpha}{\beta}\,.\label{eq:psi_0}
\end{equation}
This is the homogeneous value of the condensate, which should vanish
if the temperature is approaching the critical temperature. This can
be achieved by assuming an effective phenomenological temperature
dependence for the prefactor $\alpha$ of the following form:
\[
\alpha\to\alpha_{0}\left(T-T_{c}\right)\,,
\]
with the actual temperature $T$ and the critical temperature $T_{c}$.
Note that $\alpha$ is negative in the superconducting phase. This
smooth transition of the order parameter indicates a second order
phase transition. Since the free energy has to be bounded from below
in order to allow for a finite energy ground state, we have to demand
that $\nicefrac{\alpha_{0}}{\beta}<0$. We will refer to this fact
as "boundedness of the potential". 

As a next step we want to investigate how small perturbations of the homogeneous
condensate behave. For this purpose, we are computing the change of
$\vf$ and $\mathbf{A}$ perpendicular to the surface of a plane in $y,z$-direction separating
a superconducting half-space from a normal conducting one. In the
normal conducting half-space, we apply a constant magnetic field in
the $z$-direction. In this simple setup, the problem becomes effectively
one dimensional. In order to obtain analytical results, we investigate
the behavior of the condensate and the gauge field far away from the
boundary by introducing perturbations around the equilibrium value,
\begin{align}
\vf & =\vf_{0}+\delta\vf(x)\,,\\
\mathbf{A} & =\delta A(x)\hat{\mathbf{e}}_{y}\,.
\end{align}
Due to the geometry of the system we have assumed that all perturbations
solely depend on the direction pointing out of the phase-separation
plane, and that the magnetic field in the superconductor is parallel
to the external magnetic field. The first component of Eq.~(\ref{eq:GLeom2}) shows immediately that the phase of the condensate is constant. This can be seen by separating modulus and phase of $\vf$. Therefore, we can assume the perturbations
to be real and normalize the condensate per $\vf_{0}$, $f=\frac{\delta\vf}{\vf_{0}}$.
We now linearize the EOMs in $f$ and $\delta A$ and obtain
\begin{align}
f^{''}+4m\alpha f & =0\,,\\
\delta A^{''} & =\frac{4\pi q^{2}\vf_{0}^{2}}{m}\delta A\,,
\end{align}
where we have used Eq.~(\ref{eq:psi_0}) in order to eliminate the
constant terms and $\beta$. The equations are solved by
\begin{align}
f & =c_{1}e^{-\sqrt{4m\left|\alpha\right|}x}=c_{1}e^{-\frac{x}{\xi}}\,,\\
\delta A & =c_{2}e^{-\sqrt{\nicefrac{4\pi q^{2}\vf_{0}^{2}}{m}}x}=c_{2}e^{-\frac{x}{\ell}}\,,
\end{align}
where $c_{1}$ and $c_{2}$ are integration constants and we dropped
the exponentially growing solutions. We see that the perturbations
fall off with characteristic length scales $\xi$, called the \textbf{Ginzburg-Landau
coherence length}, and $\ell$, the \textbf{London penetration depth}.
The coherence length is consequently the length scale at which the
condensate varies. The penetration depth is the crucial ingredient
for the Meissner effect: computing the curl of $\mathbf{A}$ allows
us to determine the magnetic field in the superconductor,
\begin{equation}
\mathbf{B}=\mathbf{B}_{0}e^{-\frac{x}{\ell}}\,,
\end{equation}
where $B_{0}$ is the magnetic field at the surface, i.e. the external
field. We see that the field is not constant in the superconductor,
but drops exponentially away from the surface. Note that in the majority
of the literature the London depth is denoted by $\lambda$, which we reserve for the self-coupling constants of our field-theoretical model.

The exact form of the length scales will differ slightly in a relativistic
setup, but their meaning will be unaltered. The temperature dependence
of $\xi$ and $\ell$ can be obtained by entering the temperature dependence
of $\alpha$, which yields
\begin{align}
\xi & =\frac{1}{\sqrt{4m\left|\alpha(T)\right|}}\propto\frac{1}{1-t}\,,\\
\ell & =\frac{\sqrt{m}}{2\sqrt{\pi}q_{s}\vf_{0}}\propto\frac{1}{1-t}\,,
\end{align}
with the reduced temperature $t=\frac{T}{T_{c}}$. Note that, like
for all second-order phase transitions, the coherence length diverges at the transition
at $T=T_{c}$. In the next chapter I am going to show that the ratio
of these characteristic length scales determines the behavior of superconductors
in external magnetic fields.

\section{Behavior of Superconductors in Magnetic Fields}
\label{sec:SCmagnetic}
In this section, we are going to explore in more detail the behavior
of ordinary superconductors in external magnetic fields. Although
calculations are done within Ginzburg-Landau theory, the physical
consequences are qualitatively correct. Most of the following arguments
can be found in the standard textbooks, e.g.~Ref.~\cite{tinkham2004introduction}.
Additionally, parts of this section are simplified versions of the
calculations performed in Refs.~\cite{Haber:2016ljn,Haber:2017kth,Haber:2017oqb}.
In the previous section I have shown that the Meissner effect prevents
magnetic fields to penetrate the superconductor. In the next chapter
we are going to see how this can be explained in a more field-theoretical
framework. However, in this section I am going to explain that the
situation is actually more complicated and argue that the behavior
of a SC in an external magnetic field crucially depends on the ratio
of the London penetration depth $\ell$ and the Ginzburg-Landau coherence
length $\xi$. This dimensionless quantity is called the Ginzburg-Landau
parameter $\kappa,$
\begin{equation}
\kappa=\frac{\ell}{\xi}\,.
\end{equation}
Most materials in the lab have constant values of $\kappa$, which
allows us to characterize materials by their Ginzburg-Landau parameter.
Neutron stars on the other hand show a density dependent $\kappa,$
as we will later argue. In the non-relativistic setup used in this
introductory part, $\kappa$ is given by
\begin{equation}
\kappa^{2}=\frac{m^{2}\beta}{\pi q_{s}^{2}}\,,
\end{equation}
which is indeed dimensionless. 

In order to understand the $\kappa-$dependence of the magnetic response,
we have to look at positive and negative contributions to the free
energy of the system. Note that a "positive" contribution in energy
tends to disfavor a certain phase since the phase with the lowest
free energy is realized in thermodynamic equilibrium. 

The total magnetic field $\mathbf{B}$ in any material is given by
the sum of the external or "free" magnetic field $\mathbf{H}$
and the induced magnetization $\mathbf{M}$, in our case produced
by the supercurrent $\mathbf{j}_{s}$. 
\begin{equation}
\mathbf{B}=\mathbf{H}+4\pi\mathbf{M}\,,
\end{equation}
which means in order to reach $\boldsymbol{B}=0$ in the SC, the external
field is always balanced by a energetically costly magnetization.
On the other hand, the gain of condensation energy, which is a negative
contribution since it is the ground state of the system at $T=H=0$,
outweighs the cost of the magnetization up to a certain magnetic field.
If there are no other relevant contributions to the free energy, we
can compute the critical magnetic field $H_{c}$ by comparing the
energy of the SC phase with the normal conducting phase. Since we
work at a fixed external magnetic field $\mathbf{H}$, we have to
take the interplay between the external magnetic field and the resulting
total $\mathbf{B}$-field into account. In other words, we perform
a Legendre transformation from the Helmholtz free energy $F$ to the
Gibbs free energy ${\cal G}$,
\begin{equation}
\label{eq:Gibbs_def}
{\cal G}=F-\frac{\mathbf{H}}{4\pi}\cdot\int d^{3}r\,\mathbf{B}\,.
\end{equation}
The Gibbs free energy includes the work done by the "generator"
of the magnetic field, essentially by taking the magnetization effectively
into account. 

In the SC the Meissner effect eliminates the magnetic field, therefore
$\mathbf{B}=0$, whereas in the normal conducting phase we ignore
any possible magnetization of the normal conducting matter and set
$\mathbf{B}=\mathbf{H}$. Remember that $F_{N}$ is defined at $\mathbf{B}=0$,
thus we have to add the contribution of the magnetic field in the
normal phase which is given by $\nicefrac{\mathbf{B}^{2}}{8\pi}$.
Starting from Eq.~(\ref{eq:F_GL_single}), we find for the normal
conducting and the superconducting phase 
\begin{align}
\frac{{\cal G}_{N}}{V} & =\frac{F_{N}}{V}-\frac{H^{2}}{4\pi}+\frac{H^{2}}{8\pi}\,,\\
\frac{{\cal G}_{SC}}{V} & =\frac{F_{GL}}{V}\,.
\end{align}
The critical magnetic field $H_{c}$ can be found by setting ${\cal G}_{N}={\cal G}_{SC}$
and $H=H_{c}$,
\begin{equation}
H_{c}=\sqrt{\frac{8\pi}{V}\left(F_{N}-F_{GL}\right)}\,.
\end{equation}
At this field, the system undergoes a first-order phase transition
from the Meissner phase to the normal conducting phase, since the
condensate jumps from its homogeneous value to zero. Additionally,
the magnetic field $B$ immediately jumps from zero to $B=H$. This
kind of behavior was later termed \textbf{type-I superconductivity}
by Abrikosov. The given expression is very general and will hold in
more complicated settings, but can be evaluated in the presented case
by inserting the definition of $F_{GL}$, which yields
\begin{equation}\label{eq:HC_simple}
H_{c}=\sqrt{4\pi\vf_{0}^{2}\alpha}\,.
\end{equation}
To summarize the discussion above, we define the critical magnetic
field $H_{c}$ as in Ref.~\cite{Haber:2017kth}:
\begin{itemize}
\item \textit{Definition: }The critical magnetic field $H_{c}$
is the magnetic field at which the Gibbs free energies of the superconducting
phase in the Meissner state and the normal-conducting phase are identical,
resulting in a first-order phase transition between them.
\end{itemize}
First-order phase transitions have an interesting property: they can
lead to \textbf{mixed phases}. The mixed phase interpolates between
the two separate phases by mixing phase $a$ and $b$ on a macroscopic
scale. The geometry of the mixing depends on the surface energy between
$a$ and $b$. For example, one can imagine bubbles of phase $a$
immersed in an otherwise homogeneous phase $b$, or long strands or
sheets. In neutron stars, the transition from an ion lattice in the
crust to a neutron superfluid is thought to undergo various mixed
phases with different geometries, which are termed "nuclear pasta"
due to the pasta-like shape \cite{Ravenhall:1983uh,Caplan:2016uvu}.
In a type-I superconductor, the mixed phase consists of macroscopic
normal conducting regions within a superconducting material. The geometry
largely depends on the geometry of the superconducting sample itself
and the surface tension between the phases, because the magnetic field
will surpass the critical magnetic field $H_{c}$ in some regions
of e.g.~a spherical probe earlier than in others. The mixed phase
now interpolates between $B=0$ in the Meissner and $B=H$ in the
normal phase, in the sense that over the sample averaged magnetic
field $\left\langle B\right\rangle =\nicefrac{1}{V}\int_{V}d^{3}r\,B$
rises from zero to $H_{c}$. Thermodynamically speaking,
this accounts to fixing the averaged magnetic field $\left\langle B\right\rangle $
in the calculation instead of the external field $H$. These mixed
phases have been already discussed in the early works of Landau \cite{landau_intermed}
and in the textbook of London \cite{London_book}, for a lengthy pedagogical
discussion see Ref.~\cite{tinkham2004introduction}.

\begin{figure}[t]
\begin{centering}
\hbox{\includegraphics[width=0.45\textwidth]{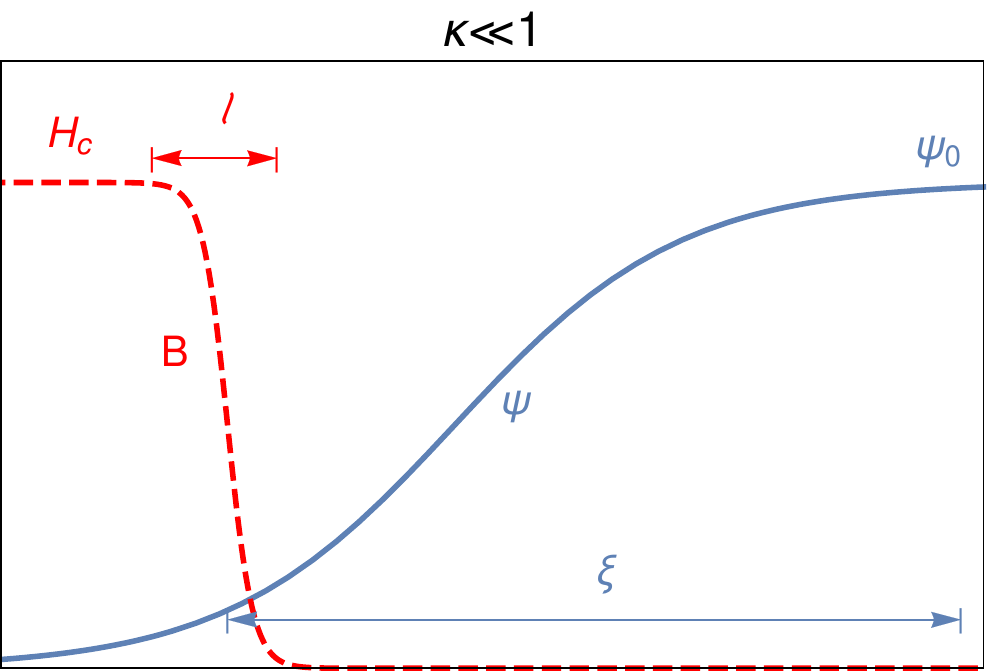}\hspace*{0.1\textwidth}\includegraphics[width=0.45\textwidth]{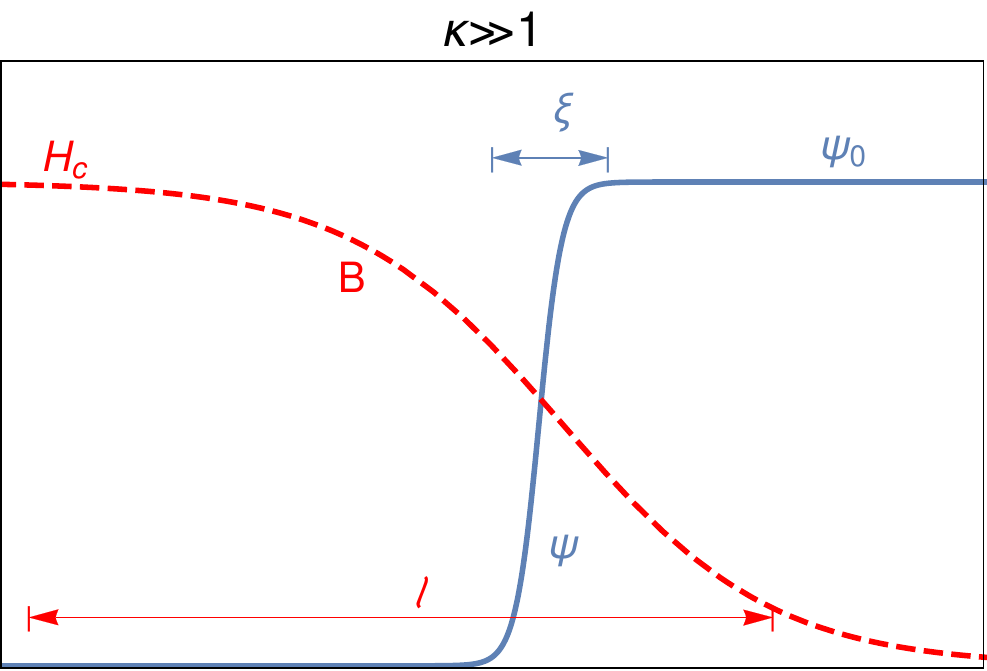}}
\par\end{centering}
\caption{\label{fig:surface_kappa}For $\kappa\ll1$ the magnetic field is
quickly damped leading to a high cost in magnetization energy whereas
the condensate has not recovered yet, leading to a positive surface
energy. For $\kappa\gg1$ the short coherence length leads to a high
amount of condensation energy whereas a small $\ell$ reduces the
cost for expelling the field, leading to a negative surface energy. }
\end{figure}

\begin{figure}[t]
\begin{centering}
\includegraphics[width=0.45\textwidth]{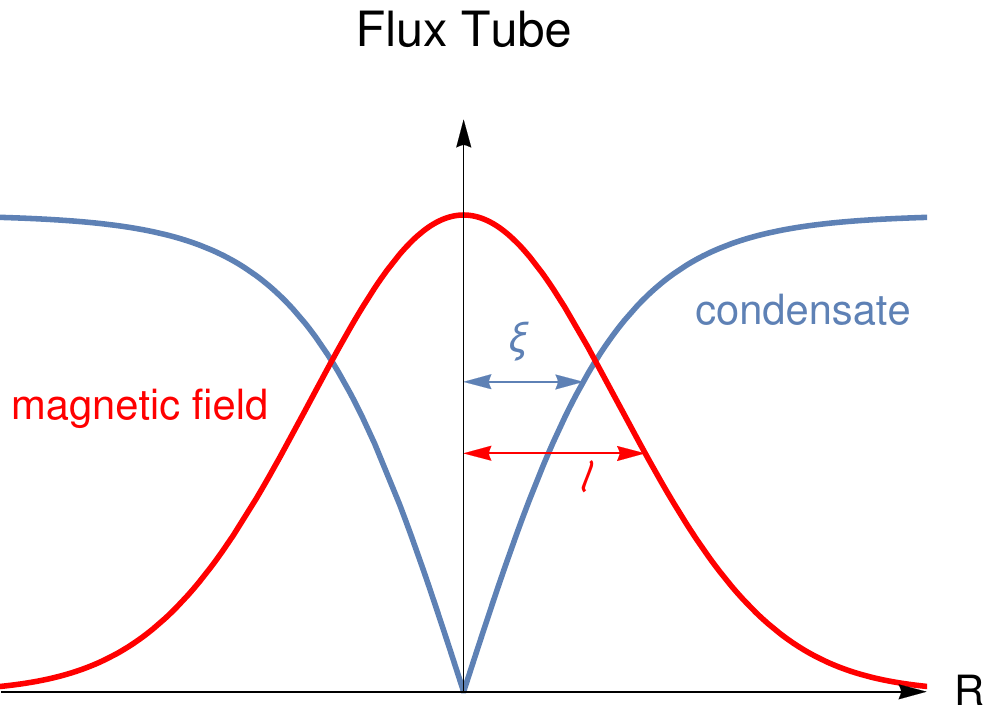}
\par\end{centering}
\caption{\label{fig:FT_schematic}Schematic plot of a flux tube. The condensate
goes to zero in the center and recovers at a scale $r\sim\xi$, whereas
the magnetic field is maximal in the center and falls of at a scale
$r\sim\ell$.}
\end{figure}

What we have not discussed so far are other possible lowering contributions
to the energy. One possible contribution which can diminish or raise
the free energy is the surface energy or tension between the normal
conducting and the superconducting phase, which is of course irrelevant
for an infinite superconductor in the Meissner phase. However, we
could imagine the magnetic field partially penetrating the SC in order
to reduce the energetic cost of the magnetization. For this to happen,
the surface energy should be negative. One can intuitively see that
the surface energy depends on the ratio of $\ell$ and $\xi$. In each panel of
Fig.~\ref{fig:surface_kappa}, we can see the transition from the normal
phase on the left to the superconducting phase on the right for two extreme
values of $\kappa$.
For $\kappa\ll1$, the magnetic field is quickly damped
leading to a relatively high cost in magnetization energy, since more volume is in the Meissner state. The condensate only
recovers on a larger scale $\xi$, which means a slow gain of condensation
energy, leading to a positive surface energy. For $\kappa\gg1$, the
short coherence length leads to a high amount of condensation energy
whereas a small $\ell$ reduces the cost for expelling the field,
leading to a negative surface energy. The exact transition from a
negative to a positive surface energy happens at a critical $\kappa$
of 
\begin{equation}
\kappa_{c}=\frac{1}{\sqrt{2}}\,,
\end{equation}
which can be found in a numerical calculation of the surface energy. A negative surface
energy suggests that a possible mixed phase seeks to maximize the surface between the superconducting and the
normal conducting phase, thus splitting into smaller and smaller regions
of separate normal conducting layers until this process is counteracted
by a positive energy contribution. In this discussion, the gradient
term in the free energy provides such an energy penalty: if the condensate
has to vary to quickly from zero in the normal phase to $\vf_{0}$
in the SC phase, the gain of surface tension will be compensated entirely.
We have already seen that the characteristic length scale of superconductivity
is given by the coherence length, therefore we can expect that the
normal conducting regions will have a typical size of $\xi$. Furthermore,
we have argued that the magnetic field is shielded in the Meissner
phase, such that the "defect" in the superconductor has to reach
the surface in order to allow the magnetic field to penetrate the
superconductor. A suitable solution seems to be a cylinder of normal conducting
matter with a typical radius $r\sim\xi$ parallel to the external
magnetic field. These magnetic defects of superconductors are called
\textbf{flux tubes} and are the defining property of \textbf{type-II
superconductivity}, first discussed by Abrikosov in 1957 \cite{abrikosov_magnetic}. The magnetic field in the flux tube is locally
created by a ring-shaped supercurrent and is maximal in the center
of the flux tube before it declines exponentially with the London
penetration depth. A schematic plot of the radial profile of a flux
tube can be seen in Fig.~\ref{fig:FT_schematic}. 

\subsection{Type-II Superconductivity \& Flux Tubes\label{subsec:flux_tubes}}

Due to their importance for this thesis, I am going to discuss flux
tubes in more detail in this subsection using GL theory. One of the
most intriguing properties of flux tubes is the fact that the magnetic
flux in the vortex is quantized, i.e.~it can only take integer values
of the fundamental magnetic flux quantum $\Phi_{0}$ . This can be
made plausible by looking at the definition of the supercurrent in
Eq.~(\ref{eq:GL_superj}) and of the magnetic flux itself,
\begin{equation}
\Phi=\int_{A}\mathbf{B}\cdot d\mathbf{A}=\int_{A}\,\left(\nabla\times\mathbf{A}\right)\cdot d\mathbf{A}=\oint_{\p A}\mathbf{A}\cdot d\mathbf{s},
\end{equation}
where $A$ is the area and $\p A$ the boundary of the area we are considering
and we have used Stokes' theorem. Assume that we want to compute the
flux of a single flux tube in an otherwise infinite homogeneous superconductor.
First we express $\vf$ in the equivalent form
\begin{equation}
\vf=\frac{1}{\sqrt{2}}\rho e^{i\psi}\,,
\end{equation}
with the absolute value $\rho$, which we will call the condensate
from now on, and the phase $\psi$. Instead of the two complex variables
$\vf$ and $\vf^{*}$ we now use two real ones. Since we need to evaluate the current infinitely far away from the flux tube,
where the condensate is homogeneous, we can neglect gradients of $\rho$
and obtain 
\begin{equation}
\mathbf{j}_{s}=\frac{q_{s}}{2m}\rho^{2}\left(\nabla\psi-q\mathbf{A}\right)\,.
\end{equation}
From the Maxwell equations we know that every current creates a magnetic
field, but we have argued that we can only have a magnetic field in
the flux tube, not in the homogeneous superconductor. Consequently,
$\mathbf{j}_{s}=0$ at $r=\infty$, with the radial distance from
the flux tube center $r$. This translates to a relation of the gradient
of the phase and the vector potential, $\mathbf{A}=\nicefrac{1}{q_{s}}\nabla\psi$.
This means that a rotation of the phase is counteracted by a corresponding
change of the gauge field. Inserting this relation into the definition
of the flux yields
\begin{equation}
\Phi=\oint_{\p A}\mathbf{A}\cdot d\mathbf{s}=\frac{1}{q_{s}}\oint_{\p A}\nabla\psi\cdot d\mathbf{s}=\frac{1}{q_{s}}\Delta\psi_{tot}\,,
\end{equation}
with the total change of the phase $\Delta\psi_{tot}$. Uniqueness
of the order parameter demands that the phase can only change by integer
multiples of $2\pi$, which yields
\begin{equation}
\Phi=\frac{2\pi n}{q_{s}}=n\Phi_{0}\,,
\end{equation}
with the winding number $n$ and the fundamental flux quantum
\begin{equation}
\Phi_{0}=\frac{2\pi}{q_{s}}\,.
\end{equation}

This picture provides us with further insight into the nature of flux
tubes: for winding number $n=1$, the phase of the condensate has
to rotate by $2\pi$ around a closed circle around the flux tube.
Now assume that we shrink this circle. The phase has to rotate "faster
and faster" around the tube, comparable to the relation of the velocity
of a rotating string at fixed angular velocity, which leads to a higher
and higher contribution of the increasing gradient of the phase to
the free energy, which has, among other positive contributions, terms
$F_{GL}\propto\rho^{2}\left(\nabla\psi\right)^{2}$. Shrinking the radius
of our path around the flux tube shows that we would receive an infinite
contribution from this gradient term if the condensate $\rho$ does
not vanish in an area around the center of the flux tube. This also
allows us to anticipate the radial flux tube profile of flux tubes
with higher winding numbers: for $n>1$ the phase has to wind around
the flux tube even faster, therefore we expect a larger area where
the condensate vanishes. We are going to confirm this expectation
by analytic and numerical calculations. 

Another important property of flux tubes is topological stability.
We have just shown that we cannot smoothly contract a circle around
a flux tube into a point, since the condensate vanishes in the center. We have "drilled" a hole (the flux
tube) into the vacuum (the homogeneous superconductor), and
the flux tube solution is homotopically distinct from the pure vacuum
solution. As a direct consequence, the time dependent Ginzburg-Landau
equations of motion have no solution which smoothly interpolates between
a static solution with a flux tube with winding number $n$ to a solution
with a different winding number, especially not to the completely
homogeneous solution with $n=0$. Changing the total winding of the
system requires a simultaneous change of the solution at every point
in space, which means that it requires an infinite amount of energy
in an infinite system. This is why flux tubes are called topological
stable. They cannot simply decay, in the lab they can only leave
the system by creeping out of the sample. Note that this argument
holds for the total winding of an infinite system. Locally, single
flux tubes might decay into multiple flux tubes with smaller winding
$n$ or may merge, which will be associated with a certain energy
cost.

It is important to note that we have made a simplifying assumption
which is true for the discussed case: the existence of a single gauge
field. In field-theoretical terms, electromagnetism is a $U(1)$ gauge
theory, which leads to a single gauge field ($\dim U(1)=1$). If we
additionally consider the strong force, we deal with an additional
eight color-gauge fields, since QCD is a $SU(3)$ gauge theory ($\dim SU(N)=N^{2}-1$).
Roughly speaking, the winding of the phase can be compensated by rotating several
gauge fields, which allows flux tubes associated with the electromagnetic
gauge field for instance to "unwind". From
a topological point of view, the discussion above can be described
as computing homotopy groups. We will resume to this more mathematical
description later when we discuss the symmetry breaking patterns for
(color-) superconductivity.

As a next step I will compute the critical magnetic field $H_{c1}$
starting from which flux tubes can penetrate the system. We are going
to use the following definition again taken from Ref.~\cite{Haber:2017kth}:
\begin{itemize}
\item \textit{Definition: }The critical magnetic field $H_{c1}$ is the
magnetic field at which it becomes energetically favorable to put
a \textit{single flux tube} into the superconductor in the Meissner
phase, resulting in a second-order phase transition from the Meissner
phase into the flux tube phase. $H_{c1}$ is an \textit{upper} bound
for this transition because there can be a first-order transition
at some smaller $H$, i.e., it can be favorable to directly form a
flux tube lattice with a finite distance between the flux tubes instead
of adding a single flux tube. We call this first-order critical field
$H_{c1}'$. 
\end{itemize}
We will discuss a possible first order transition, which depends on
the interaction between the flux tubes themselves, later on. Consequently,
the Gibbs free energy of the flux tube phase ${\cal G}_{\circlearrowleft}$
has only one additional contribution $F_{\circlearrowleft}$, the
free energy of a single flux tube.
\begin{align}
{\cal G}_{\cl} & =F_{GL}+F_{\cl}-\frac{1}{4\pi}\mathbf{H}\cdot\int d^{3}r\,\mathbf{B}\,,\nonumber \\
 & =F_{GL}+F_{\cl}-\frac{Hn\Phi_{0}}{4\pi}L\,,
\end{align}
where we have used 
\begin{equation}
\int d^{3}r\,B=n\Phi_{0}L\,,
\end{equation}
for a flux tube with length $L$. The critical magnetic field can
now be computed by equating ${\cal G}_{\cl}$ with the Gibbs free
energy of the SC in the Meissner state, which is $F_{GL}$ (remember
$B=0$ in the Meissner state):
\begin{equation}\label{eq:Hc1_single}
H_{c1}=\frac{2q_{s}}{n}\frac{F_{\cl}}{L}
\end{equation}
 The single flux tube free energy $F_{\cl}$ depends on the radial
profile of the flux tubes and the behavior of the gauge fields which
can only be computed numerically. However, some analytic solutions
can be found for either small or large distances from the flux tube
center $r$, and in the extreme type-II regime, $\kappa\gg1$. In
order to derive the corresponding equations of motion we use cylindrical
coordinates and the following ansatz for the condensate and the gauge
field:
\begin{align}
\vf & =\frac{1}{\sqrt{2}}\rho(r)e^{i\psi(\theta)}=\frac{1}{\sqrt{2}}f(r)\rho_{0}e^{i\psi(\theta)}\,,\label{eq:ft_ansatz}\\
\mathbf{A} & =\frac{na(r)}{q_{s}r}\hat{\mathbf{e}}_{\theta}\Rightarrow\mathbf{B}(r)=\frac{n}{q_{s}r}\frac{\p a(r)}{\p r}\hat{\mathbf{e}}_{z}\,,
\end{align}
where we will call $f(r)$ and $a(r)$ the profile function of the
flux tube and the gauge field, respectively, and $\hat{\mathbf{e}}_{\theta}$
and $\hat{\mathbf{e}}_{z}$ are the unit vectors in $\theta-$ and
$z-$direction. Due to the symmetry of the setup we assumed a purely
radial dependence of the profile functions. The phase of the condensate
around a cylindrical flux tube can only depend on the angle we have
rotated around the tube, not the distance, thus $\psi=\psi(\theta)$.
In the ansatz of the gauge field we have already anticipated the appearance
of the winding number $n$, however one can always redefine $a(r)$
and absorb any prefactors. For a numerical investigation the particular
ansatz facilitates the boundary conditions. The homogeneous value
of the condensate is given by $\rho_{0}^{2}=\frac{1}{2}\left|\vf_{0}\right|^{2}=-\nicefrac{\alpha}{2\beta}$
and the magnetic field is computed by taking the curl of the gauge
field in cylindrical coordinates. By inserting this ansatz into the
free energy Eq.~(\ref{eq:F_GL_single}) and varying it with respect
to $\psi$, $f$ and $a$ we obtain the EOMs for a single flux tube.
For a detailed derivation see App.~\ref{App:single_FT}. The free
energy itself can be written down as
\begin{equation}
\frac{F_{\cl}}{L}=\pi\rho_{0}^{2}\int_{0}^{\infty}dR\,R\left\{ \frac{n^{2}\kappa^{2}a^{'2}}{R^{2}}+f^{'2}+f^{2}\frac{n^{2}\left(1-a\right)^{2}}{R^{2}}+\frac{1}{2}\left(1-f^{2}\right)^{2}\right\} \,,
\end{equation}
which leads to the EOMs
\begin{align}
f^{''}+\frac{f^{'}}{R}+f\left[\left(1-f^{2}\right)-\frac{n^{2}}{R^{2}}\left(1-a\right)^{2}\right] & =0\,,\label{eq:sFT_EOM1}\\
a^{''}-\frac{a^{'}}{R} & =-\frac{f^{2}}{\kappa^{2}}\left(1-a\right)\,,\label{eq:sFT_EOM2}
\end{align}
where we have introduced the dimensionless coordinate 
\begin{equation}
R=\frac{r}{\xi}\,.
\end{equation}
The boundary conditions for the profile functions are given by 
\begin{equation}
f(0)=0\,,\qquad f(\infty)=1\,,
\end{equation}
due to our normalization construction and
\begin{equation}
a(0)=0\,,\qquad a(\infty)=1\,,
\end{equation}
which is in agreement with Eq.~(\ref{eq:sFT_EOM1}) for large $R$, where the profile
function of the condensate $f$ becomes flat and approaches $1$.
Therefore, nearly all terms cancel and we are left with $1-a\left(\infty\right)=0$.
This justifies the ansatz for the gauge field a posteriori.

For a numerical calculation and some analytic estimations it is useful
to derive the solutions for the equations of motions in the limits
$R\ll1$ and $R\gg1$, i.e.~for (physical) distances smaller and
larger than the coherence length. 

\subsubsection*{Solution for $R\ll1$}

For small values of $R$, we neglect all terms that do not contain
a derivative or are not inverse proportional to $R$,

\begin{eqnarray}\label{eq:core_sol}
f^{''}+\frac{f^{'}}{R}-\frac{n^{2}(1-a)^{2}}{R^{2}}f & = & 0\,,\\
a^{''}-\frac{a^{'}}{R} & = & 0\,,
\end{eqnarray}
The equations for the gauge field can in principle be solved by a
constant. This would lead to the trivial case without magnetic field.
However, the equation is of the Euler form and can therefore be solved
by an ansatz $a(R)=cR^{\lambda}$, which leads to 
\begin{eqnarray}
\lambda\left(\lambda-1\right)R^{\lambda-2}-\lambda R^{\lambda-2} & = & 0\,,\\
\lambda & = & 2\,.\nonumber 
\end{eqnarray}
We have therefore shown that $a\propto R^{2}$ for $R\ll1$ independently
of $n$. Plugging this result into the remaining equation and dropping
the term proportional to $n^{2}f$ and $n^{2}fR$ leads to
\begin{eqnarray*}
f^{''}+\frac{f'}{R}-\frac{n^{2}(1-cR^{2})^{2}f}{R^{2}} & = & 0\,,\\
f^{''}+\frac{f^{'}}{R}-n^{2}\frac{f}{R^{2}} & = & 0\,.
\end{eqnarray*}
This equation is again an Euler differential equation and solved by
the same ansatz $f\propto R^{\gamma}$ leading to 
\begin{eqnarray*}
\gamma & = & \pm n\,.
\end{eqnarray*}
With the boundary condition $f(0)=0$ (the condensate vanishes in
the center) we can deduce that $f\propto r^{n}$ for $R\ll1$ as stated
in Ref.~\cite{Alford:2007np} on p.~5. This confirms the earlier
statement that the region of vanishing condensate will be larger with
higher winding numbers $n$, since the profiles will start flatter
and flatter in the center.

\subsubsection*{Solution for $R\gg1$}

For distances much larger than the coherence length, the condensate
and the gauge field approach their asymptotic value. Therefore, we
write 
\begin{equation}
a(R)=1+Rv(R)\,,\qquad f(R)=1+u(R)\,,
\end{equation}
and linearize the equations of motion in the perturbations $v$ and
$u$, leading to 
\begin{align}\label{eq:FTeom_single}
0 & \simeq R^{2}v^{''}+Rv^{'}-\left(1+\frac{R^{2}}{\kappa^{2}}\right)v\,,\\
\Delta u & \simeq2u\,,
\end{align}
with the Laplace operator in cylindrical coordinates $\Delta f=\nabla\cdot\nabla f$.
The equations decouple and are solved by the modified Bessel functions
of the second kind $I_{\alpha}$ and $K_{\alpha}$, which are defined
as the two linear independent solutions to the differential equation
\begin{equation}
x^{2}\frac{d^{2}y}{dx^{2}}+x\frac{dy}{dx}-\left(x^{2}+\alpha^{2}\right)y=0\,,
\end{equation}
where $I_{\alpha}$/$K_{\alpha}$ is exponentially growing/decaying
for large $x$. By making the variable transformation $x=\nicefrac{R^{2}}{\kappa^{2}}$
or $x^{2}=2R^{2}$ (note that all factors of $\kappa$ or
$2$ then cancel) in the equation for $v$ and $u$ and realizing
that $\alpha=1$ in the first and $\alpha=0$ in the second case,
we can write down the solutions as
\begin{align}
a(R) & =1+CRv(R)=1+CRK_{1}\left(R^{2}/\kappa^{2}\right)=1+\tilde{C}rK_{1}\left(r^{2}/\ell^{2}\right)\,,\\
f(R) & =1+Du(R)=1+DK_{0}\left(R^{2}\right)=1+DK_{0}\left(r^{2}/\xi^{2}\right)\,,
\end{align}
with the numerical constants $C$ and $D$. In the first line, the
additional factor of $\xi$ has been absorbed into the redefinition
of the constant $C$. By switching back to the dimensionful coordinates
$r=R\xi$ we can see again the physical meaning of the length scales
$\ell$ and $\xi$: the deviation from the homogeneous solution falls
of exponentially with the characteristic length $\xi$ for the condensate
and $\ell$ for the gauge field. 

\subsection{Flux Tube Lattice}\label{sec:ft_lattice}

\begin{figure}[t]

\begin{centering}
\subfloat[\label{fig:Schematic-phase-diagram}Schematic phase diagram of a superconductor
in a magnetic field as a function of $\kappa$. Solid lines indicate
first order, dashed lines second-order phase transitions.]{\begin{centering}
\includegraphics[width=0.38\textwidth]{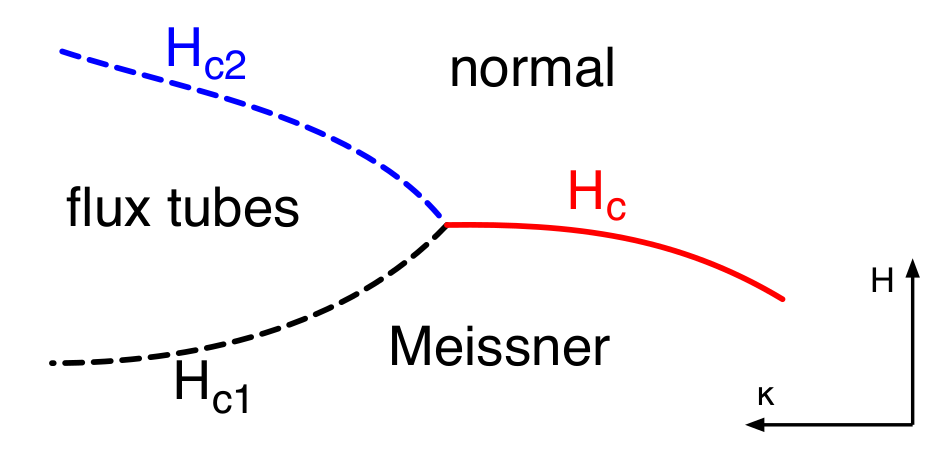}
\par\end{centering}
}\hfill{}\subfloat[\label{fig:interaction_potential_sch}Schematic plot of the flux tube
- flux tube interaction potential as function of the distance from
the flux tube center.]{\begin{centering}
\includegraphics[width=0.52\textwidth]{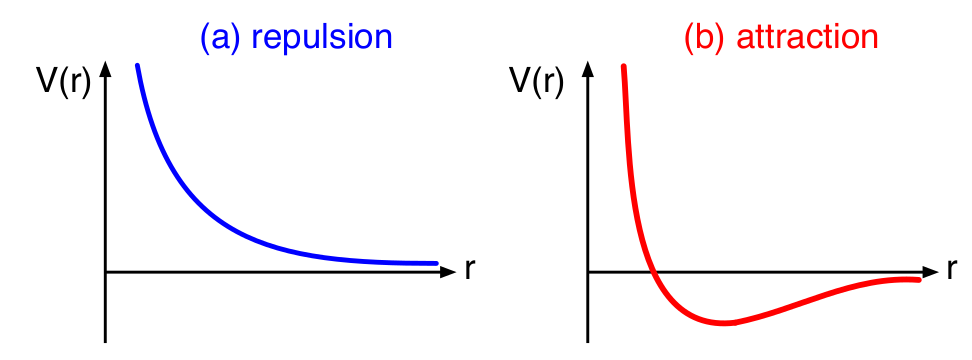}
\par\end{centering}
}
\par\end{centering}
\caption{In a simple superconductor, the transition from ${\color{blue}{\color{blue}a}}$
to ${\color{red}b}$ happens at $\kappa^{2}=1/2$ and indicates the transition from type-I to type-II behavior, where the three critical magnetic fields intersect, see left panel. The minimum of
the attractive potential defines the flux tube lattice spacing $R_{0}$.}

\end{figure}

So far we have only discussed single flux tube solutions. In a bulk
superconductor, we expect an entire \textbf{lattice of flux tubes}
to form, where the average distance $R_{0}$ between two flux tubes
depends on the external magnetic field. The average magnetic field
inside the superconductor can then be calculated from the flux tube
lattice density $\nu$. With increasing external field and decreasing
distance between the flux tubes, the cores of the tubes will start
to overlap and "eat up" the superconductor. Whereas close to $H_{c1}$
we can model the superconductor as a lattice of flux tubes of the
form computed in the single flux tube limit from Eqs.~(\ref{eq:sFT_EOM1}-\ref{eq:sFT_EOM2}),
the actual solution will deviate strongly with increasing flux tube
density and has to be computed numerically in a full $2D$ simulation.
This type of transition, where a (topological) defect successively
destroys a phase is very common and leads to a second-order phase
transition, where the (average) condensate becomes arbitrarily small
close to the transition. We can derive the maximal magnetic field
a superconductor can sustain by searching for the solution to the
linearized EOMs which allows for the highest external magnetic field.
A convenient gauge choice leading to a field in $z-$direction is
given by 
\begin{equation}
A_{y}=Hx\,.
\end{equation}
Inserting this into the GL-EOM Eq.~(\ref{eq:GLeom1}) and linearizing
the equation in $\vf$ leads to 
\begin{equation}
\left[-\Delta+\frac{4\pi i}{\Phi_{0}}Hx\p_{y}+\left(\frac{2\pi H}{\Phi_{0}^{2}}\right)^{2}x^{2}\right]\vf=\frac{1}{\xi^{2}}\vf\,,
\end{equation}
where we used the definitions of the flux quantum $\Phi_{0}$ and
the coherence length $\xi$ to rewrite some of the parameters. Since
the equation only explicitly depends on the variable $x$, we make
the following ansatz:
\begin{equation}
\vf=e^{ik_{y}y}e^{ik_{z}z}f(x)\,,
\end{equation}
which consists of plane waves with momentum $k_{y}$ and $k_{z}$
in the corresponding directions and a scalar function $f(x)$. The EOM
for $f(x)$ then reads
\begin{equation}
-f^{''}(x)+\left(\frac{2\pi H}{\Phi_{0}}\right)^{2}\left(x-x_{0}\right)^{2}f(x)=\left(\frac{1}{\xi^{2}}-k_{z}^{2}\right)f(x)\,,
\end{equation}
where we defined $x_{0}=k_{y}\Phi_{0}/(2\pi H)$. This equation can
immediately be solved by realizing its equivalence to the Schrödinger
equation for the quantum harmonic oscillator, 
\begin{equation}
-\psi^{''}+m^{2}\omega^{2}u^{2}\psi=2mE\psi\,,
\end{equation}
where we formally compare the prefactors 
\begin{equation}
u\eqhat x-x_{0}\,,\qquad2mE\eqhat\left(\frac{1}{\xi^{2}}-k_{z}^{2}\right)\,,\qquad\omega\eqhat\frac{2\pi H}{m\Phi_{0}}\,.
\end{equation}
The solution for the eigenvalues of the Schrödinger equations are
known to be $E_{n}=\omega\left(n+\tfrac{1}{2}\right)$, leaving
us with 
\begin{equation}
H=\frac{\Phi_{0}}{2\pi(2n+1)}\left(\frac{1}{\xi^{2}}-k_{z}^{2}\right)\,.
\end{equation}
The maximum of this expression can be found by setting $n=k_{z}=0$.
The resulting value is called the second critical magnetic field $H_{c2}$,
\begin{equation}
H_{c2}=\frac{\Phi_{0}}{2\pi\xi^{2}}\,.
\end{equation}
By inserting the definitions of $H_{c}$, $\Phi_{0}$
and $\kappa$ we can compare the critical magnetic fields and find
that 
\begin{equation}\label{eq:HcHc2_single}
H_{c2}=\sqrt{2}\kappa H_{c}\,.
\end{equation}
A numerical calculation in Chap.~\ref{chap:phasesHC} will show that actually all
critical magnetic fields of a single superconductor will meet at $\kappa^{2}=1/2$.
Consequently, we have just discovered a new criterion for the transition
from type-I to type-II superconductivity: the intersection
of the critical magnetic fields. We will see that this criterion
becomes more complicated in the multicomponent systems, nevertheless
the analytic intersection of $H_{c}$ and $H_{c2}$ will serve as
a guide line to find the type I/type II transition region. A schematic
plot of the critical magnetic fields of a superconductor can be seen
in Fig.~\ref{fig:Schematic-phase-diagram}, where dashed lines indicate second order and solid lines
first-order phase transitions.

In principle, we have not proven yet that the superconductor undergoes
a second-order phase transition from the flux tube phase (which is
also called Shubnikov phase \cite{shubnikov}) to the normal phase, since we have not
compared the free energies of the phases. However, it is pedagogically
not very instructive to perform this calculation here and repeat it
for the coupled case we are going to investigate in full detail later
on, from which the single superconductor result can be extracted trivially.
We will see that the superconducting phase is actually preferred over
the normal phase right below $H_{c2}$, which proves that the method
presented above actually produces the correct critical magnetic field
under the assumption of a second-order phase transition. It does not
exclude the possibility of a first order transition at a higher critical
magnetic field, since the entire discussion is based on the linearized
equations of motion. It seems unlikely though to find a first order
transition to the normal phase below $H_{c2}$ because we know that
there is a region in parameter space close to $H_{c2}$ where the
linearized equations are valid and the superconducting phase with
a small average condensate is preferred. It is physically unreasonable
(but not impossible) to go from a Meissner state to the flux tube
state, then per first order transition to the normal phase and for
even higher magnetic fields back into the superconducting phase, before
finally reaching the normal phase again via second-order phase transition
at $H_{c2}$. Let us summarize the discussion above by the following definition of $H_{c2}$.
\begin{itemize}
\item \textit{Definition:} Suppose there is a second-order phase transition
between the superconductor in the flux tube phase and the normal-conducting
phase, such that the equations of motion can be linearized in the
charged condensate. Then, the critical magnetic field $H_{c2}$ is
the maximal magnetic field allowed by the equations of motion. $H_{c2}$
is a \textit{lower} bound for the actual transition from the flux
tube phase to the normal-conducting phase because it does not exclude
a first-order transition at some larger $H$. We call the critical
field for such a first-order transition $H_{c2}'$. 
\end{itemize}
In order to discuss the first order transitions and the lattice structure,
the interaction between the flux tubes themselves within the flux
tube lattice is crucial. For a sparse lattice, where the single flux
tube approximation is still useful and we only have to deal with a
two-body interaction, the interaction energy can be calculated. I
will do so in the later parts of this thesis, and once again just
mention the most important result for the simple superconductor system
here, since it can be extracted easily from the complete calculation.
The \textbf{flux tube - flux tube interaction} has two major contributions: 
\begin{enumerate}
\item \textit{Attractive contribution: }for energetic reasons, flux tubes
prefer to overlap, i.e.~minimize the distance between the separate
tubes. That way, for a fixed number of tubes, which means for fixed
magnetic flux, the amount of lost (negative) condensation energy becomes
minimal. 
\item \textit{Repulsive contribution}: the magnetic field of one flux tube
might partially reach the next flux tube, where it interacts with
its supercurrent. Using basic vector analysis
and the fact that the magnetic fields of the separate flux tubes are
aligned, we find that the resulting Lorentz force $\mathbf{F}_{L}$
is repulsive: 
\begin{equation}
\mathbf{F}_{L}(\mathbf{r})=\mathbf{j}_{2}(\mathbf{r}-\mathbf{r}_{2})\times\mathbf{B}_{1}(\mathbf{r}-\mathbf{r}_{1})\,,
\end{equation}
where $\mathbf{j}_{2}$ is the current of the second flux tube, $\mathbf{B}_{1}$
the magnetic field of the first tube and $\mathbf{r}_{1}/\mathbf{r}_{2}$
are the position vectors of the first/second flux tube. 
\end{enumerate}
Depending on the value of $\kappa$, one term or the other is dominating.
In the simple system discussed here, for $\kappa^{2}>1/2$, which
means in the flux tube regime of the phase diagram, the magnetic penetration
length $\ell$ is rather large compared to the coherence length $\xi$,
leading to the dominance of the magnetic contribution of the interaction.
Thus, the interaction is purely repulsive. The flux tube lattice is
stabilized by the external magnetic field, which forces magnetic flux
into the superconductor. At the transition, the interaction becomes
attractive at long distances but irrelevant, since in the type-I regime
flux tubes never exist. For very short distances, the interaction
however stays repulsive, because of the strong contribution of the
magnetic field. Due to this interplay, the interaction potential develops
a non-trivial minimum at a finite flux tube separation $R_{0}$. We
will see that due to the interaction of the superconductor with a
second component, this effect can become important. A schematic plot
of the interaction potential is shown in Fig.~\ref{fig:interaction_potential_sch}.
As explained in Ref.~\cite{Haber:2017kth}, such a non-trivial
minimum, if existent in the type-I regime, can have very interesting
consequences. Recall that $H_{c1}$ is the magnetic field at which
the phase with a single flux tube is preferred over the phase with
complete field expulsion. In other words, at $H_{c1}$ the flux tube
density is zero and increases continuously with rising magnetic field, while the flux tube distance
decreases starting from infinity at $H_{c1}$. If the interaction
at infinite distances is attractive, the flux tubes do not \textquotedbl{}want\textquotedbl{}
to form an array with arbitrarily small density. Assuming that the
interaction always becomes repulsive at short range, there is a minimum
in the flux tube - flux tube potential, which corresponds to a favored
distance between the flux tubes. The negative contribution of the
interaction energy in this scenario provides an extra sort of "motivation"
for the superconductor to form an entire lattice of flux tubes at
magnetic fields even lower than $H_{c1}$. The transition from the
Meissner phase to the flux tube phase occurs at a critical field even
lower than $H_{c1}$, which we call $H_{c1}'$. At this field, the flux
tube density jumps from zero to a nonzero, finite value. An instructive analogy
is the onset of nuclear matter as a function of the baryon chemical
potential $\mu_{B}$. If the nucleon - nucleon potential was purely
repulsive, there would be a second-order onset at the baryon mass,
$\mu_{c}=m_{B}$. In reality, there is a binding energy $E_{b}$,
and the baryon onset is a first-order transition at a lower chemical
potential $\mu_{c}'=m_{B}-E_{b}$. Here, the role of the chemical
potential is played by the external field $H$, the role of the nucleons
is played by the flux tubes with mass per unit length $H_{c1}=2q_{s}F_{\circlearrowleft}/(nL)$,
and the binding energy is generated by the attractive interaction
between the flux tubes. In the textbook of Tinkham, Ref.~\cite{tinkham2004introduction},
only a simple interaction model, where only the magnetic contribution
is taken into account, is discussed in some detail. 

\begin{figure}[t]
\begin{centering}
\subfloat[\label{fig:ft_latt_exp}Hexagonal flux tube lattice in a superconductor.
The hexagonal unit cell with lattice spacing $R_{0}$ is shown in
red.]{\begin{centering}
\includegraphics[width=0.35\textwidth]{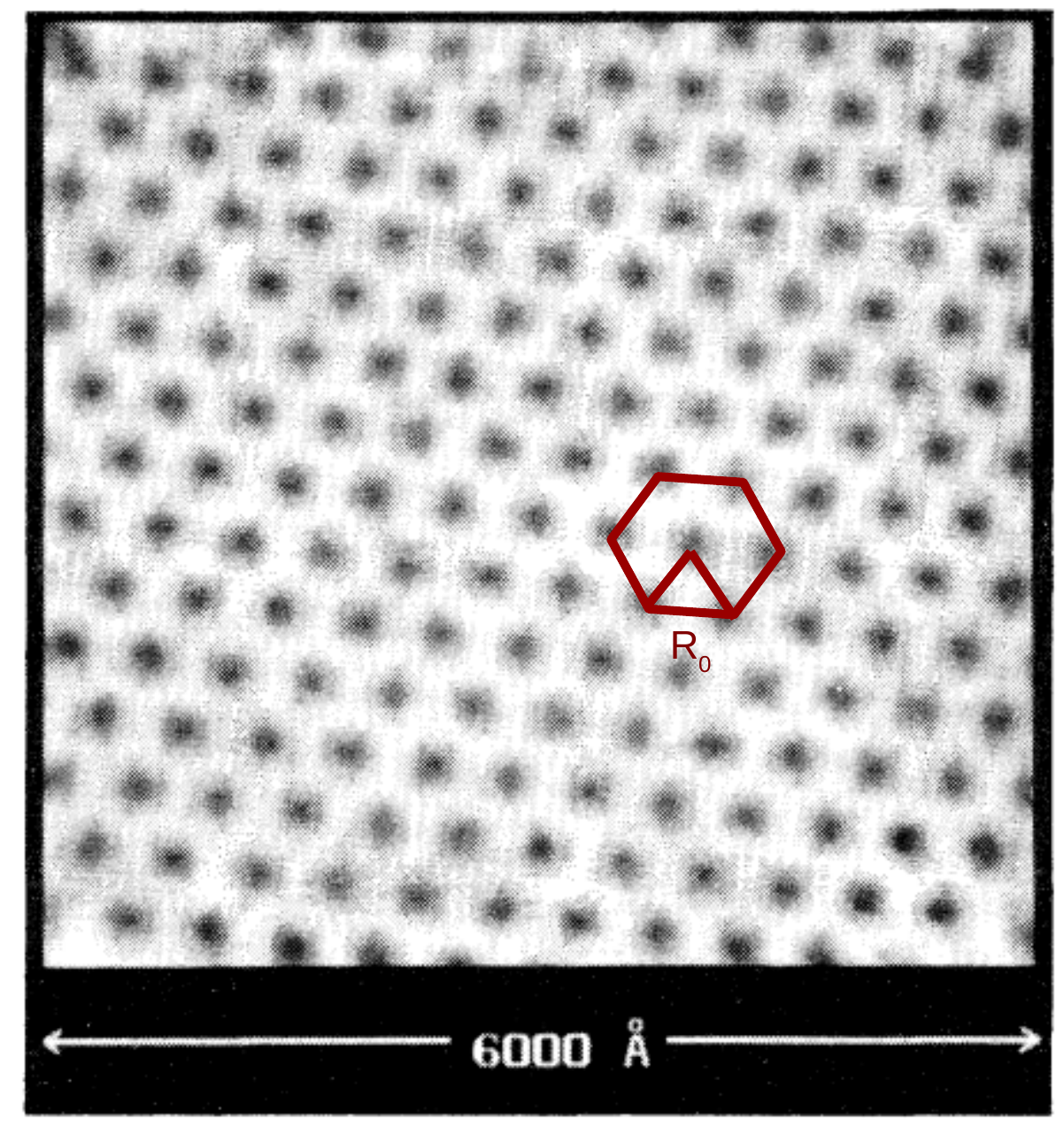}
\par\end{centering}
}\hfill{}\subfloat[\label{fig:interaction_potential_sch-1}Schematic flux tube lattice
forms (square and hexagonal)]{\begin{centering}
\includegraphics[width=0.52\textwidth]{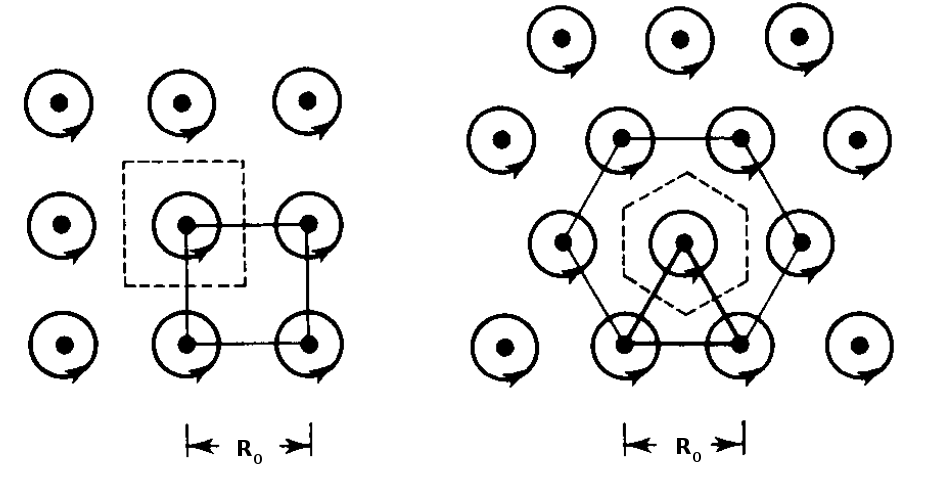}
\par\end{centering}
}
\par\end{centering}
\caption{The left panel shows an experimentally measured flux tube lattice
in $\text{NbSe}_{2}$ in an magnetic field of $1$ Tesla, modified
from Ref.~\cite{ft_lattice}. The right panel shows the different
unit cells in a square lattice and a hexagonal lattice, taken and
modified from Ref.~\cite{tinkham2004introduction}.}
\end{figure}

Finally, we have to discuss the geometric form of the lattice. A repulsive
interaction between the flux tubes suggest a hexagonal lattice of
flux tubes, since at a fixed magnetic flux the average distance between
the tubes is bigger than in a square lattice. This lattice structure
is also known as hexagonal close-packed (hcp). A numerical computation
of the geometrical lattice structure close to $H_{c2}$ discussed
in Ref.~\cite{tinkham2004introduction} reveals that this is indeed
the case, which has been confirmed experimentally, as can be seen
in Fig.~\ref{fig:ft_latt_exp}.

However, the difference between the free energy of a square lattice
and the hexagonal lattice is very small, leading Abrikosov originally to the
wrong conclusion, which was corrected by Kleiner et.~al.~in Ref.~\cite{PhysRev.133.A1226}
in 1964. Although the form of the lattice might change with the magnetic
field or with the onset of the attractive interaction, the expected
deviations from the hexagonal lattice is small, based on the argumentation
above. Therefore, for the rest of this thesis we are going to assume
a hexagonal lattice.
\newpage
To summarize this section, we have seen various different criteria
for the transition of the type-I to the type-II regime as a function
of $\kappa$. In a single superconductor, all these criteria lead
to the same "critical" value of $\kappa^{2}=1/2$.
\begin{enumerate}
\item The intersection of the three critical magnetic fields
$H_{c}$, $H_{c1}$ and $H_{c2}$ in one common point.
\item The change of the sign of surface tension between the normal
and the superconducting phase.
\item The change of the interaction between the flux tubes
from purely repulsive to long-range attractive.
\end{enumerate}
In Part.~\ref{par:SCSF}, I will show that in more complex systems not only the
value of the critical $\kappa$ will change but also that these criteria
are not identical anymore. 

\chapter{Superfluidity and Superconductivity in Quantum Field Theory}
\label{chap:sf_qft}
In this chapter I am going to show how superfluidity and superconductivity
can be described within the framework of quantum field theory and
how we can compute the Ginzburg-Landau potential discussed extensively
in Sec.~\ref{sec:Ginzburg-Landau-Theory} as tree-level potential
of a particular scalar field theory model. Most of the material presented here is basic
textbook material and discussed extensively for instance in Ref.~\cite{superbook},
which I will follow more or less directly. However, since in the mentioned
reference one can find a very comprehensive but highly pedagogical introduction
into the topic, I will keep this section short and only explain the
necessary points for understanding the research presented in the later
parts of this work.

\section{Scalar Field Theory and Noether's Theorem}
\label{sec:noether}
In order to describe superfluidity on a more microscopic level,
we use a rather simple model of a bosonic field with a contact self-interaction.
The version of the model presented here is known as $\phi^{4}$-theory
and serves as a standard toy-model for countless textbooks introducing
the concepts of quantum field theory \cite{Srednicki:2007qs,zee_qft-nut}. We will see that we can derive
the Ginzburg-Landau free energy from it directly. However, starting from a more
fundamental level allows us to go beyond GL-theory and introducing
temperature, interactions with other fluids, entrainment interactions
and more effects consistently. The Lagrangian of the theory
is given by 
\begin{equation}
\L=\p_{\mu}\vf^{*}\p^{\mu}\vf-m^{2}\left|\vf\right|^{2}-\lambda\left|\vf\right|^{4}\,,\label{eq:Lag-phi4}
\end{equation}
where $m$ is the particle mass and $\lambda$ the (self-)coupling
constant. We can use this bosonic model for spin-$0$ particles to
some extend to describe a superfluid of fermion Cooper pairs as well.
In this case, all parameters refer to the properties of the Cooper
pair, not of the particles themselves. In this approximation, we neglect
any influence on and effects of the constituents of the Cooper pairs.
Additionally, the model is only valid as long as the Cooper pairs
stay intact, because the fermionic nature of the
single particles is not reflected in the bosonic description. 

We can easily check that the theory is invariant under global $U(1)$ rotations
of the field $\vf$, i.e.~we can rescale the field by a complex phase,
\begin{equation}
\vf\to e^{-i\alpha}\vf\,,
\end{equation}
without any change of the Lagrangian. The parameter $\alpha\in\mathbb{R}$
is constant in space and time, since the symmetry is global. For
local symmetries, $\alpha$ becomes spacetime dependent. An important
feature of global symmetries is the emergence of conserved quantities.
This connection was found by the German mathematician Emmy Noether
in 1915 and published 1918, and can be stated as follows:
\begin{itemize}
\item \textbf{Noether's Theorem}: To every continuous global symmetry of
the action corresponds a conserved quantity. 
\end{itemize}
Very often, the Lagrangian is used as a starting point, however in
some cases only the action, which is the physically relevant quantity
and given by 
\begin{equation}
S=\int dt\,\L\,,
\end{equation}
is invariant. In field theory, the conserved quantity can be expressed
as a four-current $j^{\mu}$ that obeys a continuity equation, 
\begin{equation}
\p_{\mu}j^{\mu}=\p_{t}j^{0}+\p_{i}j^{i}=0\,.
\end{equation}
The corresponding conserved charge can be obtained by integrating
the zero-component of the current over space,
\begin{equation}
Q=\int_{\mathbb{R}^3}d^{3}x\,j^{0}\,.
\end{equation}
In order to compute the current, we assume the transformation parameter
$\alpha$ to be spacetime dependent, $\alpha\to\alpha(x)$, vary
the Lagrangian w.r.t.~the derivative of $\alpha$ and set $\alpha=0$
in the end.
\bea
\vf &\to&\vf^{'}=e^{i\alpha(x)}\vf\,,\\[2ex]
\L^{'}&=& \left(-i\p_{\mu}\alpha\vf^{*}+\p_{\mu}\vf^{*}\right)\left(i\p^{\mu}\alpha\vf+\p^{\mu}\vf\right)-m^{2}\left|\vf\right|^{2}-\lambda\left|\vf\right|^{4}\nonumber \\
&=& \p_{\mu}\alpha\p^{\mu}\alpha\left|\vf\right|^{2}-i\p_{\mu}\alpha\vf^{*}\p^{\mu}\vf+i\p_{\mu}\vf^{*}\p^{\mu}\alpha\vf+\p_{\mu}\vf^{*}\p^{\mu}\vf-m^{2}\left|\vf\right|^{2}-\lambda\left|\vf\right|^{4} \, .\label{eq:Lprime}
\eea
The current can now be computed as 
\begin{align}
j^{\mu} & =\frac{\p\L}{\p\left(\p_{\mu}\alpha\right)}\left|_{\alpha=0}\right.\\
 & =\left\{ 2\p^{\mu}\alpha\left|\vf\right|^{2}-i\vf^{*}\p^{\mu}\vf+i\p^{\mu}\vf^{*}\vf\right\} \left|_{\alpha=0}\right.\nonumber 
\end{align}
which finally yields 
\begin{equation}
j^{\mu}=i\left(\vf\p^{\mu}\vf^{*}-\vf^{*}\p^{\mu}\vf\right)\,.\label{eq:cons_curr_1fl}
\end{equation}
We can easily check that this current is actually conserved by using
the equations of motions for $\vf$ and $\vf^{*}$:
\begin{align}
\p_{\mu}\p^{\mu}\vf & =-\vf\left(m^{2}+\lambda\left|\vf\right|^{2}\right)\,,\\
\p_{\mu}\p^{\mu}\vf^{*} & =-\vf^{*}\left(m^{2}+\lambda\left|\vf\right|^{2}\right)\,,
\end{align}
which has to be plugged into 
\begin{equation}
\p_{\mu}j^{\mu}=i\left(\vf\p_{\mu}\p^{\mu}\vf^{*}-\vf^{*}\p_{\mu}\p^{\mu}\vf\right)\,,
\end{equation}
which proves our calculation. A conserved current allows
us to properly introduce a chemical potential. This procedure is explained
in detail in the textbook of Kapusta and Gale, Ref.~\cite{kapusta_tft},
Chap.~2.4. The chemical potential enters on the level of the Hamiltonian
density as 
\begin{equation}
{\cal H}-\mu{\cal N}\,,
\end{equation}
such that the energy needed to add one particle to the system is subtracted
by the new term and the thermodynamic potential stays constant. The
charge density ${\cal N}$ is nothing but the zeroth component of
the conserved current, 
\begin{equation}
{\cal N}=j^{0}=i\left(\vf\p^{0}\vf^{*}-\vf^{*}\p^{0}\vf\right)\,.
\end{equation}
The Hamiltonian density can be obtained from the Lagrangian density
by a Legendre transformation, for which we need to compute the conjugate
momenta to the fields. It can be shown that, after switching back
to the Lagrangian and integrating out purely quadratic terms in the
conjugate momenta, the chemical potential enters like the zeroth component
of a gauge field,
\begin{equation}
{\cal L}=\left|\left(\p_{0}-i\mu\right)\vf\right|^{2}-\left|\nabla\vf\right|^{2}-m^{2}\left|\vf\right|^{2}-\lambda\left|\vf\right|^{4}\,.
\end{equation}
This is equivalent to introducing it as the time derivative of a complex
phase after switching to polar coordinates. For that purpose we write
\begin{equation}
\vf=\frac{1}{\sqrt{2}}\phi e^{i\psi}\,,
\end{equation}
which yields, after putting it into the original Lagrangian without chemical potential
in Eq.~(\ref{eq:Lag-phi4}) and neglecting gradients of the phase,
\begin{equation}
{\cal L}=\frac{1}{2}\left|\left(\p_{0}-i\p_{0}\psi\right)\phi\right|^{2}-\frac{1}{2}\left|\nabla\phi\right|^{2}-\frac{m^{2}}{2}\phi^{2}-\frac{\lambda}{4}\phi^{4}\,.
\end{equation}
By setting $\p_{0}\psi=\mu$ we recover the Lagrangian at finite chemical
potential. We will proof later on that both procedures are still valid
in the multi-fluid calculation. 

Earlier, it was argued that Bose-Einstein condensation is a crucial
ingredient for superfluidity and conductivity. In order to incorporate
the BEC, we separate the condensate,
which is the expectation value (or vacuum expectation value, short:
vev) of the complex field $\left\langle \vf\right\rangle$, from
the fluctuations which we will for simplicity still refer to by $\vf$:
\begin{equation}
\vf(x)\to\left\langle \vf\right\rangle (x)+\vf(x)\,,\label{eq:fluctuations}
\end{equation}
which essentially amounts to shifting the field by its vev. This
form is a sense an ansatz for the condensate. Depending on the external parameters,
the vev will take on a finite value or not. In contrast to the condensate,
the fluctuations are a dynamical field, i.e.~we have to perform a
functional integration over the fluctuations in the path integral
of the partition function. The condensate on the other hand is a classical
field which has to be determined via the equations of motion. In order
to obtain analytic results, sometimes it is justified and useful (e.g.~for
small temperatures) to expand the partition function in the fluctuations
up to second order in the path integral, which then can be solved
analytically. As a next step, we write the complex vev in polar coordinates
in terms of its modulus $\rho$ and phase $\psi$,
\begin{equation}
\left\langle \vf\right\rangle =\frac{1}{\sqrt{2}}\rho(x)e^{i\psi(x)}\,.
\end{equation}
For the moment, we separate the fluctuations from the condensate and
neglect them in order to compute
\begin{align}
{\cal L} & ={\cal L}^{(0)}+\text{fluct.\,,}\\
{\cal L}^{(0)} & =\frac{1}{2}\p_{\mu}\rho\p^{\mu}\rho+\frac{\rho^{2}}{2}\left(\p_{\mu}\psi\p^{\mu}\psi-m^{2}\right)-\frac{\lambda}{4}\rho^{4}\,,
\end{align}
where the superscript $i$ in ${\cal L}^{(i)}$ denotes the order
of the fluctuations taken into account. From $\L^{(0)}$ we compute
the equations of motion for the condensate and the phase,
\begin{align}
\p_{\mu}\p^{\mu}\rho & =\rho\left(p^{2}-m^{2}-\lambda\rho^{2}\right)\,,\label{eq:EOM-cond-sf}\\
\p_{\mu}\left(\rho^{2}\p^{\mu}\psi\right) & =0\,,
\end{align}
where we have introduced the Lorentz scalar $p^{2}=\p_{\mu}\psi\p^{\mu}\psi$.
By rewriting the conserved current in Eq.~(\ref{eq:cons_curr_1fl})
using the new parametrization, we see that the last equation is nothing
but the conservation equation for the current,
\begin{equation}
\p_{\mu}j^{\mu}=\p_{\mu}\left(\rho^{2}\p^{\mu}\psi\right)=0\,.
\end{equation}
We have obtained it from the Euler Lagrange equations by varying the Lagrangian w.r.t.\ $\p_\mu \psi$,
\be
j^\mu=\frac{\p\L}{\p\left(\p_\mu\psi\right)}\, ,
\ee
which we expect from Noether's theorem. For a single field, we find
\be \label{eq:jmuQFT}
j^\mu=\rho^2\p^\mu\psi \, .
\ee
The current in general is computed by a variation of the Lagrangian (or more generally the generalized pressure) with respect to its conjugate momentum $p^\mu$, which we found to be equal to $\p^\mu\psi$.\footnote{This explains the choice of the nomenclature $p^2=\p_\mu\psi\p^\mu\psi$ as the square of the four-vector $p^\mu$.} This actually serves as the definition of the conjugate momentum. In this simple setup, we see that the current is four-parallel to its conjugate momentum.
From this form of the current, it is possible to gain a better understanding
of the phase of the condensate in hydrodynamic terms. In general, the current can be written
as the charge density $n$ it is transporting times the four-velocity
of the fluid,
\begin{equation}
j^{\mu}=nv^{\mu}\,,
\end{equation}
where 
\begin{equation}
v^{\mu}=\gamma\left(1,\mathbf{v}_{s}\right)\,,\qquad\gamma=\frac{1}{\sqrt{1-v_{s}^{2}}}\,,\label{eq:four-vel}
\end{equation}
with the superfluid (three-) velocity $\mathbf{v}_{s}$. Due to $v^{\mu}v_{\mu}=1$
per definition and use of the mostly-minus metric convention, we find
that 
\begin{equation}
n=\sqrt{j_{\mu}j^{\mu}}=\rho^{2}p^{2},
\end{equation}
where we used 
\begin{equation}
j^{\mu}=\rho^{2}\p^{\mu}\psi\,,\qquad p^{2}=\p_{\mu}\psi\p^{\mu}\psi\,.
\end{equation}
Consequently, 
\begin{equation}
v^{\mu}=\frac{j^{\mu}}{n}=\frac{\p^{\mu}\psi}{p}\,.
\end{equation}
We find that the macroscopic four-velocity is directly related to
the gradient of the microscopic phase of the condensate. The \textbf{superfluid
velocity }now follows from Eq.~(\ref{eq:four-vel}):
\begin{equation}
\mathbf{v}_{s}=-\frac{\nabla\psi}{\p_{0}\psi}\,.\label{eq:superfluid-velocity}
\end{equation}
In general, every gradient field is curl free. Therefore, if $\p_0 \psi$ is constant, we find
\begin{equation}
\nabla\times\mathbf{v}_{s}=0\,.
\end{equation}
However, for instance when computing hydrodynamic modes, this is not necessarily the case. In a non-relativistic setup, the superfluid velocity is proportional to the gradient divided by the mass and thus always curl free. In a relativistic calculation, the actual gradient should be used as a variable instead. The existence of such a curl-free variable is responsible for the existence of superfluid vortices: only
the vortex, where the condensate vanishes in the center as it is the
case of a superconducting flux tube, can carry an angular momentum
imposed from the outside. The result of Eq.~(\ref{eq:superfluid-velocity})
helps us to understand the role of the Lorentz-scalar $p$. We have
argued above that the chemical potential can be introduced as the
time derivative of the phase. Thus,
\begin{equation}
p=\sqrt{\p_{\mu}\psi\p^{\mu}\psi}=\sqrt{\left(\p_{0}\psi\right)^{2}-\left(\nabla\psi\right)^{2}}=\mu\sqrt{1-v_{s}^{2}}\,.
\end{equation}
This is the Lorentz transformation of the chemical potential, therefore
$\p_{0}\psi=\mu$ is the chemical potential in the lab-frame, where
the fluid moves with the velocity $\mathbf{v}_{s}$, whereas $p$
is the chemical potential in the rest-frame of the fluid! Note that
neither the gradient nor the time derivative of the phase are determined
by the equations of motion, they are external thermodynamic parameters
set by the physical environment.

\begin{figure}[t]
\begin{centering}
\subfloat[]{\begin{centering}
\includegraphics[width=0.45\textwidth]{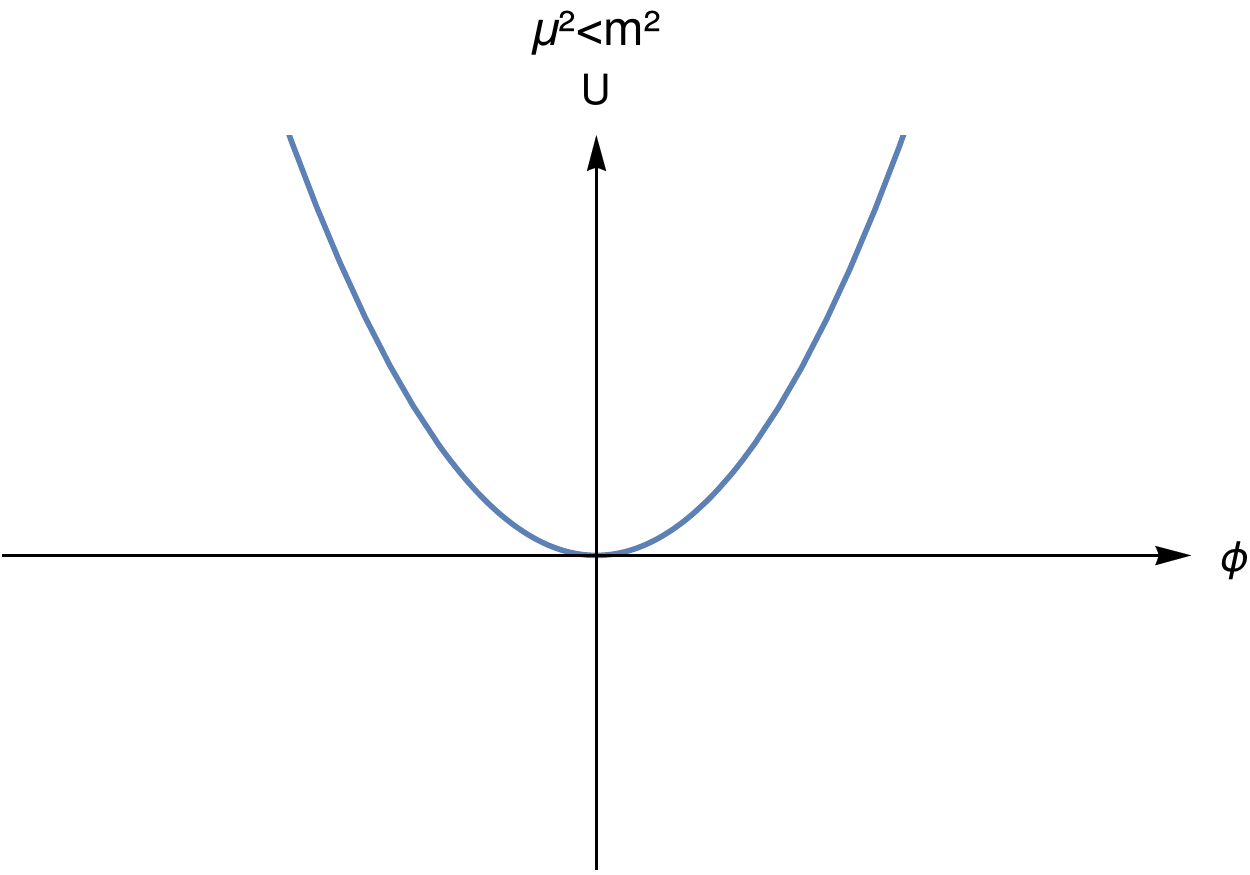}
\par\end{centering}
}\hfill{}\subfloat[]{\begin{centering}
\includegraphics[width=0.45\textwidth]{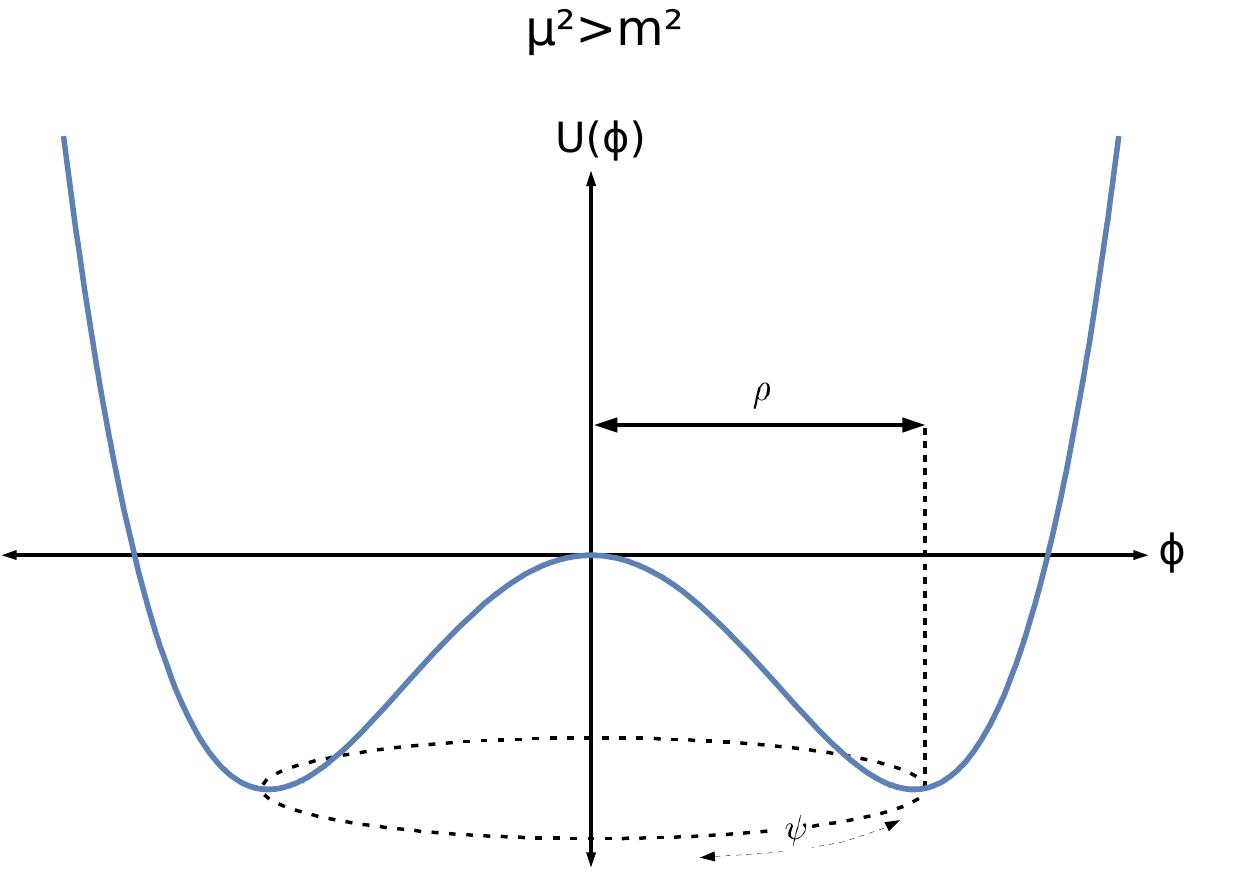}
\par\end{centering}
}
\par\end{centering}
\caption{Form of the potential as a function of the complex field $\phi$ before
(left panel) and after (right panel) spontaneous symmetry breaking.
Due to the finite value of the condensate , which is determined by
the minimum of the potential, the $U(1)$ symmetry is broken for $\mu^{2}>m^{2}$. The magnitude $\rho$ and the phase of the condesate $\psi$ are shown as well.\label{fig:SSB}}
\end{figure}

Going back to the equations of motion Eqs.~(\ref{eq:EOM-cond-sf}),
we still have to solve the first equation. Let us assume for the moment
that the condensate is static and homogeneous, then we find 
\begin{equation}
\rho\left(p^{2}-m^{2}-\lambda\rho^{2}\right)=0\,,
\end{equation}
which has two solutions,
\begin{equation}
\rho=0\,,\qquad\mathrm{and}\qquad\rho^{2}=\frac{p^{2}-m^{2}}{\lambda}\,,
\end{equation}
where the second solution can only exist for 
\begin{equation}
p^{2}>m^{2}\,.
\end{equation}
This is the usual requirement for the chemical potential: as long
as it is smaller than the mass $m$, the minimum of the potential,
which is given by 
\begin{equation}
U=-\L^{(0)}\,,
\end{equation}
is found at $\rho=0$, i.e.~there is no condensation. On the one
hand, this further proves our interpretation of $p$ to be correct.
On the other hand, we have just discovered another fundamental property
of superfluidity: \textbf{spontaneous symmetry breaking}. In
order to have condensation, the prefactor of the quadratic term has
to be negative ($p^{2}>m^{2}$). In this case, the potential takes
on the form of a Mexican hat or the bottom of a wine bottle. The $U(1)$
symmetry of the Lagrangian manifests itself in the rotational symmetry
of the potential in the space of real and imaginary part of $\phi$. However, if $\rho\ne0$, the ground state of the
system breaks this invariance spontaneously. SSB thus refers to a
system where the action obeys a symmetry that the ground state of
the system does not. The condensate plays the role of the order parameter,
since it is responsible for the breaking of the symmetry. A graphic
illustration of this discussion can be found in Fig.~\ref{fig:SSB}.

Another conserved quantity, which follows from Noether's theorem for
translational invariance of the system, is the stress-energy tensor $T^{\mu\nu}$. By using the gravitational definition, which is valid
for curved space-times as well and manifestly symmetric, we compute
\begin{align}
T^{\mu\nu} & =\frac{2}{\sqrt{-g}}\frac{\delta\left(\sqrt{-g}\L\right)}{\delta g_{\mu\nu}}=2\frac{\delta\L}{\delta g_{\mu\nu}}-g^{\mu\nu}\L\,,\\
T^{\mu\nu} & =\p^{\mu}\rho\p^{\nu}\rho+\rho^{2}\p^{\mu}\psi\p^{\nu}\psi-g^{\mu\nu}\L^{(0)}\,,\label{TmunuQFT}
\end{align}
where $\sqrt{-g}\equiv\sqrt{-\det g^{\mu\nu}}=2$ for Minkowski space.
By using the equations of motion for the condensate and the conserved
current, it is straightforward to show that 
\begin{equation}
\p_{\mu}T^{\mu\nu}=0\,,
\end{equation}
which will be fundamentally important for the hydrodynamic description
of superfluidity. Note that for a homogeneous and static condensate
both conservation equations are fulfilled trivially.
\section{Fluctuation Propagator and Goldstone Mode\label{sec:Fluc-prop_SF}}
\textbf{Goldstone's theorem }tells us that systems which undergo spontaneous
symmetry breaking of continuous global symmetries inherently possess a number
of massless (or gapless) modes. The number of modes depends on the
number of broken generators. Usually, the dispersion relation of the
Goldstone mode is linear in the origin at $k=0$. In this case, the
number of modes is equal to the number of broken generators. This
massless excitation is attributed to perturbations around the vev along the circle
at the bottom of the Mexican hat potential. Since I have already discussed
the phenomenological aspect of the Goldstone mode in the introductory
chapter on superfluidity, see Sec.~\ref{sec:Landau's-Critical-Velocity},
we will now derive the dispersion relation within our field-theoretical
model. There are two basic ways to compute the excitation energies
of the system, both of which we are going to use at some point. The
direct way consists of introducing harmonic perturbations for the
modulus and the phase of the condensate in the equations of motion.
\begin{equation}
\rho\to\rho+\delta\rho e^{i\left(\omega t-\mathbf{k}\cdot\mathbf{x}\right)}\,,\qquad\psi\to\psi+\delta\psi e^{i\left(\omega t-\mathbf{k}\cdot\mathbf{x}\right)}\,,
\end{equation}
 Afterwards, linearizing in $\delta\rho$ and $\delta\psi$ yields
a condition for $\omega$ from which the excitations can be extracted.
Here, we are going to proceed by calculating the fluctuation propagator from which
the dispersion relations can be extracted by calculating the zeros
of its determinant. Going back to Eq.~(\ref{eq:fluctuations}), we
slightly modify the fluctuations by giving them the same phase as
the condensate, i.e.~we write
\begin{equation}
\vf=\frac{1}{\sqrt{2}}\left(\rho+\vf_{1}+i\vf_{2}\right)e^{i\psi}\,,
\end{equation}
where $\vf_{1}$ and $\vf_{2}$ are the real and imaginary part of
the fluctuation. This change of basis leads to a diagonal tree-level
propagator in momentum space. We now decompose the Lagrangian in orders
of the fluctuations, from zeroth to fourth order:
\[
{\cal L}={\cal L}^{(0)}+{\cal L}^{(1)}+{\cal L}^{(2)}+{\cal L}^{(3)}+{\cal L}^{(4)}\,,
\]
where
\begin{align}
{\cal L}^{(1)} & =\varphi_{1}\Big[\rho(p^{2}-m^{2}-\lambda\rho^{2})-\p_{\mu}\p^{\mu}\rho\Big]-\frac{\varphi_{2}}{\rho}\partial_{\mu}(\rho^{2}\partial^{\mu}\psi)+\partial_{\mu}(\varphi_{1}\partial^{\mu}\rho+\varphi_{2}\rho\partial^{\mu}\psi)\,,\\
{\cal L}^{(2)} & =\frac{1}{2}\Big[\partial_{\mu}\varphi_{1}\partial^{\mu}\varphi_{1}+\partial_{\mu}\varphi_{2}\partial^{\mu}\varphi_{2}+(\varphi_{1}{}^{2}+\varphi_{2}{}^{2})(p^{2}-m^{2})\nonumber \\
 & +2\partial_{\mu}\psi(\varphi_{1}\partial^{\mu}\varphi_{2}-\varphi_{2}\partial^{\mu}\varphi_{1})-\lambda\rho^{2}(3\varphi_{1}^{2}+\varphi_{2}{}^{2})\Big]\,,\\
{\cal L}^{(3)} & =-\lambda\rho\varphi_{1}(\varphi_{1}^{2}+\varphi_{2}^{2})\,,\\
{\cal L}^{(4)} & =-\frac{\lambda}{4}(\varphi_{1}{}^{2}+\varphi_{2}{}^{2})^{2}\,.
\end{align}
Computing the terms linear in the perturbations accounts to basically
rederiving the equations of motion for the condensate up to a total
derivative term which does not contribute to the action. Therefore the
linear contributions vanish. As an approximation, we neglect the fluctuations
in third and forth order and calculate the fluctuation propagator
from the quadratic contributions. This calculation is performed in
Fourier space, thus we introduce the dimensionless Fourier transformed
fields as
\begin{equation}
\vf_{i}(X)=\frac{1}{\sqrt{TV}}\sum_{K}e^{-iK\cdot X}\vf_{i}(K)\,,
\end{equation}
with the Minkowski four-product $K\cdot X=k_{0}x_{0}-\mathbf{k}\cdot\mathbf{x}$.
Note that we use the imaginary time formalism of thermal field theory,
where $x_{0}=-i\tau$ with $\tau\in[0,\beta]$ and the inverse temperature
$\beta=\nicefrac{1}{T}$, and $k_{0}=-i\omega_{n}$. The bosonic Matsubara
frequencies are given by 
\begin{equation}
\omega_{n}=2\pi nT\,,\qquad n\in\mathbb{Z}\,.
\end{equation}
In this formalism, the Minkowski product becomes essentially Euclidean,
\be
K\cdot X=-\left(\omega_{n}\tau+\mathbf{k}\cdot\mathbf{x}\right) \, .
\ee
The grand-canonical thermodynamic potential density or short grand potential $\Omega$ is the
negative of the pressure, $\Omega=-P$, and can be computed from the
partition function $Z$ 
\begin{equation}
Z=\text{Tr}e^{-\beta\hat{H}}\,,
\end{equation}
where $\hat{H}$ is the Hamiltonian operator, via 
\begin{equation}\label{eq:OmegaZ}
\Omega=-\frac{T}{V}\ln Z\,.
\end{equation}
 The partition function is calculated in the path-integral formalism,
\begin{equation}
Z=N\int{\cal D}\vf_{1}{\cal D}\vf_{2}e^{S}\,,
\end{equation}
with the action 
\begin{equation}
S=\int_{0}^{\beta}d\tau\int d^{3}x\,\L\,,
\end{equation}
and a potentially divergent normalization constant $N$. In the free
energy, due to the properties of the logarithm, this constant will
enter additively into the free energy, which we can then renormalize
by subtracting this constant. The zeroth-order contribution is trivial:
due to the assumption that the condensate is homogeneous and static,
the integral over the Lagrangian $\L^{(0)}$ simply adds a factor $\nicefrac{V}{T}$.
There are no dynamic fields, thus the functional integration ${\cal D}\vf_{i}$
is trivial as well and can be absorbed into the normalization $N$.
As a result, we obtain 
\begin{equation}
\Omega^{(0)}=\frac{T}{V}\ln e^{-\frac{V}{T}\L^{(0)}}=U\,,
\end{equation}
where we used $\L^{(0)}=-U$ with the tree-level potential $U$. On
a diagrammatic level, the tree-level potential does not include any
loops, only "trees", hence the name. For the quadratic contributions,
we perform the calculation of the propagator in some more detail in
the appendix, see App.~\ref{App:exc_SF}. As a result, we find
\begin{equation}
S^{(2)}=-\frac{1}{2}\sum_{K}\left[\vf_{1}(-K),\vf_{2}(-K)\right]\frac{D^{-1}(K)}{T^{2}}\left(\begin{array}{c}
\vf_{1}(K)\\
\vf_{2}(K)
\end{array}\right)\,,
\end{equation}
with the inverse propagator 
\begin{equation}
D^{-1}(K)=\left(\begin{array}{cc}
-K^{2}-p^{2}+m^{2}+3\lambda\rho^{2} & -2iK_{\mu}\p^{\mu}\psi\\
2iK_{\mu}\p^{\mu}\psi & -K^{2}-p^{2}+m^{2}+\lambda\rho^{2}
\end{array}\right)\,.
\end{equation}
For simplicity, we are going to proceed without superflow by setting
$\nabla\psi=0$, which allows us to write $p^{2}=\mu^{2}$ and $K_{\mu}\p^{\mu}\psi=k_{0}\mu$.
In this case, the dispersion relations are isotropic and can be derived
analytically. Before we actually compute the dispersions we proceed
slightly more generally and compute the contribution of $S^{(2)}$ to
the free energy density $\Omega$. The functional integration of a
Gaussian integral can be done exactly and is proportional to the determinant
of the inverse propagator. Together with the zeroth-order contribution
we obtain
\begin{equation}
\Omega=U+\frac{1}{2}\frac{T}{V}\ln\det\frac{D^{-1}(K)}{T^{2}}=U+\frac{1}{2}\frac{T}{V}\text{Tr}\ln\frac{D^{-1}(K)}{T^{2}}\,,
\end{equation}
where we have used the matrix identity $\ln\det A=\text{Tr\ensuremath{\ln}}A$
for any square matrix $A$. The dispersion relations are given by the zeros
of the determinant, of which there are two different ones in the isotropic
case, $\epsilon_{k}^{\pm}$, and their negatives. Thus we decompose
the determinant into its zeros by writing $\ensuremath{\det\,D^{-1}=[k_{0}^{2}-(\epsilon_{k}^{+})^{2}][k_{0}^{2}-(\epsilon_{k}^{-})^{2}]}$,
which yields 
\begin{align*}
\frac{1}{2}\frac{T}{V}\ln\det\frac{D^{-1}(K)}{T^{2}} & =\frac{1}{2}\frac{T}{V}\ln\prod_{K}\frac{[(\epsilon_{k}^{+})^{2}-k_{0}^{2}][(\epsilon_{k}^{-})^{2}-k_{0}^{2}]}{T^{4}}\\
 & =\frac{1}{2}\frac{T}{V}\sum_{K}\left[\ln\frac{(\epsilon_{k}^{+})^{2}-k_{0}^{2}}{T^{2}}+\ln\frac{(\epsilon_{k}^{-})^{2}-k_{0}^{2}}{T^{2}}\right]\\
 & =\sum_{e=\pm}\int\frac{d^{3}k}{(2\pi)^{3}}\left[\frac{\epsilon_{k}^{e}}{2}+T\ln\left(1-e^{-\epsilon_{k}^{e}/T}\right)\right]\,.
\end{align*}
In the last step we have performed the summation over the Matsubara
frequencies as described in detail in Ref.~\cite{kapusta_tft}, and
taken the thermodynamic limit $V\to\infty$, where we have to replace
\begin{equation}
\frac{1}{V}\sum_{K}\to\int\frac{d^{3}k}{\left(2\pi\right)^{3}}\,.
\end{equation}
The integral over the first term is in general divergent and represents
an infinite contribution of the vacuum energy at $\mu=T=0$. In principle,
any proper renormalization procedure renders this contribution finite. One way is to subtract the vacuum free energy at $\mu=T=0$, which is proportional to $\epsilon_k|_{\mu=T=0}$. However, here we proceed by simply dropping the entire term $\epsilon_k/2$, which leaves us with 
\begin{equation}
\label{eq:T-potential}
\Omega=U+\sum_{e=\pm}\int\frac{d^{3}k}{(2\pi)^{3}}T\ln\left(1-e^{-\epsilon_{k}^{e}/T}\right)\,.
\end{equation}
This is sometimes called the no-sea approximation. Note that we subtract more than the pure vacuum contribution here, since $\epsilon_k$ depends on the full condensate which implicitly depends on the temperature and the chemical potential.
Finally, we are computing the dispersion by solving $\det D^{-1}(K)=0$
for $k_{0}$, which results in 
\begin{equation}
\epsilon_{k}^{\pm}=\sqrt{k^{2}+m^{2}+2\lambda\rho^{2}+\mu^{2}\mp\sqrt{4\mu^{2}(k^{2}+m^{2}+2\lambda\rho^{2})+\lambda^{2}\rho^{4}}}\,.
\end{equation}
We can now discuss some interesting limiting cases. By setting the condensate
to zero, $\rho=0$, we obtain the dispersion relations of free bosons,
\begin{equation}
\epsilon_{k}^{\pm}=\sqrt{k^{2}+m^{2}}\mp\mu\,,
\end{equation}
where the sign distinguishes between particles and anti-particles.
On the other hand, inserting the result for the condensate at vanishing
temperature allows us to compute 
\begin{equation}
\epsilon_{k}^{\pm}=\sqrt{k^{2}+(3\mu^{2}-m^{2})\mp\sqrt{4\mu^{2}k^{2}+(3\mu^{2}-m^{2})^{2}}}\,.
\end{equation}
Finally, we have found the Goldstone mode: $\epsilon_{k}^{+}$ vanishes
for $k=0$ and behaves linearly for small $k$, for which we approximate
\begin{equation}
\epsilon_{k}^{+}=\sqrt{\frac{\mu^{2}-m^{2}}{3\mu^{2}-m^{2}}}\,k+{\cal O}(k^{3})\,.
\end{equation}

\begin{figure}[t]
\begin{centering}
\includegraphics{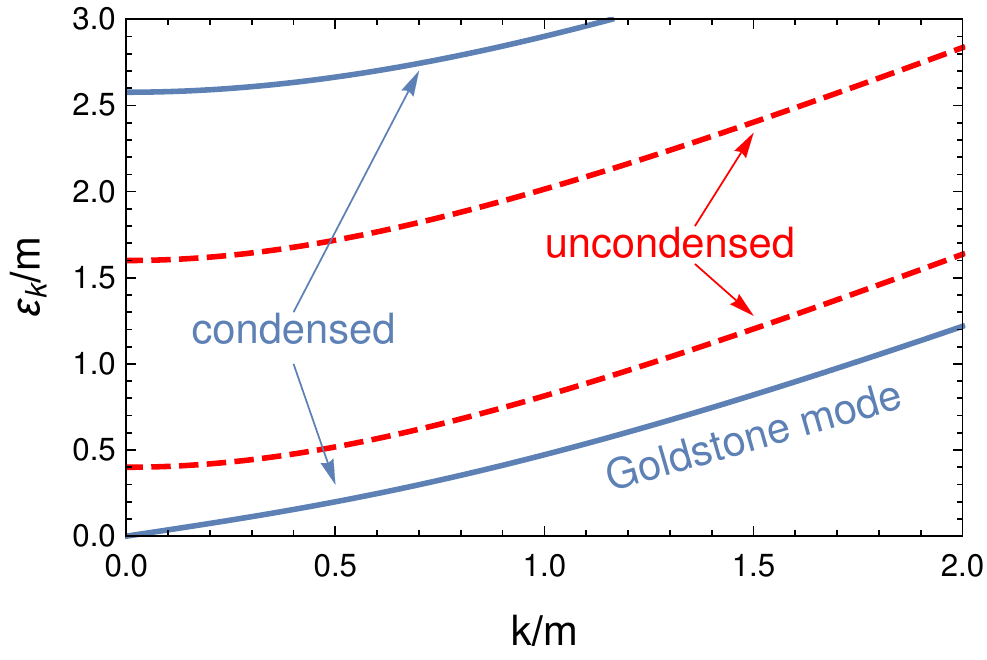}
\par\end{centering}
\caption{\label{fig:gold_disp}The dispersion relations for the condensed and
uncondensed case are plotted. The lowest mode behaves linear in the
origin and is called Goldstone mode, which results from the spontaneous
breaking of the global $U(1)$ symmetry by the condensate. Reproduced from Ref.~\cite{superbook}. }

\end{figure}

The second mode is gapped and thus called the massive or sometimes the Higgs
mode and given by 
\begin{equation}
\epsilon_{k}^{-}=\sqrt{2}\sqrt{3\mu^{2}-m^{2}}+\frac{1}{2\sqrt{2}}\frac{5\mu^{2}-m^{2}}{(3\mu^{2}-m^{2})^{3/2}}\,k^{2}+{\cal O}(k^{4})\,.
\end{equation}
A comparison of the dispersions in the condensed and uncondensed case
is shown in Fig.~\ref{fig:gold_disp}.

\section{Massive Gauge Boson}
\label{sec:massive_boson}

In order to describe a superconductor in the field-theoretical formalism,
we have to incorporate the charge of the field and the photon, which
interacts with the charged field. The photon is described by a gauge
field, as it is known from classical electrodynamics. In order to
incorporate a gauge field into our Lagrangian, we promote the global
$U(1)$ symmetry to a local gauge symmetry. Due to the derivative
in the kinetic term of ${\cal L}$ in Eq.~(\ref{eq:Lag-phi4}), the
system is a priori not invariant under this transformation, as we
have seen in the derivation of the Noether current in Eq.~(\ref{eq:Lprime}).
By adding a vector field which transforms in such a way that the extra
terms are canceled, we can achieve local gauge invariance. The gauge
field $A_{\mu}$ is coupled to the scalar field with charge $q$ via
the covariant derivative $D_{\mu}=\p_{\mu}-iqA_{\mu}$. The transformation
description is then given by
\begin{equation}
\vf\to e^{-i\alpha(X)}\vf\,,\qquad A_{\mu}\to A_{\mu}-\frac{1}{q}\p_{\mu}\alpha\,.
\end{equation}
The gauged Lagrangian now reads
\begin{equation}
\L=\left(D^{\mu}\vf\right)^{*}D_{\mu}\vf-m^{2}\left|\vf\right|^{2}-\lambda\left|\vf\right|^{4}-\frac{1}{16\pi}F^{\mu\nu}F_{\mu\nu}\,,\label{eq:phi^4-gauged}
\end{equation}
where the kinetic term of the gauge field is given by the standard
Yang-Mills Lagrangian $\L_{YM}$,
\begin{equation}
\L_{YM}=-\frac{1}{16\pi}F^{\mu\nu}F_{\mu\nu}\,,
\end{equation}
with the field strength tensor 
\begin{equation}
F_{\mu\nu}=\p_{\mu}A_{\nu}-\p_{\nu}A_{\mu}\,.
\end{equation}
This theory is often called scalar quantum electrodynamics (scalar
QED). The chemical potential can be still introduced like the zero-component
of a gauge field, i.e.~we can further modify the covariant derivative
to
\begin{equation}
D_{\mu}=\p_{\mu}-iqA_{\mu}+\delta_{\mu0}\mu
\end{equation}
Let us use polar coordinates again for the field and separate the fluctuations from the condensate,
\begin{equation}
\vf=\frac{1}{\sqrt{2}}\left(\rho_{0}+\vf\right)e^{i\psi}\,,
\end{equation}
where we assume that the condensate is independent of spacetime and
takes its homogeneous value 
\begin{equation}
\rho_{0}^{2}=\frac{\mu^{2}-m^{2}}{\lambda}\,,
\end{equation}
which we derived earlier. As a reminder and for comparison we quickly
repeat the exercise without gauge field and find
\begin{equation}
\L=\frac{1}{2}\p_{\mu}\vf\p^{\mu}\vf+\frac{\left(\rho_{0}+\vf\right)^{2}}{2}\p_{\mu}\psi\p^{\mu}\psi-\lambda\rho_{0}^{2}\vf^{2}-\sqrt{\lambda}\left(\mu^{2}-m^{2}\right)\vf^{3}-\frac{\lambda}{4}\vf^{4}+\frac{\lambda\rho_{0}^{4}}{4}\,.\label{eq:higgs_L}
\end{equation}
Note that $\p^{\mu}\psi$ is the phase of the fluctuations and therefore
a dynamic field (we have introduced the chemical potential directly
instead of as the phase of the condensate). We immediately see that
there is a massive mode $\vf$ with mass term $-\lambda\rho_{0}^{2}$
and a massless mode $\psi$, for which there are only a kinetic term and
interaction terms, but no quadratic mass term. This is the Goldstone
mode which we have computed explicitly in the last section. In the
gauged case, we incorporate an intermediate step in which we use polar
coordinates for the complete, not yet shifted field of the form $\vf=\frac{1}{\sqrt{2}}\phi e^{i\psi}$,
and compute 
\begin{align}
\L & =\frac{1}{2}\p^{\mu}\phi\p_{\mu}\phi+\frac{q^{2}\phi^{2}}{2}\left(A_{\mu}-\frac{1}{q}\p_{\mu}\psi\right)\left(A^{\mu}-\frac{1}{q}\p^{\mu}\psi\right)\nonumber \\
 & +\frac{\left(\mu^{2}-m^{2}\right)}{2}\phi^{2}-\frac{\lambda}{4}\phi^{4}-\frac{1}{16\pi}F^{\mu\nu}F_{\mu\nu}\,.
\end{align}
The expressions in the parentheses, which couple the gauge field
and the mode $\psi$, are gauge invariant. This means, a redefined
gauge field of the form 
\begin{equation}
B_{\mu}\equiv A_{\mu}-\frac{1}{q}\p_{\mu}\psi
\end{equation}
is invariant under gauge transformations, because the phase transforms as $\psi\to\psi-\alpha$ which cancels the transformation of $A_\mu$.
This procedure of redefinition is equivalent to using unitary gauge,
where one sets $\alpha(X)=\psi(X)$, such that the phase cancels.
As before, we shift the field by its vev (but keep the common phase), $\phi\to\rho+\vf$
and obtain
\bea
\L  &=&\frac{1}{2}\p^{\mu}\vf\p^{\mu}\vf+\frac{q^{2}\rho^{2}}{2}B^{\mu}B_{\mu}+q^{2}\rho\vf B_{\mu}B^{\mu}+\frac{q^{2}}{2}\vf^{2}B^{\mu}B_{\mu}-\left(\mu^{2}-m^{2}\right)\vf^{2}\non[2ex]
 &&-\sqrt{\lambda}\left(\mu^{2}-m^{2}\right)\vf^{3}-\frac{\lambda}{4}\vf^{4}+\frac{\left(\mu^{2}-m^{2}\right)^{2}}{4\lambda}-\frac{1}{16\pi}F_{\mu\nu}F^{\mu\nu}\,,
\eea
where $F^{\mu\nu}$ is unchanged since $\p_{\mu}A_{\nu}-\p_{\nu}A_{\mu}=\p_{\mu}B_{\nu}-\p_{\nu}B_{\mu}$.\footnote{Note that this relies on the interchangeability of derivatives, $[\p_{\mu},\p_{\nu}]\psi=0$,
which is not true if we deal with flux tubes where $\psi$ has a finite
winding, in which case $[\p_{\mu},\p_{\nu}]\psi$ has to be interpreted
as a distribution around the flux tube center. }By comparing to Eq.~(\ref{eq:higgs_L}), we see a very important
difference: the gapless mode is not present in the gauged case, it
has been "eaten" by the gauge field. The gauge field on the other
hand has become massive, where the mass is proportional to the vev
of the scalar field. This mechanism is known as the Higgs mechanism
and was first described using this very model presented here in Ref.~\cite{Higgs_orig}.
Note that in the literature, this is often referred to as spontaneous
symmetry breaking of a gauge symmetry. However, Elitzur's theorem,
which was proposed and shown to be correct in a lattice formulation
in Ref.~\cite{Elitzur} in 1975, tells us that local symmetries can
not be spontaneously broken. In the Higgs mechanism, only the global
part of the symmetry is broken by the condensate and leads to a conserved
current. Applying Noether's theorem to local symmetries leads to vanishing
conserved charges. The fact that the ground state of the theory appears
to be not invariant under local transformations is due to the gauge
fixing rather than to spontaneous breaking of the gauge symmetry.
The gauge invariant field $B_{\mu}$ should not be interpreted as
a gauge field, because it transforms differently under the local gauge
transformation, it stays the same. For a discussion of this topic see Ref.~\cite{Friederich:2011xs}.

Although the Goldstone mode disappeared, the number of degrees of
freedom is the same in both cases before and after the condensation.
In the superfluid case without gauge field, we have:
\begin{itemize}
\item Before condensation: two massive degrees of freedom coming from the
two scalar fields $\vf_{1}$ and and $\vf_{2}$ that form the complex
field $\vf=\vf_{1}+i\vf_{2}$.
\item After condensation: one massive mode with a mass term proportional
to the condensate and the gapless Goldstone mode $\psi$
\end{itemize}
If we include the gauge field, we start with two additional degrees
of freedom coming from the gauge field.
\begin{itemize}
\item Before condensation: two massive degrees of freedom from the complex
field $\vf$ and the two degrees of freedom from the massless gauge
field corresponding to the two polarizations of the photon, making a total of four.
\item After condensation: only one massive mode because of the "eating"
of the Goldstone mode, but three degrees of freedom from the now massive
gauge field, which allows for longitudinal and transverse modes. 
\end{itemize}

\section{Ginzburg-Landau Theory from Scalar Field Theory}

We have now all the ingredients to show how the GL free energy emerges
as the tree-level potential from the Lagrangian of scalar $\phi^{4}-$theory.
We can directly start from the gauged version of the theory since
simply setting the charge to zero allows us to convert the superconductor
to a superfluid. The starting point is therefore the Lagrangian given
in Eq.~(\ref{eq:phi^4-gauged}). We now ignore any fluctuations and
temporal derivatives of the condensate $\rho$, but account for an
inhomogeneous $\rho$ as we prepare for vortices and flux tubes.
Consequently, we replace the complex fields by
\begin{equation}
\vf=\frac{1}{\sqrt{2}}\rho e^{i\psi}\,,
\end{equation}
and introduce the chemical potential as the time derivative of the
phase $\psi$. After doing so, we obtain the tree level potential
$U$ by taking the negative of the resulting Lagrangian density, $U=-\L_{\vf\to\left\langle \vf\right\rangle }$,
\begin{equation}
U=\frac{1}{2}\left|\left(-i\nabla-q\mathbf{A}\right)\rho e^{i\psi}\right|^{2}-\frac{\mu^{2}-m^{2}}{2}\rho^{2}+\frac{\lambda}{4}\rho^{4}+\frac{B^{2}}{8\pi}\,,
\end{equation}
where we have used that $F_{\mu\nu}F^{\mu\nu}=2\left(\mathbf{B}^{2}-\mathbf{E}^{2}\right)$
in general and set the electrical field $\mathbf{E}$ to zero. By comparison
to the Ginzburg-Landau free energy in Eq.~(\ref{eq:F_GL_single}),
we see that we have just obtained the same result as a limiting case,
namely the tree-level potential, from scalar field theory. The coefficients
$\alpha$ and $\beta$ basically just have been replaced by the chemical
potential, the mass and the self-coupling constant $\lambda$. Additionally,
we started from a fully relativistic setup, which explains the different
prefactor of the gradient-energy, which has no mass dependence in
the relativistic case. Although this approach has so far no advantage
over starting directly from the Ginzburg-Landau free energy, it will
allow us, especially in the more complicated two-fluid setup in the
next chapter, to introduce temperature in a consistent way.

\chapter{Two-Component Model}
\section{Motivation and Methods}
The main goal of this thesis is to explore the behavior of \textbf{multicomponent
(super)fluids and superconductors} at finite superflow and in external
magnetic fields. In the interior of neutron stars, this situation
occurs in many places: First, in the crust, where superfluid neutrons
interact with the ion lattice that partially can be described as a
normal fluid. Second, in the outer core, where neutrons as well as
protons can form Cooper pair condensates, giving rise to a two-fluid
system of a superfluid and a superconductor \cite{Bogolyubov1958,MIGDAL1959655,Sedrakian:2006xm,Page2014,Haber:2016ljn}.
If the inner core consists of nuclear matter, hyperons can be present and form Cooper pairs as
well. Then, even more superfluid components can exist \cite{Gusakov:2009kc}.
Additionally, a deconfined quark phase in the inner core is possible.
Such a color-superconductor can exist in many different states, depending
on the quarks which participate in the formation of the Cooper pairs.
The most symmetric phase is described by the the color-flavor locked
(CFL) phase \cite{Alford:1998mk,Alford:2007xm}, which is effectively
described by a three-component superconductor. The CFL-phase as well
as the color-spin locked phase \cite{Schafer:2000tw,Schmitt:2004et}
are additionally superfluids, and kaon condensation in CFL may lead
to a two-component superfluid. In each case, temperature effects will
add an additional fluid, which renders dense neutron star matter a
complicated multi-fluid system. Although I will keep all results as
generic as possible in order to apply the presented calculations to
other, more down-to-earth systems, this is the environment we have
in mind whenever we have to assign numeric values to the parameters
of the model. Neutron stars also serve as primary motivation to include,
besides the simpler density coupling, a derivative or entrainment
coupling between the two components, which leads to the so-called
Andreev-Bashkin effect \cite{1976JETP...42..164A,2007JETP..105..135S},
which will be described in more detail later on. For a calculation of the strength of this coupling in dense nuclear matter see for instance Ref.\ \cite{Chamel:2006rc}.

\subsection*{Hydrodynamic Instabilities}

At vanishing magnetic field but finite uniform superflow, we will mainly focus on the discussion of hydrodynamic
instabilities in two-fluid systems. These are phenomenologically
interesting for instance for pulsar glitches, which can be observed
as sudden jumps in the rotation frequency of the star that are commonly
explained by a collective unpinning of superfluid vortices from the
ion lattice in the inner crust of the neutron star. For an elaborated
pedagogical explanation see the introductory part of this work and references therein. Hydrodynamic
instabilities are one candidate for triggering such a collective effect
\cite{2004MNRAS.354..101A,Peralta:2006um,Haskell:2015jra}. In this
scenario, the role of the second (normal) fluid, besides the neutron
superfluid, is played by the lattice of ions, not unlike an atomic
superfluid in an optical lattice \cite{PhysRevA.64.061603}. Recently
it has been argued that also the superfluids in the core of the star
might be important for the glitch mechanism, because entrainment effects
between the superfluid neutrons and the lattice in the inner crust
\cite{pethick2010superfluid,Chamel:2012pk} reduce the efficiency
of the transfer of angular momentum from the superfluid to the crust
\cite{Andersson:2012iu}, and thus an analysis of hydrodynamic instabilities
in a two-superfluid system of neutrons and protons is of phenomenological
interest. For simplicity however, we will neglect a possible charge
of one or both of the components, whether it is a normal or a superfluid.
This part is entirely based on our publication Ref.~\cite{Haber:2015exa}.
I want to emphasize that the presented results are obtained in an
idealized situation and are thus not directly applicable to the actual
physics inside a compact star. For instance, by neglecting the possible
charge of the fluids, any effects of electromagnetism are ignored,
which would be necessary to describe the coupled system of neutrons,
protons and electrons, unless protons and electrons can be viewed
as a single, neutral fluid. Also, our starting point is a bosonic
effective theory while the superfluids in a neutron star are mostly
of fermionic nature which condense via the Cooper mechanism. Additionally,
any effects of rotation or a magnetic field, i.e.~superfluid vortices
or superconducting flux tubes, are ignored in the study of hydrodynamic
instabilities. 

Although our main motivation comes from the study of neutron stars,
the described model is much more general. Two-fluid systems
where at least one component is a superfluid are realized in many
different contexts. For instance, any superfluid at nonzero temperature
is such a system because it can be described in terms of a superfluid
and a normal fluid within the two-fluid picture of superfluidity,
which was first described in Refs.~\cite{tisza38,landau41}, for
a pedagogical introduction see Chap.~2.3 of Ref.~\cite{superbook}
or \cite{Stetina:2015exa}. Systems with two superfluid components
can be realized in the laboratory by mixtures of two different species
at sufficiently low temperatures. Examples are $^{3}$He-$^{4}$He
mixtures, where experimental attempts towards simultaneous superfluidity
of both components have been made \cite{2002JLTP..129..531T,PhysRevB.85.134529},
and superfluid Bose-Fermi mixtures of ultra-cold atomic gases, which
have been realized recently in the laboratory as well, see Refs.~\cite{2014Sci...345.1035F,PhysRevLett.115.265303}.
We can in principle directly infer all limits, from the ultra-relativistic
to the non-relativistic limit, directly from our results, which makes this study quite versatile.
While liquid helium and ultra-cold gases are of course most conveniently
described in a non-relativistic framework, a relativistic treatment
is desirable in the astrophysical context. On the microscopic level,
this is mandatory for quark matter and for sufficiently dense nuclear
matter in the core of the star, whereas the high mass of the nucleons
allows for a non-relativistic treatment at lower densities
in the crust. Under some circumstances, for instance in rapidly rotating
neutron stars, fluid velocities can assume sizable fractions of the
speed of light, such that also on the hydrodynamic level relativistic
corrections may become important. The non-relativistic limit can always
be taken straightforwardly by increasing the mass of the constituent
fluid particles and/or by decreasing the fluid velocities, such that
the presented results can also be applied to superfluids in the laboratory. 

In all mentioned systems, a counterflow between the fluids can be
created experimentally or, in the case of neutron stars, will necessarily
occur. It is well known from plasma physics that this may lead to
certain dynamical instabilities, called "two-stream instabilities"
(or sometimes "counterflow instabilities"). Such an instability
manifests itself in a nonzero imaginary part of a sound velocity,
where the magnitude of the imaginary part determines the time scale
on which the given mode becomes unstable. In this work, the critical velocity of two-fluid systems at which the two-stream
instability sets in is computed. The two-stream instability can even occur in
a single superfluid at nonzero temperature \cite{Schmitt:2013nva},
in mixtures of two superfluids \cite{PhysRevA.63.063612,2004MNRAS.354..101A,2011PhRvA..83f3602I,2015EPJD...69..126A,PhysRevLett.115.265303},
and in a superfluid immersed in a lattice \cite{PhysRevA.64.061603}.
In each case, it is interesting to address the relation between this
dynamical instability and Landau's critical velocity, where the quasiparticle
energy of the Goldstone mode becomes negative (for studies of Landau's
critical velocity in a two-fluid system see Refs.~\cite{2003JETPL..78..574A,2006JETP..103..944A,2008JLTP..150..612K},
for a simple derivation and explanation see Chap.~\ref{sec:Landau's-Critical-Velocity}).
We shall thus also compute the onset of this energetic instability
and in particular ask the question whether an \textit{energetic} instability
is a necessary condition for the two-fluid system to become \textit{dynamically}
unstable. For the case of two superfluids, we will start from a globally
$U(1)\times U(1)$ symmetric Lagrangian for two complex scalar fields.
For this investigation, both inter-coupling terms are taken into account:
the non-derivative coupling as well as the derivative coupling, the
latter giving rise to entrainment between the two fluids. We restrict
ourselves to uniform superfluid velocities, but will allow for arbitrary
angles between the directions of the counterflow and the sound mode,
thus being able to analyze the full angular dependence of the instability.
In the zero-temperature approximation which is applied for this calculation,
the sound modes are identical to the two Goldstone modes that arise
from spontaneous breaking of the underlying global symmetry group,
and we study them through the bosonic propagator in the condensed
phase and through linearized two-fluid hydrodynamics. For both computation
methods, I will present the same calculation for the simple one fluid
case for pedagogical reasons. For the calculation of the Goldstone
mode from the fluctuation propagator, see Chap.~\ref{sec:Fluc-prop_SF}.
On top of that, we shall compare our results for the two-component
superfluid with the case where one or both of the superfluids is replaced
by a normal, ideal fluid. Even though we always neglect dissipation,
there is an important difference between a superfluid and a normal
fluid. In a superfluid, density and velocity oscillations are not
completely independent because they are both related to the phase
of the condensate. As a consequence, there is a constraint to the
hydrodynamic equations, and only longitudinal modes are allowed. We
will discuss additional solutions that occur in the presence of one
or two normal fluids and point out an interesting manifestation of
the two-stream instability in the presence of entrainment for the
case of two normal fluids, which is completely absent if at least
one of the fluids is a superfluid. 

\subsection*{External Magnetic Fields}

\begin{figure}[t]
\centering
\includegraphics[width=\textwidth]{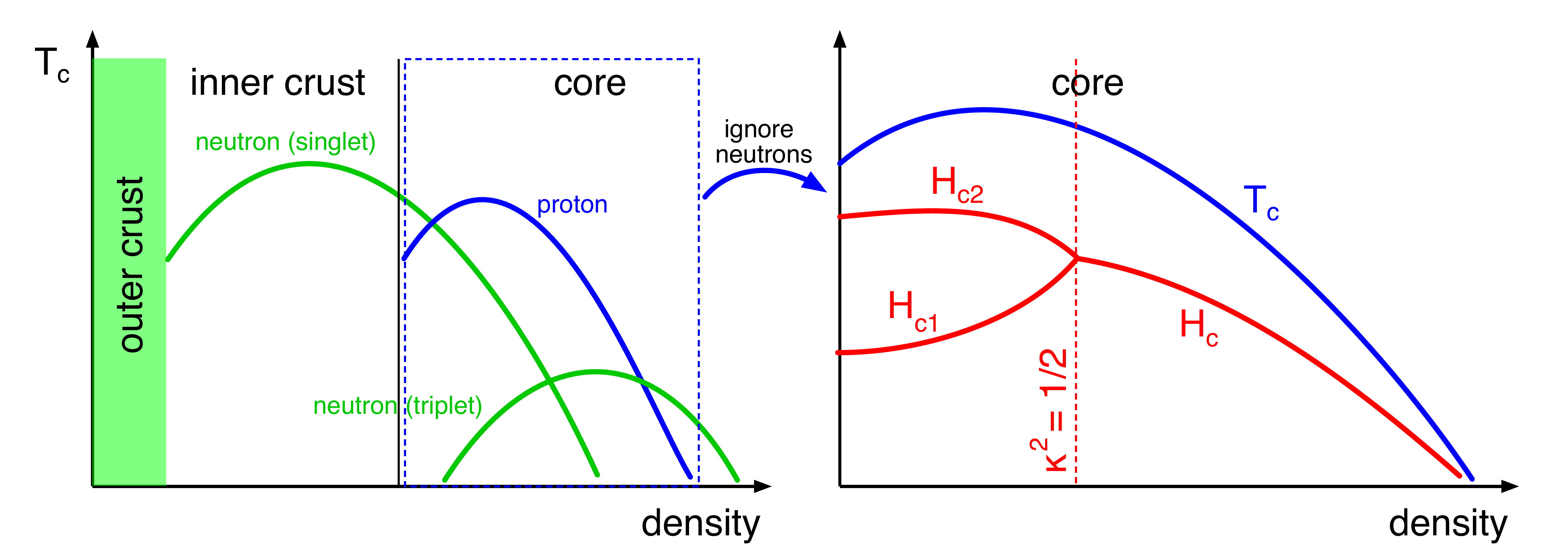}

\caption{\label{Fig:schematic_gaps}Schematic view of the interior of a compact star. (Reproduced with modifications from Refs.\ \cite{Glampedakis:2010sk,Graber:2016imq}, published in Ref.~\cite{Haber:2016ljn}) Left panel: critical temperatures for proton superconductivity 
and neutron superfluidity, where Cooper pairing can occur in the ${}^1S_0$ (singlet) or ${}^3P_2$ (triplet) channels. Outer and inner crust contain a lattice of ions. In the inner crust, a neutron superfluid is immersed in this 
lattice. In the core, where the density exceeds nuclear saturation density, neutron superfluidity is expected to coexist with a proton superconductor. Right panel: critical magnetic fields for the proton superconductor, 
as expected naively from a system without a coexisting superfluid, explained in detail for the single superconductor case earlier in this thesis. 
} 
\label{fig:schematic} 
\end{figure}

At finite external magnetic field, big emphasis will be put on the\textbf{
transition} from the \textbf{type-I to the type-II regime} of superconductivity
in the presence of a superfluid. As a simplification, we will assume that there is no counterflow between the the components when there are no flux tubes. In a flux tube, the supercurrent generates a flow between the two components, however this flow is a result of the external magnetic field we apply and is not a free thermodynamic parameter in our setup. The goal of this investigation is to study the critical magnetic fields for the flux
tube lattice in a two-component system, where the superconductor is
coupled to a superfluid. As explained, in an astrophysical context this is expected to exist in the core of neutron stars.
Microscopic calculations -- which have to be taken with care at these extreme baryon number densities -- suggest that the proton superconductor turns from type II to type I as the density increases, i.e., as we
move further towards the center of the star. In other words, a neutron star has a spatially varying $\kappa$, and the transition from type-II to type-I superconductivity might be 
realized as a function of the radius of the star \cite{Glampedakis:2010sk}. For a schematic plot describing the behavior of the neutron singlet, neutron triplet and proton gaps as a function of density in the star and the expected schematic critical magnetic fields if the coupling to the neutrons is neglected can be seen in Fig.~\ref{Fig:schematic_gaps}.The details of all presented curves are poorly known at large densities: it is not clear whether singlet and triplet neutron pairing indeed coexist in a certain density regime, and the neutron triplet gap is rather weak, leading to small critical temperatures for condensation in some regions. It is therefore conceivable that some shells in the outer core are neither superfluid nor superconducting. For a recent calculation of the neutron singlet and triplet gaps within chiral effective theory see for instance Ref.~\cite{Drischler:2016cpy}. Additionally, the density might not become large enough to indeed realize a type-I superconductor. Also, the possible transition to a quark matter phase could cut off the shown nuclear matter phases somewhere in the core. 

However, the possible interface between the type-I and type-II superconducting phase might affect the evolution of the magnetic field in the star and is thus of potential relevance to observations. Even if this interface is not realized, it is important to understand the magnetic properties of the flux tube phase in the presence of the neutron superfluid.  

The magnitudes of the energy gaps, which are varying non-monotonically along the profile of the star, reach a maximum of the order of $1\, \mathrm{MeV}$ 
at intermediate densities and are much smaller at higher densities deep in the core \cite{wambach1993quasiparticle}.
Therefore, the critical temperatures, which can be as high as $T_c \sim 10^{10}\,\mathrm{ K}$, are very small in certain regions of the star. And, the critical magnetic fields 
for proton superconductivity, at their maximum about $H_c \sim 10^{16}\, \mathrm{ G}$ -- larger than the largest measured surface fields -- become very small as well. (A very feeble 
superconducting pairing gap is neither robust against temperature nor against a magnetic field.) This motivates us to study the behavior of the superconductor at magnetic fields
close to the critical fields, and it motivates us to include temperature. 
For actual predictions in the astrophysical context, the coefficients of our effective model should be made density-dependent, using results from more microscopic calculations. In the present work we mainly focus on deriving general results and only mimic the situation of dense nuclear matter by varying our parameters in a way that is  reminiscent of the situation in a neutron star.

Some of the multicomponent systems mentioned before also consist of at least one charged component. For instance, a charged hyperon condensate in coexistence with a proton superconductor can possibly form a two-superconductor system. 
In the color-flavor locked (CFL) phase, the pairing of all quarks is usually described by a single gap function. This is different in the presence of a magnetic field, and the study of color-magnetic flux tubes \cite{Iida:2004if} or domain walls \cite{Giannakis:2003am} in a Ginzburg-Landau approach shows striking similarities with the investigated two-component system. This motivates us to use our knowledge in the last part of this thesis to invest these phenomena in quark matter in more detail. In coexistence with a kaon condensate \cite{Bedaque:2001je,Alford:2007xm}, the CFL phase  couples a color superconductor with a superfluid and represents another interesting system to which our results can be potentially applied directly.

In the context of laboratory systems, most approaches use neutral atoms, which then form superfluids. However, at least for a single
atomic species, the coupling to a "synthetic magnetic field" has been realized, including the observation of analogues of magnetic flux 
tubes \cite{2009Natur.462..628L,2011RvMP...83.1523D,2014RPPh...77l6401G}. Therefore, 
future experiments may well allow for the creation of a laboratory version of a coupled superconductor/superfluid system. 

Systems of two superconducting components have been discussed in the 
literature \cite{Carlstrom:2010wn,Brandt2011,2012arXiv1206.6786B,Wu:2015sqk} and can be realized in the form of two-band superconductors, or even in liquid metallic hydrogen \cite{babaev2004superconductor}.

The presented study is also related to so-called type-1.5 superconductivity, predicted to occur in systems with two superconducting components \cite{PhysRevB.72.180502,PhysRevLett.105.067003,Carlstrom:2010wn}.  Although in the two-component system we are investigating only one component is 
charged, we shall find very similar effects, for instance the possibility of flux tube clusters. 


From a technical standpoint, we again consider a system of two complex
scalar fields but add an abelian gauge field, with the two scalar
fields coupled to each other as before and one of them coupled to
the gauge field -- the neutral scalar field is then indirectly coupled
to the gauge field through the charged scalar field. Various aspects
of this system will be discussed, such as the effect of different
forms of the coupling between the scalar fields (density coupling
vs.~derivative coupling), effects of nonzero temperature,
and the interaction between magnetic flux tubes. I will show that
the transition region between type-I and type-II behavior is altered
drastically because of the presence of the superfluid, and one of
the main results will be the topology of the phase diagram in this
region. 

The calculations are based on a Ginzburg-Landau free energy for two
condensates, in the same way I have discussed it at length in earlier
chapters. As shown, it is possible to start from a field-theoretical
Lagrangian from which we compute the thermal fluctuations of the system.
This is necessary in order to systematically generalize the standard
temperature-dependent coefficients of the Ginzburg-Landau potential
to the situation of two coupled fields. As we do at finite superflow,
we shall work in a relativistic formalism, but the main results hold
for non-relativistic systems as well because we only consider the
static limit, where the equations of motion are identical. The coupled
equations of motion for the two condensates and the gauge field --
which yield the profile and energy of a single flux tube -- are
computed numerically using the Newton-SOR method explained in App.~\ref{App:num_methods}.
Nevertheless, where possible, simple analytical results are derived.
For instance, for the computation of the free energy of a flux tube
array, an approximation valid for sparse arrays is employed, which
is based on the numerical solution for a single flux tube. This is
sufficient to derive certain aspects of the phase structure. For a
complete study of the phase diagram, a fully numerical calculation
has to be carried out, which is beyond the scope of this thesis. The obtained results should provide guidance and
physical insights that can support and complement such a numerical
calculation in future studies. 

\subsection{Structure}
In the following section, I will introduce the general model in the more general case which includes gauged fields and all cross-coupling terms. From the derived expression, the case of two superfluids can be recovered by setting the charges to zero. After calculating the critical temperatures in the homogeneous case (i.e.\ without magnetic field or superflow), which includes a computation of the fluctuation propagator, I will present various phase diagrams. All possible thermodynamic phases of the system as a function of temperature and chemical potential, including finite superflow, are explained. In order to do so, we have to properly introduce the chemical potential for the two-component system in the presence of a derivative coupling. 
In the chapters following the general introduction of the model, we will first focus on the computation of hydrodynamic instabilities. Since these calculations are partially performed in the framework of relativistic hydrodynamics, we will shortly discuss the use of hydrodynamics as an effective theory and its limitations, and give a short introduction into the topic. Afterwards, the sound modes are computed from the hydrodynamic equations, which are derived in detail, for the three cases of two superfluids, one superfluid and one normal fluid, and two normal fluids. Since we neglect any dissipative effects, our normal fluid is an ideal fluid. These results are then used to discuss extensively various (hydrodynamic) instabilities and their relation.

Since the proton component in the core actually forms a superconductor, we will then go on to explore the behavior of the two-component, partially charged system, in an external magnetic field. The expressions for the critical magnetic fields $H_c$, $H_{c1}$, and $H_{c2}$ for our two-component system are derived. We will see that the flux tube - flux tube interaction can enable the existence of first-order phase transitions. Our numerical results, most of them in the form of phase diagrams, 
are presented consequently, together with a discussion of the type-I/type-II transition region. Finally, possible mixed phases are discussed, where flux tube clusters form within the superfluid or the Meissner phase.

\section{Two-Fluid Model}
In order to describe a system of two interacting superconductors, we start from a very similar Lagrangian as presented in Chap.~\ref{chap:sf_qft}, especially Eq.~(\ref{eq:Lag-phi4}). Of course we have to double the field content to account for the second field. Having a nucleonic system in mind, we want the two fields to interact directly with each other and not only via the coupling to a common $U(1)$ gauge field. In order to preserve the symmetry of the two fields and therefore superfluidity, we construct the interaction term in such a way that it does not break the symmetry explicitly. Since we allow the two fields to have separate masses and self-coupling constants, the symmetry of the system is given by 
\be
U(1)\times U(1) \, ,
\ee
instead of a maximally possible $SU(2)$ symmetry. Depending on the charges of the fields, the symmetry is either completely global (for $q_1=q_2=0$ and therefore no gauge field), partially local (for one non-vanishing charge) or completely local for two gauged fields. The index $i=1,2$ now denotes the first (e.g.\ the "neutron") or the second (e.g.\ the "proton") field, not the real or complex part of the complex scalar field.  The Lagrangian is then given by
\be \label{L}
{\cal L}={\cal L}_1 + {\cal L}_2 + {\cal L}_\mathrm{ int} + {\cal L}_\mathrm{ YM}\, , 
\ee
where  
\begin{subequations}
\bea 
{\cal L}_i&=&D_\mu\varphi_i(D^\mu \varphi_i)^*-m_i^2|\varphi_i|^2-\lambda_i|\varphi_i|^4  \, , \qquad i=1,2 \, , \\[2ex]
{\cal L}_\mathrm{ int} &=& 2h|\varphi_1|^2|\varphi_2|^2-\frac{g_1}{2}\Big[\varphi_1\varphi_2(D_\mu\varphi_1)^*(D^\mu\varphi_2)^*+\mathrm{ c.c.}\Big]\non[2ex]&-&\frac{g_2}{2}\Big[\varphi_1\varphi_2^*(D_\mu\varphi_1)^*D^\mu\varphi_2+\mathrm{ c.c.}\Big]  \, , \\[2ex]
 {\cal L}_\mathrm{ YM}&=&-\frac{F_{\mu\nu}F^{\mu\nu}}{16\pi} \, ,
\eea
\end{subequations}
with the covariant derivative $D_\mu \varphi_i = (\partial_\mu+iq_iA_\mu)\varphi_i$, where $A_\mu$ is the gauge field and $q_1$, $q_2$ the electric charges, with the complex scalar fields $\varphi_1$, $\varphi_2$, the mass parameters  $m_i\ge 0$, the self-coupling constants $\lambda_i>0$, and the field strength tensor $F_{\mu\nu}=\p_\mu A_\nu-\p_\nu A_\mu$. Since the Lagrangian is a real quantity, the complex conjugate of the derivative-interaction term is added and denoted by c.c.\ . We have included two types of cross-couplings between the fields: a density coupling with dimensionless coupling constant $h$, and a derivative coupling which allows for two different structures with coupling constants $g_1$ and $g_2$ of mass dimension $-2$. Due to this derivative coupling, the model is non-renormalizable and an ultra-violet cutoff is required in general. However, in a Ginzburg-Landau-like study we are only interested in an effective potential for which the only occurring momentum integral is made finite by nonzero temperature. Therefore, the non-renormalizability will not play any role in the following. This derivative coupling is responsible for the entrainment effect. Loosely speaking, entrainment between two fluids denotes the ability of one fluid to drag the other fluid with it. In a more field-theoretical definition, entrainment leads to the fact that the conserved currents of each field are not four-parallel to their respective conjugate momenta, but receive a contribution from the conjugate momentum of the other fluid. This is not automatically the case for all derivative couplings, and never occurs in systems with a simple density coupling, which we have introduced with the coupling constant $h$. I will discuss this in more detail when we derive the hydrdynamic equations from the conserved currents later on. The density coupling earns its name due to its mix of the squares of the two fields, which are proportional to the particle density of the corresponding field (in the uncoupled case). For the moment, let us compute the conserved currents which arise due to the symmetry using Noether's theorem and the procedure described in Sec.~\ref{sec:noether}: \begin{subequations}\label{j1j2a}
\bea 
j_1^\mu &=& i(\varphi_1 D^\mu\varphi_1^*-\varphi_1^*D^\mu\varphi_1) + g|\varphi_1|^2i(\varphi_2 D^\mu\varphi_2^*-\varphi_2^*D^\mu\varphi_2) \, , \\[2ex] 
j_2^\mu &=& i(\varphi_2 D^\mu\varphi_2^*-\varphi_2^*D^\mu\varphi_2) + g|\varphi_2|^2i(\varphi_1 D^\mu\varphi_1^*-\varphi_1^* D^\mu\varphi_1) \, ,
\eea  
\end{subequations}
where the difference of the two entrainment couplings is abbreviated by
\be
g \equiv \frac{g_{1}- g_{2}}{2} \, . 
\ee
Later we shall also need the sum of them,
\be
G \equiv \frac{g_{1}+ g_{2}}{2} \, .
\ee
\subsection{Chemical Potentials with Derivative Couplings}
The global part of the symmetry allows us to introduce a chemical potential for each field separately. The chemical potentials $\mu_1$ and $\mu_2$ are introduced in the usual way, they can be formally included in the Lagrangian as temporal components of the gauge fields
in the covariant derivatives, $q_iA_0\to -\mu_i$. It is important to note that they have to be included in the covariant derivatives of the coupling terms as well, which is described in Ref.~\cite{Haber:2015exa} and I will show now. For simplicity, we will neglect the charges and the gauge fields for the moment, since they do not change the nature of the proof. The main idea of the proof is to introduce 
a compact matrix notation in which the derivation then proceeds very similarly to the standard scenario without derivative coupling, presented in Chap.~\ref{chap:sf_qft}. As an aside, I shall point out a complication arising from the functional integration in the path integral over the canonical momenta, which produces a nontrivial field-dependent factor in the presence of a derivative coupling.

Let us introduce a new notation for the real and imaginary parts of the complex fields in order to reserve the index $i=1,2$ to indicate the field,
\be \label{ReIm}
\varphi_i = \frac{1}{\sqrt{2}}(\varphi'_i+i\varphi''_i) \, .
\ee
Here, we do not have to separate the condensates and the fluctuations. In this basis, the Lagrangian becomes
\bea
\L &=& \sum_{i=1,2}\left[\frac{1}{2}\partial_\mu\varphi'_i\partial^\mu\varphi'_i+\frac{1}{2}\partial_\mu\varphi''_i\partial^\mu\varphi''_i
-\frac{m_i^2}{2}(\varphi_i'^2+\varphi_i''^2)-\frac{\lambda_i}{4}(\varphi_i'^2+\varphi_i''^2)^2\right]\non[2ex]
&&+ \frac{h}{2}(\varphi_1'^2+\varphi_1''^2)(\varphi_2'^2+\varphi_2''^2)-\frac{g_1}{4}\Big[(\varphi_1'\varphi_2'-\varphi_1''\varphi_2'')(\partial_\mu\varphi_1'\partial^\mu\varphi_2'-
\partial_\mu\varphi_1''\partial^\mu\varphi_2'')\non[2ex]
&&+(\varphi_1'\varphi_2''+\varphi_1''\varphi_2')(\partial_\mu\varphi_1'\partial^\mu\varphi_2''+
\partial_\mu\varphi_1''\partial^\mu\varphi_2')\Big]-\frac{g_2}{4}\Big[(\varphi_1'\varphi_2'+\varphi_1''\varphi_2'')\non[2ex]
&&(\partial_\mu\varphi_1'\partial^\mu\varphi_2'+\partial_\mu\varphi_1''\partial^\mu\varphi_2'')+(\varphi_1'\varphi_2''-\varphi_1''\varphi_2')(\partial_\mu\varphi_1'\partial^\mu\varphi_2''-
\partial_\mu\varphi_1''\partial^\mu\varphi_2')\Big] \, .
\eea
By introducing the vector $\vec{\varphi}$ through 
\be
\vec{\varphi} = \left(\begin{array}{c} \vec{\varphi}_1 \\[1ex] \vec{\varphi}_2 \end{array}\right) \, , \qquad 
\vec{\varphi}_i = \left(\begin{array}{c} \varphi_i' \\[1ex] \varphi_i'' \end{array}\right) \, , 
\ee
we can write the Lagrangian in the compact form
\be \label{L1}
{\cal L} = \frac{1}{2}\Big[\partial_\mu\vec{\varphi}^T Q^{-1} \partial^\mu\vec{\varphi} - \vec{\varphi}^T(m^2+Y)\vec{\varphi}\Big]
\, ,
\ee
with $m^2 = \mathrm{ diag}(m_1^2,m_2^2)$. Here we have abbreviated 
\be 
Q^{-1} \equiv \left(\begin{array}{cc} 1 & A \\[1ex] A^T & 1 \end{array}\right) \, , \qquad 
Y \equiv \frac{1}{2} \left(\begin{array}{cc} \lambda_1 Y_{11} & -hY_{12} \\[1ex] -hY_{21} & \lambda_2 Y_{22} \end{array}\right) \, ,
\ee
where 
\be
A \equiv -\frac{1}{2}(GY_{12} +g\tau_2Y_{12}\tau_2) \, , \qquad \tau_2=\left(\begin{array}{cc} 0 & -1 \\ 1 & 0 \end{array}\right) \,  , \qquad 
Y_{ij} \equiv \left(\begin{array}{cc} \varphi_i'\varphi_j' & \varphi_i'\varphi_j'' \\[1ex] \varphi_i''\varphi_j' & \varphi_i''\varphi_j'' \end{array}\right) \, .
\ee
The compact form (\ref{L1}) seems to suggest that the Lagrangian is quadratic in the fields, which would allow us to perform the path integral analytically as an effectively Gaussian integral. This is of course not true, one has to keep in mind 
that the matrices $Q^{-1}$ and $Y$ contain terms quadratic in the fields too. The point was to write the Lagrangian in a form that shows the derivative terms explicitly, and absorb all remaining structure in the most compact way. This facilitates the introduction of the conjugate momenta and the chemical potentials, and also makes the structure of the Lagrangian very transparent: $Y$ contains the non-derivative 
self-couplings (diagonal terms, proportional to $\lambda_i$), and the non-derivative cross-coupling (off-diagonal terms, proportional to 
$h$), while $Q^{-1}$ contains the derivative cross-couplings (off-diagonal terms, proportional to $g_i$) and the kinetic terms (diagonal). We do not include derivative self-couplings, which would occur diagonally in $Q^{-1}$. 

The canonical momenta conjugate to the fields are defined as 
\bea
\pi_i'&=&\frac{\partial {\cal L}}{\partial(\partial_0\varphi_i')}  \, , \qquad  \pi_i''=\frac{\partial {\cal L}}{\partial(\partial_0\varphi_i'')}  \, , 
\eea
which can be compactly written as 
\be\label{pidphi}
\vec{\pi} = Q^{-1} \,\partial^0\vec{\varphi} \, , 
\ee
with the vector $\vec{\pi}$ defined analogously to $\vec{\varphi}$. Below we shall need the inverse relation $\partial^0\vec{\varphi} = Q\vec{\pi}$ with 
\be \label{Qinv}
Q = \left(\begin{array}{cc} (1-AA^T)^{-1} & -A(1-A^TA)^{-1} \\[1ex] -A^T(1-AA^T)^{-1} & (1-A^TA)^{-1} \end{array}\right) \, .
\ee
We can now exactly follow the procedure explained for the single field. The Hamiltonian is given by the Legendre transform of ${\cal L}$ with respect to the pair of variables $(\partial^0\vec{\varphi}, \vec{\pi}$),
\bea
{\cal H} &=& \vec{\pi}^T\partial^0\vec{\varphi}-{\cal L} = \frac{1}{2}\Big[\vec{\pi}^T Q\vec{\pi}+\nabla\vec{\varphi}^T\cdot Q^{-1}\nabla\vec{\varphi}
+\vec{\varphi}^T(m^2+Y)\vec{\varphi}\Big]  \, . 
\eea
Introducing two chemical potentials for the two conserved charges amounts to a shifted Hamiltonian ${\cal H}-\mu_1{\cal N}_1-\mu_2{\cal N}_2$ with the two charge densities 
given by the temporal components of the currents, ${\cal N}_i = j_i^0$. The currents (\ref{j1j2a}) can be written as 
\be
j_1^\mu = -\vec{\varphi}_1^T\tau_2\left(\partial^\mu\vec{\varphi}_1-\frac{g}{2}\tau_2Y_{12}\tau_2\partial^\mu\vec{\varphi}_2\right) \, , \qquad
j_2^\mu = -\vec{\varphi}_2^T\tau_2\left(\partial^\mu\vec{\varphi}_2-\frac{g}{2}\tau_2Y_{21}\tau_2\partial^\mu\vec{\varphi}_1\right) \, . 
\ee
Remember that we have set $q_1=q_2=0$, therefore we have to replace all covariant derivatives in the original expression of the currents by ordinary partial derivatives.
Now, using the temporal component of these expressions and inserting $\partial^0\vec{\varphi} = Q\vec{\pi}$ with $Q$ from Eq.\ (\ref{Qinv}), one computes
the remarkably simple relations
\begin{subequations}
\bea
j_1^0 &=& -\vec{\varphi}_1^T\tau_2\vec{\pi}_1 = \varphi_1'\pi_1''-\varphi_1''\pi_1'  \, ,\\[2ex]
j_2^0 &=& -\vec{\varphi}_2^T\tau_2\vec{\pi}_2 = \varphi_2'\pi_2''-\varphi_2''\pi_2'  \, ,
\eea
\end{subequations}
and thus $\mu_1j_1^0 + \mu_2j_2^0 = - \vec{\varphi}^T\mu\tau_2\vec{\pi}$ with $\mu=\mathrm{ diag}(\mu_1,\mu_2)$. 
(Note that $\tau_2$ is a matrix in the space of real and imaginary parts, while $\mu$ is a matrix in the space of fields 1 and 2.) 
The partition function is
\be
Z = \int {\cal D}\varphi_1{\cal D}\varphi_2{\cal D}\pi_1{\cal D}\pi_2\,\exp\left[-\int_X\left({\cal H}-\mu_1{\cal N}_1 - \mu_2{\cal N}_1
-\vec{\pi}^T\partial_0\vec{\varphi}\right)\right]  \, ,
\ee
with the abbreviation
\be
\int_X\equiv \int_0^{1/T} d\tau \int d^3x \, ,
\ee
where $\tau$ is the imaginary time. 
Therefore, we compute 
\bea \label{Lmu}
&&-\left({\cal H}-\mu_1j_1^0 - \mu_2j_2^0-\vec{\pi}^T\partial_0\vec{\varphi}\right) =\non[2ex]
&&-\frac{1}{2}\vec{\pi}^TQ\vec{\pi} + (\partial_0\vec{\varphi}+\mu\tau_2\vec{\varphi})^T\vec{\pi}
-\frac{1}{2}\left[\nabla\vec{\varphi}^T\cdot Q^{-1}\nabla\vec{\varphi}+\vec{\varphi}^T(m^2+Y)\vec{\varphi}\right] = \non[2ex]
&&-\frac{1}{2}\vec{\Pi}^TQ\vec{\Pi} +\frac{1}{2}\Big[(D_\mu\vec{\varphi})^T Q^{-1} D^\mu\vec{\varphi} - \vec{\varphi}^T(m^2+Y)\vec{\varphi}\Big] \, ,
\eea
with the shifted momenta $\vec{\Pi} = \vec{\pi}-Q^{-1}(\partial_0\vec{\varphi}+\mu\tau_2\vec{\varphi})$. 
The second term in the third line is the "new" Lagrangian; it is identical to the original Lagrangian (\ref{L1}), but with the derivatives replaced by the 
"covariant" derivatives $D^\mu = \partial^\mu+\delta^\mu_0 \mu\tau_2$. Consequently, the chemical potentials are effectively introduced by adding to all derivatives
in the Lagrangian the chemical potential, not just in the kinetic terms as one might expect. They can thus equivalently be introduced in the phase of the condensates.  

The integration over the shifted momenta can now easily be 
performed. In the presence of a derivative coupling, this integration induces a nontrivial, i.e., field-dependent, factor in the integrand of the 
partition function, 
\be
Z =  \int {\cal D}\varphi_1{\cal D}\varphi_2\,(\mathrm{ det}\,Q)^{-1/2}\exp\left\{\frac{1}{2}\int_X
\Big[(D_\mu\vec{\varphi})^T Q^{-1} D^\mu\vec{\varphi} - \vec{\varphi}^T(m^2+Y)\vec{\varphi}\Big]\right\} \, , 
\ee
with 
\be
\mathrm{ det}\,Q = \frac{1}{\left(1-\frac{G^2}{4}|\varphi_1|^2|\varphi_2|^2\right)\left(1-\frac{g^2}{4}|\varphi_1|^2|\varphi_2|^2\right)} \, .
\ee
If we expand $\mathrm{ det}\,Q$ around the condensates $\rho_1$, $\rho_2$ 
and only keep the lowest order contribution we can write   
\bea \label{ZQ0}
Z \simeq   [\mathrm{ det}\,Q^{(0)}]^{-1/2} \int {\cal D}\varphi_1{\cal D}\varphi_2\,\exp\left[-\frac{V}{T}U+ \frac{1}{2}\int_X {\L}^{(2)} +\ldots \right] \, ,  
\eea
where
\be \label{Q0}
\mathrm{det}\,Q^{(0)}  = \frac{1}{\left(1-\frac{G^2}{4}\rho_1^2\rho_2^2\right)\left(1-\frac{g^2}{4}\rho_1^2\rho_2^2\right)} \, ,
\ee
and where we have also employed the expansion in the exponent, keeping terms up to second order in the fluctuations. 
\section{Tree-level Potential and Phase Diagrams}
\label{sec:phases}
Whenever we perform a calculation, e.g.\ if we compute the hydrodynamic instabilities or if we analyze the interaction of the superconductor with the superfluid, we have to choose some numerical values for the thermodynamic parameters, which are the two chemical potentials, the temperature, and the two superflows $\mu_i,\ T,$ and $\nabla\psi_i$, and the coupling constants $\lambda_i,\ h,\ g$ and $G$ as well as for the masses $m_i$. However, we have to check whether the given set of parameters leads to condensation in the first place, i.e.\ if we are in the superfluid respectively superconducting phase at all. As we have learned, a single field condenses whenever $\mu_i>m_i$. In the coupled case, this becomes more complicated. Consequently, the study of the phase structure is a crucial first step. For this purpose, we compute the tree-level potential and the homogeneous solutions to the equations of motion at vanishing magnetic field. At first, we will work in the zero temperature approximation before computing finite temperature effects from the fluctuations. Once again, we separate the fluctuations from the expectation value, 
\be\label{shift}
\varphi_i = \langle\varphi_i\rangle + \mathrm{fluctuations} \, ,
\ee
and neglect the fluctuations for the moment by setting them to zero. This means that the fields are solely given by their expectation values $\langle\varphi_i\rangle$ (the "condensates"), which we parameterize once again by their modulus $\rho_i$ and their phase $\psi_i$,
\be \label{rhopsi}
\langle\varphi_i\rangle = \frac{\rho_i}{\sqrt{2}}e^{i\psi_i} \, .
\ee
Since we are interested in a superconductor coupled to a superfluid, we assume only one of the fields to be charged, say field 1, and the second to be neutral, 
\be
q_1\equiv q \, , \qquad q_2=0 \, .
\ee

Because we restrict ourselves to uniform fluid velocities, $\partial^\mu\psi_i = \mathrm{const}$, the phases $\psi_i$ depend linearly on time and space, with $\partial_0\psi_i = \mu_i$ and $\nabla\psi_i =-\mu_i \mathbf{v}_i$. As I have proven, if the chemical potentials $\mu_i$ are introduced correctly via $\mathcal{H} - \mu_1\mathcal{N}_1 - \mu_2\mathcal{N}_2$, where $\mathcal{H}$ is the Hamiltonian and $\mathcal{N}_i = j^0_i$ are the charge densities, then it is equivalent to introduce the chemical potential just like a background temporal gauge field, $\partial_0\varphi_i\to (\partial_0-i\mu_i)\varphi_i$. Consequently, we can indeed introduce the chemical potentials directly in the phase of the condensates. The same arguments can be applied to the superfluid velocity, or more precisely to the spatial components of the momentum conjugate to the current, such that we can write more generally $\mathcal{H}-p_{1\mu}j_1^\mu-p_{2\mu}j_2^\mu$, and the vectors $\mathbf{p}_i$ can be viewed as spatial components of a background gauge field. 
Moreover, we are only interested in static solutions and thus drop all time derivatives of the condensates. Then, the zero-temperature tree-level potential 
$U=-{\cal L}_{\varphi_i\to\langle \varphi_i \rangle}$ is
\bea \label{Ux}
U(\mathbf{r}) &=& \frac{(\nabla\rho_1)^2}{2}+\frac{(\nabla\rho_2)^2}{2}- \frac{p_1^2-(\nabla\psi_1-q\mathbf{A})^2-m_{1}^2}{2}\rho_1^2-
\frac{p_2^2-(\nabla\psi_2)^2-m_{2}^2}{2}\rho_2^2+\frac{\lambda_1}{4}\rho_1^4 \non[2ex]&&+\frac{\lambda_2}{4}\rho_2^4 -\frac{h+gp_{12}}{2}\rho_1^2\rho_2^2 -\frac{G}{2}\rho_1\rho_2\nabla\rho_1\cdot\nabla\rho_2+\frac{g}{2}\rho_1^2\rho_2^2(\nabla\psi_1-q\mathbf{A})\cdot\nabla\psi_2 +\frac{B^2}{8\pi} \, ,\non[0.5ex]
\hfill
\eea
where we have reduced the Yang-Mills contribution to a purely magnetic term, $\mathbf{B} = \nabla\times\mathbf{A}$, and used the abbreviations
\be
p_i^2 \equiv \partial_\mu\psi_i\partial^\mu\psi_i = \mu_i^2(1-v_i^2) \, , \qquad p_{12}^2 \equiv \partial_\mu\psi_1\partial^\mu\psi_2 = \mu_1\mu_2(1-\mathbf{v}_1\cdot\mathbf{v}_2)
\, .
\ee
As explained in Sec.~\ref{sec:noether}, $p_1$ and $p_2$ are the chemical potentials in the rest frames of the fluids. 
Here we keep the general notation $p$ introduced there, deviating slightly from the notation in Refs.\ \cite{2013PhRvD..87f5001A,Schmitt:2013nva}, where 
$\partial_\mu\psi\partial^\mu\psi$ was instead denoted by $\sigma^2$.

The potential $U$ needs to be bounded from below, otherwise there is no ground state with finite energy. This requires $\lambda_1,\lambda_2>0$ (which we shall always assume and is already true in the single field case)
and 
\be
h+gp_{12}^2<\sqrt{\lambda_1\lambda_2} \, .
\ee 
In particular, the potential is bounded for 
arbitrary negative values of $h+gp_{12}^2$. Notice that the boundedness of the potential depends on the chemical potentials, which enter $p_{12}$ (and also on the fluid velocities). This is not a problem as long as we identify the unbounded region and always work with externally fixed chemical potentials in the bounded region. Our first goal is to find the phase structure of the model within a uniform ansatz at vanishing magnetic field, i.e.\ we restrict ourselves to uniform condensates as well, $\nabla\rho_i=0$. Note that, as a direct consequence, the derivative coupling $G$ drops out of the potential at $T=0$ and plays therefore no role in the coming discussion. For the phase structure, we now need to minimize the resulting potential $U$
with respect to the condensates $\rho_1$ and $\rho_2$,
\be
0 = \frac{\partial U}{\partial\rho_1} = \frac{\partial U}{\partial\rho_2} \, . 
\ee
We now identify the four different phases that are solutions of these equations and that are distinguished by their residual symmetry group.
Let us first briefly discuss the trivial situation without coupling, $g=0$, and without any velocities, $\mathbf{v}_1=\mathbf{v}_2=0$. Bose-Einstein condensation occurs for chemical potentials larger than the mass of the bosons. Therefore, in an uncoupled system, there is no condensate for $\mu_1<m_1$, $\mu_2<m_2$, there is exactly one condensate if exactly one of the chemical potentials becomes larger than the corresponding mass, and there are 
two condensates if both chemical potentials are larger than the corresponding masses, $\mu_1>m_1$, $\mu_2>m_2$. A coupling between the two condensates can disfavor or favor coexistence of the two condensates, depending on the sign of the coupling constant. At finite coupling, the following solutions are possible:
\begin{itemize}
\item In the normal phase ("NOR"), neither the charged nor the neutral field condenses and the residual group is $U(1)\times U(1)$.
\be \label{NOR}
\rho_1=\rho_2=0 \, , \qquad U_\mathrm{ NOR}=0 \, .
\ee
\end{itemize}
The other solutions are determined by the equations 
\begin{subequations}
\bea
p_1^2-m_1^2-\lambda_1\rho_1^2 + (h+gp_{12}^2)\rho_2^2 &=& 0 \, , \\[2ex]
p_2^2-m_2^2-\lambda_2\rho_2^2 + (h+gp_{12}^2)\rho_1^2 &=& 0 \, . 
\eea
\end{subequations}
 \begin{itemize}
 
\item In the (pure) superconductor ("SC"), only the charged field forms a condensate, whereas the condensate of the other field is zero, and only one global $U(1)$ remains unbroken. The transition from the normal to the condensed phase happens when the corresponding chemical potential reaches the mass of the field.
\bea 
\rho_1^2 &=& \rho_\mathrm{ SC}^2 \equiv\frac{p_1^2-m_{1}^2}{\lambda_1}  \, , \qquad  \rho_2=0\, , \qquad U_\mathrm{SC}=-\frac{\lambda_1\rho_\mathrm{ SC}^4}{4} \, .\label{eq:rhoSC}
\eea
Whenever we neglect the charge of this field as well, especially while discussing the hydrodynamic instabilities, we equivalently call this phase superfluid one phase, "SF$_1$". 
\newpage
\item If only the second field condenses, we are in a pure superfluid phase ("SF"), while the charged fields remains uncondensed and the (local) $U(1)$ remains unbroken, 
\bea 
\rho_2^2 &=&  \rho_\mathrm{ SF}^2 \equiv \frac{p_2^2-m_{2}^2}{\lambda_2} \, , \qquad  \rho_1=0\, , \qquad U_\mathrm{ SF}=-\frac{\lambda_2\rho_\mathrm{ SF}^4}{4} \, .\label{eq:rhoSF} 
\eea
Once again, for the discussion of hydrodynamic instabilities, an index  will be added to distinguish the two superfluid phases from each other. Therefore, this phase will be equivalently  called the SF$_2$ phase.
\item In the coexistence phase ("COE"), both condensates exist simultaneously, and the symmetry is broken down to $\mathbf{1}$. Without coupling, the coexistence phase is realized if and only if 
both chemical potentials are larger than the corresponding masses. The inter-species couplings, the density coupling $h$ and the entrainment coupling $g$, favor ($h>0$ or $g>0$) or disfavor ($h<0$ or $g<0$ in our convention) the COE phase. The second derivative coupling plays no role at $T=0$ since it drops out of the tree-level potential.\footnote{In a system of neutrons and protons inside a neutron star, the results of Ref.\ \cite{Chamel:2006rc} suggest that the entrainment 
coupling $g$ is negative. This can be seen 
by rewriting the homogeneous version of the free energy (\ref{Ux}) as
\be
U = U(\mathbf{v}_1=\mathbf{v}_2=0) + \frac{\mu_1^2\rho_1^2}{2}v_1^2+\frac{\mu_2^2\rho_2^2}{2}v_2^2 +\frac{g\mu_1\mu_2\rho_1^2\rho_2^2}{2}\mathbf{v}_1\cdot\mathbf{v}_2 \, .
\ee
By comparing this expression with the non-relativistic version in Ref.\ \cite{Chamel:2006rc} and using the results from the fermionic microscopic theory therein, 
we conclude $g<0$ (with $|g|$ depending on the baryon density). In view of the results presented here it is thus an interesting question whether  
the entrainment coupling between neutrons and protons may forbid the coexistence of both condensates, in particular under circumstances (i.e., at a given temperature and 
baryon density) in which each of the condensates would be allowed to exist on its own.} The condensates and the free energy density are 
\end{itemize}
\bea \label{COE}
&&\rho_{01}^2 = \frac{\lambda_2(p_1^2-m_1^2)+(h+gp_{12}^2)(p_2^2-m_2^2)}{\lambda_1\lambda_2-(h+gp_{12}^2)^2} \, , 
\quad \rho_{02}^2 = \frac{\lambda_1(p_2^2-m_2^2)+(h+gp_{12}^2)(p_1^2-m_1^2)}{\lambda_1\lambda_2-(h+gp_{12}^2)^2} \, , \non[2ex]
&&U_\mathrm{ COE} = - \frac{\lambda_1(p_2^2-m_2^2)^2+\lambda_2(p_1^2-m_1^2)^2+2(h+gp_{12}^2)(p_1^2-m_1^2)(p_2^2-m_2^2)}{4[\lambda_1\lambda_2-(h+gp_{12}^2)^2]}\, . \label{UCOE}
\eea

The ground state is then found by determining the global minimum of $U$ depending on the free thermodynamic parameters, which include the superflow.

\begin{figure} [t]
\begin{center}
\hbox{\includegraphics[width=0.5\textwidth]{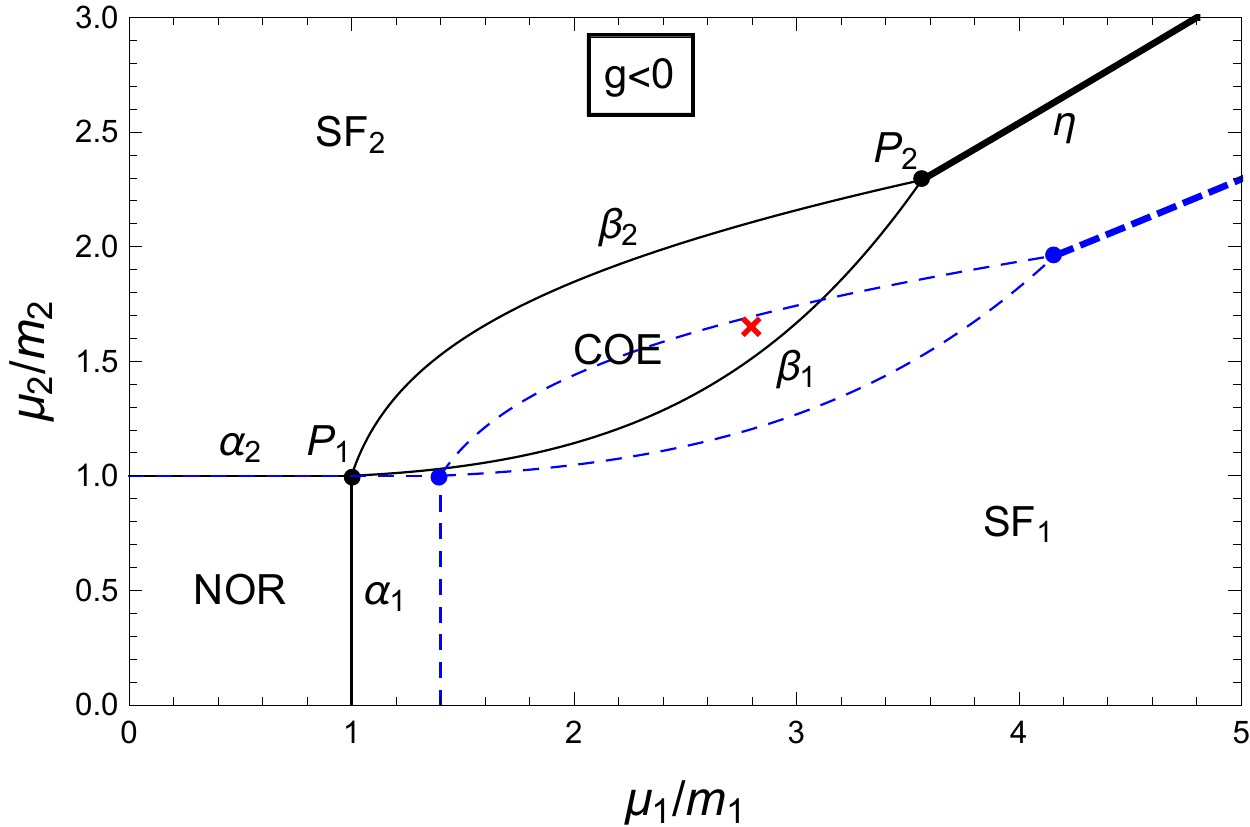}\includegraphics[width=0.5\textwidth]{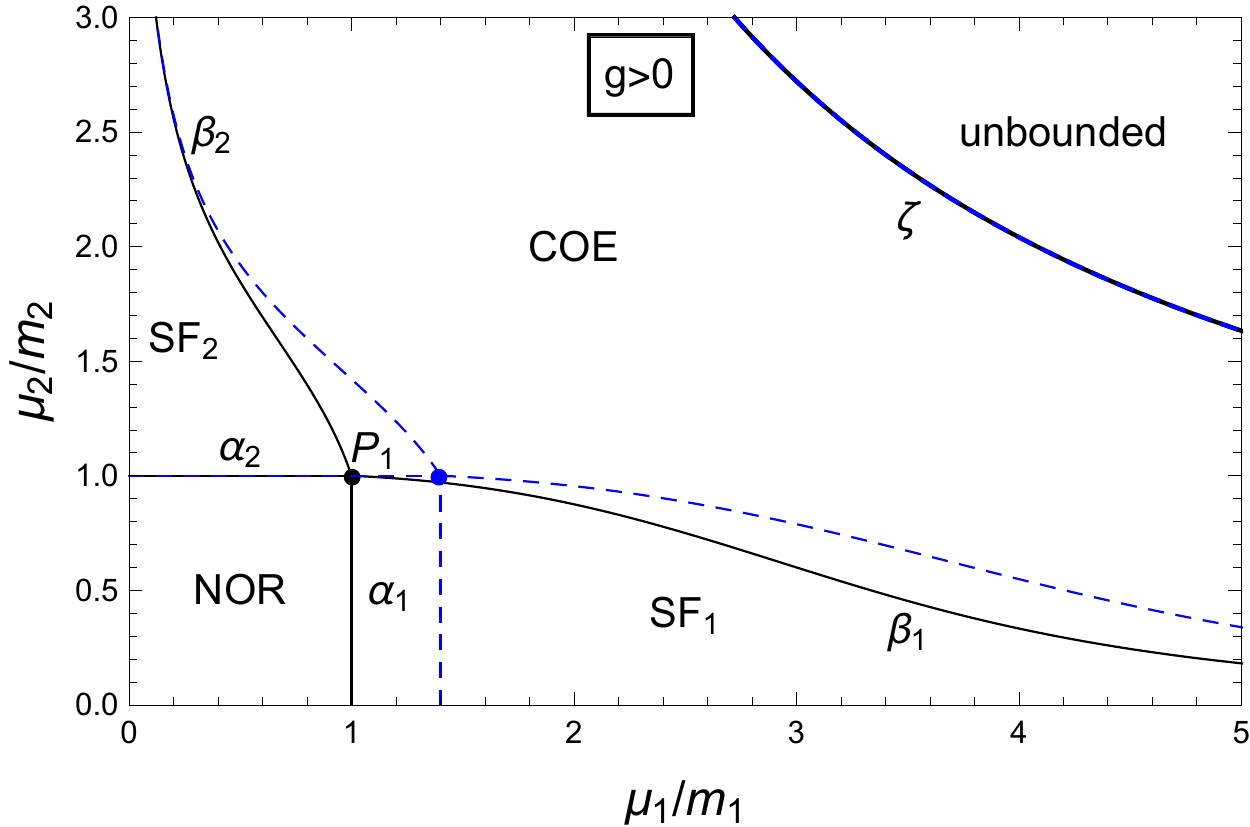}}
\caption{Phase diagrams for set 1) in the plane of the two chemical potentials $\mu_1$, $\mu_2$ for two different signs of the entrainment coupling $g$ 
with the same magnitude $|g|=0.03/(m_1 m_2)$ and vanishing non-entrainment coupling $h=0$. The mass ratio is chosen to be $m_2/m_1 = 1.5$ and the self-coupling 
constants are set to $\lambda_1=0.3$, $\lambda_2=0.2$. Solid (black) lines correspond to vanishing fluid velocities, $v_1=v_2=0$, while dashed (blue) lines 
correspond to $v_1=0.7$, $v_2=0$ (in units of the speed of light). 
Thin lines are second-order phase transitions, while the thick lines in the left panel are phase transitions of first order. In 
the upper right corner of the right panel the potential is unbounded from below. The lines and points are labeled, and their expressions are given in Table \ref{table0}. For later discussions, a point within the COE phase is chosen, marked here with a (red) cross. }
\label{fig:phases1}
\end{center}
\end{figure}

In principle, one can now draw several phase diagrams and study the influence of several parameters and coupling-constants simultaneously. In order to tame the parameters space, we are going to focus on two main combinations, summarized in Tab.~\ref{tab:parasets}: 
\begin{itemize}
\item[1)] Finite entrainment coupling $g$ and finite superflow $\mathbf{v}_i$, but vanishing density coupling, $h=0$. This particular choice of parameters will be mostly used in the discussion of the hydrodynamic instabilities. We will see that the effect of finite temperature mainly reduces to altering the chosen parameters, therefore we proceed mostly with $T=0$.
\item[2)] Finite density coupling $h$ but no entrainment coupling, $g=0$, and both fluids at rest in the homogeneous phases. This set of parameters will be used for the examination of our two-component system in an external magnetic field, where a superflow can be induced by flux tubes but is not an external thermodynamic parameter in our setup. Since finite temperatures and magnetic fields both weaken the condensate, we will compute $T>0$ effects, where the derivative coupling $G$ enters the temperature dependent condensates. 
\end{itemize}
In both cases, we have studied the influence of the other parameters as well and found no qualitative differences of the results, although the phase structure becomes slightly richer. Therefore, a simple set of parameter that features all interesting results is chosen.
\begin{table}
\begin{tabular}{|c||c|c|c|c|c|}
\hline 
set & thermod.~var.\ & phases & couplings & neglected & charges\tabularnewline
\hline 
1) & $T,\,\mu_{i},\,\nabla\psi_{i}$ & $\text{SF}_{1},\text{SF}_{2}$, COE, NOR & $\lambda_{1},\,\lambda_{2},\,g$ & $h=0$, $T=0$ & $q_{1}=q_{2}=0$\tabularnewline
\hline 
2) & $T,\,\mu_{i},\,H$ & SC, SF, COE, NOR & $\lambda_{1},\,\lambda_{2},\,h\,,G$ & $g=0$ & $q_{1}\equiv q,\,q_{2}=0$\tabularnewline
\hline 
\end{tabular}
\caption{\label{tab:parasets}Important parameter combinations for the investigation of hydrodynamic instabilities (1) and flux tubes (2). The first column shows the free external parameters, where $i=1,2$.}
\end{table}
Let us start by discussing the first set of parameters by looking at the phase diagram in Fig.~\ref{fig:phases1}.
\begin{figure} [t]
\begin{center}
\hbox{\includegraphics[width=0.5\textwidth]{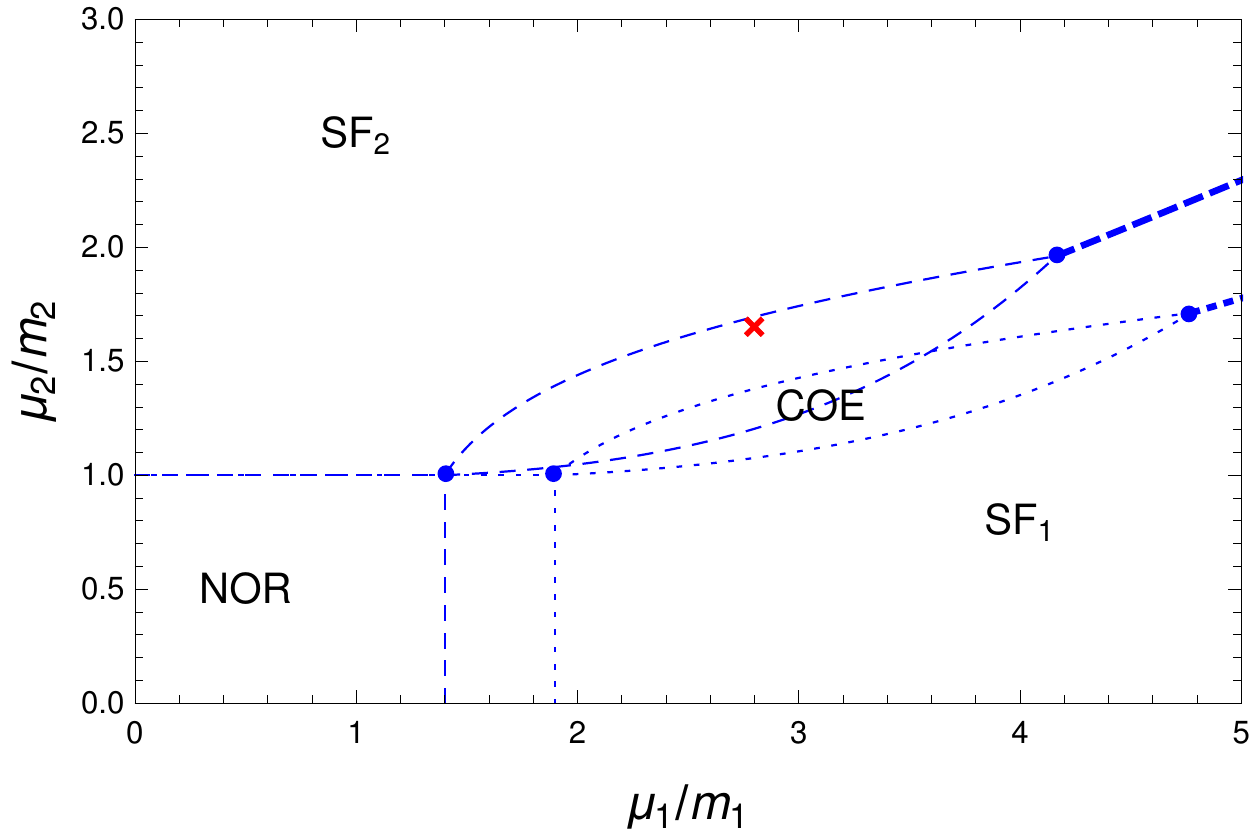}\includegraphics[width=0.5\textwidth]{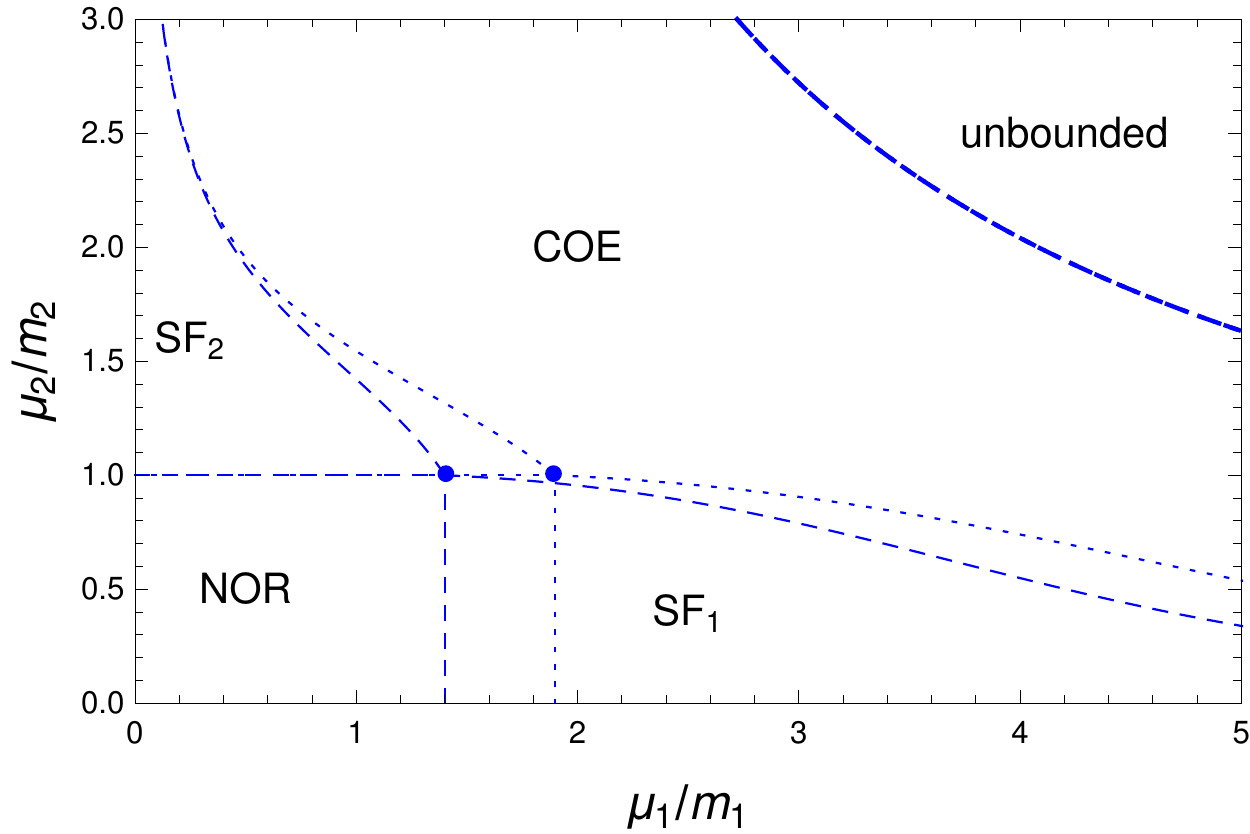}}
\caption{Same phase plot as in Fig.~\ref{fig:phases1}, but for $v_1=0.7$, $v_2=0$ (blue, dashed lines, as before) and even higher velocities $v_1=0.85$, $v_2=0$. We see that the chosen point in the $\mu_1$-$\mu_2$ plane (red cross) "leaves" the COE-phase due to the higher velocity.}
\label{fig:phases1a}
\end{center}
\end{figure}
\begin{table*}[t]
\centering
\resizebox{\columnwidth}{!}{%
\begin{tabular}{|c||c|} 
\hline
\rule[-1.5ex]{0em}{4ex} 
 $\alpha_i$ & $\displaystyle{\mu_{i} = \gamma_i m_{i}}$ \\[1ex] \hline
\rule[-1.5ex]{0em}{8ex} 
 $\;\;\beta_{1/2}\;\;$ & $\displaystyle{\mu_{2/1} = \gamma_{2/1}^2\left[\sqrt{\frac{g^2\mu_{1/2}^2}{4}\rho_{\mathrm{SF}_1/\mathrm{SF}_2}^4(1-\mathbf{v}_1\cdot\mathbf{v}_2)^2+\frac{m_{2/1}^2}{\gamma_{2/1}^2}}
-\frac{g\mu_{1/2}}{2}\rho_{\mathrm{SF}_1/\mathrm{SF}_2}^2(1-\mathbf{v}_1\cdot\mathbf{v}_2)\right]}$ \\[4ex] \hline
\rule[-1.5ex]{0em}{6ex}
 $\eta$ & $\displaystyle{\mu_1 = \gamma_1\sqrt{m_1^2+\sqrt{\lambda_1\lambda_2}\rho_{\mathrm{SF}_2}^2}}$ \\[2ex] \hline
\rule[-1.5ex]{0em}{6ex}
 $\zeta$ & $\displaystyle{\mu_2 = \frac{\sqrt{\lambda_1\lambda_2}}{g\mu_1(1-\mathbf{v}_1\cdot\mathbf{v}_2)}}$ \\[3ex] \hline\hline
\rule[-1.5ex]{0em}{4ex}
 $P_1$ & $\displaystyle{(\gamma_1m_1,\gamma_2 m_2)}$ \\[1ex] \hline
\rule[-1.5ex]{0em}{7ex}
 $P_2$ & $\;\;(\mu_1,\mu_2)$ with $\displaystyle{\mu_{1/2}^2 = \frac{\gamma_{1/2}^2}{2\sqrt{\lambda_{2/1}}}\Bigg[\sqrt{\left(\sqrt{\lambda_2}m_1^2-\sqrt{\lambda_1}m_2^2\right)^2
+\frac{4(\lambda_1\lambda_2)^{3/2}}{g^2\gamma_1^2\gamma_2^2(1-\mathbf{v}_1\cdot\mathbf{v}_2)^2}}}$\\ &
$ \pm(\sqrt{\lambda_2}m_1^2-\sqrt{\lambda_1}m_2^2)\Bigg]\;\;$ \\[3ex] \hline
\end{tabular}}
\caption{Expressions for the phase transition lines $\alpha_1$, $\alpha_2$, $\beta_1$, $\beta_2$ (second order), $\eta$ (first order), the critical line $\zeta$ for the unboundedness 
of the potential,
and the critical points $P_1$ and $P_2$ in the phase diagrams of Fig.~\ref{fig:phases1} (as in the phase diagrams, the non-entrainment coupling is set to $h=0$ for simplicity, see Tab.~\ref{tab:parasets}). 
The Lorentz factors are  $\gamma_i = 1/\sqrt{1-v_i^2}$ and the condensates in the absence of a second field $\rho_{\mathrm{SF}_i}^2=(p_i^2-m_i^2)/\lambda_i$, $i=1,2$. 
}
\label{table0}
\end{table*}

Without loss of generality we can restrict ourselves to $\mu_1,\mu_2>0$ because the chemical potentials enter 
the free energy $U$ only quadratically and through the combination $g\mu_1\mu_2$, and $g$ only enters in this combination. All phase transition lines and critical points are 
given by simple analytical expressions, see Table\ \ref{table0}. We see in the left panel that the region for the COE phase gets squeezed by the entrainment coupling. Let us examine the various phase transitions by 
following a horizontal line in the phase diagram: for $1<\mu_2/m_2\lesssim 2.3$ we start in the superfluid phase SF or $\mathrm{SF}_2$ phase depending on whether we set both charges to zero. Besides altering the nomenclature of the phases, the charge $q$ does not alter the phase structure at vanishing temperature in any way. Upon increasing $\mu_1$, the line $\beta_2$ is reached. At this line
both condensates change continuously, in particular the condensate of field 1 becomes nonzero. 
This second-order phase transition line is found from $U_{\mathrm{SF}} = U_\mathrm{COE}$. It is easy to check that it is identical to the line where $\rho_{01}$ from 
Eq.~(\ref{COE}) is zero. By further increasing $\mu_1$ we leave the COE phase through a second-order phase transition line $\beta_1$ and reach the superconducting phase SC (alternatively $\mathrm{SF}_1$). 
For $\mu_2/m_2\gtrsim 2.3$, we do not reach the COE phase by increasing $\mu_1$, although there is a region where the COE phase is allowed, i.e., where both $\rho_{01}$ and $\rho_{02}$ from 
Eq.\ (\ref{COE}) assume real nonzero values. However, the free energy of this state turns out to be larger than the free energies of the two phases with a single non-vanishing condensate.
Therefore, there is a direct first-order phase transition line $\eta$ between $\mathrm{SF}_1$ and $\mathrm{SF}_2$. For larger values of $|g|$ the region of the COE phase becomes smaller, 
and beyond a critical value the COE phase completely disappears from the phase diagram. 
This critical value is reached when the points $P_1$ and $P_2$ coincide, and it is given by
\be
g = -\frac{\sqrt{\lambda_1\lambda_2}}{m_1m_2}\frac{\sqrt{(1-v_1^2)(1-v_2^2)}}{1-\mathbf{v}_1\cdot\mathbf{v}_2} \, . 
\ee
The right panel shows the scenario where the coupling is in favor of the COE phase. In this case, there is a region for large chemical potentials 
where the tree-level potential becomes unbounded. 

In each panel the phase structure for the case of zero velocities and for the case where the velocity of superfluid 1 is nonzero is shown. The effect of the nonzero velocity 
on the $\mathrm{SF}_1$ phase is 
very simple: condensation occurs if $p_1>m_1$, i.e., if the chemical potential measured in the rest frame of the fluid
is sufficiently large. We work, however, with fixed $\mu_1$, i.e., we fix the chemical potential measured in the frame where the fluid moves with velocity $\mathbf{v}_1$. 
Therefore, the chemical potential relevant for condensation, $p_1 = \mu_1\sqrt{1-v_1^2}$, is reduced by a nonzero $\mathbf{v}_1$ through a standard Lorentz factor, and thus a nonzero 
velocity effectively disfavors condensation of the given field. The effect of the velocity on the COE phase is a bit more complicated, in this case there is no frame in which 
the velocity dependence can be eliminated. For either sign of the coupling $g$, a nonzero velocity reduces the region of the COE phase, i.e., there is a parameter region in which, 
for zero velocity, the COE phase is preferred, but which is taken over by a single-condensate phase or the NOR phase at nonzero velocity. For negative values of the entrainment 
coupling $g$, 
see left panel of Fig.~\ref{fig:phases1}, a nonzero velocity can also work in favor of the COE phase:
there are points in the single-superfluid phase $\mathrm{SF}_1$ at zero $\mathbf{v}_1$ which undergo a phase transition to the COE phase at nonzero $\mathbf{v}_1$. This can be seen in more detail in Fig.~\ref{fig:phases1a}, where the phase boundaries at even higher velocities are shown.

It has to be emphasized that the discussion in this section has been on a purely thermodynamic level, in the sense that the velocities have been treated as external 
parameters in the same way as the chemical potentials. This is possible when the velocity fields are constant in space and time. It is very natural from the point of view
of the covariant formalism since chemical potential and superfluid velocity are different components of the same four-vector, the conjugate momentum $\partial^\mu\psi$.
This "generalized" thermodynamics can be carried further to compute Landau's critical velocity from "generalized" susceptibilities 
\cite{2003JETPL..78..574A,2008JLTP..150..612K}. We shall discuss this connection later on.

\begin{figure} [t]
\begin{center}
\hbox{\includegraphics[width=0.5\textwidth]{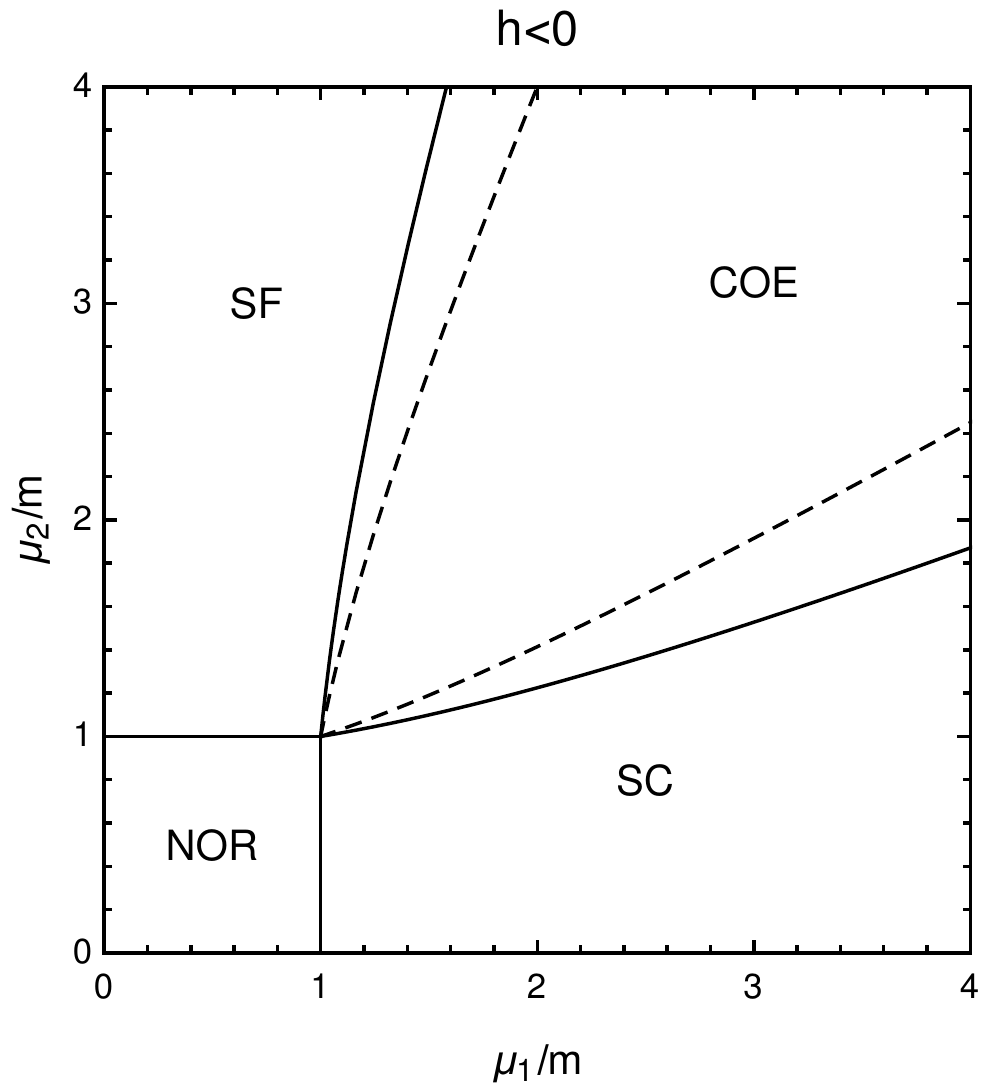}\includegraphics[width=0.5\textwidth]{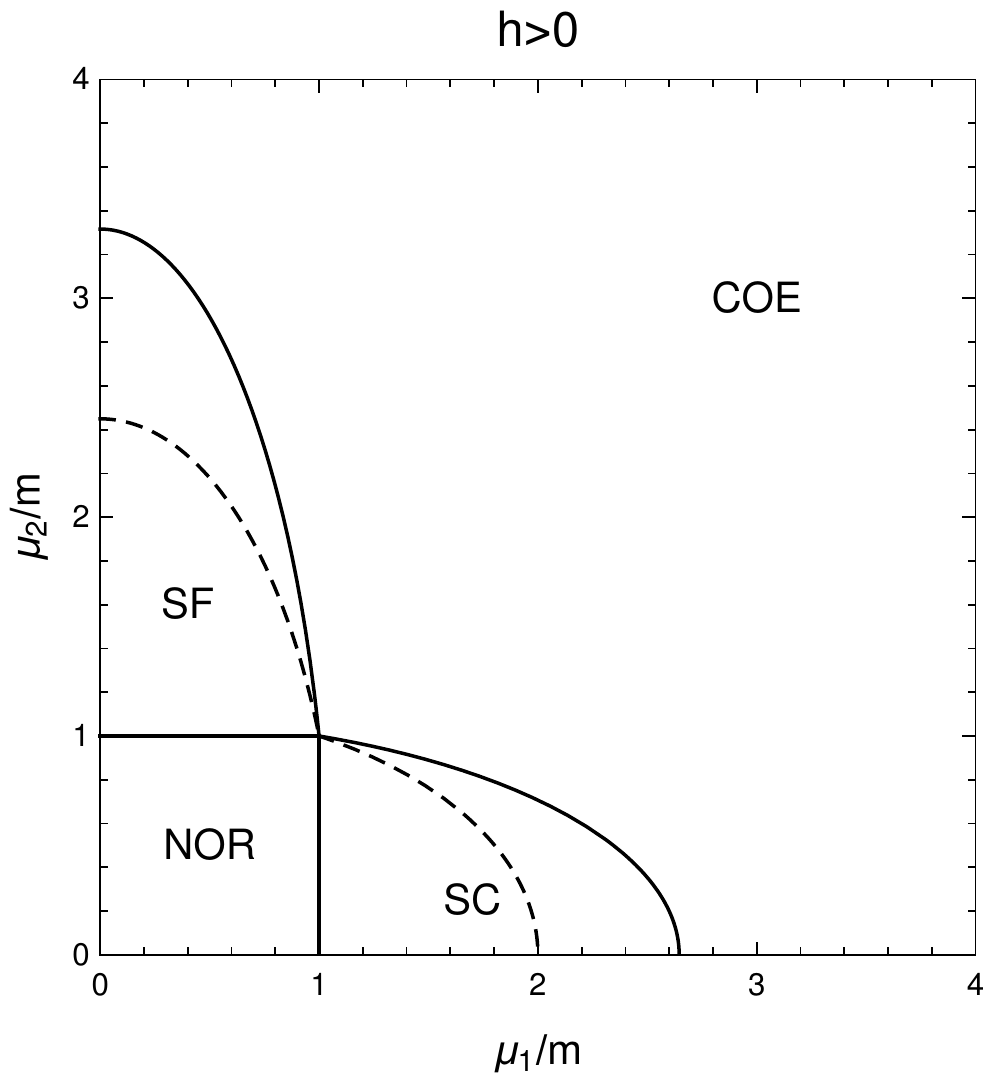}}
\caption{Phase diagrams for set 2) of Tab.~\ref{tab:parasets} in the plane of the two chemical potentials $\mu_1$, $\mu_2$ for two different signs and absolute values of the density coupling $h$ with magnitude $|h|=0.1$ (dashed lines) and $|h|=0.05$ (solid lines) and vanishing entrainment coupling, $g=0$. The masses are equal, $m_1=m_2=m$, and the self-coupling constants are $\lambda_1=0.3$, $\lambda_2=0.5$. All phase transitions are first order. }
\label{fig:phases2}
\end{center}
\end{figure}

Let us now turn to the second set of parameters and examine the phase structure for finite density coupling, shown in Fig.~\ref{fig:phases2}.
 As it has been the case with the entrainment coupling $g$, the inter-species coupling $h$ can facilitate or hinder the formation of the coexistence phase. Depending on the sign, increasing the magnitude of the coupling will further decrease ($h<0$) or increase ($h>0$) the coexistence phase. For large negative values, the coexistence phase can vanish entirely, which happens at a value of 
 \be
 h=-\sqrt{\lambda_1\lambda_2} \, .
 \ee
The phase transition lines are slightly simpler in this setup and are given by $\mu_2$ as a function of $\mu_1$:
\bea
\mathrm{SF}\to\mathrm{COE}:\qquad\mu_2&=&\sqrt{m_2^2+h/\lambda_1\left(m_1^2-\mu_1^2\right)}\, ,\\
\mathrm{COE}\to\mathrm{SC}:\qquad\mu_2&=&\sqrt{m_2^2+\lambda_2/h\left(m_1^2-\mu_1^2\right)}\, ,  
\eea
where the first equation denotes the transition from the SF-phase to the COE phase, and the second one the transition from the COE-phase to the SC-phase.
In order to include a finite temperature, we have to compute the fluctuation propagator and the excitation energies, which I am going to present in the next section.
\section{Introducing Temperature}
We intend to include temperature $T$ into the potential (\ref{Ux}) in an effective way. In Ginzburg-Landau models this is usually done by introducing temperature dependent coefficients, with a $T$-dependence that is strictly valid only close to the critical temperature. In the two-component system, the form of these coefficients is not obvious because we have two fields and hence (at least) two critical temperatures. We thus proceed by introducing temperature in the underlying field theory and derive an effective potential. This will be done in a high-temperature approximation, assuming the condensates to be uniform, and without background magnetic field. 
In order to calculate finite temperature effects, we therefore have to compute the fluctuation propagator at tree-level as it has been shown in Sec.~\ref{sec:Fluc-prop_SF} for a single field. As argued, in the case of two superfluids, the $T=0$ approximation for a neutron star environment is quite reasonable. However, a high magnetic field can weaken the \textit{superconducting} gap, leaving the superconductor "vulnerable" to even small temperatures. Therefore, we are going to focus on the superconductor-superfluid setup during this discussion, which means that we set the superflow in the homogeneous phases to zero and simply work with the chemical potentials $\mu_i$, which replace the frame dependent $p_i$. Taking into account the gauged nature of the charged field is essential, since the excitations are quite different due to the "eating" of the Goldstone mode, as discussed in Sec.~\ref{sec:massive_boson}. Later on we will need the fluctuation propagator for the two superfluid setup as well, which is very similar from a technical point of view.
 
Neglecting zero-temperature quantum corrections, the one-loop potential is 
\be
\label{eq:LOfiniteT}
\Omega(\mu_1,\mu_2,T) = U+T\sum_{i=1}^{6}\int\frac{d^3k}{(2\pi)^3}\ln\big(1-e^{-\epsilon_{ki}/T}\big) \, ,
\ee
where the sum is taken now over all six quasiparticle excitations $\epsilon_{ki}$ instead of just two in the case of a single scalar field, where we have derived the latter formula, see Eq.~(\ref{eq:T-potential}).  Without condensation, each of the complex scalar fields yields 2 excitations 
(both massive if $m_i>0$), corresponding to particle and anti-particle excitations, while the gauge field has two massless excitations, corresponding to the two possible polarizations of massless photons. These are 6 modes in total. 
In the coexistence phase both scalar fields condense. As a consequence, there is one Goldstone mode from the 
neutral field and one would-be Goldstone boson from the charged field, which becomes a third mode of the now massive gauge field. Together with the two massive modes from the scalar fields and the two original modes of the gauge field -- which are now massive as well -- these are again 6 modes. 
The excitations $\epsilon_{ki}$ are computed from the tree-level propagator. Their expressions are very complicated, but for the high-$T$ approximation we only need their behavior 
at large momenta. All details of this calculation are deferred to appendix \ref{app:excitations}. From a field-theoretical perspective our high-$T$ approximation is very crude, and for a
quantitative evaluation of the model for all temperatures more sophisticated methods are needed, such as the two-particle irreducible formalism \cite{Alford:2013koa} or functional 
renormalization group techniques \cite{Fejos:2016wza}. These methods are beyond the scope of the present work because, firstly, if applied to our present context of magnetic flux tubes and their interactions, they would render the calculation much more complicated and purely numerical methods would be required. Secondly, having in mind the application of our model to nuclear matter, the next step towards a more sophisticated description should probably be to employ a fermionic model, rather than improving the bosonic one (note for instance that our bosonic system has well-defined quasiparticle excitations for all energies, while a fermionic one has 
a continuous spectral density for energies larger than twice the energy gap from Cooper pairing). 

\begin{figure} [t]
\begin{center}
\hbox{\includegraphics[width=0.5\textwidth]{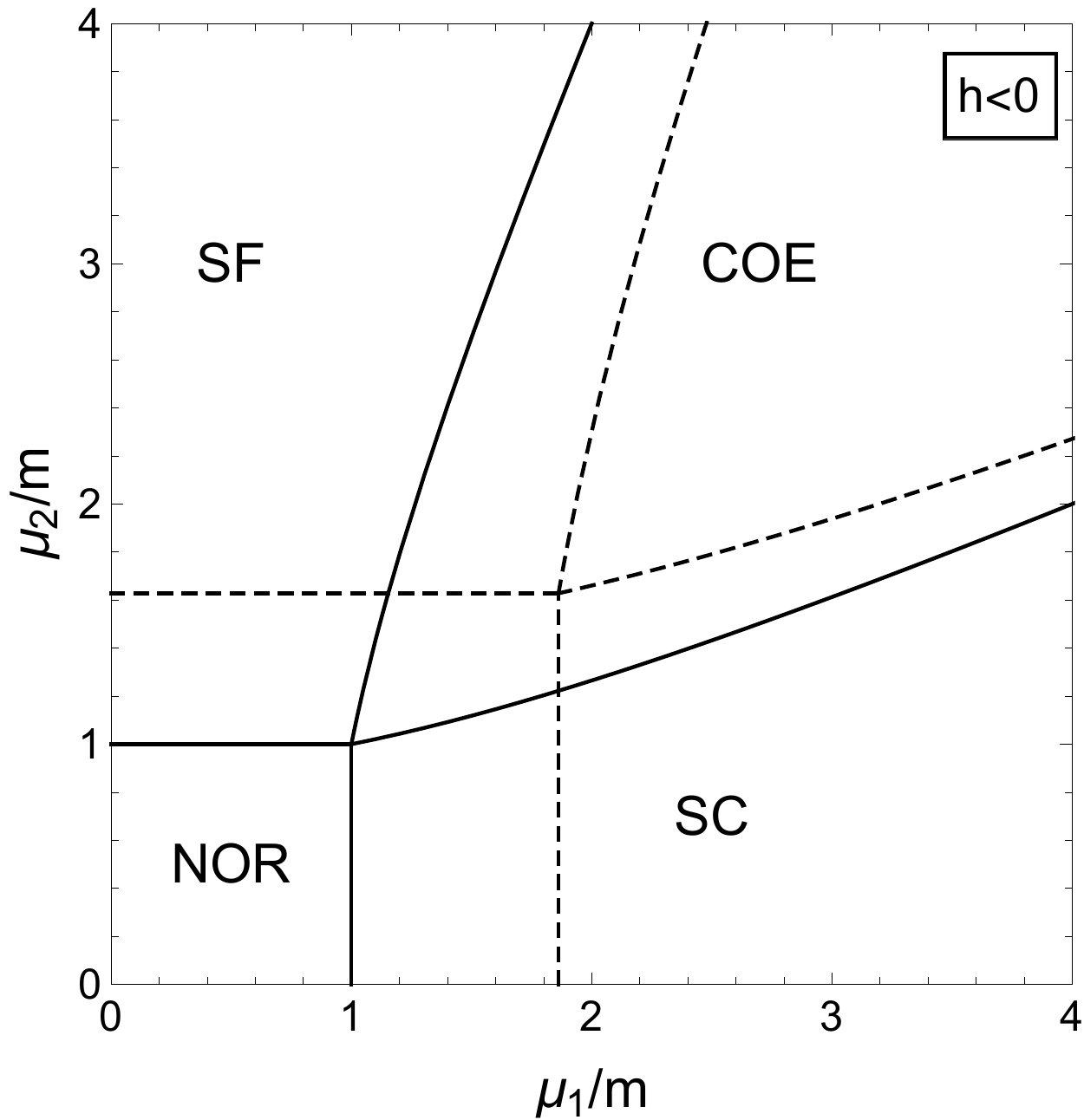}\includegraphics[width=0.5\textwidth]{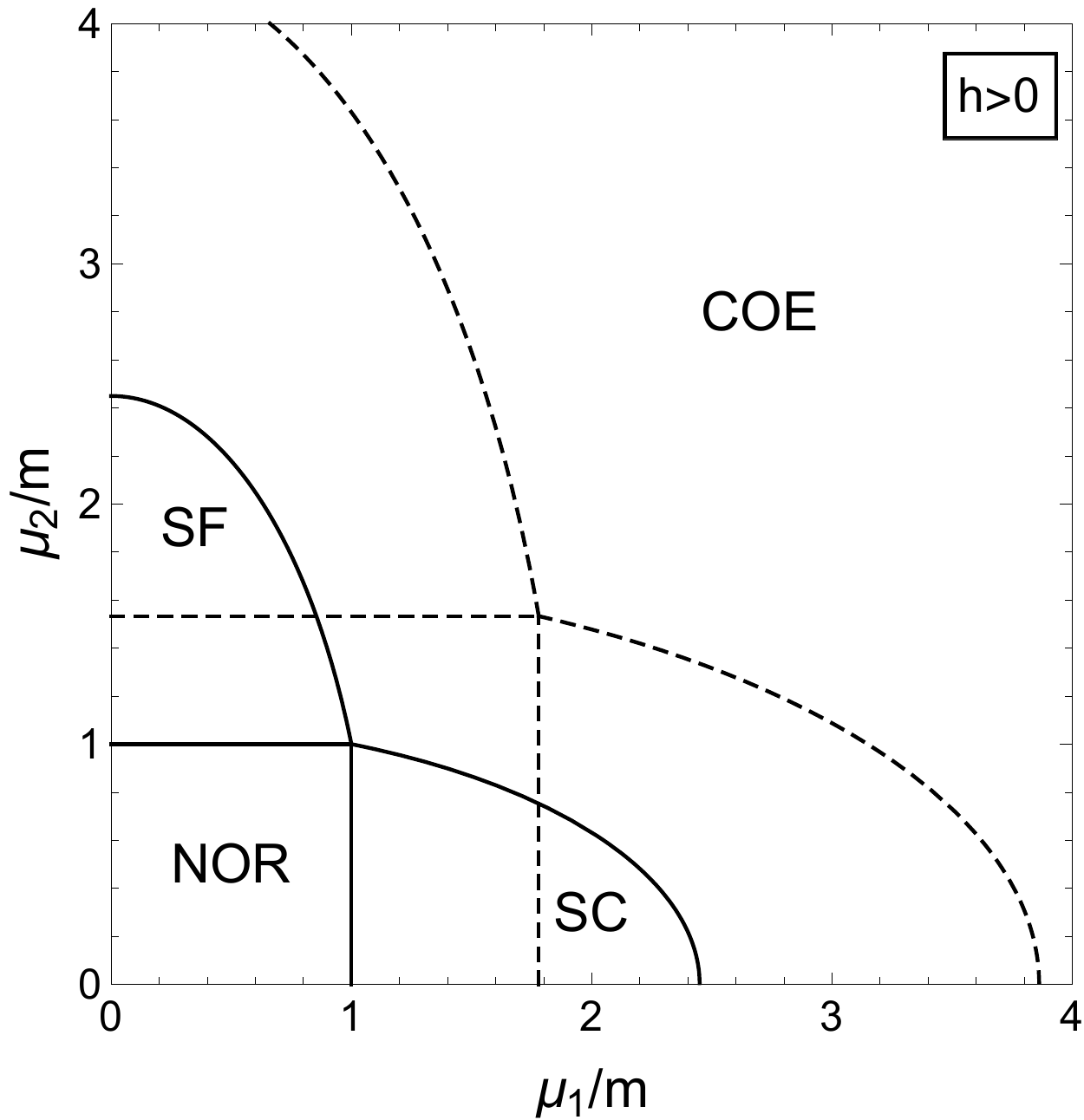}}
\caption{Phase diagrams in the plane of the two chemical potentials $\mu_1$, $\mu_2$ with same parameters as in Fig.~\ref{fig:phases2}for $|h|=0.1$ and $T=0$ (solid lines) and $T=3m$ (dashed lines). Additionally, we set $G=0$, which would alter the effective value of $h$.}
\label{fig:phases2T}
\end{center}
\end{figure}

We also simplify the result by only keeping the leading order contribution from the derivative coupling $G$.  As a result, all temperature corrections can be absorbed into thermal masses and a thermal density coupling, and we can work with the effective potential
\bea \label{UxT}
U(\mathbf{r}) &\simeq& \frac{(\nabla\rho_1)^2}{2}+\frac{(\nabla\rho_2)^2}{2}- \frac{\mu_1^2-(\nabla\psi_1-q\mathbf{A})^2-m_{1,T}^2}{2}\rho_1^2-
\frac{\mu_2^2-(\nabla\psi_2)^2-m_{2,T}^2}{2}\rho_2^2\non[2ex]&&+\frac{\lambda_1}{4}\rho_1^4 +\frac{\lambda_2}{4}\rho_2^4 
-\frac{h_T}{2}\rho_1^2\rho_2^2 -\frac{G}{2}\rho_1\rho_2\nabla\rho_1\cdot\nabla\rho_2 +\frac{B^2}{8\pi} \, ,
\eea
where 
 \begin{subequations}\label{thermal_mass}
\bea 
m_{1,T}^2 &=& m_1^2+\frac{2\lambda_1-h+6\pi q^2}{6}T^2\, ,\\[2ex]
m_{2,T}^2 &=& m_2^2+\frac{2\lambda_2-h}{6}T^2\, , \\[2ex]
h_T&=& h\left(1+\frac{GT^2}{6}\right) \, .
 \eea
 \end{subequations}
For the following, we can thus simply take Eqs.\ (\ref{NOR}) -- (\ref{COE}) and replace the masses and the density coupling by their thermal generalizations. 
From these results we are able to compute the critical temperatures. This can be done by comparing the now temperature dependent free energies of the various phases and determine the value of $T$ at which they are equal. As expected, the temperature dependent condensates vanish at these critical values, indicating a second-order phase transition. Therefore, it is sufficient to solve for the temperature at which the condensates go to zero. In the presence of a derivative coupling $G$ the resulting expressions are very lengthy and not very insightful. Therefore, we set $G=0$ for the moment, such that the only effect of temperature is a modification of the 
masses $m_1$ and $m_2$, not $h$. Inserting the thermal masses into Eqs.\ (\ref{COE}), we compute the
$T$-dependent condensates by minimization,
\be
\rho_{0i}^2(T)=\rho_{0i}^2(T=0)\left(1-\frac{T^2}{T_{ci}^2}\right) \, , 
\ee
where the critical temperatures $T_{c1}$ and $T_{c2}$ indicate the phase transitions to the SF and SC phases, 
 \begin{subequations} \label{TcCOE}
 \bea
T_{c1}^2&=&\frac{6(\lambda_1\lambda_2-h^2)}{\lambda_2(2\lambda_1+h+6\pi q^2)-h^2} \rho_{01}^2(T=0) \, , \label{Tc1} \\[2ex]
T_{c2}^2&=&\frac{6(\lambda_1\lambda_2-h^2)}{\lambda_1(2\lambda_2+h)-h(h-6\pi q^2)} \rho_{02}^2(T=0) \, .\label{Tc2}
\eea
\end{subequations}
In the limit $h=0$,  Eq.\ (\ref{Tc1}) reduces to the well-known result for a single charged field, see for instance Eq.\ (4.24) in Ref.\ \cite{Kapusta:1981aa} (in this 
reference Heaviside-Lorentz units are used, i.e., our charge $q$ has to be divided by $\sqrt{4\pi}$ to match that result exactly). If we set $h=0$ in Eq.\ (\ref{Tc2}) 
the result becomes independent of the charge $q$, as it should be because field 2 is neutral and couples to the gauge field only indirectly through field 1.
For the transition from the SC to the NOR and the SF to the NOR phase, we find
\be\label{eq:TcSCSF}
T_c(\mathrm{SC}\to\mathrm{NOR})=\frac{\sqrt{6}\sqrt{m_1^2-\mu_1^2}}{\sqrt{h-2\left(3\pi q^2+\lambda_1\right)}}\, ,\qquad T_c(\mathrm{SF}\to\mathrm{NOR})=\frac{\sqrt{6}\sqrt{m_2^2-\mu_2^2}}{\sqrt{h-2\lambda_2}}\, .
\ee
The critical temperatures (\ref{TcCOE}) and their more complicated versions with nonzero $G$ are interesting in themselves. For instance, they can be used 
to analyze systematically in which regions of parameter space the COE phase is superseded by the SF phase at high temperature (i.e., the charged condensate 
melts first, $T_{c1}<T_{c2}$) or by the SC phase (i.e., the neutral condensate melts first, $T_{c2}<T_{c1}$). Or, they can be used to 
identify regions in the parameter space where one or both critical temperatures squared become negative, indicating that one or both condensates "refuse" to melt.
This interesting observation -- although it may 
be an artifact of our approximation -- has been pointed out previously in the literature, see for instance appendix C in Ref.\ \cite{Alford:2007qa} and references therein. 
Here we shall not further analyze the critical temperatures. None of the parameter sets used in the following 
show this unusual behavior, i.e., we choose parameters such that $T_{c1}$ and $T_{c2}$ exist. 
The result of our finite temperature calculation can be seen for instance in Fig.~\ref{fig:phases2T}.

\begin{figure} [t]
\begin{center}
\hbox{\includegraphics[width=0.5\textwidth]{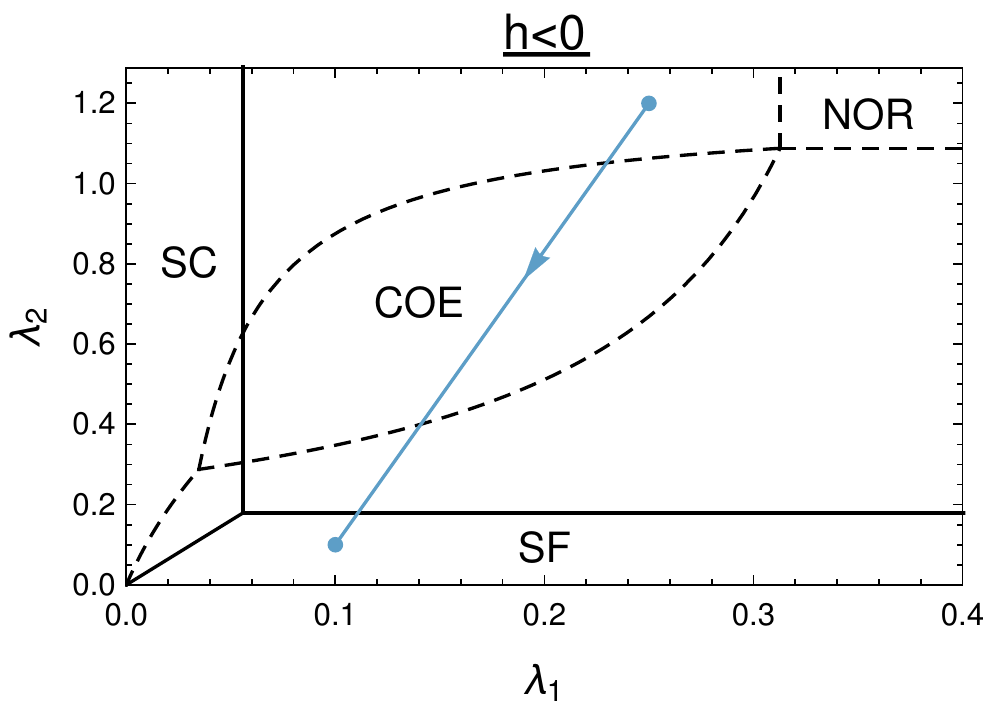}\includegraphics[width=0.5\textwidth]{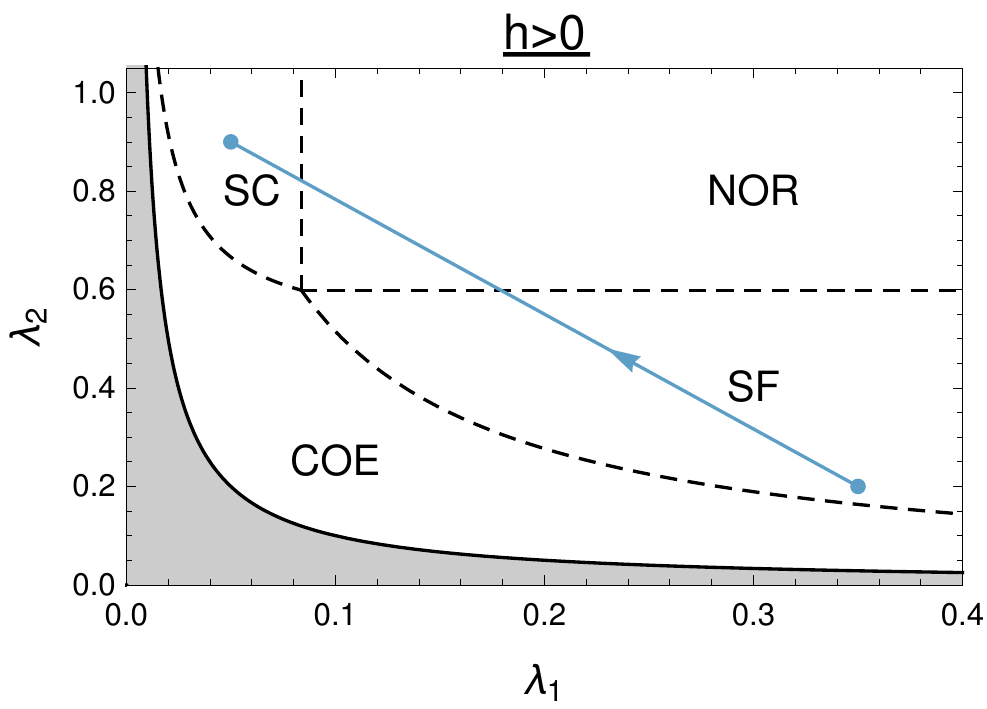}}
\caption{ Phases in the $\lambda_1$-$\lambda_2$-plane at $T=0$ [solid (black) curves] and $T>0$ [dashed (black) curves], at vanishing magnetic
 field and superflow. The shaded region in the right panel has to be excluded because there the potential is unbounded from below, $h>\sqrt{\lambda_1\lambda_2}$.  While all derivative couplings are set to zero, $g=G=0$, the results are shown for two different signs of $h=\pm|0.1|$. The other parameters are $m_1=m_2\equiv m$, 
 $\mu_1=1.5m$, $\mu_2=1.8m$ for both panels, and $T=2.43m$ (left), while $T=3.5m$ (right). The (blue) paths in both panels 
define the effective parameter $\alpha$, see Eq.\ (\ref{alpha}),  with
  $\vec{\lambda}_\mathrm{ start}=(0.25,1.2)$, $\vec{\lambda}_\mathrm{ end}=(0.1,0.1)$ for $h<0$ and 
 $\vec{\lambda}_\mathrm{ start}=(0.35,0.2)$, $\vec{\lambda}_\mathrm{ end}=(0.05,0.9)$ for $h>0$.
\label{fig:l1l2}}
\end{center}
\end{figure}

\begin{figure} [t]
\begin{center}
\hbox{\includegraphics[width=0.5\textwidth]{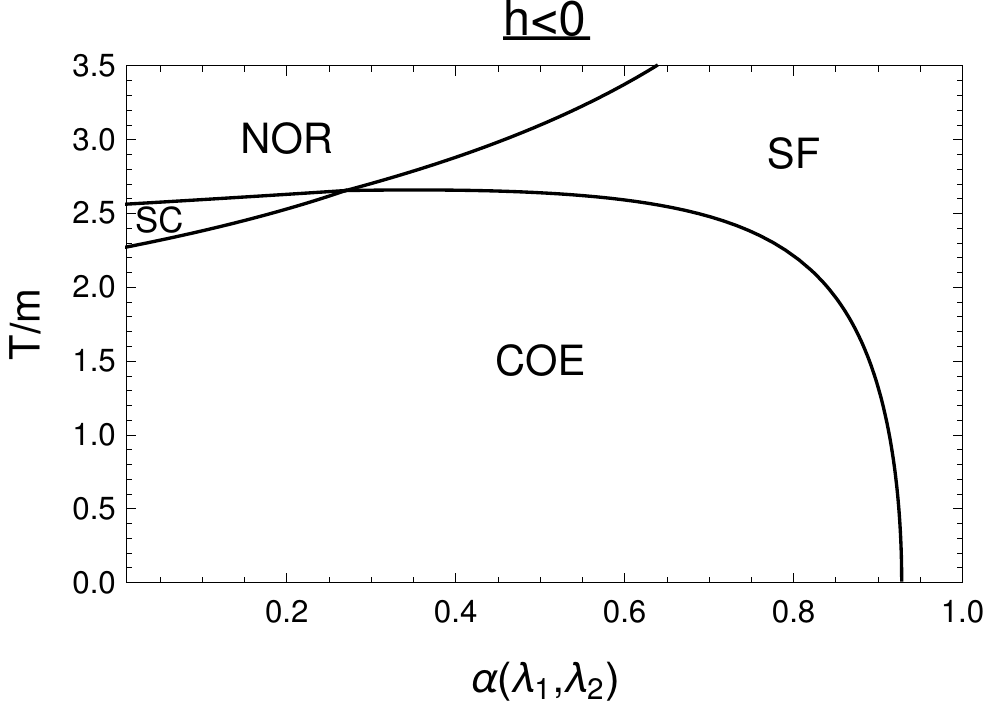}\includegraphics[width=0.5\textwidth]{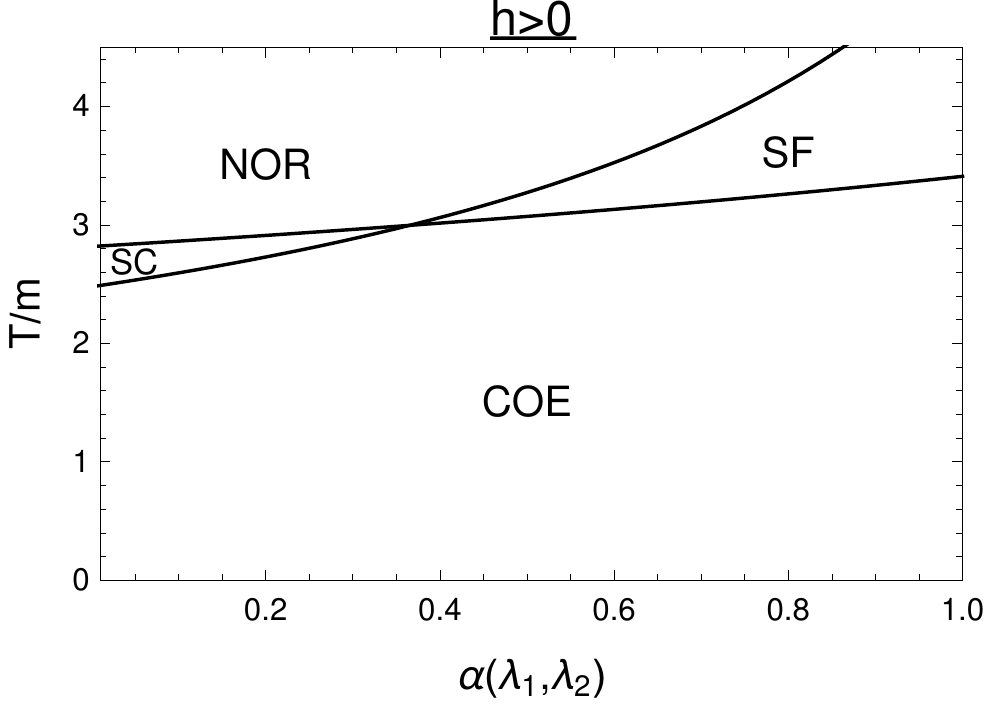}}
\caption{ Critical temperatures as a function of the parameter $\alpha$ as defined in Fig.~\ref{fig:l1l2}. All parameters are taken from the same plot. In the left panel, we find that there is an upper limit for $\alpha$ beyond which the COE phase cannot exist, which is in agreement with the definition of the parameter in Fig.~\ref{fig:l1l2}, where the end point lies in the SF phase for $T=0$. \label{fig:Tc}}
\end{center}
\end{figure}

There, the phase diagram in the plane of the chemical potentials is shown identically to Fig.~\ref{fig:phases2}. Since the masses are effectively increased by the final temperature, we see a delayed onset of all condensed phases compared to the solid $T=0$ lines. 
Alternatively, one can draw the phase diagram as a function of the self-coupling constants $\lambda_1$ and $\lambda_2$. A lot of effects that we are going to discuss depend on these couplings, for instance the critical temperatures which we want to plot here. In a first attempt to "tame" the parameter space, we parametrize a simple path in the $\lambda_1-\lambda_2$ plane by $\alpha\in [0,1]$, which is defined by 
\bea \label{alpha}
\vec{\lambda}  = \vec{\lambda}_\mathrm{start}+\alpha(\vec{\lambda}_\mathrm{end}-\vec{\lambda}_\mathrm{ start})  \, ,
\eea
with $\vec{\lambda}=(\lambda_1,\lambda_2)$. This parameterization is reminiscent of how these parameters are expected to vary as a function of increasing density in a neutron star. This effective parameter will be used especially for the discussion of critical magnetic fields and discussed in more detail later on. For the moment, it allows us to vary at least two parameters simultaneously while keeping the others fixed. A phase diagram in the plane of $\lambda_1$-$\lambda_2$ is shown in Fig.~\ref{fig:l1l2}, where the definition of the parameter $\alpha$ is visualized as well. The results are shown at vanishing and finite temperature.
Interestingly, the normal region cannot be reached by increasing the self-couplings at finite temperature for either sign of the inter-species coupling, only at finite temperatures the NOR region appears. For the NOR phase to be preferred (which is chosen to have free energy equals to zero), the COE phase has to become positive. The only positive terms in the potential (\ref{UxT}) dependent on $\lambda_1$ and $\lambda_2$ are the quartic self-coupling terms, which are, in leading order at $T=0$, still proportional to $\lambda_i^{-1}$. At finite temperature, the additional $\lambda_i-$dependence of the thermal masses then drive the transition to the normal region. This argument is independent of the sign of $h$.
For positive values of $h$, which strengthen the COE phase, the system stays in the COE phase at $T=0$ as long as the potential is bounded from below. The unbounded region where $h>\sqrt{\lambda_1\lambda_2}$, is shaded in grey. This of course requires that the other parameters are chosen in such a way that the system is in the COE phase in the first place. At finite temperatures, once again the thermal masses lead to a more complex phase structure, by not only enabling the transition to the normal but also to the two single-condensate phases.
Finally, we plot the critical temperatures given by Eqs.~(\ref{eq:TcSCSF}) and Eqs.~(\ref{TcCOE}) as a function of the newly introduced effective parameter $\alpha$ in Fig.~\ref{fig:Tc}.

\part{Instabilities in Two-Component (Super)fluids}
After introducing the model and its phase structure in the last part, we will now focus on the derivation of hydrodynamic instabilities of the two-fluid system. In order to do so, we employ a field-theoretical approach for the system of two superfluids, where we compute the dispersion relation of the Goldstone mode, from which we can deduce the speed of sound. The sound modes indicate various instabilities, like the two-stream and the Landau instability. Additionally, we take a hydrodynamic approach to compute the speed of sound of said system, as well as of a mix of an ideal (i.e.\ dissipationless) fluid and a superfluid or two ideal fluids.

For centuries, \textbf{hydrodynamics} was understood as an application of Newton's laws to a continuous deformable medium \cite{Schaefer:2014awa}. Nowadays, we have a much more general understanding of hydrodynamics as an effective theory of non-equilibrium, long-wavelength, low-frequency dynamics of many-body systems. These many body systems include quarks and gluons in the QGP, as well as the sea of neutrons and protons in a compact star, or a color superconductor which might exist in the core of such a star.

Hydrodynamics is a quite useful description since it is a coarse-grained theory. In many-body systems, the shear amount of particles makes it nearly impossible to follow and compute the microscopic details of the system. However, when the system contains sufficiently many particles, thermodynamic concepts start to apply, like the notion of the static equilibrium. Such systems can be described by a few quantities, like temperature, collective velocity (e.g.\ the superflow) and chemical potential, which determine the thermodynamic quantities such as the energy density, the pressure and so on \cite{Jeon:2015dfa}. Hydrodynamics now allows us to study deviations from this equilibrium. The validity of hydrodynamics is based on the existence of two different time or length scales. On the one hand, there is the microscopic time scale which describes how long it takes for a local disturbance to relax. This can equivalently  be described by the scattering length or mean free path $l$, since particles with short mean free paths scatter more often and can therefore equilibrate faster. The second time scale is the macroscopic time scale of the observation, or in terms of lengths the size of the observed system, $L$. For a hydrodynamic description, we require 
\be 
l \ll L \, .
\ee
The ratio of these two numbers is called the Knudsen number Kn, so equivalently we could demand
\be
\mathrm{Kn} \equiv \frac{l}{L} \ll 1 \, .
\ee
Only then it is meaningful to talk about fluid elements, which are macroscopically small but large on a microscopic scale such that each fluid element is in local equilibrium. If one considers such a fluid cell around a position $\mathbf{x}$ at a given time $t$, then it becomes meaningful to describe the system with a scalar temperature field $T(\mathbf{x},t)$, and a velocity vector field $v^\mu(\mathbf{x},t)$ as well as with a local chemical potential $\mu(\mathbf{x},t)$. These quantities are related to the conjugate variables of the conserved quantities like the energy, the momentum and the charge density. Consequently, we will use their conservation equations to study the dynamics of the fluid with input from the underlying microscopic theory \cite{Jeon:2015dfa}.

The existence of a second particle species does not automatically allow for  a two-fluid description. We now have to deal with four length scales, the system size $L$, the two separate mean free paths $l_1$ and $l_2$, and the inter-species mean free path $l_{12}$. On the one hand, we have to require that both mean free paths are shorter than the system size in order to enable a hydrodynamic description. On the other hand, we need a clear separation of the two individual mean free paths and $l_{12}$,
\be 
l_1,\, l_2 \ll L \qquad \mathrm{and} \qquad l_1,~l_2 \ll l_{12} \lesssim L \, .
\ee  
Otherwise, the two particle species interact microscopically with each other and one should instead describe them as a single fluid consisting of two particle species. The inter-species mean free path can be of the order of the system size or smaller, but not larger, because then the two fluids would essentially not interact.
For a long introduction to relativistic hydrodynamics see the reviews in \cite{Kovtun:2012rj,Schaefer:2014awa,Jeon:2015dfa}, Chap.~2 of Ref.~\cite{superbook}, and for a review on its use in nuclear collisions see for instance Ref.~\cite{Romatschke:2017ejr}. For a connection of hydrodynamics and field theory see Refs.~\cite{2013PhRvD..87f5001A, Stetina:2015exa}.
In the following chapter I will give a short introduction into single- and two-fluid hydrodynamics and introduce the necessary quantities for the discussion of instabilities in a two-fluid system. Consequently, the sound modes of a single fluid are computed, before we turn to the more complicate two-fluid systems. Additionally, the dispersion relations of the Goldstone modes are computed in order to study energetic and dynamic instabilities in more detail.
\chapter{Sound Modes from Hydrodynamic Equations}
In this chapter we derive the sound modes in the presence of nonzero fluid velocities from the hydrodynamic equations. 
As a warm-up exercise, and in order to establish the notation, we will discuss a single fluid before turning to two coupled fluids. The derivation is general in the sense that it holds for normal fluids as well as for superfluids. This enables us to discuss the three cases of two superfluids, one superfluid and one normal fluid, and two normal fluids from the final result. This chapter largely follows the Ref.~\cite{Haber:2015exa} and is expanded with the help of Ref.~\cite{superbook}.
\section{Single Fluid}
The starting point of our calculation are the conservation equations for the conserved current and the energy-momentum tensor,
\be\label{conserve}
\p_\mu j^\mu=0\, , \qquad \p_\mu T^{\mu\nu}=0\, .
\ee
In contrast to our discussion of these quantities in quantum field theory, we are now looking for expressions in terms of thermodynamic quantities instead of microscopic ones. Largely, we will be able to relate these later on. For the current, we can write as usual 
\be
j^\mu=nv^\mu \, ,
\ee
with the associated conserved charge density $n$ and the relativistic four-velocity $v^\mu=\gamma(1,\mathbf{v})$. In the rest frame of the fluid, where all thermodynamic quantities are defined, the energy-momentum  tensor of an ideal fluid is known to be
\be
T^{\mu\nu}_0= \mathrm{diag}(\epsilon,\, P,\, P,\, P) \, ,
\ee
with the energy density $\epsilon$ and the pressure $P$. We can now Lorentz transform this tensor with a general Lorentz transformation $\Lambda^{\mu}{}_{\;\nu}$, 
\be 
T^{\mu\nu}=\Lambda^{\mu}{}_{\alpha}\Lambda^\nu{}_\beta T^{\alpha\beta}_0 \, ,
\ee  
to obtain the energy-momentum tensor in the frame where the fluid moves with the velocity $v^\mu$:
\be\label{eq:Tmunu_ideal}
T^{\mu\nu}=(\epsilon+P)v^\mu v^\nu - g^{\mu\nu}P \, ,
\ee
with the Minkowski metric $g^{\mu\nu}$. Although not specified here, all quantities may in general depend on spacetime. The conservation laws for energy and momentum now become
\be
\p_\mu T^{\mu \nu} = -\p_\nu P + \p_\mu \left[\left(\epsilon+P\right)v^\mu v^\nu \right] \, .
 \ee
One can gain some insight by separating the spatial and temporal components of these equations. For $\nu=0$ we obtain
\be
\frac{\p P}{\p t}-\frac{\p}{\p t}\left(\frac{\epsilon+P}{1-\mathbf{v}^2}\right)-\nabla\cdot\left[\frac{\left(\epsilon+P\right)\mathbf{v}}{1-\mathbf{v}^2}\right]=0 \, ,
\ee
whilst for $\nu=i$ we find
\bea
0&=&\nabla_iP+\frac{\p}{\p t}\left[\frac{\left(\epsilon+P\right)v_i}{1-\mathbf{v}^2}\right]+\nabla_j\cdot\left[\frac{\left(\epsilon+P\right)v_j v_i}{1-\mathbf{v}^2}\right]\\[2ex]
&=& \nabla_i P + v_i\frac{\p P}{\p t}+\frac{\epsilon+P}{1-\mathbf{v}^2}\left(\frac{\p v_i}{\p t}+v_j\cdot\nabla_j v_i\right) \, ,
\eea
where the Latin indices denote the components of the corresponding three-vector in order to clarify which objects have to be contracted. In the second line we used the temporal equation in order to eliminate the time derivatives of $\epsilon+P$. The last equation can be further transformed into
\be
\frac{\p\mathbf{v}}{\p t}+\left(\mathbf{v}\cdot\nabla\right)\mathbf{v}=-\frac{1-\mathbf{v}^2}{\epsilon+P}\left(\nabla P +\mathbf{v}\frac{\p P}{\p t}\right) \, ,
\ee
which is nothing but the relativistic version of the Euler equation.
From the current conservation we directly compute
\be
\p_\mu j^\mu = 0 \qquad \Rightarrow \qquad \frac{\p}{\p t}\left(\frac{n}{\sqrt{1-\mathbf{v}^2}}\right)+\nabla\cdot\left(\frac{n\mathbf{v}}{\sqrt{1-\mathbf{v}^2}}\right)=0 \, ,
\ee
which is the relativistic version of the continuity equation. For a derivation of the non-relativistic limit see for instance Chap.~2 of Ref.~\cite{superbook}.

In principle, we could use these equations and introduce harmonic fluctuations of the chemical potential and the velocity in oder to compute the sound modes of the ideal fluid. By relating the chemical potential and the superflow via the phase of the condensate we would obtain the sound modes of a superfluid. However, we will introduce a more general notation at this point, which will allow us to work with the same variables later on in the multi-fluid case.

As a first step, we use the thermodynamic relation\footnote{Note that in most textbooks, the relation is given with $\mu$ as the chemical potential because the different frames of reference are not taken into account. Since we defined $\mu$ as  the chemical potential in the "lab frame", and all thermodynamic relations are defined in the rest frame of the fluid, we have to use the boosted potential $p$.} at zero temperature
\be
\epsilon+P=p n \, ,
\ee
to rewrite the energy-momentum tensor in the following way:
\be
T^{\mu\nu}=j^\mu p^\nu - g^{\mu\nu}P \, ,
\ee
which leads to the conservation law of the form
\be
\p_\mu T^{\mu\nu}=j^\mu\p_\mu p^\nu-\p^\nu P \, ,
\ee
where we have used that the current is conserved.
We can compare the energy-momentum tensor to our field theoretical result from Eq.~(\ref{TmunuQFT}). Together with the result for the current from Eq~(\ref{eq:jmuQFT}), we can read off that the pressure at $T=0$ for a single superfluid is given by the tree-level Lagrangian. This is of course expected, because the grand canonical potential  is nothing but the negative of the pressure, $\Omega=-P$. $\Omega$ itself at vanishing temperature is given by the negative of the tree-level Lagrangian, canceling the other minus sign. Therefore we can use
\be
\p^\nu P = \frac{\p P}{\p p^\mu}\p^\nu p^\mu= \frac{\p \L^{(0)}}{\p p^\mu}\p^\nu p^\mu=j_\mu \p^\nu p^\mu \, ,
\ee
where we have used the definition of the conjugate momentum to replace the derivative of $\L$ with the conserved current. The conservation equation now becomes
\be
\p_\mu T^{\mu\nu}=j_\mu\left(\p^\mu p^\nu-\p^\nu p^\mu\right) = j_\mu \omega^{\mu\nu} \, ,
\ee
 where we have introduced the vorticity
 \be
\omega^{\mu\nu}\equiv\p^\mu p^\nu-\p^\nu p^\mu \, .
\ee
For a superfluid we see that the vorticity is identically zero since $p^\mu=\p^\mu\psi$ and the commutativity of the four-gradients.

We want to generalize this discussion to the case of mixtures of ideal and superfluids. In these cases, we might not always be able to start from a microscopic theory, or the pressure itself does not depend on all conjugate momenta of the conserved currents. For instance in the case of a single superfluid at finite temperature, the entropy current $s^\mu$ is conserved as well and leads to the existence of a second sound mode, but does not have a conjugate momentum in the pressure. Thus, we introduce the generalized pressure $\Psi$, such that all conserved currents can be computed by
\be
j^\mu_i=\frac{\p\Psi}{\p p_{i,\mu}}  \, ,
\ee
with $i$ indicating the $i-$th conserved current. In the limit of a single fluid, we know that  $\Psi$ is equal to the pressure in the fluid rest frame. Consequently, we demand that $\Psi$ is a Lorentz scalar as well, which means that it depends on the squares and four-products of the conjugate momenta $p^\mu_i$. For two fluids for instance, we can write
\be
\Psi=\Psi\left(p_1^2,p_2^2,p_{12} \right) \, ,
\ee
with $p_{12}=p_1^\mu p_2^\nu g_{\mu\nu}$. In general, we can decompose the energy-momentum tensor as
\be \label{eq:tmunu_i}
T^{\mu\nu}=\sum_i j^\mu_i p^\nu_i-g^{\mu\nu}\Psi \, ,
\ee
where $i$ runs over all conserved currents and their conjugate momenta. Additionally, we introduce the generalized energy density $\Lambda$, which is sometimes also called master function, via the relation
\be\label{eq:gen_dens}
\Lambda+\Psi=\sum_i j^\mu_i p^\nu_i \, ,
\ee
analogous to the thermodynamic relation $\epsilon+P=pn$. For a more elaborate discussion of these quantities see Refs.~\cite{Alford:2012mv,2013PhRvD..87f5001A,Stetina:2015exa}, and Ref.~\cite{superbook} for a more pedagogical approach. Equipped with this setup, we can turn to the computation of the speed of sound of a single ideal fluid.

For a single fluid, we write
\be\label{Tmunu}
T^{\mu\nu} = j^\mu p^\nu -g^{\mu\nu}\Psi   \, .
\ee
As the generalized pressure only depends on one conjugated momentum squared, the Lorentz scalar $p^2=p_\mu p^\mu$, we use the chain rule to write
\be \label{jdef}
j^\mu = \frac{\partial \Psi}{\partial p_\mu} = 2\frac{\partial \Psi}{\partial p^2} p^\mu \, . 
\ee
With this relation between the conjugate momentum and the current it 
is obvious that the stress-energy tensor is symmetric under exchange of the indicies $\nu$ and $\mu$. In Chap.~\ref{chap:sf_qft} we have seen that the superfluid velocity is proportional to the derivative of the phase of the condensate, which is the conjugate momentum to the current. We can repeat the argument more generally for the ideal fluid as well. As a reminder, the Lorentz scalar $p$ is the chemical potential measured in the rest frame of the fluid. We write $j^\mu = nv^\mu$, where $n$ is the charge density measured in the rest frame of the fluid. Since $n^2=j^2$, we compute 
\be \label{n}
n = 2\frac{\partial \Psi}{\partial p^2} p \, . 
\ee
The generalized energy density can also be written as $\Lambda = T^\mu_{\;\;\mu}+3\Psi$. 
Using Eqs.\ (\ref{Tmunu}) and (\ref{n}), we consistently recover $\Lambda + \Psi = p_\mu j^\mu = pn$ from Eq.~(\ref{eq:gen_dens}).

The current can now be written as $j^\mu = n p^\mu/p$, from which we conclude that the fluid velocity is 
\be \label{vmu}
v^\mu = \frac{p^\mu}{p}  \, .
\ee
Using $v^\mu = \gamma(1,\mathbf{v})$, where $\gamma=1/\sqrt{1-\mathbf{v}^2}$ is the usual Lorentz factor, the three-velocity becomes $\mathbf{v}=\mathbf{p}/\mu$, with  
$\mu = p_0$ being the chemical potential measured in the frame in which the fluid moves with velocity $\mathbf{v}$. 
In that frame, the charge density is $j^0 = n\mu/p$. With the help of Eqs.\ (\ref{jdef}), (\ref{n}), (\ref{vmu}) and the relation $pn=\Lambda + \Psi$, we see that the stress-energy tensor
(\ref{Tmunu}) can be written in the familiar form $T^{\mu\nu}=(\Lambda+\Psi)v^\mu v^\nu-g^{\mu\nu}\Psi$, which can be obtained from Eq.~(\ref{eq:Tmunu_ideal}) by replacing all quantities by the corresponding "generalized" form.  

We write the hydrodynamic equations (\ref{conserve}) as 
\begin{subequations} \label{jT}
\bea
0&=&\partial_\mu j^\mu = \frac{n}{p}\left[g_{\mu\nu}+\left(\frac{1}{c^2}-1\right)\frac{p_\mu p_\nu}{p^2}\right]\partial^\mu p^\nu \, , \\[2ex]
0&=&\partial_\mu T^{\mu\nu} = j^\mu(\partial_\mu p^\nu -\partial^\nu p_\mu) \, , \label{jT2}
\eea
\end{subequations} 
where, in the second relation, we have used $\partial_\mu j^\mu=0$, and we have introduced the speed of sound $c$ in the rest frame of the fluid, 
\be
c^2 = \frac{n}{p}\left(\frac{\partial n}{\partial p}\right)^{-1} \, . 
\ee

To compute the sound modes, we need to treat the temporal and spatial components of the momentum $p^\mu = (\mu, \mathbf{p})$ separately. Each component is assumed to fluctuate
harmonically about its equilibrium value with frequency $\omega$ and wave vector $\mathbf{k}$,
\be \label{osc}
\mu(\mathbf{x},t) = \mu +\delta\mu \,e^{i(\omega t-\mathbf{k}\cdot\mathbf{x})} \, , \qquad  \mathbf{p}(\mathbf{x},t) = \mathbf{p} +\delta\mathbf{p} \,e^{i(\omega t-\mathbf{k}\cdot\mathbf{x})} \, , 
\ee
where $\delta\mu$ and $\delta\mathbf{p}$ are the fluctuations which will be kept to linear order. Then, Eqs.\ (\ref{jT}) become 
\begin{subequations} \label{hydrolin}
\bea
0 &=& \frac{n}{p}\left[\omega\delta\mu-\mathbf{k}\cdot\delta\mathbf{p}+\frac{\mu\omega_v}{p^2}\left(\frac{1}{c^2}-1\right)(\mu\delta\mu-\mathbf{p}\cdot\delta\mathbf{p})\right]   \, , \label{hydrolin1} \\[2ex]
0 &=& \frac{n}{p}\mathbf{p}\cdot(\omega\delta\mathbf{p}-\mathbf{k}\delta\mu)  \, , \label{hydrolin2}\\[2ex]
0 &=& \frac{n}{p}\left[\mu\omega_v\delta\mathbf{p}-\mathbf{k}(\mu\delta\mu - \mathbf{p}\cdot\delta\mathbf{p})\right] \, ,\label{hydrolin3}
\eea
\end{subequations}
where $\omega_v \equiv \omega - \mathbf{v}\cdot\mathbf{k}$. We have decomposed 
Eq.\ (\ref{jT2}) into its temporal component (\ref{hydrolin2}) and its spatial components (\ref{hydrolin3}).

We know that in the case of a superfluid, the conjugate momentum can be written as the gradient of 
a phase $\psi\in [0,2\pi]$, $p^\mu = \partial^\mu \psi$. This has very interesting implications for the nature of the sound modes. By remembering that
\be \label{super}
\mu = \partial_0\psi \, , \qquad \mathbf{v} = -\frac{\nabla\psi}{\mu} \, ,
\ee
we find that, together with $\partial_0\mathbf{p}=-\nabla\partial_0\psi=-\nabla\mu$ and $\mathbf{p}=\mu\mathbf{v}$, that the fluctuations in the chemical potential and the superfluid velocity are not independent. Instead, they are both related to the phase. As a consequence, only longitudinal modes are allowed where $\delta\mathbf{p}$ oscillates in the direction of the wave vector $\mathbf{k}$, 
$\omega\delta\mathbf{p}=\mathbf{k}\delta\mu$, and Eqs.\ (\ref{hydrolin2}) and (\ref{hydrolin3}) are 
automatically fulfilled. We have argued this before, when we have seen that for the very same reason the vorticity of a superfluid is zero and the conservation of the energy-momentum tensor therefore trivial. We nevertheless wrote down the conservation equation for the harmonic perturbations in order to cover the case of an ideal fluid as well. It only remains to solve the continuity equation which yields 
a quadratic polynomial in $\omega$. Since in the given approximation all modes are linear in momentum, we can write $\omega = uk$ with the angular-dependent sound velocity 
\be \label{usingle}
u = \frac{(1-c^2)v\cos\theta \pm c\sqrt{1-v^2}\sqrt{1-c^2v^2-(1-c^2)v^2\cos^2\theta}}{1-c^2v^2} \, .
\ee
Here, $v\equiv |\mathbf{v}|$ is the modulus of the three-velocity and $\theta$ the angle between the directions of the fluid velocity and the propagation of the 
sound mode, $\mathbf{k}\cdot\mathbf{p}=\mu vk\cos\theta$.
As a first check, we set $v=0$ and directly obtain that the sound
velocity is equal to the sound velocity defined in the fluid rest
frame, $u=c$. As a further check one can set $c=1$. If the sound
velocity in the fluid rest frame already equals the speed of light,
it has to be the same in all frame of references, i.e.~$u=1$.
\begin{eqnarray}
\left(1-v^{2}\right)u^{2}+\left(v^{2}-1\right) & = & 0\,,\\
u & = & 1\,.
\end{eqnarray}
 
For a given angle $\theta \in [0,\pi]$, we consider only the upper sign
in Eq.\ (\ref{usingle}), such that the sound speed is positive for small velocities $v$. Since we shall later be interested in instabilities, we may already ask at this
point when the speed of sound becomes negative. This happens at the critical velocity $v=c$, where $u$ starts to become negative in the upstream direction $\theta=\pi$.
We also see that $u$ never becomes complex because the arguments of either of the square roots in the numerator only become negative for unphysical
velocities larger than the speed of light, $v>1$.   

In the case of a normal fluid, the same longitudinal modes are found,
but $\mathbf{p}$ is now allowed to oscillate in transverse directions with respect to $\mathbf{k}$. This yields the additional mode $u = \mathbf{v}\cdot\hat{\mathbf{k}}$ with the conditions for the fluctuations  
$\mathbf{p}\cdot\delta\mathbf{p} = \mu\delta\mu$ and $\mu \mathbf{k}\cdot\delta\mathbf{p} = \mathbf{k}\cdot\mathbf{p}\,\delta\mu$. 

\section{Two Fluids}
\label{sec:hydro}
We now generalize the results to the case of two coupled fluids. According to Eq.~(\ref{eq:tmunu_i}), we write the stress-energy tensor as
\be
T^{\mu\nu} = j_1^\mu p_1^\nu +j_2^\mu p_2^\nu -g^{\mu\nu} \Psi \, . 
\ee
Now $\Psi$ is a function of \textbf{all} Lorentz scalars that can be constructed from the two conjugate momenta $p_1^\mu$, $p_2^\mu$, i.e., $\Psi = \Psi(p_1^2,p_2^2,p_{12}^2)$.
The currents are defined as in Eq.\ (\ref{jdef}) and are slightly more complicated than before, since the conjugate momentum of each current now appears in the mixed scalar $p_{12}$ as well. This yields  
\begin{subequations} \label{j1j2}
\bea
j_1^\mu &=& {\cal B}_1 p_1^\mu + {\cal A}\, p_2^\mu \, , \\[2ex]
j_2^\mu &=& {\cal A}\, p_1^\mu + {\cal B}_2 p_2^\mu \, , 
\eea
\end{subequations}
where we have abbreviated the various derivatives with
\be
{\cal B}_i \equiv 2\frac{\partial \Psi}{\partial p_i^2} \, , \qquad 
{\cal A} \equiv \frac{\partial \Psi}{\partial p_{12}^2} \, ,
\ee
with the fluid index $i=1,2$.
In general, the currents are not four-parallel to their own conjugate momentum anymore, but, if $\Psi$ depends on $p_{12}^2$, receive a contribution from the 
conjugate momentum of the other fluid. This effect is the entrainment effect we have discussed before and ${\cal A}$ is called entrainment coefficient. To guarantee the symmetry of the stress-energy tensor, there can only be one single entrainment coefficient, appearing
in both currents. The conservation equations now read
\begin{subequations}
\bea
\partial_\mu j_1^\mu &=& \Big[{\cal B}_1 g_{\mu\nu}+b_1 p_{1\mu}p_{1\nu}+a_1(p_{1\mu} p_{2\nu}+p_{2\mu}p_{1\nu})+a_{12} p_{2\mu}p_{2\nu}\Big]\partial^\mu p_1^\nu \non[2ex]
&&+\Big[{\cal A} g_{\mu\nu}+a_1p_{1\mu}p_{1\nu}+d\,p_{1\mu} p_{2\nu}+a_{12}p_{2\mu}p_{1\nu}+a_2p_{2\mu}p_{2\nu}\Big]\partial^\mu p_2^\nu\, , \label{cons1}\\[2ex]
\partial_\mu j_2^\mu &=& (1\leftrightarrow 2) \, , \label{cons2} \\[2ex]
\partial_\mu T^{\mu\nu} &=& j_1^\mu(\partial_\mu p_1^\nu-\partial^\nu p_{1\mu})+j_2^\mu(\partial_\mu p_2^\nu-\partial^\nu p_{2\mu}) \, ,
\eea
\end{subequations}
where Eq.\ (\ref{cons2}) is obtained from Eq.\ (\ref{cons1}) by exchanging the indices $1\leftrightarrow 2$ ($a_{21}=a_{12}$), and we have abbreviated second derivatives by
\bea
b_i &\equiv& 4\frac{\partial^2 \Psi}{\partial (p_i^2)^2}\, , \qquad a_i \equiv 2\frac{\partial^2 \Psi}{\partial p_i^2\partial p_{12}^2} \, , 
\qquad a_{12} \equiv \frac{\partial^2 \Psi}{\partial (p_{12}^2)^2} \, , \qquad d\equiv 4\frac{\partial^2 \Psi}{\partial p_1^2\partial p_2^2} \, .
\eea
Here, $b_1$, $b_2$ are susceptibilities that are given by the properties of each fluid separately, while $d$ describes a non-entrainment coupling between the two fluids, 
and the coefficients $a_1$, $a_2$, $a_{12}$ are only nonzero in the presence of an entrainment coupling.
Again we introduce fluctuations in the temporal and spatial components of the conjugate momenta, as given in Eq.\ (\ref{osc}). The conservation equations then become
\begin{subequations}
\bea
0&=& \Big[\omega {\cal B}_1+\omega_{v_1}\mu_1(\mu_1b_1+\mu_2a_1)+\omega_{v_2}\mu_2(\mu_1a_1+\mu_2a_{12})\Big]\delta\mu_1 \non[2ex]
&&+\Big[\omega {\cal A}+\omega_{v_1}\mu_1(\mu_1a_1+\mu_2d)+\omega_{v_2}\mu_2(\mu_1 a_{12}+\mu_2 a_2)\Big]\delta\mu_2 \non[2ex]
&&-(\omega_{v_1}\mu_1 b_1+\omega_{v_2}\mu_2 a_1)\mathbf{p}_1\cdot\delta\mathbf{p}_1 
-(\omega_{v_1}\mu_1 d +\omega_{v_2}\mu_2 a_2)\mathbf{p}_2\cdot\delta\mathbf{p}_2 -{\cal B}_1\mathbf{k}\cdot\delta\mathbf{p}_1-{\cal A}\mathbf{k}\cdot\delta\mathbf{p}_2\non[2ex]
&&-(\omega_{v_1}\mu_1 a_1+\omega_{v_2}\mu_2 a_{12})\mathbf{p}_2\cdot\delta\mathbf{p}_1
-(\omega_{v_1}\mu_1 a_1+\omega_{v_2}\mu_2 a_{12})\mathbf{p}_1\cdot\delta\mathbf{p}_2  \, , \label{one}\\[2ex]
0&=& (1\leftrightarrow 2) \, , \label{two}\\[2ex]
0&=& (\mathbf{p}_1{\cal B}_1+\mathbf{p}_2 {\cal A})\cdot(\omega\delta\mathbf{p}_1-\mathbf{k}\delta\mu_1) + (\mathbf{p}_2{\cal B}_2+\mathbf{p}_1 {\cal A})\cdot(\omega\delta\mathbf{p}_2-\mathbf{k}\delta\mu_2) 
\, , \label{2fl3}\\[2ex]
0&=& (\omega_{v_1}\mu_1 {\cal B}_1+\omega_{v_2}\mu_2{\cal A})(\omega\delta\mathbf{p}_1-\mathbf{k}\delta\mu_1)+
(\omega_{v_2}\mu_2 {\cal B}_2+\omega_{v_1}\mu_1{\cal A})(\omega\delta\mathbf{p}_2-\mathbf{k}\delta\mu_2) \, , \label{2fl4}
\eea
\end{subequations}
where we have inserted $\partial_\mu T^{\mu 0}=0$ (\ref{2fl3}) into the spatial components $\partial_\mu T^{\mu \ell}$ ($\ell=1,2,3$) to obtain Eq.\ (\ref{2fl4}).

\subsection{Super-Super}

If both fluids are superfluids, the relation of the chemical potential and the fluid velocity via the gradient of the phase dictates $\omega\delta\mathbf{p}_i=\mathbf{k}\delta\mu_i$. The conservation of energy and momentum 
is automatically fulfilled in this case, and the two continuity equations read
\begin{subequations} \label{dmu1dmu2}
\bea
0&=& \left[{\cal B}_1(\omega^2-k^2)+\mu_1^2b_1\omega_{v_1}^2+2\mu_1\mu_2a_1\omega_{v_1}\omega_{v_2}+\mu_2^2a_{12}\omega_{v_2}^2\right]\delta\mu_1 \non[2ex]
&&+\left[{\cal A}(\omega^2-k^2)+\mu_1^2a_1\omega_{v_1}^2+\mu_1\mu_2(a_{12}+d)\omega_{v_1}\omega_{v_2}+\mu_2^2a_2\omega_{v_2}^2\right]\delta\mu_2 \, , \\[2ex]
0&=& \left[{\cal A}(\omega^2-k^2)+\mu_2^2a_2\omega_{v_2}^2+\mu_1\mu_2(a_{12}+d)\omega_{v_1}\omega_{v_2}+\mu_1^2a_1\omega_{v_1}^2\right]\delta\mu_1\non[2ex]
&&+\left[{\cal B}_2(\omega^2-k^2)+\mu_2^2b_2\omega_{v_2}^2+2\mu_1\mu_2a_2\omega_{v_1}\omega_{v_2}+\mu_1^2a_{12}\omega_{v_1}^2\right]\delta\mu_2 \, .
\eea
\end{subequations}
After some rearrangements this can be compactly written as
\be
(u^2\chi_2 + u\chi_1+\chi_0)\delta\mu = 0  \, , 
\ee
with $u=\omega/k$, the vector $\delta \mu = (\delta\mu_1,\delta\mu_2)$, and the $2\times 2$ matrices $\chi_2$, $\chi_1$, $\chi_0$ whose entries are given by 
\be
(\chi_2)_{ij} = \frac{\partial^2\Psi}{\partial\mu_i\partial\mu_j} \, , \qquad  
(\chi_1)_{ij} = \left(\frac{\partial^2\Psi}{\partial\mu_i\partial p_{j\ell}}+\frac{\partial^2\Psi}{\partial\mu_j\partial p_{i\ell}}\right)\hat{k}_\ell \, , \qquad 
(\chi_0)_{ij} = \frac{\partial^2\Psi}{\partial p_{i\ell}\partial p_{jm}} \hat{k}_\ell\hat{k}_m \, , 
\ee
with $i,j = 1,2$, and spatial indices $\ell,m=1,2,3$. (The indices $i,j$ are reserved for the two different 
fluid species, which should not lead to any confusion since spatial indices will not appear explicitly from here on.) 
The sound velocity $u$ is then determined from 
\be \label{detu}
\mathrm{ det}\,(u^2\chi_2 + u\chi_1+\chi_0) = 0 \, , 
\ee
which is a quartic polynomial in $u$ with analytical, but very complicated, solutions. We shall discuss and interpret these solutions in Chap.~\ref{chap:instabilities}.

\subsection{Super-Normal}

Let us now assume that one of the fluids is a normal fluid. Say fluid 1 is a superfluid, $\omega\delta\mathbf{p}_1=\mathbf{k}\delta\mu_1$, while we make no assumptions about 
$\delta\mathbf{p}_2$. Of course, we still find the same modes as for the two superfluids since the normal fluid can also accommodate the longitudinal oscillations of a superfluid. 
An additional mode is found if we enforce a transverse mode by requiring $\omega\delta\mathbf{p}_2\neq\mathbf{k}\delta\mu_2$. Then, Eq.\ (\ref{2fl4}) yields the mode 
\be \label{vnk1}
u = \frac{\mathbf{v}_2\mu_2{\cal B}_2+\mathbf{v}_1\mu_1{\cal A}}{\mu_2{\cal B}_2+\mu_1{\cal A}}\cdot\hat{\mathbf{k}} \, .
\ee
This is the generalization of the mode $u = \mathbf{v}\cdot\hat{\mathbf{k}}$ 
mentioned in the discussion of the single normal fluid. We may apply this expression to a single superfluid at nonzero temperature.
In this case, there are two currents, the conserved charge current $j_1^\mu = j^\mu$ and the entropy current $j_2^\mu= s^\mu$ 
(which is also conserved if we neglect dissipation). Their conjugate momenta 
are $p_1^\mu=\partial^\mu \psi$, where $\psi$ is the phase of the condensate, and $p_2^\mu=\Theta^\mu$, whose temporal component is the temperature, $\Theta_0=T$, measured in the 
normal-fluid rest frame, where $\mathbf{s}=0$. Analogously to Eq.\ (\ref{j1j2}) we can write \cite{Carter:1995if,2013PhRvD..87f5001A,superbook} 
\begin{subequations} 
\bea
j^\mu &=& {\cal B} \partial^\mu \psi  + {\cal A} \Theta^\mu  \, , \\[2ex]
s^\mu &=& {\cal A}\partial^\mu \psi + {\cal C} \Theta^\mu \, . 
\eea
\end{subequations}
Consequently, we can identify $\mathbf{v}_2\mu_2{\cal B}_2+\mathbf{v}_1\mu_1{\cal A}\to \mathbf{s}$ and $\mu_2{\cal B}_2+\mu_1{\cal A}\to s_0$. Note that $\mu_2$, the temporal component
of the conjugate momentum $p_2^\mu$ corresponds to the temperature $T$, and the velocity $\mathbf{v}_2$ corresponds to $\mathbf{\Theta}/T$. The four-velocity of the normal fluid is defined 
by $v_n^\mu = s^\mu/s$, which yields the three-velocity $\mathbf{v}_n = \mathbf{s}/s_0$. (If normal fluid and superfluid velocities are used as 
independent hydrodynamic variables, one works in a "mixed" representation with respect to currents and momenta:
 while the superfluid velocity corresponds to the \textit{momentum} of one fluid, $v_s^\mu = \partial^\mu\psi$, the normal fluid 
velocity corresponds to the \textit{current} of the other fluid, $v_n^\mu = s^\mu/s$.) Inserting all this into Eq.\ (\ref{vnk1}) yields
\be
u = \mathbf{v}_n\cdot\hat{\mathbf{k}} \, . 
\ee
This is in exact agreement with Ref.\  \cite{PhysRevB.77.144515}, where this mode has been discussed in the non-relativistic context. (In Refs.\ \cite{2013PhRvD..87f5001A,Alford:2013koa},
where sound modes in a relativistic superfluid at nonzero temperatures were studied within a field-theoretical setup, this mode was not mentioned because the calculation 
was performed in the rest frame of the normal fluid.)

\subsection{Normal-Normal}

In the case of two normal fluids, we make no assumptions about the fluctuations $\delta\mathbf{p}_i$, $\delta\mu_i$. Let us first suppose there was no coupling between the 
two fluids at all, i.e., ${\cal A}=a_1=a_2=a_{12}=d=0$. Without loss of generality, we can work in the rest frame of one of the fluids, 
say $\mathbf{p}_2=0$ and thus $\omega_{v_2}=\omega$. Then, 
we can express $\mathbf{p}_1\cdot \delta\mathbf{p}_1$, $\mathbf{k}\cdot\delta\mathbf{p}_1$, $\mathbf{k}\cdot\delta\mathbf{p}_2$ in terms of $\delta\mu_1$ and $\delta\mu_2$ to 
obtain 
\be
0=\mu_1\omega_{v_1}[{\cal B}_1(\omega^2-k^2)+\omega_{v_1}^2\mu_1^2b_1]\delta\mu_1 + \mu_2\omega[{\cal B}_2(\omega^2-k^2)+\omega^2\mu_2^2b_2]\delta\mu_2 \, .
\ee
This equation yields the separate modes of the two fluids: by setting $\delta\mu_2=0$ we find the modes for fluid 1, and by setting $\delta\mu_1=0$ we find the modes for fluid 2. With 
\be
{\cal B}_i=\frac{n_i}{p_i} \, , \qquad b_i=\frac{n_i}{p_i^3}\left(\frac{1}{c_i^2}-1\right) \qquad (i=1,2) \, , 
\ee
we recover the modes discussed above for the single fluid. 

Now let us switch on the coupling between the fluids. The simplest 
case is to neglect any entrainment, $a_1=a_2=a_{12}=0$, but keep a nonzero non-entrainment coupling, $d\neq 0$. For a compact notation we define the  
"mixed susceptibilities"
\be 
\Delta_{1/2} \equiv \frac{p_{2/1}}{n_{1/2}}\frac{\partial n_{1/2}}{\partial p_{2/1}}\, , 
\ee
(such that $d = \Delta_1B_1/p_2^2=\Delta_2B_2/p_1^2$), which are only nonvanishing for nonzero coupling $d$. 
Then, for $\delta\mu_2=0$ we find the modes $u=v_1\cos\theta$, and 
\be \label{dmu20}
\delta\mu_2=0:\qquad u = \frac{(1-c_1^2)v_1\cos\theta\pm c_1\sqrt{1-v_1^2}\sqrt{1-(1+\Delta_1)v_1^2[c_1^2+(1-c_1^2)\cos^2\theta]+\Delta_1c_1^2}}{1-c_1^2[v_1^2(1+\Delta_1)-\Delta_1]} 
\, . 
\ee
As for the case of a single fluid, we may ask whether and when the speed of sound turns negative. And, in contrast to the single fluid, $u$ may even become complex. The critical 
velocities for these two instabilities (to be discussed in detail in Sec.\ \ref{chap:instabilities}) are, respectively,
\be \label{dmu20vcr}
\delta\mu_2=0:\qquad  v_{\mathrm{ c}}^< = \frac{c_1}{\sqrt{c_1^2\sin^2\theta+\cos^2\theta}} \, , \qquad v_{\mathrm{ c}}^> = v_{\mathrm{ c}}^<\frac{\sqrt{1+\Delta_1 c_1^2}}{c_1\sqrt{1+\Delta_1}}
\ge v_{\mathrm{ c}}^<
 \, ,
\ee
For $\delta\mu_1=0$ we have 
\be \label{dmu10}
\delta\mu_1=0:\qquad u = \frac{\Delta_2 c_2^2v_1\cos\theta\pm c_2\sqrt{1-v_1^2}\sqrt{1-v_1^2+\Delta_2(c_2^2-v_1^2\cos^2\theta)}}{1-v_1^2+\Delta_2c_2^2} \, .
\ee
Again, we can easily compute the critical velocities,
\be \label{dmu10vcr}
\delta\mu_1=0:\qquad v_{\mathrm{ c}}^< = \frac{1}{\sqrt{1+\Delta_2\cos^2\theta}} \, , \qquad v_{\mathrm{ c}}^> = v_{\mathrm{ c}}^<\sqrt{1+\Delta_2c_2^2} \ge v_{\mathrm{ c}}^<
\, .
\ee
In both cases, it is obvious (and we have indicated it by our choice of notation), that the critical velocity for a negative sound velocity is smaller than or equal to that for a 
complex sound velocity. We shall come back to this observation and also discuss the general case with entrainment 
at the end of Sec.\ \ref{chap:instabilities}.

\chapter{Quasiparticle Propagator and Goldstone Modes}
\label{chap:goldstone}
At vanishing temperature, the speed of sound of a superfluid can equivalently be extracted from the slope of the Goldstone mode which results from the spontaneous breaking of the $U(1)\times U(1)$ symmetry. We have performed this calculation before, even in the presence of a gauge field when introducing finite temperatures in our microscopical model. Here, we neglect any possible charge of the fields and corresponding gauge fields, but have to work at finite superflow. Here, in the main text, I simply present the propagator in momentum space and then directly proceed with evaluating the excitation energies. In appendix \ref{app:exc_SFSF}, some details of the derivation are laid out.

The inverse propagator obtained from the Lagrangian given in Eq.~(\ref{L}) can be written as 
\be
S^{-1} = \left(\begin{array}{cc} S^{-1}_{11} & S^{-1}_{12} \\ S^{-1}_{21} & S^{-1}_{22} \end{array}\right) \, , 
\ee
where
\begin{subequations} \label{Sinv}
\bea
S^{-1}_{11/22} &=& \left(\begin{array}{cc} -K^2+\lambda_{1/2}(3\rho_{1/2}^2-\rho_{\mathrm{SF}1/\mathrm{SF}2}^2)  & 
2iK\cdot\partial\psi_{1/2} 
\\ -2iK\cdot\partial\psi_{1/2} & 
-K^2+\lambda_{1/2}(\rho_{1/2}^2-\rho_{\mathrm{SF}1/\mathrm{SF}2}^2)   \end{array}\right) \non[2ex]
&&-\left(\begin{array}{cc}  (h+gp_{12}^2)\rho_{2/1}^2  & 
-ig\rho_{2/1}^2 K\cdot\partial\psi_{2/1} 
\\ ig\rho_{2/1}^2 K\cdot\partial\psi_{2/1} & 
 (h+gp_{12}^2)\rho_{2/1}^2  \end{array}\right) \, , \allowdisplaybreaks\\[2ex]
S^{-1}_{12/21} &=& \frac{\rho_1\rho_2}{2}\left(\begin{array}{cc} GK^2-4(h+gp_{12}^2)\;\;  & 
2igK\cdot\partial\psi_{1/2} \\ -2igK\cdot\partial\psi_{2/1} & 
-gK^2  \end{array}\right) \, ,
\eea
\end{subequations}
with the four-momentum $K=(k_0,\mathbf{k})$, the four-product $K\cdot\partial\psi_{1/2} = K_\mu \partial^\mu\psi_{1/2}$, and the condensates in the SF$_1$/SF$_2$ phase from Eqs.~(\ref{eq:rhoSC},\ref{eq:rhoSF}). Note that here also the sum of the two derivative coupling constants $G$ appears, while in the tree-level potential only their difference $g$ had entered.
This form of the propagator is general and holds for all possible phases. We are only interested in the excitation energies of the COE phase, where both condensates are nonzero. Therefore, we set $\rho_i = \rho_{0i}$ with the condensates in the COE phase from Eq.~(\ref{COE})
and use \mbox{$\lambda_{1/2}(\rho_{01/02}^2-\rho_{\mathrm{SF}_1/\mathrm{SF}_2}^2) -(h+gp_{12}^2)\rho_{02/01}^2 =0$}, to simplify
\be
S^{-1}_{11/22} = \left(\begin{array}{cc} -K^2+2\lambda_{1/2}\rho_{01/02}^2\;\;  & 
2iK\cdot\partial\psi_{1/2} 
\\ -2iK\cdot\partial\psi_{1/2} & 
-K^2  \end{array}\right)
+
ig\rho_{02/01}^2K\cdot\partial\psi_{2/1}\left(\begin{array}{cc} 0  & 
1 \\ -1 & 0  \end{array}\right) \, .
\ee
The excitation energies are given by $\mathrm{ det}\,S^{-1}=0$. Due to the symmetry of the determinant under $K\to -K$, the excitations come in 4 pairs: if $k_0=\epsilon_{r,\mathbf{k}}$ is a zero, then also 
$k_0=-\epsilon_{r,-\mathbf{k}}$, ($r=1,\ldots ,4$). In appendix \ref{app:exc_SFSF} it is shown that only one energy of each pair has to be kept in order to compute the thermodynamic properties of the system. In the COE phase, the solutions of $\mathrm{ det}\, S^{-1}=0$ correspond to the two Goldstone modes and two massive modes.   
For the simplest scenario, let us set the superfluid velocities, the mass parameters, and the non-entrainment coupling to zero, $\nabla\psi_1=\nabla\psi_2=m_1=m_2=h=0$.
Then, the energy of the massive modes has the form $\epsilon_{\mathbf{k}} = M + {\cal O}(k^2)$ with the two masses 
\be
M = \sqrt{6}\left[\mu_{1/2} + \frac{\mu_{2/1}^3}{2\lambda_{2/1}}\,g + {\cal O}(g^2)\right] \, .
\ee
These modes are of no further interest to us since we shall focus on the low-energy properties of the system. Also, we should recall that most of the real-world 
superfluids we have in mind 
are of fermionic nature. Therefore, our bosonic approach can at best be a low-energy effective description. In a fermionic superfluid there is a massless mode too because of the 
Goldstone theorem, but typically there is no stable massive mode. There rather is a continuum of states for energies 
larger than twice the fermionic pairing gap \cite{Fukushima:2005gt,2015JPSJ...84d4003Y}, and thus at these energies our effective bosonic description breaks down. 

The energies of the Goldstone modes have the form $\epsilon_{\mathbf{k}} = u k + {\cal O}(k^3)$, where
\be
u  = \frac{1}{\sqrt{3}}\left[1\pm\frac{\mu_1\mu_2}{2\sqrt{\lambda_1\lambda_2}}\,g + {\cal O}(g^2)\right] \, .
\ee
We see that the effect of a small entrainment coupling is to split the two Goldstone modes, with one mode becoming faster and one mode becoming slower. One can 
check that the behavior of a non-entrainment coupling $h$ is different: if we set the entrainment couplings to zero, $G=g=0$, but keep a nonzero $h$, we find that one mode remains unperturbed by 
the coupling and the other acquires a larger speed. 

After these preparations we can now consider nonzero fluid velocities and discuss the resulting instabilities. 

\chapter{Dynamical and Energetic Instabilities}
\label{chap:instabilities}
 
\section{Results for Sound Modes and Identification of Instabilities}

Having identified the regions in parameter space where both species 1 and 2 become superfluid and having derived the quasiparticle propagator $S$ for this
phase, we now compute the excitation energies numerically. We focus on the two Goldstone modes and do not discuss the massive modes any further. 
For small momenta, the dispersion relations of the Goldstone modes are linear, $\epsilon_{\mathbf{k}} = u k$, 
and their slopes $u$ can also be computed from the linearized hydrodynamic equations, i.e., if we are only interested in the low-momentum behavior we may alternatively employ 
Eq.\ (\ref{detu}) instead of $\mathrm{ det}\,S^{-1}=0$. In general, this is not true. For instance, in a single superfluid at nonzero temperature, there is 
only one Goldstone mode, but there are two different sound modes, usually called first and second sound. Only for small temperatures, the Goldstone mode is well 
approximated by first sound, in general neither first nor second sound corresponds to the Goldstone mode. In our zero-temperature approximation of two coupled superfluids, 
we have "two first sounds" which coincide with the Goldstone modes at low momentum.

\begin{figure}[t]
\begin{center}
\hbox{\includegraphics[width=0.28\textwidth]{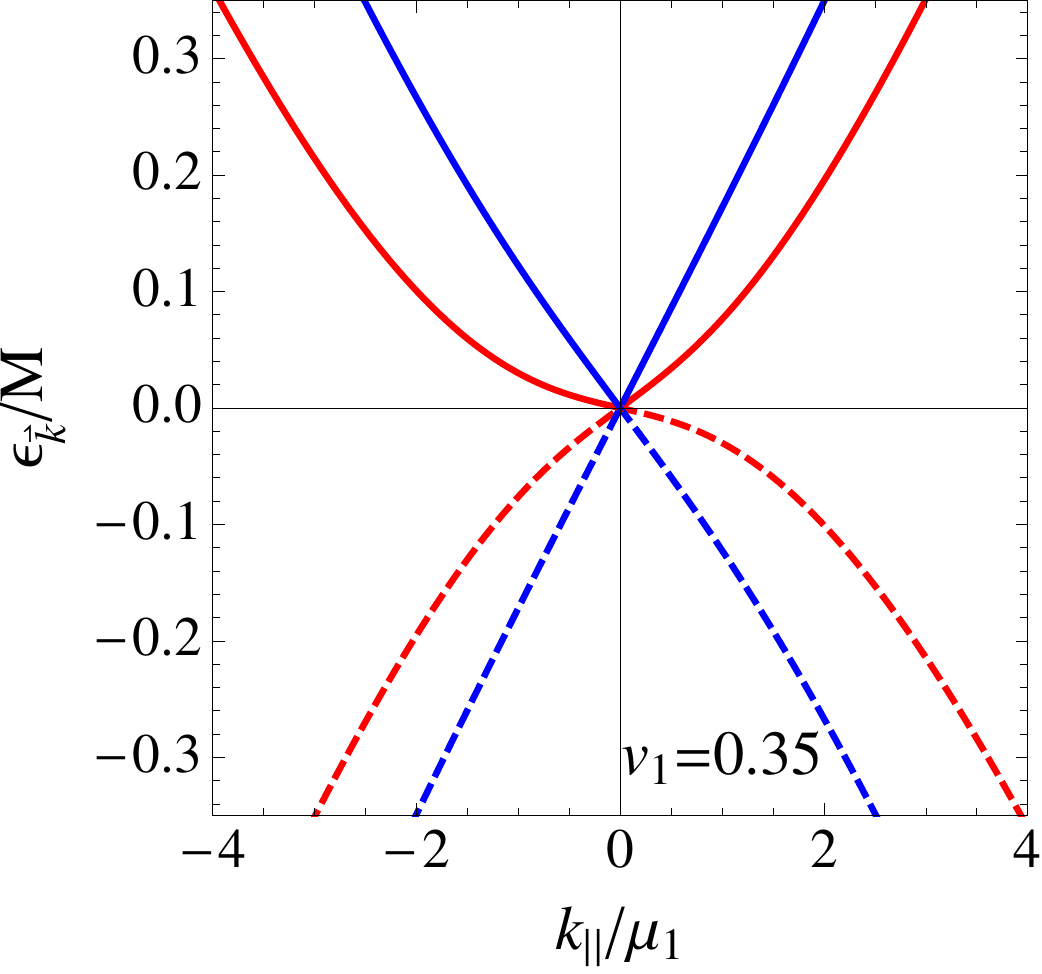}\includegraphics[width=0.235\textwidth]{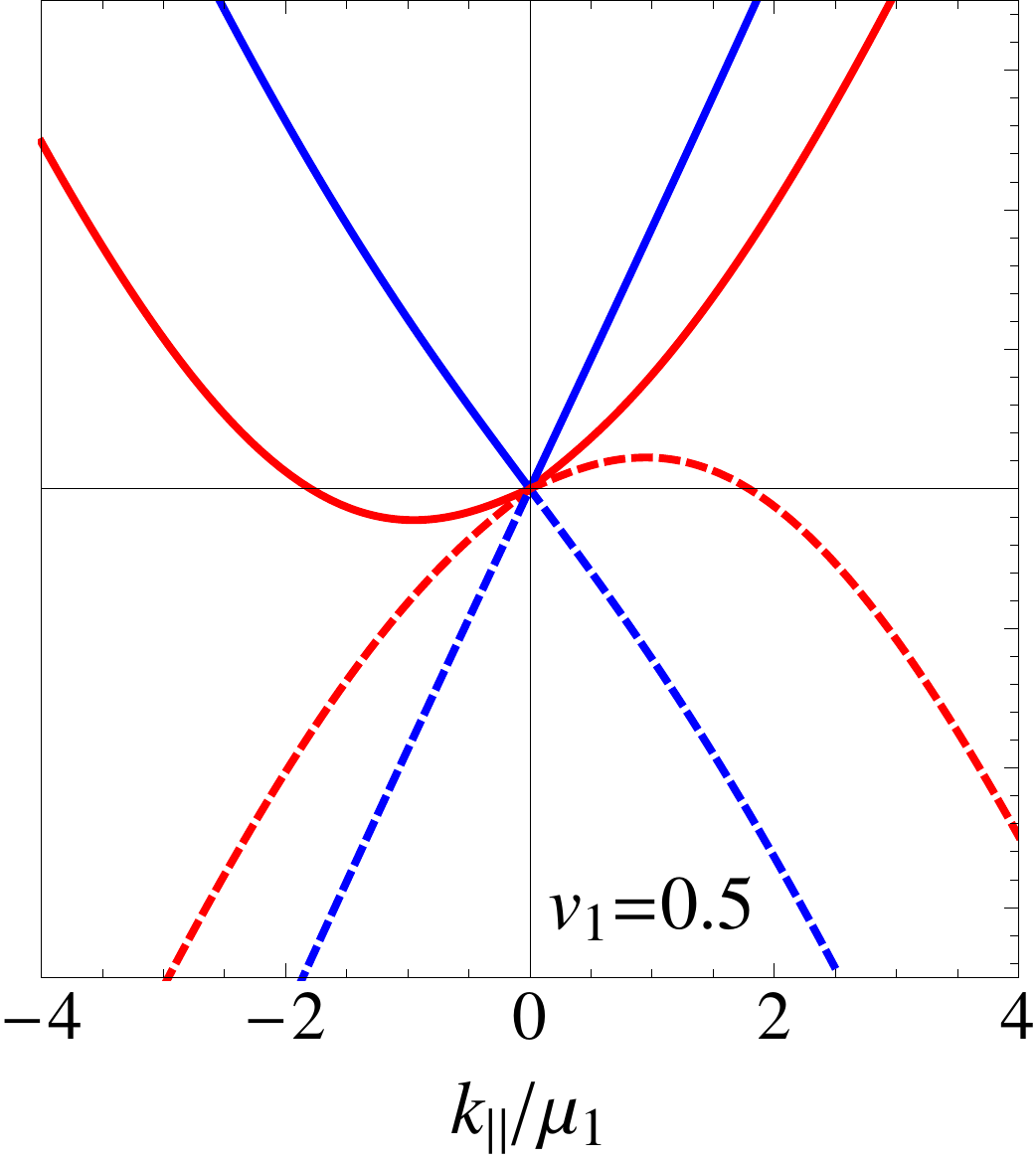}
\includegraphics[width=0.235\textwidth]{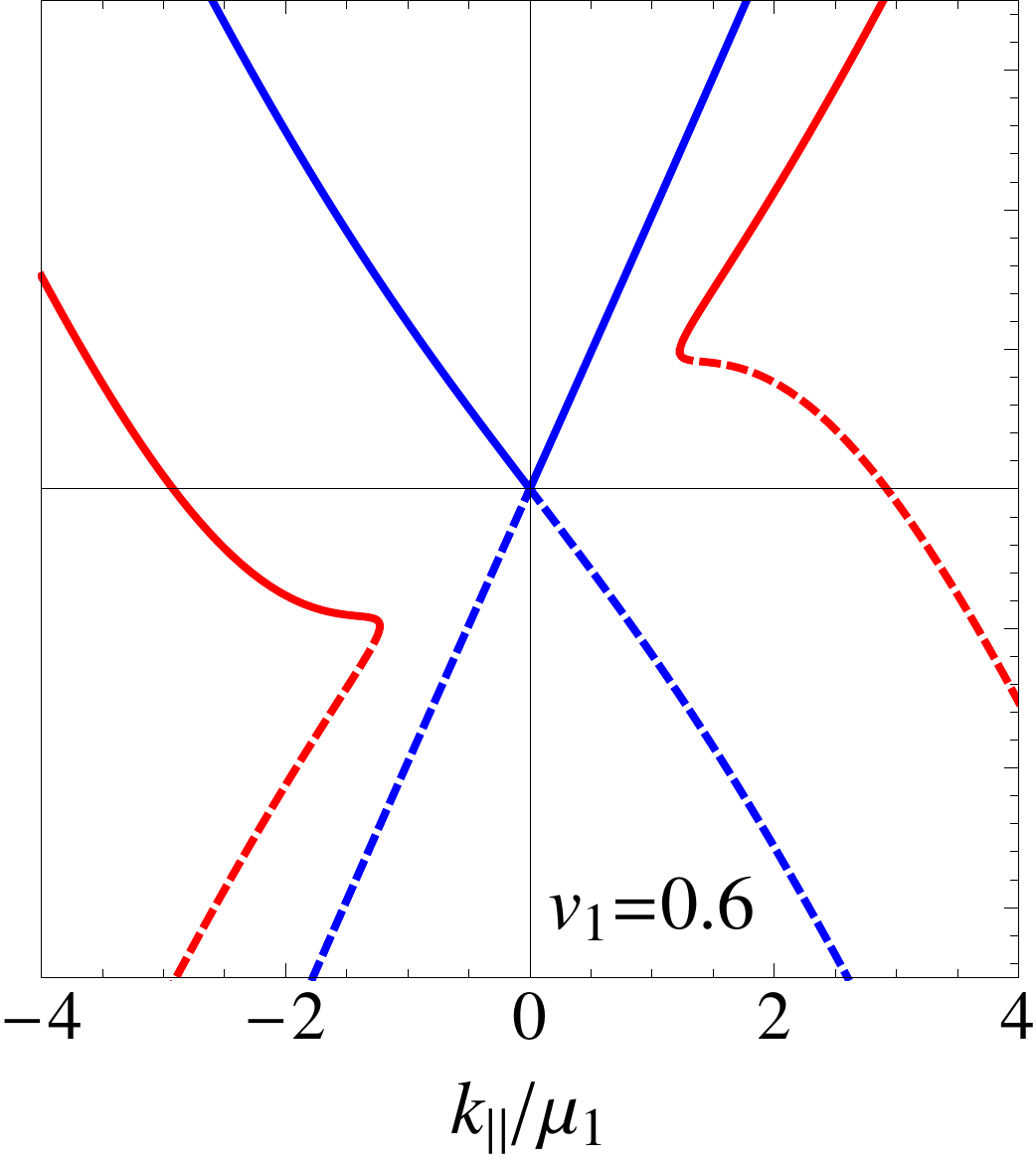}\includegraphics[width=0.235\textwidth]{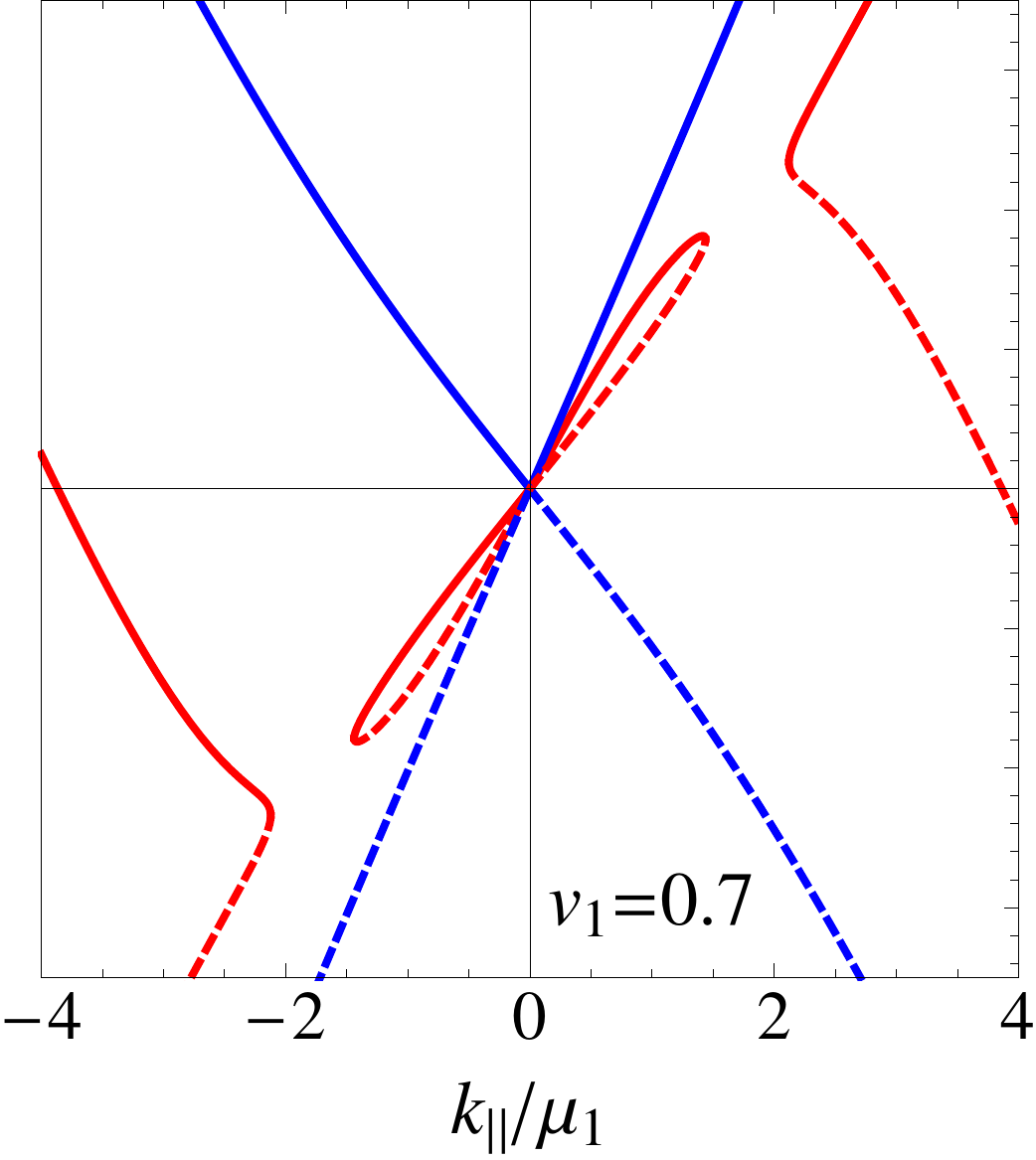}}
\caption{Goldstone dispersion relations with the parameters of the phase diagram in the left panel of Fig.\ \ref{fig:phases1} with $\mu_1/m_1=2.8$, $\mu_2/m_2=1.67$ (marked point in that 
phase diagram), and four different velocities $v_1$, parallel ($k_{||}>0$) and anti-parallel ($k_{||}<0$) to 
$\mathbf{v}_1$. 
The excitation energy $\epsilon_{\mathbf{k}}$ is normalized to the mass $M$ of the lighter of the two massive modes. For small momenta the dispersion relations are linear, their 
slope is shown in Fig.~\ref{figpolar} for all angles. For a given $k_{||}$ 
all four massless solutions of $\mathrm{ det}\,S^{-1}=0$, including the negative "mirror branches" (dashed lines), are presented. 
From the second panel on, the branches that were positive for vanishing superflow (solid lines) acquire negative energies for certain momenta. 
In the third and fourth panels there are gaps in the curves for certain momenta where $\epsilon_{\mathbf{k}}$ is complex, 
indicating a dynamical instability. }
\label{figdisp}
\end{center}
\end{figure}
\begin{figure}[t]
\begin{center}
\hbox{\includegraphics[width=0.27\textwidth]{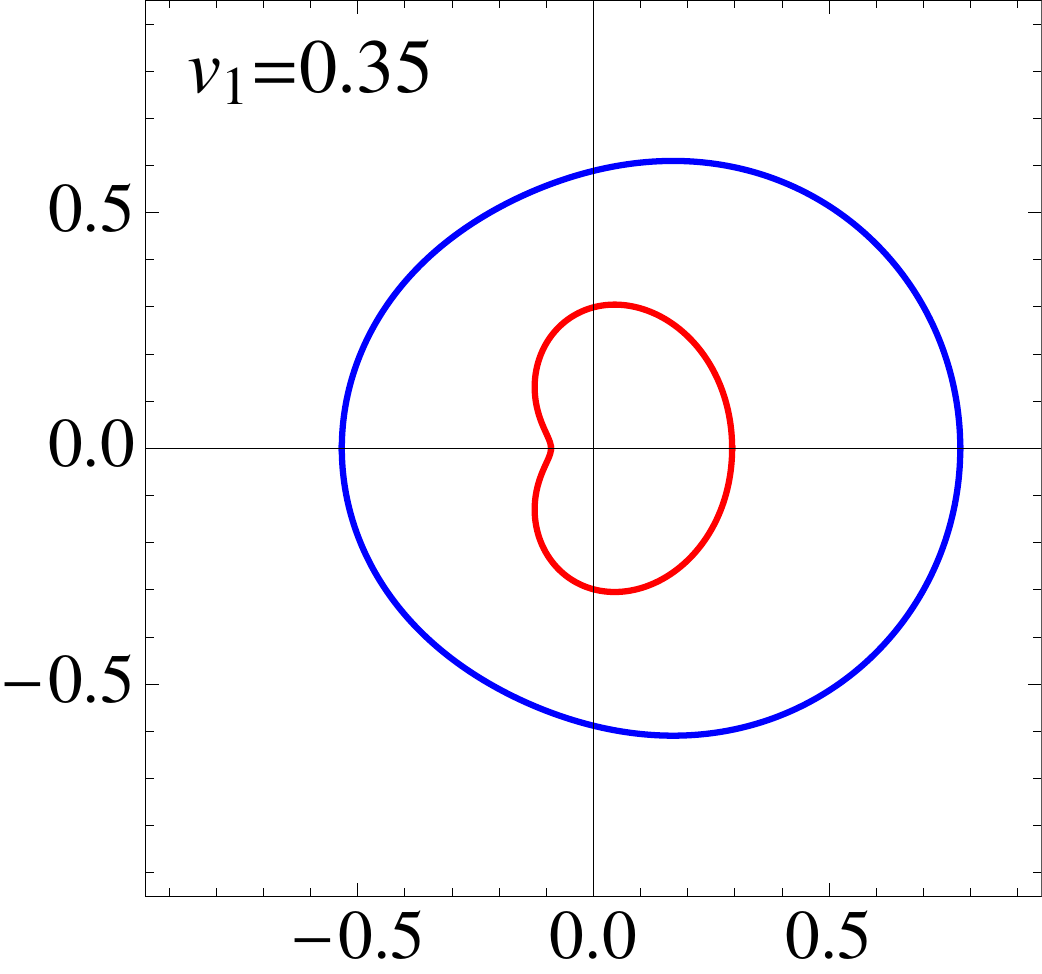}\,\includegraphics[width=0.235\textwidth]{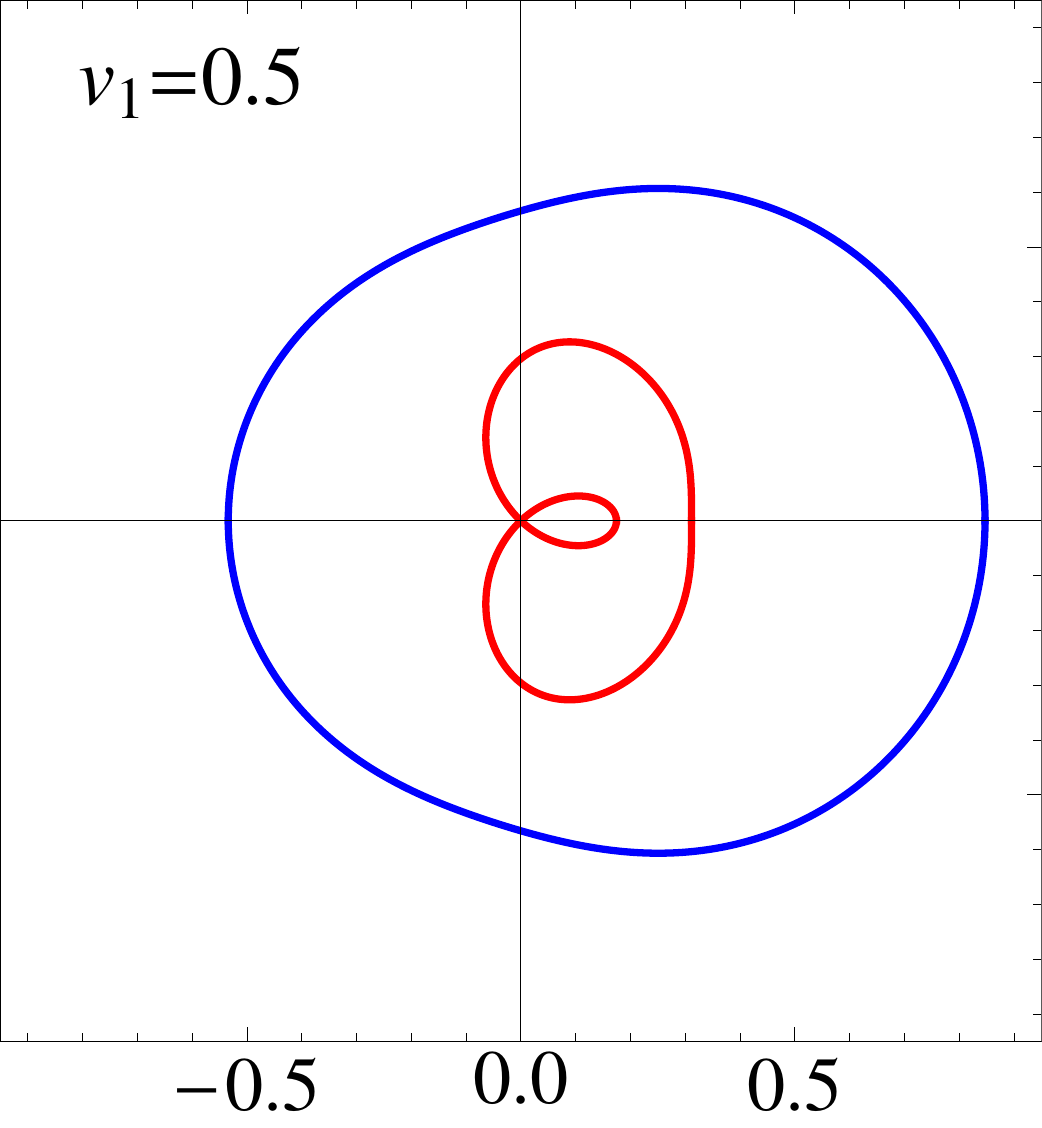}
\includegraphics[width=0.235\textwidth]{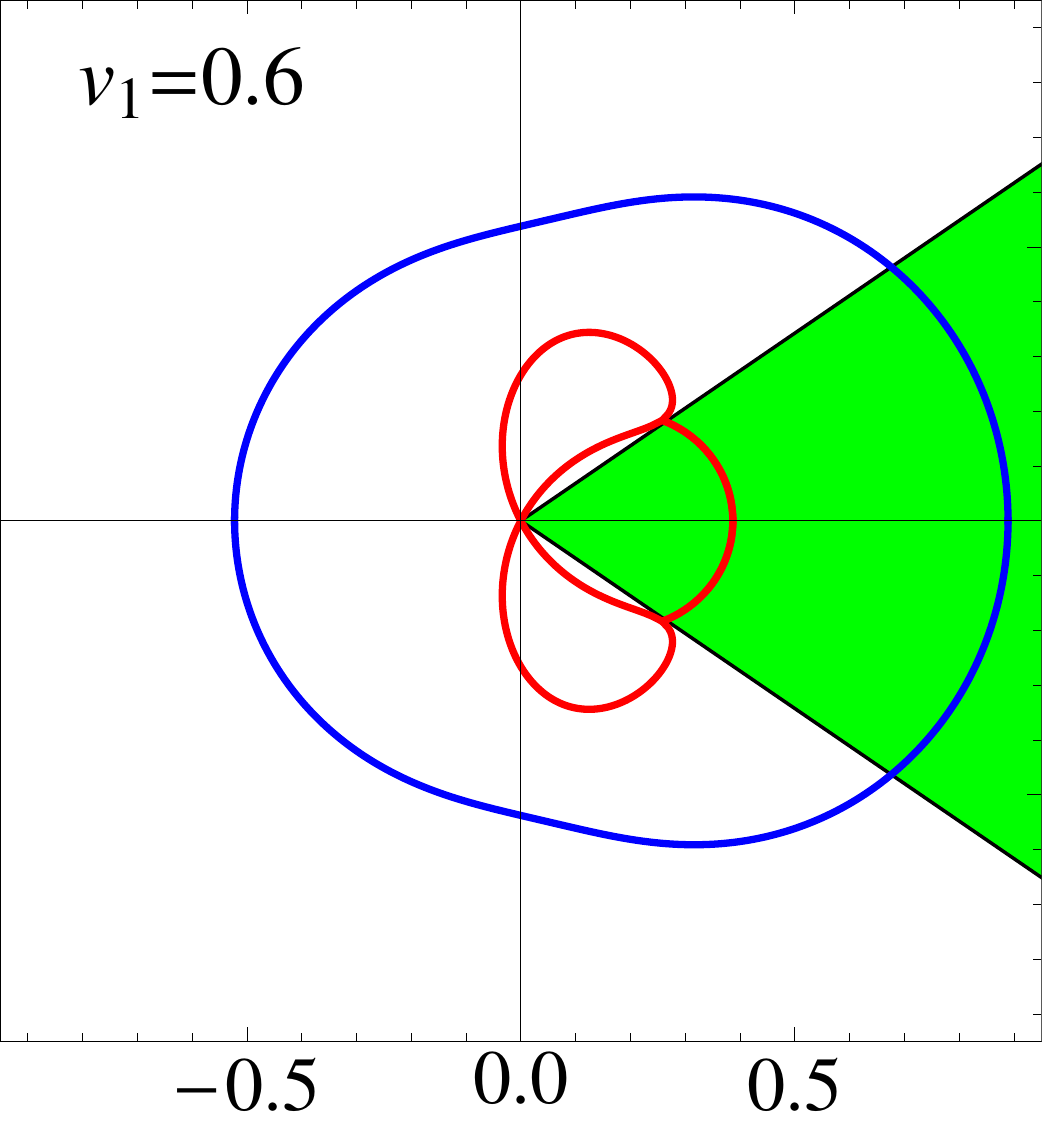}\,\includegraphics[width=0.235\textwidth]{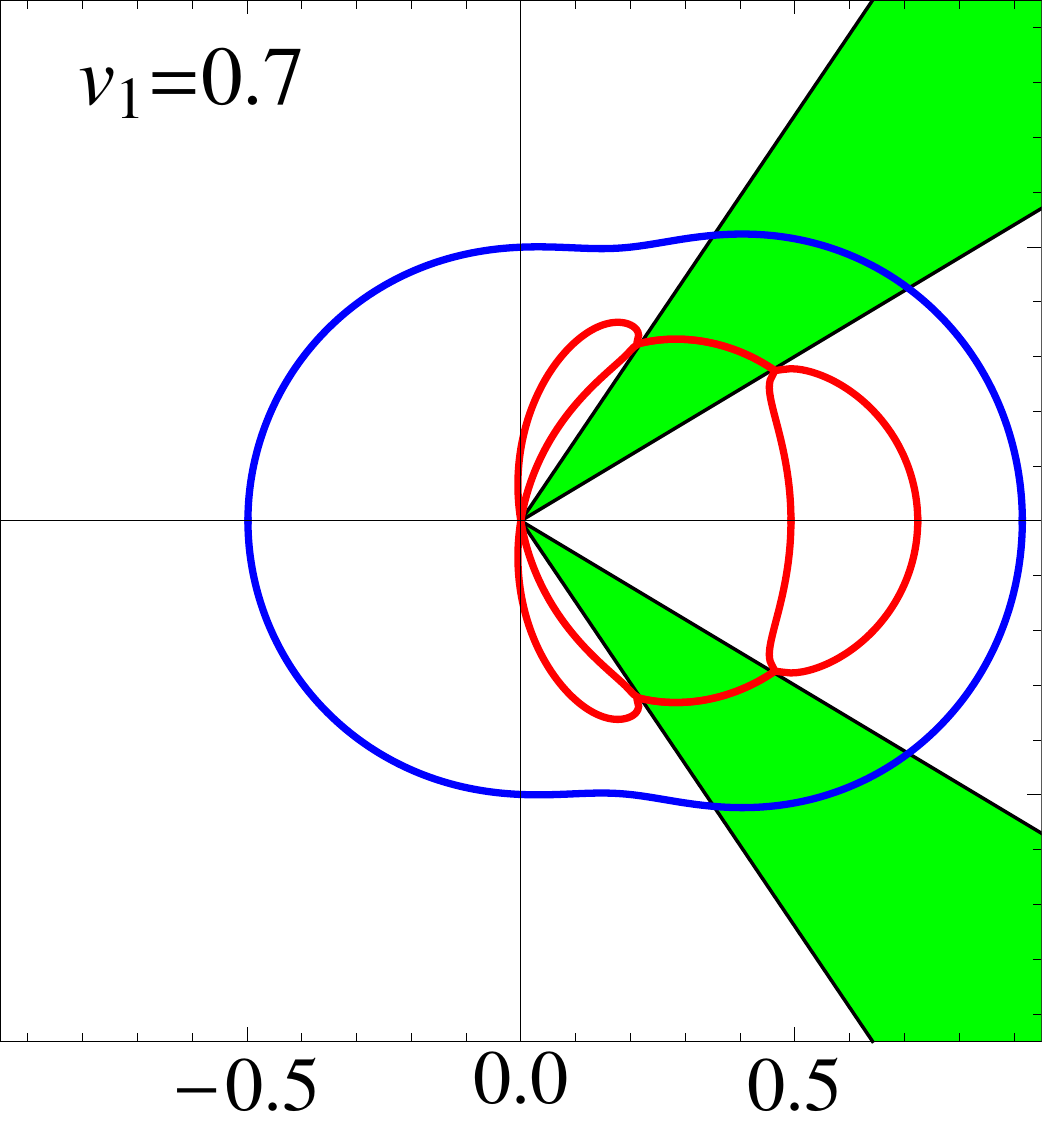}}
\caption{Real part of the sound speeds with the parameters of Fig.~\ref{figdisp} for all angles between the wave vector $\mathbf{k}$ and the velocity $\mathbf{v}_1$, 
and four different magnitudes $v_1$ (the distance from the origin to the curve is the speed of 
sound for a given angle). The velocity $\mathbf{v}_1$ points to the right. In the two last panels there are unstable directions for which the sound speed becomes complex and the 
real parts of two branches coincide, indicated by the shaded (green) areas.}
\label{figpolar}
\end{center}
\end{figure}

As discussed, the microscopic model contains several parameters, and a complete survey of the parameter space is very tedious and not necessary for our purpose. Of course, many details depend on the 
specific choice of the parameters. For instance, the difference in the phase structure if we allow for a non-entrainment coupling $h$ instead of the entrainment coupling $g$, becomes apparent by comparing Fig.~\ref{fig:phases1} with Fig.~\ref{fig:phases1a}. The topology of the phase diagrams could be even richer if we had allowed for both couplings simultaneously.
However, the phase structure is not our main concern, we are rather interested in the instabilities at nonzero fluid velocities. 
To this end, we consider a particular point of the phase diagram in the left panel of Fig.~\ref{fig:phases1}, marked with a red cross. This means that for the moment we restrict ourselves to the case of a pure entrainment coupling, setting $h=0$. The behavior of this point 
in terms of dynamical and energetic instabilities is generic in a sense that we will discuss in Sec.~\ref{sec:analysis}. 
In fact, for the two-component superfluid, the following discussion about dynamical and energetic instabilities would be quantitatively the same for a non-entrainment coupling, 
entrainment is \textit{not} crucial. Therefore, (the non-relativistic limit of) the results will also be relevant for dilute atomic gases, where entrainment is believed to be negligible.
For a system of two normal fluids, however, entrainment does make a qualitative 
difference for the dynamical instability, see Fig.~\ref{fignn} and discussion at the end of Sec.~\ref{sec:analysis}.
The entrainment terms themselves contain two different coupling constants and in principle we can choose both of them independently. 
We have seen that for the tree-level potential only the combination $g = (g_1-g_2)/2$ 
matters, while in the propagator both constants enter separately. One can check, however, that for the linear, low-energy part of the dispersion, again only $g$
matters, such that we can keep working with the single entrainment parameter $g$. Only in Fig.~\ref{figdisp} we show dispersion relations that go beyond linear order in momentum,
and in this case we have made the choice $g_2=0$, such that $g=G$ (again with quantitative, but for our conclusions irrelevant, changes if $g_1$ and $g_2$ are chosen 
differently). 

In Fig.~\ref{figdisp} we show the dispersion relations of the two Goldstone modes for four different values of $v_1$, with $v_2=0$, i.e., 
the calculation is done in the rest frame of superfluid 2. All four massless solutions of $\mathrm{det}\,S^{-1}=0$ are plotted. As explained in the previous chapter, for each solution $\epsilon_{r,\mathbf{k}}$ (solid lines),  
$-\epsilon_{r,-\mathbf{k}}$ is also a solution
(dashed lines). Several observations are obvious from the four panels. First of all we see the trivial effect that 
a nonzero superflow (here of superfluid 1) leads to anisotropic dispersion relations, in particular to different sound speeds parallel and anti-parallel to the superflow
(first panel). Beyond a certain value of the superflow, negative excitation energies appear for small momenta (second panel), before the energies become 
complex at small momenta (third panel), and this complex region moves to larger momenta (fourth panel). Complementary information for the same parameter set is shown 
in Fig.~\ref{figpolar}. The four panels in this figure show less in the sense that only the sound speeds are plotted (i.e., the slope of the Goldstone dispersion at small 
momenta), but they show more in the sense that these speeds are shown for \textit{all} angles between the direction of the sound wave and the superflow. Also, 
in Fig.~\ref{figpolar} we have restricted ourselves to positive excitations, i.e., only the branches of the upper half of Fig.~\ref{figdisp} are shown in Fig.~\ref{figpolar}.
For instance, for $v_1=0.5$, there is a branch with negative energy in the upstream direction (anti-parallel to $\mathbf{v}_1$), see the lower (red) solid line in the second panel of 
Fig.\ \ref{figdisp}. At the same time, the "mirror branch" in the downstream direction has acquired positive energy, the upper (red) dashed curve in the same panel. The latter is shown 
as a solid curve in the second panel of Fig.\ \ref{figpolar}, which also shows that this branch exists for all angles in the half-space $\mathbf{k}\cdot\mathbf{v}_1>0$.

\begin{figure} [t]
\begin{center}
\hbox{\includegraphics[width=0.5\textwidth]{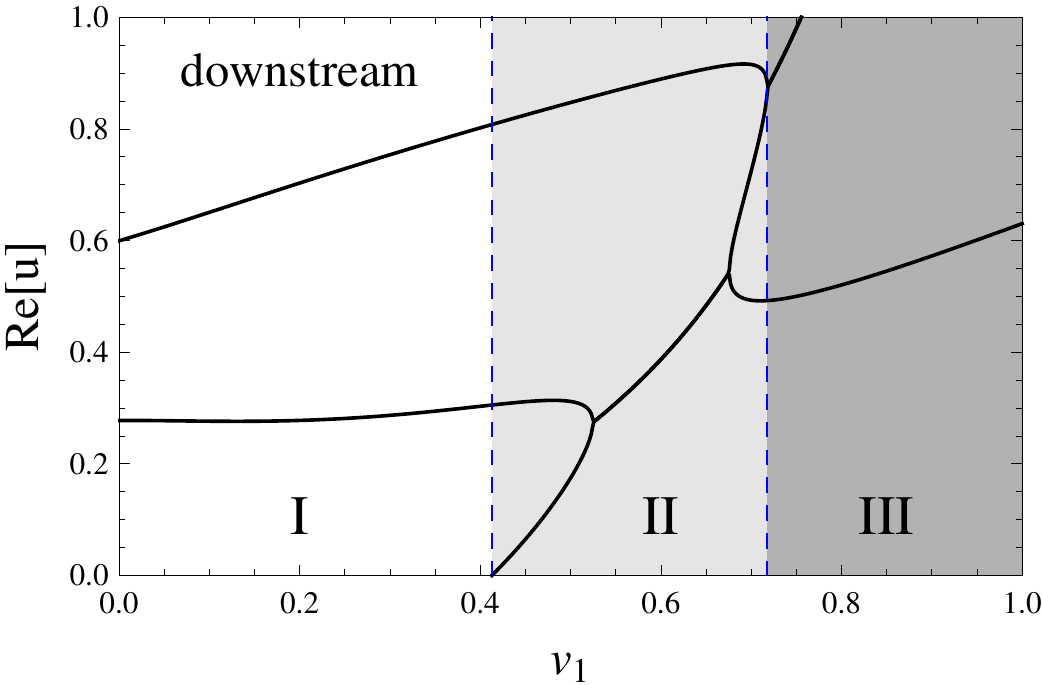}\includegraphics[width=0.5\textwidth]{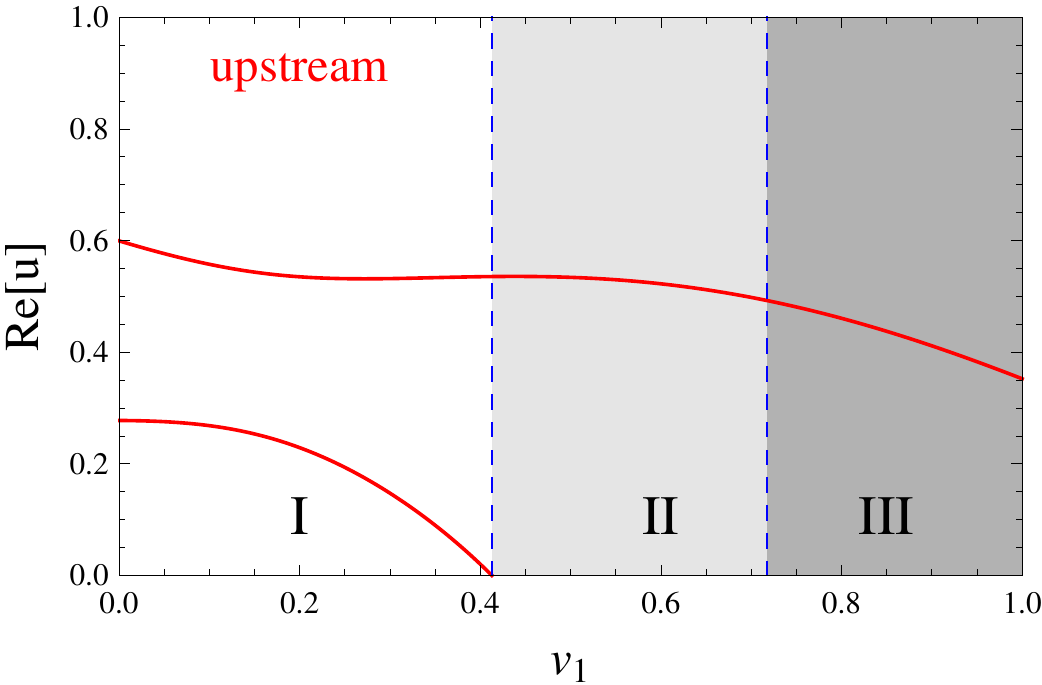}}
\caption{Real part of the sound speeds parallel ("downstream") and anti-parallel ("upstream") to the velocity of superfluid 1, for the parameters of Figs.\ \ref{figdisp} and
\ref{figpolar}. Region I: stable; region II: energetically unstable and containing a dynamically unstable region; region III: single-superfluid phase $\mathrm{ SF}_2$ preferred.}
\label{figUpDown}
\end{center}
\end{figure}

The most relevant information of Figs.\ \ref{figdisp} and \ref{figpolar} is extracted in Fig.\ \ref{figUpDown}, which we now use to discuss the various instabilities. 

\subsection{Energetic Instability}
The transition from region I to region II is defined by the point where the excitation energy of one of the 
Goldstone modes becomes negative. Such a point is well known in the theory of superfluidity. It exists even for a single superfluid and the corresponding critical velocity
is nothing but Landau's critical velocity. Compared to Landau's original argument, our calculation is a generalization to a two-component superfluid system and to the relativistic 
case. Since in our system the onset of negative excitation energies is equivalent to the sound speed becoming zero, we can also use Eq.\ (\ref{detu})
to compute the critical velocity. From this equation we easily read off that $u=0$ occurs for $\mathrm{ det}\, \chi_0=0$  
which, in turn, occurs at the point where one of the eigenvalues of $\chi_0$ changes its sign; 
both eigenvalues are positive in region I, one eigenvalue is negative in region II. 
Consequently, in our approximation, where the quasiparticle modes and the sound modes coincide, 
Landau's critical velocity is a manifestation of the negativity of the "current susceptibility", 
the second derivative of the pressure with respect to the spatial components of the conjugate 
four-momentum $p^\mu=\partial^\mu\psi$. A more common susceptibility is the "number susceptibility", the 
second derivative with respect to the chemical potential, which is the \textit{temporal} component of the 
conjugate four-momentum. In this case, the negativity implies that the density decreases
by increasing the corresponding chemical potential, which indicates an instability. This is completely analogous to the spatial components: here, a negative susceptibility 
implies that the three-current decreases by increasing the corresponding velocity, 
again indicating an unstable situation. In general, one has to check both stability criteria 
separately; Landau's critical velocity does not necessarily coincide with the onset of a negative current susceptibility. An example is a single superfluid at nonzero temperature, 
where neither of the two sound modes coincides with the quasiparticle mode and thus the connection between the quasiparticle energy and the 
susceptibility cannot be made.

\begin{figure}[t]
\begin{center}
\hbox{\includegraphics[width=0.485\textwidth]{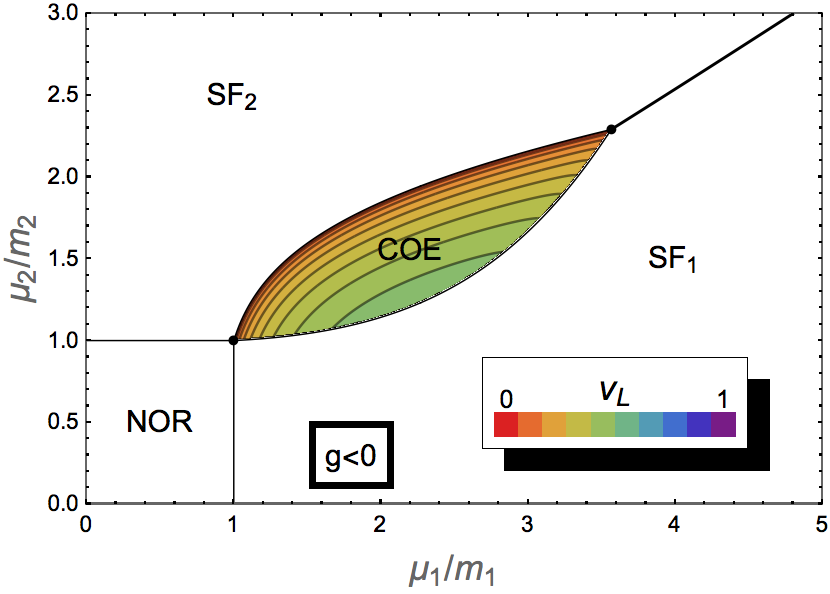}\includegraphics[width=0.495\textwidth]{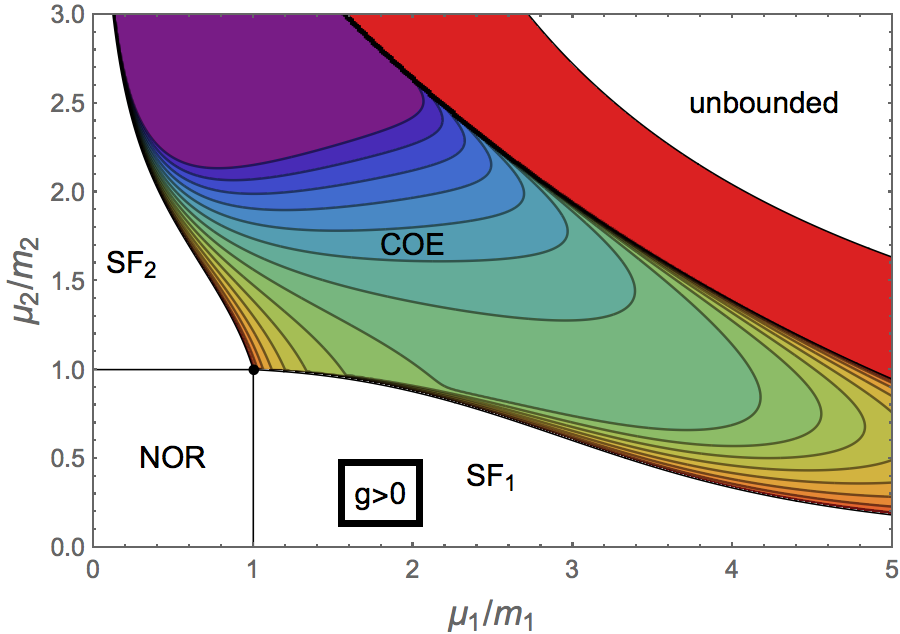}}
\caption{Landau's critical velocity $v_L$ ("energetic instability") in the COE phase of the two phase diagrams from Fig.\ \ref{fig:phases1}, i.e., for negative (left panel) 
and positive (right panel) values of the entrainment coupling $g$. 
The asymmetry between the two superfluids arises because we work, without loss of generality, in the rest frame of superfluid 2, i.e., the critical velocity
refers to $\mathbf{v}_1$. 
In the right panel, the COE region is separated from the region where the tree-level potential is unbounded from below
by a band where $v_L=0$, i.e., states in this band are energetically unstable already in the static case although they appear stable within our uniform ansatz.
The critical velocity for the two-stream instability is, where it exists, larger than Landau's critical velocity throughout the phase diagram.  
The phases where only one field condenses, $\mathrm{ SF}_1$ and $\mathrm{ SF}_2$, also become unstable beyond certain critical velocities, but these are not shown here.}
\label{figVc}
\end{center}
\end{figure}

In Fig.~\ref{figVc} Landau's critical velocity $v_L$ for all values of the chemical potentials for which the COE phase is preferred in the absence of any fluid 
velocity is shown. We see that 
for negative values of the entrainment coupling \textit{all} states that where stable at $\mathbf{v}_1=0$ become energetically unstable at some critical velocity. This is different 
for positive values of the entrainment coupling where there is a region in the phase diagram with no instability, indicated by $v_L=1$. This case is also interesting because we find a region 
with vanishing critical velocity, i.e., an unstable COE state which appeared to be stable in the calculation based on the tree-level potential. 
This means that the COE phase may well be the global 
minimum of the potential $U$ within our ansatz of a uniform condensate, but it may be a saddle point if an anisotropic or inhomogeneous condensate is allowed for. 

Besides the negativity of the susceptibility, we can describe the energetic instability also in the following intuitive way, using the picture of the sound waves: 
if a sound mode propagates "upstream", it is natural to expect that it is slowed down compared to the situation without superflow. 
Landau's critical velocity
is the point at which the mode is slowed down so much that it comes to rest, and for larger velocities it appears to propagate in the "wrong" direction. 
It then shows up as an additional 
downstream mode, which is manifest in all Figs.~(\ref{figdisp}) -- \ref{figUpDown}, for instance in 
the left panel of Fig.~\ref{figUpDown} where
at the transition between regions I and II a third mode appears in the downstream direction.  
Rephrased in this way, the energetic instability encountered here is analogous to the 
Chandrasekhar-Friedman-Schutz (CFS) instability \cite{PhysRevLett.24.611,1978ApJ...221..937F} (including the $r$-mode instability \cite{Andersson:1997xt}, explained in Sec.~\ref{sec:QMinCS}) known from 
astrophysics\footnote{In the astrophysical literature such an instability is called \textit{secular} instability. 
Here we use the term \textit{energetic} instability synonymously, following the condensed matter literature, see for instance Ref.\ \cite{PhysRevA.87.063610}.}: 
in that case, certain oscillatory modes of a neutron star can be "dragged" by the rotation of the star, such that from a distant observer they appear to propagate, say, 
counter-clockwise, 
while in the co-rotating frame they propagate clockwise. This is exactly the same kind of behavior as we observe here, we just have to replace the oscillatory mode of the 
neutron star by the sound mode and the angular velocity of the star by the velocity of the superfluid. It is very instructive to develop this analogy a bit further: in a rotating neutron 
star, the instability is realized due to the emission of gravitational waves, which are able to transfer angular momentum from the system. In the same way, as already pointed out
by Landau in his original argument, the negative excitation energies lead to an instability if there is a mechanism that can exchange momentum with the superfluid, 
for example the presence of the walls of a capillary in which the superfluid flows. The exchange of angular momentum or momentum is crucial for the instability to set in,
and knowledge about how this exchange works is required to determine the 
time scale on which the instability operates. This is in contrast to the \textit{dynamical} instability, where a time scale is inherent and which we discuss next.

\subsection{Dynamical Instability} 
In a sub-region of region II, two of the sound speeds acquire an imaginary part of opposite sign and equal magnitude, while their real parts 
coincide (since the polynomial from which the sound speeds are computed has real coefficients, for any given solution also the complex conjugate is a solution). This means that the 
amplitude of one of the modes is damped, while it increases exponentially for the other one, with a time scale given by the magnitude of the imaginary part. 
This kind of instability is well-known in two-fluid or multi-fluid plasmas \cite{Buneman:1959zz,1963PhRvL..10..279F,2001AmJPh..69.1262A}, 
and is termed two-stream instability or counterflow instability. 
In plasma physics, usually the fluids are electrically charged, and the calculation of the sound modes is somewhat different because the hydrodynamic equations
become coupled to Maxwell's equations. This provides an "indirect" coupling between the two fluids, while we have coupled the two fluids "directly" through a coupling 
term in the Lagrangian. In the context of two-component 
superfluids the two-stream instability has been discussed in a non-relativistic context \cite{2004MNRAS.354..101A,PhysRevA.87.063610,2015EPJD...69..126A} and, without reference to superfluidity, 
in a relativistic context \cite{Samuelsson:2009up}. It has also been discussed in a single, relativistic superfluid at nonzero temperature \cite{Schmitt:2013nva}. 
For a truly dynamical discussion of the two-stream instability one has to go beyond linearized hydrodynamics \cite{Hawke:2013haa} and possibly take into account the formation of
vortex rings and turbulence \cite{2011PhRvA..83f3602I}.  
Our present calculation only yields the time scale of the exponential growth at the onset of the instability.

\subsection{Phase Transition to Single-Superfluid Phase}   
As the phase diagram in the left panel of Fig.\ \ref{fig:phases1} shows, a point within the COE phase with fixed chemical potentials 
will simply leave the COE region beyond a critical velocity. Therefore, in region III a different phase, even within our simple uniform ansatz,
is preferred, in this case the phase $\mathrm{ SF}_2$, where only field 2 forms a condensate. This instability is also seen in the sound modes because the COE 
phase ceases to be a local minimum at that point. In other words, the phase transition is of second order.

\section{Further Analysis of Instabilities and Comparison to Normal Fluids} 
\label{sec:analysis}

In the results shown so far, the dynamical instability occurs in an energetically unstable region, i.e., a complex sound speed only occurs if there is already a negative excitation 
energy. In other words, the two branches that merge in the left panel of Fig.~\ref{figUpDown} are not the two "original" downstream modes, but one downstream mode and the original 
upstream mode that has changed its direction. Two questions arise immediately: 

\begin{enumerate}

\item[$(i)$] Does the occurrence of complex sound speeds have any physical meaning if they occur in an 
energetically unstable region? 

\item[$(ii)$] Is this behavior generic or can a dynamical instability arise even in the absence of an energetic instability? 

\end{enumerate}

As for point $(i)$, we will give a brief qualitative discussion, while for point $(ii)$ we will give a definite answer within the presented approach.

$(i)$ The problem that seems to arise is that an energetically unstable system will choose a different configuration, either another equilibrium state with lower free energy than the
one that exhibits the negative excitation energies, or it will refuse to be in equilibrium altogether. In either case, the two-stream instability we have observed may well be absent 
in the new configuration, simply because the calculation in the energetically unstable state is not valid since this state is not realized. 
As mentioned above, the energetic instability
is, in our approximation, identical to a negative current susceptibility. 
Negative number susceptibilities are well-known indicators of an instability, for instance for Cooper-paired systems  
with mismatched Fermi surfaces \cite{Gubankova:2006gj,Deng:2006ed,Huang:2006kr}, where the resolution may be phase separation, i.e., a spatially inhomogeneous 
state with paired regions separated from unpaired regions. In our present calculation, the negative susceptibility may as well be cured by an inhomogeneous state which we 
have not included into our ansatz, for example in the form of stratification of the two superfluid components \cite{2004LaPhL...1...50Y,2009JPhCS.150c2057M} or  
a crystalline structure of the condensates \cite{Landea:2014naa}. While these solutions to the problem concern equilibrium states, the fate of an energetically unstable state 
in a real physical system, be it in a neutron star or in ultra-cold atoms, may be more complicated. As mentioned above, the realization of the energetic (or secular, in the 
astrophysical terminology) instability depends on a mechanism that is able to transfer momentum to and from the system. If such a mechanism is absent or operates on a large
time scale it is thus conceivable (depending on the actual physical situation) 
that the two-component superfluid becomes unstable only at the larger critical velocity where the 
two-stream instability sets in. 

\begin{figure} [t]
\begin{center}
\hbox{\includegraphics[width=0.355\textwidth]{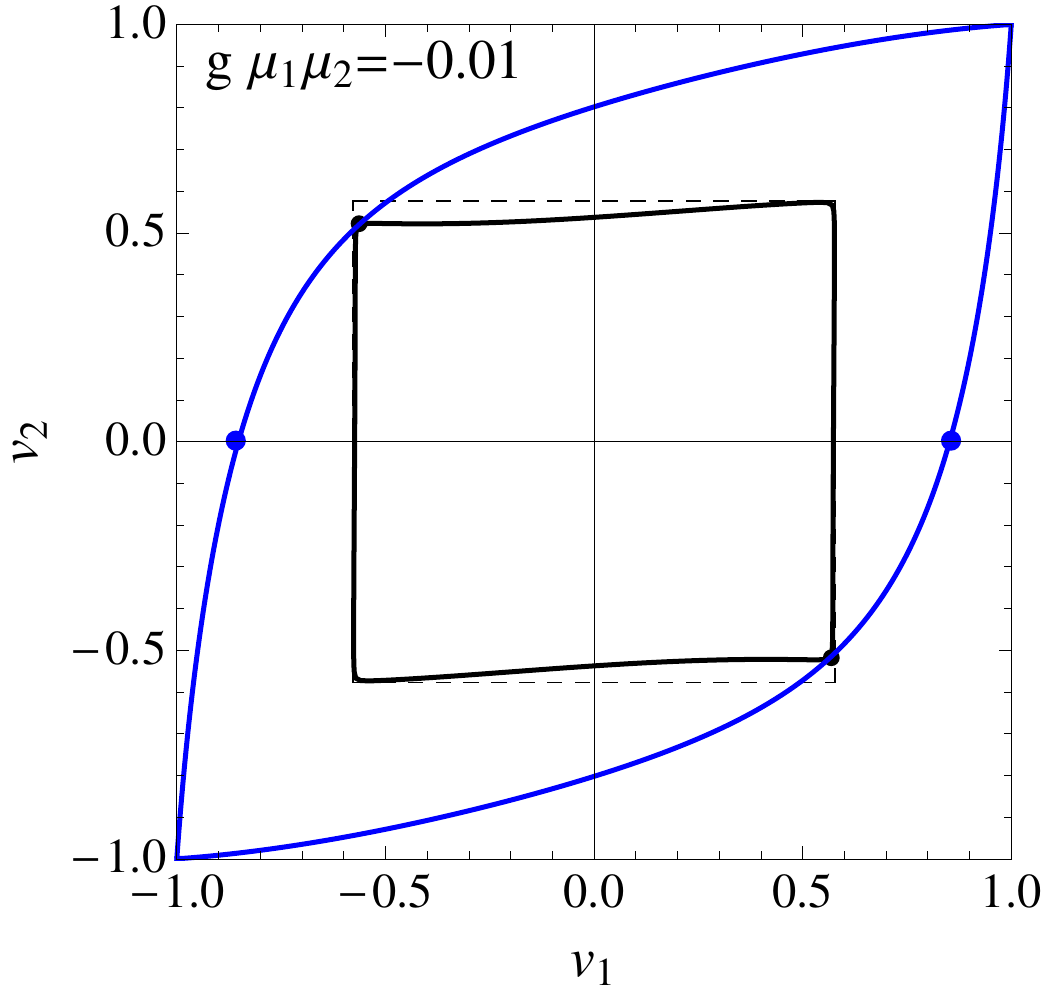}\includegraphics[width=0.31\textwidth]{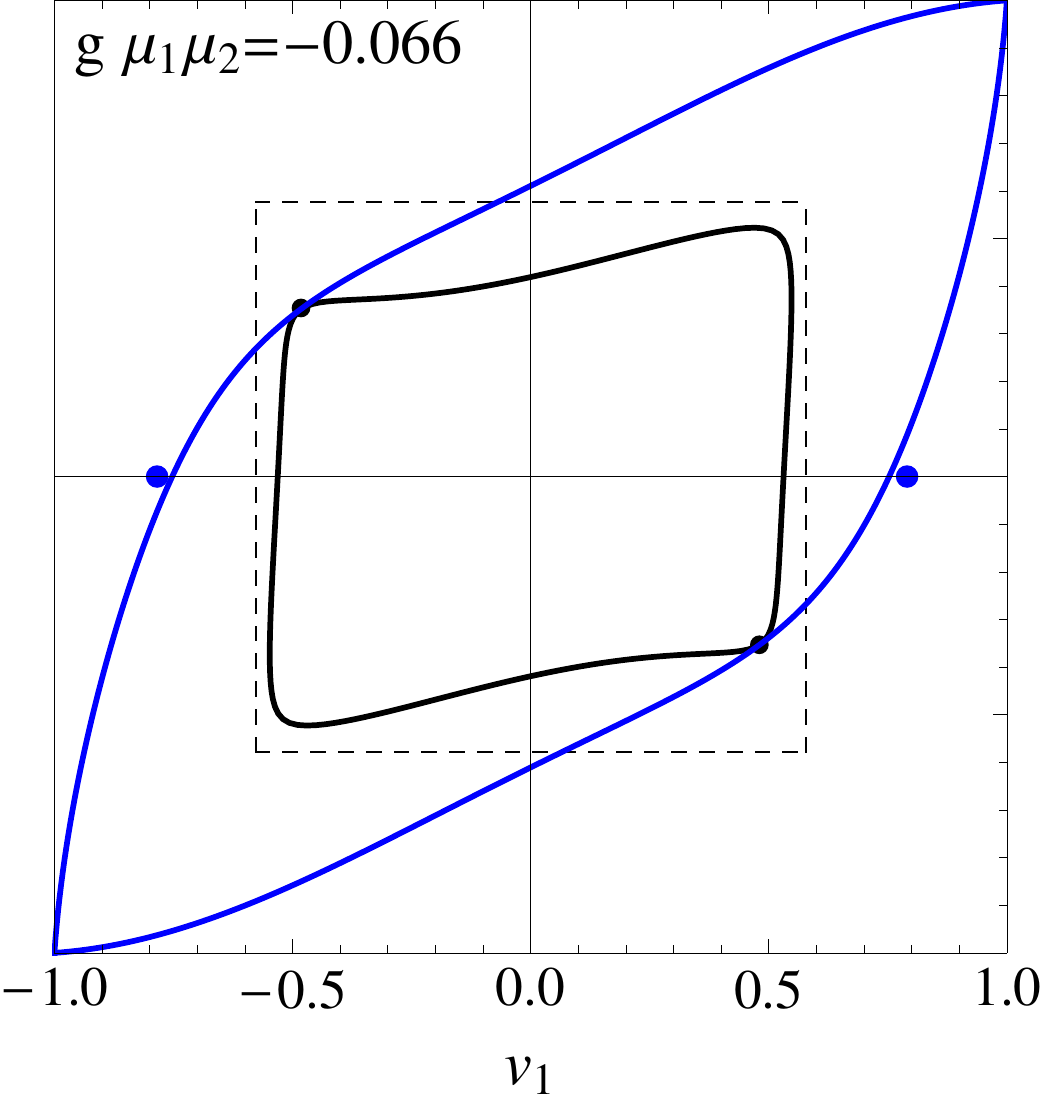}
\includegraphics[width=0.31\textwidth]{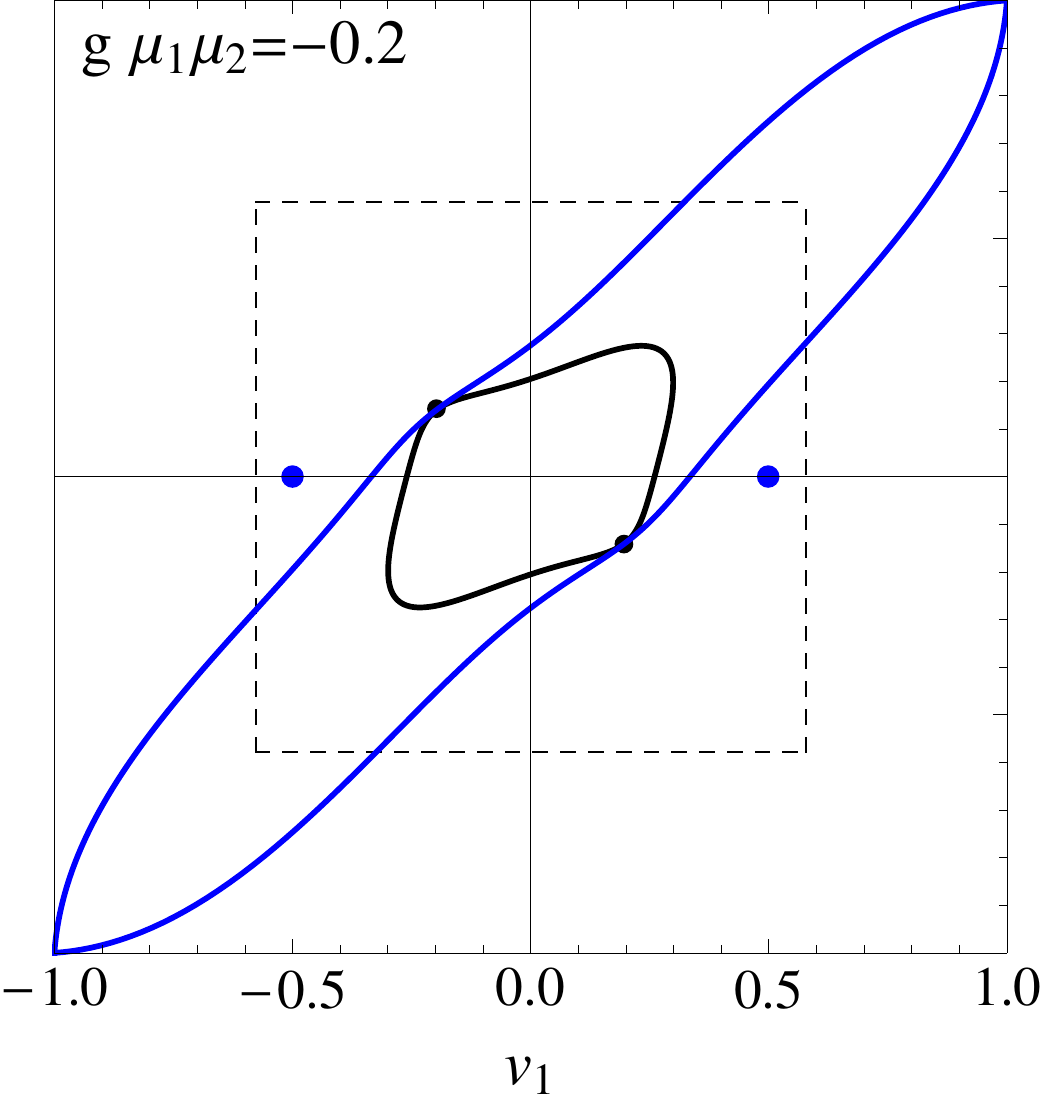}}
\caption{Landau's critical velocity $v_L$ [inner (black) solid curve] and critical velocity for the two-stream instability $v_\mathrm{ two-stream}$ 
[outer (blue) solid curve] for two superfluids moving with velocities $v_1$ and $v_2$
parallel [$\mathrm{ sgn}(v_1v_2)>0$] or anti-parallel [$\mathrm{ sgn}(v_1v_2)<0$] to each other for negative entrainment coupling $g$ of three different strengths, and $m_1=m_2=0$, 
$\lambda_1=0.3$, $\lambda_2=0.2$. (Angles in this plot correspond to different ratios $v_1/v_2$, not to angles between $\mathbf{v}_1$ and $\mathbf{v}_2$, which are always aligned or
anti-aligned.)  The dashed square shows Landau's critical velocity $1/\sqrt{3}$ in the absence of a coupling between the fluids. The large (blue) dots on the horizontal axis mark the 
analytical result for small coupling strength for the onset of the two-stream instability (\ref{v2stream}), while the small (black) dots mark the points where both 
critical velocities coincide.} 
\label{figv1v2}
\end{center}
\end{figure}

$(ii)$
First of all we recall that the negative excitation energies and Landau's critical velocity are frame dependent. So far we have worked in the rest frame of one of the 
two superfluids. But one can well imagine that there is a third meaningful reference frame, for instance the rest frame of the entropy fluid if we consider 
nonzero temperatures, which is the most convenient rest frame in a field-theoretical calculation \cite{2013PhRvD..87f5001A}. 
Therefore, for a general analysis of Landau's critical velocity, we should allow both fluids to move with nonzero velocities $\mathbf{v}_1$ and $\mathbf{v}_2$. 
(In Landau's original argument for a single superfluid a second reference frame is given by the capillary.) Of course, the 
calculation becomes very tedious if we allow for three different vectors with arbitrary directions: $\mathbf{v}_1$, $\mathbf{v}_2$, and the wave vector $\mathbf{k}$. Therefore,
we restrict ourselves to the case where all these vectors are aligned with each other. As we have seen in Fig.\ \ref{figpolar}, it is the downstream direction where energetic and
dynamical instabilities set in "first" (i.e., for the smallest velocities). Therefore, it \textit{is} a restriction to consider only aligned $\mathbf{v}_1$, $\mathbf{v}_2$, 
but once they are aligned it is no further restriction to align $\mathbf{k}$ too if we are interested in the critical velocities.
 
In Fig.\ \ref{figv1v2} we plot Landau's critical velocity $v_L$ and the critical velocity for the two-stream instability $v_\mathrm{ two-stream}$
for arbitrary values of $v_1$ and $v_2$, and three different values of the entrainment coupling. If the fluids were uncoupled, Landau's critical velocity for each of the fluids
is $1/\sqrt{3}$ in the ultra-relativistic limit (for simplicity, we have set the mass parameters to zero in this plot), 
irrespective of the velocity of the other fluid. This is indicated by the dashed square. A nonzero 
entrainment coupling reduces Landau's critical velocity, and it tends to do more so if the fluids move in opposite directions, where $\mathrm{ sgn}\,(v_1v_2)<0$. (This is 
different if we choose the opposite sign for the coupling constant, $g>0$: in this case Landau's critical velocity is \textit{enhanced} for anti-aligned flow.) This is a similar effect
as in the phase diagrams of Fig.\ \ref{fig:phases1}: an entrainment coupling $g<0$ disfavors the COE phase, it is stable only in a smaller region compared to the uncoupled case
in the $\mu_1$-$\mu_2$ plane 
(Fig.\ \ref{fig:phases1}) or in the $v_1$-$v_2$ plane (Fig.\ \ref{figv1v2}). The behavior of $v_\mathrm{ two-stream}$,  the outer instability curve, 
is easy to understand in the following way: one might expect the two-stream instability to depend only on the relative velocity of the two superfluids. So, if relativistic effects were 
neglected, one might expect two straight lines given by $v_2-v_1=\pm \mathrm{ const}$. The actual curves are different for two reasons: first, relativistic effects bend the 
curves in the $v_1$-$v_2$ plane according to the relativistic addition of velocities and, second, the velocities also have a nontrivial effect on the condensates of the superfluids,
as already discussed in the context of the phase diagram in the $\mu_1$-$\mu_2$ plane. 

For $v_2=0$ and small values of the entrainment coupling, one can find a simple analytical expression for the critical velocity.
The two stream instability occurs if the nature of the roots of the
polynomial for the sound modes change. In the stable region, there
are four real solutions, where two can be disregarded. At the onset
of the instability, two of the solutions merge. Since a quartic polynomial
always has four complex solutions, the merged solution becomes complex,
where the complex conjugated number is also a solution. The nature
of the solutions can be determined from the discriminant of quartic
polynomials, with the coefficients $a$ trough $e$ of the quartic function
$au^{4}+bu{}^{3}+cu^{2}+du+e=0$,

\begin{align}\Delta= & 256a^{3}e^{3}-192a^{2}bde^{2}-128a^{2}c^{2}e^{2}+144a^{2}cd^{2}e-27a^{2}d^{4}\\ \nonumber
 & +144ab^{2}ce^{2}-6ab^{2}d^{2}e-80abc^{2}de+18abcd^{3}+16ac^{4}e\\ \nonumber
 & -4ac^{3}d^{2}-27b^{4}e^{2}+18b^{3}cde-4b^{3}d^{3}-4b^{2}c^{3}e+b^{2}c^{2}d^{2} \, .
\end{align}

If $\Delta<0$, the equation has two real and two complex conjugate
roots, i.e. the system is two-stream unstable. If $\Delta>0$, the
function has four real (or four complex) solutions and no two-stream
instability occurs \cite{1922}. The coefficients of the quartic function
that enable us to derive the sound modes depend on the velocity of the
fluids also in a polynomial way. Therefore, it is possible to compute
an approximation for the critical velocity by calculating the power
series of the discriminant $\Delta$ in second order of the entrainment
coupling $g$, solve it for $v_{1}$ (for simplicity we are working
in the second fluid rest frame, i.e. $\mathbf{v}_{2}=0$), and to expand
it again to first order in $g$. In the non-relativistic
case without entrainment, the critical velocity is given by the sum
of the sound speeds of the two separate fluids, $v_{\mathrm{two-stream}}=c_{1}+c_{2}$.
In the ultra-relativistic case ($m_{1}=m_{2}=0)$, we know that the
sound speeds are given by the conformal limit, $c_{1}=c_{2}=\frac{1}{\sqrt{3}}$.
Using the relativistic velocity-addition formula, we compute as the
lowest order approximation,
\begin{equation}
v_{\mathrm{two-stream}}\approx\frac{c_{1}+c_{2}}{1+c_{1}c_{2}}=\frac{\sqrt{3}}{2}\,.
\end{equation}
The entrainment coupling $g$ reduces or enhances this result, depending on the sign of the coupling constant. The first correction in the entrainment coupling is given by
\begin{equation}\label{v2stream}
v_{\mathrm{two-stream}}=\frac{\sqrt{3}}{2}\left(1+g\frac{\lambda_{2}\mu_{1}^{4}+16\lambda_{1}\mu_{2}^{4}}{32\lambda_{1}\lambda_{2}\mu_{1}\mu_{2}}+{\cal O}\left(h^{2}\right)\right)\,,
\end{equation}
which shows good agreement with the full numerical result. 

We have indicated the value (\ref{v2stream}) in Fig.~\ref{figv1v2} as (blue) dots on the horizontal axis and see that the approximation becomes worse for larger couplings.  
In interpreting Eq.~(\ref{v2stream}) we have to keep in mind that we have determined the critical velocity at fixed 
$\mu_1$, $\mu_2$, i.e., we have fixed the chemical potentials in the "lab frame", not in the respective rest frames of the fluids. Therefore, increasing $v_1$ while keeping $v_2=0$ affects the condensate 1 and not the 
condensate 2 (in addition to making the two fluids move with respect to each other). This asymmetry is manifest in the term proportional to $g$ in Eq.\ (\ref{v2stream}), and it becomes manifest even in the limit 
$g\to 0$ if we include nonzero masses $m_1$, $m_2$. We have checked that if we work with fixed chemical potentials in the rest frames of the superfluids $p_1$, $p_2$, the critical velocity $v_\mathrm{ two-stream}$ does become
symmetric in the two fluids, and the $g\to 0$ limit is the relativistic sum of the two Landau critical velocities for all masses $m_1$, $m_2$. This includes the non-relativistic limit, in agreement with Ref.\ \cite{2015EPJD...69..126A}. 

The main conclusion from Fig.~\ref{figv1v2} regarding the above question $(ii)$ is obvious: $v_L\le v_\mathrm{ two-stream}$ for all ratios $v_2/v_1$,  
and there exists one ratio $v_2/v_1$ where $v_L = v_\mathrm{ two-stream}$. 
In other words, the scenario shown in Fig.~\ref{figUpDown}\textit{is} generic, we have not found any region 
in the parameter space where the two-stream instability sets in at smaller velocities than the energetic instability. A general, rigorous proof of this statement is difficult
because the sound modes are solutions of quartic equations. Thus, strictly speaking, we have not rigorously proven this
statement, but, besides the results shown in Fig.~\ref{figv1v2} we have checked many other parameter sets, including a different sign of the entrainment coupling and including
a non-entrainment coupling. The situation we are asking for is a merger of the two original downstream modes, i.e., the two curves in region I in the left panel of 
Fig.~\ref{figUpDown}. We have in particular looked at parameter sets where the curves of these modes cross in the absence of any coupling. But, if a coupling is switched on, 
no imaginary part
but rather an "avoided crossing" develops at this point. Therefore, in all cases we have considered, the qualitative conclusion about the order of the two critical velocities remains. 

At this point, let us come back to the modes that we have discussed in Sec.~\ref{sec:hydro}. We pointed out that a two-fluid system allows for a richer spectrum of massless modes if 
one or both of the fluids are normal fluids rather than superfluids. In a superfluid, the oscillations are constrained because chemical potential and superfluid velocity are both 
related to the phase of the condensate. As a consequence, only longitudinal oscillations are allowed. 
If exactly one of the fluids is a normal fluid, one additional transverse mode appears, given in Eq.~(\ref{vnk1}). This mode has a fixed form, for all possible couplings between the 
fluids, and thus does not couple to the other, longitudinal modes. Therefore, the discussion of energetic and dynamical instabilities reduces to exactly the same modes as discussed
in this section for the two-component superfluid. Of course, for a specific discussion
we need the generalized pressure 
as a function of the Lorentz scalars $p_1^2$, $p_1^2$, $p_{12}^2$, of which the microscopic theory here only provides one specific example. An example for a system of one normal fluid 
and one superfluid is a single superfluid at nonzero temperature, for which the generalized pressure has been derived from a microscopic model  
in the low-temperature approximation, keeping terms of order $T^4$ \cite{Carter:1995if,2013PhRvD..87f5001A,superbook}, 
see for instance Eq.\ (4.45) in Ref.\ \cite{superbook}. In its regime of validity, i.e., low temperatures compared to the 
chemical potential and also small superfluid velocities, there is no dynamical instability, at least not in an energetically stable 
regime \cite{2013PhRvD..87f5001A}. At arbitrary temperatures below the 
critical temperature a covariant form of the generalized pressure, based on a microscopic model, is unknown to the best of our knowledge, 
and one has to apply more complicated methods. Within the self-consistent 2-particle-irreducible 
formalism, a two-stream instability was indeed found at nonzero temperatures, 
remarkably in an energetically \textit{stable} regime, for velocities slightly below Landau's critical velocity \cite{Schmitt:2013nva}.

\begin{figure} [t]
\begin{center}
\hbox{\includegraphics[width=0.5\textwidth]{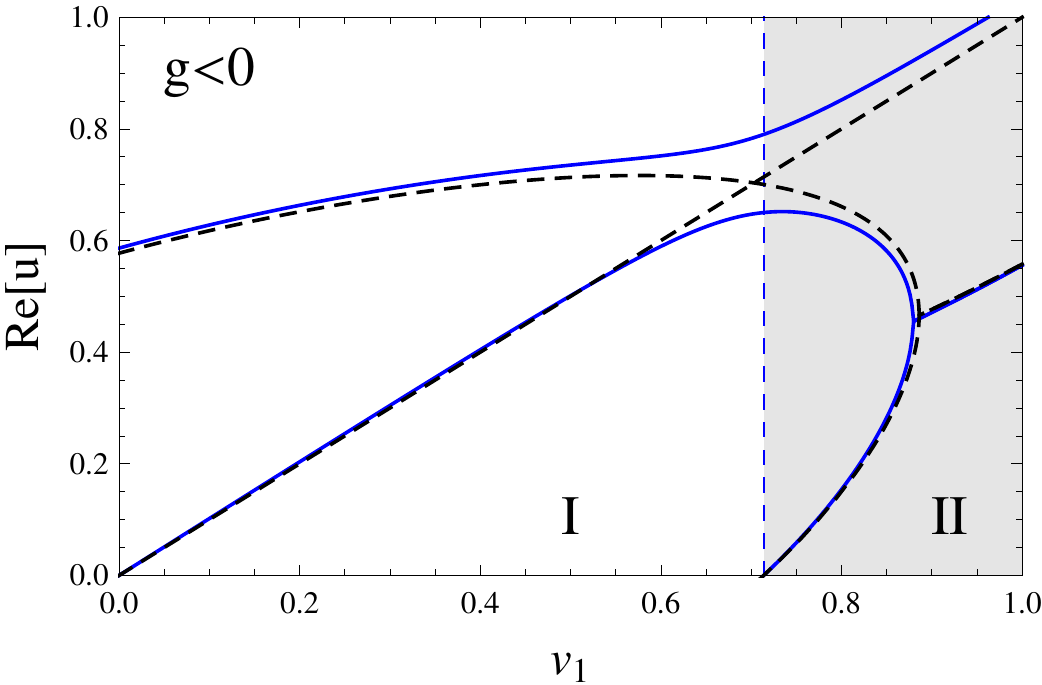}\includegraphics[width=0.5\textwidth]{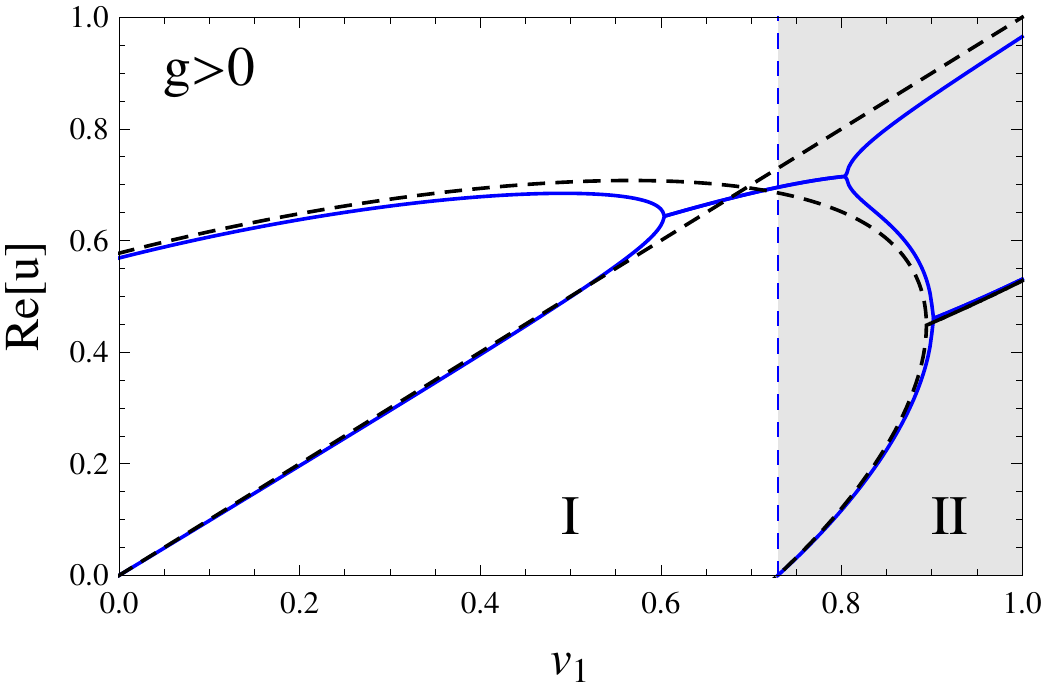}}
\caption{Real part of the sound speeds for transverse modes in the downstream direction, in a system of two \textit{normal} fluids with entrainment [solid (blue) lines]. 
Regions I and II are energetically stable and unstable regions, respectively. For a positive 
entrainment coupling, there is a two-stream instability in an energetically stable regime,  
which does not occur in the system of two coupled superfluids, where the two-stream
instability is always in region II. Here
we have used the equation of state of the superfluid system and chose the ultrarelativistic limit, $m_1=m_2=0$, the ratio of chemical potentials $\mu_2/\mu_1=1.5$, 
the self-couplings $\lambda_1=0.3$, $\lambda_2=0.2$, the non-entrainment coupling $h=0.05$, and the entrainment coupling $g\mu_1\mu_2=-0.003$ (left panel) and 
$g\mu_1\mu_2=+0.003$ (right panel). The dashed (black) lines are the results with the same parameters, but in the absence of entrainment, $g=0$.}
\label{fignn}
\end{center}
\end{figure}

For the case of two normal fluids, we discuss two intriguing 
observations. Firstly, in the absence of entrainment, but
in the presence of a non-entrainment coupling, we have found sound modes whose speed has a very simple analytical form, see Eqs.~(\ref{dmu20}) and (\ref{dmu10}). Just like
the two-component superfluid, they show an energetic instability and a dynamical instability, with corresponding critical velocities given 
by Eqs.\ (\ref{dmu20vcr}) and (\ref{dmu10vcr}). As these equations make obvious, the critical velocity for the energetic instability is always smaller than 
that of the dynamical instability. We have thus found an example for a set of (transverse) 
modes where there is a very simple proof that for all values of the coupling constant and for 
all equations of state for each single fluid the dynamical instability can only occur if the system is already energetically unstable. 
Secondly, let us discuss the modes (\ref{dmu20}) and (\ref{dmu10}) in the 
presence of entrainment. In this case, their analytical form is extremely 
complicated, so we again resort to a numerical evaluation. To this end, we need to specify an equation of state, and we simply use the pressure from our 
microscopic model of the complex fields, i.e., $P=-U_\mathrm{ COE}$ with $U_\mathrm{ COE}$ from Eq.\ (\ref{COE}). This is one particular example where both 
fluids enter symmetrically, and we simply reinterpret the pressure in the sense that the conjugate four-momentum is not related to any phase, thus allowing for 
non-longitudinal oscillations. The results for a certain choice of parameters
is shown in Fig.~\ref{fignn}. The dashed (black) lines show the result in the absence of entrainment: these are the modes from Eq.\ (\ref{dmu20}) just 
discussed, with one mode becoming negative (in the upstream direction, not shown in the figure) and, at a larger velocity, two modes that merge and acquire complex 
values.
This is very similar to the superfluid case in Fig.~\ref{figUpDown}.
But, additionally, there is the transverse mode $u=v_1\cos\theta$ which, in the absence of entrainment, is unaffected by the other modes. In the presence of 
entrainment, the modes couple. Mathematically speaking, the polynomial does not factorize in this case.  
In particular, for a certain sign of the entrainment coupling constant, in our convention $g>0$, there is a two-stream instability 
due to the presence of this mode\footnote{We may also consider the generalization of the modes (\ref{dmu10}) in the presence of entrainment, i.e., the modes that are continuously 
connected to the uncoupled oscillations of the fluid at rest. In this case, it is the zero mode $u=0$ instead of the mode $u=v_1\cos\theta$ which, in the presence of entrainment, 
couples to the modes (\ref{dmu10}), also resulting in a dynamical instability.}. 
This is remarkable because of two reasons: it is an example where entrainment, as opposed to a non-entrainment coupling, 
 leads to a qualitative difference regarding the dynamical
instability. (In the case of the two-superfluid system, although we have only shown results with entrainment coupling, the qualitative conclusions would not have changed had we 
worked with a non-entrainment coupling.) And, it is an example for a two-stream instability occurring in an energetically stable regime: none of the modes has negative energy at the 
point where two of the modes become complex. 
\chapter{Summary: Hydrodynamic Instabilities in Two-Fluid Systems}
\label{sec:summary1}

In this part of the thesis, we have investigated quasiparticle excitations in a relativistic two-component superfluid, by starting from a bosonic field-theoretical model for two complex
scalar fields and  including an inter-species derivative coupling that gives rise to entrainment between the two fluids. We have focused on the simplest hydrodynamic
situation where the fluid velocities are uniform in space and time and have restricted ourselves to zero temperature and zero magnetic field. 
As a preparation, in Sec.~\ref{sec:phases} the phase structure of the system in an "extended" grand canonical ensemble was computed, where besides the two chemical potentials also the two fluid velocities serve as externally given parameters. This is very natural because 
both quantities are different components of the same four-momentum conjugate to the conserved current. By including fluctuations about the condensates, 
we have computed the dispersion relations
of the two Goldstone modes in the phase where both complex fields condense. Besides this microscopic approach, a general derivation of the sound modes 
from two-fluid relativistic hydrodynamics in the linear approximation was presented. While this approach reproduces the Goldstone dispersion relations of the two-component superfluid at low energy, 
it also allows for a study of the sound modes in two-component \textit{normal} fluids. 

Our main focus has been on the instabilities that arise at nonzero fluid velocities. We have systematically analyzed energetic and dynamical instabilities.
The energetic instability manifests itself in negative excitation energies beyond a certain critical velocity. This critical velocity - or rather a critical surface in the space of both chemical potentials and fluid velocities
-- is a generalization of Landau's critical velocity for a single superfluid. With the help of the hydrodynamic equations, we have seen that this 
instability is, within the used zero-temperature approximation, 
equivalent to a negative eigenvalue of the "current susceptibility" matrix, the matrix of second derivatives of the pressure with 
respect to the fluid velocities (analogous to the number susceptibilities which are second derivatives with respect to the chemical potentials). In a certain parameter 
range this instability even occurs in the limit of vanishing fluid velocities. Besides the energetic instability we have also 
computed the critical velocity for the two-stream instability. This dynamical instability manifests itself in a complex sound velocity, whose imaginary part determines the time scale 
on which the unstable mode grows. 

As a result of this analysis, we have found that, in the case of a two-component superfluid  
at zero temperature, the dynamical instability can only occur in the presence of an energetic
instability. In other words, one mode must acquire negative energy before it can couple to another mode to develop an exponentially growing amplitude. 
Our numerical evaluation suggests that this is a general result, holding throughout the parameter space and for both kinds of couplings we have considered, 
non-entrainment and entrainment couplings. We have complemented this
result by an analysis of two normal fluids, which allow for non-longitudinal oscillations. This is in contrast to a superfluid, where a constraint on the oscillations is given
by the relation of both chemical potential and superfluid velocity to the phase of the condensate. Most importantly, we
have found a set of transverse modes in a two-component normal fluid that couple to each other if entrainment is present and that show a two-stream instability in the \textit{absence} of 
an energetic instability. 

An obvious extension is to include electromagnetism, i.e., to promote one or both of the global symmetry groups of our Lagrangian to a local one. It would be interesting to see whether this changes any of the conclusions regarding the dynamical instability, especially in view of its possible relevance for pulsar glitches. However, already for vanishing fluid velocities, an external magnetic field will lead to interesting effects, like a possible formation of flux tubes. The interaction of the remaining superfluid component with the superconductor and the resulting phase structure at zero superflow but in a magnetic field will be the topic of the next part of this thesis.

\part{Critical Magnetic Fields of a Superconductor Coupled to a Superfluid}
\label{par:SCSF}
After a thorough investigation of instabilities in the multicomponent system, we now turn our attention to the effect of electromagnetism. For that purpose, we take into account the electrical charge of one of the fields. For simplicity we will ignore any potential relative velocities of the fluids at vanishing magnetic field, which means that all thermodynamic quantities are now unambiguously defined in the common rest frame of the fluid and the heat bath. As a further simplification, we will set the entrainment coupling $g$, which enters into the phase structure, to zero, $g=0$. However, we will partially compute the effect of the second entrainment coupling $G$.
 As we have discussed in some detail in the introductory part of this work, a type-II superconductor in an external magnetic field can form an array of flux tubes. This is in contrast to type-I superconductors, where the magnetic field is either completely expelled or completely destroys the superconducting state, but never penetrates partially through quantized flux tubes. The Ginzburg-Landau parameter $\kappa$  predicts whether a superconductor is of type I or of type II. In standard systems, the boundary between type-I and type-II superconductors is given by
\be 
\kappa_c=\frac{1}{\sqrt{2}} \, .
\ee
As one result, we will see that the different criteria to distinguish between the two types are shifted and become ambiguous in the presence of the superfluid component.
The goal of this part is to study the critical magnetic fields for the flux tube lattice in the two-component system. Since the two scalar fields are coupled to each other and one of them is coupled to the gauge field ,the neutral scalar field is coupled indirectly to the gauge field as well. Various aspects of this system will be discussed, such as the effect of different forms of the coupling between the scalar fields (density coupling vs.\ derivative coupling), effects of nonzero temperature, and the interaction between magnetic flux tubes. 

As discussed extensively in Sec.~\ref{sec:ft_lattice}, the properties of the type-II state are driven by flux tubes and their interactions. Magnetic flux tubes in a proton superconductor in neutron stars have already been studied in the literature, usually with a bigger emphasis on phenomenological consequences. However, more microscopic approaches often do not include a consistent treatment of both components as performed here, and rather put together separate results from the proton superconductor and the neutron superfluid (which may be a good approximation for certain quantities because of the small proton fraction in neutral, $\beta$-equilibrated nuclear matter). Studies relevant to this thesis that do include both components within a single model can be found in
Refs.~\cite{Alpar:1984zz,Alford:2007np,2015arXiv150400570K,Sinha:2015bva}. In Ref.\ \cite{Alford:2007np}, flux tube profiles and energies are computed. These results are largely reproduced and utilized in the following chapters. The calculation of the interaction between flux tubes is performed within an approximation valid for large flux tube separations, based on old literature for a single-component superconductor \cite{Kramer:1971zza}; for a different method leading to the same result see Ref.~\cite{Speight:1996px}.
Extensions to a system of a superconductor coupled to a superfluid can be found in Refs.\ \cite{Buckley:2003zf,Buckley:2004ca}, where the results were restricted to the symmetric situation of approximately equal self-coupling and cross-coupling strengths of the scalar fields (which is unrealistic for neutron star matter \cite{Alford:2005ku}), and no derivative cross-coupling was taken into account. Interactions between flux tubes have also been computed, based on the same approximation, in the context of cosmic strings for one-component \cite{Bettencourt:1994kf} and two-component \cite{MacKenzie:2003jp} systems.

Special emphasis will also be put on the transition region between type-I and type-II behavior, because this region is changed qualitatively by the presence of the superfluid, and one of the main results will be the topology of the phase diagram in this region.

This part follows closely our publications published in Ref.~\cite{Haber:2016ljn} and Ref.~\cite{Haber:2017kth}. 

\chapter{Critical Magnetic Fields}
\label{chap:critical_fields}
In order to compute the critical magnetic fields in the system, we follow in principle the procedure presented in Chap.~\ref{chap:SF_SC_gen}, and especially in Sec.~\ref{sec:SCmagnetic}. The major difference consists of the relativistic treatment here and obviously the coupling to the second, neutral, field. For that purpose, we compute the Gibbs free energy $\mathcal{G}$ defined in Eq.~(\ref{eq:Gibbs_def}). The free energy $F$ can be computed from the potential (\ref{UxT}) by integrating over space,
\be
F = \int d^3r \,U(\mathbf{r}) \, .
\ee
We are interested in the phase structure at fixed external (and homogeneous) magnetic field of the form $\mathbf{H} = H\mathbf{e}_z$. As a reminder, the correct thermodynamic potential for this scenario, the Gibbs free energy, takes the form
\be \label{Gibbsdef}
{\cal G} = F - \frac{\mathbf{H}}{4\pi}\cdot\int d^3r \, \mathbf{B} \, .
\ee
To determine the complete phase diagram, we would in principle have to compute the Gibbs free energy for all possible phases at each point in the phase space given by the thermodynamic variables
$(\mu_1,\mu_2,T,H)$. The possible phases are the NOR, SF, SC, and COE phases listed in Sec.~\ref{sec:phases}. Note that the notation of the phases now accommodates the difference between the charged and the neutral scalar field by relabeling the phases SF$_1$ and SF$_2$ by SC and SF respectively. For the phases that are superconducting (SC and COE), we have to distinguish the Meissner phase, in which the magnetic field is completely expelled, $\mathbf{B}=0$, from the flux tube phase, where a lattice of magnetic flux tubes is formed, admitting part of the applied magnetic field in the superconductor. We shall simplify this problem by not computing the Gibbs free energy for the flux tube phase in full generality, which would require us to determine the spatial profile of the condensate and the magnetic field, including the preferred lattice structure, fully dynamically. 
Instead -- following the usual textbook treatment presented in Sec.~\ref{sec:SCmagnetic} -- we shall compute the critical magnetic fields $H_{c1}$, $H_{c2}$, and $H_c$, although they do not provide complete information of the phase diagram, not even for a single-component superconductor. To interpret their meaning for the phase diagram (in particular in our two-component system)
it is important to precisely recall how they are computed, and thus I will shortly repeat the definition of the corresponding critical magnetic field before we compute them for the coupled system. 
In general, when we speak of the superconducting phase, this can be either the COE or the SC phase, while the normal-conducting phase can either be NOR or SF.
The concrete calculations will always be done for the most interesting case, where both charged and neutral condensates exist in the superconducting phase (COE)
and the normal conductor is the pure superfluid (SF).  The critical magnetic fields for the transition between the COE and NOR and 
between the SC and NOR phases are not needed for our main results, but can be computed analogously. The latter appears 
to be the standard textbook scenario. However, in our two-component system it is conceivable that in the SC phase a neutral condensate is induced in the center of a flux tube \cite{Forgacs:2016ndn,Forgacs:2016iva}. Therefore,  
the pure superconductor SC might acquire a superfluid admixture, which can affect  the critical magnetic fields for the transition to the completely uncondensed phase (NOR). In the present calculation, we shall only consider flux tube
solutions that approach the COE phase, not the SC phase, far away from the center of the flux tube.

\section{Critical Magnetic Field \texorpdfstring{$H_c$}{Hc}}
\label{sec:Hc}
In Sec.~\ref{sec:SCmagnetic}, we have learned how to compute the critical magnetic fields in a superconductor. We can use the same definitions to derive the critical fields in the two-component system. The first-order phase transition from the Meissner phase to the normal-conducting, i.e.~ the SF phase, can be obtained by a comparison of the corresponding free energies.
The Gibbs free energy of the COE phase with complete expulsion of the magnetic field is
\be \label{GibbsCOE}
{\cal G}_\mathrm{ COE} = VU_\mathrm{ COE} \, ,
\ee
where $V$ is the total volume of the system and $U_\mathrm{ COE}$ is the free energy density from Eq.~(\ref{UCOE}), where we set $g=0$ and neglect the superflow. Therefore, we can replace the conjugate momenta to the conserved currents $p_i$ with the chemical potentials $\mu_1$ and $\mu_2$. We further incorporate finite temperature effects by replacing the masses $m_1$, $m_2$ and the coupling $h$  by their thermal generalizations $m_{1,T}$, $m_{2,T}$, $h_T$, presented in Eqs.~(\ref{thermal_mass}). We neglect any magnetization in the normal-conducting phases, and thus $\mathbf{B}=\mathbf{H}$ in the SF phase, which yields the Gibbs free energy 
\be \label{GibbsSF}
{\cal G}_\mathrm{ SF} = V\left(U_\mathrm{ SF} -\frac{H^2}{8\pi}\right) \, ,
\ee
with $U_\mathrm{ SF}$ from Eq.\ (\ref{eq:rhoSF}). Note that the $H^2$ term is a sum of the magnetic energy $\propto B^2$ and the term $\propto HB$ in the Legendre transformation 
from the free energy $F$ to the Gibbs free energy ${\cal G}$. Therefore, the critical magnetic field, defined by ${\cal G}_\mathrm{ COE} = {\cal G}_\mathrm{ SF}$, becomes
\be \label{Hc}
H_c = \sqrt{8\pi(U_\mathrm{ SF}-U_\mathrm{ COE})} =  2\pi q\sqrt{2}\kappa\sqrt{1-\frac{h_T^2}{\lambda_1\lambda_2}} \rho^2_{01}\, ,
\ee
where $\rho_{01}$ is the value of the charged condensate in the COE phase. Here we have introduced the relativistic version of the Ginzburg-Landau parameter 
\be 
\kappa=\frac{\ell}{\xi}=\sqrt{\frac{\lambda_1}{4\pi q^2}} \, ,
\ee
with the magnetic penetration depth $\ell$  and the coherence length $\xi$, 
\be\label{ellxi}
\ell=\frac{1}{\sqrt{4\pi q^2}\rho_{01}} \, , \qquad \xi=\frac{1}{\sqrt{\lambda_1}\rho_{01}} \, .
\ee

\section{Critical Magnetic Field \texorpdfstring{$H_{c2}$}{Hc2}}
\label{sec:Hc2}
The procedure to compute the second critical magnetic field in the two-component system, which separates the flux tube phase from the SF phase, is equivalent to the one component case presented in Sec.~\ref{sec:SCmagnetic}. Therefore, we use the same definition and nomenclature introduced before.
By the definition of a second-order phase transition, as we approach $H_{c2}$, the \textit{charged} condensate approaches zero and the neutral condensate approaches the condensate of the SF phase.
For magnetic fields $H$ close to and smaller than $H_{c2}$, we can write the condensates and the gauge field as their values at $H_{c2}$ plus small perturbations.
Then, for the calculation of $H_{c2}$, we compute the the in the charged condensate linearized equations. The critical magnetic field is the maximally field allowed by these equations. 

We are also interested in checking whether and in which parameter regime 
the flux tube phase is energetically preferred just below $H_{c2}$. This is done within the same calculation, but taking 
into account higher order terms in the equations of motion and the free energy. This calculation is somewhat lengthy 
and is explained in appendix \ref{app:Hc2}. Here only the results are summarized.
The critical magnetic field becomes
\be\label{HcHc2}
H_{c2} = \frac{1}{q\xi^2}\left(1-\frac{h_T^2}{\lambda_1\lambda_2}\right)=\sqrt{2}\kappa\sqrt{1-\frac{h_T^2}{\lambda_1\lambda_2}}H_c\, ,
\ee
where the second expression relates $H_{c2}$ to $H_c$ by using Eq.~(\ref{Hc}). At zero temperature, $H_{c2}$ 
does not depend on the gradient coupling $G$. However, the difference in Gibbs free energies between the superconducting and the normal-conducting phases does depend on $G$, see Eq.~(\ref{DeltaGfull}). For $G=0$ we have
\be\label{DeltaG}
\frac{{\cal G}_\mathrm{ COE}}{V} = \frac{{\cal G}_\mathrm{ SF}}{V} + \lambda_1\langle\bar{\varphi}_1^4\rangle\left[\frac{1}{2\kappa^2}-1+\frac{h^2}{\lambda_1\lambda_2}
{\cal I}_1(p)\right] \, ,
\ee
where $\langle\bar{\varphi}_1^4\rangle $ is the spatial average of the charged condensate (\ref{eq:SEsol}), where 
\be
p^2 = \frac{2\lambda_2\rho_\mathrm{ SF}^2}{qH_{c2}} \, ,
\ee 
and where 
\be
{\cal I}_1(p) \equiv \frac{pe^{p^2/4}}{2\sqrt{2}} \int_{-\infty}^\infty dt\,e^{-t^2}\left\{
e^{pt}\left[1-\mathrm{ erf}\left(\frac{p}{2}+t\right)\right]+e^{-pt}\left[1-\mathrm{ erf}\left(\frac{p}{2}-t\right)\right]\right\} \, ,
\ee
with the error function erf.

In the limit of a single superconductor, $h = 0$, we recover the standard result that we have found earlier in Eq.~(\ref{eq:HcHc2_single}), where the critical fields $H_c$ and $H_{c2}$ coincide 
at $\kappa^2=1/2$, and Eq.\ (\ref{DeltaG}) shows that the flux tube phase is preferred, ${\cal G}_\mathrm{ COE}<{\cal G}_\mathrm{ SF}$, if and only if $\kappa^2>1/2$. We did not carry out this calculation for the simple case in the introduction, since it can easily be obtained from the complete calculation of the two-component model by setting $h=0$.
In the coupled system the situation is more complicated. Now, from Eq.~(\ref{HcHc2}) we see that $H_c$ and $H_{c2}$ coincide at a larger value of $\kappa$ 
(since $h^2<\lambda_1\lambda_2$ to ensure the boundedness  of the  potential for $h>0$ and to ensure the existence of the COE phase for $h<0$, the square root is always real and smaller than $1$). This appears to take away phase space from the flux tube phase. However, from Eq.\ (\ref{DeltaG}) we see that the difference in Gibbs free energies between the COE and the SF phases changes sign at a different point,
and this point is given not just by the coupling constant $h$, but also depends on $p$, i.e., on the magnitude of the neutral condensate $\rho_\mathrm{ SF}$ compared to the 
square root of the critical magnetic field $H_{c2}$. 
Despite this dependence we can make a general statement: we find $0\le {\cal I}_1(p) <1$, and thus the factor ${\cal I}_1(p)$ weakens the 
effect of the term $h^2/(\lambda_1\lambda_2)$. At the value of $\kappa$ where $H_c$ and $H_{c2}$ are equal, the superconducting phase is preferred and -- for all $p$ --
remains preferred along $H_{c2}$, until the smaller $\kappa$ defined through Eq.~(\ref{DeltaG}) is reached. This observation is indicative of 
the complications at the transition between type-I and type-II superconductivity in the two-component system, and we shall find further discrepancies to the standard scenario when we compute the critical field $H_{c1}$.

Anticipating the numerical results in Sec.~\ref{chap:phasesHC}, let us comment on a possible first-order phase transition at $H'_{c2}$, as mentioned in the definition at the 
beginning of this chapter. Suppose we are in a parameter region where the flux tube phase is favored just below $H_{c2}$, i.e., let $\kappa$ be larger than the critical $\kappa$ 
defined through Eq.~(\ref{DeltaG}). Then, any phase transition from the flux tube phase to the normal phase at a critical field smaller than $H_{c2}$ is practically excluded because we know that the 
system prefers to be in the flux tube phase just below $H_{c2}$ (here we ignore the very exotic possibility that the system quits the flux tube phase and then re-enters it below $H_{c2}$). A phase transition at a critical magnetic field larger than $H_{c2}$ -- instead of the one at $H_{c2}$ -- is however possible. This phase transition must be of first order 
because by definition $H_{c2}$ is the largest magnetic field at which a second-order transition may occur. Putting these arguments together leads to the conclusion that $H_{c2}$ is a lower bound for the transition from the flux tube phase to the normal phase, possibly replaced by a first order transition at $H'_{c2}>H_{c2}$. The numerical results I will present
will indeed suggest such a first-order phase transition. However, we shall find $H'_{c2}<H_{c2}$, which, as we will explain, is an artifact of the approximation applied for the interaction between flux tubes.  Nevertheless, the result will allow us to speculate about the correct critical field $H'_{c2}$, obtained in a more complete calculation that goes beyond this approximation.

\section{Critical Magnetic Field \texorpdfstring{$H_{c1}$}{Hc1}}
\label{sec:Hc1}
The last critical magnetic field that we have to compute is the one responsible for the second-order phase transition from the Meissner to the flux-tube state. The slightly unconventional order in which the critical fields are presented can be explained easily. Whereas the other two fields can be computed analytically within the used approximations, $H_{c1}$ requires a numerical computation of the single flux tube energy. We have now to perform the energy comparison between the Gibbs free energy for the COE phase with a single magnetic flux tube, $\mathcal{G}_\mathrm{ COE}^{\circlearrowleft}$, and the one completely without flux tubes. Thus we compute, according to the definition (\ref{Gibbsdef}), 
\be \label{Gibbsfl}
\mathcal{G}_\mathrm{ COE}^{\circlearrowleft}=VU_\mathrm{ COE}+F_{\circlearrowleft}-\frac{Hn\Phi_0}{4\pi}  L \, ,
\ee
where $F_{\circlearrowleft}$ is the free energy of the flux tube, and where we have used the quantization properties of the flux tubes,
\be \label{drB}
\int d^3r \, B = n\Phi_0 L \, ,
\ee
with the winding number $n$ of the flux tube, the length of the flux tube $L$, and the fundamental flux quantum $\Phi_0 = 2\pi/q$. Placing a single flux tube into the two-fluid system results in a loss in (negative) condensation energy, and thus the free energy increases. However, at fixed magnetic field $H$, there is an energy gain from allowing magnetic flux into the system. For a more detailed discussion see Sec.~\ref{sec:SCmagnetic}.
As a consequence, there is a competition between these two contributions of opposite sign in Eq.~(\ref{Gibbsfl}). 
At the critical point, the two contributions exactly cancel each other, 
\be 	\label{Hc1}
H_{c1}=\frac{2q}{n}\frac{F_{\circlearrowleft}}{L} \,,
\ee
which is identical as for the single superconductor, see Eq.~(\ref{eq:Hc1_single}),but with an altered expression for the single flux tube energy $F_{\circlearrowleft}$. The calculation of $H_{c1}$ thus amounts to the calculation of the free energy of a single flux tube $F_{\circlearrowleft}$, for which we can largely follow Ref.~\cite{Alford:2007np}. 
We work in cylindrical coordinates, $\mathbf{r}=(r,z,\theta)$, and make the following, radially symmetric, ansatz for the condensates, 
\be 
\rho_i(r)=\rho_{0i} f_i(r)\, , \qquad \psi_1(\theta) = n\theta \, , \qquad \psi_2=0 \, ,
 \ee
and the gauge field
\be
\mathbf{A}(r)=\frac{na(r)}{qr}\mathbf{e}_{\theta}\quad \Rightarrow \qquad \mathbf{B}(r)=\frac{n}{qr}\frac{\partial a}{\partial r}\mathbf{e}_z \, .
\ee 
The profile functions $f_i$ and $a$ have to be computed numerically. Their boundary conditions are $f_{i}(\infty)=a(\infty)=1$, $f_1(0)=0$, 
and $\partial_rf_2(\infty)=\partial_r a(\infty)=0$, such that the condensates approach their homogeneous values $\rho_{0i}$ and the magnetic field vanishes far 
away from the center of the flux tube. The values of the neutral condensate and the gauge field at the center of the flux tube are determined dynamically.
We have set the winding number of the neutral condensate to zero because the flux tube does not induce a superfluid 
vortex \cite{Alford:2007np}. 

We insert our ansatz into the potential (\ref{UxT}) and separate the potential of the homogeneous COE phase,
\be
U(\mathbf{r}) =  U_{\circlearrowleft}(\mathbf{r}) + U_\mathrm{ COE} \, ,
\ee
with
\be
U_\mathrm{ COE} = -\frac{\mu_1^2-m_{1,T}^2}{2}\rho_{01}^2 -\frac{\mu_2^2-m_{2,T}^2}{2}\rho_{02}^2 +\frac{\lambda_1}{4}\rho_{01}^4 +\frac{\lambda_2}{4}\rho_{02}^4 
-\frac{h_T}{2}\rho_{01}^2\rho_{02}^2 \, .
\ee
To write the free energy of the flux tube in a convenient form, we introduce the dimensionless variable
\be
R=\frac{r}{\xi} \, ,
\ee
abbreviate the dimensionless gradient coupling by 
\be
\Gamma\equiv G\rho_{01}\rho_{02} \, ,
\ee
and the ratio of neutral over charged condensate by 
\be
x\equiv\frac{\rho_{02}}{\rho_{01}} \, .
\ee
It is also useful to write $\mu_1^2-m_{1,T}^2 = \lambda_1\rho_\mathrm{ SC}^2 = \lambda_1\rho_{01}^2-h_T\rho_{02}^2$ and $\mu_2^2-m_{2,T}^2 = \lambda_2\rho_\mathrm{ SF}^2 = \lambda_2\rho_{02}^2-h_T\rho_{01}^2$, which follows from Eq.\ (\ref{COE}). Then, we obtain the free energy per unit length
\bea \label{Efl}
\frac{F_{\circlearrowleft}}{L} &=& \frac{1}{L}\int d^3r\, U_{\circlearrowleft}(\mathbf{r}) \\[2ex]
&=&  \pi \rho_{01}^2 \int_0^\infty dR\,R\left\{\frac{n^2\kappa^2 a'^2}{R^2}+f_1'^2+f_1^2\frac{n^2(1-a)^2}{R^2}+\frac{(1-f_1^2)^2}{2}\right. \non[2ex]
&& \left. +x^2
\left[f_2'^2 + \frac{\lambda_2}{\lambda_1}x^2\frac{(1-f_2^2)^2}{2}\right]-\frac{h_T}{\lambda_1} x^2 (1-f_1^2)(1-f_2^2)-\Gamma x f_1f_2f_1'f_2'
 \right\} \, ,\nonumber 
\eea
where prime denotes derivative with respect to $R$.
This yields the following, coupled differential equations of motion for $a$, $f_1$, $f_2$, 

\begin{subequations} \label{eomA}
\bea
0\hspace{-.2cm}&=&\hspace{-.2cm}a''-\frac{a'}{ R}+\frac{f_1^2}{\kappa^2}\left(1-a\right)  \, , \label{eoma}\\[2ex]
0\hspace{-.2cm}&=&\hspace{-.2cm}f_1''+\frac{f_1'}{ R}+ f_1\left[1-f_1^2-\frac{n^2(1-a)^2}{R^2}\right] -\frac{h_T}{\lambda_1} x^2 f_1(1-f_2^2)
-\frac{\Gamma x}{2}f_1\left[f_2'^2+f_2\left(f_2''+\frac{f_2'}{ R}\right)\right]  \, ,\non[0.5ex]\hfill\\[0.5ex] 
0\hspace{-.2cm}&=&\hspace{-.2cm}f_2''+\frac{f_2'}{R}+ f_2\frac{\lambda_2}{\lambda_1}x^2\left(1-f_2^2\right)- \frac{h_T}{\lambda_1} f_2\left(1-f_1^2\right)-\frac{\Gamma}{2x}
f_2\left[f_1'^2+f_1\left(f_1''+\frac{f_1'}{ R}\right)\right] \, . 
\eea
\end{subequations}
We solve these equations numerically with a successive over-relaxation method. The profiles themselves have been discussed in detail in 
Ref.\ \cite{Alford:2007np}\footnote{Eqs.\ (\ref{eomA}) are identical to Eqs.\ (16) in Ref.\ \cite{Alford:2007np}  if we identify
\be
\frac{\Gamma}{2} \leftrightarrow \sigma \, , \qquad x \leftrightarrow \frac{\langle\phi_n\rangle}{\langle\phi_p\rangle} \, , \qquad \frac{h_T}{\lambda_1} 
\leftrightarrow -\frac{a_{pn}}{a_{pp}}  \, , \qquad \frac{\lambda_2}{\lambda_1} \leftrightarrow \frac{a_{nn}}{a_{pp}} \, . \nonumber
\ee}, additionally we will discuss them later in this section. For the moment, we continue with the asymptotic solution, which will be needed later.   
\subsection{Solutions for \texorpdfstring{$R\gg 1$}{R>>1}}
Analogously to Sec.~\ref{sec:SCmagnetic}, we now compute the behavior of the profile functions far away from the core of the flux tube. In this region, all profile functions are close to one. Therefore, we decompose the profile functions into an asymptotic part and small deviations from the asymptotic values, 
\bea \label{aff}
a(R) &=& 1+ Rv(R) \, , \qquad 
f_1(R) = 1+ u_1(R) \, , \qquad 
f_2(R) = 1+u_2(R) \, ,
\eea
such that we can linearize the profile equations (\ref{eomA}) in $v$, $u_1$, and $u_2$, 
\begin{subequations} \label{vuM}
\bea
0&\simeq& R^2v''+Rv'-\left(1+\frac{R^2}{\kappa^2}\right) v    \, , \label{vuM1}\\[2ex]
\Delta  u  &\simeq& M u  \, .
\eea
\end{subequations}
We notice that the equations are nearly identical to Eqs.~(\ref{eq:FTeom_single}) for the single superconductor. Especially the equaion for the gauge field is identical, but the equations for the two profiles here mix via the following vector and matrix, 
\be
u \equiv \left(\begin{array}{c} u_1 \\[2ex] u_2 \end{array}\right) \, , \qquad 
M \equiv 2\left(\begin{array}{cc} 1 & -\frac{\Gamma x}{2} \\[2ex] -\frac{\Gamma}{2x} & 1 \end{array}\right)^{-1} \left(\begin{array}{cc} 1 & -\frac{h_T}{\lambda_1}x^2  \\[2ex] -\frac{h_T}{\lambda_1} & \frac{\lambda_2}{\lambda_1}x^2 \end{array}\right) \, .
\ee
However, in order to solve them analogously we can decouple the equations for $u_1$ and $u_2$ by diagonalizing $M$, 
\be
\mathrm{ diag}\,(\nu_+,\nu_-) = U^{-1}MU \, , \qquad U =  \left(\begin{array}{cc} \gamma_+ & \gamma_- \\[2ex] 1 & 1 \end{array}\right) \, , 
\ee
where $\nu_\pm$ are the eigenvalues of $M$ and $(\gamma_\pm,1)$ its eigenvectors, given by
\bea \label{nugam}
\nu_\pm = \frac{\lambda_1+\lambda_2 x^2-h_T\Gamma x\pm{\cal Q}}{\lambda_1(1-\Gamma^2/4)} \, , \qquad \gamma_\pm = \frac{x(\lambda_1-\lambda_2 x^2\pm {\cal Q})}{\lambda_1\Gamma -2h_T x} \, ,
\eea
where ${\cal Q} \equiv[(\lambda_1-\lambda_2 x^2)^2-2h_T\Gamma x(\lambda_1+\lambda_2 x^2)+x^2(4h_T^2+\Gamma^2\lambda_1\lambda_2)]^{1/2}$. 
This yields two uncoupled equations for $\tilde{u}_1$ and $\tilde{u}_2$, where $\tilde{u}=U^{-1}u$, which we solve with the boundary condition 
$\tilde{u}_1(\infty)=\tilde{u}_2(\infty)=0$ (which leaves one integration constant from each equation undetermined). We undo the rotation with $u=U\tilde{u}$, and, together with the
solution to Eq.\ (\ref{vuM1}), insert the 
result into Eq.\ (\ref{aff}) to obtain the asymptotic solutions
\begin{subequations}
\label{asympsol}
\bea
a(R) &\simeq& 1+C R K_1(R/\kappa) \, , \label{Qasym} \\[2ex]
f_1(R) &\simeq& 1+ D_+ \gamma_+K_0(\sqrt{\nu_+}R)+ D_- \gamma_-K_0(\sqrt{\nu_-}R)   \, , \label{f1asym} \\[2ex]
f_2(R) &\simeq& 1+ D_+ K_0(\sqrt{\nu_+}R) + D_-K_0(\sqrt{\nu_-}R)  \, , \label{f2asym}
\eea
\end{subequations}
where $K_0$ and $K_1$ are again the modified Bessel functions of the second kind, and the constants $C$, $D_+$, $D_-$ can only be determined numerically by solving the full equations of motion, including the boundary conditions at $R=0$. Comparing these asymptotic solutions to the full numerical solutions to determine the numerical constants turns out to be rather difficult and extremely dependent on the matching points. For our calculations we therefore mostly use the full numerical results.
In deriving the linearized equations (\ref{vuM}), we have not only used $u_1, u_2, v\ll 1$, but also $v^2\ll u_1, u_2$, which implies $e^{-2R/\kappa} \ll e^{-\sqrt{\nu_\pm}R}$. This assumption is violated if $\kappa$ is sufficiently large compared to $1/\sqrt{\nu_{\pm}}$ (compared
to $1/\sqrt{2}$ in a single superconductor), which means deep in the type-II regime. Later, when we use the solutions of  
the linearized equations for the interactions between flux tubes, we are only interested in the transition region between type-I and type-II behavior, where $1/\kappa \simeq \sqrt{\nu_\pm}$. Thus the linearization is a valid approximation for our purpose. 

\subsection{Solutions for \texorpdfstring{$R\ll 1$}{R<<1}}
\label{sec:Rll1}
For small values of $R$, we neglect all terms in the equations of motion which do not contain a derivative or are not inversely proportional to $R$.
\begin{eqnarray}
f_1''+\frac{f_1'}{R}-\frac{n^{2}(1-a)^{2}f_1}{R^{2}} & = & 0\,,\\
a''-\frac{a'}{R} & = & 0\,,\\
f_2''+\frac{f_2'}{R} & = & 0\,.
\end{eqnarray}
The equations for the gauge field and the neutral condensate can be
solved either trivially by a constant or a polynomial. The two equations for the charged profile function $f_1$ and the neutral one $f_2$ completely decouple. The neutral field cannot be influenced by the magnetic field in this limit, thus we do not expect deviations from the trivial solution of the neutral condensate, and we deduce that $f_2(r)=\mathrm{const.}$\ is the correct solution. The other two equations are identical to Eqs.~(\ref{eq:core_sol}), therefore we can immediately write
$a\propto R^{2}$ and $f\propto r^{n}$ for $R\ll1$ as stated in Ref.~\cite{Alford:2007np} on p.~5.

\section{Flux Tube Profiles}
\label{sec:ft_disc}
\begin{figure} [t]
\begin{center}
\hbox{\includegraphics[width=0.5\textwidth]{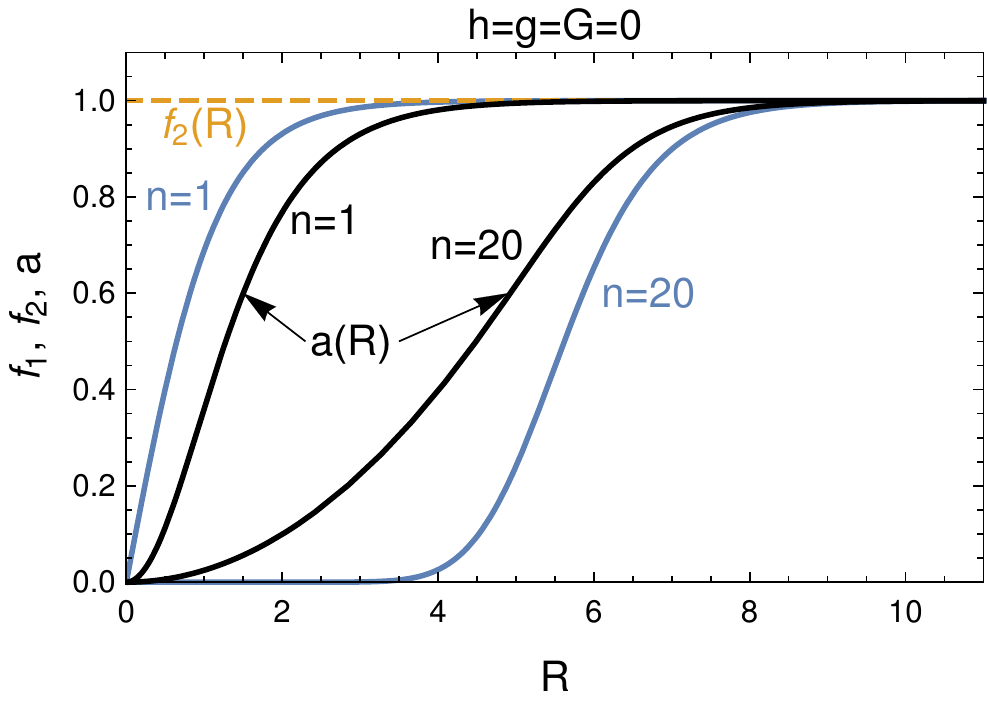}\includegraphics[width=0.49\textwidth]{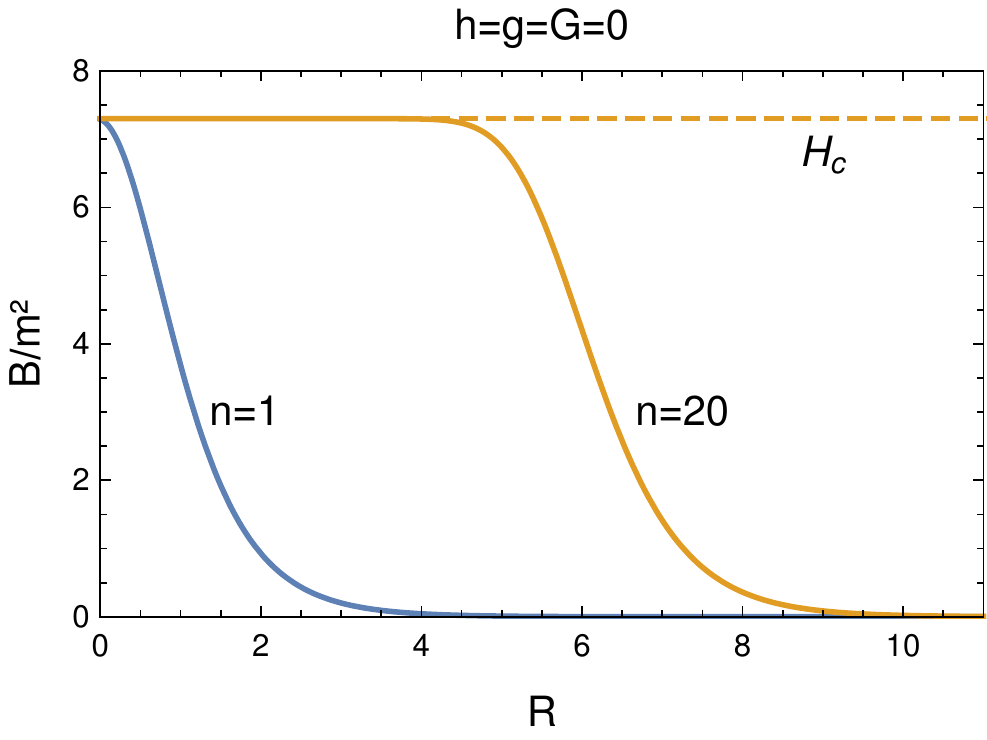}}
\caption{\textit{Left:}Flux tube and gauge field profile functions $f_1(R)$ (blue), $f_2(R)$ and $a(R)$ as a function of the dimensionless radial coordinate $R$ for winding numbers $n=1$ and $n=20$ at vanishing interspecies coupling. The neutral condensate stays constant, whereas the flux tube profile of the charged condensate shows the $R^n$ behavior for different winding numbers.\\
\textit{Right:} For large $n$, the magnetic field approaches $H_{c}$ in the core of the flux tube and then drops exponentially due to the Meissner effect. All numerical parameters can be found in Tab.~(\ref{tab:para1}).
\label{fig:single_tube}}
\end{center}
\end{figure}

\begin{figure} [t]
\begin{center}
\hbox{\includegraphics[width=0.5\textwidth]{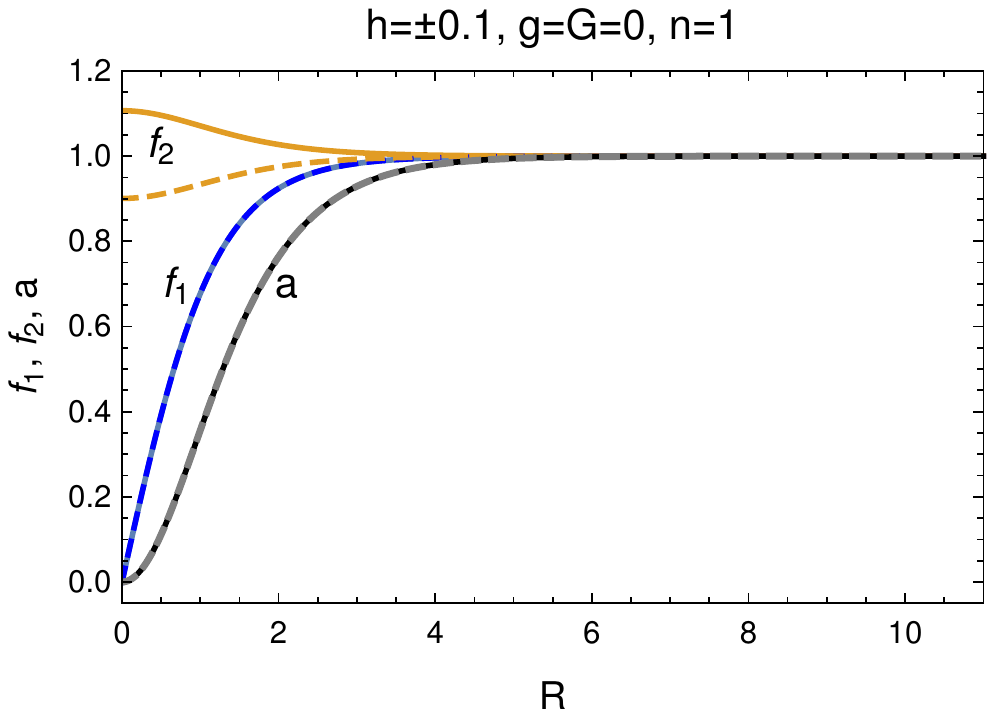}\includegraphics[width=0.5\textwidth]{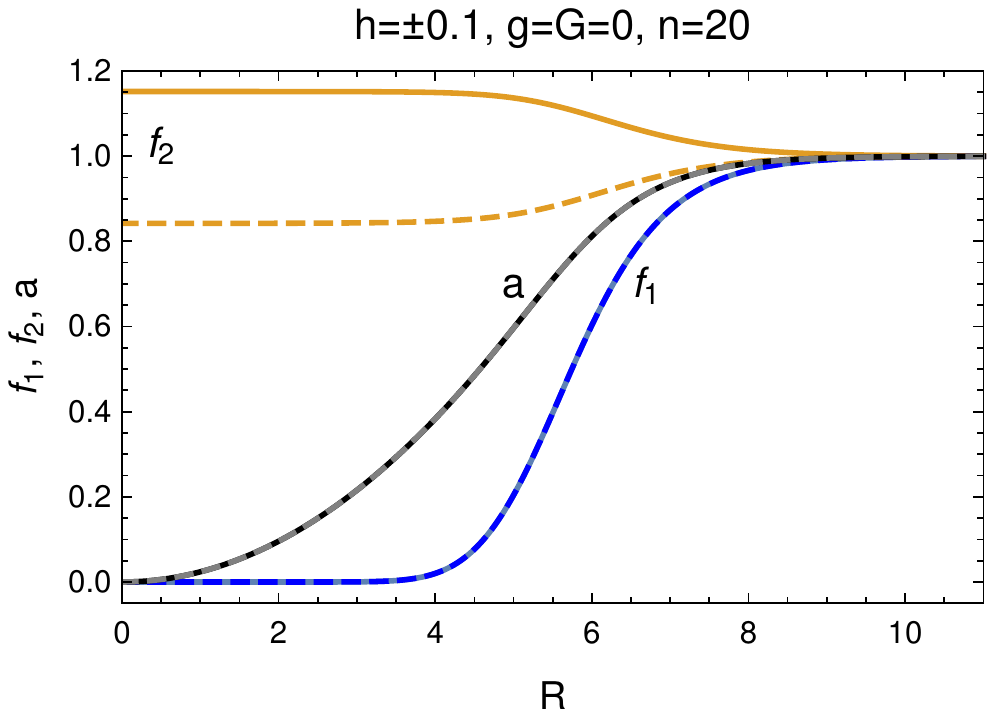}}
\caption{Flux tube and gauge field profile functions $f_1(R)$ (blue), $f_2(R)$ (orange) and $a(R)$ (black) for winding numbers $n=1$ and $n=20$ at different signs of $h=\pm0.1$. The profiles for $h>0$ are plotted with dashed lines, for $h<0$ solid lines are used. All other parameters are taken from Tab.~(\ref{tab:para1}). The neutral condensate is either enhanced or diminished in the core. For large $n$ the neutral condensate approaches the value of the uncoupled condensate $\rho_{\mathrm{SF}}$ from Eq.~(\ref{eq:rhoSF}).
\label{fig:coupled1}}
\end{center}
\end{figure}

\begin{table}
\begin{center}
\begin{tabular}{|c|c|c|c|c|c|c|}
\hline 
$m_{1}$ & $m_{2}$ & $\mu_{1}$ & $\mu_{2}$ & $\lambda_{1}$ & $\lambda_{2}$ & $T$\tabularnewline
\hline 
\hline 
m & m & $1.5$ m & $1.8$ m & $0.184$ & $0.716$ & 0\tabularnewline
\hline 
\end{tabular}
\caption{Numerical parameters for the flux tube profiles and magnetic fields in Sec.~\ref{sec:ft_disc}.\label{tab:para1}}
\end{center}
\end{table}
Although the profile functions have been studied in great detail in Ref.~\cite{Alford:2007np}, we will discuss them for some parameters in order to obtain some intuitive feeling for the behavior of the multicomponent system. In order to do so, we solve the coupled equations derived in Eqs.~(\ref{eomA}) using the numerical procedure explained in App.~\ref{App:num_methods}. The relaxation itself is carried out using the Fortran code presented in the same appendix, whereas the input and the data processing is carried out externally with the help of Wolfram Mathematica.  As a first step we reproduce the schematic flux tube presented in Fig.~\ref{fig:FT_schematic} by setting all couplings to zero. The result is presented in Fig.~\ref{fig:single_tube}.
As expected, the uncharged condensate, which is now completely uncoupled from the magnetic field, stays constant at $f_2=1$, which is a result of our normalization. The behavior of the charged condensate depends on the winding number. In agreement with our investigation in Sec.~\ref{sec:Rll1} we find a larger normal-conducting flux tube core, which also results in a broader plateau-like structure of the magnetic field, which is shown in the right panel of the same plot. We see that the magnetic field in the core for large winding numbers saturates at the critical field for the transition from the Meissner to the normal state in the type-I regime (indicated by the horizontal dashed line). Since we have already discussed the structure of a single flux tube, this calculation merely serves as a conformation that the code we use for the numerical investigation is able to produce reasonable results, therefore we turn on the cross-coupling $h$ in a next step. For the moment, we still set the derivative couplings $g$ and $G$ to zero but investigate different signs of $h$. The results are shown in Fig.~\ref{fig:coupled1}.

\begin{figure} [t]
\begin{center}
\includegraphics[width=0.5\textwidth]{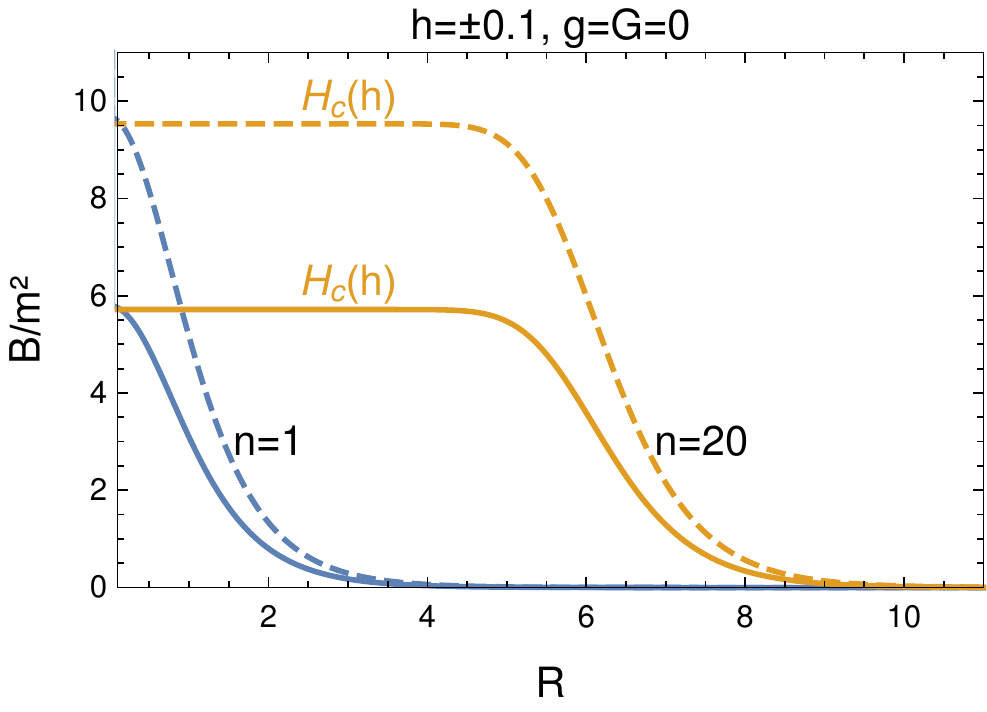}
\caption{Magnetic field corresponding to the profile functions plotted in Fig.~\ref{fig:coupled1} with the parameters from Tab.~(\ref{tab:para1}). Solid lines correspond to $h<0$ whereas dashed lines indicate $h>0$. The magnetic field in the core, which approaches $H_c$ for large $n$, depends on the external parameters. The field is plotted in units of the mass parameter $m^2$.
\label{fig:bcoupled1}}
\end{center}
\end{figure}

\begin{figure} [t]
\begin{center}
\hbox{\includegraphics[width=0.5\textwidth]{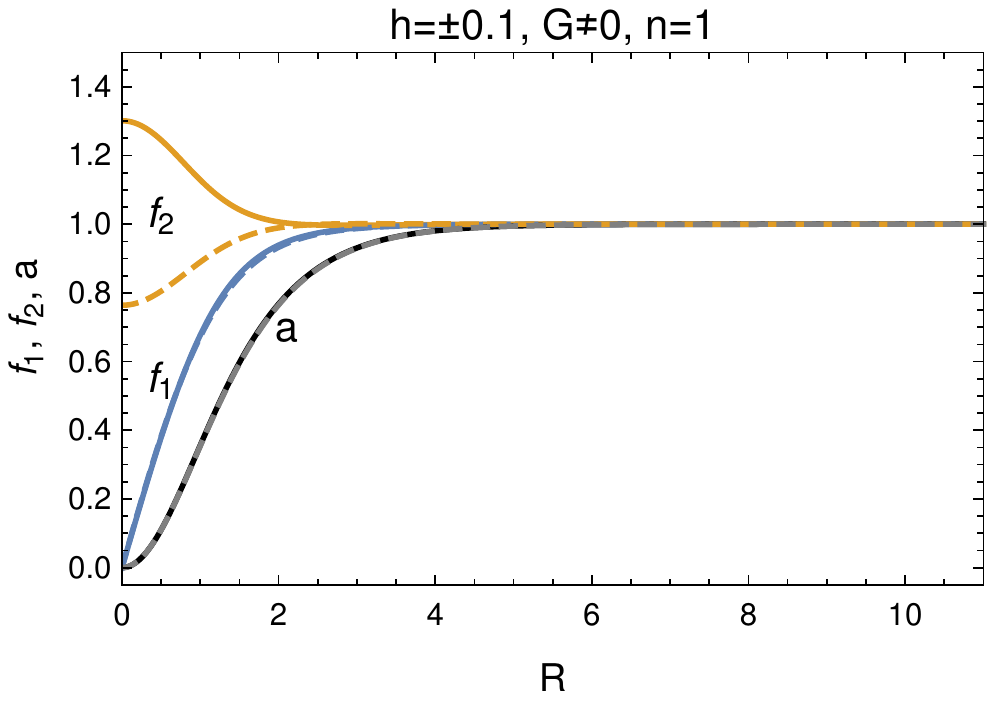}\includegraphics[width=0.5\textwidth]{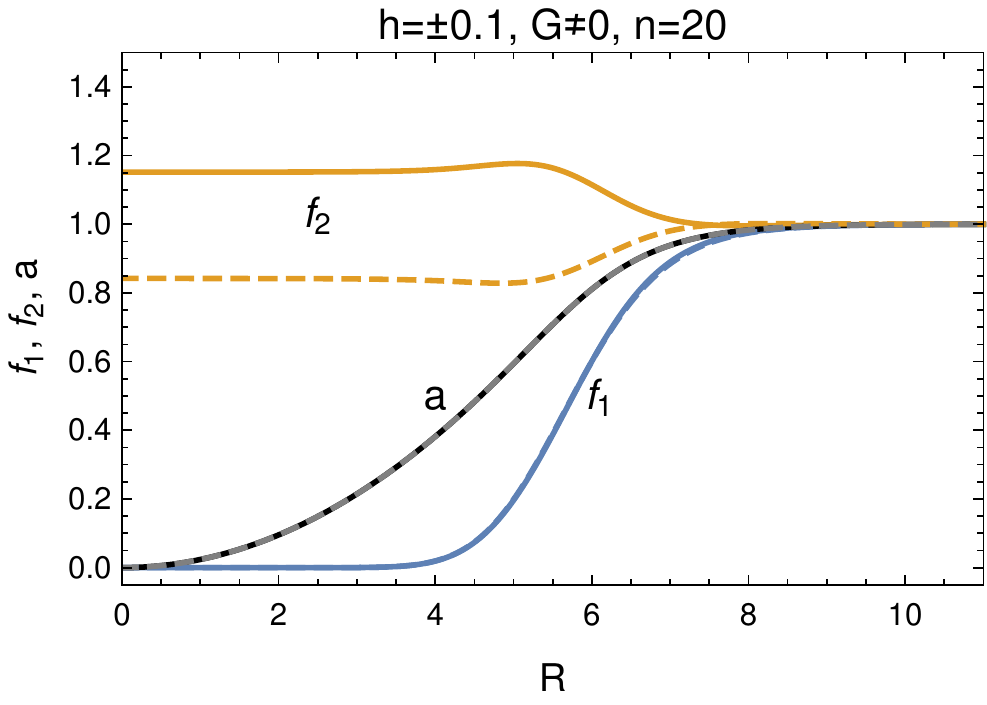}}
\caption{Flux tube and gauge field profile functions $f_1(R)$ (blue), $f_2(R)$ (orange) and $a(R)$ (black) for winding numbers $n=1$ and $n=20$. The solid lines correspond to $h=-0.1$ and $G=-0.4/m^2$ or equivalently $\Gamma\approx-1.442$. For the dashed lines, $h=0.1$ and $G=0.2/m^2$, which corresponds to $\Gamma\approx1.274$, since for higher values of $G$ the code turns unstable, for the remaining parameters see Tab.~(\ref{tab:para1}).
\label{fig:coupledG}}
\end{center}
\end{figure}

In these plots, the profiles are shown for $h=\pm 0.1$. For negative values of $h$, the two condensates repel each other, therefore the density of the neutral condensate in the core, where the charged condensate is destroyed, is enhanced. The neutral condensate is "pushed" into the core by the charged one. By changing the sign of the interaction, the two condensates energetically prefer to occur together, leading to a decreased value of the second condensate in the core. The profile functions $f_1(R)$ and $a(R)$ seem to be nearly identical for both values of $h$, which is a result of the \textit{nearly} symmetric splitting of the neutral condensate around the unperturbed value $f_2=1$. Note however that this is a result of the chosen parameters and is not exact. As expected, the region where the charged condensates vanishes as well as the enhanced or diminished plateau of the neutral condensate grows with the winding number. The slope of the profile function for the gauge field $a(r)$ decreases, which leads to a plateau in the magnetic field which is basically given by the derivative of $a(R)$. This behavior is shown in Fig.~\ref{fig:bcoupled1}.
As a final step we additionally turn on the derivative coupling $G$. Since the magnetic field does not change qualitatively we focus on the profile functions of the two condensates. In the profile functions, the effect of the derivative coupling becomes clearly visible. Whenever the gradient of the charged condensate is significant, clear features in the neutral condensate can be seen, for details see Fig.~\ref{fig:coupledG}.

\begin{figure} [t]
\begin{center}
\includegraphics[width=0.5\textwidth]{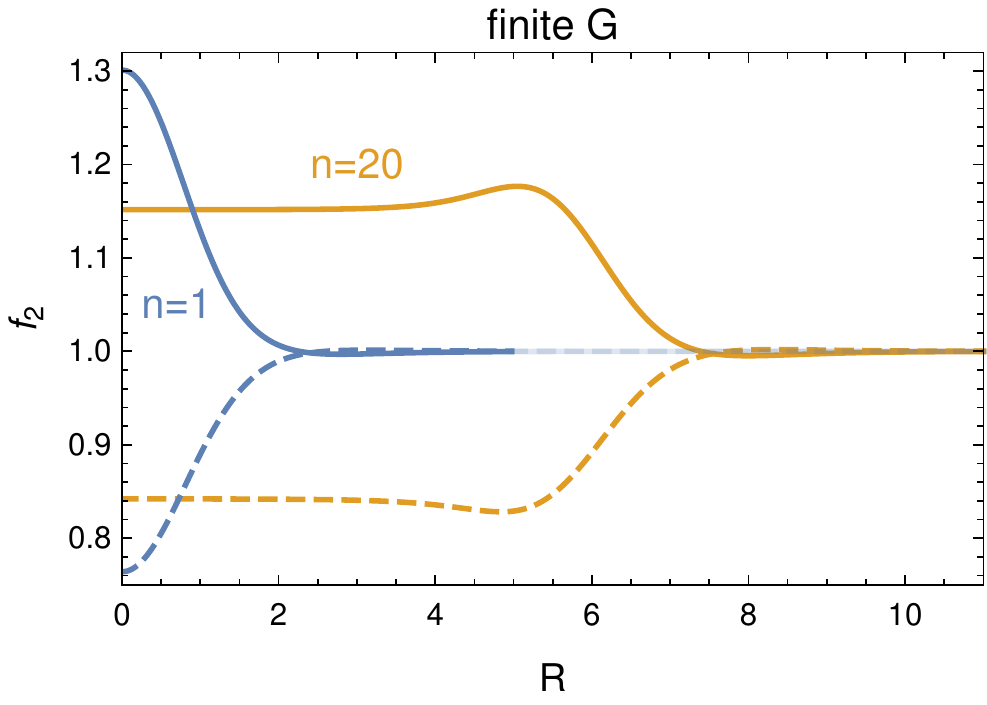}
\caption{Zoom in of the profile function of the neutral condensate $f_2(R)$ with finite derivative coupling $G$ from Fig.~\ref{fig:coupledG}.
\label{fig:zoomG}}
\end{center}
\end{figure}

Generally, a slight extra bump can be seen in the neutral condensate. In Fig.~\ref{fig:zoomG}, a zoom in for the neutral condensate is presented.
\chapter{Flux Tube -- Flux Tube Interaction and First-Order Phase Transitions}
\label{chap:inter}
After the discussion of the single flux tube solutions in the latter chapter, we now turn our attention to the interaction between the flux tubes themselves. If the phase transitions from the Meissner phase to the flux tube phase and from the flux tube phase to the normal-conducting phase were of second order we would be done. 
The critical magnetic fields of the previous chapters would be sufficient to determine the phase structure. We shall see, however, that, due to the presence of the superfluid, 
first-order phase transitions become possible. To this end, we compute the Gibbs free energy of the entire flux tube lattice, rather than only of a single flux tube.
We shall do so in an approximation of flux tube distances much larger than the width of a flux tube. 

We generalize the Gibbs free energy (\ref{Gibbsfl}) to a system with flux tube area density $\nu$ and add a term that takes into account the interaction between the flux tubes,
\be \label{Gibbs_lattice}
\frac{\mathcal{G}_\mathrm{ COE}^{\circlearrowleft\circlearrowleft}}{V}\simeq U_\mathrm{ COE}+\frac{n\nu}{2q}(H_{c1}-H)+\frac{t\nu}{2}\frac{F_\mathrm{ int}^{\circlearrowleft}(R_0)}{L}\, ,
\ee
where we have eliminated $F_{\circlearrowleft}$ in favor of $H_{c1}$ with the help of Eq.\ (\ref{Hc1}), and where we have employed the nearest-neighbor approximation for the interaction 
term with the number of nearest neighbors $t$, and the dimensionless lattice constant $R_0$. For a hexagonal lattice, which we shall use in our explicit calculation, 
$t=6$ and $\nu=2/(\sqrt{3}R_0^2)$. The interaction energy $F_\mathrm{ int}^{\circlearrowleft}(R_0)$ is defined by writing the 
total free energy of two flux tubes with distance $R_0$, say 
flux tubes $(a)$ and $(b)$, in terms of the free energy of the flux tubes in isolation plus the interaction energy, 
\be \label{Fintdef}
F_{\circlearrowleft}^{(a)+(b)}=F_{\circlearrowleft}^{(a)}+F_{\circlearrowleft}^{(b)}+F_\mathrm{ int}^{\circlearrowleft}(R_0) \, .
\ee 
We calculate $F_\mathrm{ int}^{\circlearrowleft}(R_0)$ in appendix \ref{app:fl_int} in an approximation that is valid for large $R_0$. This calculation makes use of the method 
first employed in Ref.\cite{Kramer:1971zza},  adapted to the two-component system with gradient coupling here. All related references mentioned throughout the introduction and this chapter are based on this method or an equivalent one, and our results reproduce the ones of those references in various limits. The result is given by
\bea \label{Fint}
\frac{F_\mathrm{ int}^{\circlearrowleft}(R_0)}{L} &\simeq& 2\rho_{01}^2 R_0 \int_{R_0/2}^\infty \frac{dR}{\sqrt{R^2-(R_0/2)^2}}\bigg\{\frac{\kappa^2 n^2 a'(1-a)}{R^2}-(1-f_1)f_1'-x^2(1-f_2)f_2'\non[2ex]
&&+\frac{\Gamma x}{4}(f_1+f_2+f_1f_2-1)[(1-f_1)f_2'+(1-f_2)f_1']\bigg\} \, . 
\eea
As explained in the appendix in more detail, the integration can be reduced to an integral over the plane  that separates the two Wigner-Seitz cells, which, in this simple 
setup, are two half-spaces. Since the integration along the direction of the flux tubes is trivial, we are left with a one-dimensional integral. As a consequence of the
approximation, only the profile functions of a single flux tube appear in the integrand. In the derivation we have also assumed the asymptotic values of the condensates to be identical to the homogeneous values in the Meissner phase, $\rho_{01}$ and $\rho_{02}$. If the flux tube density increases, the condensate will not be able to recover fully to its homogeneous value, indicating the break down of our approximation. The coherence length $\xi$ can therefore be used as an estimate of the applicability of the sparse lattice approximation. Alternatively, the interaction energy between the flux tubes can be computed by introducing source terms for the asymptotic solutions. Although this approach allows for a simple integration and yields the same result when expressed in terms of the Bessel functions obtained from the asymptotic solutions, the first approach is better suitable for a numerical calculation. It is actually easier to work with the full numeric solutions of the profile functions instead of the asymptotic solutions, due to the complication arising from fixing the numerical constants to the full solution. In the source term approach, a result as a function of the profile functions cannot be obtained, since the dependence on the source terms only vanishes due to the final integration procedure.
We shall later insert our numerical solutions $f_1$, $f_2$, and $a$ into Eq.\ (\ref{Fint}) 
to compute the Gibbs free energy  numerically. Before we do so we extract some simple
analytical results with the help of the asymptotic solutions (\ref{asympsol}). Inserting them into Eq.\ (\ref{Fint}) yields a lengthy expression which is not 
very instructive, especially due to the terms proportional to the gradient coupling. In appendix \ref{app:asymp} it is shown that a simple expression can be extracted, even including
the gradient coupling, if we restrict ourselves to the leading order contribution at large distances. Here we proceed with the simpler case of vanishing gradient 
coupling, $\Gamma=0$, to obtain straightforwardly 
\be \label{Fintasymp}
\frac{F_\mathrm{ int}^{\circlearrowleft}(R_0)}{L} \simeq2\pi\rho_{01}^2\Big[\kappa^2 n^2 C^2 K_0(R_0/\kappa) - D_+^2(\gamma_+^2+x^2) K_0(R_0\sqrt{\nu_+})- D_-^2(\gamma_-^2+x^2) K_0(R_0\sqrt{\nu_-})\Big] \, ,
\ee
where we have used $\gamma_+\gamma_-+x^2=0$ for $\Gamma=0$, which follows from Eqs.\ (\ref{nugam}), the derivatives $K_1'(x) = -K_0(x) - K_1(x)/x$, $K_0'(x) = -K_1(x)$, and 
the integral 
\be \label{Kint}
\int_{R_0/2}^\infty\frac{dR\, K_0(\alpha R)K_1(\alpha R)}{\sqrt{R^2-(R_0/2)^2}} = \frac{\pi  K_0(\alpha R_0)}{\alpha R_0} \, .
\ee
The result (\ref{Fintasymp}) shows that there is a positive contribution, which makes the flux tubes repel each other due to their magnetic fields, and there is a negative contribution, 
which makes the flux tubes attract each other due to the lower loss of (negative) condensation energy if the flux tubes overlap. Let us first see how the case of a single 
superfluid is recovered by switching off the coupling $h$.  (Since we have set $\Gamma=0$, there is no temperature dependence left in $h_T$ and we drop the subscript $T$ in 
this discussion.) As $h\to 0$, the quantities $\nu_\pm$ and $\gamma_\pm$ go to different limits, depending on the sign of $\lambda_1-\lambda_2x^2$. If $\lambda_2 x^2>\lambda_1$, we have $\gamma_+\sim h$ and $\gamma_-\sim h^{-1}$. Numerically, we find that while $\gamma_-$ diverges, 
the product $D_-\gamma_-$ goes to a finite value. Moreover, $D_+$ goes to zero, such that the attractive terms reduce to $- D_-^2\gamma_-^2 K_0(R_0\sqrt{2})$ since $\nu_-\to 2$ for $h\to 0$. In particular, all dependence on $x$, which contains the neutral condensate, has disappeared, as it should be. If, on the other hand, 
$\lambda_2 x^2<\lambda_1$, we see from Eqs.\ (\ref{nugam}) that now $\gamma_+\sim h^{-1}$ and $\gamma_-\sim h$, and it is the other term, $- D_+^2\gamma_+^2 K_0(R_0\sqrt{2})$, which survives, again reproducing the correct result of a single superconductor. The result can be used to find the sign of the interaction at $R_0\to \infty$, 
i.e., to determine whether the flux tubes repel or attract each other at large distances.
Since the Bessel functions fall off exponentially for large $R_0$, we simply compare the arguments of the Bessel functions of the negative and positive contributions. 
For the single superconductor, the long-distance flux tube interaction is thus attractive for $\kappa^2<1/2$ and repulsive for $\kappa^2>1/2$, i.e., the sign change 
appears exactly at the point where $H_c=H_{c2}$. 

Going back to the full expression (\ref{Fintasymp}) for the two-component system, we compare $\nu_-$ with $1/\kappa^2$, because
$\nu_-<\nu_+$, i.e., the term proportional to $K_0(R_0\sqrt{\nu_-})$ is less suppressed for $R_0\to\infty$. Therefore, the point at which the long-range interaction changes from repulsive to attractive is given by 
\bea\label{sign}
\frac{1}{\kappa^2} &=& 1+\frac{\lambda_2}{\lambda_1}x^2 
-\sqrt{\left(1-\frac{\lambda_2}{\lambda_1} x^2\right)^2 + \frac{4h^2 x^2}{\lambda_1^2}} \non[2ex]
&=& \frac{H_{c2}^2}{\kappa^2 H_c^2} \left[1-\frac{h^2}{\lambda_2^2 x^2}+{\cal O}\left(\frac{1}{x^4}\right)\right] \, .
\eea
By comparing Eq.~(\ref{sign}) with Eq.~(\ref{HcHc2}), we see that in the two-component system the long-distance interaction changes its sign at a point \textbf{different} from $H_c=H_{c2}$. In Sec.~\ref{sec:ft_lattice} we learned that this is an indication for the transition from the type-I to the type-II regime, which, for a single superconductor, is equivalent to the intersection point of the critical fields.
This change in the coupled system is made particularly obvious in the second line of Eq.\ (\ref{sign}), where we have expanded the result for large values of $x$, i.e., for large values of the 
neutral condensate compared to the charged one,  $\rho_{02}/\rho_{01} \gg 1$. This limit is interesting for the interior of neutron stars, where protons are expected to 
contribute only about 10\% to the total baryon number 
density\footnote{In Ref.\ \cite{Buckley:2004ca}, the limit $x \gg1 $ was considered ($n_1/n_2 \ll 1$ in the notation of that reference), and it was argued that the critical 
$\kappa$'s for $H_c=H_{c2}$ and the sign change of the long-range interaction are identical, in agreement with the leading-order contribution of our Eq.\ (\ref{sign}). Ref.\ \cite{Buckley:2004ca}
only considered the near-symmetric situation $\lambda_1=\lambda_2\equiv\lambda$, $h=-\lambda+\delta\lambda$ with $0<\delta\lambda\ll\lambda$ (notice that $h<0$ here). 
In this case, our results show that $H_c=H_{c2}$ occurs at $\kappa^2 \simeq \frac{\lambda}{4\delta\lambda}$ and the sign change in the long-range interaction energy at $\kappa^2\simeq\frac{\lambda}{4\delta\lambda}\frac{1+x^2}{x^2}$. Consequently, even in the near-symmetric situation the two 
critical $\kappa$'s are different and only become identical in the limit $x\gg 1$.
}. 
From Eq.\ (\ref{sign}) we recover $\kappa^2=1/2$ for $h=0$, but only if $\lambda_2 x^2>\lambda_1$. 
The reason is that the limits $R_0\to \infty$ and $h\to 0$ do not commute in general: 
in deriving Eq.\ (\ref{sign}) we have fixed $h$ at a nonzero value and let $R_0\to\infty$, while in our above discussion of the single superconductor, we fixed $R_0$ while first letting 
$h\to 0$.

An attractive long-distance interaction between the flux tubes can have very interesting consequences. Recall that $H_{c1}$ is the magnetic field at which the phase with 
a single flux tube is preferred over the phase with complete field expulsion. 
In other words, at $H_{c1}$ the flux tube density is zero and increases continuously, while the flux tube distance decreases continuously from 
infinity at $H_{c1}$. If the interaction at infinite distances is attractive, the flux tubes do not "want" to form an array with arbitrarily small density. Assuming that the interaction 
always becomes repulsive at short range [which our numerical results confirm if we extrapolate Eq.\ (\ref{Fint}) down to lower distances], there is a minimum in the flux tube - flux tube potential, which corresponds to a favored distance between the flux tubes. For a schematic plot see Fig.~\ref{fig:interaction_potential_sch}. As a consequence, the transition from the Meissner phase to 
the flux tube phase occurs at a critical field lower than $H_{c1}$, which we call $H_{c1}'$, at which the flux tube density jumps from 
zero to a nonzero value. 

In the single-component system, this first-order phase transition is not realized because it occurs in the type-I regime.
More precisely, if we were to continue $H_{c1}$ into the type-I regime, then, at $H_{c1}$, it does not matter that the flux tube phase is made more favorable by an attractive interaction 
because the normal-conducting phase is the ground state (under the assumption that the gain in Gibbs free energy is not sufficient to overcome the 
difference to the normal phase). In the two-component system, however, the attractive interaction may exist in the regime where, at $H_{c1}$, the Meissner phase (and the phase with a single flux tube) is already preferred over the normal phase. Hence, any \textit{arbitrarily small} binding energy will lead to a first-order phase transition at $H_{c1}'<H_{c1}$. 
As we move along $H_{c1}$ towards smaller values of $\kappa$, i.e., towards the type-I regime, we hit the critical point given by Eq.~(\ref{sign}), where  
the second-order transition turns into a first-order transition. Since our approximation is accurate for infinitesimally small flux tube densities, our prediction for this point is exact. If we 
then keep moving along $H_{c1}'$, the flux tube density at the transition increases and our results have to be taken with care.

We can directly compute $H_{c1}'$ by equating the Gibbs free energy of the flux tube phase (\ref{Gibbs_lattice}) 
to the Gibbs free energy of the Meissner phase (\ref{GibbsCOE}). In the flux tube phase we have to find the preferred flux tube distance $R_0$ (or, equivalently, the
preferred flux tube density $\nu$), which is given by minimizing the Gibbs free energy. Hence, we compute $H_{c1}'$ by solving the coupled equations
\be\label{Hcprime}
{\cal G}^{\circlearrowleft\circlearrowleft}_\mathrm{ COE} = {\cal G}_\mathrm{ COE} \, , \qquad \frac{\partial {\cal G}^{\circlearrowleft\circlearrowleft}_\mathrm{ COE}}{\partial R_0} = 0 
\ee
for $H$ and $R_0$.
These equations can be solved in the following elegant way:

We start by writing down the interaction energy Eq.~(\ref{Fint}) (for $\Gamma=0$) in a general form ,
\be
\frac{F_\mathrm{ int}^{\circlearrowleft}(R_0)}{L} = 2\rho_{01}^2R_0\int_1^\infty\frac{d\rho}{\sqrt{\rho^2-1}}{\cal F}\left(\frac{R_0\rho}{2}\right) \, , 
\ee
with the integrand
\be
{\cal F}(R) = \frac{\kappa^2n^2 a'(1-a)}{R^2}-(1-f_1)f_1'-x^2(1-f_2)f_2' \, ,
\ee
and where we have introduced the new variable $\rho=2R_0R$ in order to eliminate the dependence of the lower boundary of the integral on $R_0$. Using this, we can write 
\bea \label{dnu2}
0 = \frac{\partial {\cal G}^{\circlearrowleft\circlearrowleft}_\mathrm{ COE}}{\partial \nu} = \frac{n}{2q}[H_{c1}(n)-H] +\frac{t\rho_{01}^2 R_0}{2}\int_1^\infty \frac{d\rho}{\sqrt{\rho^2-1}}\left[{\cal F}\left(\frac{R_0\rho}{2}\right)-\frac{R_0\rho}{2}{\cal F}'\left(\frac{R_0\rho}{2}\right)
\right] \, . \non[0.5ex]
\hfill
\eea
This equation can be trivially solved for $H$. Then we insert the result into the first equation of Eqs.~(\ref{Hcprime}), where the Gibbs free energy of the lattice depends on $H$. The resulting equation becomes very simple,
\be
0 = \int_1^\infty \frac{d\rho}{\sqrt{\rho^2-1}}\left[{\cal F}\left(\frac{R_0\rho}{2}\right)+\frac{R_0\rho}{2}{\cal F}'\left(\frac{R_0\rho}{2}\right)\right] \, .
\ee
This equation can be solved numerically for $R_0$, which is computed in such a way that it represents the energetically favored flux tube spacing at the \textbf{onset}, which then in turn is inserted into the expression for $H$ to compute the corresponding critical magnetic field for the first-order phase transition.  

We may use the same method to compute a potential first-order phase transition from the flux tube phase to the normal-conducting phase, i.e., in the free energy comparison
we replace ${\cal G}_\mathrm{ COE}$ with ${\cal G}_\mathrm{ SF}$ from Eq.\ (\ref{GibbsSF}) and compute the resulting critical field $H_{c2}'$. 

\chapter{Phase Diagrams}
\label{chap:phasesHC}

\section{Taming the Parameter Space}

The results in the previous chapters have shown that the presence of the superfluid 
affects the transition from type-I to type-II superconductivity in a qualitative way, and we will make these results now more concrete by discussing the phase diagram of our model. 
To this end, we need to locate this transition in the parameter space. As discussed, we have to deal with a large number of parameters, 
$m_1$, $m_2$, $\lambda_1$, $\lambda_2$, $q$, $h$, $G$, and the external thermodynamic parameters $T$, $H$, $\mu_1$, $\mu_2$. As we have done it in the discussion of the homogeneous phase structure, we set $m_1=m_2\equiv m$ and $q=2e$ reminiscent of a neutron and proton condensate, and express all dimensionful quantities in units of $m$. Many interesting 
results can already be obtained with a density coupling alone, and we shall therefore set the gradient coupling to zero, $G=0$, which implies $h_T=h$, for all numerical results. 
This leaves us with the 3 coupling constants $\lambda_1$, $\lambda_2$, $h$, plus 4 thermodynamic parameters. 
If we take the condition $H_{c2}=H_c$ as an indication for the location of the type-I/type-II transition, then Eq.\ (\ref{HcHc2}) shows that the transition is, for $G=0$ and fixed $q$, given by a surface in the $\lambda_1$-$\lambda_2$-$h$-space. (This surface is independent of $\mu_1$, $\mu_2$, and $T$, but these parameters of course determine the favored phase, and thus, if embedded in the larger parameter space, not everywhere on that surface the COE phase is the preferred phase at $H=0$.) 
Therefore, the phase diagrams in Fig.\ \ref{fig:phases2}, where $\lambda_1$, $\lambda_2$, and $h$ are fixed, are not very useful for our present purpose, and it is more suitable to start from the $\lambda_1$-$\lambda_2$ plane, 
where, for a given cross-coupling $h$, we obtain a nontrivial curve $H=H_{c2}$. These phase diagrams were presented in Fig.~\ref{fig:l1l2}. Two phase diagrams in the $\lambda_1$-$\lambda_2$ plane at vanishing magnetic field 
are shown in Fig.~\ref{fig:l1l2}, one for positive and one for negative cross-coupling $h$. The chemical potentials are chosen to be larger than the common 
mass parameter, $\mu_i>m$, in which case it is always possible to find negative and positive values of $h$ such that at sufficiently low 
$T$ and $H$ there is a region in the phase diagram where the COE phase is preferred, cf.\ Fig.~\ref{fig:phases2}.

In the interior of a neutron star, as we move towards the center and thus increase the total baryon number, the system will take some complicated path in our multi-dimensional 
parameter space, under the assumption that the model describes dense nuclear matter reasonably well. Here we do not attempt to construct this path. But, we keep in mind
that nuclear matter is expected to cross the critical surface $H=H_{c2}$ if we move to sufficiently large densities. Therefore, we now choose a path with this property. 
Starting from the diagrams in Fig.\ \ref{fig:l1l2}, the simplest way to do this is to choose a path in the $\lambda_1$-$\lambda_2$ plane with all other parameters held fixed. We have already used this parametrization for the discussion of the finite temperature effects, where we parametrized the path by $\alpha\in [0,1]$, which is defined in Eq.~(\ref{alpha}). In Fig.~\ref{fig:l1l2} the paths for positive and negative $h$ that we shall use in the following are shown. Both paths cross from a type-II region for small $\alpha$ into a type-I region for large $\alpha$. In a very crude way, $\alpha$ plays the role of the baryon density in a neutron star. 
Since our paths are chosen such that $\lambda_1$ decreases along them and the charge $q$ is fixed, 
the Ginzburg-Landau parameter $\kappa$ decreases as $\alpha$ increases.

\begin{figure} [t]
\begin{center}
\hbox{\includegraphics[width=0.49\textwidth]{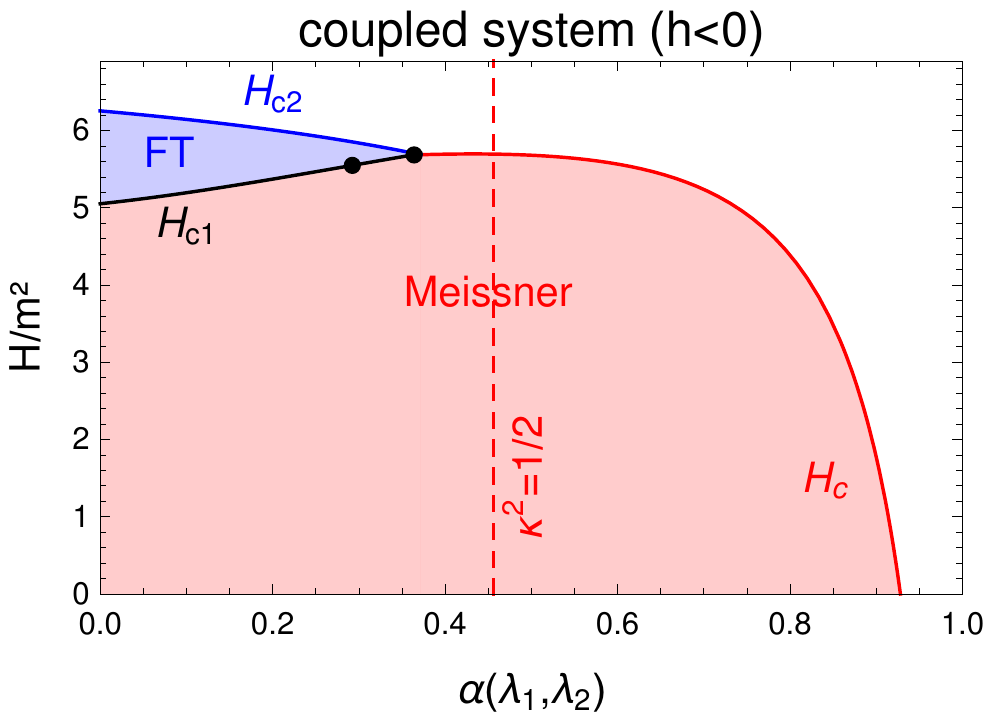}\includegraphics[width=0.5\textwidth]{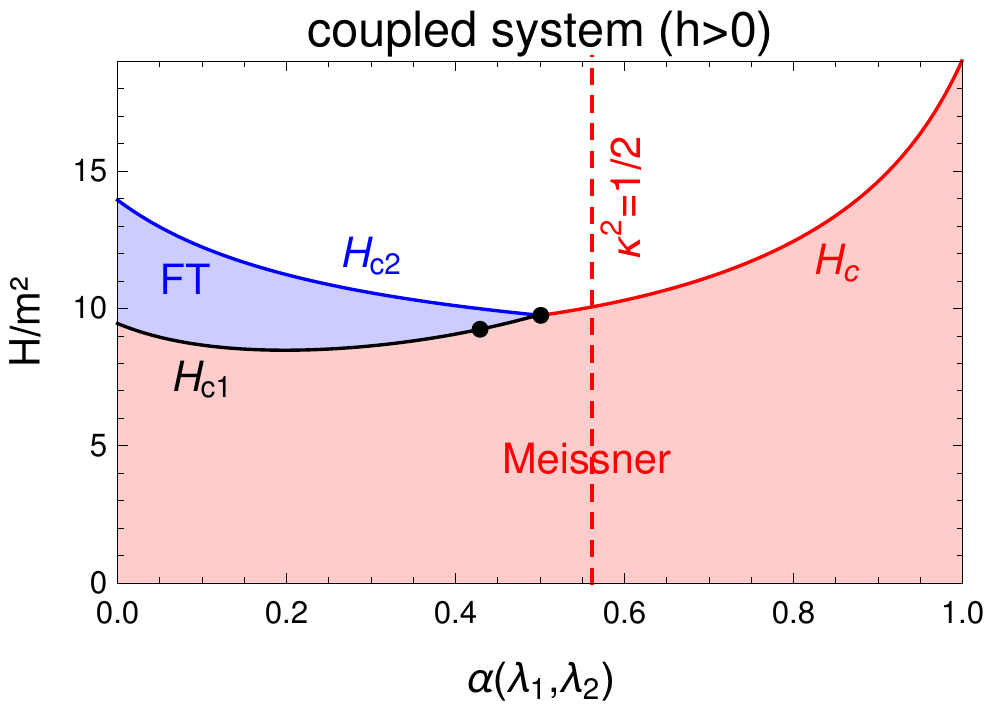}}
\caption{Zero-temperature critical magnetic fields along the paths defined in Fig.~\ref{fig:l1l2}. The magnetic fields are given in units of $m^2$. The black dots on the $H_{c1}$-curves represent the onset of the first-order phase transition.  The three critical magnetic fields do not intersect in a single point although they appear to do so in 
these plots, see Fig.~\ref{fig:zoom} for a zoom-in. The red-shaded area represents the Meissner phase, where as the blue-shaded area is the flux tube (FT) phase. The vertical dashed (red) line shows where the transition from type-I to type-II behavior occurs in the uncoupled system. The numerical parameters are $h=\pm0.1$, $m_1=m_2=m$, $\mu_1=1.5m$ and $\mu_2=1.8m$, the derivative couplings and temperature are zero, $g=G=0,\ T=0$.}
\label{fig:phasesl1l2}
\end{center}
\end{figure}

\section{Critical Magnetic Fields}

In Fig.~\ref{fig:phasesl1l2} we show the zero-temperature critical magnetic fields $H_c$, $H_{c2}$, and $H_{c1}$, computed as explained in Secs.\ \ref{sec:Hc} -- \ref{sec:Hc1}. The horizontal axis is given by $\alpha$. In principle, we can use the model straightforwardly to determine the phases in the entire $\alpha$-$H$-$T$-space. As a rough guide to this three-dimensional space 
notice that increasing the magnetic 
field at fixed $T$ will eventually destroy the charged condensate, i.e., if $H$ is sufficiently large only the SF and NOR phases 
survive, while increasing the temperature at fixed $H$ will eventually destroy all condensates, i.e., at sufficiently large $T$ only the NOR phase survives. For the critical temperatures see Fig.~\ref{fig:Tc}.
Working out the details of the entire phase space might be interesting, but it is tedious and not necessary for the main purpose of this thesis. 
Nevertheless, this possibility makes our model very useful for nuclear matter inside a neutron star. For instance, comparing Fig.~\ref{fig:phasesl1l2} with Fig.~1 in Ref.~\cite{Glampedakis:2010sk}, we see that our results are -- on the one hand -- 
a toy version of more concrete calculations of dense nuclear matter, but -- on the other hand -- more sophisticated because they include all possible phases in a consistent way, not relying on any result within a single-fluid system. 

Here we proceed with the discussion of the critical magnetic fields, and for the remainder of this part we shall restrict ourselves to zero temperature.

\section{Type-I/type-II Transition Region}

At first sight, the phase structure in Fig.\ \ref{fig:phasesl1l2} regarding the critical magnetic fields looks as expected from a single superconductor, only with a critical $\kappa$ that is shifted from the standard value. But, we already know from Sec.\ \ref{sec:inter} that the point at which the second-order onset of flux tubes turns into a first-order transition is different from the point where $H_{c}$ and $H_{c2}$ intersect. We have marked this point in both plots of Fig.\ \ref{fig:phasesl1l2}. Moreover, 
in the presence of the superfluid, the three critical magnetic fields do not intersect in a single point. This is only visible 
on a smaller scale, and we discuss this transition region in detail now. With respect to that region, there is no qualitative difference between the two parameter sets chosen in 
Fig.\ \ref{fig:phasesl1l2}, and therefore we will restrict ourselves to the set with $h<0$.

\begin{figure} [t]
\begin{center}
\hbox{\includegraphics[width=0.5\textwidth]{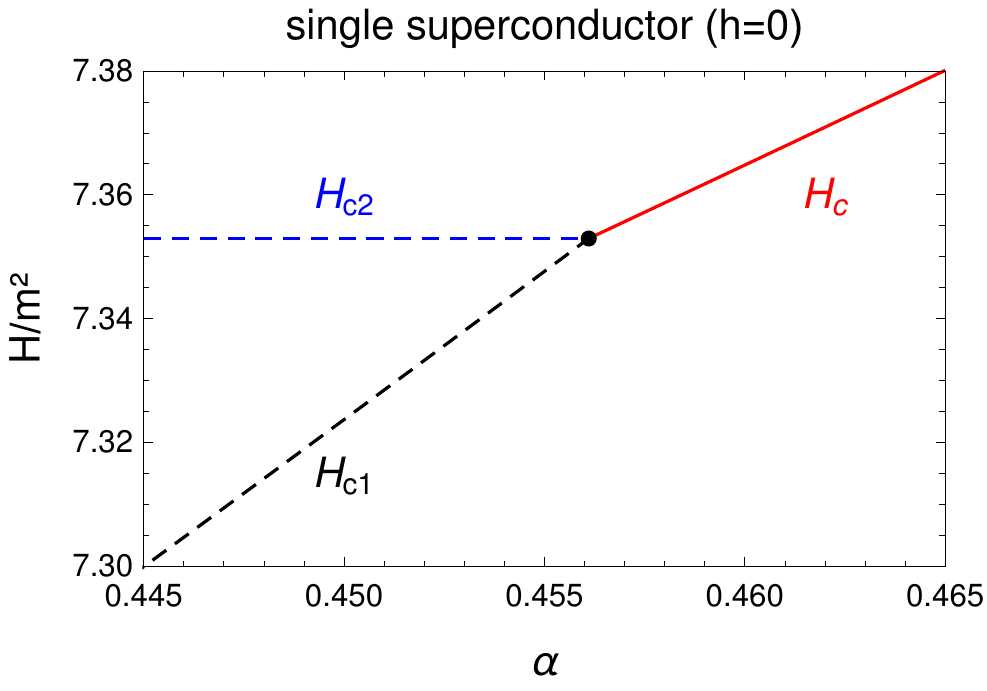}\includegraphics[width=0.5\textwidth]{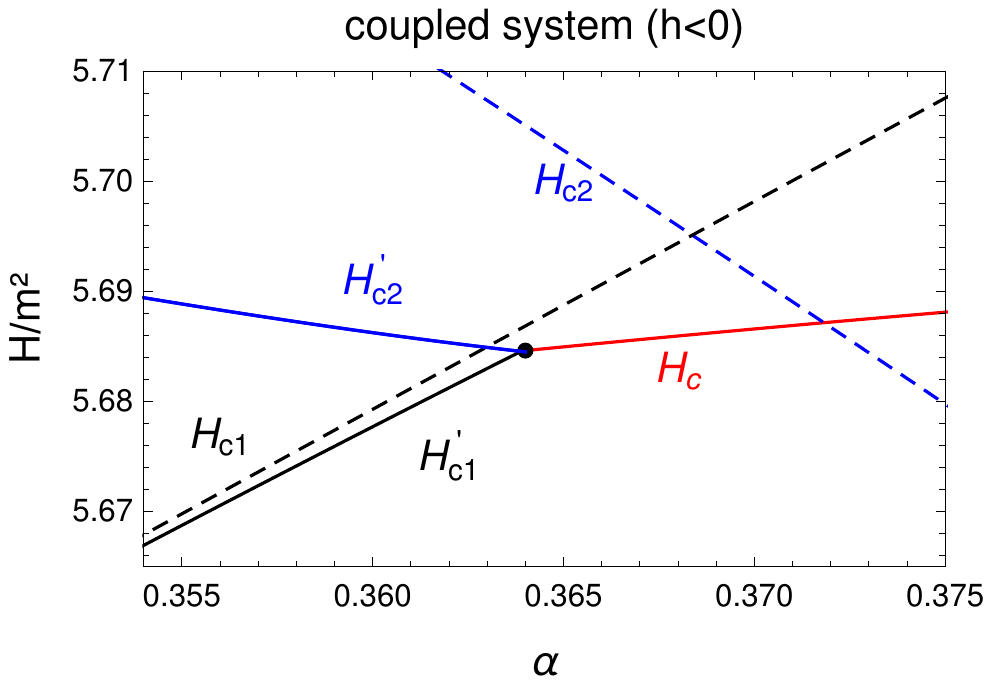}}
\caption{Critical magnetic fields in the type-I/type-II transition region as a function of the parameter $\alpha$ for a single superconductor, 
$h=0$ (left panel), and a superconductor coupled to a superfluid with negative density coupling, $h<0$ (right panel). All other parameters are taken from
 Fig.~\ref{fig:phasesl1l2}, i.e., the right panel is a zoom-in to the transition region of the left panel of Fig.~\ref{fig:phasesl1l2}. Solid (dashed) lines are first (second) order phase transitions.}
\label{fig:zoom}
\end{center}
\end{figure}

In Fig.\ \ref{fig:zoom}, the critical magnetic fields in the region that covers their intersection point(s) is presented. In the left panel, we have, for comparison, set the coupling to the 
superfluid to zero, $h=0$, with all other parameters held fixed. As a result, we obtain the expected phase structure of an ordinary superconductor, as schematically predicted in Fig.~\ref{fig:Schematic-phase-diagram}. All three critical magnetic 
fields intersect at one point -- which can be viewed as a check for our numerical calculation of $H_{c1}$ -- and this point corresponds to $\kappa^2=1/2$. 
For magnetic fields smaller than $H_{c}$ and $H_{c1}$ the superconductor expels the magnetic field completely, and magnetic fields larger than $H_{c}$ and $H_{c2}$ penetrate the system and superconductivity breaks down. In the open "wedge" between $H_{c1}$ and $H_{c2}$, an array of flux tubes 
(with varying flux tube density) is expected to exist, with second-order phase transitions at $H_{c1}$ and $H_{c2}$.

\begin{figure} [h]
\begin{center}
\hbox{\includegraphics[width=0.48\textwidth]{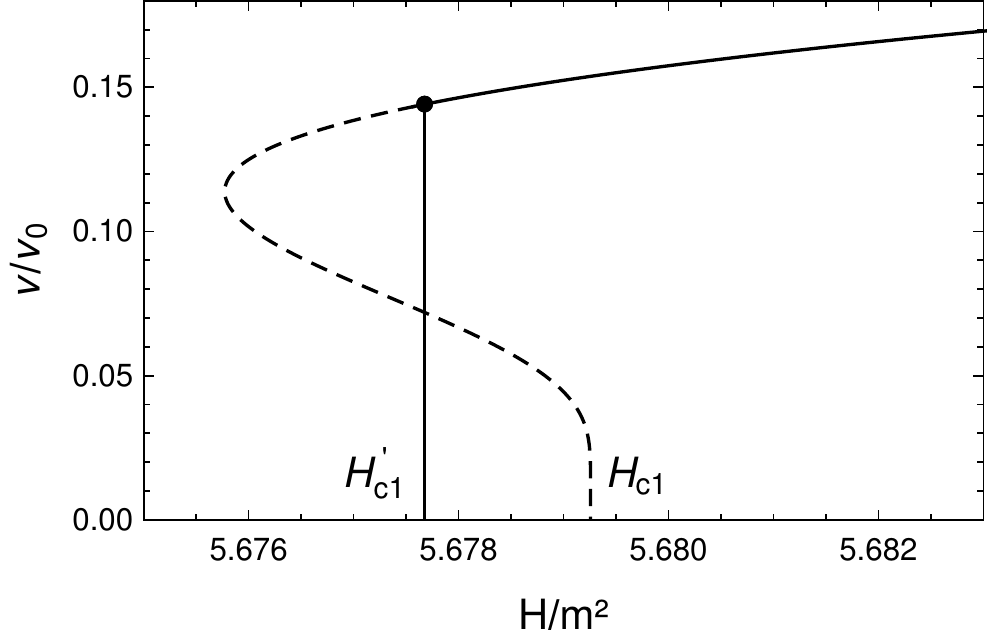}\includegraphics[width=0.52\textwidth]{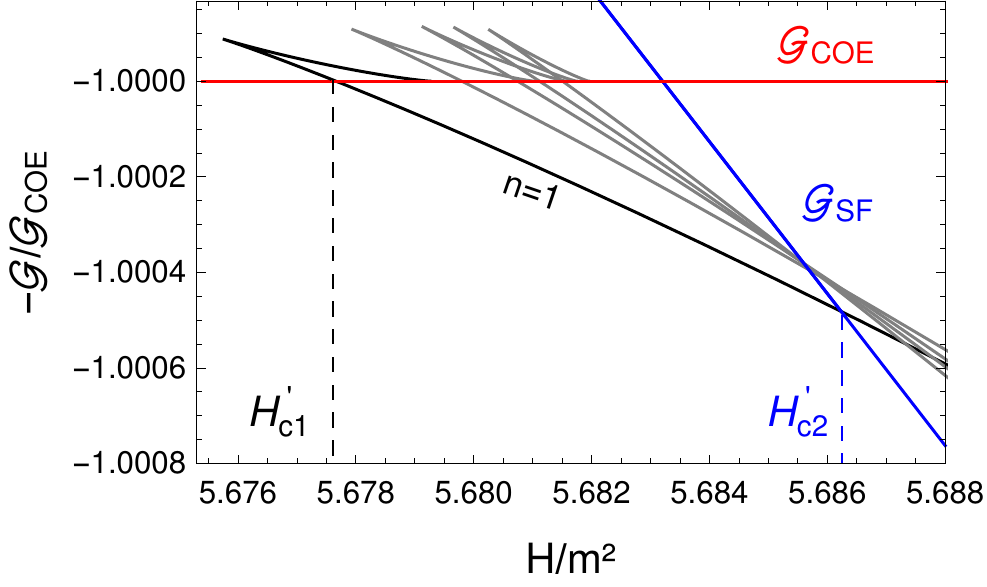}}
\caption{Left panel: flux tube density as a function of $H$ with the parameters of the right panel of Fig.\ \ref{fig:zoom} and $\alpha=0.360$, in units of $\nu_0=1/(\pi\xi^2)$. 
The dashed line shows the 
unstable and metastable part of the solution and is not realized, i.e., the density jumps at $H=H_{c1}'$ from zero to a finite value indicated by the black dot. For $\nu\to 0$,
the dashed line approaches the mass per unit length of the flux tube, i.e., the "would-be" second-order transition $H_{c1}$. 
Right panel: Gibbs free energies as a function of the external magnetic field $H$ for the Meissner, flux tube, and normal-conducting phases, including higher winding numbers, 
$n=2,4,6,10$, which are energetically disfavored.}
\label{fig:Hnu}
\end{center}
\end{figure}

In the right panel we zoom in to the critical region of the left panel of Fig.~\ref{fig:phasesl1l2}. 
From our analytical results we know the following.
\begin{itemize}

 \item[(i)]The critical magnetic fields $H_c$ and $H_{c2}$ intersect at a point given by 
Eq.\ (\ref{HcHc2}), which corresponds for the chosen parameters to $\alpha\simeq 0.37182$. 

\item[(ii)]Just below the curve $H_{c2}$ the flux tube phase is energetically favored over
the normal-conducting phase (not necessarily over the Meissner phase) for all $\alpha < 0.38265$, as we can compute from Eq.\ (\ref{DeltaG}). This point is beyond the right end of the scale shown in Fig.~\ref{fig:zoom}.

\item[(iii)]
The second-order phase transition from the Meissner phase to the flux tube phase turns into a first-order transition at the point given by Eq.~(\ref{sign}), here 
$\alpha\simeq 0.29236$, which is  beyond the left end of the  scale of the plot. In the single superconductor, these three $\alpha$'s (or $\kappa$'s) coincide. Had we only computed 
$H_c$, $H_{c2}$, and $H_{c1}$, we would have obtained a puzzling collection of potential phase transition lines. However, together with the first-order phase transitions
$H_{c1}'$ and $H_{c2}'$, computed from Eq.\ (\ref{Hcprime}), a consistent picture of the phase structure emerges. 
\end{itemize}
Before we comment on this structure, we make the behavior 
at  $H_{c1}'$ more explicit by plotting the flux tube density $\nu$ and the Gibbs free energies in Fig.\ \ref{fig:Hnu}. 
The right panel of this figure includes the results for higher winding numbers. We see that they are energetically disfavored for the parameter set chosen here. 
In Ref.~\cite{Alford:2007np} it was shown that higher winding numbers become important if the magnetic flux, instead of the external field $H$, is fixed. We have confirmed that our numerical procedure presented here indeed reproduce that observation, but we have not checked systematically whether and for which parameters flux tubes with higher winding numbers are favored 
in an externally given magnetic field $H$. 

\begin{figure} [t]
\begin{center}
\includegraphics[width=\textwidth]{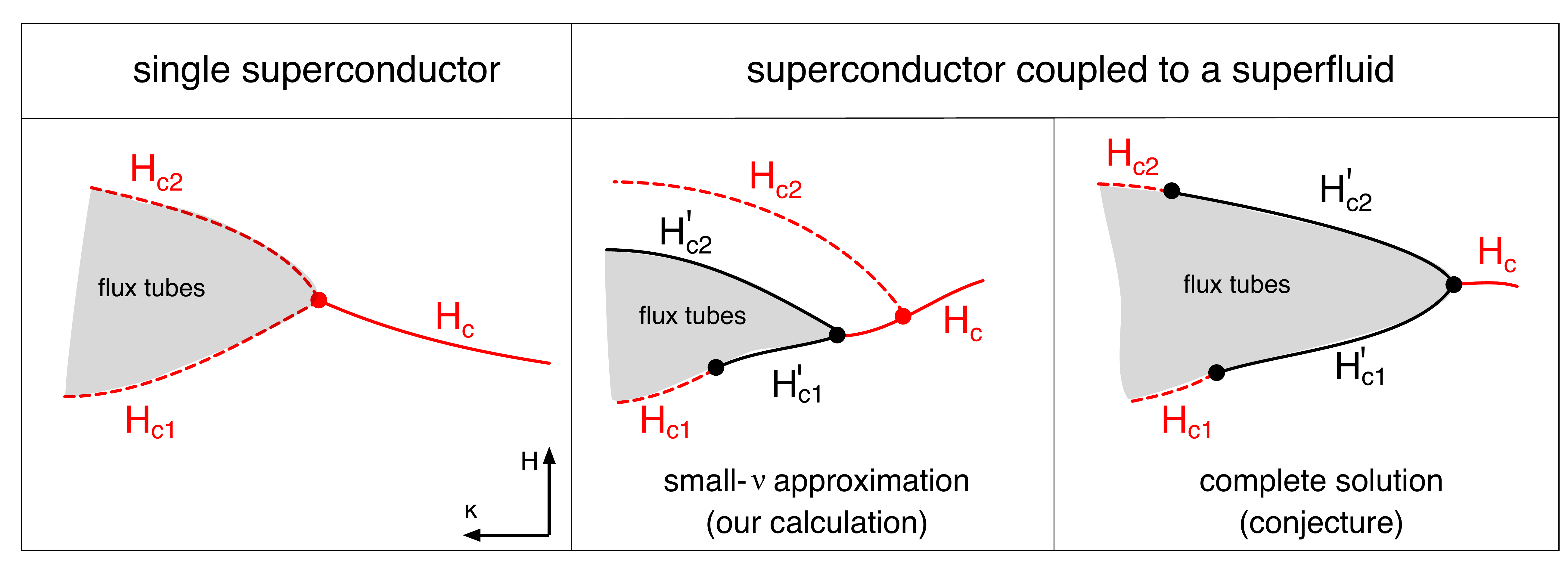}
\caption{Schematic phase structures for a single superconductor and our two-component system in the type-I/type-II transition region. 
Solid (dashed) lines are first (second) order phase 
transitions. The approximation of small flux tube densities $\nu$ rigorously predicts the critical point at which $H_{c1}$ becomes first order. By extrapolating this approximation to compute also the upper critical field  -- where $\nu$ is not small --  one arrives at the inconsistent diagram shown in the middle panel: the first-order transition $H_{c2}'$, computed  from the small-$\nu$ approximation, must not be smaller than $H_{c2}$ ($H_{c2}$ is a rigorous result, independent of the approximation). 
The conjectured phase structure in the right panel is the simplest one consistent with all presented results, including a possible critical point between $H_{c2}$ and $H_{c2}'$.  
 }
\label{fig:HHH}
\end{center}
\end{figure}

The most straightforward interpretation of the right panel of Fig.~\ref{fig:zoom} is to simply ignore the second-order phase transition curves. Then, the topology of the critical region is 
the same as in the left panel, only with first-order instead of second-order transitions at the boundaries of the flux tube phase (with $H_{c1}'$ turning into a second-order phase transition at $\alpha\simeq 0.29236$). However, this cannot be the complete picture. The reason is that after we
have left the flux tube phase through $H_{c2}'$ and keep increasing $H$ we reach $H_{c2}$, and we know that there should be flux tubes just below $H_{c2}$ for 
all $\alpha < 0.38265$. In other words, our result contradicts the observation that $H_{c2}$ is a lower bound for the transition from the flux tube phase to the normal-conducting phase, as explained at the end of Sec.~\ref{sec:Hc2}. 
This contradiction is resolved by remembering the regime of validity of the sparse flux tube lattice approximation for the free energy of the flux tube lattice. This approximation is accurate where $H_{c1}$ turns into $H_{c1}'$ because the distance between the flux tubes is infinitely large at this critical point. 
As we move along $H_{c1}'$ upon increasing $\alpha$, and then along $H_{c2}'$ upon decreasing $\alpha$, the approximation becomes worse and worse. Within the present calculation it is thus not possible to determine the phase structure unambiguously, but it is easy to guess a simple topology of the type-I/type-II transition region that is consistent 
with all the presented results and takes into account the shortcomings of the approximation. This conjectured phase structure is shown in Fig.~\ref{fig:HHH}. 

The motivation for the conjecture is as follows. The existence of the first-order line $H'_{c1}$ and its starting point is predicted rigorously in our approach. Let us
move along that line assuming that we go beyond the approximation and know the complete result. As we move towards large $\alpha$, we will deviate from the line predicted by the sparse lattice approximation. At some value of $\alpha$, we will intersect the curve $H_c$. In order to resolve the contradiction of the resulting 
phase structure, we expect this intersection to occur "on the other side" of the intersection between $H_{c2}$ and $H_{c}$.
This implies that our approximation underestimates the binding energy of the flux tubes, i.e., the flux tube phase is expected to be more favored in the full numerical result. A simple reason -- other than the inconsistency of the phase structure -- why the approximation distorts the full result in this, and not the other, direction, cannot be found trivially within the presented model. Now, at the new, correct, 
intersection of $H'_{c1}$ and $H_c$, there must necessarily be a third line attached, namely $H_{c2}'$ (just like in the results of the sparse lattice approximation). The reason is that if we cross $H_{c1}'$ we end up in the flux tube phase and if 
we cross $H_c$ we end up in the normal-conducting phase, and these two phases must be separated by a phase transition line. This critical field $H_{c2}'$ might be 
larger than $H_{c2}$ for all $\alpha$ (below the $\alpha$ of the triple point where $H'_{c1}$, $H'_{c2}$ and $H_c$ intersect) or $H_{c2}'$ might merge with $H_{c2}$, leading to 
an additional critical point. The latter is the scenario shown in the right panel of Fig.~\ref{fig:HHH}. One might ask whether $H_{c1}'$ and $H_c$ intersect exactly at the point where $H_c$
and the second-order line $H_{c2}$ intersect. In this case, the entire upper critical line would be of second order and given by $H_{c2}$. However, this 
seems to require some fine-tuning of the interaction between the flux tubes since the second-order line $H_{c2}$ does not know anything about this interaction.

\chapter{Flux Tube Clusters}

\begin{figure} [t]
\begin{center}
\hbox{\includegraphics[width=0.5\textwidth]{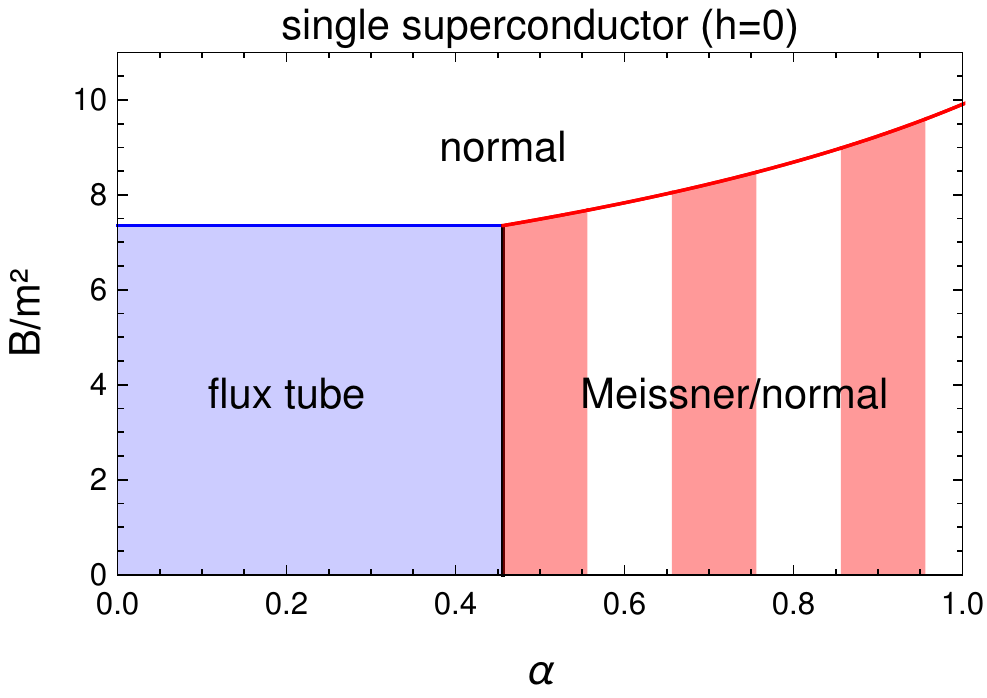}\includegraphics[width=0.5\textwidth]{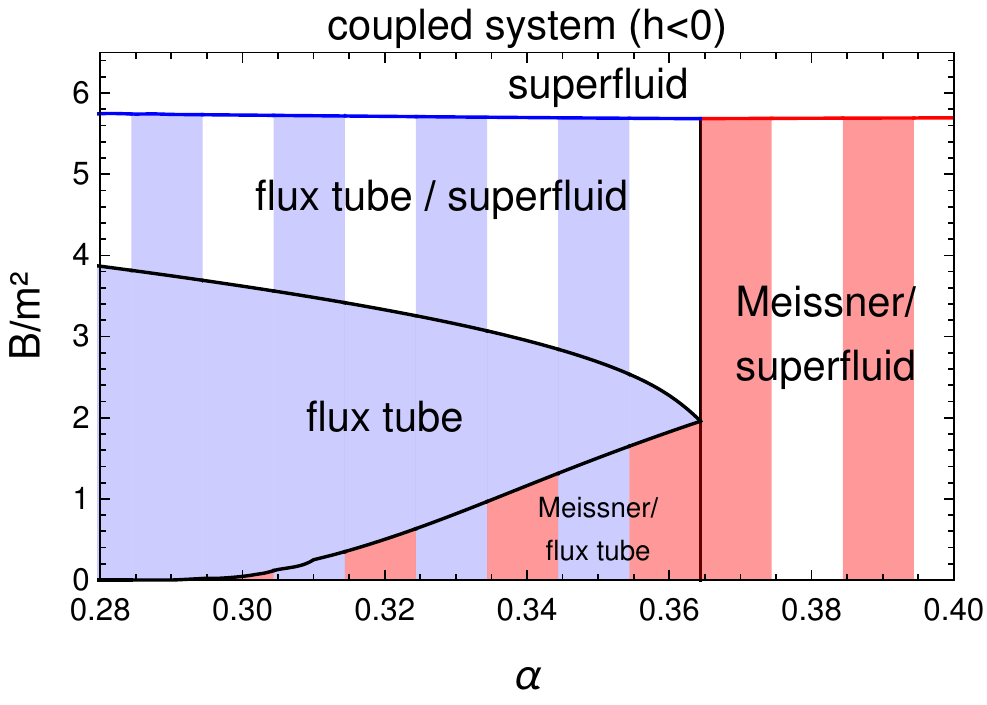}}
\caption{Phases in the $B$-$\alpha$ plane, computed with the parameters
and from the results of Fig.~\ref{fig:zoom}. In a single superconductor (left panel), the magnetic field penetrates in the form of a flux tube array ("flux tube"), 
through macroscopic regions in a mixed phase ("Meissner/normal") or homogeneously and space filling ("normal"). In a superconductor coupled to a superfluid 
(right panel), it can also penetrate in the form of flux tube clusters, either in 
a mixture with field-free regions ("Meissner/flux tube") or in a mixture with normal-conducting regions ("flux tube/superfluid").
The "Meissner/flux tube" phase is, for the chosen parameters, only possible for $\alpha> 0.29236$ (where the 
phase transition in the $H$-$\alpha$ plane is of first order).
 }
\label{fig:Balfa}
\end{center}
\end{figure}

First-order phase transitions in general open up the possibility of mixed phases. In a first order transition, the order parameter, in our case for instance the flux tube density $\nu$ or the internal magnetic field $B$, jumps instead of smoothly changing. In a second-order phase transition, the flux tube density starts at zero at $H_{c1}$, which is not possible in the first-order transition, where it jumps at the onset of the flux tube (lattice) phase. This jump $\Delta\nu$ can be "softened" by allowing for mixed phases, where the \textbf{averaged} order parameter changes smoothly. This means we expect clusters of flux tubes within either the Meissner or the pure superfluid phase. More technically speaking, the first-order phase transitions with $H$ as an external variable translate into mixed phases if we fix the magnetic field $B$ (spatially averaged) 
instead. Again, this can be illustrated by the analogy to the onset of baryonic matter at small temperatures as a function of $\mu_B$. This onset 
is a first-order transition  with a discontinuity in  the
baryon number density $n_B$. If we instead probe this onset with fixed $n_B$ (spatially averaged), we pass through a region of mixed phases, for example nuclei 
in a periodic lattice, until we reach the saturation density. This mixed phase of clustered nuclear matter in the vacuum (mixed with some electrons) is the basic form of matter that our everyday worlds is composed of. 
Further mixed phases are realized in the outer regions of a neutron star and called "nuclear pasta" \cite{RevModPhys.89.041002}, and it would be an intriguing manifestation  
if the mixed flux tube phases discussed here are realized in the core of the star. Each first-order transition in $H$ yields two critical magnetic fields $B$ which 
we compute as follows. At $H_{c1}'$, the lower critical field is $B=0$, and the upper critical field is $\langle B\rangle=\Phi_0 \nu$ [using Eq.\ (\ref{drB})], where $\nu$ is the 
numerically computed flux tube area density as we approach the first-order transition from above; 
at $H_{c2}'$, the lower critical field is $\langle B\rangle=\Phi_0 \nu$, with $\nu$ now being the numerically computed density as we approach  
the first-order transition from below, while the upper critical field is $B=H_{c2}$; at $H_c$, the lower critical field is $B=0$, and the upper one is $B=H_c$. This calculation is performed using the parameters of Fig.~\ref{fig:zoom}. 
As discussed for the $H$-$\alpha$ phase diagrams above, also for the $B$-$\alpha$ phase structure we do not expect the sparse lattice approximation to yield quantitatively reliable results where the flux tube density is large. Therefore, these results 
reflect the topology of the $B$-$\alpha$ phase diagram correctly, but the precise location of the phase transition lines cannot be determined within this approach. The phase diagrams for the single superconductor and the two-component system are shown in Fig.~\ref{fig:Balfa}.
In a single superconductor, there is only one possible mixed phase: macroscopic regions in which the magnetic field penetrates, mixed with regions in which the magnetic field remains expelled \cite{tinkham2004introduction}. The geometric structure of these regions depends on the details of the system such as the surface tension, and it is beyond the scope of this paper to determine them, however this mixed state is experimentally well known. 
In the two-component system, two additional mixed phases are possible, both of which contain flux tube clusters. (Unrelated to the first-order phase transitions pointed out here, flux tube clusters have been suggested to exist in neutron stars in the vicinity of superfluid neutron vortices \cite{1995ApJ...447..305S}.) 
Firstly, at $H_{c1}'$, flux tube clusters are immersed in a field-free superconducting region, as predicted for "type-1.5 superconductivity" \cite{PhysRevB.72.180502}. Secondly, at $H_{c2}'$, there is a mixed phase of flux tubes with 
the normal-conducting phase, i.e., superconducting regions that enclose flux tubes and that are themselves surrounded by completely normal-conducting regions. Once again, the exact form of these clusters is hard to determine and left for future research.

\chapter{Summary: Critical Magnetic Fields of a Superconductor Coupled to a Superfluid}
\label{chap:summary}

We have seen that the coupling to a superfluid can have profound effects on the magnetic properties of a superconductor. As for the investigation of the hydrodynamic instabilities in the multicomponent system, we have started from our microscopic 
model for two complex scalar fields, coupled to each other via density and gradient coupling terms, with one of the fields being electrically charged. 
By computing the thermal excitations of the system we have derived a Ginzburg-Landau-like effective potential 
for the charged and neutral condensates and the gauge field. This potential has then been evaluated at nonzero temperatures and external magnetic fields, 
computing the two condensates dynamically for all 4 possible phases: condensation of both fields (superconductor + superfluid), condensation of only one field 
(pure superconductor or pure superfluid), or no condensation. The structure of the resulting phase diagram in the multi-dimensional parameter space, 
with the main focus on
the transition region between type-I and type-II superconductivity, has been discussed. To this end, the critical magnetic fields $H_c$, $H_{c2}$
(analytically) and $H_{c1}$ (numerically, based on the profile functions of a magnetic flux tube) have been computed. In contrast to the standard scenario of a single superconductor, these
three magnetic fields do not intersect in a single point if the superconductor coexists with a superfluid. The phase structure around these intersection points is (at least partially) 
resolved by computing the first-order phase transitions $H_{c2}'$ and $H_{c1}'$. This has been done by employing a simple approximation for the free energy of a flux tube array
that is valid for large flux tube distances and that effectively reduces the calculation to solving the equations of motion for a single flux tube. The 
new critical fields $H_c$, $H_{c2}'$, $H_{c1}'$ \textit{do} intersect in a single point, restoring the topology of the transition region, with (segments of) the second-order transition lines
replaced by first-order transitions. In particular, we have identified a new critical point -- and derived an analytical expression for its location -- where the 
second-order flux tube onset $H_{c1}$ turns into a first order transition $H_{c1}'$. The presence of the first-order transitions allows for mixed phases with flux tube clusters, very similar to a type-1.5 superconductor, which consists of two charged fields coupled indirectly through the gauge field. 

There are several possible improvements and extensions. The approximation for the flux tube array can be improved for instance by determining dynamically 
the values of the condensates far away from the flux tubes instead of using the values of the homogeneous phase. To settle the precise location of the phase transition lines, 
it would be interesting to perform a brute force numerical calculation of the free energy of the flux tube phase, for which our results are a valuable guidance. First numerical investigations in Ref.~\cite{david_alex} suggest the correctness of the proposed structure of the full phase diagram. There are several other interesting aspects of our model which were mentioned but not worked out in detail. For instance, 
one could perform a more systematic study of the effect of the derivative coupling, which has been included in all the analytical results, but set to zero in the final numerical results
of the phase diagrams. Or one could perform a more detailed study of flux tubes with higher winding numbers, which turned out to be energetically disfavored for the 
parameter regime we have studied, but which are known to potentially play a role in the two-component system. 

The setup and obtained results are applicable to dense nuclear matter in the core of neutron stars. However, for more direct predictions one must fit the model parameters, such as the density coupling and gradient coupling, to values predicted for nuclear matter and eventually compute the phase structure as a function of the baryon number density rather than of an abstract 
model parameter.

\part{Magnetic Defects in Color Superconductivity}

\chapter{Color Superconductivity}
\label{chap:color}
We have argued in Sec.~\ref{sec:QMinCS} that nuclear matter at high baryon density undergoes a phase transition to deconfined quark matter. It is unclear at which densities this transition occurs, however it might be within the conditions present in compact stars. Most likely, possible quark matter in compact stars will exist in a color-superconducting state.
Since quarks are fermions, they undergo Cooper pair condensation instead of direct Bose-Einstein condensation. This possibility was first mentioned before the invention of the current theory of QCD itself in Refs.~\cite{Ivanenko:1969gs,Ivanenko:1969bs}, but only later it was shown that such a quark matter phase exists at sufficiently high densities \cite{Baym:1976yu}. The actual study of quark Cooper pairing started in the late seventies of the $20^{th}$ century in works of Barrois \cite{Barrois:1977xd,Barrois:1979pv} and Frautschi \cite{Frautschi:1978rz}, where the term "color superconductivity" originates. Bailin and Love already classified many possible pairing patterns of color superconductivity in the works Ref.~\cite{Bailin:1979nh, Bailin:1983bm}. Color superconductivity became of phenomenological interested due to the prediction of large pairing gaps in the quasi-particle spectrum, where models using a contact interaction of quarks predict a gap of $\Delta\approx10-100$ MeV \cite{Alford:1998mk}. Note that in contrast to the standard BCS-gap, which is exponentially suppressed by the square of the coupling, the color-superconducting gap is parametrically larger,
\be
\Delta_{\mathrm{BCS}}\propto\mathrm{exp}\left(-\frac{4\pi^2 M^2}{g^2\mu^2}\right) \, , \qquad \Delta_{\mathrm{CFL}} \propto \mathrm{exp}\left(-\frac{3\pi^2}{\sqrt{2}g_s}\right) \, , 
\ee
where $M$ and $g$ are the mass of the Cooper pair and the coupling constant of the BCS Hamiltonian and $g_s$ is the (strong) QCD coupling constant. This particular form of the gap makes it inaccessible for perturbative methods, since in a Taylor expansion around $g_s=0$, every term of the expansion vanishes.
For a general review on color superconductivity see Ref.~\cite{Alford:2007xm} and references therein.

The attractive interaction, which is necessary for Cooper pairs to form, is provided by the strong interaction. At (asymptotically) high densities, this interaction is dominated by one-gluon exchange \cite{Schafer:1999jg}. We have seen in Fig.~\ref{fig:quarks} that quarks come in six different "flavors", however in compact stars the high masses of the charm, bottom and top quarks which are all on the GeV scale, allow us to ignore their contributions. The chemical potential in compact stars is most likely only high enough to populate the lighter three flavors, consisting of up, down and strange. Note that the bare mass of the strange quark at $m_s\approx95$ MeV is already comparably heavier than the mass of up and down quarks at a few MeV. Additionally, quarks come in three different color charges, which are denoted by red, green and blue. Considering that a Cooper pair can roughly be written as the diquark expectation value 
\be
\left< \psi \psi \right> \, ,
\ee
with the quark spinor $\psi$, we can see that there is a plethora of possible pairing structures in color, flavor and spin space. Before discussing this structure in a more detailed, technical manner we quickly recapitulate the symmetry properties of QCD.

\subsubsection*{Symmetries of QCD}
\begin{itemize}
\item \textit{local symmetry:} The local gauge group of QCD is given by $SU(N_c)$, where the number of colors in QCD is given by $N_c=3$. Since the dimension of $SU(N)$ is given by the dimension of the group, $\mathrm{dim}(SU(N))=N^2-1$, $SU(3)$ is generated by the eight generators $T_a$, $a=1,...,8$, which can be represented by the Gell-Mann matrices $\lambda_a$ via $T_a=\frac{\lambda_a}{2}$. The Gell-Mann matrices can be found in App.~\ref{app:gell-mann}. The generators follow the Lie-Algebra
\be
[T_a,T_b]=if^{abc}T_c \, ,
\ee  
where $f^{abc}$ are the structure constants of $SU(3)$.
As a consequence, we have to deal with eight self-interacting gauge fields $A^\mu_a$ which represent the eight gluons. In contrast, in the discussion of electromagnetic superconductivity, we had to deal with a single gauge field since the gauge group of quantum electrodynamics is given by $U(1)$.
The resulting field strength tensor, which includes the gluon self-interaction, is given by
\be \label{eq:Fgluon}
F^{\mu\nu}_{a}=\p^\mu A^\nu_a-\p^\nu A^\mu_a+g_s f^{abc}A^\mu_b A^\nu_c \, ,
\ee
with the strong coupling constant $g_s$ and where the gauge fields contain the corresponding generator and therefore do not commute in general.
\item \textit{global symmetries}: In the case of three \textit{massless} quark flavors, QCD exhibits a global $U(N_f)\times U(N_f)$ symmetry with the number of flavors $N_f=3$. This can be decomposed into
\be
SU(3)_L\times SU(3)_R \times U(1)_B \times U(1)_A \, .
\ee
The groups $SU(3)_R$ and $SU(3)_L$ represent the chiral symmetry of (massless) three-flavor QCD: the Lagrangian is invariant under separate rotations of left- and right-handed quarks. For (asymptotically) high chemical potentials, the approximation of   massless quarks becomes exact. For lower chemical potentials, the masses break the chiral symmetry explicitly because left- and right-handed spinors are mixed by the mass terms. In the case of equally massive quarks, i.e.\ a mass matrix which is proportional to the unitary matrix in flavor space, the chiral symmetry is broken down to simultaneous rotations of left- and right-handed particles,
\be 
SU(3)_L\times SU(3)_R \to SU(3)_{R+L} \, .
\ee
But even if we take all quark masses for all quark flavors into account, the strong interaction conserves flavors; different quarks cannot transform into each other. In this case, the flavor symmetry reduces to separate $U(1)$ groups for each quark flavor. In principle, this would allow us to introduce a separate chemical potential for each quark flavor. However, in compact stars the weak interaction, which does not conserve flavor, leads to beta-equilibrium. In the limit of equal (or vanishing as here) quark masses, beta-equilibrium is in accordance with the use of a single chemical potential $\mu_q$.

Additionally, the (approximate) chiral symmetry can be \textit{spontaneously} broken by a chiral condensate $\left<\bar{\psi}\psi\right>$.
The conservation of baryon number $U(1)_B$ is exact and will play an important role in the discussion of superfluidity in quark matter. The axial symmetry $U(1)_A$ is anomalous, which means that it is conserved classically via Noether's theorem but broken by quantum effects.
\end{itemize}

\section{Order Parameter of Color Superconductivity}
For Cooper's theorem to work, we need an attractive interaction between the fermions we want to pair. Following Chap.~4 of Ref.~\cite{Schmitt:2010pn}, we write down the pairing process in terms of representations of the quarks, which live in the fundamental representation of the color gauge group $SU(3)$. Consequently, we can write
\be 
[3]_c\otimes[3]_c = [\bar{3}]^A_c\oplus[6]_c^S \, .
\ee
This basically states that the two quarks in the fundamental representation on the left-hand side interact in an antisymmetric (A) anti-triplet channel and a symmetric (S) sextet channel. The attractive interaction takes place in the anti-triplet. This can be made plausible by the following argument: consider an overall color neutral baryon, which consists per definition of three quarks. It can be thought for instance of consisting of a diquark of a red and a green quark, and an additional blue quark. The combination of red and green gives anti-blue, so it lives in the anti-triplet. In combination with the third, blue quark the baryon becomes color neutral. Since it is effectively a bound state of the diquark and the blue quark, the interaction in this channel must be attractive. Therefore, we expect quarks of different color to pair. Since the Cooper pair condensate is a quark quark condensate, we have no possibility to form a color-neutral object. As a consequence, the order parameter is color charged. This is why we speak of color superconductivity.

It is most likely that the pairing for high densities takes place in the anti-symmetric spin zero channel, although under certain circumstances spin-$1$ pairing is feasible \cite{Schmitt:2004hg}. Since the overall wave function has to be anti-symmetric as well, the pairing should take place in the anti-symmetric flavor channel. Let us consider the chiral symmetry to be exact for the moment, then both $SU(3)_f$ groups with $f=L,R$ lead to the same representations as described for the color gauge group. We can therefore write down the color-flavor structure of the Cooper pair as
\be
\left<\psi\psi\right>\in [\bar{3}]^A_c\otimes [\bar{3}]^A_f \, .
\ee
As a final ingredient, we have to discuss the Dirac structure of the Cooper pair, which defines the pairing pattern even further. For instance, a Cooper pair of the form $\left<\psi C \gamma_5\psi\right>$,  where $C=i\gamma^2\gamma^0$ is the charge conjugation operator with the gamma matrices $\gamma_i$, leads to even-parity, spin singlet pairing. In principle, many different structures are possible, for a general discussion see Ref.~\cite{Pisarski:1999av}. In order to find the correct structure for a given set of thermodynamic parameters, one has to compute the thermodynamical potential and minimize it with respect to the gap for a given Dirac structure. The potential which has the lowest free energy after inserting the result for the gap wins. We will omit this computation, since it is not relevant for the following discussion, and work with the presented structure.
Let us denote the basis matrices in color and flavor space, where we need three matrices for both spaces, by $(J^A)^{\alpha\beta}$ and $(I_B)_{ij}$ with the color indices $A,\alpha,\beta\le 3$ and the flavor indices $B,i,j\le 3$. Due to their anti-symmetric properties, the epsilon tensor turns out to be a viable candidate, thus we write
\be
(J^A)^{\alpha\beta}=-i\varepsilon^{\alpha\beta A} \, \qquad \mathrm{and} \qquad (I_B)_{ij}=-i\varepsilon_{ijB} \, .
\ee
Using this basis, we write
\be
\left<\psi_i^\alpha C \gamma_5\psi_j^\beta\right>\propto\varepsilon^{\alpha\beta A}\varepsilon_{ijB}\Phi_A^B \, ,
\ee
where the $3\times3$ matrix $\Phi$ determines the specific color-superconducting phase. Our goal in the next chapter will be to write down a Ginzburg-Landau type potential using this matrix.

In the following we are going to distinguish between two different order parameters. Assume for the moment that we are looking for the ground state of three flavor quark matter at $T=0$ for asymptotically large baryon chemical potential. In this scenario, it is reasonable to neglect all three quark masses, leading to an exact chiral symmetry. As an educated guess, we are looking for the most symmetric pairing pattern, which means we assume
\be
\phi_A^B\propto \delta_A^B \Rightarrow \left<\psi_i^\alpha C \gamma_5\psi_j^\beta\right>\propto\varepsilon^{\alpha\beta A}\varepsilon_{ijA}\, .
\ee
This is the order parameter of the \textbf{color-flavor locked} phase, short CFL \cite{Alford:1998mk}.
By checking for which color and flavor indices the order parameter is non-vanishing, we can read off the CFL pairing pattern, which is given by $rd-gu,\,bu-rs,\,bd-gs,\,ru-gd-bs$, where $rd-gu$ denotes the pairing of a red down with a green up quark and so on. All colors are "locked" to  certain flavors, i.e.\ there is a one-to-one correspondence between the three colors and the three flavors. The CFL order parameter is the only one where all quarks participate in the pairing, which leads to the biggest gain in condensation energy, confirming our claim that CFL presents the ground state of QCD under the discussed conditions.
In terms of symmetries, the CFL phase is still symmetric under a combined rotation in color and flavor space. The symmetry breaking pattern is thus given by
\be\label{eq:SSB_CFL}
[SU(3)_c]\times SU(3)_R\times SU(3)_L \times U(1)_B \to SU(3)_{c+L+R}\times \mathbb{Z}_2 \, ,
\ee
where square brackets denote local symmetries. We can see that CFL breaks the global $U(1)_B$ as well, meaning that the CFL phase is, besides a color superconductor, a superfluid as well. Additionally, chiral symmetry is broken due to the locking of color and flavor and not by the formation of a chiral condensate, which is a particle-\textit{anti-}particle condensate. This gives rise to eight (pseudo-) Goldstone bosons, the meson octet, which includes for instance kaons and pions. The local $SU(3)_c$ is completely broken, therefore we expect eight massive gauge bosons. But is CFL an electrical superconductor as well? To answer this question we have to investigate the fate of the generator of electromagnetism, which we denote by $Q$ and which acts in flavor space. In the unbroken phase, it is contained  in the chiral groups, 
\be 
[U(1)_Q] \subset SU(3)_R\times SU(3)_L \, ,
\ee
since it can be built by generators of this larger group. However, after the SSB, only a different generator $\tilde{Q}\subset SU(3)_{c+L+R}$ remains unbroken, which is a linear combination of the original charge generator and (at least) one gluon generator. \footnote{In most cases this linear combination only consists of the charge operator and the eighth gluon generator, however this statement is gauge dependent and one can in principle work in a more complicated basis where there is an admixture of the third gluon as well.} Speaking in different terms, there is a linear combination of a gluon (in CFL normally the eighth) with the photon that remains massless. The orthogonal combination $\tilde{T}_8$ becomes, together with the remaining seven gluons, massive. This phenomenon is called \textbf{rotated electromagnetism}. Every diquark in the CFL phase is neutral with respect to $\tilde{Q}$, but differently charged under $\tilde{T}_8$. Whether $\tilde{Q}$ is predominately a photon or a gluon depends on the mixing angle. For CFL, it turns out that the mixing angle is rather small, making the CFL phase almost transparent for ordinary light. This is plausible since the relative size of the coupling constants plays a crucial role in determining the angle, and the strong coupling in neutron star environments is much larger than the electromagnetic coupling, $g\gg e$. Only the orthogonal generator, which is a small admixture of the photon to a gluon, becomes massive. The answer whether CFL can be considered a electronic superconductor therefore is yes, but only "a little bit". Let us assume a sphere of quark matter in the CFL phase, like the core of a compact star, in an external electromagnetic field. Most of the field, given by the mixing angle, freely penetrates the sphere. However, a small part of the field is, depending on the type of superconductivity present and the strength of the field, either expelled due to the Meissner effect or forms $\tilde{Q}$-flux tubes. We will discuss rotated electromagnetism in more (technical) detail later on, where we will determine the mixing angle for CFL for instance. 
For a more detailed discussion on the symmetry breaking pattern of CFL and its topological defects, see Ref.~\cite{Eto:2013hoa}.

\begin{figure}[t]
\begin{center}
\hbox{\includegraphics[width=0.33\textwidth]{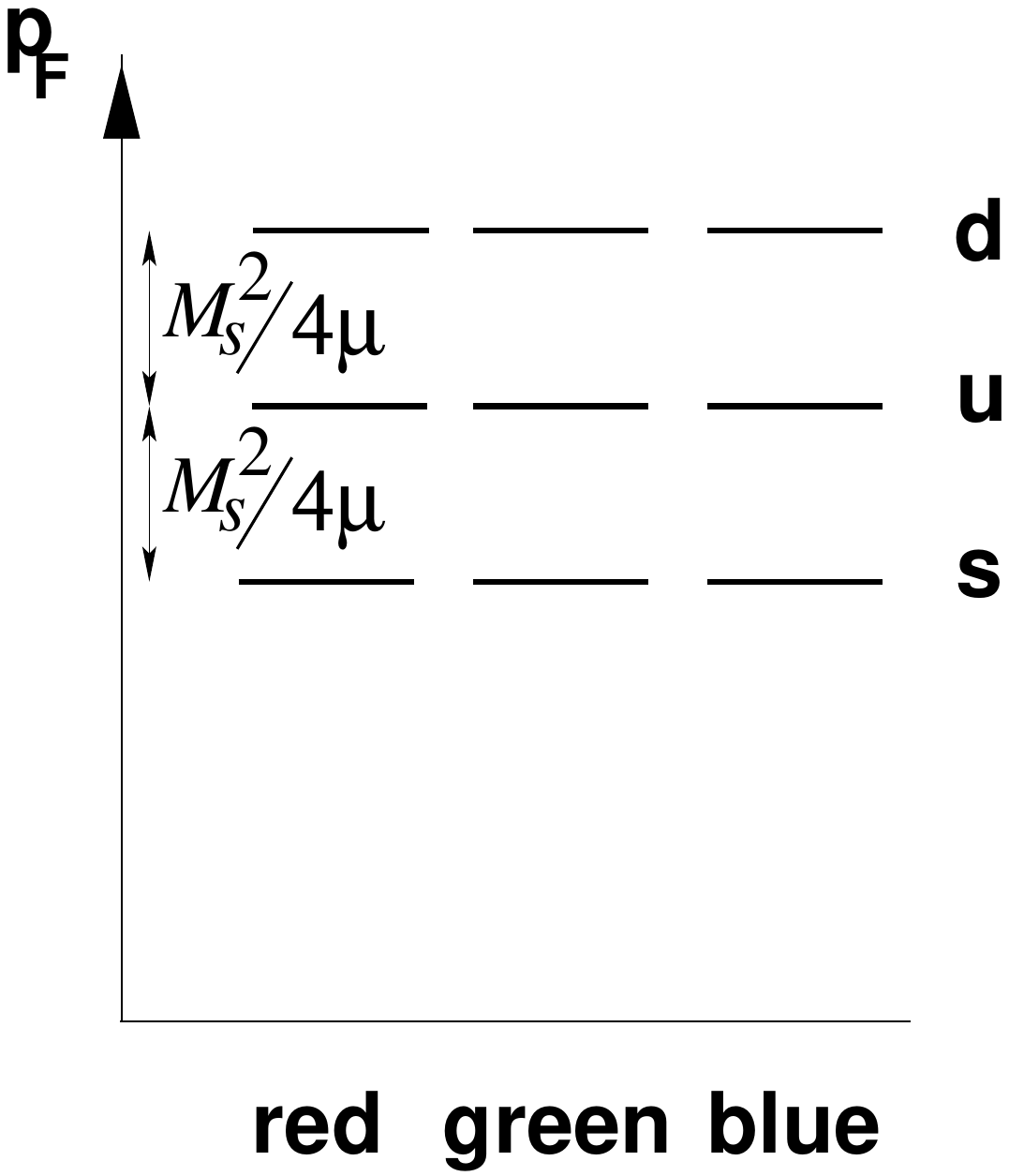}\includegraphics[width=0.33\textwidth]{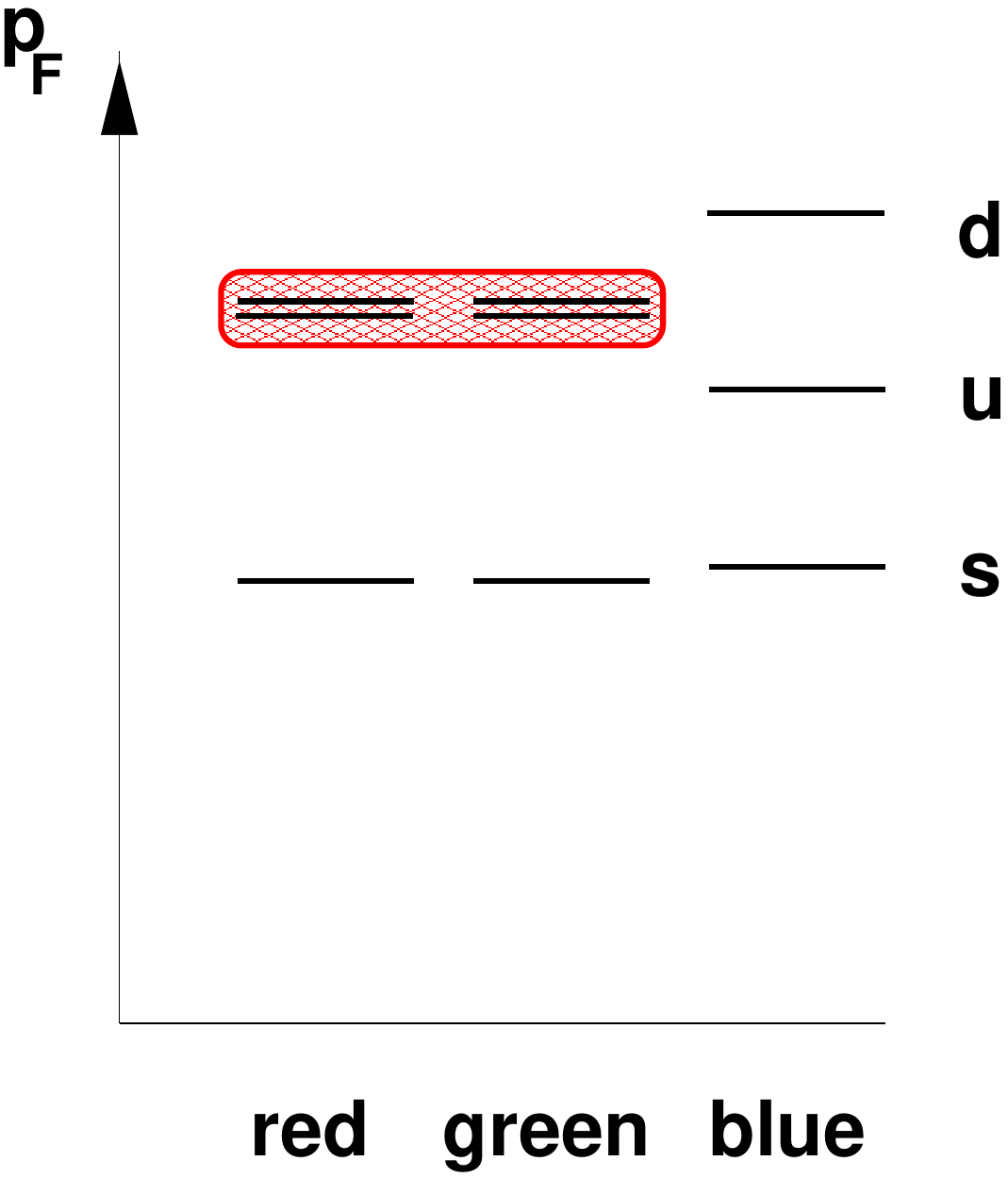}
\includegraphics[width=0.33\textwidth]{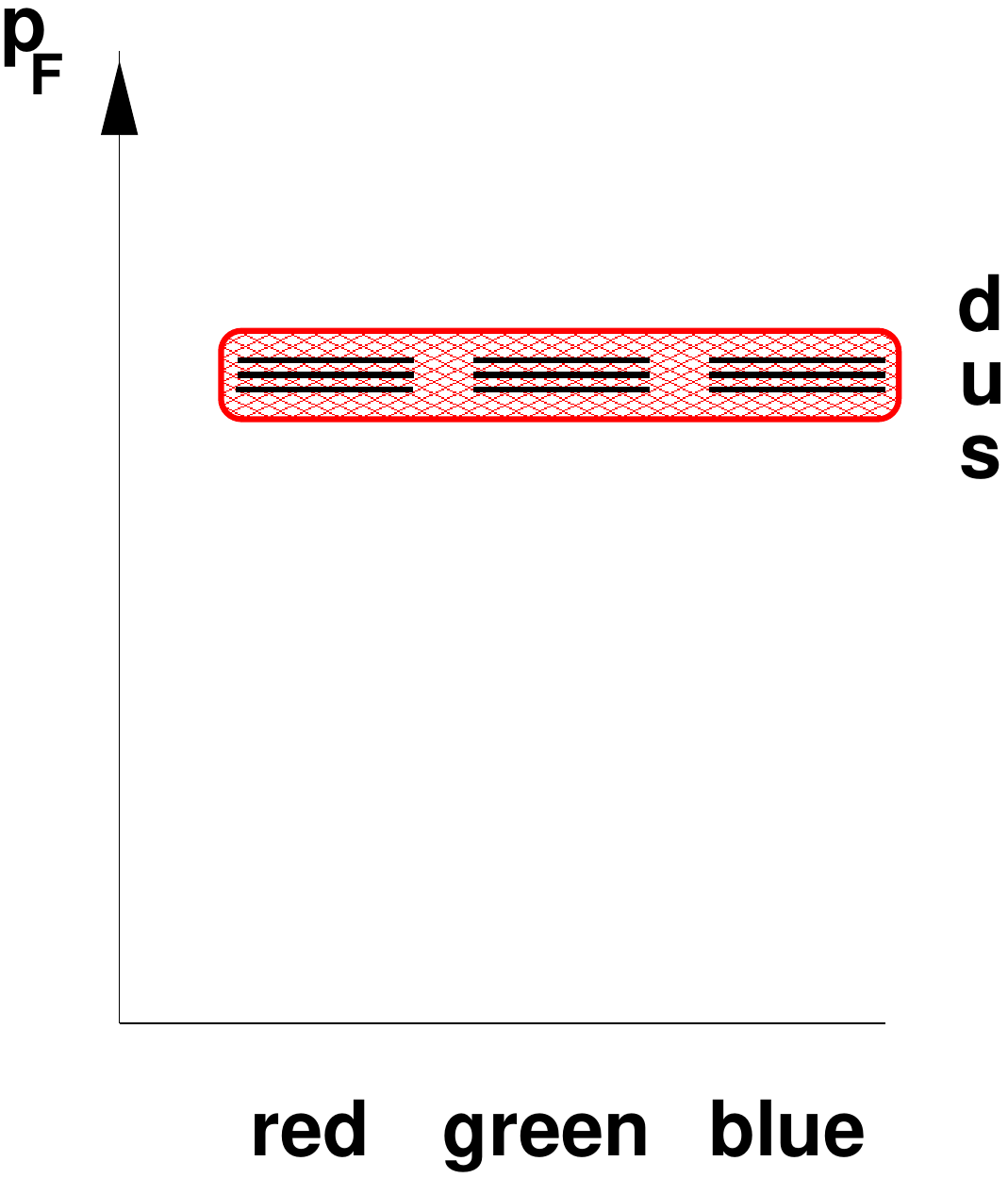}}
\caption{ Pairing pattern and splitting of Fermi surfaces $p_F$ in unpaired and paired quark matter in the 2SC and CFL phase, taken from Ref.~\cite{Alford:2007xm}. It shows quarks of which flavor and color participate in the pairing and the formation of the common Fermi sphere for Cooper pairing of quarks living in split Fermi spheres due to the strange quark mass, here denoted by $M_s$.}
\label{fig:splitting}
\end{center}
\end{figure}

In the entire discussion above, we have assumed that we can neglect all quark masses, including the mass of the strange quark which is, upon neglecting medium effects, $m_s\approx 95$ MeV. However, for smaller chemical potential the strange quark mass induces a stress on the Cooper pair. Roughly speaking, the mass of the strange quark leads to a splitting of the Fermi spheres of the three flavors by an amount of $\Delta p_F \approx m_s^2/4\mu$. Since Cooper pairing at small interactions occurs on the Fermi surface, as an intermediate step a common Fermi sphere is created, which costs energy. If the energy cost of the creation of this common Fermi sphere becomes too large, other color-superconducting phases can take over. The splitting of the Fermi spheres and the pairing in common Fermi spheres is schematically shown in Fig.~\ref{fig:splitting} taken from Ref.~\cite{Alford:2007xm}. 

A logical candidate is a phase where the strange quark does not participate in the pairing. The simplest possible order parameter for such a phase is given by 
\be
\phi^A_B \propto \delta^A_3 \delta^3_B \qquad  \Rightarrow \qquad \left<\psi_i^\alpha C\gamma_5 \psi_j^\beta\right>\propto \varepsilon_{ij3}\varepsilon^{\alpha\beta 3} \, , 
\ee 
which is called the \textbf{2SC phase}, where the strange quark and all blue quarks remain unpaired \cite{Bailin:1979nh,Bailin:1983bm,Alford:1997zt}. As in the CFL phase, rotated electromagnetism is present, where the mixing angle of the 2SC phase slightly differs from the one in the CFL phase. The symmetry breaking pattern, assuming massless up and down quarks, is found to be
\be
[SU(3)_c]\times SU(2)_L\times SU(2)_R \times U(1)_B\times U(1)_S
\to [SU(2)_{rg}]\times SU(2)_L\times SU(2)_R\times U(1)_{\tilde{B}} \times U(1)_S  \, .
\ee
Interestingly, no global symmetry is broken, since the baryon symmetry survives as $\tilde{B}$, which is a linear combination of the original generator with the broken diagonal $T_8$ color generator. For a more detailed discussion see Ref.~\cite{Alford:2007xm}. Consequently, 2SC is not a superfluid. 

\begin{figure} [t]
\begin{center}
\includegraphics[width=0.7\textwidth]{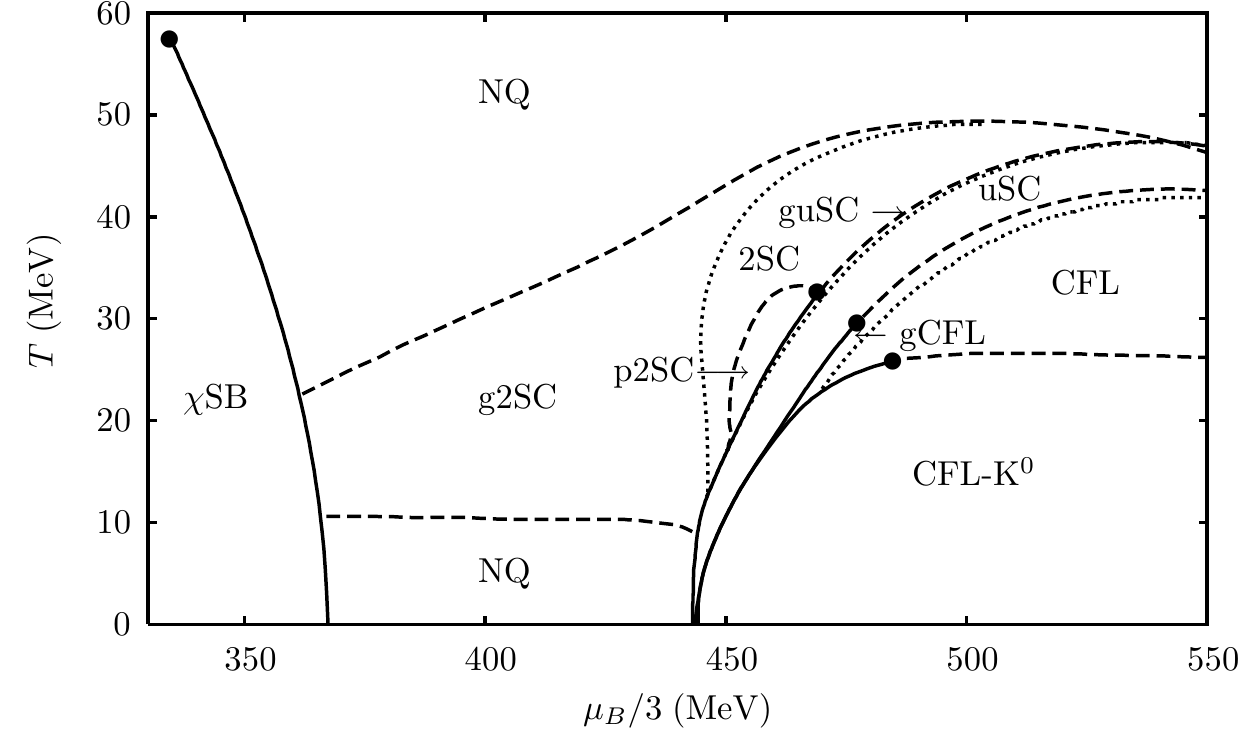}
\caption{Phase diagram of electrically and color neutral quark matter in weak equilibrium, in the plane of baryon chemical
potential and temperature, taken from Ref.~\cite{Warringa:2006dk}, where an explanation of the depicted phases can be found in Tab.~I. Here, first-order phase transitions are denoted by a solid line and second-order phase transitions by a dashed
line. This phase diagram shows part of the rich phase structure of color superconductivity.}
\label{fig:NJL_phases}
\end{center}
\end{figure}

Although we have only discussed two particular phases, it is clear that the rich structure of color superconductivity allows for a much more diverse phase diagram of color-superconducting phases. One possibility to gain further insight into the pairing of quark matter is to use Nambu--Jona-Lasinio type models, where the gluon interaction is only effectively taken into account via a direct quark interaction \cite{NJL,1961PhRv..124..246N}. An extensive review of the use of NJL models in quark matter can be found in Ref.~\cite{Buballa:2003qv}. In Ref.~\cite{Warringa:2006dk}, a phase diagram including only homogeneous phases is presented, which is shown in Fig.~\ref{fig:NJL_phases}. For a description of the depicted phases see Tab.~I of the original reference.

 An alternative approach to such calculations is the use of perturbative QCD methods, which use the asymptotic freedom of QCD \cite{Gross:1973id} to treat the problem at weak coupling. We will use results obtained within this framework later on, however its validity is probably limited to extremely high chemical potentials, $\mu \gtrsim 10^6$ MeV, whereas chemical potentials expected in compact stars are roughly at $\mu_q \approx 400$ MeV. 
\chapter{Magnetic Defects in Color Superconductivity}
We already know that ordinary superconductivity can be destroyed by an external magnetic field: either partially, by the formation of magnetic flux tubes if the superconductor is of type II, or completely,
if the external field is sufficiently large \cite{abrikosov_magnetic,tinkham2004introduction,RevModPhys.82.109}. Now, we want to investigate the fate of \textit{color} superconductivity in three-flavor 
quark matter in the presence of an ordinary external magnetic field, with an emphasis on the magnetic defects created in type-II color superconductors. The entire following discussion until the end of this thesis was originally published in Ref.~\cite{Haber:2017oqb}.

We have made plausible in the latter chapter that at the highest densities, three-flavor quark matter is in the color-flavor locked (CFL) phase, where all quarks participate in Cooper 
pairing. In this phase, all Cooper pairs are neutral with respect to a certain combination of the electromagnetic gauge field and the eighth gluon gauge field. This phenomenon is what we introduced as rotated electromagnetism. The corresponding magnetic field, which we call $\tilde{B}$, can penetrate the CFL phase, while the magnetic field corresponding to the orthogonal combination, termed $\tilde{B}_8$, and the fields corresponding 
to the other seven gluons are expelled due to the Meissner effect. Since an ordinary magnetic field $B$ has a $\tilde{B}_8$ component it will eventually destroy the CFL phase and, in the type-II regime for intermediate field strengths, will lead to the formation of magnetic flux tubes that carry $\tilde{B}_8$ flux. 

\section{Method and Main Ideas}
The main idea of this part is to write down an effective, Ginzburg-Landau style theory for the diquark condensate, based on the order parameter discussed in Chap.~\ref{chap:color}. Then, we have a theory of an up to three-component superconductor, for which we can use our knowledge from the first parts of this thesis.
Magnetic flux tubes in CFL have been studied before in Refs.~\cite{Iida:2002ev,Iida:2004if} within a Ginzburg-Landau 
approach \cite{Bailin:1983bm,Iida:2000ha,Iida:2001pg}, including an analysis of whether CFL is a type-I or type-II superconductor. 
This question was also
addressed within the same approach in Ref.\ \cite{Giannakis:2003am}, by calculating the surface energy. In these works, CFL was effectively described as a 
two-component superconductor, where the two components have different charges with respect to the rotated color gauge field $\tilde{A}_8^\mu$. 
In the following, the same Ginzburg-Landau approach is employed, but improved in several ways. Firstly, we make use of our recently gained understanding about two-component superconductivity in the previous part of this thesis, in particular the unconventional behavior of such systems in the type-I/type-II transition region. Secondly, we show that the 
CFL phase is, upon increasing the magnetic field but still for massless quarks, superseded by the  2SC phase (except for very small values of the strong coupling constant), which is indicative of the kind of flux tubes that develop in CFL. Thirdly, we will see that a new kind 
of flux tubes is energetically preferred in the parameter regime that is relevant for applications to compact stars. This new flux tube configuration is found by allowing all three diagonal components of the order parameter to be different, in contrast to the two-component approach in the 
literature. The total winding of the three components is minimized by setting the winding number of one component to zero, 
resulting in a CFL flux tube with a 2SC-like core. By computing the critical magnetic field at which flux tubes start to populate the system, 
we shall demonstrate that this configuration is favored over the previously discussed CFL flux tubes with an unpaired core. 

We will also study flux tubes in 2SC itself. Since the 2SC phase is a single-component superconductor, the flux tube configuration considered in Ref.\ \cite{Alford:2010qf} appears to be unique, analogous to ordinary superconductors. However, our general setup allows us to check whether additional color-flavor components of the order parameter are induced in the core of a 2SC flux tube. We find that this is indeed the case. These new flux tube solutions can reduce their energy by increasing their winding number and thus their radius, eventually resulting in a domain wall rather than a one-dimensional string. 

By using the purely bosonic Ginzburg-Landau theory we neglect any effect of the charges of the constituents of the Cooper pairs, which are now quarks instead of nucleons, and a fermionic approach would have to be used to go beyond this 
approximation \cite{Ferrer:2005vd,Ferrer:2006vw,Noronha:2007wg,Fukushima:2007fc}. Moreover, the same restrictions of this approach apply, like the fact that a Ginzburg-Landau approach is strictly speaking only valid for small condensates, for instance for temperatures close to the critical temperature. For the main numerical results, we do not 
investigate the complete parameter space of the Ginzburg-Landau potential or introduce effective parameters like we have done for the two-component system, but rather restrict ourselves  to  the 
weak-coupling form of the parameters. Then, we extrapolate the results to larger values of the coupling, which are expected in an astrophysical environment. We also work in the simplified scenario of vanishing quark masses, and 
it remains to be seen how our results are modified if the strange quark mass is taken into account; mass terms were included in the Ginzburg-Landau approach in 
Refs.\ \cite{Iida:2003cc,Iida:2004cj,Schmitt:2010pf}. 

\section{Relation to Superfluid Vortices in CFL}

All flux tubes we discuss in detail have, as an imposed restriction, a vanishing baryon circulation far away from the flux tube. 
In other words, the flux tubes we are interested in can only be induced by a magnetic field, not by rotation. Flux tubes 
that do have baryon circulation, in particular the so-called semi-superfluid vortices, have been discussed extensively in the literature, 
for instance in Refs.~\cite{Balachandran:2005ev,Eto:2009kg,Vinci:2012mc,Alford:2016dco}, for a review see Ref.~\cite{Eto:2013hoa}.
These vortices, just like the vortices in an ordinary superfluid, have a logarithmically divergent energy, and a finite system or a lattice of vortices is required to regularize this divergence. This means that in an numerical investigation, the energy of the vortex will depend logarithmically on the upper bound of the integration.

The flux tubes we discuss here, just like the flux tubes in an ordinary 
superconductor, do not show this divergence and their energy is finite even in an infinite volume. This would make a comparison to the vortices above on a pure energy level difficult. To put our discussion into a wider context, we shall briefly discuss how all line defects, with and without baryon circulation, with and without color-magnetic flux,  are  obtained by choosing different triples of winding numbers of the three order parameter components. 

In contrast to the CFL vortices, the flux tubes discussed here are not protected by 
topology. This can be seen by computing the first homotopy group $\pi_1$, since flux tubes are string-like defects.  \cite{Eto:2013hoa}. This means that configurations with different windings are continuously connected. In particular, the configurations we consider are continuously connected to the zero-winding configuration (not unlike the so-called "semilocal cosmic strings" \cite{Vachaspati:1991dz}), i.e., they can be unwound into "nothing" 
without encountering a discontinuity.
Since such a discontinuity typically translates into an energy barrier, one might 
question the stability of the objects we consider in this paper. However, the main result of our calculation is a critical magnetic field at which 
the flux tube is energetically preferred over the configuration without flux tube. Therefore, even though we do not explicitly prove  local stability by introducing fluctuations about the flux tube state, the magnetic field stabilizes the flux tube and by comparing free energies 
we establish  global stability. (The chosen ansatz is not completely general in color-flavor space, i.e., while we will prove that the flux tube cannot 
decay into "nothing" at a sufficiently large magnetic field, we can, strictly speaking, not exclude that it decays into more exotic 
color-magnetic flux tubes.) 

\section{Astrophysical Implications of Color Superconductivity}
We have shortly discussed the phenomenological relevance of quark matter in general in the introductory part of this thesis, see Sec.~\ref{sec:QMinCS}.
\textbf{Color-magnetic defects} in CFL and 2SC quark matter are very interesting as well for the phenomenology of quark stars or neutron stars with a quark matter core. The critical magnetic fields we compute here -- as already suggested from previous work -- are most likely too large to be reached in compact stars. Nevertheless, there might be other mechanisms to create magnetic defects in quark matter. As argued in Ref.\ \cite{Alford:2010qf}, flux tubes can form if quark matter is cooled into a color-superconducting phase at a given, approximately constant magnetic field. It is then a dynamic question how and on which time 
scale the magnetic field is expelled from the system. A full dynamical simulation of the expulsion of the magnetic field is extremely 
complicated and most likely involves the formation of flux tubes or domain walls, see for instance Ref.\ \cite{1991PhRvL..66.3071L} for such a study in the context of ordinary superconductors. While the results I am going to present only concern equilibrium configurations, they 
show to the very least that new defects, so far overlooked in the literature, should be taken into account in this discussion.  

It has been argued that the color-flux tubes thus created
support a deformation of the rotating star ("color-magnetic mountains"). This deformation gives rise to a continuous emission of gravitational waves because of the misalignment of rotational and magnetic axes \cite{Glampedakis:2012qp}.
(A different mechanism in quark matter to support continuous gravitational waves is the formation of a crystalline phase  \cite{Lin:2007rz,Haskell:2007sh,Knippel:2009st,Anglani:2013gfu}.)
The larger energy (and the only slightly smaller number) of the color-magnetic flux tubes compared to  flux tubes in superconducting nuclear matter makes this mechanism particularly efficient and the resulting gravitational waves potentially detectable. The following calculation
provides a quantitative, numerical  calculation of the flux tube energy, putting the estimates used in Ref.~\cite{Glampedakis:2012qp}
on solid ground. It also slightly changes this estimate due to the new flux tube configuration, although this change is small 
compared to the uncertainties involved in the estimate of the ellipticity of the star, which is necessary for the emission of gravitational waves. 

\section{Structure}
The final part of this thesis is organized as follows. In Chap.~\ref{chap:GL} we introduce the Ginzburg-Landau potential which is expanded to color superconductivity, and a general ansatz for the order parameter. Then,  as a necessary preparation for the study of the flux tubes, in Chap.~\ref{chap:hom} we discuss the homogeneous phases and the phase diagram at nonzero external magnetic field, just as we did in the two-component system. We turn to the CFL flux tubes in Chap.~\ref{chap:CFL}, with a classification 
of the flux tubes and their radial profiles shown in Sec.~\ref{sec:circ}. In Chap.~\ref{chap:2SC} we discuss 2SC flux tubes and domain 
walls and present the corresponding profiles in Sec.~\ref{sec:profiles}. The main results, putting together the phase diagram of the homogeneous phases with the critical fields for the magnetic defects, are discussed in Chap.~\ref{chap:results_color}, and I end with a brief summary and outlook of this part, including further astrophysical implications, in Chap.~\ref{chap:summary_color}.
In this part I use Heaviside-Lorentz units for the gauge fields, in which the 
elementary charge is $e=\sqrt{4\pi\alpha}\simeq 0.3$. These are the units used in the most closely related literature about the 
CFL phase, for instance Ref.\ \cite{Giannakis:2003am}. Note, however, that Gaussian units are used in the rest of this thesis and in other literature on multicomponent superconductors, for instance in our publication Ref.\ \cite{Haber:2017kth}.

\chapter{Setup for Color Superconductivity}
\label{chap:GL}

\section{General Form of Ginzburg-Landau Potential}

We have seen that the order parameter $\Psi$ for spin-zero Cooper pairing of three-flavor, three-color quark matter is an anti-triplet  
in color and flavor space, $\Psi \in [\bar{3}]_c\otimes [\bar{3}]_f$. We can thus introduce the components $\Phi_{ij}$ of the order parameter in the presented basis $(J_i)_{jk} = -i\epsilon_{ijk}$ in color space and $(I_i)_{jk} = -i\epsilon_{ijk}$ in flavor space introduced in Chap.~\ref{chap:color} via 
\be
\Psi =  \Phi_{ij} J_i\otimes I_j \, . 
\ee
Later, we shall only work with the $3\times 3$ matrix $\Phi$, not with the $9\times 9$ tensor $\Psi$, and simply refer to $\Phi$ as the order parameter. In general, there are two order parameters $\Psi_L$ and $\Psi_R$ for pairing in the left-handed and right-handed sectors. They are different for instance if kaon condensation is considered \cite{Schmitt:2010pf}. Here we assume $\Psi_L=\Psi_R\equiv \Psi$.
The Ginzburg-Landau potential up to quartic order in $\Psi$ is \cite{Giannakis:2003am}
\bea \label{U}
U &=& -3\Big\{\Tr[(D_0\Psi)^\dag(D_0\Psi)]-u^2\Tr[(D_i\Psi)^\dag(D_i\Psi)]\Big\}+k\Tr[\Psi^\dag\Psi]\non[2ex]
&&+\frac{l_1}{2}\Tr[(\Psi^\dag\Psi)^2]+\frac{l_2}{2}(\Tr[\Psi^\dag\Psi])^2 
+\frac{1}{4}F_{\mu\nu}^aF_a^{\mu\nu}+\frac{1}{4}F_{\mu\nu}F^{\mu\nu} \, , 
\eea
where $u^2 = \frac{1}{3}$, with the gluon field strength tensors  $F_{\mu\nu}^a$ from Eq.~(\ref{eq:Fgluon}) with $a=1,\ldots,8$,  the color gauge fields $A_\mu^a$, the strong coupling constant $g$,  and the SU(3) structure constants $f^{abc}$, and  
$F_{\mu\nu}=\partial_\mu A_\nu-\partial_\nu A_\mu$ is the electromagnetic field strength tensor with the electromagnetic gauge field $A_\mu$. The parameters $k$, $l_1$, $l_2$ 
can be computed in the weak-coupling limit from perturbation theory \cite{Iida:2000ha}. The covariant derivative is 
\bea\label{DPsi}
D_\mu \Psi &=& \partial_\mu \Psi +igA_\mu^a\Phi_{ij}(T_a J_i+J_i T_a^T)\otimes I_j +ie A_\mu \Phi_{ij} J_i\otimes (QI_j+I_j Q^T) 
\, ,
\eea
where $T_a=\lambda_a/2$, with the Gell-Mann matrices $\lambda_a$,  such that $\Tr[T^aT^b]=\frac{1}{2}\delta^{ab}$,  where $e$ is the elementary electric charge, and where $Q=\mathrm{ diag}(q_1,q_2,q_3)$ is the 
$U(1)$ charge generator in flavor space with the individual electric charges of the quarks $q_1$, $q_2$, $q_3$.

For simplicity, we shall work in the massless limit throughout the thesis, such that flavor symmetry is only 
broken by the electric charges, not by the quark masses. In particular, there is no distinction between $d$ and $s$ quarks in this approximation. We can write the covariant derivative as  
\bea
D_\mu \Psi 
&=& (D_\mu\Phi)_{ij} J_i\otimes I_j\, ,
\eea
with
\be \label{Dphi}
D_\mu\Phi = \partial_\mu\Phi -igA_\mu^a T_a^T\Phi+ieA_\mu \Phi \bar{Q} \, ,
\ee
where we have used $T_a J_i+J_i T_a^T = -(T_a)_{ij}J_j$ and  $QI_j+I_j Q^T = \bar{Q}_{jk} I_k$ with 
$\bar{Q}= \mathrm{ diag}(q_2+q_3,q_1+q_3,q_1+q_2)$. Since the electric charges of $u$, $d$ and $s$ quarks add up to zero, we have 
$\bar{Q} = -Q$, and thus it is not strictly necessary to  introduce the notation $\bar{Q}$. But, 
one should keep in mind that the relevant charge matrix contains the 
charges of Cooper pairs, not of individual quarks, as the notation $-Q$ instead of $\bar{Q}$ would have suggested.  
We can now perform the trace over the 9-dimensional color-flavor space in Eq.\ (\ref{U}) and write the Ginzburg-Landau potential in terms of $\Phi$, 
\bea \label{UPhi}
U &=& -12\Big\{\Tr[(D_0\Phi)^\dag(D_0\Phi)]-u^2\Tr[(D_i\Phi)^\dag(D_i\Phi)]\Big\}+4k \Tr[\Phi^\dag\Phi]+l_1\Tr[(\Phi^\dag\Phi)^2]\non[2ex] 
&&+(l_1+8l_2) (\Tr[\Phi^\dag\Phi])^2 +\frac{1}{4}F_{\mu\nu}^aF_a^{\mu\nu}+\frac{1}{4}F_{\mu\nu}F^{\mu\nu}
 \, , 
\eea
where the following properties of the color matrices are used:
\be
\Tr[J_iJ_j]=2\delta_{ij} \, ,\qquad \Tr[J_iJ_jJ_kJ_\ell] = \delta_{ij}\delta_{k\ell}+\delta_{i\ell}\delta_{jk} \, ,   
\ee
(and the same for the flavor matrices $I_i$) and, as a consequence,
\begin{subequations}
\bea
\Tr[(D_\mu\Psi)^\dag(D^\mu\Psi)] &=& 4\Tr[(D_\mu\Phi)^\dag(D^\mu\Phi)] \\[2ex]
\Tr[\Psi^\dag\Psi]&=& 4\Tr[\Phi^\dag\Phi] \\[2ex]
\Tr[(\Psi^\dag\Psi)^2] &=& 2(\Tr[\Phi^\dag\Phi])^2 + 2\Tr[(\Phi^\dag\Phi)^2] \, .
\eea
\end{subequations}

All traces in the final result are now taken over the 3-dimensional order parameter space.

\section{Superfluid Velocity}

Superfluid vortices are characterized by a nonzero circulation around the vortex. We shall see that line defects in CFL 
can carry magnetic flux \textit{ and} baryon circulation. Therefore, we first derive a general expression for the superfluid velocity, which 
can then be used to compute the baryon circulation for particular flux tube solutions, see Sec.~\ref{sec:circ}. The superfluid velocity is computed in analogy to the case of a scalar field \cite{2013PhRvD..87f5001A}; for a derivation in the context of CFL see 
Refs.\ \cite{Iida:2001pg,Eto:2013hoa}.
We first introduce an overall phase $\psi$ associated with 
baryon number conservation $U(1)_B$,
\be
\Phi = e^{i\psi}\Delta \, .
\ee
This allows us to compute the baryon four-current via 
\bea
j^\mu = -\frac{\partial U}{\partial (\partial_\mu\psi)} \, . 
\eea
We find 
\bea
j^0&=& 12 i\Tr[(D^0\Phi)^\dag\Phi-\Phi^\dag(D^0\Phi)] \, , \qquad j^i= 12 u^2 i\Tr[(D^i\Phi)^\dag\Phi-\Phi^\dag(D^i\Phi)] \, .
\eea
The superfluid four-velocity $v^\mu$ is defined through $j^\mu = n_s v^\mu$ 
with the superfluid density $n_s$ and $v_\mu v^\mu=1$, as described in the introductory part of the thesis. 
With $v^\mu=\gamma(1,\mathbf{v}_s)$, the components of the superfluid three-velocity 
$\mathbf{v}_s$ become
\be \label{superv}
(\mathbf{v}_s)_i = \frac{j^i}{j^0} = \frac{u^2}{4\mu_q}\frac{i\Tr[(D^i\Phi)^\dag\Phi-\Phi^\dag(D^i\Phi)]}{\Tr[\Phi^\dag\Phi]} \, , 
\ee
where we have assumed $\Delta$ to be time-independent, set the temporal components of the gauge fields to zero, \mbox{$A_0=A^a_0=0$}, 
and introduced the quark chemical potential $\mu_q$ through the time dependence of the phase, $\partial_0 \psi= 2\mu_q$, where 
the factor 2 arises from the diquark nature of the order parameter.

\section{Ansatz and Gibbs Free Energy}

We evaluate the potential (\ref{UPhi}) for the diagonal order parameter $\Phi = \mathrm{ diag}(\phi_1,\phi_2,\phi_3)$, with the complex 
scalar fields $\phi_1$, $\phi_2$, $\phi_3$. Allowing all three diagonal components to be different is a more general ansatz than used in the literature before. It is not the most general ansatz because the reduced symmetry due to the electric charges of the quarks
(and quark masses if they were taken into account) does not allow to rotate an arbitrary order parameter matrix into an equivalent diagonal form. 

Due to rotated electromagnetism, the eighth gluon and the photon mix, which can for instance be seen in a microscopic calculation of the gauge boson polarization tensor \cite{Schmitt:2003aa}. This mixing can also be derived within the Ginzburg-Landau approach by computing the 
magnetic fields in the CFL phase in the presence of an externally applied magnetic field, which we will do in Chap.~\ref{chap:hom}. 
 We anticipate this mixing by defining the rotated gauge fields 
\begin{subequations} \label{mix0}
\bea
\tilde{A}_\mu^8 &=& \cos\theta \,A_\mu^8+\sin\theta\, A_\mu \, , \\[2ex]
\tilde{A}_\mu &=& -\sin\theta \,A_\mu^8+\cos\theta\, A_\mu \, , 
\eea
\end{subequations}
with the mixing angle given by
\be \label{costheta}
\cos\theta = \frac{\sqrt{3}g}{\sqrt{3g^2+4e^2}} \, ,  \qquad \sin\theta = - \frac{2e}{\sqrt{3g^2+4e^2}} \, .
\ee
The derivation of the CFL mixing angle is presented in App.~\ref{app:rotated}. In the new rotated basis, the magnetic field $\tilde{B}_8$ is the one which experiences a Meissner effect in the CFL phase whereas the magnetic field 
$\tilde{B}$ penetrates the CFL phase unperturbed, if the quark flavors in the charge matrix are ordered $(d,s,u)$, such that $Q=\mathrm{ diag}(-1/3,-1/3,2/3)$ is proportional to $T_8$. If the order $(u,d,s)$ is used, the mixing between the gauge fields involves $A_\mu^3$ \cite{Iida:2001pg}. We shall work with the more convenient order $(d,s,u)$ in the CFL phase, but change to $(u,d,s)$ in Chap.~\ref{chap:2SC}, where 
we discuss magnetic defects in the 2SC phase. 

For our diagonal order parameter, it is consistent with the non-abelian Maxwell equations
\be
\p_\mu F^{a\mu\nu}+gf^{abc}A_\mu^b F^{c\mu\nu}=j^{a\nu} \, ,
\ee
to set all gauge fields corresponding to the non-diagonal 
$SU(3)$ generators to zero, $A_1^\mu=A_2^\mu=A_4^\mu=A_5^\mu=A_6^\mu=A_7^\mu=0$. The current on the right hand side of Maxwell's equations can be computed from the terms in the potential Eq.~(\ref{UPhi}) that mix gauge and scalar fields, which are all contained in
\bea
\hspace*{-3cm}&&\Tr\left[\left(D_\mu\Phi\right)^\dagger\left(D^\mu\Phi\right)\right] = \left(\p_\mu+i\tilde{g}_8\tilde{A}^8_\mu+i\frac{g}{2}A_\mu^3 \right)\phi_1^*\left(\p_\mu-i\tilde{g}_8\tilde{A}^8_\mu-i\frac{g}{2}A_\mu^3 \right)\phi_1 \hspace*{2cm} \non[2ex]
&&+\left(\p_\mu+i\tilde{g}_8\tilde{A}^8_\mu-i\frac{g}{2}A_\mu^3 \right)\phi_2^*\left(\p^\mu-i\tilde{g}_8\tilde{A}_8^\mu+i\frac{g}{2}A^\mu_3 \right)\phi_2\non[2ex]
&&+\left(\p_\mu-2i\tilde{g}_8\tilde{A}^8_\mu\right)\phi_3^*\left(\p^\mu+2i\tilde{g}_8\tilde{A}_8^\mu\right)\phi_3 +\frac{g^2}{4}\Big[\left(A_\mu^1 A_1^\mu+A_\mu^2 A_2^\mu\right)\left(|\phi_1|^2+|\phi_2|^2\right)\non[2ex]
&&+\left(A_\mu^4 A_4^\mu+A_\mu^5 A_5^\mu\right)\left(|\phi_1|^2+|\phi_3|^2\right)+\left(A_\mu^6 A_6^\mu+A_\mu^7 A_7^\mu\right)\left(|\phi_2|^2+|\phi_3|^2\right)\Big] \, ,
\eea
where we have denoted the coupling to the rotated color field $\tilde{\mathbf{A}}_8$ by
\be \label{g8}
\tilde{g}_8 \equiv \frac{g}{2\sqrt{3}\cos\theta}  \, .
\ee
The currents $j^{a\mu}$ are now obtained by variation of these terms with respect to $A_\mu^a$.

Additionally, we set all electric fields to zero, and only keep 
the magnetic fields $\mathbf{B}_3 = \nabla\times\mathbf{A}_3$, $\tilde{\mathbf{B}}_8 = \nabla\times\tilde{\mathbf{A}}_8$, and $\tilde{\mathbf{B}} = \nabla\times\tilde{\mathbf{A}}$.  We also ignore all time dependence since we are only interested in equilibrium configurations. 
Putting all this together yields 
the potential 
\be \label{UUB}
U = U_0 + \frac{\tilde{\mathbf{B}}^2}{2} \, , 
\ee
with 
\bea \label{U123}
U_0 &=& \left|\left(\nabla+i\frac{g}{2}\mathbf{A}_3+i\tilde{g}_8\tilde{\mathbf{A}}_8\right)\phi_1\right|^2
+\left|\left(\nabla-i\frac{g}{2}\mathbf{A}_3+i\tilde{g}_8\tilde{\mathbf{A}}_8\right)\phi_2\right|^2+\left|\left(\nabla-2i\tilde{g}_8\tilde{\mathbf{A}}_8\right)\phi_3\right|^2 \non[2ex]
&&-\mu^2(|\phi_1|^2+|\phi_2|^2+|\phi_3|^2)+\lambda(|\phi_1|^4+|\phi_2|^4+|\phi_3|^4)\non[2ex]
&&-2h(|\phi_1|^2|\phi_2|^2+|\phi_1|^2|\phi_3|^2+|\phi_2|^2|\phi_3|^2)+\frac{\mathbf{B}_3^2}{2}+ \frac{\tilde{\mathbf{B}}_8^2}{2}  \, . 
\eea
We have separated the rotated field $\tilde{\mathbf{B}}$ because all scalar fields are neutral with respect to the corresponding charge, and the only contribution
is the trivial $\tilde{\mathbf{ B}}^2$ term. 
Furthermore, we have introduced the new Ginzburg-Landau parameters 
\begin{subequations} \label{weak123}
\bea
\mu^2 &=& -k \simeq \frac{48\pi^2}{7\zeta(3)}T_c(T_c-T) \, , \\[2ex]
\lambda&=&\frac{l_1}{8}+\frac{l_2}{2}  \simeq\frac{72\pi^4}{7\zeta(3)}\frac{T_c^2}{\mu_q^2}  \, ,\\[2ex]
h&=&-\left(\frac{l_1}{16}+\frac{l_2}{2}\right)  \simeq-\frac{36\pi^4}{7\zeta(3)}\frac{T_c^2}{\mu_q^2} \, .
\eea
\end{subequations} 
In the last expression of each line, the weak-coupling results have been used\footnote{Here, the convention of Ref.\ \cite{Giannakis:2003am} is used. To compare with 
Refs.\ \cite{Iida:2002ev,Iida:2004if,Eto:2013hoa}, the order parameter has to be rescaled as 
\be
\Phi \to \sqrt{\frac{3}{7\zeta(3)}}\frac{\pi^2T_c}{2\mu_q} \Phi\, .\nonumber 
\ee
}
  with the temperature $T$
and the critical temperature for color superconductivity $T_c$. 
(At weak coupling, although the relation between the 
critical temperature and the zero-temperature gap differs from phase to phase \cite{Schmitt:2002sc}, the absolute values of the critical temperatures of CFL and 2SC are the same.) The potential (\ref{U123}) describes three massless bosonic fields which have 
the same chemical potential $\mu$, the same self-interaction given by $\lambda$, interact pairwise with the same coupling constant $h$, and have different charges with respect to the three gauge fields. (In comparison, the model of the neutron-proton system contained only two massive scalar fields but with different chemical potentials and different self-couplings, including derivative coupling terms between the fields.)
For $\phi_1=\phi_2$ the system is
 neutral  with respect to $A^\mu_3$ at every point in space and we recover the potential used in Ref.\  \cite{Giannakis:2003am}. Since we allow for $\phi_1\neq \phi_2$, we must keep 
$\mathbf{A}_3$.

We are interested in the phase structure in an externally given homogeneous magnetic field $\mathbf{H}$, which, without 
loss of generality, we align with the $z$-direction, $\mathbf{H}=H\mathbf{e}_z$ with $H\ge 0$. Therefore, 
we need to consider the Gibbs free energy 
\be
G = \int d^3\mathbf{r}\,(U-\mathbf{H}\cdot\mathbf{B})  = \int d^3\mathbf{r} \left[U_0+\frac{\tilde{\mathbf{ B}}^2}{2}-\mathbf{H}\cdot(\tilde{\mathbf{ B}}\cos\theta+\tilde{\mathbf{ B}}_8\sin\theta)\right] \, .
\ee
Since $\tilde{\mathbf{A}}$ does not couple to the three condensates, its equation of motion is trivially fulfilled by any constant $\tilde{\mathbf{B}}$ and 
the Gibbs free energy is minimized by $\tilde{\mathbf{ B}}=\tilde{B}\mathbf{e}_z$ with
\be
\tilde{B} = H\cos\theta \,, 
\ee
such that we can write the Gibbs free energy density as 
\be \label{GV}
\frac{G}{V} = -\frac{H^2\cos^2\theta}{2}+ \frac{1}{V}\int d^3\mathbf{r} \left(U_0-\mathbf{H}\cdot\tilde{\mathbf{ B}}_8\sin\theta\right) \, ,
\ee
 where $V$ is the total volume of the system.

\section{Strategy of the Calculation}
The calculation we are aiming for is very similar to the one carried out in the calculation of the flux tubes in a superconductor coupled to a superfluid, with the obvious complication of an extra scalar field and extra gauge fields. Our first aim is to identify the region in parameter space where magnetic flux tubes form, therefore we need to compute the three critical magnetic fields $H_c$, $H_{c1}$ and $H_{c2}$. The critical field $H_c$ follows again from a simple comparison of Gibbs free energies of the homogeneous phases. The critical field $H_{c2}$ is calculated in the same manner as for the simpler system, by linearizing the equations of motion. Once again, only $H_{c1}$ requires a fully numerical calculation, except for approximations that are valid only in the deep type-II regime. Therefore, as a first location for the transition from type-I to type-II behavior we compute $H_c$ and $H_{c2}$ and determine 
the point at which $H_c=H_{c2}$. In a one-component system, this yields a critical value for the Ginzburg-Landau 
parameter $\kappa=\kappa_c=1/\sqrt{2}$, where $\kappa$ is the ratio
of magnetic penetration depth and coherence length. However, we have already seen that due to the coupling to other fields, this is not exact anymore. Note that we have now to distinguish different coherence lengths and London penetration lengths for the various condensates and magnetic fields. 

This makes the situation more complicated in a color superconductor, which was already realized in Refs.\ \cite{Giannakis:2003am,Iida:2004if}, where it was pointed out that various criteria for type-I/type-II behavior do not coincide, i.e., do not yield a single critical $\kappa$, as we have seen in the two-component system as well. 
Moreover, in our present three-component system there is not simply a single superconducting phase and critical fields for the transition to the normal-conducting phase. 
Instead, we need to compute the critical fields for all possible transitions between the CFL, 2SC, and unpaired phases. 
The strategy is thus as follows. Analogously to the neutron-proton system, we start with the homogeneous phases to construct a phase diagram at nonzero external 
magnetic field $H$. This corresponds to computing the various critical fields $H_c$. Then, we  compute the critical fields 
$H_{c2}$, and the intersection where $H_c=H_{c2}$ will give us an idea (although not a precise location, because of the multicomponent structure) for the transition between type-I and type-II 
behavior. The resulting phase diagram is then used as a foundation for the calculation of the flux tube profiles and energies, which 
is done in the type-II regime. For the color-superconducting system, we will not attempt to resolve the details of the type-I/type-II transition region as we did before. Instead we will focus on possible types of magnetic defects in the CFL and the 2SC phase.

\chapter{Homogeneous Phases}
\label{chap:hom}
In this chapter we classify the homogeneous phases without magnetic field, as we have done it in similar fashion for the two-component system. Once again, we write the complex scalar fields as 
\be \label{rhopsi_c}
\phi_i = \frac{\rho_i}{\sqrt{2}}e^{i\psi_i} \, , 
\ee
where the index $i$ is now extended to $i=1,2,3$.
In this section, we only consider homogeneous solutions, $\nabla\rho_i=\nabla\psi_i=0$ (then, the phases $\psi_i$ do not play any 
role). In this case, our ansatz for the gauge fields is $\mathbf{A}_3=xB_3\mathbf{e}_y$, $\tilde{\mathbf{A}}_8=x\tilde{B}_8 \mathbf{e}_y$, such that the magnetic fields, given by the curl of the corresponding vector potentials, are homogeneous and parallel to the externally applied field $\mathbf{H}$. Then, the potential 
from Eq.\ (\ref{U123}) becomes
\bea \label{U0hom}
U_0 &=& \frac{B_3^2}{2}+\frac{\tilde{B}_8^2}{2}-\frac{\mu^2}{2}(\rho_1^2+\rho_2^2+\rho_3^2)+\frac{\lambda}{4}(\rho_1^4+\rho_2^4+\rho_3^4)-\frac{h}{2}(\rho_1^2\rho_2^2+\rho_1^2\rho_3^2+\rho_2^2\rho_3^2) \non[2ex]
&& +\frac{x^2\rho_1^2}{2}\left(\frac{g}{2}B_3+\tilde{g}_8\tilde{B}_8\right)^2+\frac{x^2\rho_2^2}{2}\left(-\frac{g}{2}B_3+\tilde{g}_8\tilde{B}_8\right)^2
+\frac{x^2\rho_3^2}{2} \left(2\tilde{g}_8\tilde{B}_8\right)^2 \, .
\eea
The equations of motion for $\mathbf{A}_3$ and $\tilde{\mathbf{A}}_8$ are
\begin{subequations} \label{A3A8}
\bea
0&=& \rho_1^2\left(\frac{g}{2}B_3+\tilde{g}_8\tilde{B}_8\right)-\rho_2^2\left(-\frac{g}{2}B_3+\tilde{g}_8\tilde{B}_8\right) \, , \\[2ex]
0&=& \rho_1^2\left(\frac{g}{2}B_3+\tilde{g}_8\tilde{B}_8\right)+\rho_2^2\left(-\frac{g}{2}B_3+\tilde{g}_8\tilde{B}_8\right)+4\rho_3^2\tilde{g}_8\tilde{B}_8  \, , 
\eea
\end{subequations}
and the equations of motion for the condensates $\rho_i$ are
\begin{subequations} \label{rho123}
\bea
0&=&\rho_1\left[\lambda\rho_1^2-h(\rho_2^2+\rho_3^2)-\mu^2+x^2\left(\frac{g}{2}B_3+\tilde{g}_8\tilde{B}_8\right)^2\right] \, , \label{rho1}\\[2ex]
0&=&\rho_2\left[\lambda\rho_2^2-h(\rho_1^2+\rho_3^2)-\mu^2+x^2\left(-\frac{g}{2}B_3+\tilde{g}_8\tilde{B}_8\right)^2\right] \, ,\label{rho2} \\[2ex]
0&=&\rho_3\left[\lambda\rho_3^2-h(\rho_1^2+\rho_2^2)-\mu^2+x^2\left(2\tilde{g}_8\tilde{B}_8\right)^2\right] \, .
\eea
\end{subequations}
Since in this section the condensates and magnetic fields are constant in space by assumption, the terms proportional to $x^2$ and the $x$-independent terms in Eqs.\ (\ref{rho123}) must vanish separately. As a consequence, the terms proportional to $x^2$ in the potential (\ref{U0hom}) vanish as well. This must be the case because  otherwise the free energy, obtained by integrating $U_0$ over space, would become infinite. We conclude that any given combination of nonzero condensates yields a condition for the magnetic fields. 
We discuss all possible combinations now.

\begin{itemize}

\item If all three condensates are nonzero, Eqs.\ (\ref{rho123}) show that $B_3=\tilde{B}_8=0$ [which trivially fulfills Eqs.\ (\ref{A3A8})]. This is the CFL solution, and Eqs.\  (\ref{rho123}) yield 
\be \label{rho0}
\rho_1^2=\rho_2^2=\rho_3^3 = \frac{\mu^2}{\lambda(1-2\eta)} \equiv \rho_\mathrm{ CFL}^2 \, ,
\ee
where we have abbreviated the ratio of the cross-coupling constant to the self-coupling constant by
\be
\eta\equiv \frac{h}{\lambda} \, .
\ee
To ensure the boundedness of the potential, we must have $\eta<0.5$ (including all negative values), which also ensures 
$\rho_\mathrm{ CFL}^2\ge 0$. 
With the weak-coupling results from Eq.\ (\ref{weak123}), $\eta=-0.5$. 
The Gibbs free energy density of the homogeneous CFL phase is now computed with the help of  Eqs.\ (\ref{GV}) and (\ref{U0hom}),
\be \label{GCFL}
\frac{G_\mathrm{ CFL}}{V} = -\frac{H^2\cos^2\theta}{2} + U_\mathrm{ CFL}  \, , 
\ee
where 
\be \label{UCFL}
U_\mathrm{ CFL} = -\frac{3\mu^4}{4\lambda(1-2\eta)} \, .
\ee

\item If exactly one of the condensates vanishes, we also have $B_3=\tilde{B}_8=0$ in all three possible phases. The two non-vanishing condensates 
are identical, $\rho^2 = \mu^2/[\lambda (1-\eta)]$, and $U_0 = -\mu^4/[2\lambda(1-\eta)]$. We thus conclude that these phases are preferred over the CFL phase if and only if $\eta<-1$, for arbitrary magnetic field $H$. However, we shall see that in this regime the 2SC phase or the completely unpaired phase (to be discussed next) are preferred. Therefore, the phases in which exactly one of the three condensates is zero never occur and we  will ignore them from now on. 

\item If two of the condensates vanish, we have the following possible phases:

\begin{enumerate}

\item[$(i)$] $\rho_1=\rho_3=0$ ("2SC$_\mathrm{ ud}$"). If we label the three color components as usual by (red, green, blue), this phase 
corresponds to Cooper pairing of red and blue up quarks with 
blue and red down quarks, respectively. In this case, Eqs.\ (\ref{A3A8})
yield a relation between $B_3$ and $\tilde{B}_8$, and Eq.\ (\ref{rho2}) yields the value for the nonzero condensate,
\be \label{rho2SC}
\rho_2^2 = \frac{\mu^2}{\lambda} \equiv \rho_\mathrm{ 2SC}^2\, .
\ee 
Eliminating one of the magnetic fields, say $B_3$ in favor of $\tilde{B}_8$, in the Gibbs free energy (\ref{GV}) and minimizing the resulting expression with respect to $\tilde{B}_8$ yields
\be \label{2SC1}
\qquad B_3 = \frac{\sqrt{3}\sin\theta\cos\theta}{1+3\cos^2\theta} H \, , \qquad \tilde{B}_8 = \frac{3\sin\theta\cos^2\theta}{1+3\cos^2\theta} H \, , 
\ee
where we have used Eq.\ (\ref{g8}). The Gibbs free energy density becomes 
\be \label{G2SC}
\frac{G_{\mathrm{ 2SC}_\mathrm{ ud}}}{V} = -\frac{H^2\cos^2\theta}{2} -\frac{H^2}{2}\frac{3\sin^2\theta\cos^2\theta}{1+3\cos^2\theta}+ U_{\mathrm{ 2SC}} \, , 
\ee
where
\be \label{U2SC}
U_{\mathrm{ 2SC}} = -\frac{\mu^4}{4\lambda} \, .
\ee
Up to a relabeling of the colors because of the chosen flavor convention which corresponds to $Q=\mathrm{ diag}(1/3,1/3,-2/3)$, this phase is the phase  commonly termed 2SC in the literature. 
In the 2SC phase, we expect a Meissner effect for a certain combination of the photon 
and the eighth gluon, just like in CFL \cite{Schmitt:2003aa}. However, the result (\ref{G2SC}) shows that both $B_3$ and $\tilde{B}_8$ are nonzero. The reason is that 
the 2SC phase has a different mixing angle. Since we are interested in comparing the free energies of the different phases, we obviously have to work within the same basis for all phases. The use of the CFL mixing angle, together with the chosen convention for the charge matrix $Q$, 
therefore leads to a seemingly complicated result for the 2SC phase. The mixing angle of the 2SC phase can be recovered from these results by writing the Gibbs free energy (\ref{G2SC}) in the same form as the one for CFL (\ref{GCFL}),
\be \label{Gudhom}
\frac{G_{\mathrm{ 2SC}_\mathrm{ ud}}}{V} = -\frac{H^2\cos^2\vartheta_1}{2} + U_{\mathrm{ 2SC}} \, , 
\ee
where
\be
\cos^2\vartheta_1 = \frac{3g^2}{3g^2+e^2} \, .
\ee
(In Chap.~\ref{chap:2SC}, where we discuss defects in 2SC, we shall use an additional rotation given by $\vartheta_2$, hence the
notation $\vartheta_1$.)

\item[$(ii)$] $\rho_2=\rho_3=0$ ("2SC$_\mathrm{ us}$"). This phase corresponds to green/blue and up/strange pairing. 
The only difference to the 2SC$_\mathrm{ ud}$ phase is that $B_3$ has opposite sign, i.e., now $\mathbf{B}_3$
and $\tilde{\mathbf{ B}}_8$ are anti-parallel, not parallel. In particular, the Gibbs free energies are identical, because $B_3$ enters quadratically. 
This is expected since we work in the massless limit and thus interchanging $d$ with $s$ quarks should not change any physics.

\item[$(iii)$]  $\rho_1=\rho_2=0$ ("2SC$_\mathrm{ ds}$"). This phase  corresponds to red/green and down/strange pairing and is genuinely different from the usual 2SC phase -- even in the massless limit -- because now quarks with the 
same electric charge pair. In this case, we find $B_3=\tilde{B}_8=0$, $\rho_3^2=\mu^2/\lambda$, and
\be
\frac{G_{\mathrm{ 2SC}_\mathrm{ ds}}}{V} = -\frac{H^2\cos^2\theta}{2} + U_{\mathrm{ 2SC}} \,  .
\ee
\end{enumerate}
Without magnetic field, these three phases have the same free energy and are preferred over the CFL phase for $\eta < -1$.
In the presence of a magnetic field, the Gibbs free energy of the 2SC$_\mathrm{ ds}$ phase is always larger than that of the 2SC$_\mathrm{ ud}$ and 2SC$_\mathrm{ us}$ phases. Therefore, we no longer need to consider the 2SC$_\mathrm{ ds}$ phase and use the term 2SC for both 2SC$_\mathrm{ ud}$ and 2SC$_\mathrm{ us}$ in the present section. (In Chap.~\ref{chap:2SC} we will come back to the definitions of 2SC$_\mathrm{ ud}$ and 2SC$_\mathrm{ us}$ because we will discuss domain walls that interpolate between these two order parameters.)

\item Finally, in the completely unpaired phase ("NOR"), where $\rho_1=\rho_2=\rho_3=0$, we find 
\be
B_3=0 \, , \qquad \tilde{B}_8 = H\sin\theta \, , 
\ee
and the Gibbs free energy density is
\be
\frac{G_\mathrm{ NOR}}{V}=-\frac{H^2}{2} \, .
\ee

\end{itemize}

\begin{figure} [t]
\begin{center}
\hbox{\includegraphics[width=0.5\textwidth]{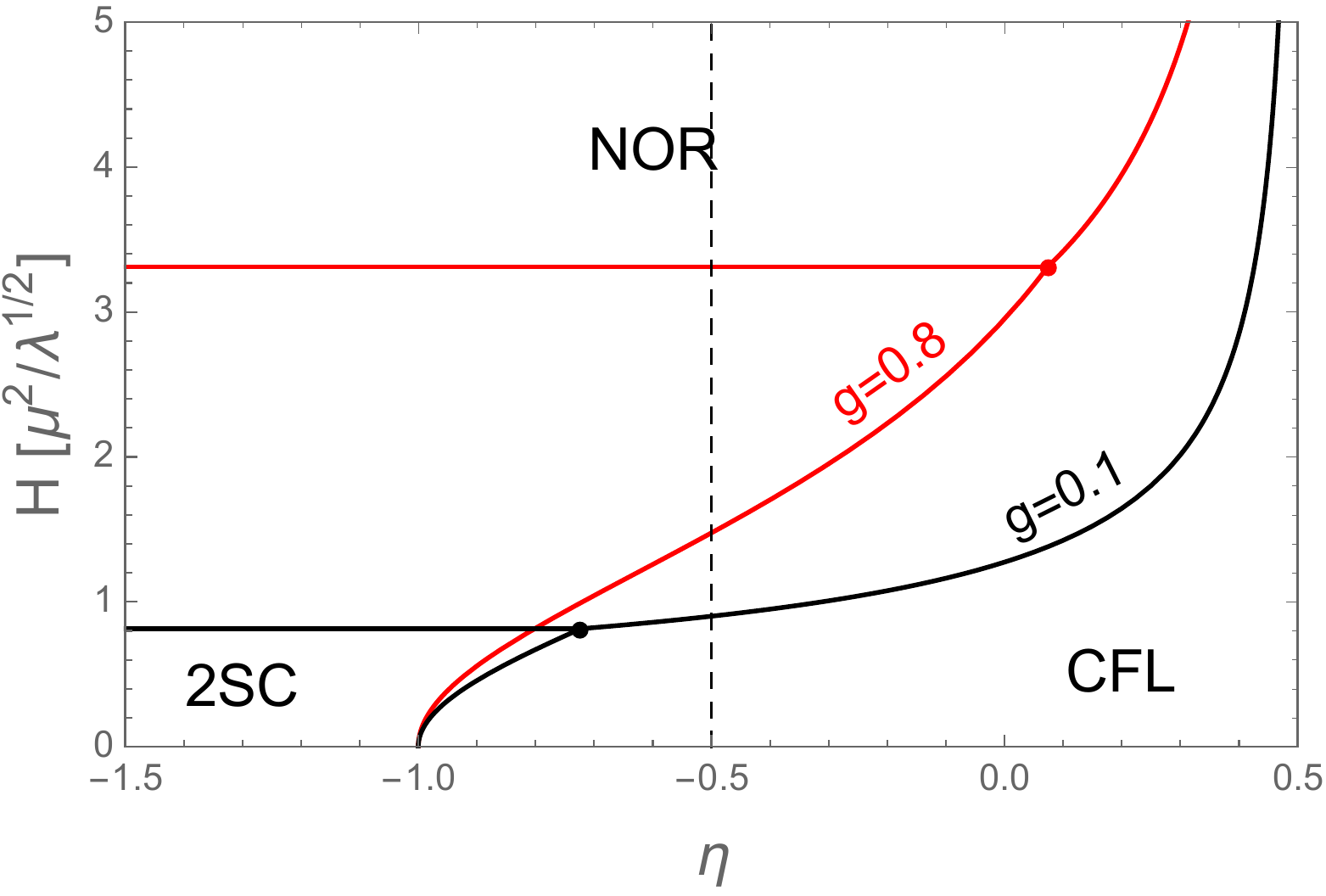}\includegraphics[width=0.5\textwidth]{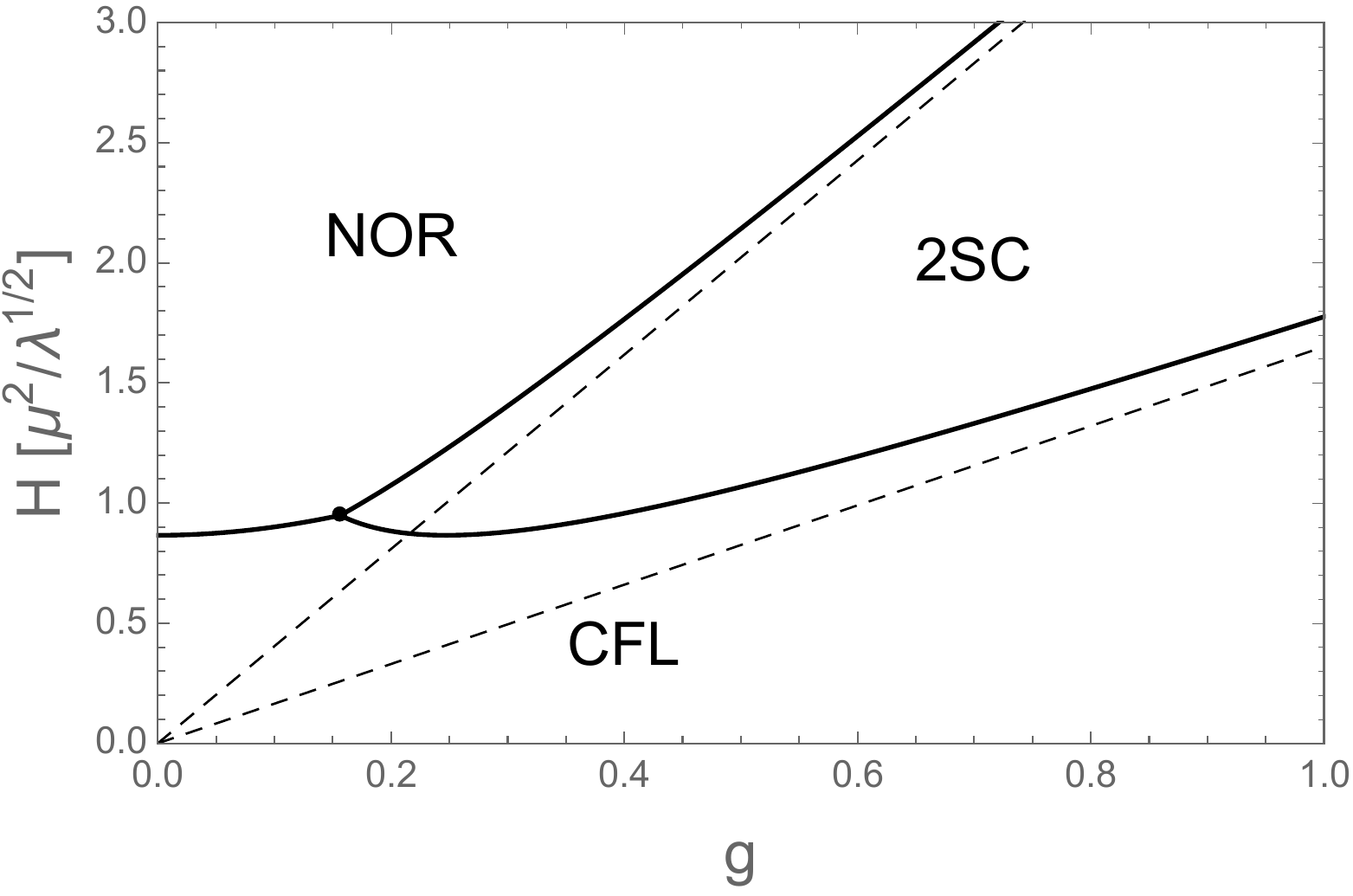}}
\caption{\textit{Left panel:} phases in the plane of external magnetic field $H$ and ratio of cross-coupling to self-coupling $\eta=h/\lambda$. The solid lines are the critical fields $H_c$ from Eq.\ (\ref{Hcs}) for two different values of the strong coupling constant $g$. The vertical dashed line 
indicates the weak-coupling value $\eta=-0.5$. The horizontal scale terminates at the maximum value $\eta=0.5$, beyond which 
the Ginzburg-Landau potential becomes unbounded from below.\newline
\textit{Right panel:} phases for $\eta=-0.5$. The critical point
where all three phases meet is given by $(g,H)=(2e/\sqrt{15},3/\sqrt{10}\,\mu^2/\sqrt{\lambda})$. For $g\to 0$, the critical 
field between CFL and NOR phases goes to $\sqrt{3}/2\, \mu^2/\sqrt{\lambda}$. The dashed lines are the critical fields for $g\gg e$.}
\label{fig:Heta}
\end{center}
\end{figure}

\section{Critical Fields \texorpdfstring{$H_{c}$}{Hc}}

With these results one can easily compute the critical magnetic fields of the phase transitions between CFL, 2SC, and NOR phases by comparing the corresponding free energies,
\be \label{Hcs}
\frac{H_c^2}{\mu^4/\lambda} = \left\{\begin{array}{cc} \displaystyle{\frac{3g^2+e^2}{2e^2}} & \mbox{2SC/NOR} \\[4ex] 
\displaystyle{\frac{3}{2(1-2\eta)}\frac{3g^2+4e^2}{4e^2}} & \mbox{CFL/NOR} \\[4ex]
\displaystyle{\frac{1+\eta}{1-2\eta}\frac{(3g^2+e^2)(3g^2+4e^2)}{9e^2g^2}} & \mbox{2SC/CFL} \end{array}\right. \, .
\ee
We plot the critical fields in the phase diagrams of Fig.~\ref{fig:Heta}. In the chosen units for the magnetic field, the phase structure
only depends on $\eta$ and the strong coupling 
constant $g$ (the electromagnetic coupling constant $e$ is held fixed). This will no longer be
true when we discuss the type-I/type-II transition in the
subsequent chapters. This transition  
depends also on $\lambda$ separately, i.e., on the ratio $T_c/\mu_q$. To avoid a multi-dimensional study of the parameter space, we shall thus later 
restrict ourselves to the weak-coupling results of the Ginzburg-Landau parameters, which imply $\eta=-0.5$, and extrapolate  these results 
to large values of $g$.  This is already done in the 
right panel of Fig.~\ref{fig:Heta}, which means that the left panel of this 
figure is the only plot where $\eta$ is kept general. 

We see that at zero magnetic field and weak coupling CFL is preferred over 2SC, which is well known and remains true if a small strange quark mass together with the conditions of color and electric neutrality are taken into account \cite{Alford:2002kj}. 
If $\eta$ is kept general, there is a regime where 2SC is preferred, even for vanishing magnetic field. This can be 
understood within the three-component picture, having in mind that $\eta=h/\lambda$ with $\lambda>0$: a negative coupling
$h$ implies repulsion between the three components. If this repulsion is sufficiently large, the condensates no longer "want" to coexist and the 2SC phase becomes preferred.  

In the presence of a magnetic field $H$, the Gibbs free energy can be lowered by admitting this field into the system. 
This situation becomes slightly more complicated due to the rotated electromagnetism. In CFL, part of the magnetic field 
is already admitted because it is $\tilde{B}_8$, not $B$, that is completely expelled from the superconductor. Admitting 
a larger $B$ field can be achieved by breaking 
all condensates (now the entire applied magnetic field penetrates, $H=B$, but all condensation energy is lost) or by first going to the "intermediate" 2SC phase, where some condensation energy is maintained. Both scenarios are realized, as the right panel shows: 
for small values of the strong coupling constant, the CFL phase is directly superseded by the unpaired phase, while for all 
$g>2e/\sqrt{15}$ the 2SC phase appears between CFL and NOR.

\section{Critical Fields \texorpdfstring{$H_{c2}$}{Hc2}}

Next, we compute the critical field $H_{c2}$ for all three phase transitions given in 
Eq.~(\ref{Hcs}). We follow the standard procedure as we did for the single superconductor and the two-component system, to compute these fields , which becomes slightly more complicated 
for the CFL/2SC transition, where we can follow the two-component treatment. 
The  equations of motion for the complex fields are computed from Eq.\ (\ref{U123}), 
\begin{subequations}
\bea
\left[\left(\nabla+i\frac{g}{2}\mathbf{A}_3+i\tilde{g}_8\tilde{\mathbf{A}}_8\right)^2+\mu^2-2\lambda|\phi_1|^2+2h(|\phi_2|^2+|\phi_3|^2)\right] \phi_1 &=& 0 \, , \\[2ex]
\left[\left(\nabla-i\frac{g}{2}\mathbf{A}_3+i\tilde{g}_8\tilde{\mathbf{A}}_8\right)^2+\mu^2-2\lambda|\phi_2|^2+2h(|\phi_1|^2+|\phi_3|^2)\right] \phi_2 &=& 0 \, , \\[2ex]
\left[\left(\nabla-2i\tilde{g}_8\tilde{\mathbf{A}}_8\right)^2+\mu^2-2\lambda|\phi_3|^2+2h(|\phi_1|^2+|\phi_2|^2)\right] \phi_3 &=& 0 \, .
\eea
\end{subequations}
We discuss the three phase transitions separately.

\begin{itemize}

\item The simplest case is the transition between 2SC and NOR, where $\phi_1=\phi_3=0$ in both phases. 
We linearize in $\phi_2$ and set $\mathbf{A}_3=0$ because $B_3=0$ in the unpaired phase.
This leaves the single equation 
\be
\left[\left(\nabla+i\tilde{g}_8\tilde{\mathbf{A}}_8\right)^2+\mu^2\right]\phi_2 = 0 \, .
\ee
With the usual argument \cite{tinkham2004introduction} this gives a maximal field $\tilde{B}_8=-\mu^2/\tilde{g}_8$.  
Since in the normal phase $\tilde{B}_8=H\sin\theta$, the critical field is
 \be
H_{c2} = \frac{3\mu^2}{e} \qquad \mbox{(for 2SC/NOR)} \, .
\ee
At the 2SC/NOR transition, the system is an ordinary single-component superconductor, and we expect an ordinary type-I/type-II transition at exactly $H_c=H_{c2}$. This can be confirmed by the numerical calculation of $H_{c1}$ for ordinary 2SC flux tubes, see Fig.\ \ref{fig:phases} in Chap.~\ref{chap:results_color}. Therefore, using Eq.\ (\ref{Hcs}) and the 
weak-coupling expression for $\mu$ from Eq.\ (\ref{weak123}), 2SC flux tubes appear for
\be
\frac{T_c}{\mu_q}>\frac{\sqrt{7\zeta(3)}}{12\sqrt{3}\pi^2}\sqrt{g^2+\frac{e^2}{3}} \simeq 0.014 \sqrt{g^2+\frac{e^2}{3}} \, .
\ee
This standard type-I/type-II transition is expected to occur 
at $\kappa_\mathrm{ 2SC}^2=1/2$. As a check, we may thus define the corresponding Ginzburg-Landau parameter a posteriori, 
\be
\kappa_\mathrm{ 2SC}^2 = \frac{72\pi^4}{7\zeta(3)}\frac{3}{g^2+\frac{e^2}{3}} \frac{T_c^2}{\mu_q^2} \, , 
\ee
which is in exact agreement with Eq.\ (112) of Ref.\ \cite{Iida:2002ev}.

\item For the transition between CFL and NOR phases we linearize in all three condensates and set $\mathbf{A}_3=0$,
because in the phase above $H_{c2}$ all condensates and $B_3$ vanish. This leads to the three equations  
\bea
\left[\left(\nabla+i\tilde{g}_8\tilde{\mathbf{A}}_8\right)^2+\mu^2\right] \phi_1 &=& 0 \, ,\non[2ex] 
\left[\left(\nabla+i\tilde{g}_8\tilde{\mathbf{A}}_8\right)^2+\mu^2\right] \phi_2 &=& 0 \, , \\
\left[\left(\nabla-2i\tilde{g}_8\tilde{\mathbf{A}}_8\right)^2+\mu^2\right] \phi_3 &=& 0 \, .\nonumber
\eea
The first two equations give a maximal field $\tilde{B}_8=-\mu^2/\tilde{g}_8$, 
which we use to compute $H_{c2}$, such that at least one of the condensates is nonzero below $H_{c2}$. This definition of $H_{c2}$ for the CFL/NOR transition agrees with Ref.\ \cite{Iida:2004if}, and  we find the same critical field as for the 2SC/NOR transition,
 \be
H_{c2} = \frac{3\mu^2}{e}  \qquad \mbox{(for CFL/NOR)} \, .
\ee
As an estimate for the location of the type-I/type-II transition we again use the point $H_c=H_{c2}$, although in this case the critical 
region is expected to look more complicated because CFL is a multicomponent system. We find that CFL flux tubes appear (if the next phase up in $H$ is the NOR phase) for
\be \label{Tmucond}
\frac{T_c}{\mu_q} > \frac{\sqrt{7\zeta(3)}}{24\pi^2}\frac{\sqrt{g^2+\frac{4}{3}e^2}}{\sqrt{1-2\eta}} \simeq 8.7\times 10^{-3}\sqrt{g^2+\frac{4}{3}e^2} \, ,
\ee
where, for the numerical estimate, $\eta=-0.5$ is used. 
As Fig.\ \ref{fig:Heta} demonstrates, the CFL/NOR transition is only relevant for $g<2e/\sqrt{15}\simeq  0.16$, where one would expect the weak-coupling results to be applicable. 
Hence, in this regime, $T_c/\mu_q \propto \exp(-\mathrm{ const}/g)$ is exponentially suppressed and it seems very unlikely that the type-II regime is realized.

\item For the transition between CFL and 2SC, without loss of generality, the 2SC$_\mathrm{ ud}$ phase is used.  In this phase, $\phi_1=\phi_3=0$ and
thus we can linearize in $\phi_1$ and $\phi_3$ (but not in $\phi_2$). Moreover, in 2SC$_\mathrm{ ud}$ we have $g\mathbf{A}_3=2\tilde{g}_8\tilde{\mathbf{A}}_8$,
which follows from  Eq.\ (\ref{2SC1}). This relation is used to eliminate $\mathbf{A}_3$ and we arrive at the two equations
\be
\left[\left(\nabla\pm2i\tilde{g}_8\tilde{\mathbf{A}}_8\right)^2+\mu^2+2h|\phi_2|^2\right] \phi_{1/3} = 0 \, ,
\ee
and the homogeneous solution for the second condensate $|\phi_2|^2 = \mu^2/(2\lambda)$. With $\tilde{\mathbf{A}}_8 = x\tilde{B}_8\mathbf{e}_y$
this becomes 
\be
\mu^2(1+\eta)\phi_{1/3}=\left[-\Delta\mp 2i(2\tilde{g}_8\tilde{B}_8)x\partial_y+(2\tilde{g}_8\tilde{B}_8)^2x^2\right]\phi_{1/3} \, ,
\ee
where $\Delta = \partial_x^2+\partial_y^2+\partial_z^2$. As for the standard scenario, this equation has the form of the Schr{\"o}dinger 
equation for the harmonic oscillator, and we can compute the critical field in the usual way from the lowest eigenvalue
\cite{tinkham2004introduction,Haber:2017kth}. The result is
\be
H_{c2} = \frac{2\mu^2(1+\eta)(3g^2+e^2)}{3eg^2} \qquad \mbox{(for CFL/2SC)} \, .
\ee 
Again, we can determine the point $H_c=H_{c2}$, which suggests type-II behavior for
\be \label{Tcmu1}
\frac{T_c}{\mu_q} > \frac{\sqrt{14\zeta(3)}}{24\pi^2\sqrt{1-2\eta}\sqrt{1+\eta}}\frac{g\sqrt{3g^2+4e^2}}{\sqrt{3g^2+e^2}} \simeq 0.017 g\sqrt{\frac{3g^2+4e^2}{3g^2+e^2}} \,.
\ee
If we use the critical temperature for CFL from perturbative calculations \cite{Alford:2007xm,Schmitt:2002sc}, 
\be
T_c = 2^{1/3}\frac{e^\gamma}{\pi}\Delta_0 \, , 
\ee
with the Euler-Mascheroni constant $\gamma$ and the zero-temperature gap
\be
\Delta_0 = \mu_q b \exp\left(-\frac{3\pi^2}{\sqrt{2}g}\right) \, , \qquad 
b\equiv 512\pi^4\left(\frac{2}{g^2 N_f}\right)^{5/2}e^{-\frac{\pi^2+4}{8}}2^{-1/3} \, ,
\ee
and extrapolate the resulting ratio $T_c/\mu_q$ to large values of the coupling, we find that the criterion (\ref{Tcmu1}) for type-II behavior is not fulfilled for any $g$.
Thus, if we take Eq.\ (\ref{Tcmu1}) as the relevant criterion, we have to assume that strong-coupling effects, not captured by the extrapolation of the weak-coupling result, drive $T_c$ sufficiently large to allow for type-II behavior. 
As model calculations suggest, $T_c/\mu_q \gtrsim 0.06$ [choosing $g = 3.5$ in Eq.\ (\ref{Tcmu1}), which is plausible
for interiors of neutron stars] is not unrealistically large. We note, however, that the multicomponent nature of CFL suggests that flux tubes can appear for \textit{smaller} values of $T_c/\mu_q$ due to a possible first-order onset of flux tubes that
increases the region in the phase diagram where a lattice of flux tubes is preferred \cite{Haber:2017kth}. The exact calculation of the modified critical $T_c/\mu_q$ would require a numerical study of the flux tube lattice, and it is conceivable that even the extrapolated weak-coupling result allows for type-II behavior.

\end{itemize}

\chapter{CFL Flux Tubes}
\label{chap:CFL}

We now turn to the flux tube solutions in the CFL phase. The first step is the formulation of  the equations of motion in the most 
general way (within our diagonal ansatz for the gap matrix). This allows us to discuss the various possible flux tube 
configurations, compare their profiles and free energies, and determine the energetically 
most preferred flux tube configuration  by computing the critical fields $H_{c1}$. 

\section{Equations of Motion and Flux Tube Energy for CFL}

Having in mind a single, straight flux tube, we assume cylindrical symmetry and work in cylindrical coordinates $\mathbf{r} = (r,\varphi,z)$.  We write the modulus and the phase of the condensates from Eq.\ (\ref{rhopsi_c}) as  ($i=1,2,3$),
\be \label{fn}
\rho_i(\mathbf{r}) = f_i(r)\rho_\mathrm{ CFL} \, , \qquad \psi_i(\mathbf{r}) = n_i\varphi \, , 
\ee
with the CFL condensate in the homogeneous phase $\rho_\mathrm{ CFL}$ from Eq.\ (\ref{rho0}) and  
dimensionless functions $f_i(r)$. Single-valuedness of the order parameter requires $n_i\in \mathbb{Z}$. These are the winding numbers, for which there is a priori no additional condition, in particular they can be chosen independently of each other.  We will see that this choice determines the properties of the flux tube. 
For the gauge fields, we make the ansatz 
\be
\mathbf{A}_3(\mathbf{r}) = \frac{a_3(r)}{r}\mathbf{e}_\varphi  \, , \qquad \tilde{\mathbf{A}}_8(\mathbf{r}) = \frac{\tilde{a}_8(r)}{r}\mathbf{e}_\varphi  \, ,
\ee
with the dimensionless functions $a_3(r)$ and $\tilde{a}_8(r)$. This yields magnetic fields in the $z$ direction, 
\be \label{B38}
\mathbf{B}_3(r) = \frac{1}{r}\frac{\partial a_3}{\partial r}\mathbf{e}_z \, , \qquad \tilde{\mathbf{ B}}_8(r) = \frac{1}{r}\frac{\partial \tilde{a}_8}{\partial r}\mathbf{e}_z
 \, .
 \ee
After eliminating $\mu$ in favor of $\rho_\mathrm{ CFL}$ with the help of Eq.\ (\ref{rho0}), we can write the potential (\ref{U123}) as
\bea \label{U00}
U_0 &=& U_\mathrm{ CFL} + U_{\circlearrowleft} \, , 
\eea
with $U_\mathrm{ CFL}$ from Eq.~(\ref{UCFL}) and the free energy density of the flux tube
\bea \label{Uflux}
U_{\circlearrowleft}&=&\frac{\lambda\rho_\mathrm{ CFL}^4}{2}\left\{\frac{\lambda(a_3'^2+\tilde{a}_8'^2)}{R^2} + f_1'^2+f_2'^2+f_3'^2 +\frac{(1-f_1^2)^2}{2}+\frac{(1-f_2^2)^2}{2}+\frac{(1-f_3^2)^2}{2} \right.\non[2ex]
&& \left.-\eta\Big[(1-f_1^2)(1-f_2^2)+(1-f_1^2)(1-f_3^2)+(1-f_2^2)(1-f_3^2)\Big] \right.\non[2ex]
&& \left.+f_1^2\frac{{\cal N}_1^2}{R^2}+f_2^2\frac{{\cal N}_2^2}{R^2}+f_3^2\frac{{\cal N}_3^2}{R^2} \right\} \, , 
\eea
where we have introduced the new dimensionless coordinate 
\be
R=r\sqrt{\lambda}\,\rho_\mathrm{ CFL} \, , 
\ee
have denoted derivatives with respect to $R$ by a prime, and have abbreviated
\be \label{N123}
{\cal N}_1 \equiv n_1+\frac{g}{2}a_3+\tilde{g}_8\tilde{a}_8 \, , \qquad {\cal N}_2 \equiv n_2-\frac{g}{2}a_3+\tilde{g}_8\tilde{a}_8
\, , \qquad {\cal N}_3\equiv n_3-2\tilde{g}_8\tilde{a}_8 \, .
\ee
Consequently, the equations of motion for the gauge fields become
\begin{subequations} \label{a3a8}
\bea
a_3''-\frac{a_3'}{R}&=&\frac{g}{2\lambda}\left(f_1^2{\cal N}_1-f_2^2{\cal N}_2\right) \, , \\[2ex]
\tilde{a}_8''-\frac{\tilde{a}_8'}{R}&=&\frac{\tilde{g}_8}{\lambda}\left(f_1^2{\cal N}_1+f_2^2{\cal N}_2
-2f_3^2{\cal N}_3\right) \, ,
\eea
\end{subequations}
and the equations of motion for the condensates are 
\begin{subequations} \label{f123}
\bea
0&=& f_1''+\frac{f_1'}{R}+f_1(1-f_1^2)-f_1\frac{{\cal N}_1^2}{R^2} -\eta f_1(2-f_2^2-f_3^2) \, , \\[2ex]
0&=& f_2''+\frac{f_2'}{R}+f_2(1-f_2^2)-f_2\frac{{\cal N}_2^2}{R^2} -\eta f_2(2-f_1^2-f_3^2) \, , \\[2ex]
0&=& f_3''+\frac{f_3'}{R}+f_3(1-f_3^2)-f_3\frac{{\cal N}_3^2}{R^2} -\eta f_3(2-f_1^2-f_2^2) \, . 
\eea
\end{subequations}
The boundary values of the scalar fields are as follows. Far away from the flux tube, the system is in the 
CFL phase, such that $f_i(\infty)=1$. In the origin, the scalar fields  vanish
if the respective component has nonzero winding, $f_i(0)=0$ if $n_i\neq 0$. Otherwise, we require $f_i'(0)=0$ as a boundary condition, and  $f_i(0)$ must be determined dynamically.
For the gauge fields, we use Eqs.\ (\ref{a3a8}) to determine their values at infinity. Assuming $a_3'(\infty)=a_3''(\infty)=\tilde{a}_8'(\infty)=\tilde{a}_8''(\infty)=0$, we find 
\be \label{ainf}
a_3(\infty) = \frac{n_2-n_1}{g} \, , \qquad \tilde{a}_8(\infty)=\frac{2n_3-n_1-n_2}{6\tilde{g}_8} \, .
\ee
In the origin we then have to require $a_3(0)=a_8(0)=0$, which follows from the equations of motion evaluated for small $R$. We solve the coupled differential equations (\ref{a3a8}) and (\ref{f123}) numerically 
with the help of a successive over-relaxation method to obtain the profiles of the flux tubes. The flux tube energy $F_\circlearrowleft$ 
per unit length is then 
obtained by inserting the result into Eq.\ (\ref{Uflux}) and integrating over space. We write the result as 
\bea \label{FL}
\frac{F_\circlearrowleft}{L} &=& \frac{1}{L}\int d^3\mathbf{r}\, U_\circlearrowleft 
= \pi\rho_\mathrm{ CFL}^2 \,{\cal I}_\circlearrowleft \, , 
\eea
where $L$ is the length of the flux tube in the $z$-direction, and 
\bea \label{Icircle}
{\cal I}_\circlearrowleft\equiv \int_0^\infty dR\,R\left[\frac{\lambda(a_3'^2+\tilde{a}_8'^2)}{R^2}+\frac{1-f_1^4}{2}
+\frac{1-f_2^4}{2}+\frac{1-f_3^4}{2} -\eta(3-f_1^2f_2^2-f_1^2f_3^2-f_2^2f_3^2)\right]\, , \hspace*{-1cm}\non[0.5ex]
\hspace{-5cm}
\eea
where partial integration and the equations of motion (\ref{f123}) have been used. 

\section{Critical Field \texorpdfstring{$H_{c1}$}{Hc1}} 
\label{sec:Hc1_CFL}

To determine the critical magnetic field $H_{c1}$ we need to compute the Gibbs free energy of the CFL phase 
in the presence of a flux tube. We insert the energy density $U_0$ from Eq.\ (\ref{U00}) with the notation introduced in 
Eq.\ (\ref{FL}) into the general form of the Gibbs free energy (\ref{GV}). Furthermore, 
we use 
\be \label{drB_CFL}
\int d^3\mathbf{r}\, \tilde{B}_8 = 2\pi L \tilde{a}_8(\infty) \, , 
\ee
which follows directly from the form of the magnetic field in Eq.\ (\ref{B38}) and the boundary condition $\tilde{a}_8(0)=0$. Recall that
we have defined $\tilde{\mathbf{ B}}_8 = \tilde{B}_8 \mathbf{e}_z$, i.e., $\tilde{B}_8$ is the $z$-component, not the modulus, of $\tilde{\mathbf{ B}}_8$. Therefore, $\mathbf{H}\cdot\tilde{\mathbf{B}}_8 =H \tilde{B}_8$ with $H$ being non-negative by assumption and the sign of $\tilde{B}_8$ 
indicating whether $\tilde{\mathbf{ B}}_8$ is aligned or anti-aligned with $\mathbf{H}$. 

This yields the Gibbs free energy density 
\be
\frac{G}{V} = -\frac{H^2\cos^2\theta}{2} +U_\mathrm{ CFL} +\frac{L}{V}\left[\frac{F_{\circlearrowleft}}{L}-2\pi \tilde{a}_8(\infty) H \sin\theta \right] \, .
\ee
It is favorable to place a single flux tube into the system if this reduces the free energy of the homogeneous CFL phase (\ref{GCFL}), i.e., if the expression in the square brackets becomes negative. By definition, 
this occurs at the critical magnetic field $H_{c1}$. Writing this critical field in the same units as the critical fields in Fig.\ \ref{fig:Heta}, we find 
\bea \label{Hc1_CFL}
\frac{H_{c1}}{\mu^2/\sqrt{\lambda}}  &=& \frac{(3g^2+4e^2)\,{\cal I}_\circlearrowleft}{4e\sqrt{\lambda}(1-2\eta)(n_1+n_2-2n_3)}
\, ,
\eea
where we have used Eqs.\ (\ref{costheta}), (\ref{g8}), (\ref{rho0}), (\ref{ainf}), and (\ref{FL}). Note that the critical field is proportional to the flux tube energy per winding number $n_1+n_2-2n_3$. In general, the expression on the right-hand side can be positive or negative, but we have assumed $H$ to be 
positive and hence $H_{c1}$ must be positive. We have $1-2\eta>0$ for all allowed values 
of $\eta$ and ${\cal I}_\circlearrowleft>0$ [which we always find to be the case, although it is not manifest from
Eq.\ (\ref{Icircle}) since $f_i(r)>1$ is possible]. Therefore,
the winding numbers must be chosen such that $n_1+n_2-2n_3>0$, which can be understood as follows. 
If $n_1+n_2-2n_3>0$, we have $\tilde{a}_8(\infty)<0$ because of Eq.\ (\ref{ainf}). 
Hence, due to $\tilde{a}_8(0)=0$ and Eq.\ (\ref{B38}), and assuming  $\tilde{a}_8(r)$ to be a monotonic function of $r$, $\tilde{\mathbf{ B}}_8$ is \textit{anti-parallel} to $\mathbf{H}$ for all $r$.  Therefore,  $\tilde{\mathbf{ B}}_8\sin\theta$, which is the contribution to $\mathbf{B}$, is \textit{parallel} to $\mathbf{H}$ because $\sin\theta<0$, as it should be. 

\section{Asymptotic Behavior}
\label{sec:asym}

It is useful to determine the point at which the long-range interaction between two flux tubes 
changes from repulsive to attractive in color superconductors as well. 
To compute the interaction between flux tubes in this more complicated setting, we first need to discuss the asymptotic behavior of the flux tube profiles. 
Far away from the center of the flux tube, i.e., for large $R$, we use the ansatz for the gauge fields $a_3(R) = a_3(\infty) + R v_3(R)$, $\tilde{a}_8(R) = \tilde{a}_8(\infty) + R \tilde{v}_8(R)$ and for the scalar fields 
$f_i(R)=1+u_i(R)$ ($i=1,2,3$). We assume $n_1+n_2+n_3=0$. This is equivalent to a vanishing baryon circulation far away from the 
flux tube, as will be discussed in detail in Sec.\ \ref{sec:circ}.  

We linearize the equations of motion (\ref{a3a8}) and (\ref{f123}) in the functions $v_3, \tilde{v}_8, u_1, u_2, u_3$. 
The equations for the gauge fields then yield decoupled equations for $v_3$ and $\tilde{v}_8$, 
\begin{subequations}
\bea
v_3''+\frac{v_3'}{R} &\simeq & \left(1+\frac{R^2}{\kappa_3^2}\right) \frac{v_3}{R^2} \, , \\[2ex]
\tilde{v}_8''+\frac{\tilde{v}_8'}{R} &\simeq & \left(1+\frac{R^2}{\tilde{\kappa}_8^2}\right) \frac{\tilde{v}_8}{R^2} \, , 
\eea
\end{subequations}
where we have used Eq.\ (\ref{ainf}), and where
\bea \label{kappa38}
\kappa_3^2 \equiv \frac{2\lambda}{g^2} \, , \qquad \tilde{\kappa}_8^2\equiv\frac{\lambda}{6\tilde{g}_8^2} \, .
\eea
We can see that the two gauge fields come with different Ginzburg-Landau parameter, since their magnetic penetration length depends on the charge, which is different. The solutions of these equations are 
\begin{subequations} \label{asymp}
\bea
v_3(R) &=& c_3K_1(R/\kappa_3) \, , \\[2ex]
\tilde{v}_8(R) &=& \tilde{c}_8K_1(R/\tilde{\kappa}_8) \, , 
\eea
\end{subequations}
where $K_n$ are the modified Bessel functions of the second kind and $c_3$ and $\tilde{c}_8$ are integration constants which can 
only be determined numerically. By canceling the coherence length from the two different $\kappa$ parameters with the one of the dimensionless radial variable $R$, we find the expected exponential behavior of the gauge fields with the characteristic scale set by the penetration depth.
The linearized equations for the scalar fields are 
\begin{subequations} \label{asympf_eq}
\bea
0&\simeq& u_1''+\frac{u_1'}{R}-2u_1+2\eta(u_2+u_3) \, , \\[2ex]
0&\simeq& u_2''+\frac{u_2'}{R}-2u_2+2\eta(u_1+u_3) \, , \\[2ex]
0&\simeq& u_3''+\frac{u_3'}{R}-2u_3+2\eta(u_1+u_2) \, .
\eea
\end{subequations}
We solve these coupled equations as in the two-component system by diagonalizing the mixing matrix. Hence, we first write them as
\bea
\Delta u = M u \, , \qquad M\equiv 2\left(\begin{array}{ccc}1&-\eta&-\eta\\-\eta&1&-\eta\\-\eta&-\eta&1\end{array}\right) \, , \qquad u \equiv \left(\begin{array}{c} u_1\\u_2\\u_3\end{array}\right) \, ,
\eea
where $\Delta$ is the Laplacian in cylindrical coordinates.
Then, we diagonalize this system of equations, 
\be
\Delta\tilde{u} = (U^{-1}MU) \tilde{u} \, , 
\ee
with $\tilde{u}=U^{-1}u$ and 
\be
U = \left(\begin{array}{ccc} 1&-1&-1\\1&0&1\\1&1&0 \end{array}\right) \, , \qquad U^{-1}MU = \left(\begin{array}{ccc}\nu_1&0&0\\0&\nu_2&0\\0&0&\nu_2\end{array}\right) \, ,
\ee
where the eigenvalues of $M$ are denoted by 
\be
\nu_1\equiv 2(1-2\eta) \, , \qquad \nu_2\equiv 2(1+\eta) \, .
\ee
Solving the uncoupled equations and then undoing the rotation yields the 
asymptotic solutions 
\begin{subequations} \label{asympf}
\bea
u_1(R)&=&d_1K_0(\sqrt{\nu_1}R)-(d_2+d_3)K_0(\sqrt{\nu_2}R) \, , \\[2ex]
u_2(R)&=&d_1K_0(\sqrt{\nu_1}R)+d_3K_0(\sqrt{\nu_2}R) \, , \\[2ex]
u_3(R)&=&d_1K_0(\sqrt{\nu_1}R)+d_2K_0(\sqrt{\nu_2}R) \, , 
\eea
\end{subequations}
with  integration constants $d_1$, $d_2$, $d_3$. From Fig.\ \ref{fig:Heta} we know that the CFL phase only exists for $-1<\eta<0.5$. For values outside that regime the 2SC phase is preferred (large negative values of $\eta$), or the Ginzburg-Landau potential is unbounded from below 
(large positive values). Therefore, both eigenvalues $\nu_1$ and $\nu_2$ are positive in the relevant regime and the square roots
in Eqs.\ (\ref{asympf}) are real. 

We have thus found that all gauge fields and scalar fields fall off exponentially for 
$R\to\infty$, which guarantees the finiteness of the free energy of the flux tube configuration and justifies the boundary conditions used above for the gauge fields. This is not the case if the baryon circulation is nonzero, $n_1+n_2+n_3\neq 0$, where, as suggested from 
ordinary superfluid vortices, at least one of the fields falls off with a power law \cite{Eto:2009kg}.

\section{Interaction Between Flux Tubes}
\label{sec:inter}

We can now use the asymptotic solutions to compute the interaction between two flux tubes at large distances. 
This calculation has been explained in detail for a two-component system in App.~\ref{app:fl_int}, based on 
well-known approximations for a one-component superconductor \cite{Kramer:1971zza}.
 The extension to the present case with three scalar components 
and two gauge fields is straightforward, although somewhat tedious. The interaction energy $F^\circlearrowleft_\mathrm{ int}(R_0)$
between two flux tubes, say flux tube $(a)$ and flux tube $(b)$, whose centers are in a distance $R_0$ from each other, is again defined as 
\be
F^{(a)+(b)}= F^{(a)}+ F^{(b)}+ F^\circlearrowleft_\mathrm{ int}(R_0) \, ,
\ee
where $F^{(a)+(b)}$ is the total free energy of the two flux tubes, $F^{(a)}$ is the free energy of flux tube $(a)$ in the absence of flux 
tube $(b)$, and vice versa for $F^{(b)}$.  A brief sketch of the calculation is given in appendix \ref{app:inter_color}. The result 
for the interaction energy per unit length is 
\bea \label{Fint2}
\frac{F_\mathrm{ int}^{\circlearrowleft}}{L} &=& 2\pi \rho_\mathrm{ CFL}^2\Big[\frac{\kappa_3^2g^2c_3^2}{2}K_0(R_0/\kappa_3)
+6\tilde{\kappa}_8^2\tilde{g}_8^2\tilde{c}_8^2K_0(R_0/\tilde{\kappa}_8)-3d_1^2K_0(\sqrt{\nu_1}R_0)\non[2ex] &&-2(d_2^2+d_3^2+d_2d_3)K_0(\sqrt{\nu_2}R_0)\Big] \, . \;\;
\eea
This is in agreement with Eq.\ (46) in Ref.\ \cite{Iida:2004if}, where the term proportional to $K_0(R_0/\kappa_3)$ was absent because 
only flux tubes without $B_3$-flux were considered. 
There are positive (repulsive) contributions from the gauge fields and negative (attractive) contributions from the scalar 
fields. For $\eta < 0$ we have $\nu_2<\nu_1$, and thus the long-distance behavior of the attractive contribution is dominated by $K_0(\sqrt{\nu_2}R_0)$ [note that $2(d_2^2+d_3^2+d_2d_3)=(d_2+d_3)^2+d_2^2+d_3^2>0$]. Since at weak coupling $\eta=-0.5$, we shall focus on this case. 
For the repulsive part we notice that always $\kappa_3 >\tilde{\kappa}_8$, such that, if there is a nonzero $B_3$-flux, the dominant contribution is 
given by $K_0(R_0/\kappa_3)$. Then, the interaction is attractive for $\sqrt{\nu_2}<1/\kappa_3$. If the $B_3$-flux vanishes, the contribution containing $\kappa_3$ does not exist and the 
interaction is attractive for  $\sqrt{\nu_2}<1/\tilde{\kappa}_8$. Inserting the definitions for $\kappa_3$ and $\tilde{\kappa}_8$ from Eq.\ (\ref{kappa38}),
we find that the interaction is repulsive for 
\bea \label{repul}
\frac{T_c}{\mu_q} > \left\{\begin{array}{cc} \displaystyle{\frac{\sqrt{7\zeta(3)}}{12\pi^2\sqrt{2(1+\eta)}}\sqrt{g^2+\frac{4}{3}e^2}\simeq 0.025\sqrt{g^2+\frac{4}{3}e^2}} & \;\;\;\;\mbox{for $B_3=0$} \\ [4ex]
\displaystyle{\frac{\sqrt{7\zeta(3)}}{12\pi^2\sqrt{2(1+\eta)}}g\simeq 0.025 g } & \;\;\;\;\mbox{for $B_3\neq0$} \end{array}\right. \, , 
\eea
where, for the numerical approximation, we have inserted the weak-coupling result $\eta=-0.5$. We shall make use of these results in our discussion of the phase diagram in Chap.~\ref{chap:results_color}.

\section{Baryon Circulation and Magnetic Flux}
\label{sec:circ}

In general, the flux tubes described by Eqs.\ (\ref{a3a8}) and (\ref{f123})  have nonzero baryon circulation $\Gamma$ and 
nonzero magnetic fluxes $\Phi_3$ and $\tilde{\Phi}_8$. These three quantities are used in the following to discuss the properties of the possible flux tube 
configurations.

The baryon circulation is computed by inserting our ansatz for the order parameter into the superfluid velocity (\ref{superv}) to obtain
\bea \label{VS}
\mathbf{v}_s 
&=& \frac{1}{6\mu_q}\frac{\rho_1^2 n_1+\rho_2^2n_2+\rho_3^2n_3+\tilde{g}_8\tilde{a}_8(\rho_1^2+\rho_2^2-2\rho_3^2)+\frac{g}{2}a_3(\rho_1^2-\rho_2^2)}{\rho_1^2+\rho_2^2+\rho_3^2} \frac{\mathbf{e}_\theta}{r}
\, ,
\eea
where we have used $u^2=1/3$. Then, the baryon circulation around a CFL flux tube along a circle at infinity becomes
\be
\Gamma = \oint d \bm{\ell}\cdot\mathbf{v}_s = \frac{\pi}{3\mu_q}\frac{n_1+n_2+n_3}{3} \, ,
\ee
where we have used that far away from the flux tube the condensates assume their homogeneous CFL values and become identical, 
$\rho_1=\rho_2=\rho_3$. 
Consequently, the CFL flux tube has vanishing baryon circulation if the three winding numbers add up to zero. In particular, 
the gauge fields have dropped out of the result. This is different from an ordinary flux tube in a single-component superconductor, where
the circulation can only vanish due to a cancellation between the winding number and the gauge field, as can be seen by setting $\rho_1=\rho_2=0$ in Eq.~(\ref{VS}). 

The magnetic fluxes are
\begin{subequations}\label{Phi3Phi8}
\bea
\Phi_3 &=& \oint d\bm{\ell} \cdot \mathbf{A}_3 = 2\pi a_3(\infty) = \frac{2\pi}{g}(n_2-n_1)  \, , \\[2ex]
\tilde{\Phi}_8 &=& \oint d\bm{\ell} \cdot \tilde{\mathbf{A}}_8 = 2\pi \tilde{a}_8(\infty) = \frac{\pi}{\tilde{g}_8}\frac{2n_3-n_1-n_2}{3}  \, . 
\label{Phi8}
\eea
\end{subequations} 
We can now classify all possible flux tubes by their three winding numbers and use the baryon circulation and the color-magnetic 
fluxes to understand their main properties. In Table \ref{tab:n1n2n3} the most important configurations that are expected to 
appear in CFL in the presence of an externally imposed  rotation and/or an externally imposed  magnetic field are listed. One point 
of this table is to demonstrate that the CFL line defects considered so far in the literature and the new configurations discussed 
here are all defined by a particular choice of the triple of winding numbers.  (We recall that the three-component nature of our 
system is a consequence of the diagonal ansatz of the gap matrix. In principle, more components might appear through non-diagonal gap matrices, which would induce additional color magnetic fields.  To our knowledge, such configurations have not been studied in the literature.) 

\begin{table}[t]
\centering
\resizebox{\columnwidth}{!}{%
\begin{tabular}{|c|c||c|c|c|} 
\hline
\rule[-1.5ex]{0em}{4ex} 
 CFL line defect  &$\;\;$ $(n_1,n_2,n_3)$ $\;\;$ & $\;\;$ $\Gamma\; [\pi/3\mu_q]$ $\;\;$& $\;\;$$\Phi_3\; [2\pi/g]$ $\;\;$
  & $\;\;$$\tilde{\Phi}_8\; [\pi/\tilde{g}_8]$ $\;\;$ \\[1ex] \hline\hline
\rule[-1.5ex]{0em}{6ex} 
$\;\;$ $T_{111}$ (global vortex \cite{Forbes:2001gj})$\;\;$& $(n,n,n)$ & $n$ & 0 & 0\\[2ex] \hline
\rule[-1.5ex]{0em}{6ex} 
$T_{001}$ (semi-superfluid vortex, "$M_1$" \cite{Balachandran:2005ev}) & $(0,0,n)$ & $\displaystyle{\frac{n}{3}}$ & 0 & $\displaystyle{\frac{2n}{3}}$ \\[2ex] \hline
\rule[-1.5ex]{0em}{6ex} 
$T_{110}$ (semi-superfluid vortex, "$M_2$" \cite{Balachandran:2005ev}) & $(n,n,0)$ & $\displaystyle{\frac{2n}{3}}$ & 0 & $\displaystyle{-\frac{2n}{3}}$ \\[2ex] \hline
\rule[-1.5ex]{0em}{6ex} 
$T_{112}$ (magnetic flux tube \cite{Iida:2004if}) & $(n,n,-2n)$ & 0 & 0 & $-2n$  \\[2ex] \hline
\rule[-1.5ex]{0em}{6ex} 
$\;\;$ $T_{101}$ (magnetic flux tube)$\;\;$ & $(n,0,-n)$ & 0 & $-n$ & $-n$  \\[2ex] \hline
\end{tabular}}
\caption{Line defects in CFL, classified by the  winding numbers of the three components of the order parameter, 
$n\in \mathbb{Z}$,  from which baryon number circulation $\Gamma$ and color-magnetic fluxes $\Phi_3$ and $\tilde{\Phi}_8$ are obtained.}
\label{tab:n1n2n3}

\end{table}

If an external rotation is applied to CFL, vortices with nonzero 
baryon circulation must be formed. This has been discussed in detail in the literature. For instance, it has been found that the global vortex $T_{111}$ (which has no color-magnetic flux) is unstable with respect to decay into three so-called semi-superfluid vortices
\cite{Balachandran:2005ev,Alford:2016dco}. 
Each semi-superfluid vortex has nonzero color-magnetic fluxes, but a triple of vortices $T_{100}$, $T_{010}$, $T_{001}$
(in an obvious generalization of the notation introduced in Tab.~(\ref{tab:n1n2n3})) is color neutral. We do not discuss rotationally induced 
vortices here. We rather focus on configurations with vanishing baryon circulation $\Gamma$ and non-vanishing 
magnetic flux $\tilde{\Phi}_8$,
\begin{subequations} \label{constraints}
\bea
n_1+n_2+n_3&=&0 \, , \\[2ex]
n_1+n_2-2n_3& >& 0 \, .
\eea
\end{subequations}
These are flux tubes that are formed in the type-II regime of CFL if an external (ordinary) 
magnetic field is applied, but no rotation. 
In the interior of a neutron star, there is nonzero rotation \textit{and} a nonzero magnetic field, i.e., the total magnetic flux and the 
total angular momentum must be nonzero. We know that the rotational axis and the magnetic field axis are, at least for some neutron stars,  not aligned, otherwise we would not observe them as pulsars. This suggests that, if there is a CFL core in the pulsar,  magnetic flux and baryon circulation are 
not maintained by a single species of flux tubes. Therefore, it appears that purely magnetic flux tubes, without circulation, are necessary.  

\begin{figure} [t]
\begin{center}
\hbox{\includegraphics[width=0.5\textwidth]{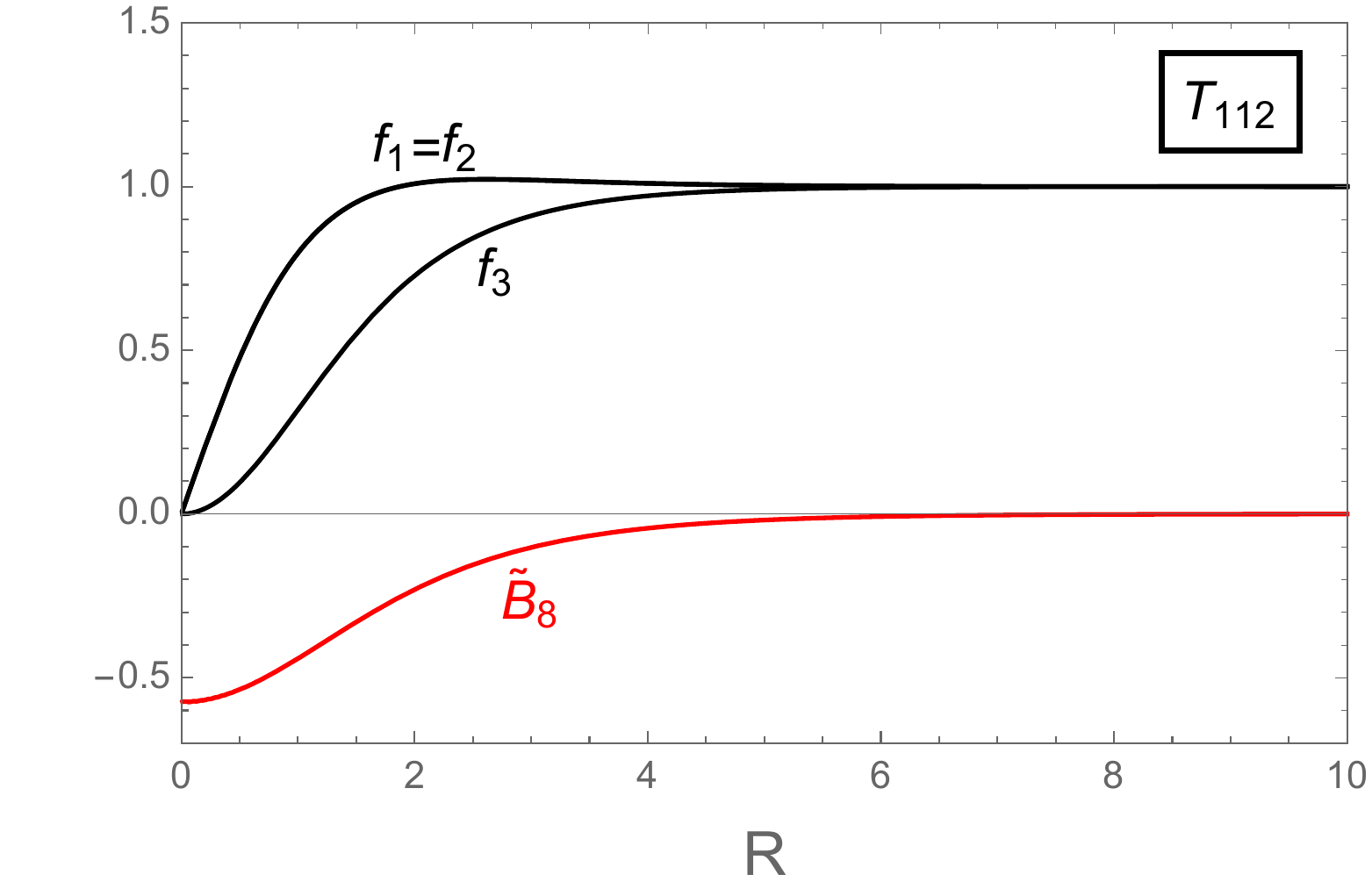}\includegraphics[width=0.5\textwidth]{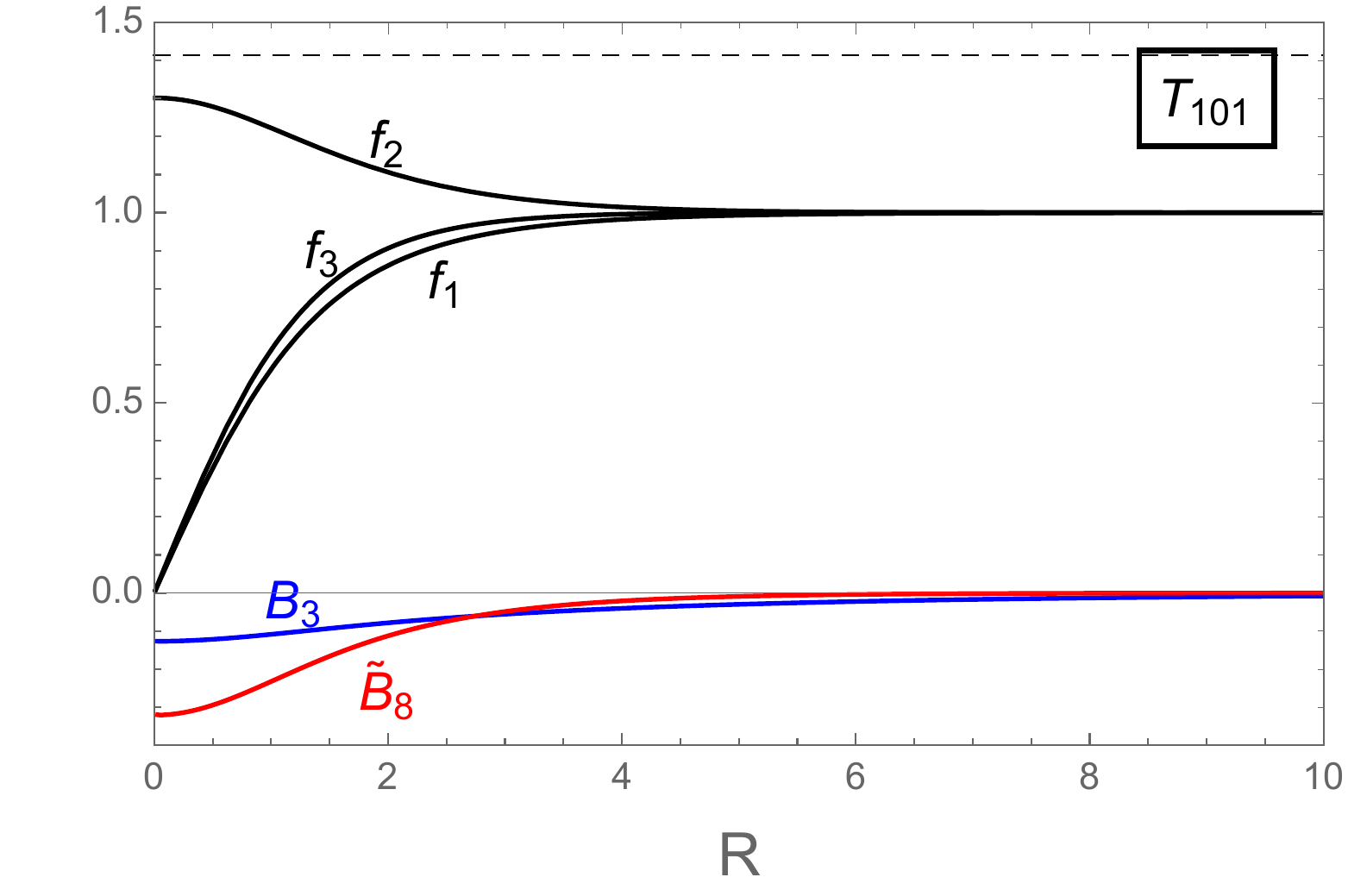}}
\caption{Dimensionless condensates $f_1$, $f_2$, $f_3$ and magnetic fields $B_3$ and $\tilde{B}_8$ in units 
of $\mu^2/\sqrt{\lambda}$ for the CFL flux tubes $T_{112}$ (left, where $B_3=0$) and $T_{101}$ (right) with $n=1$ at the same coupling $g=0.1$ and $\eta=-0.5$, 
$T_c/\mu_q\simeq 0.012$, as a function of the dimensionless radial coordinate $R=r\sqrt{\lambda}\,\rho_\mathrm{ CFL}$. The horizontal dashed line in the right panel marks the homogeneous 2SC condensate $f_2=\sqrt{1-2\eta}$ ($f_2$
is rescaled with the CFL condensate). If we increase the winding, $n_1, -n_3\to \infty$, the condensate $f_2$  approaches this value. The negative sign of $\tilde{B}_8$ ensures that $\tilde{B}_8\sin\theta \,\mathbf{e}_z$ is aligned with the 
magnetic field $\mathbf{H}$. At the relatively small value of $g$ chosen here, the $B_3$ field in the right panel falls off on a larger length 
scale than the $B_8$ field, $\kappa_3/\tilde{\kappa}_8=\sqrt{1+4e^2/(3g^2)}\simeq 3.6$. }
\label{fig:profilesCFL}
\end{center}
\end{figure}

Within the two constraints (\ref{constraints})  we are interested in the energetically most preferred 
flux tube. In the previous literature, only the flux tube $T_{112}$ was discussed, but there are obviously 
infinitely many more possibilities to choose  winding numbers that fulfill the constraints (\ref{constraints}). 
One can systematically study all possibilities: for instance, define the length (squared) of the vector $(n_1,n_2,n_3)$ by 
$N^2\equiv n_1^2+n_2^2+n_3^2$, then choose an $N_0$ and solve the equations of motion for all vectors $(n_1,n_2,n_3)$ 
that fulfill Eqs.\ (\ref{constraints}) and whose length is smaller than $N_0$. This can easily be automatized with a computer. 
Such a calculation, including the comparison of the free energies of the different flux tubes (for a certain choice of the 
Ginzburg-Landau parameters) suggests that the obvious expectation is fulfilled: unless we are in the type-I regime, where flux tubes are never preferred, configurations with 
a small "total winding" $N$ tend to be favored. Therefore, we rather focus exclusively 
on the two configurations with the smallest $N$, namely $T_{112}$ and $T_{101}$.  

The price one has to pay for minimizing 
the total winding in $T_{101}$ compared to $T_{112}$ is a nonzero $B_3$ field. This gives an energy cost due to the $B_3^2$ term in the 
free energy. (Presumably this is the reason why this flux tube has so far been ignored in the literature.) However, one of the 
scalar fields has zero winding and thus it is allowed to remain nonzero in the center of the flux tube. 
Moreover, the negative sign of the effective coupling constant $h$ (using the weak-coupling result) suggests that the scalar components interact repulsively with each other. Hence, if 
$\rho_1$ and $\rho_3$ go to zero, $\rho_2$ does not only not vanish, but is even expected to be enhanced in the center of the flux tube. This implies a gain in condensation energy and is exactly what the numerical result will show. 

There is another way of understanding the difference between $T_{101}$ and $T_{112}$. 
If, in the configuration $T_{112}$, the winding $n$ is increased, the flux tube gets wider and the completely unpaired phase 
in the center of the tube grows until eventually CFL has been replaced by the NOR phase. As a consequence, $H_{c1}$ approaches $H_c$ for $n\to \infty$. (In the type-I regime, $H_{c1}\to H_c$ from above, and in the type-II regime from below.) This suggests that, in the absence of flux tubes, there is a transition from the CFL to the NOR phase. However, we have seen in 
Chap.~\ref{chap:hom} that there is a parameter regime where CFL is, upon increasing $H$, replaced by 2SC, not by the NOR phase. 
The configuration $T_{101}$ accounts for this transition: now, if the winding $n$ is sent to infinity, the second component survives and one 
arrives in the 2SC phase (more precisely, the 2SC$_\mathrm{ ud}$ phase). This suggests that where there is a transition from CFL to 2SC, 
the configuration $T_{101}$ should be favored. 

We will thus refer to $T_{112}$ as a "CFL flux tube with a NOR core" and to $T_{101}$
as a "CFL flux tube with a 2SC core", keeping in mind that this is a simplifying terminology for the fully dynamically computed flux tube profiles. 
The profiles of both configurations are shown in Fig.~\ref{fig:profilesCFL} for the coupling constant $g=0.1$ and the ratio $T_c/\mu_q$ at which the critical fields $H_{c1}$ of both configurations turn out to be identical. 
We shall compare the critical magnetic field $H_{c1}$ for both kind of flux tubes more systematically in Chap.~\ref{chap:results_color}.

\section{Physical Units and Numerical Estimates}
\label{sec:units}

As already pointed out in Refs.~\cite{Iida:2001pg,Iida:2004if}, the critical magnetic fields associated with the (partial) breaking of color superconductivity
are extremely large. The main reason is that color superconductors -- in an astrophysical environment where $g\gg e$ -- admit 
a large part of the externally applied magnetic field because the massless gauge boson is almost identical to the photon, 
with a small admixture of one of the gluons. Therefore, breaking the superconductor, or partially breaking it through the formation 
of  magnetic defects, requires an enormously large ordinary magnetic field. In all our results, the magnetic fields are given 
in units of $\mu^2/\sqrt{\lambda}$, which is very convenient since it minimizes the number of parameters to be specified. To translate this into physical units we use the definitions (\ref{weak123}) and find
\be \label{HinG}
\frac{\mu^2}{\sqrt{\lambda}}  \simeq 1.597\times 10^{19}(1-t)\mu_{q400}^2\frac{T_c}{\mu_q}\, G \, ,
\ee 
where $t\equiv T/T_c$ and $\mu_{q400}\equiv \mu_q/(400\, \mathrm{ MeV})$. Although the ratio $T_c/\mu_q$ is exponentially small at weak coupling, this is certainly not true in the interior of neutron stars. 
Therefore, Eq.~(\ref{HinG}) shows that the critical magnetic fields (for instance in Fig.\ \ref{fig:Heta})  are 
much larger than the measured magnetic fields at the surface of the star, which are at most of the order of
$10^{16} \,  \mathrm{ G}$. Magnetic fields in the interior that are several orders of magnitude larger seem unlikely, although not inconceivable,
 given the estimate of maximal magnetic fields in a quark matter core of the order of $10^{20} \, \mathrm{ G}$ \cite{Ferrer:2010wz}. 
As we shall see later, the new flux tube solution $T_{101}$ has a smaller critical field $H_{c1}$ compared to $T_{112}$,
but this decrease does  not change the order of magnitude estimate of the critical field strength.

We may also estimate the width of the flux tubes in physical units. From the 
asymptotic solutions of the CFL flux tubes (\ref{asympf}) and the definition of the dimensionless radial coordinate 
$R=r \sqrt{\lambda}\,\rho_\mathrm{ CFL}$ we read off the coherence length $\xi$. This is the length scale on which all three 
condensates approach their homogeneous values. Again using Eqs.\ (\ref{weak123}) we find 
\be
\xi^{-1} = \sqrt{\lambda}\,\rho_\mathrm{ CFL}  \simeq 10.76 \frac{T_c}{\mu_q}\sqrt{1-t}\,\mu_{q400} \, \mathrm{ fm}^{-1} \, .
\ee
For a numerical estimate, let us set $T_c\simeq 40\, \mathrm{ MeV}$, such that $T_c/\mu_q \simeq 0.1$. Judging from 
model calculations and extrapolations from the perturbative result, this is a large, but conceivable, critical temperature. 
Then, setting $T=0$, we find that $\xi \simeq 0.93 \, \mathrm{ fm}$. 
The magnetic penetration depth $\ell$ is obtained from the asymptotic solution 
(\ref{asymp}). We have to distinguish between the penetration depths of $B_3$ and $\tilde{B}_8$, which become identical only for 
$g\gg e$,
\begin{subequations}
\bea
\ell_3^{-1} &=& \frac{g\rho_\mathrm{ CFL}}{\sqrt{2}} \simeq 0.37 g\sqrt{1-t}\,\mu_{q400} \, \mathrm{ fm}^{-1} \, , \\[2ex]
\ell_8^{-1} &=&\sqrt{6}\tilde{g}_8\rho_\mathrm{ CFL} \simeq 0.37 \sqrt{g^2+\frac{4e^2}{3}} \sqrt{1-t}\,\mu_{q400} \, \mathrm{ fm}^{-1}  \, .
\eea
\end{subequations}
With $T=0$ and $g\simeq 3.5$ we find $\ell_3\simeq \ell_8 \simeq 0.77\, \mathrm{ fm}$. 

Finally, we write the energy of the flux tube per unit length from Eq.\ (\ref{FL}) as 
\be
\frac{F_{\circlearrowleft}}{L} = \pi\rho_{\mathrm{CFL}}^2 {\cal I}_{\circlearrowleft} \simeq 1.378\times 10^{9}(1-t)\mu_{q400}^2\,{\cal I}_{\circlearrowleft}\,\frac{\mathrm{erg}}{\mathrm{cm}} \, , 
\ee
where ${\cal I}_{\circlearrowleft}$ has to be computed numerically. For instance, with $g=3.5$ and $T_c/\mu_q=0.1$ we find
for the $T_{112}$ tube ${\cal I}_{\circlearrowleft} \simeq 5.9$ and for the $T_{101}$ tube ${\cal I}_{\circlearrowleft} \simeq 2.5$, both 
with $n=1$, in rough agreement with the simple estimates used in Ref.\ \cite{Glampedakis:2012qp}, which yield $F_{\circlearrowleft}/L\simeq 1.5 \times 10^{10}\mu_{q400}^2\,\mathrm{ erg}/\mathrm{ cm}$.

\chapter{Magnetic Defects in the 2SC Phase}
\label{chap:2SC}

At first sight, color-magnetic flux tubes in 2SC (= flux tubes that approach the 2SC phase at infinity) are 
less exotic than their counterparts in CFL because 2SC is a single-component
superconductor, which means that only one of the scalar fields in the Ginzburg-Landau potential is nonzero. In an ordinary 2SC flux tube, which 
we will refer to as\footnote{To distinguish 2SC flux tubes from CFL flux tubes, we denote them by $S$, instead of $T$. The 2SC domain wall will be denoted by $D$.} $S_1$, this component has a nonzero winding and vanishes in the center of the tube \cite{Alford:2010qf}. 
One may ask, however, whether the other two components are induced inside the flux tube, similarly to the flux tubes discussed in 
Refs.\ \cite{Forgacs:2016ndn,Forgacs:2016iva}, or like the enhancement of the neutral condensate in the two-component system. We shall investigate this 
possibility by considering 2SC flux tubes within the full three-component calculation. The result suggests the 
existence of domain walls, which will emerge as the infinite-radius limit of the flux tubes.

In the 2SC phase, we work with $Q=\mathrm{ diag}(2/3,-1/3,-1/3)$. This amounts to an ordering of the quark flavors as $(u,d,s)$. Then, the usual 2SC phase with 
up/down pairing, 2SC$_\mathrm{ ud}$,  is given by a nonzero condensate $\rho_3$. Since we work in the massless limit, this phase is
equivalent to the 2SC$_\mathrm{ us}$ phase, where only $\rho_2$ is nonzero\footnote{Recall that in all preceding sections we used $Q=\mathrm{ diag}(-1/3,-1/3,2/3)$, which is more convenient for CFL, and thus the 2SC$_\mathrm{ us}$ and 2SC$_\mathrm{ ud}$ phases were given by 
a nonzero $\rho_1$ and $\rho_2$, respectively.}. For the magnetic defects in 2SC, it is convenient to introduce the following rotated fields\footnote{In many aspects, the 2SC calculation is analogous to the CFL calculation, and it is helpful to reflect this in the notation. Therefore, the notation 
$\tilde{A}_\mu^8$ and 
$\tilde{A}_\mu$ is used again, although these fields are different from the rotated fields in the CFL calculation. Since the CFL mixing will not appear from now on, this should not lead to any confusion.},
\be
\left(\begin{array}{c} \tilde{A}_\mu^3 \\ \tilde{A}_\mu^8 \\ \tilde{A}_\mu \end{array}\right) =  
\left(\begin{array}{ccc} \cos\vartheta_2 & 0 &\sin\vartheta_2 \\ 0&1&0 \\ -\sin\vartheta_2&0&\cos\vartheta_2  \end{array}\right)\left(\begin{array}{ccc} 1 & 0 &0 \\ 0& \cos\vartheta_1 & \sin\vartheta_1 \\ 0 & -\sin\vartheta_1 & \cos\vartheta_1   \end{array}\right)\left(\begin{array}{c} A_\mu^3 \\ A_\mu^8 \\ A_\mu \end{array}\right)  \, ,
\ee
with 
\begin{subequations}
\bea
\sin\vartheta_1 &=& \frac{e}{\sqrt{3g^2+e^2}} \, , \qquad \cos\vartheta_1 = \frac{\sqrt{3}g}{\sqrt{3g^2+e^2}} \, , \\[2ex]
\sin\vartheta_2 &=& \frac{\sqrt{3}e}{\sqrt{3g^2+4e^2}} \, , \qquad \cos\vartheta_2 = \frac{\sqrt{3g^2+e^2}}{\sqrt{3g^2+4e^2}} \, .
\eea
\end{subequations}
This two-fold rotation is motivated as follows. If we were interested in the homogeneous 2SC$_\mathrm{ ud}$ 
phase, given by a nonzero $\rho_3$, the gauge field $A_\mu^3$ 
would play no role and applying the rotation given by $\vartheta_1$  yields a magnetic field that is expelled, $\tilde{B}_8$, and the orthogonal combination that penetrates the 2SC phase. This is well-known, see for instance Ref.\ \cite{Schmitt:2003aa}. 
Here, however,  we are interested in keeping all condensates. One finds that $\rho_1$ and $\rho_2$ are charged under all 
three gauge fields that are obtained from this first rotation. The second rotation,  given by $\vartheta_2$, simplifies the situation by
creating a field, namely $\tilde{A}_\mu$, under which \textit{all three} condensates are neutral, while leaving $\tilde{A}_\mu^8$ unchanged.  
This is useful because it eliminates $\tilde{A}_\mu$ from the calculation of the flux tube and domain wall profiles, and we only have to 
deal with two gauge fields in the numerical calculation. This derivation is laid out in more detail in App.~\ref{app:rotated}, where the rotation is carried out in two intermediate steps for pedagogical reasons.

The Ginzburg-Landau potential in terms of the new rotated fields is obtained by starting from the potential given by Eqs.\ (\ref{UUB}) and (\ref{U123}), undoing the CFL rotation and applying the 2SC rotations, or by re-starting from the original potential (\ref{UPhi}). In either case, one derives
\be
U = \frac{\tilde{B}^2}{2} + U_0 \, , 
\ee
with 
\bea \label{U02SC}
U_0 &=& \frac{\tilde{\mathbf{ B}}_3^2}{2}+ \frac{\tilde{\mathbf{ B}}_8^2}{2} +\frac{(\nabla\rho_1)^2}{2}+\frac{(\nabla\rho_2)^2}{2}+\frac{(\nabla\rho_3)^2}{2}\\[2ex]&&-\frac{\mu^2}{2}(\rho_1^2+\rho_2^2+\rho_3^2)+\frac{\lambda}{4}(\rho_1^4+\rho_2^4+\rho_3^4)-\frac{h}{2}(\rho_1^2\rho_2^2+\rho_2^2\rho_3^2+\rho_1^2\rho_3^2)\non[2ex]
&&+\left(\nabla\psi_1+\tilde{q}_3\tilde{\mathbf{A}}_3+\tilde{q}_{81}\tilde{\mathbf{A}}_8\right)^2\frac{\rho_1^2}{2}
+\left(\nabla\psi_2-\tilde{q}_3\tilde{\mathbf{A}}_3+\tilde{q}_{82}\tilde{\mathbf{A}}_8\right)^2\frac{\rho_2^2}{2}+\left(\nabla\psi_3+\tilde{q}_{83}\tilde{\mathbf{A}}_8\right)^2\frac{\rho_3^2}{2} \nonumber \,  ,
\eea
where we have written the scalar fields in terms of their moduli and phases according to Eq.\ (\ref{rhopsi_c}), and where 
we have abbreviated 
\be
\tilde{q}_{81}\equiv \frac{3g^2+4e^2}{6\sqrt{3g^2+e^2}} \, , \qquad \tilde{q}_{82}\equiv \frac{3g^2-2e^2}{6\sqrt{3g^2+e^2}} \, , \qquad \tilde{q}_{83}\equiv \frac{\sqrt{3g^2+e^2}}{3} \, , 
\ee 
and
\be
\tilde{q}_3\equiv \frac{g}{2}\sqrt{\frac{3g^2+4e^2}{3g^2+e^2}} \, .
\ee
We can write the Gibbs free energy density as
\be
\frac{G}{V} = -\frac{H^2\cos^2\vartheta_1\cos^2\vartheta_2}{2}+\frac{1}{V}\int d^3\mathbf{r}
\left[U_0-H(\tilde{B}_3\cos\vartheta_1\sin\vartheta_2+\tilde{B}_8\sin\vartheta_1)\right] \, ,
\ee
where we have used $\tilde{B} = H\cos\vartheta_1\cos\vartheta_2$, which follows from minimizing $G$ with respect to $\tilde{B}$. 
For the homogeneous phases we repeat the calculation from Chap.~\ref{chap:hom} to find
\begin{subequations} \label{B3B823}
\bea
&\mathrm{ 2SC}_\mathrm{ ud}:& \qquad \tilde{B}_3=H\cos\vartheta_1\sin\vartheta_2 \, , \qquad \tilde{B}_8 = 0 \, ,\\[2ex]
&\mathrm{ 2SC}_\mathrm{ us}:& \qquad \tilde{B}_3=\frac{3ge(3g^2-2e^2)H}{2\sqrt{3g^2+4e^2}(3g^2+e^2)^{3/2}} \, , \qquad \tilde{B}_8 = \frac{9g^2eH}{2(3g^2+e^2)^{3/2}} \, . 
\eea
\end{subequations}

\section{Flux Tubes in 2SC}

In analogy to the CFL calculation, we write the scalar fields as 
$\rho_i(\mathbf{r}) = f_i(r)\rho_\mathrm{ 2SC}$ with the homogeneous 2SC condensate $\rho_\mathrm{ 2SC}$ from Eq.\ (\ref{rho2SC}), and introduce the winding numbers in the 
phases through $\psi_i(\mathbf{r}) = n_i\varphi$. We use the 2SC$_\mathrm{ ud}$ phase 
for our boundary condition far away from the flux tube, i.e., 
$f_1(\infty)=f_2(\infty)=0$, $f_3(\infty)=1$, while $f_i(0)=0$ if the corresponding winding number $n_i$ is nonzero.
For the gauge fields we write 
\be \label{A3A8tt}
 \tilde{\mathbf{A}}_3(\mathbf{r}) = \left[\frac{H\cos\vartheta_1\sin\vartheta_2}{2}r+\frac{\tilde{a}_3(r)}{r}\right]\mathbf{e}_\varphi \, ,
 \qquad \tilde{\mathbf{A}}_8(\mathbf{r}) = \frac{\tilde{a}_8(r)}{r}\mathbf{e}_\varphi \, ,
\ee
with $\tilde{a}_3'(\infty)=\tilde{a}_8'(\infty)=0$ and $\tilde{a}_3(0)=\tilde{a}_8(0)=0$. In contrast to the CFL flux tubes, there is a magnetic field, $\tilde{B}_3$, which is nonzero far away from the 
flux tube (in addition to the homogeneous field $\tilde{B}$, which simply penetrates the superconductor). This field 
will become inhomogeneous in the flux tube, unless the system chooses to keep $\rho_1$ and $\rho_2$ zero
everywhere. We have separated the 
homogeneous part of the $\tilde{B}_3$ field in our ansatz (\ref{A3A8tt}), such that far away from the flux tube $\tilde{a}_3$ does not contribute to the magnetic field and we have $\tilde{\mathbf{ B}}_3(\infty)=H\cos\vartheta_1\sin\vartheta_2\, \mathbf{e}_z$. 
This separation is useful, but not crucial. Alternatively,  one could have implemented the external field in the boundary condition for $\tilde{a}_3$.

Inserting our ansatz into the potential (\ref{U02SC}), we compute the Gibbs free energy density
\bea
&&U-\mathbf{H}\cdot \mathbf{B} = U_\mathrm{ 2SC} -\frac{H^2\cos^2\vartheta_1}{2} - \lambda\rho_\mathrm{ 2SC}^2H\sin\vartheta_1\frac{\tilde{a}_8'}{R} 
+ \frac{\lambda\rho_\mathrm{ 2SC}^4}{2}\left\{\frac{\lambda(\tilde{a}_3'^2+\tilde{a}_8'^2)}{R^2}+f_1'^2+f_2'^2\right.\non[2ex]
&&\left. +f_3'^2+f_1^2\left(\frac{f_1^2}{2}-1\right)+f_2^2\left(\frac{f_2^2}{2}-1\right)+\frac{(1-f_3^2)^2}{2}-\eta(f_1^2f_2^2+f_1^2f_3^2+f_2^2f_3^2) \right.\non[2ex]
&&\left. +\frac{({\cal N}_1+\Xi R^2)^2f_1^2+({\cal N}_2-\Xi R^2)^2f_2^2+{\cal N}_3^2f_3^2}{R^2}\right\} \, , 
\eea
with $U_\mathrm{ 2SC}$ from Eq.\ (\ref{U2SC}). Analogously to Chap.~\ref{chap:CFL} we have introduced the dimensionless coordinate 
$R=r\sqrt{\lambda}\,\rho_\mathrm{ 2SC}$, prime denotes derivative with respect to $R$, we have defined the dimensionless external magnetic field 
\be \label{Xi}
\Xi=\frac{\tilde{q}_3 H\cos\vartheta_1\sin\vartheta_2}{2\lambda\rho_\mathrm{ 2SC}^2} = \frac{3eg^2}{4\sqrt{\lambda}(3g^2+e^2)}\frac{H}{\mu^2/\sqrt{\lambda}} \, ,
\ee
and we have abbreviated 
\bea
{\cal N}_1 &\equiv& n_1+\tilde{q}_3\tilde{a}_3+\tilde{q}_{81}\tilde{a}_8 \, ,\qquad 
{\cal N}_2 \equiv n_2-\tilde{q}_3\tilde{a}_3+\tilde{q}_{82}\tilde{a}_8 \, ,\qquad 
{\cal N}_3 \equiv n_3+\tilde{q}_{83}\tilde{a}_8 \, , 
\eea
in analogy to Eq.\ (\ref{N123}). 

The equations of motion are obtained in a complete analogous way as in the CFL calculation, presented in App.~\ref{App:CFL_FT}, therefore we only state the result, which is given for the gauge fields by
\begin{subequations}
\bea
\tilde{a}_3''-\frac{\tilde{a}_3'}{R}&=& \frac{\tilde{q}_3}{\lambda}[({\cal N}_1+\Xi R^2)f_1^2-({\cal N}_2-\Xi R^2)f_2^2] \, , \\[2ex]
\tilde{a}_8''-\frac{\tilde{a}_8'}{R}&=& \frac{1}{\lambda}[\tilde{q}_{81}({\cal N}_1+\Xi R^2)f_1^2+\tilde{q}_{82}({\cal N}_2-\Xi R^2)f_2^2+\tilde{q}_{83}{\cal N}_3f_3^2] \, , \label{a8t}
\eea
\end{subequations}
and for the scalar fields we find
\begin{subequations}
\bea
0&=& f_1''+\frac{f_1'}{R}+f_1\left[1-f_1^2-\frac{({\cal N}_1+\Xi R^2)^2}{R^2} +\eta(f_2^2+f_3^2)\right] \, , \\[2ex]
0&=& f_2''+\frac{f_2'}{R}+f_2\left[1-f_2^2-\frac{({\cal N}_2-\Xi R^2)^2}{R^2} +\eta(f_1^2+f_3^2)\right] \, , \\[2ex]
0&=& f_3''+\frac{f_3'}{R}+f_3\left[1-f_3^2-\frac{{\cal N}_3^2}{R^2} +\eta(f_1^2+f_2^2)\right] \, . 
\eea
\end{subequations}
Evaluating Eq.\ (\ref{a8t}) at $R=\infty$ yields 
\be \label{n3a}
\tilde{a}_8(\infty) = -\frac{n_3}{\tilde{q}_{83}} \, ,
\ee
which is the usual relation for a single-component superconductor and implies vanishing baryon circulation far away from the flux tube.
There is no analogous condition for $\tilde{a}_3(\infty)$, and we determine this value dynamically in the numerical solution. 

We can write the Gibbs free energy density as 
\be \label{G2SCflux}
\frac{G}{V}=U_\mathrm{ 2SC}-\frac{H^2\cos^2\vartheta_1}{2}+\frac{L}{V}\left[\frac{F_\circlearrowleft}{L}-2\pi \tilde{a}_8(\infty)H\sin\vartheta_1\right] \, , 
\ee
where the flux tube energy per unit length, in analogy to the CFL calculation, is
\be
\frac{F_\circlearrowleft}{L} = \pi\rho_\mathrm{ 2SC}^2\,  {\cal I}_\circlearrowleft \, , 
\ee
with
\be
{\cal I}_\circlearrowleft \equiv \int_0^\infty dR\,R\left[\frac{\lambda(\tilde{a}_3'^2+\tilde{a}_8'^2)}{R^2}-\frac{f_1^4}{2}-\frac{f_2^4}{2}+\frac{1-f_3^4}{2}+\eta
(f_1^2f_2^2+f_1^2f_3^2+f_2^2f_3^2)\right] \, .
\ee
The critical magnetic field $H_{c1}$ is again calculated by setting the expression in the square brackets in Eq.\ (\ref{G2SCflux}) to zero, since the remaining terms are the Gibbs free energy density of the homogeneous 2SC phase (\ref{Gudhom}). However, this calculation is more complicated than in the CFL phase because $F_\circlearrowleft$ now depends implicitly on $H$. Therefore, instead of simply computing the free energy of the flux tube we have to solve  the following equation numerically,
\be \label{Xictube}
\Xi_{c1} +\frac{g^2 {\cal I}_\circlearrowleft(\Xi_{c1})}{8\lambda n_3} = 0 \, .
\ee
This equation has a solution $\Xi_{c1}>0$ only for $n_3<0$. This  means that, in our convention for the winding number $n_3$, 
we need $n_3<0$ to align the $B$-component of the magnetic field in the flux tube along the external magnetic field. In the simple case of the ordinary 2SC flux tube, i.e., where only the condensate $\rho_3$ is nonzero and where only the gauge field $\tilde{a}_8$ needs 
to be taken into account in the calculation of profiles, the free energy of the flux tube does not depend on the external magnetic field. In this case, it is useful to write Eq.\ (\ref{Xictube}) in the form
\be
\frac{H_{c1}}{\mu^2/\sqrt{\lambda}} = -\frac{(3g^2+e^2){\cal I}_\circlearrowleft}{6e\sqrt{\lambda}n_3}= -\frac{(3g^2+e^2)}{6e\sqrt{\lambda}n_3}\int_0^\infty dR\, R\left(\frac{1-f_3^4}{2}+\lambda\frac{\tilde{a}_8'^2}{R^2}\right)  \, , 
\ee 
where now the right-hand side directly yields the critical magnetic field.

\section{Domain Walls in 2SC}
\label{sec:domain}

The profiles of the flux tubes from the previous subsection approach the 2SC$_\mathrm{ ud}$ phase at infinity. We know that in the massless limit considered here the 2SC$_\mathrm{ us}$ phase is equivalent to the 2SC$_\mathrm{ ud}$ phase. Therefore, we can construct a 
domain wall that
approaches 2SC$_\mathrm{ us}$ far away from the wall on one side and 2SC$_\mathrm{ ud}$ on the other side. It is conceivable that the "twist" 
that changes 2SC$_\mathrm{ us}$ into 2SC$_\mathrm{ ud}$ admits a magnetic field in the wall, which leads to a gain in Gibbs free energy 
and might favor the domain wall over the homogeneous phase in the presence of an externally applied field. We shall see that 
this is indeed the case and that, in a certain parameter regime,  the domain wall solution is 
favored over the flux tubes from the previous subsection. 

Domain walls in the 2SC phase in the presence of a magnetic field were already suggested in Ref.\ \cite{Son:2007ny}. These domain 
walls are associated with the axial $U(1)_A$. This symmetry is broken due to the axial anomaly 
of QCD, but becomes an approximate symmetry at high density and is spontaneously broken by the 2SC condensate. These domain walls are perpendicular to the magnetic field 
and their width is given by the inverse of the mass of the $U(1)_A$ pseudo-Goldstone boson. This is different from the domain walls discussed here, which align themselves parallel to the magnetic field and which have finite width even though our potential does not 
include $U(1)_A$ breaking terms. The "anomalous" 
domain walls have been discussed within an effective Lagrangian for the Goldstone mode \cite{Son:2007ny}, and it would be interesting
for future work to investigate their competition or coexistence with the domain walls discussed here in a common framework.

The equations that have to be solved to compute the profile of 
the domain wall are derived as follows. Due to the geometry of the problem, we  work in cartesian coordinates 
rather than the cylindrical coordinates
used for the flux tubes. We keep the external magnetic field in the $z$-direction and, without loss of generality,  place the domain wall in the $y$-$z$-plane, such that the problem becomes one-dimensional along the $x$-axis. 
For the gauge fields, our ansatz is
\bea
\tilde{\mathbf{A}}_3(\mathbf{r}) = \left[(x-x_0)H\cos\vartheta_1\sin\vartheta_2+ \sqrt{\lambda}\rho_\mathrm{ 2SC}\tilde{a}_3(x)\right]\mathbf{e}_y \, , \qquad 
\tilde{\mathbf{A}}_8(\mathbf{r})= \sqrt{\lambda}\rho_\mathrm{ 2SC} \tilde{a}_8(x)\mathbf{e}_y \, , 
\eea
such that the magnetic fields point in the $z$-direction with $z$-components
\be \label{BBB}
\tilde{B}_3 = H\cos\vartheta_1\sin\vartheta_2 +\lambda\rho_\mathrm{ 2SC}^2\tilde{a}_3' \, , \qquad 
\tilde{B}_8=\lambda\rho_\mathrm{ 2SC}^2\tilde{a}_8' \, , 
\ee
where prime now denotes the derivative with respect to the dimensionless coordinate $X\equiv \sqrt{\lambda}\rho_\mathrm{ 2SC}\, x$. 
We have added an $x$-independent term proportional to $x_0$ to the gauge field $\tilde{\mathbf{A}}_3$. This term is irrelevant for the 
magnetic field and does not affect any physics. It is merely a useful term for the numerical evaluation because it can be used to shift
 the location of the domain wall on the $x$-axis. Since this location depends on the values of the parameters, we conveniently 
 adjust $x_0$ to keep the domain wall in the $x$-interval which we have chosen for the numerical calculation.

We set $\rho_1=0$ and introduce the dimensionless condensates as above through $\rho_i(\mathbf{r})=f_i(x)\rho_\mathrm{ 2SC}$ for $i=2,3$. 
As just explained, the phases of the condensates do not wind as we move across the wall, and thus we set $\psi_i=0$. 
One could define a new angle $\alpha$ by writing $f_1= f \cos\alpha$, $f_2= f \sin\alpha$ and solve the equations of motion for $f$ and $\alpha$, see Ref.\ \cite{Chernodub:2010sg} for a similar calculation in a two-component superconductor. This angle, which 
rotates between the two condensates, \textit{does} wind across the domain wall. 
But this change of basis is not necessary, and we
 shall stick to the variables $f_1$, $f_2$.  Then, from Eq.\ (\ref{U02SC}) we compute the Gibbs free energy density 
\bea
U-\mathbf{H}\cdot \mathbf{B} &=& U_\mathrm{ 2SC} -\frac{H^2\cos^2\vartheta_1}{2} - \lambda\rho_\mathrm{ 2SC}^2H\sin\vartheta_1\,\tilde{a}_8' 
+ \frac{\lambda\rho_\mathrm{ 2SC}^4}{2}\Big\{\lambda(\tilde{a}_3'^2+\tilde{a}_8'^2)+f_2'^2+f_3'^2 \non[2ex]
&&+\left[{\cal M}_2-2\Xi(X-X_0)\right]^2f_2^2+{\cal M}_3^2f_3^2-f_2^2-f_3^2+\frac{1}{2}(f_2^4+f_3^4)+\frac{1}{2}-\eta f_2^2f_3^2 \Big\} \, , 
\eea
with $\Xi$ from Eq.\ (\ref{Xi}), $X_0\equiv \sqrt{\lambda}\rho_\mathrm{ 2SC} \,x_0$, and
\be
{\cal M}_2\equiv -\tilde{q}_3\tilde{a}_3+\tilde{q}_{82}\tilde{a}_8 \, , \qquad 
{\cal M}_3\equiv \tilde{q}_{83}\tilde{a}_8 \, . 
\ee
The equations of motion are
\begin{subequations}
\bea
\tilde{a}_3'' &=& -\frac{\tilde{q}_3}{\lambda}\left[{\cal M}_2-2\Xi(X-X_0)\right]f_2^2 \, , \label{aX}\\[2ex]
\tilde{a}_8''&=& \frac{\tilde{q}_{82}}{\lambda} \left[{\cal M}_2-2\Xi(X-X_0)\right]f_2^2 +\frac{\tilde{q}_{83}}{\lambda}{\cal M}_3 f_3^2 \, ,
\eea
\end{subequations} 
and
\begin{subequations}
\bea
0&=& f_2'' + f_2\left\{1-f_2^2-\left[{\cal M}_2-2\Xi(X-X_0)\right]^2+\eta f_3^2\right\} \, , \label{f2X}\\[2ex]
0&=& f_3'' + f_3\left(1-f_3^2-{\cal M}_3^2+\eta f_2^2\right) \, . \label{f3X}
\eea
\end{subequations}
The boundary conditions are determined as follows. On one side far away from the domain wall, say at $X=+\infty$,  we put the 
2SC$_\mathrm{ ud}$ phase, while on the other side, at $X=-\infty$, we put the 2SC$_\mathrm{ us}$ phase. Then, the boundary conditions for the scalar fields are $f_2(+\infty)=f_3(-\infty)=0$ and $f_2(-\infty)=f_3(+\infty)=1$.
For the boundary conditions of the gauge fields we need the magnetic fields of the two phases far away from the wall 
(\ref{B3B823}) to find
\bea \label{bound2SC2}
\tilde{a}_3'(-\infty) &=& -\frac{4\tilde{q}_3\Xi}{g^2} \, , \qquad \tilde{a}_8'(-\infty) = \frac{6\Xi}{\sqrt{3g^2+e^2}} \, ,  \qquad 
\tilde{a}_3'(+\infty)=\tilde{a}_8'(+\infty)=0 \, .
\eea
Here the external field $H$ appears inevitably in the boundary conditions (in its dimensionless version $\Xi$),
while this was avoided in the case of the flux tubes by separating the $H$-dependent part in the ansatz for $\tilde{\mathbf{A}}_3$.
In addition to the boundary conditions for the derivatives, we have $\tilde{a}_8(+\infty)=0$, which follows from evaluating 
Eq.\ (\ref{f3X}) at $X=+\infty$. All other boundary values of the gauge fields must be determined dynamically.

The Gibbs free energy density becomes
\bea \label{GVdomain}
\frac{G}{V}&=&U_\mathrm{ 2SC}-\frac{H^2\cos^2\vartheta_1}{2}+\frac{A_{yz}}{V}\frac{\sqrt{\lambda}\rho_\mathrm{ 2SC}^3}{2}{\cal I}_{||} \, , 
\eea
where $A_{yz}$ is the area of the system in the plane of the domain wall, and the dimensionless energy per unit area of the domain wall is, after partial integration and using the equations of motion, 
\be
{\cal I}_{||}  \equiv \int_{-\infty}^\infty dX \left[\lambda(\tilde{a}_3'^2+\tilde{a}_8'^2)
-4\lambda\Xi\frac{\tan\vartheta_1}{\tilde{q}_3\sin\vartheta_2}\tilde{a}_8' +\frac{1}{2}(1-f_2^4-f_3^4)+\eta f_2^2f_3^2\right] \, .
\ee
As a check, we confirm that the integrand goes to zero at $X\pm\infty$: the contribution of the scalar fields is obviously zero 
at $X=\pm\infty$ because one of the two functions $f_2$ and $f_3$ goes to 0 and the other one to 1. The gauge field contribution at $X=+\infty$ is obviously zero  because all derivatives  $\tilde{a}_3'$, $\tilde{a}_8'$ vanish. At $X=-\infty$, we employ the 
boundary conditions from Eq.\ (\ref{bound2SC2}) to show that the contributions quadratic in the derivatives of the gauge field 
are exactly canceled by the term proportional to $\tilde{a}_8'$. This term comes from the
$\mathbf{H}\cdot \mathbf{B}$ term in the Gibbs free energy and was written separately in the flux tube energies in the previous sections, see for instance Eq.\ (\ref{G2SCflux}). Since here, in the case of the domain walls,  this would have required writing down a 
divergent integral [with the divergence being canceled by the divergent $\tilde{a}_8(-\infty)$],  
we have included the term linear in $\tilde{a}_8'$ into the integral.

\section{Numerical Results and Profile Functions}
\label{sec:profiles}

\begin{figure} [t]
\begin{center}
\hbox{\includegraphics[width=0.5\textwidth]{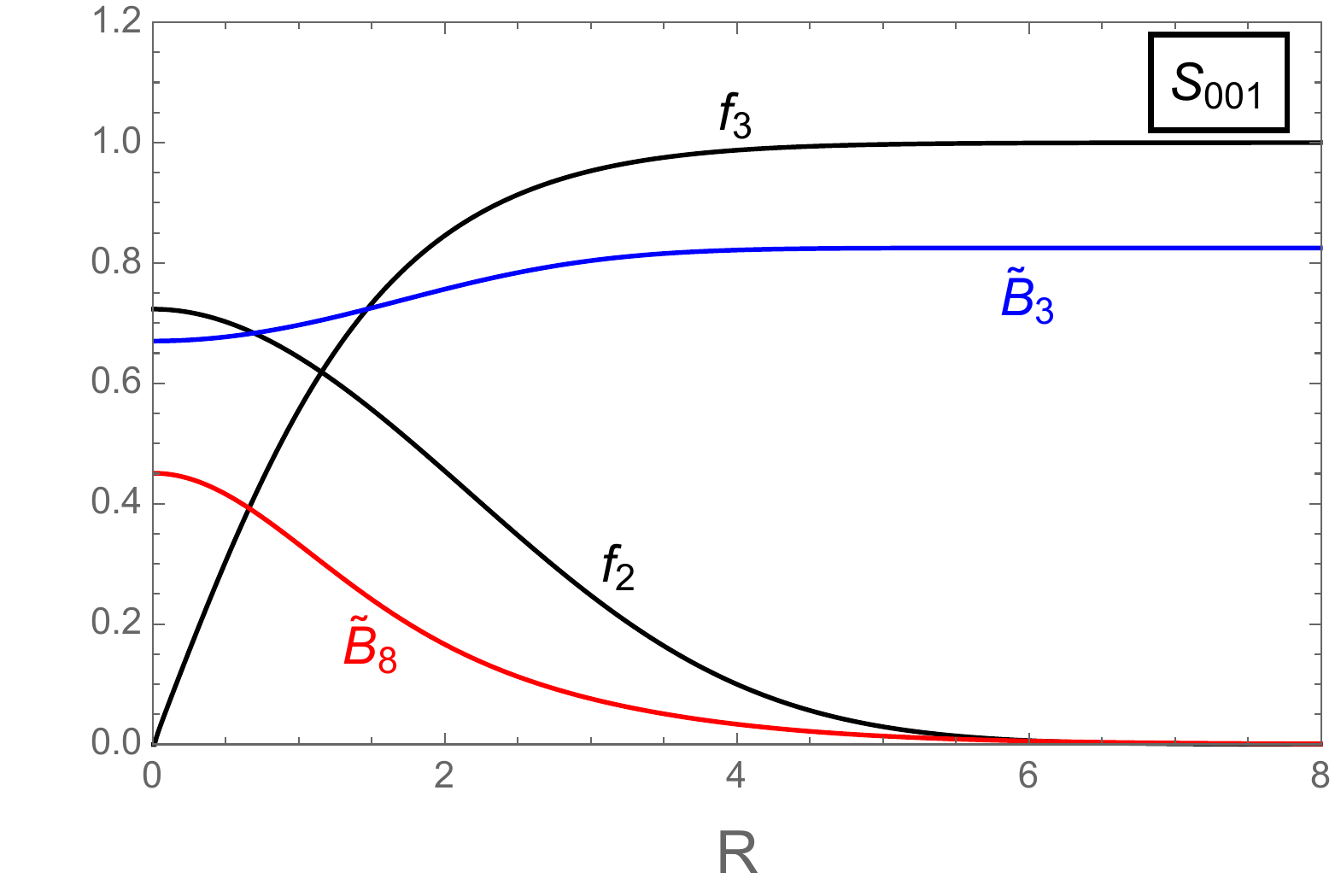}
\includegraphics[width=0.5\textwidth]{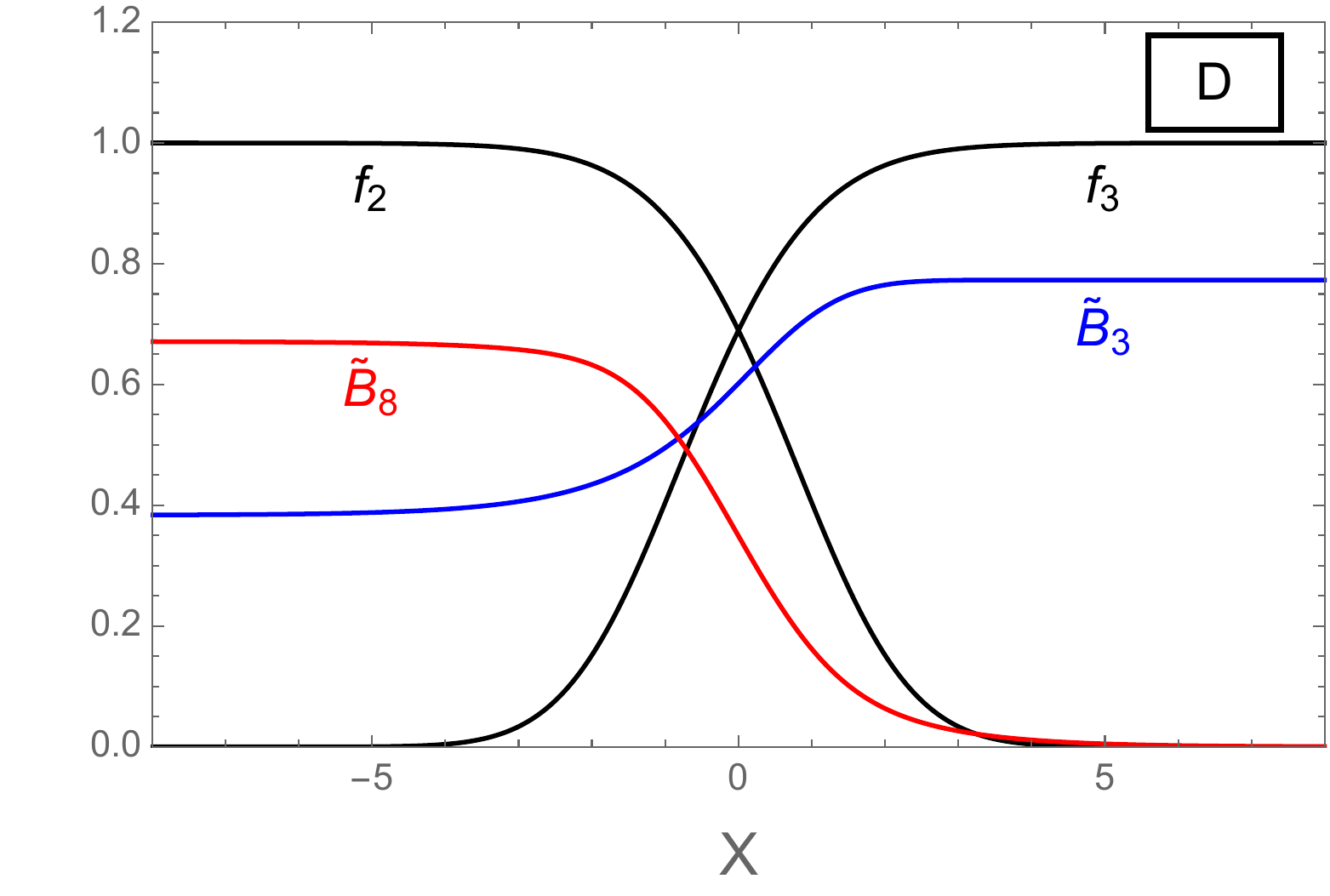}}
\caption{Profiles of the dimensionless condensates 
$f_2$, $f_3$ and the magnetic fields $\tilde{B}_3$, $\tilde{B}_8$
in units of $\mu^2/\sqrt{\lambda}$ for a 2SC flux tube (left panel, with winding number $n_3=-1$) and a 2SC domain wall (right panel). The parameters for both panels are $g=3.5$, $T_c/\mu_q\simeq 0.084$,  and the profiles are plotted at their respective critical fields $H_{c1}(S_{001})=9.59\,\mu^2/\sqrt{\lambda}$ (left) and $H_{c1}(D)=8.99\,\mu^2/\sqrt{\lambda}$ (right), see also Fig.\ \ref{fig:windings}. The dimensionless radial coordinate for the flux tube is $R=r\sqrt{\lambda}\rho_\mathrm{ 2SC}$, and the 
dimensionless cartesian coordinate $X$ for the domain wall is $X=x\sqrt{\lambda}\rho_\mathrm{ 2SC}$. We have placed the center of the 
domain wall, where $f_2=f_3$,  at the arbitrarily chosen point $X=0$.}
\label{fig:2SC}
\end{center}
\end{figure}

\begin{figure} [t]
\begin{center}
\hbox{\includegraphics[width=0.5\textwidth]{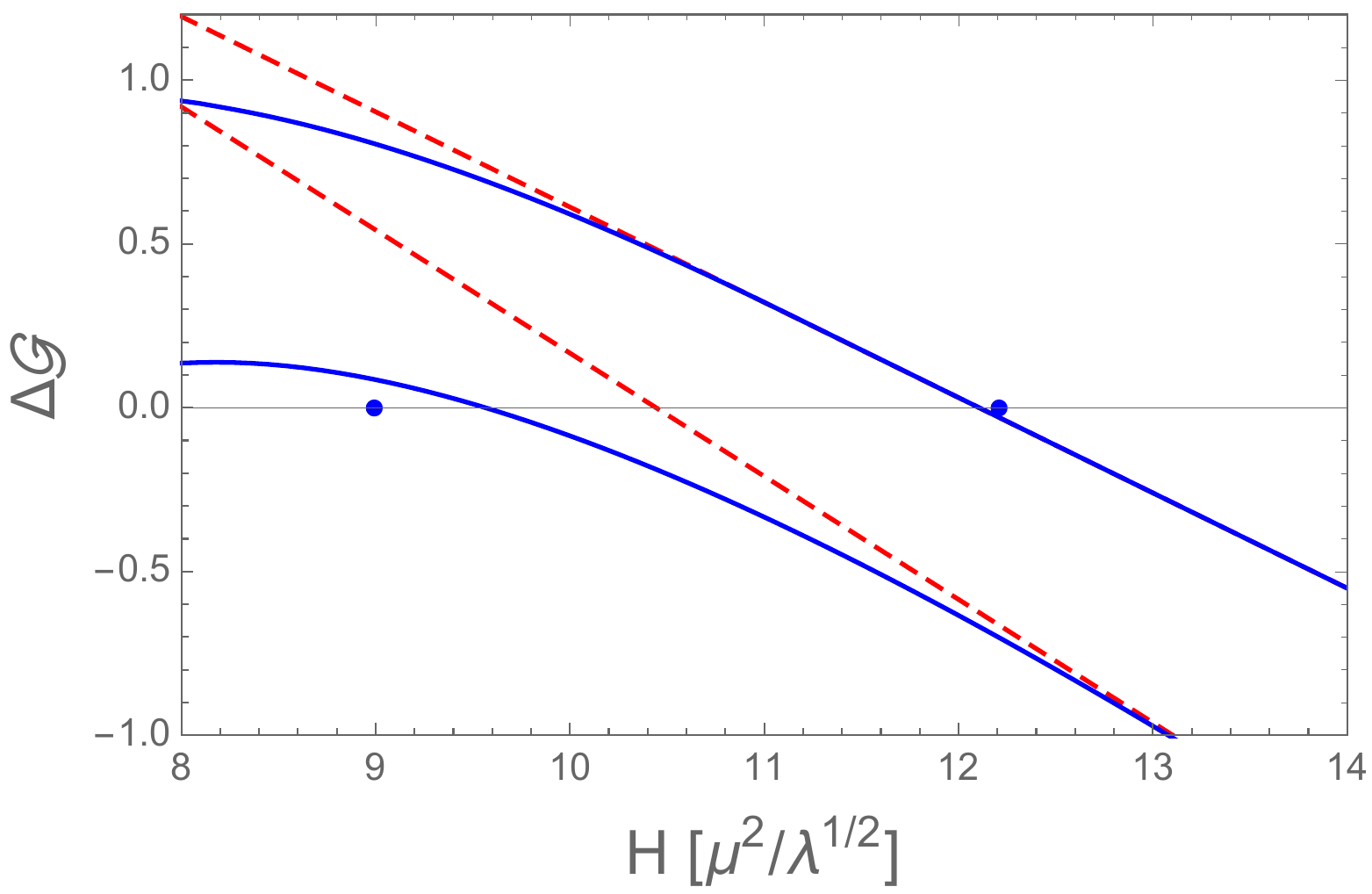}\includegraphics[width=0.5\textwidth]{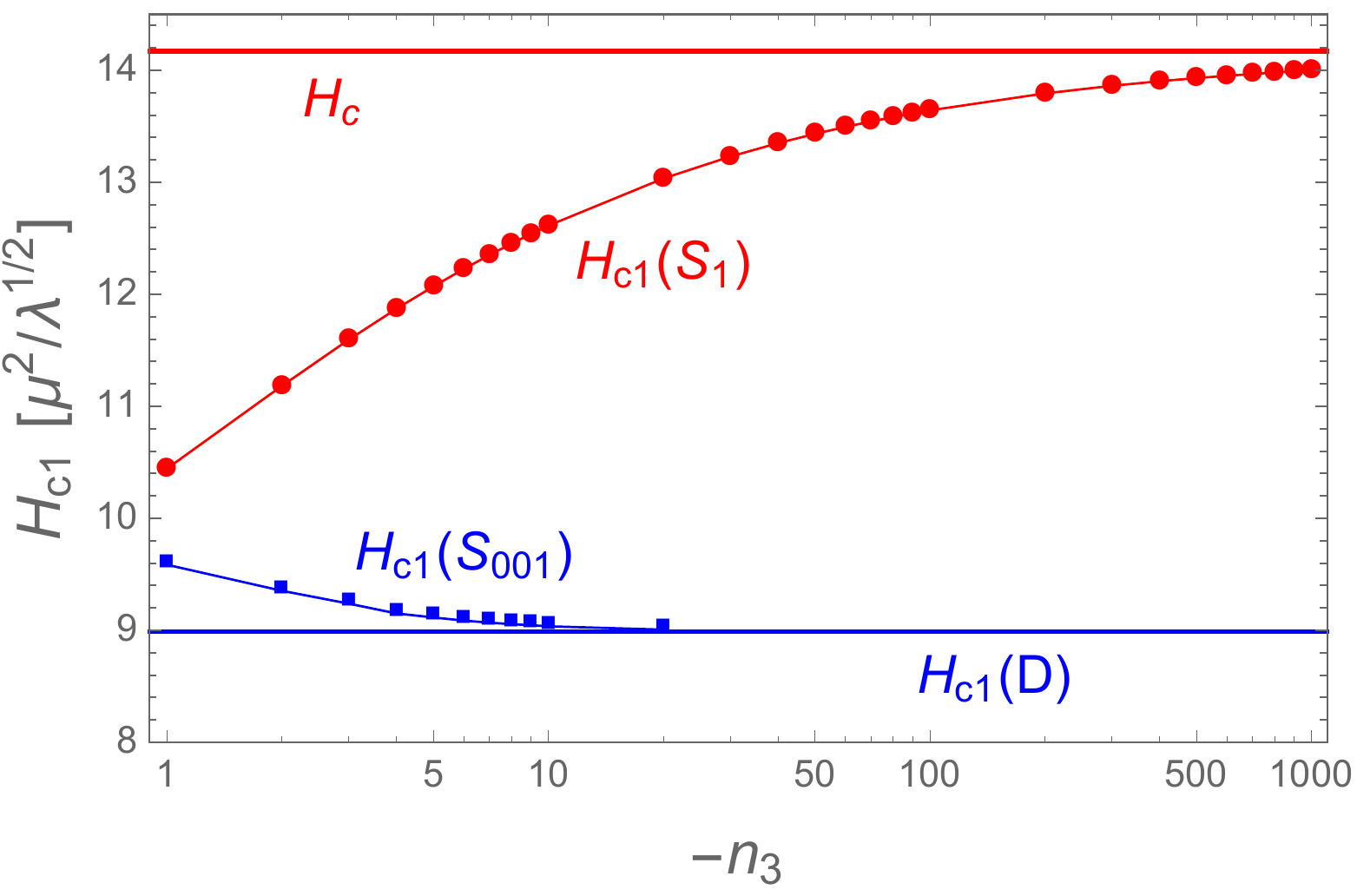}}
\caption{\textit{Left panel:} Gibbs free energy difference per unit length, in units of $\rho_\mathrm{ 2SC}^2$, between the phase with a single flux tube and the homogeneous 2SC phase ($\Delta{\cal G}=0$ defines the critical field $H_{c1}$). The solid (blue) lines
are the curves for flux tubes with an induced second condensate in the core $S_{001}$, dashed (red) lines correspond to standard flux tubes $S_1$. The lower pair of curves is computed at $T_c/\mu_q=0.084$, the upper pair at $T_c/\mu_q=0.065$, both for $g=3.5$ and winding $n_3=-1$. The two dots indicate the critical fields of the domain wall. \textit{Right panel:} Critical magnetic fields $H_{c1}$ for different winding numbers $n_3$ for $S_1$, $S_{001}$, and the domain wall $D$ for $T_c/\mu_q\simeq 0.084$, $g=3.5$. For large winding numbers, $H_{c1}(S_1)$ approaches $H_c$ from below, indicating ordinary type-II behavior, while $H_{c1}(S_{001})$ approaches  the critical field for the formation of domain walls from above. The thin lines connecting the data points are to guide the eye, only integer values of $n_3$ make sense.}
\label{fig:windings}
\end{center}
\end{figure}

The profiles for a 2SC flux tube and a 2SC domain wall are shown in 
Fig.~\ref{fig:2SC}. For all flux tube solutions discussed in the following, the winding numbers of the components that vanish far away from the flux tube are set 
to zero, $n_1=n_2=0$. A check for some selected parameter sets shows that nonzero $n_1$ and/or $n_2$ give rise to less preferred configurations, which is expected because in this case $f_1$ and/or $f_2$ must vanish in the center of the tube and can only become nonzero in an intermediate radial regime.
The left panel of the figure shows a flux tube in which one additional condensate, namely $\rho_2$, is induced in the core. 
In principle, it is possibly to find parameter regions which allow for solutions where both $\rho_1$ and $\rho_2$ become nonzero in the center of the flux tube. However, a parameter region where it is energetically favorable to place a flux tube with three nonzero condensates into the homogeneous state was not found. We shall thus ignore these configurations from now on. The configuration with two nonzero condensates, on the other hand, can become favorable over the homogeneous phase. This is shown in the left panel of Fig.~\ref{fig:windings}, where the dimensionless Gibbs free energy difference between the phase with a single flux tube and the homogeneous phase is plotted,
\be
\Delta{\cal G} \equiv \frac{G-G_{\mathrm{ 2SC}_\mathrm{ ud}}}{\rho_\mathrm{ 2SC}^2 L} = \pi\left({\cal I}_\circlearrowleft +\frac{8\lambda\Xi n_3}{g^2}\right) \, ,
\ee
with $G$ from Eq.~(\ref{G2SCflux}) and $G_{\mathrm{ 2SC}_\mathrm{ ud}}$ from Eq.\ (\ref{Gudhom}). The two pairs of curves show one example where the configuration with an induced condensate in the core is preferred at the point where $\Delta{\cal G}=0$ over the standard flux tube solution $S_1$, and one example where there is only a single condensate at $\Delta{\cal G}=0$. In the former case, it turns out that the system can further reduce its free energy by 
replacing $S_{001}$ with a domain wall, whose critical field $H_{c1}(D)$ is determined by solving ${\cal I}_{||}=0$ numerically for $\Xi$. This critical field is indicated in the left panel of Fig.~\ref{fig:windings} by a dot for both cases: $H_{c1}(D)<H_{c1}(S_{001}) < H_{c1}(S_1)$ for $T_c/\mu_q=0.084$, and $H_{c1}(S_1)<H_{c1}(D)$ for $T_c/\mu_q=0.065$.
The connection between the flux tube $S_{001}$ and the 
domain wall can be understood with the help of the right panel of Fig.~\ref{fig:windings}. Let us first explain the upper two (red) curves in this plot, which 
show the standard behavior of an ordinary type-II superconductor: the most favorable configuration is a flux tube with minimal winding number, and as we increase the winding, the 
critical field $H_{c1}$ approaches the critical field $H_c$ from below (in a type-I superconductor, it would approach it from above). This is easy to understand: as the winding is increased, the core of the flux tube becomes larger and thus the normal phase "eats up" the superconducting phase. Hence, for infinite winding, the critical field $H_{c1}$ indicates that it has now become favorable to place an infinitely large flux tube into the system, i.e., to replace the superconducting phase with the normal phase, which is nothing but the definition of $H_c$.
Similarly, the critical field for the flux tube $S_{001}$ 
approaches the critical field for the domain wall $D$: again, as we increase the winding, the phase in the core, which now approaches the 2SC$_\mathrm{ us}$ phase 
for $|n_3|\to\infty$, spreads out and "eats up" the phase far away from the flux tube, which is the 2SC$_\mathrm{ ud}$ phase. However, in contrast to the ordinary flux tube $S_1$, these two phases have the same free energy for all parameter values (in the massless limit), and there can never be a well-defined transition in the phase diagram from the homogeneous 2SC$_\mathrm{ us}$ phase to the 
homogeneous 2SC$_\mathrm{ ud}$ phase. Instead, we find that a stable domain wall forms, which interpolates between the two phases. 
While Figs.~(\ref{fig:2SC}) and (\ref{fig:windings}) only show results for specific parameters, we 
study the phase diagram more systematically in the next section.

\chapter{Phase Diagrams}
\label{chap:results_color}

\begin{figure} [t]
\begin{center}
\hbox{\includegraphics[width=0.51\textwidth]{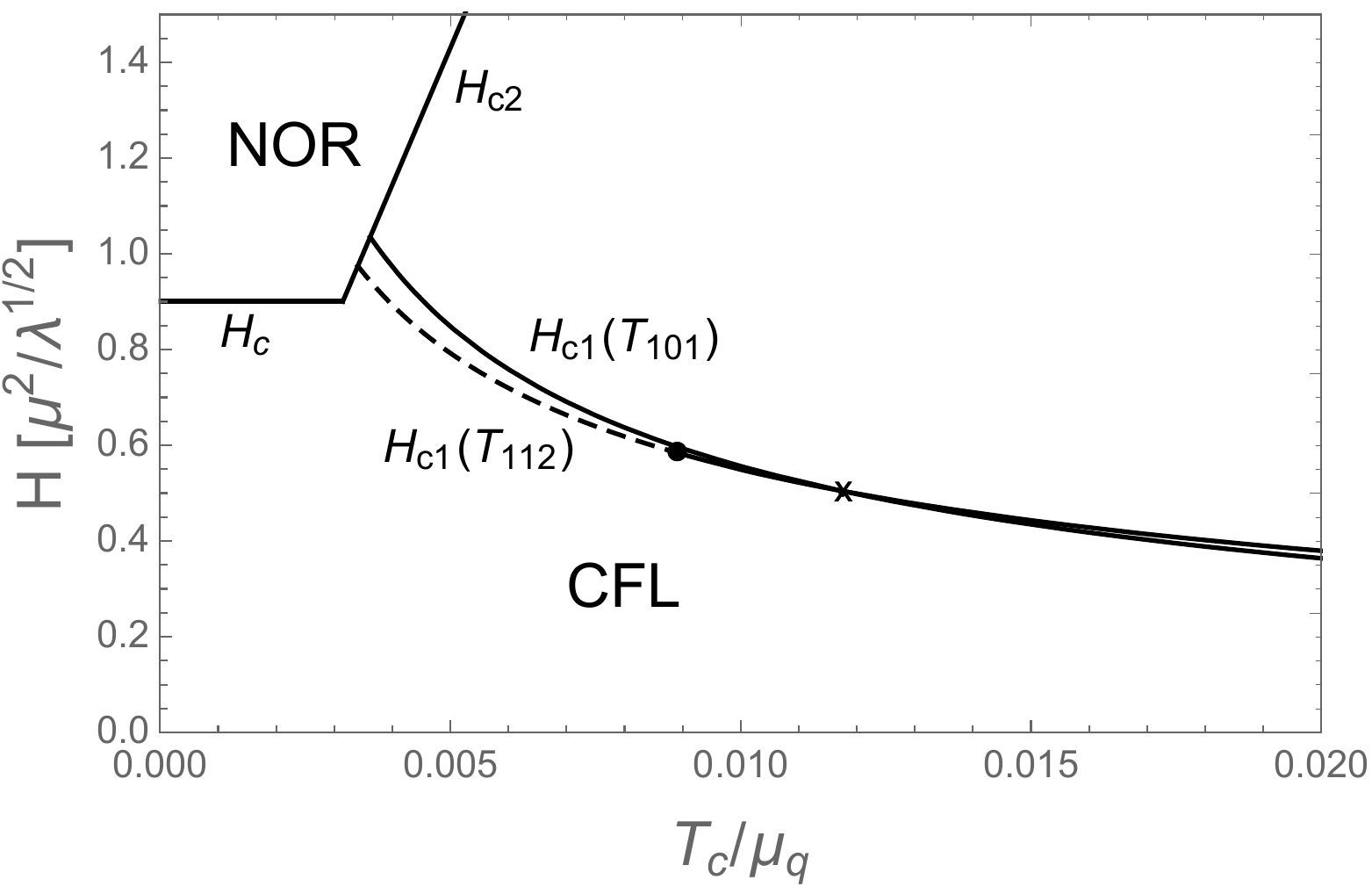}\includegraphics[width=0.49\textwidth]{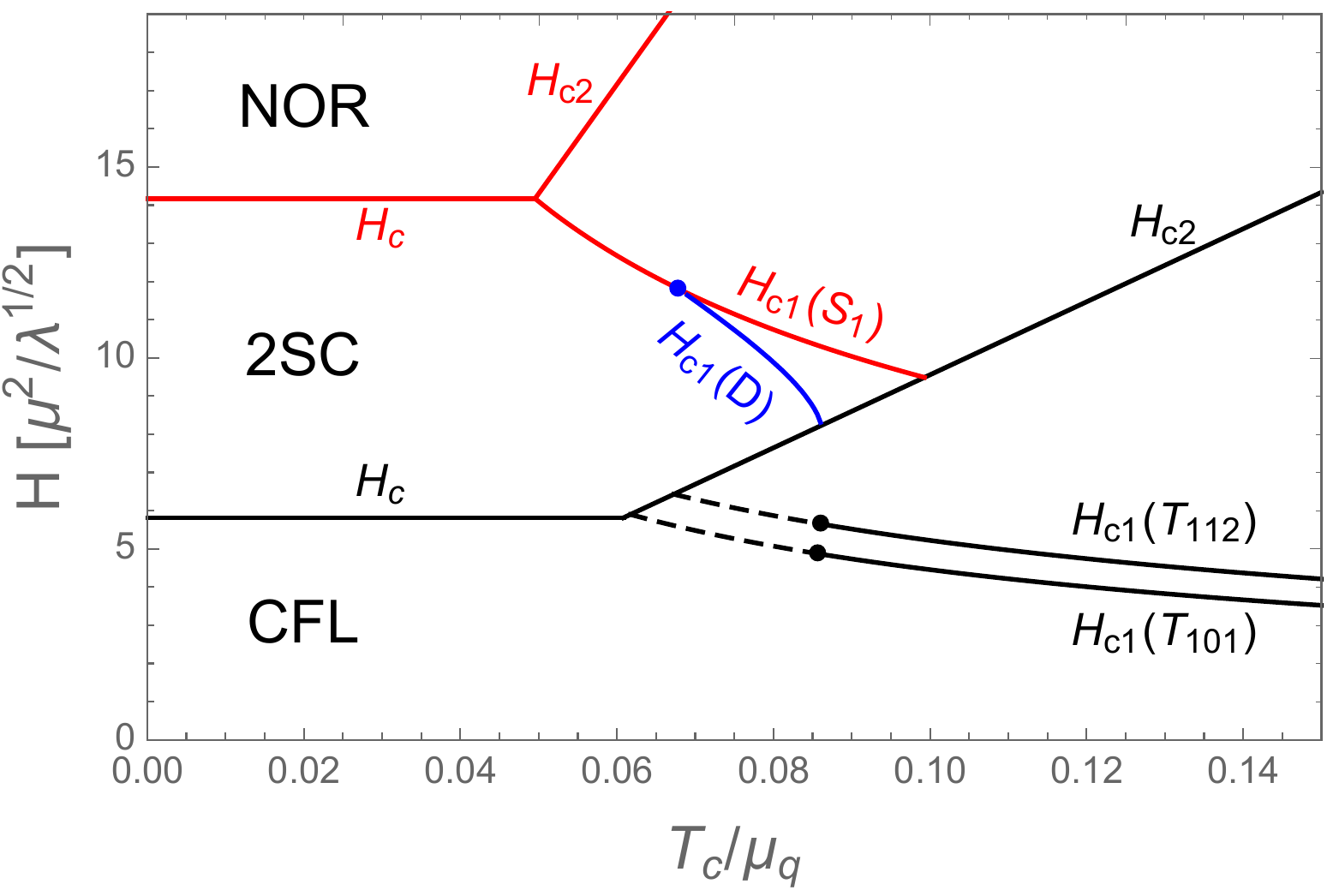}}
\caption{Critical magnetic fields for $g=0.1$ (left panel) and $g=3.5$ (right panel). For weak coupling, the 
CFL flux tube with a 2SC core $T_{101}$ is preferred over the flux tube with an unpaired core $T_{112}$  only for large 
values of $T_c/\mu_q$, while for strong coupling
this is the case for all $T_c/\mu_q$ in the type-II regime. The dots separating the 
dashed from the solid segments in $H_{c1}$ mark the transition from attractive (dashed) to repulsive (solid) long-range interaction between the flux tubes. The point marked with a cross in the left panel is  the intercept $H_{c1}(T_{101})=H_{c1}(T_{112})$. At strong coupling, where the 2SC phase appears for large magnetic fields, the 2SC domain wall $D$ is preferred over the "standard" 2SC flux tube $S_1$ for $T_c/\mu_q \gtrsim 0.07$.}
\label{fig:phases}
\end{center}
\end{figure}

Putting together the results of the previous chapters, the magnetic phase structure of color-superconducting quark matter  in the $H$-$T_c/\mu_q$-plane is shown  in Fig.~\ref{fig:phases}. The figure includes all three critical magnetic fields: $H_c$, 
indicating a first-order phase transition between homogeneous phases; $H_{c2}$, the lower boundary for the transition of a flux tube phase to a homogeneous phase; and $H_{c1}$, the field at which the system starts to form magnetic defects. 

As we have seen in Chap.~\ref{chap:hom}, for small couplings $g$ the CFL phase is directly superseded by the NOR phase as we increase the magnetic field, while the 2SC phase appears as an intermediate phase for couplings 
$g>2e/\sqrt{15}$. One example for either case is shown here, with the larger coupling chosen such that it is realistic for the interior of neutron stars (there was not any qualitative difference for other values of $g$ as long as
$g>2e/\sqrt{15}$ found). In a single-component superconductor, the critical lines $H_c$, $H_{c1}$, and $H_{c2}$ intersect in a single point, 
which marks the transition from type-I to type-II behavior, and in the type-II regime a lattice of flux tubes is expected between 
$H_{c1}$ and $H_{c2}$. This standard scenario is realized for the 2SC phase, see the intersecting (red) critical lines $H_c$, $H_{c1}(S_1)$, and $H_{c2}$ in the right panel. CFL, however, is a three-component superconductor and thus the transition region between 
type-I and type-II behavior is more complicated, see the (black) transition lines $H_c$, $H_{c1}(T_{101})$, $H_{c1}(T_{112})$,
and $H_{c2}$ in both panels which do not intersect in a single point. This was expected from our investigation of the two-component system earlier on. Along the dashed segments of the transition lines $H_{c1}$, the long-range interaction between the flux tubes is attractive, see Sec.~\ref{sec:inter}, and in this regime one expects a
first-order phase transition at some $H<H_{c1}$ as it was found in the neutron-proton system. For small coupling, the change from repulsive to 
attractive interaction occurs at different points for the $T_{101}$ and $T_{112}$ configurations (in the left panel, the $T_{101}$ tubes 
interact repulsively throughout the type-II regime). These points become identical for $g\gg e$, as we can see in the right panel and 
in Eq.~(\ref{repul}). The precise structure of this type-I/type-II transition region is not the main point of this discussion. Our investigation of the neutron-proton system suggests that flux tubes in CFL are possible also for values of $T_c/\mu_q$ smaller than indicated by the intercept of $H_c$ and $H_{c2}$. For our purpose, the main point is that for sufficiently large $T_c/\mu_q$, such that the interaction between flux tubes at long distances is repulsive,
we are in a "standard" type-II regime, and the onset of flux tubes occurs in a second-order transition.  It is this region in which 
we can compare the different critical fields $H_{c1}$ to obtain the energetically most preferred magnetic defect. 

Another complication arises in the right panel. We recall that, usually, $H_{c2}$
is the lower bound (assuming a second-order transition) for the transition of the flux tube phase to the normal-conducting phase. This 
is unproblematic in the case of the 2SC/NOR transition (upper $H_{c2}$ in the right panel). The lower $H_{c2}$ marks the transition from a CFL flux tube phase to a homogeneous 2SC phase. However, for sufficiently large $T_c/\mu_q$ we expect 2SC domain walls (or flux tubes) in the region above this $H_{c2}$. Therefore, although the curve for $H_{c2}$ is continued into the 
region of large $T_c/\mu_q$ for completeness, the actual phase transitions (possibly between different flux tube lattices or stacks of domain walls) are beyond the scope of the present approach. 

In summary, neither panel in Fig.~\ref{fig:phases} is a complete phase diagram and more complicated studies are necessary
to find all phase transition lines. But they serve the purpose to carefully locate the type-II regime where the main results presented here are valid:

\begin{itemize}
\item The CFL flux tube $T_{101}$ (which has a 2SC core) has a smaller critical magnetic field $H_{c1}$ than the flux tube $T_{112}$ (which has an unpaired core), unless the strong coupling constant is very small. This is equivalent to saying that the energy 
per unit length of $T_{101}$ is smaller. Although the configuration $T_{101}$ had never been discussed before in the literature, this result is not  surprising, because the "total winding" (for instance defined by the sum of the squares of the winding numbers $n_1$, $n_2$, $n_3$) is minimized by $T_{101}$ within the constraints of a nonzero $\tilde{B}_8$-flux and a vanishing baryon circulation. 

\item The 2SC domain wall, which interpolates between the two phases 2SC$_\mathrm{ us}$ and 2SC$_\mathrm{ ud}$, has a lower critical field $H_{c1}$ than the standard 2SC flux tube (in which two of the three condensates are identically zero) for sufficiently large $T_c/\mu_q$. Just like the flux tube, the domain wall admits 
additional $B$-flux into the system, which is the reason it can have a lower Gibbs free energy than the homogeneous phase. 

\end{itemize} 

\chapter{Summary: Color Superconductivity}
\label{chap:summary_color}

We have discussed magnetic defects -- flux tubes and domain walls -- in color-superconducting phases of dense quark matter,
using a Ginzburg-Landau approach. 
In a color superconductor, line defects can, in general, carry baryon circulation, magnetic flux, and color-magnetic flux.
We have focused  on the "pure" magnetic flux tubes, which have zero baryon circulation and thus are not induced by rotation. 
These flux tubes are not protected by topology, but can be stabilized by an external magnetic field. 
By solving the equations of motion numerically we have calculated the profiles of different kinds of flux tubes and their energy. As one of the main results, we have found a new type of CFL flux tube, which is most easily understood as a CFL flux tube with a 2SC core
(while the flux tube previously discussed in the literature has a core with unpaired quark matter). After carefully 
identifying the type-II regime, in which flux tubes are expected, we have seen that, for sufficiently large values of the strong coupling constant, the novel flux tube configuration has a smaller critical magnetic field 
than the flux tube with unpaired core. This result is supported by the observation that, in this strong-coupling regime,
 CFL is  superseded by 2SC as the magnetic field is increased, which makes the occurrence of CFL flux tubes with a 2SC core very 
plausible. (While, at small coupling, the CFL phase is superseded by the unpaired phase, and the flux tubes with unpaired core 
are favored.) Our new solution minimizes the total winding of the flux tube because one of the three condensates -- the one 
that survives in the 2SC phase -- has zero winding. The second main result presented is the discovery of magnetic domain walls in the 2SC phase, 
which emerge from 2SC flux tubes in the limit of infinite radius. The crucial ingredient, never included in the literature before, has been to allow for induced condensates in the core of the 2SC flux tubes. We have found that one of these induced condensates grows 
until it approaches the 2SC value, giving rise to a domain wall where the profiles of the condensates interpolate between two different versions of the 2SC phase. These two versions are distinguished by the pairing 
pattern ($us$ pairing vs.\ $ud$ pairing) and have the same free energy in the limit of massless quarks, in which we have worked throughout the investigation of color superconductivity. One might argue that in this limit the 2SC phase is not relevant anyway. As it was pointed out, however, the 2SC phase can be favored over the CFL phase not only if the strange quark mass is sufficiently large, but also in the case of a large magnetic field. Therefore, the 2SC domain walls do exist in a certain regime of the phase diagram, it is not necessary to artificially assume the 2SC phase to be the ground state.

Nevertheless, it would be an important extension  of the present calculation to include quark masses, and, in particular, study the fate of the 2SC domain walls in this more realistic setting.  It would also be interesting to study lattices of flux tubes or stacks of domain walls rather 
than the single, isolated magnetic defects that we have studied here. One step in this direction has been done by computing the long-distance
interaction between CFL flux tubes, but a full study of inhomogeneous phases would require more involved numerical calculations. 
It is tempting to speculate about the role of the CFL flux tubes discussed here in the interior of compact stars. If a rotating neutron star has 
a CFL core, flux tubes with nonzero baryon circulation must form, because this is how a superfluid accommodates rotation. 
Since it has been shown that color neutral vortices are disfavored, these flux tubes ("semi-superfluid vortices") have nonzero 
color-magnetic fluxes. Although the total color flux of three different semi-superfluid vortices is zero, particular arrangements of 
semi-superfluid vortices with nonzero total flux are conceivable (due to the mixing of photons and gluons, this also creates 
a nonzero flux with respect to the ordinary magnetic field). However, this would imply 
alignment of rotational and magnetic axes, which contradicts observations of pulsars because the pulsating signal that we 
observe results from the misalignment of rotation and magnetic field. One solution might be the coexistence of 
semi-superfluid vortices -- aligned with the rotational axis -- and the "pure" magnetic flux tubes considered here -- aligned with the
magnetic axis. The resulting core with CFL matter would be very complicated, not unlike a nuclear matter core where vortices from superfluid neutrons  and flux tubes from superconducting protons are expected to coexist. Another  question concerns the 
boundary between quark matter and hadronic matter. It has been discussed how the 
vortices and flux tubes of nuclear matter merge with semi-superfluid vortices \cite{Cipriani:2012hr,Alford:2018mqj,Chatterjee:2018nxe}, and it would be interesting to 
investigate this question for the non-rotational flux tubes, in particular for the flux tubes with 2SC core pointed out in this work,
 which carry an additional component of  
color-magnetic flux, on top of the flux from the rotated gluon field. Finally, it would be interesting to further investigate the influence of the color-magnetic flux tubes and domain walls on 
the emission of gravitational waves of neutron stars. Besides the already mentioned continuous emission due to color-magnetic mountains in the introduction, one could also imagine an effect 
of the color-magnetic flux tube lattice on the tidal deformability of neutron stars. This parameter is relevant for the gravitational wave emission of neutron star mergers \cite{PhysRevLett.119.161101} (similar to a possible effect of the crust of the star \cite{Penner:2011pd} or a crystalline quark matter phase in the core \cite{Lau:2017qtz}).

\part*{Conclusions and Outlook}
\addcontentsline{toc}{part}{Conclusions and Outlook}

The research carried out for this thesis is centered around the behavior of multicomponent superfluidity and superconductivity, where special emphasis was put on environments probably present in compact stars. After an introduction into these topics from a phenomenological and a technical point of view, three main topics were discussed in detail:
\begin{itemize}
\item Hydrodynamic instabilities in two-component (super-) fluids.
\item The type-I/type-II transition in a superconductor coupled to a superfluid.
\item Magnetic defects in color superconductivity.
\end{itemize}
For the first two points, a field-theoretical bosonic model of two scalar fields with quartic self-interactions and two types of interspecies coupling, a density coupling and a derivative coupling, was introduced. It was shown that the derivative coupling is responsible for the entrainment effect. Starting from this model, an effective, temperature dependent Ginzburg-Landau energy functional was derived. The Ginzburg-Landau free energy was used to derive the homogeneous phase structure at vanishing magnetic field in Sec.~\ref{sec:phases}. Various phase diagrams at finite superflow, finite temperature and as a function of the various coupling constants where plotted.
In order to examine \textbf{hydrodynamic instabilities} in a system of two superfluids, the hydrodynamic equations where derived. As an additional approach, the quasiparticle spectrum of the Goldstone mode was derived from the fluctuation propagator at tree-level. The hydrodynamic equations however allowed us to distinguish between superfluids and ideal, dissipationless but otherwise normal fluids, which allow for additional transverse modes. 
Whereas energetical instabilities manifested themselves by negative excitation energies, the onset of the two-stream instability was established to occur whenever the speed of sound becomes imaginary. It was found that for a two-component superfluid at zero temperature, this dynamical instability only occurs in an energetically unstable regime. However, for a system of two coupled normal fluids, the transverse modes can become unstable at velocities lower than the generalized version of Landau's critical velocity, which itself marks the onset of the energetic instability. These findings are of potential interest for the understanding of pulsar glitches. 

It is an open question whether the energetic instability can be resolved by the formation of an inhomogeneous superfluid, possibly through a phase separation of the two components. And, subsequently, it would be interesting to find out whether such an inhomogeneous state, if energetically stable, 
will still suffer a dynamical instability. This can be relevant not only for superfluids in the astrophysical context, but also for superfluids in the laboratory such as Bose-Fermi mixtures in cold atomic gases. 

By taking into account the charge of one of the scalar fields and including a corresponding gauge field, the influence of an external magnetic field on the two-component model was studied thoroughly. Especially the transition from type-I to type-II superconductivity was investigated. For this purpose, the critical magnetic fields of the coupled system, which represents a superconductor and a superfluid, were derived. This included a numerical computation of the single-flux tube energy. By including the interaction between the flux-tubes themselves in a sparse lattice approximation, a first order transition from the Meissner phase to a lattice of flux tubes and successively to the normal phase was discovered. It was therefore shown that the coupling to a superfluid has an immense influence on the magnetic phase structure of a superconductor. This does not only include a shift in the critical value of the Ginzburg-Landau parameter for the transition from type I to type II. In general, first-order phase transitions allow for the existence of mixed phases. By varying the averaged magnetic field in the superconductor, the possible existence of new mixed phases of flux tube cluster was demonstrated. All these effects might influence the formation of the magnetic field in the early stages of a compact star, where the critical magnetic fields become rather small due to the high temperatures present. It would be interesting to employ these results in studies of the time evolution of the magnetic field in a neutron star. Here, the ground state in equilibrium for given temperature, magnetic field and chemical potential has been computed, but for more phenomenological predictions one needs to know whether and on which time scale this ground state is reached. Simulations in Ref.~\cite{Ho:2017bia} based on calculations first performed in Ref.~\cite{baym_sfCS} yield very large time scales for the expulsion of the magnetic field, suggesting that a calculation at effectively fixed magnetic flux might be appropriate. One may also ask whether a potential phase of flux tube clusters would affect the transport properties of the core in a detectable way. 

In general, the phase structure at nonzero temperature should be studied in more detail. An improvement of the large-temperature approximation on which the Ginzburg-Landau potential was based is highly desirable, but probably requires technically more sophisticated approaches like the 2PI-formalism. Additionally, finite temperature effects add another fluid-component in the hydrodynamic treatment, given by the entropy current, which would complicate the problem drastically. It would nevertheless be promising to study dynamical instabilities in this case since for instance in superfluid neutron star matter temperature effects can be expected to be relevant. 

Although most parameters were chosen with having the physics of compact stars in mind, the described model is rather general. All the presented results can be transferred to the non-relativistic limit, where they are of potential interest for laboratory experiments as well.

Using the gained knowledge of multicomponent superconductivity, we turned our interest to \textbf{color superconductivity} in massless three-flavor quark matter. As a starting point, a Ginzburg-Landau type potential for color superconductivity consisting of three scalar fields and three (color-) gauge fields was derived. As in the two-component system, the basic phase structure was computed and revealed that the CFL phase can be superseded by the 2SC phase by increasing the magnetic field.
By using a general order parameter with separate winding numbers, it was possible to classify a large amount of magnetic defects in the color flavor locked phase known from literature \cite{Eto:2013hoa} by the triplet of winding numbers. Additionally, a new defect with minimized winding, so far unknown in the literature, was found. It was demonstrated that this new defect is energetically preferred in the parameter space expected in compact stars. An investigation of flux tubes in the 2SC phase with induced condensates in the core exposed the existence of a new defect in the 2SC phase: magnetic domain walls. Especially for the investigation of domain walls, taking into account the mass of the strange quark is a crucial next step. 

All the presented results provide a small contribution towards our understanding of dense matter in astrophysical environments. However, the models are still far from describing these physical systems accurately. At this point, the presented investigation serves as an improvement for the microscopic background of some astrophysical phenomena, like pulsar glitches and the magnetic evolution of a compact star. Obviously, fitting the model parameters, especially of the two-component model, to nuclear properties would allow us to extract numeric values for the onset of instabilities, the critical magnetic field strengths and location of the first-order phase transitions and so on. However, I have resisted this temptation in order to not overstress the accuracy of the current approach. The main goal of this thesis was to show how multicomponent superfluidity and superconductivity can be treated consistently within the approximations explained in detail. An improvement to this investigation would be the use of fermionic models instead, which might prove to be very difficult due to the more complicated energy spectrum for instance.
Within the current approach, investigating the effect of an external imposed rotation of the system appears to be the most intriguing option to render the model more realistic. The formation of vortices in the superfluid might alter our results drastically. In a first attempt, vortices parallel to the magnetic field could be investigated. However, due to the lighthouse effect, we know that in pulsars the axis of rotation and the axis of the magnetic field are not aligned. Vortices in nuclear matter can scatter on flux tubes, which unfortunately makes a general investigation of creeping vortices and flux tubes which are misaligned unfeasible within the semi-analytical approach presented in this thesis.
Also for CFL matter, including rotation can have profound effects on the presented results. Since CFL flux tubes can in principle carry angular momentum, one has to investigate the lowest lying magnetic defect at given magnetic field \textit{and} rotation. The lowest state might very well consist of separate magnetic and rotational defects, which would allow for the lighthouse effect. If it is rather found that CFL flux tubes with rotation and magnetic flux represent the true ground state under these conditions, their existence in compact stars might be questionable.

The expected results of the NICER mission and the upcoming age of gravitational-wave astronomy might allow us to make tremendous leaps in our understanding of dense nuclear and quark matter in the future. Hopefully within the next decades we can fill our gaps in our understanding of physics in these extreme environments and allow us to draw a much more accurate version of the QCD phase diagram.

\appendix
\part*{Appendix}
\addcontentsline{toc}{part}{Appendix}

\chapter{Ginzburg-Landau Equations of Motion}

Throughout this thesis, the equations of motions of a Ginzburg-Landau
style free energy are used. In order to explain the most important
steps I will derive them for the various cases in this appendix, such
that the flow of the corresponding chapters does not get interrupted
by lengthy derivations.

\section{Single Superconductor\label{app:single_SC}}

The equations of motion for a single, non-relativistic superconductor
from the Ginzburg-Landau free energy can be derived in a
rather straight forward manner. This appendix refers to Sec.~\ref{sec:Ginzburg-Landau-Theory}.
The free energy reads 
\begin{equation}
\frac{F_{GL}}{V}=\frac{F_{N}}{V}+\alpha\left|\vf\right|^{2}+\frac{\beta}{2}\left|\vf\right|^{4}+\frac{1}{2m_{s}}\left|\left(-i\mathbf{\nabla}-q_{s}\mathbf{A}\right)\vf\right|^{2}+\frac{\mathbf{B}^{2}}{8\pi}\,.
\end{equation}
The equations of motion can be derived by variation of the free energy
with respect to $\vf^{*}$ and the vector potential $\mathbf{A}$,
which leads to the Euler-Lagrange equations:
\begin{align}
\frac{\p F_{GL}}{\p\vf^{*}}-\p_{i}\frac{\p F_{GL}}{\p\left(\p_{i}\vf^{*}\right)} & =0\,,\\
\frac{\p F_{GL}}{\p A_{i}}-\p_{j}\frac{\p F_{GL}}{\p\left(\p_{j}A_{i}\right)} & =0\,.
\end{align}
The first variation is straightforward and directly leads to Eq.~(\ref{eq:GLeom1}):
\[
\alpha\vf+\beta\left|\vf\right|^{2}\vf+\frac{1}{2m_{s}}\left(-i\mathbf{\nabla}-q_{s}\mathbf{A}\right)^{2}\vf=0\,,
\]
where the square of the covariant derivative is a shorthand notation
for applying the differential operator including the vector field
two times, which also leads to derivatives of the gauge field. Calculating
the variation with respect to the vector field is best done in index
notation. We first focus in the pure magnetic field part and remember
that $\mathbf{B}=\nabla\times\mathbf{A}=\ve_{ijk}\p_{j}A_{k}$ .
\begin{align}
\mathbf{B}^{2} & =\ve_{imn}\p_{m}A_{n}\ve_{ijk}\p_{j}A_{k}\,,\\
\p_{a}\frac{\mathbf{B}^{2}}{\p\left(\p_{a}A_{b}\right)} & =\p_{a}\left(\ve_{imn}\delta_{a}^{m}\delta_{b}^{n}\ve_{ijk}\p_{j}A_{k}+\ve_{imn}\p_{m}A_{n}\ve_{ijk}\delta_{a}^{j}\delta_{b}^{k}\right)\nonumber \\
 & =2\ve_{iab}\p_{a}\ve_{ijk}\p_{j}A_{k}\nonumber \\
 & =-2\ve_{bai}\p_{a}\ve_{ijk}\p_{j}A_{k}=-2\nabla\times\mathbf{B}\,.
\end{align}
The variation of the second term containing the covariant derivative
is straight-forward since it contains no derivatives of the gauge
field and yields
\begin{equation}
\frac{q}{m}\text{Re}\left\{ \vf^{*}\left(i\mathbf{\nabla}+q_{s}\mathbf{A}\right)\vf\right\} \,.
\end{equation}
Putting these results together leads to
\begin{equation}
\frac{q}{m}\text{Re}\left\{ \vf^{*}\left(i\mathbf{\nabla}+q_{s}\mathbf{A}\right)\vf\right\} +\frac{1}{4\pi}\nabla\times\mathbf{B}=0\,,
\end{equation}
which is in agreement with Eq.~(\ref{eq:GLeom2}). Especially the
variation of the pure magnetic field term is of significance in several
chapters of this thesis.

\section{Single Flux Tube\label{App:single_FT}}

In this section I derive the equations of motion and the free energy
of a single flux tube for Sec.~\ref{subsec:flux_tubes}. We start
with the free energy from Ginzburg-Landau theory, subtract the contribution
of $F_{N}$ and insert our ansatz from Eq.~(\ref{eq:ft_ansatz})

\begin{align}
F_{\cl}+F_{0} & =L\int rdr\,d\theta\Bigg\{\alpha\rho_{0}^{2}f(r)^{2}+\frac{\beta}{8}\rho_{0}^{4}f(r)^{4}+\frac{1}{8\pi}\left(\frac{n}{q_{s}r}\frac{\p a(r)}{\p r}\right)^{2}\\
 & +\frac{\rho_{0}^{2}}{4m_{s}}\left|\left[-i\left(\mathbf{\p_{r}}\hat{\mathbf{e}}_{r}+\frac{1}{r}\p_{\theta}\mathbf{\hat{e}}_{\theta}\right)-\frac{na(r)}{r}\hat{\mathbf{e}}_{\theta}\right]f(r)e^{i\psi(\theta)}\right|^{2}\Bigg\}\,,\nonumber 
\end{align}
where we already performed the integration along the $z-$axis which
yields the length of the flux tube $L$ and $F_{0}$ is the purely
homogeneous solution which does not depend on the profile functions,
\begin{equation}
\frac{F_{0}}{V}=\frac{\alpha}{2}\rho_{0}^{2}+\frac{\beta}{4}\rho_{0}^{4}\,.
\end{equation}
As a first step we introduce dimensionless coordinates by scaling
the radial distance from the flux tube center by the coherence length
$\xi$,
\begin{equation}
R=\frac{r}{\xi},
\end{equation}
\begin{align}
F_{\cl}+F_{0} & =L\xi^{2}\int RdR\,d\theta\Bigg\{\frac{\alpha}{2}\rho_{0}^{2}f(R)^{2}+\frac{\beta}{8}\rho_{0}^{4}f(R)^{4}+\frac{1}{8\pi\xi^{4}}\left(\frac{n}{q_{s}R}\frac{\p a(R)}{\p R}\right)^{2}\\
 & +\frac{\rho_{0}^{2}}{4\xi^{2}m_{s}}\left|\left[-i\left(\mathbf{\p_{R}}\hat{\mathbf{e}}_{R}+\frac{1}{R}\p_{\theta}\mathbf{\hat{e}}_{\theta}\right)-\frac{na(R)}{R}\hat{\mathbf{e}}_{\theta}\right]f(R)e^{i\psi(\theta)}\right|^{2}\bigg\}\nonumber \\
 & =L\xi^{2}\int RdR\,d\theta\Bigg\{\frac{\alpha}{2}\rho_{0}^{2}f^{2}+\frac{\beta}{8}\rho_{0}^{4}f^{4}+\frac{1}{8\pi\xi^{4}}\left(\frac{n}{q_{s}R}a^{'}\right)^{2}\\
 & +\frac{\rho_{0}^{2}}{4\xi^{2}m_{s}}\left|\left[-if^{'}\hat{\mathbf{e}}_{R}+\frac{f}{R}\p_{\theta}\psi\mathbf{\hat{e}}_{\theta}-\frac{naf}{R}\hat{\mathbf{e}}_{\theta}\right]\right|^{2}\Bigg\}\,,\nonumber 
\end{align}
where prime denotes a derivative w.r.t.~the dimensionless coordinate
$R$ and we expressed the relevant part of the gradient in cylindrical
coordinates, 
\begin{equation}
\nabla=\p_{r}\hat{\mathbf{e}}_{r}+\frac{1}{r}\p_{\theta}\eh_{\theta}
\end{equation}
We now compute the EL-EOMs with respect to $f,$ $a$ and $\psi$.
For simplicity, we start with the equation of motion for the phase
$\psi(\theta)$, which reduces to 
\begin{equation}
\p_{\theta}^{2}\psi=0\qquad\Rightarrow\qquad\psi=a\theta+b\,,
\end{equation}
with numerical constants $a$ and $b$. Following the discussion in
Sec.~\ref{subsec:flux_tubes} on the flux quantization, we know
that the phase can only rotate by an integer multiple of $2\pi$ if
we go around the flux tube completely, therefore we can set $b=0$
w.l.o.g.~and identify $a$ with the winding number $n$,
\begin{equation}
\psi=n\theta,\qquad n\in\mathbb{N}_{0},
\end{equation}
with the set of natural numbers including zero $\mathbb{N}_{0}$.
This allows us to perform the integration w.r.t.~$d\theta$ which
simplifies the energy of the flux tube to 
\begin{align}
F_{\cl}+F_{0} & =\pi L\xi^{2}\int RdR\,\Bigg\{\alpha\rho_{0}^{2}f^{2}+\frac{\beta}{4}\rho_{0}^{4}f^{4}+\frac{1}{4\pi\xi^{4}}\left(\frac{n}{q_{s}R}a^{'}\right)^{2}\\
 & +\frac{\rho_{0}^{2}}{2\xi^{2}m_{s}}\left[\left(f^{'}\right)^{2}+\frac{n^{2}f^{2}}{R^{2}}\left(1-a\right)^{2}\right]\Bigg\}\,,\nonumber 
\end{align}
which we now use to perform the variation w.r.t.~$a$ and $f$:
\begin{align}
2\alpha\rho_{0}^{2}Rf+\beta\rho_{0}^{4}Rf^{3}+\frac{\rho_{0}^{2}}{\xi^{2}m_{s}}\left[\frac{n^{2}f}{R}\left(1-a\right)^{2}-Rf^{''}-f^{'}\right] & =0\,,\\
\frac{\rho_{0}^{2}}{m_{s}}\left[\frac{n^{2}f^{2}}{R}\left(1-a\right)\right]-\frac{n^{2}}{2\pi q_{s}^{2}\xi^{2}}\left(\frac{a^{''}}{R}-\frac{a^{'}}{R^{2}}\right) & =0\,.
\end{align}
In this step it is important to take the functional determinant of
the integral into account during the variation of the free energy.
We can further simplify things by dividing or multiplying both equations
with $R$ and using the definitions of $\xi$, $\ell$, $\kappa$
and $\rho_{0}^{2}=-\nicefrac{\alpha}{2\beta}$.

\begin{align}
f^{''}+\frac{f^{'}}{R}+f\left[\left(1-f^{2}\right)-\frac{n^{2}}{R^{2}}\left(1-a\right)^{2}\right] & =0\,,\\
a^{''}-\frac{a^{'}}{R} & =-\frac{f^{2}}{\kappa^{2}}\left(1-a\right)\,.
\end{align}
It is interesting to note that these equations are formally identical
to the relativistic version in our more complicated setup in Sec.~
taken from our Ref.~\cite{Haber:2017kth} if all couplings to the
second field and its contributions are set to zero. The free energy
of a single flux tube can now be simplified to 
\begin{equation}
\frac{F_{\cl}}{L}=\pi\rho_{0}^{2}\int_{0}^{\infty}dR\,R\left\{ \frac{n^{2}\kappa^{2}a^{'2}}{R^{2}}+f^{'2}+f^{2}\frac{n^{2}\left(1-a\right)^{2}}{R^{2}}+\frac{1}{2}\left(1-f^{2}\right)^{2}\right\} \,.
\end{equation}
\section{CFL Flux Tubes\label{App:CFL_FT}}
We calculate the equations of motion for flux tubes in the CFL phase presented in Chap.~\ref{chap:CFL}. 
Our starting point is the free energy from Eq.~(\ref{U123})
\bea
U &=&\frac{\tilde{B}^{2}}{2}+\frac{B_{3}^{2}}{2}+\frac{\tilde{B}_{8}^{2}}{2}+\left|\left(\nabla+i\gt\at_{8}+\frac{i}{2}gA_{3}\right)\phi_{1}\right|^{2}+\left|\left(\nabla+i\gt\at_{8}-\frac{i}{2}gA_{3}\right)\phi_{2}\right|^{2}\non[2ex]
&&+\left|\left(\nabla-2i\gt\at_{8}\right)\phi_{3}\right|^{2}-\mu^{2}\left(\left|\phi_{1}\right|^{2}+\left|\phi_{2}\right|^{2}+\left|\phi_{3}\right|^{2}\right)+\lambda\left(\left|\phi_{1}\right|^{4}+\left|\phi_{2}\right|^{4}+\left|\phi_{3}\right|^{4}\right)\non[2ex]
 &&-2h\left(\left|\phi_{1}\right|^{2}\left|\phi_{2}\right|^{2}+\left|\phi_{1}\right|^{2}\left|\phi_{3}\right|^{2}+\left|\phi_{2}\right|^{2}\left|\phi_{3}\right|^{2}\right)\,.\label{eq:potential}
\eea
In order to compute the equations of motion for general topological
defects we make the following ansatz in cylindrical coordinates,
\begin{align}
\phi_{i} & =\frac{1}{\sqrt{2}}\rho_{\mathrm{CFL}}f_{i}(r)e^{in_{i}\theta}\,,\\
\at_{8} & =a_{08}\frac{a_{8}(r)}{r}\hat{e}_{\theta}\,,\\
A_{3} & =a_{03}\frac{a_{3}(r)}{r}\hat{e}_{\theta}\,,
\end{align}
with the profile functions $f_{i}(r),$ the winding numbers $n_{i}$,
the asymptotic value for the condensate $\rho_{\mathrm{CFL}}$ from Eq.~(\ref{rho0}), and where $a_{08},\,a_{03}$ are kept general since they can be used to define the profile functions as dimensionless quantities, and the coordinate
angle $\theta$. Due to the high symmetry in the free energy concerning the
coupling constants, we can immediately deduce that the scalar fields
have the same asymptotic value, $\rho_{1}=\rho_{2}=\rho_{3}=\rho_{\mathrm{CFL}}$. In order to keep the following lengthy formulas more compact, we will omit the index CFL for $\rho_{\mathrm{CFL}}$ and simply write $\rho$, which is now static and constant. All the dynamics is captured in the profile functions.

Entering our ansatz for the scalar and gauge fields using the $\nabla$
and the curl operator (since $\mathbf{B}=\nabla\times\mathbf{A}$) in cylindrical coordinates\footnote{$\nabla f(r,\theta)=\partial_{r}f\hat{e}_{r}+\frac{1}{r}\partial_{\theta}f\hat{e}_{\theta}$
, $\nabla\times A=\frac{1}{r}\partial_{r}\left(rA\right)\hat{e}_{z}$ } yields
\bea
U &=&a_{03}^{2}\frac{\left(a_{3}^{'}\right)^{2}}{2r^{2}}+a_{08}^{2}\frac{\left(a_{8}^{'}\right)^{2}}{2r^{2}}+\frac{1}{2}\rho^{2}\left|f_{1}^{'}\hat{e}_{r}+\frac{in_{1}}{r}f_{1}\hat{e}_{\theta}+i\gt a_{08}\frac{a_{8}(r)}{r}\hat{e}_{\theta}f_{1}+\frac{i}{2}ga_{03}\frac{a_{3}(r)}{r}\hat{e}_{\theta}f_{1}\right|^{2}\nonumber \\[2ex]
 &&+\frac{1}{2}\rho^{2}\left|f_{2}^{'}\hat{e}_{r}+\frac{in_{2}}{r}f_{2}\hat{e}_{\theta}+i\gt a_{08}\frac{a_{8}(r)}{r}\hat{e}_{\theta}f_{2}-\frac{i}{2}ga_{03}\frac{a_{3}(r)}{r}\hat{e}_{\theta}f_{2}\right|^{2}\non[2ex]
 &&+\frac{1}{2}\rho^{2}\left|f_{3}^{'}\hat{e}_{r}+\frac{in_{3}}{r}f_{3}\hat{e}_{\theta}-2i\gt a_{08}\frac{a_{8}(r)}{r}\hat{e}_{\theta}f_{3}\right|^{2}\non[2ex]
 &&-\frac{\mu^{2}}{2}\rho^{2}\left(f_{1}^{2}+f_{2}^{2}+f_{3}^{2}\right)+\frac{\lambda}{4}\rho^{4}\left(f_{1}^{4}+f_{2}^{4}+f_{3}^{4}\right)-\frac{h}{2}\rho^{4}\left(f_{1}^{2}f_{2}^{2}+f_{1}^{2}f_{3}^{2}+f_{2}^{2}f_{3}^{2}\right)\,.\nonumber 
\eea
We now subtract $U_{\mathrm{CFL}}$ from Eq.~(\ref{UCFL}) from $U$ to get the free energy of a single
flux tube. (We could compute the EOM directly from $U$ as well, but
subtraction at this point facilitates the numerical calculation because
dimensional parameters are eliminated).
\be
U_{\circlearrowleft}=U-U_{0}\,,
\ee
which yields
\bea
U_{\circlearrowleft} &=&\frac{1}{2}\rho^{2}\Big(f_{1}^{'2}+\frac{n_{1}^{2}}{r^{2}}f_{1}^{2}+\gt^{2}a_{08}^{2}\frac{a_{8}^{2}}{r^{2}}f_{1}^{2}+\frac{g^{2}}{4}a_{03}^{2}\frac{a_{3}^{2}}{r^{2}}f_{1}^{2}+2\gt a_{08}\frac{a_{8}}{r^{2}}n_{1}f_{1}^{2}+ga_{03}\frac{a_{3}}{r^{2}}n_{1}f_{1}^{2}\non[2ex]
&&+ga_{03}\gt a_{08}\frac{a_{3}a_{8}}{r^{2}}f_{1}^{2}\Big)+\frac{1}{2}\rho^{2}\Big(f_{2}^{'2}+\frac{n_{2}^{2}}{r^{2}}f_{2}^{2}+\gt^{2}a_{08}^{2}\frac{a_{8}^{2}}{r^{2}}f_{2}^{2}+\frac{g^{2}}{4}a_{03}^{2}\frac{a_{3}^{2}}{r^{2}}f_{2}^{2}+2\gt a_{08}\frac{a_{8}}{r^{2}}n_{2}f_{2}^{2}\non[2ex]
&&-ga_{03}\frac{a_{3}}{r^{2}}n_{2}f_{2}^{2}-ga_{03}\gt a_{08}\frac{a_{3}a_{8}}{r^{2}}f_{2}^{2}\Big)+\frac{\left(a_{03}a_{3}^{'}\right)^{2}+\left(a_{08}a_{8}^{'}\right)^{2}}{2r^{2}}\non[2ex]
&&+\frac{1}{2}\rho^{2}\left(f_{3}^{'2}+\frac{n_{3}^{2}}{r^{2}}f_{3}^{2}+4\gt^{2}a_{08}^{2}\frac{a_{8}^{2}}{r^{2}}f_{3}^{2}-4\gt a_{08}\frac{a_{8}}{r^{2}}n_{3}f_{3}^{2}\right)+\frac{\mu^{2}}{2}\rho^{2}\left(3-f_{1}^{2}-f_{2}^{2}-f_{3}^{2}\right)\non[2ex]
&&-\frac{\lambda}{4}\rho^{4}\left(3-f_{1}^{4}-f_{2}^{4}-f_{3}^{4}\right)+\frac{h}{2}\rho^{4}\left(3-f_{1}^{2}f_{2}^{2}-f_{1}^{2}f_{3}^{2}-f_{2}^{2}f_{3}^{2}\right).
\eea
We now extremize the area integral of the free energy $2\pi\int dr\,rU$
w.r.t.\ the profile functions $f_{i}$ and $a_{i}$. The Euler-Lagrange
equations read
\begin{equation}
\frac{\partial\left(rU\right)}{\partial f_{i}}-\partial_{r}\frac{\partial(rU)}{\partial f_{i}^{'}}=0\,,
\end{equation}
where $f_{i}$ stands for all profile functions including the ones
of the gauge fields. This leads to five equations. 
\begingroup
\allowdisplaybreaks
\bea
\left(f_{1}^{''}+\frac{1}{r}f_{1}^{'}\right) &=&\frac{n_{1}^{2}}{r^{2}}f_{1}+\gt^{2}a_{08}^{2}\frac{a_{8}^{2}}{r^{2}}f_{1}+\frac{g^{2}}{4}a_{03}^{2}\frac{a_{3}^{2}}{r^{2}}f_{1}+2\gt a_{08}\frac{a_{8}}{r^{2}}n_{1}f_{1}+ga_{03}\frac{a_{3}}{r^{2}}n_{1}f_{1}\non[2ex]
 &&+ga_{03}\gt a_{08}\frac{a_{3}a_{8}}{r^{2}}f_{1}-\mu^{2}f_{1}+\lambda\rho^{2}f_{1}^{3}-h\rho^{2}\left(f_{2}^{2}+f_{3}^{2}\right)f_{1}\,,\\[2ex]
\left(f_{2}^{''}+\frac{1}{r}f_{2}^{'}\right) &=&\frac{n_{2}^{2}}{r^{2}}f_{2}+\gt^{2}a_{08}^{2}\frac{a_{8}^{2}}{r^{2}}f_{2}+\frac{g^{2}}{4}a_{03}^{2}\frac{a_{3}^{2}}{r^{2}}f_{2}+2\gt a_{08}\frac{a_{8}}{r^{2}}n_{2}f_{2}\non[2ex]
&&-ga_{03}\frac{a_{3}}{r^{2}}n_{2}f_{2}-ga_{03}\gt a_{08}\frac{a_{3}a_{8}}{r^{2}}f_{2}\,,\\[2ex]
\left(f_{3}^{''}+\frac{1}{r}f_{3}^{'}\right) &=&\frac{n_{3}^{2}}{r^{2}}f_{3}+4\gt^{2}a_{08}^{2}\frac{a_{8}^{2}}{r^{2}}f_{3}-4\gt a_{08}\frac{a_{8}}{r^{2}}n_{3}f_{3}\non[2ex]&&-\mu^{2}f_{3}+\lambda\rho^{2}f_{3}^{3}-h\rho^{2}\left(f_{1}^{2}+f_{2}^{2}\right)f_{3}\,,\\[2ex]
a_{03}^{2}\left(\frac{a_{3}^{''}}{r}-\frac{a_{3}^{'}}{r^{2}}\right) &=&\frac{1}{2}\rho^{2}\Bigg\{\frac{g^{2}}{2}a_{03}^{2}\frac{a_{3}}{r}f_{1}^{2}+\frac{ga_{03}}{r}n_{1}f_{1}^{2}+ga_{03}\gt a_{08}\frac{a_{8}}{r}f_{1}^{2}+\frac{g^{2}}{2}a_{03}^{2}\frac{a_{3}}{r}f_{2}^{2}\non[2ex]
&&-\frac{ga_{03}}{r}n_{2}f_{2}^2-ga_{03}\gt a_{08}\frac{a_{8}}{r}f_{2}^2\Bigg\} \,,\\[2ex]
a_{08}^{2}\left(\frac{a_{8}^{''}}{r}-\frac{a_{8}^{'}}{r^{2}}\right) &=&\frac{1}{2}\rho^{2}\Bigg\{2\gt^{2}a_{08}^{2}\frac{a_{8}}{r}f_{1}^{2}+\frac{2\gt a_{08}}{r}n_{1}f_{1}^{2}+ga_{03}\gt a_{08}\frac{a_{3}}{r}f_{1}^{2}+2\gt^{2}a_{08}^{2}\frac{a_{8}}{r}f_{2}^{2}\non[2ex]
 &&+\frac{2\gt a_{08}}{r}n_{2}f_{2}^{2}-ga_{03}\gt a_{08}\frac{a_{3}}{r}f_{2}^{2}+8\gt^{2}a_{08}^{2}\frac{a_{8}}{r}f_{3}^{2}-\frac{4\gt a_{08}}{r}n_{3}f_{3}^{2}\Bigg\}\,.\vspace*{-2cm}\non[0.5ex]
 \hfill
\eea
\endgroup
We can further simplify these equations by using $\rho_{\mathrm{CFL}}$
to eliminate $\mu$.

\begin{align}
\left(f_{1}^{''}+\frac{1}{r}f_{1}^{'}\right) & =\frac{f_{1}}{r^{2}}\left(n_{1}+\gt a_{08}a_{8}+\frac{g}{2}a_{03}a_{3}\right)^{2}-\rho^{2}\left[\lambda f_{1}\left(1-f_{1}^{2}\right)-hf_{1}\left(2-f_{2}^{2}-f_{3}^{2}\right)\right]\,,\\
\left(f_{2}^{''}+\frac{1}{r}f_{2}^{'}\right) & =\frac{f_{2}}{r^{2}}\left(n_{2}+\gt a_{08}a_{8}-\frac{g}{2}a_{03}a_{3}\right)^{2}-\rho^{2}\left[\lambda f_{2}\left(1-f_{2}^{2}\right)-hf_{2}\left(2-f_{1}^{2}-f_{3}^{2}\right)\right]\,,\\
\left(f_{3}^{''}+\frac{1}{r}f_{3}^{'}\right) & =\frac{f_{3}}{r^{2}}\left(n_{3}-2\gt a_{08}a_{8}\right)^{2}-\rho^{2}\left[\lambda f_{3}\left(1-f_{3}^{2}\right)-hf_{3}\left(2-f_{1}^{2}-f_{2}^{2}\right)\right]\,,\\
a_{03}\left(a_{3}^{''}-\frac{a_{3}^{'}}{r}\right) & =\rho^{2}\left\{ \frac{g}{2}f_{1}^{2}\left(n_{1}+\frac{g}{2}a_{03}a_{3}+\gt a_{08}a_{8}\right)-\frac{g}{2}f_{2}^{2}\left(n_{2}-\frac{g}{2}a_{03}a_{3}+\gt a_{08}a_{8}\right)\right\} \,,\\
a_{08}\left(a_{8}^{''}-\frac{a_{8}^{'}}{r}\right) & =\rho^{2}\Bigg\{\gt f_{1}^{2}\left(n_{1}+\gt a_{08}a_{8}+\frac{g}{2}a_{03}a_{3}\right)+\gt f_{2}^{2}\left(n_{2}+\gt a_{08}a_{8}-\frac{g}{2}a_{03}a_{3}\right)\nonumber \\
 & -2\gt f_{3}^{2}\left(n_{3}-2\gt a_{08}a_{8}\right)\Bigg\}\,.
\end{align}
In a last step, we introduce a dimensionless radial coordinate $r=\frac{R}{\sqrt{\lambda}\rho}$,
where $\xi=\sqrt{\lambda}\rho$ is the coherence length of the system,
and divide all equations by $\xi^{2}$. In the following, prime denotes
a derivative w.r.t. the new dimensionless variable.
\begin{align}
\left(f_{1}^{''}+\frac{1}{R}f_{1}^{'}\right) & =\frac{f_{1}}{R^{2}}\left(n_{1}+\gt a_{08}a_{8}+\frac{g}{2}a_{03}a_{3}\right)^{2}-\left[f_{1}\left(1-f_{1}^{2}\right)-\frac{h}{\lambda}f_{1}\left(2-f_{2}^{2}-f_{3}^{2}\right)\right]\,,\\
\left(f_{2}^{''}+\frac{1}{R}f_{2}^{'}\right) & =\frac{f_{2}}{R^{2}}\left(n_{2}+\gt a_{08}a_{8}-\frac{g}{2}a_{03}a_{3}\right)^{2}-\left[f_{2}\left(1-f_{2}^{2}\right)-\frac{h}{\lambda}f_{2}\left(2-f_{1}^{2}-f_{3}^{2}\right)\right]\,,\\
\left(f_{3}^{''}+\frac{1}{R}f_{3}^{'}\right) & =\frac{f_{3}}{R^{2}}\left(n_{3}-2\gt a_{08}a_{8}\right)^{2}-\left[f_{3}\left(1-f_{3}^{2}\right)-\frac{h}{\lambda}f_{3}\left(2-f_{1}^{2}-f_{2}^{2}\right)\right]\,,\\
\left(a_{3}^{''}-\frac{a_{3}^{'}}{R}\right) & =\frac{1}{\lambda a_{03}}\left\{ \frac{g}{2}f_{1}^{2}\left(n_{1}+\frac{g}{2}a_{03}a_{3}+\gt a_{08}a_{8}\right)-\frac{g}{2}f_{2}^{2}\left(n_{2}-\frac{g}{2}a_{03}a_{3}+\gt a_{08}a_{8}\right)\right\} \,,\\
\left(a_{8}^{''}-\frac{a_{8}^{'}}{R}\right) & =\frac{1}{\lambda a_{08}}\Bigg\{\gt f_{1}^{2}\left(n_{1}+\gt a_{08}a_{8}+\frac{g}{2}a_{03}a_{3}\right)+\gt f_{2}^{2}\left(n_{2}+\gt a_{08}a_{8}-\frac{g}{2}a_{03}a_{3}\right)\nonumber \\
 & -2\gt f_{3}^{2}\left(n_{3}-2\gt a_{08}a_{8}\right)\Bigg\}\,.
\end{align}
We can use the equations of motion to simplify the energy by replacing
the derivative terms and use partial integration, where the factor $R$ of the Jacobi determinant in the partial integration must not be forgotten.
We use
\begin{align}
\int dr\,r\left(f^{'}\right)^{2} & =-\int dr\,f\left(rf^{''}+f^{'}\right)\,.
\end{align}
\begin{equation}
\frac{F_{\circlearrowleft}}{L}=2\pi\int dr\,rU_{\circlearrowleft}\,,
\end{equation}
and find
\begingroup
\allowdisplaybreaks
\begin{align}
&\frac{F_{\circlearrowleft}}{L} =2\pi\int dr\,\Bigg\{ a_{03}^{2}\frac{\left(a_{3}^{'}\right)^{2}}{2r}+a_{08}^{2}\frac{\left(a_{8}^{'}\right)^{2}}{2r}+\frac{1}{2}\rho^{2}\Big[-rf_{1}\Big(\frac{f_{1}}{r^{2}}\left(n_{1}+\gt a_{08}a_{8}+\frac{g}{2}a_{03}a_{3}\right)^{2}\nonumber\\
&-\rho^{2}\left[\lambda f_{1}\left(1-f_{1}^{2}\right)-hf_{1}\left(2-f_{2}^{2}-f_{3}^{2}\right)\right]\Big)+\frac{f_{1}^{2}}{r}\left(n_{1}+\frac{g}{2}a_{03}a_{3}+\gt a_{08}a_{8}\right)^{2}\Big]\nonumber\\
 & +\frac{1}{2}\rho^{2}\Big[-rf_{2}\left(\frac{f_{2}}{r^{2}}\left(n_{2}+\gt a_{08}a_{8}-\frac{g}{2}a_{03}a_{3}\right)^{2}-\rho^{2}\left[\lambda f_{2}\left(1-f_{2}^{2}\right)-hf_{2}\left(2-f_{1}^{2}-f_{3}^{2}\right)\right]\right)\nonumber\\
&+\frac{f_{2}^{2}}{r}\left(n_{2}-\frac{g}{2}a_{03}a_{3}+\gt a_{08}a_{8}\right)^{2}\Big]+\frac{1}{2}\rho^{2}\Big[-rf_{3}\Big(\frac{f_{3}}{r^{2}}\left(n_{3}-2\gt a_{08}a_{8}\right)^{2}-\rho^{2}[\lambda f_{3}\left(1-f_{3}^{2}\right)\nonumber\\
 &-hf_{3}\left(2-f_{1}^{2}-f_{2}^{2}\right)]\Big)+\frac{f_{3}^{2}}{r}\left(n_{3}-2\gt a_{08}a_{8}\right)^{2}\Big]\nonumber\\
& +r\left[\frac{\mu^{2}}{2}\rho^{2}\left(3-f_{1}^{2}-f_{2}^{2}-f_{3}^{2}\right)-\frac{\lambda}{4}\rho^{4}\left(3-f_{1}^{4}-f_{2}^{4}-f_{3}^{4}\right)+\frac{h}{2}\rho^{4}\left(3-f_{1}^{2}f_{2}^{2}-f_{1}^{2}f_{3}^{2}-f_{2}^{2}f_{3}^{2}\right)\right]\Bigg\}.
\end{align}
\endgroup
We see that a lot of terms cancel, which leaves us with a comparably compact
form,
\begingroup
\allowdisplaybreaks
\begin{align}
&\frac{F_{\circlearrowleft}}{L} = 2\pi\int dr\,\Bigg\{ a_{03}^{2}\frac{\left(a_{3}^{'}\right)^{2}}{2r}+a_{08}^{2}\frac{\left(a_{8}^{'}\right)^{2}}{2r}+\frac{1}{2}\rho^{4}rf_{1}\left[\lambda f_{1}\left(1-f_{1}^{2}\right)-hf_{1}\left(2-f_{2}^{2}-f_{3}^{2}\right)\right] \nonumber \\
 & +\frac{1}{2}\rho^{4}rf_{2}\left[\lambda f_{2}\left(1-f_{2}^{2}\right)-hf_{2}\left(2-f_{1}^{2}-f_{3}^{2}\right)\right] +\frac{1}{2}\rho^{4}rf_{3}\left[\lambda f_{3}\left(1-f_{3}^{2}\right)-hf_{3}\left(2-f_{1}^{2}-f_{2}^{2}\right)\right]\nonumber \\
 & +r\left[\frac{\mu^{2}}{2}\rho^{2}\left(3-f_{1}^{2}-f_{2}^{2}-f_{3}^{2}\right)-\frac{\lambda}{4}\rho^{4}\left(3-f_{1}^{4}-f_{2}^{4}-f_{3}^{4}\right)+\frac{h}{2}\rho^{4}\left(3-f_{1}^{2}f_{2}^{2}-f_{1}^{2}f_{3}^{2}-f_{2}^{2}f_{3}^{2}\right)\right]\hspace{-.2cm}\Bigg\}\, .
\end{align}
\endgroup
Replacing $\mu^{2}=\rho^{2}\left(\lambda-2h\right)$ and switching to dimensionless variables leads to
\begin{align}
\frac{F_{\circlearrowleft}}{L}  =\rho^{2}_{\mathrm{CFL}}\pi\int dR\,R&\Bigg\{ \lambda\left(\frac{a_{03}a_{3}^{'}}{R}\right)^{2}+\lambda\left(\frac{a_{08}a_{8}^{'}}{R}\right)^{2}+\frac{1}{2}\left(1-f_{1}^{4}\right)+\frac{1}{2}\left(1-f_{2}^{4}\right)+\frac{1}{2}\left(1-f_{3}^{4}\right)\non[2ex]&-\frac{h}{\lambda}\left(3-f_{1}^{2}f_{2}^{2}-f_{1}^{2}f_{3}^{2}-f_{2}^{2}f_{3}^{2}\right)\Bigg\} \, ,
\end{align}
which is identical to Eq.~(\ref{FL}) upon setting $a_{08}=a_{03}=1$.

\chapter{Derivation of the Propagator}
\label{app:excitations}
In this appendix the derivation of the tree-level propagator of the
various systems is presented in some detail.

\section{Single Field\label{App:exc_SF}}

The excitation energies of the system described by the Lagrangian
in Eq.~(\ref{eq:Lag-phi4}) can be computed as follows.

We start by computing the fluctuation propagator as described
in Sec.~\ref{sec:Fluc-prop_SF}. Starting from the action containing
the quadratic contributions of the fluctuations, we write
\begin{equation}
S^{(2)}=\int_{0}^{\beta}d\tau\int d^{3}x\,\L^{(2)}\,,
\end{equation}
and use the Fourier transformed fields 
\begin{equation}
\vf_{i}(X)=\frac{1}{\sqrt{TV}}\sum_{K}e^{-iK\cdot X}\vf_{i}(K)\,,
\end{equation}
to compute the Lagrangian in Fourier space. A priori, we allow different
momenta $K_{\mu}$ and $K_{\mu}^{'}$ in the transformation whenever
two fields are multiplied with each other.
\begin{align}
S^{(2)}  =\ &\frac{1}{2TV}\sum_{K}\sum_{K^{'}}\int d\tau\int d^{3}x\,e^{-i\left(K+K^{'}\right)\cdot X}\Bigg\{-K_{\mu}K^{\mu'}\varphi_{1}(K)\varphi_{1}(K^{'})\nonumber \\
&-K_{\mu}K^{\mu'}\varphi_{2}(K)\varphi_{2}(K^{'})+\left[\varphi_{1}(K)\varphi_{1}(K^{'})+\varphi_{2}(K)\varphi_{2}(K^{'})\right](p^{2}-m^{2})\nonumber \\
 & +2\partial_{\mu}\psi\left[-i\varphi_{1}(K)K^{\mu'}\varphi_{2}(K^{'})+i\varphi_{2}(K)K^{\mu}\varphi_{1}(K^{'})\right]\non[2ex]
 & -\lambda\rho^{2}\left[3\varphi_{1}(K)\varphi_{1}(K^{'})+\varphi_{2}(K)\varphi_{2}(K^{'})\right]\Bigg\}\,.
\end{align}
We use that the Fourier transform of the delta symbol
is given by 
\begin{equation}
\delta_{K,K^{'}}=\frac{T}{V}\int d\tau\int d^{3}x\,e^{-iX\cdot(K-K^{'})}\,,
\end{equation}
to perform the integrations over $d\tau$ and $d^{3}x$ and perform
the summation over $K^{'}$ which leads to $K=-K^{'}$. We can simplify
the calculation by introducing an internal $2-$dimensional vector
space in $\vf_{1}$ and $\vf_{2}$ such that 
\begin{equation}
S^{(2)}=-\frac{1}{2}\sum_{K}\left[\vf_{1}(-K)\vf_{2}(-K)\right]\frac{D^{-1}(K)}{T^{2}}\left(\begin{array}{c}
\vf_{1}(K)\\
\vf_{2}(K)
\end{array}\right)\,,
\end{equation}
where the inverse propagator is given by 
\begin{equation}
D^{-1}(K)=\left(\begin{array}{cc}
-K^{2}-p^{2}+m^{2}+3\lambda\rho^{2} & -2iK_{\mu}\p^{\mu}\psi\\
2iK_{\mu}\p^{\mu}\psi & -K^{2}-p^{2}+m^{2}+\lambda\rho^{2}
\end{array}\right)\,.
\end{equation}
In principle, there is some arbitrariness in the off-diagonal components,
but we we choose a maximally symmetric version.
The excitations from this propagator are computed in the main text.

\section{Single Charged Field}
In this section I show the derivation of the propagator for a single charged scalar field coupled to an abelian gauge field, in the next section this derivation will be extended to the two-fields case. Elements of this derivation can be found in discussions 
of the standard abelian Higgs model, see for instance Chap.~85 of Ref.~\cite{Srednicki:2007qs}. The Lagrangian in Heaviside-Lorentz units is given by
\bea \label{Lagsingle}
{\cal L}  &=& -\frac{1}{16\pi}F_{\mu\nu}F^{\mu\nu} +[\partial_\mu + i(qA_\mu-\delta_{\mu}^0\mu)]\varphi[\partial^\mu - i(qA^\mu-\delta^{\mu}_0\mu)]\varphi^* -m^2|\varphi|^2 -\lambda |\varphi|^4  \non[2ex]
&=& -\frac{1}{16\pi}F_{\mu\nu}F^{\mu\nu}+\partial_\mu\varphi\partial^\mu\varphi^* + i(q A_\mu-\delta_{0\mu}\mu)(\varphi\partial^\mu\varphi^*-\varphi^*\partial^\mu\varphi) \non[2ex]
&&+(\mu^2-m^2+q^2A_\mu A^\mu -2\mu qA_0)|\varphi|^2 -\lambda|\varphi|^4 \, .
\eea
We now parametrize the field in terms of its real and imaginary part, $\vf=\frac{1}{\sqrt{2}}(\phi+i\chi)$.
\bea
\L&=& -\frac{1}{16\pi}F_{\mu\nu}F^{\mu\nu}+\frac{1}{2}\partial_\mu\phi\partial^\mu\phi + \frac{1}{2}\partial_\mu\chi\partial^\mu\chi + (qA_\mu-\delta_{0\mu}\mu)(\phi\partial^\mu\chi-\chi\partial^\mu\phi) 
 \non[2ex]
 &&+\frac{1}{2}(\phi^2+\chi^2)(\mu^2-m^2+q^2A_\mu A^\mu-2\mu qA_0) -\frac{\lambda}{4}(\phi^2+\chi^2)^2 \, ,
\eea

Next, we separate the real, spacetime-independent condensate by replacing $\phi \to \rho + \phi$, where we use the residual global $U(1)$ freedom to chose a real value for the condensate. Then, we collect the terms of second order in the fluctuations
$\phi$, $\chi$, $A_\mu$,
\bea
{\cal L}^{(2)} &=& -\frac{1}{16\pi}F_{\mu\nu}F^{\mu\nu}+\frac{1}{2}\partial_\mu\phi\partial^\mu\phi + \frac{1}{2}\partial_\mu\chi\partial^\mu\chi +\rho q A_\mu\partial^\mu\chi+\frac{\rho^2q^2}{2}A_\mu A^\mu
\\&&-\mu(\phi\partial_0\chi-\chi\partial_0\phi) -2\mu\rho qA_0\phi+\frac{1}{2}(\phi^2+\chi^2)(\mu^2-m^2-\lambda\rho^2)-\lambda\rho^2\phi^2 \, . \nonumber
\eea 
Here, terms of higher order in the fluctuations are neglected, and the prefactors of the terms linear in $\phi$ and $\chi$ are canceled as before by applying the equations of motions obtained from the tree-level potential. We see that the gauge field acquires a mass term $\frac{1}{2}\rho^2q^2A_{\mu}{\mu}$, which is called the abelian Higgs mechanism \cite{1964PhRvL..13..508H,1964PhRvL..13..321E}, see Sec.~\ref{sec:massive_boson}.
In order to compute the dispersions we need to add a gauge fixing term,
\be
{\cal L}_\mathrm{ gf} = -\frac{(\partial_\mu A^\mu)^2}{2\xi} \, .
\ee
In general, together with the gauge fixing term, ghost fields should be included in order to cancel unphysical modes and all influence of the gauge parameter $\xi$ in the end. Although ghost fields are trivial for abelian gauge fields which interact with scalar or fermionic fields (the only become important for non-abelian gauge theories like QCD), this is not longer the case as soon as the field forms a condensate.
We also replace $A_0\to iA_0$ because we work in the imaginary time formalism. The remaining procedure is analogous to the derivation without gauge field. By writing down the fluctuations of all fields in Fourier space, we can read off the inverse propagator in momentum space, where we introduce the Fourier transformed scalar fields via 
\be
\phi(X) = \frac{1}{\sqrt{TV}}\sum_K e^{-iK\cdot X}\phi(K) \, , \qquad \chi(X) = \frac{1}{\sqrt{TV}}\sum_K e^{-iK\cdot X}\chi(K) \, ,
\ee
and the gauge field
\be
A_{\mu}(X) = \frac{1}{\sqrt{TV}}\sum_K e^{-iK\cdot X}A_{\mu}(K) \, ,
\ee
with the spacetime four-vector $X=(-i\tau,\mathbf{x}) $ and four-momentum $K=(k_0,\mathbf{k})$, where the zeroth component of the momentum vector $k_0$ contains the the bosonic Matsubara frequencies $\omega_n$. The second-order terms in the fluctuations can then be written as  
\be
\int_X {\cal L}^{(2)}= -\frac{1}{2}\sum_K \Xi(-K)^T\frac{S^{-1}(K)}{T^2}\Xi(K) \, , 
\ee
with the abbreviation
\be
\Xi(K) = \left(\begin{array}{c} \phi(K) \\ \chi(K) \\ A_{\mu}(K) \end{array}\right) \, ,
\ee
and the propagator
\bea
S^{-1} = \left(\begin{array}{cc} S_0^{-1} & I(K) \\ [2ex] I^T(-K) & D^{-1} \end{array}\right)\, ,
\eea
The inverse propagator for the scalar field is found to be
\bea
S_0^{-1} = \left(\begin{array}{cc} -K^2-(\mu^2-m^2)+3\lambda\rho^2 & 2ik_0\mu \\ [2ex] -2ik_0\mu & -K^2-(\mu^2-m^2)+\lambda\rho^2 \end{array}\right) \, .
\eea
The gauge boson propagator, including the mass term induced by the condensate, reads
\bea \label{gaugeprop}
D^{-1}(K) = \sigma
\left(\begin{array}{cccc} \frac{\zeta(K,\rho)}{\sigma}+ k_0^2 & -i k_0k_1 & -i k_0k_2 &-i k_0k_3 \\[2ex]
-i k_0k_1 & \frac{\zeta(K,\rho)}{\sigma}- k_1^2 & - k_1k_2 &- k_1k_3 \\[2ex]
-i k_0k_2 & - k_1k_2 & \frac{\zeta(K,\rho)}{\sigma}- k_2^2 & - k_2k_3 \\[2ex]
-i k_0k_3 & - k_1k_3 & - k_2k_3 & \frac{\zeta(K,\rho)}{\sigma}-k_3^2\end{array}\right) \, ,
\eea
where $\sigma\equiv 1-1/\xi$ and $\zeta(K,\rho)\equiv-K^2+4\pi q^2\rho^2$
and the off-diagonal blocks that couple the scalar field to the gauge field are given by
\be
I(K) =\sqrt{4\pi}q\rho \left(\begin{array}{cccc} 2i\mu  & 0&0&0 \\[2ex] -k_0 & ik_1 & ik_2 & ik_3 \\[2ex] 0&0&0&0 \\[2ex] 0&0&0&0 \end{array}\right) \, .
\ee
\section{Two Scalar Fields}
\label{app:exc_SFSF}
For the computation of the sound modes of the two superfluid system, we compute the slope of the Goldstone mode. Once again, we introduce fluctuations of the form
\be
\vf_i=\frac{e^{i\psi_i}}{\sqrt{2}}\left(\rho_i+\phi_i+i\chi_i\right) \, ,
\ee
into the Lagrangian Eq.~(\ref{L}) as in the section before, but set all charges to zero and ignore the gauge field. Note the different notion compared to Ref.~\cite{Haber:2015exa}, where this derivation was originally performed. Doing so, we obtain
\bea
\L^{(2)} \hspace*{-.2cm}&=&\hspace*{-.2cm}\frac{1}{2}\sum_{i=1,2}\Big[(\partial\phi_i)^2+(\partial\chi_i)^2 +|\phi_i|^2(p_i^2-m_i^2)
+2\partial\psi_i\cdot(\phi_i\partial\chi_i-\chi_i\partial\phi_i)-\lambda_i\rho_i^2(3\phi_i^2+\chi_i^2)\Big] \non[2ex]
&&\hspace*{-.2cm}+\frac{h+gp_{12}^2}{2}\Big(\rho_1^2|\phi_2|^2+\rho_2^2|\phi_1|^2+4\rho_1\rho_2\phi_1\phi_2\Big)
-\frac{\rho_1\rho_2}{2}\Big(G\partial\phi_1\cdot\partial\phi_2-g\partial\chi_1\cdot\partial\chi_2\Big) \non[2ex]
&&\hspace*{-.2cm}+\frac{g}{2}\Big[\rho_1^2\partial\psi_1\cdot(\phi_2\partial\chi_2-\chi_2\partial\phi_2)
+\rho_2^2\partial\psi_2\cdot(\phi_1\partial\chi_1-\chi_1\partial\phi_1)\non[2ex]&&\hspace*{-.2cm}+2\rho_1\rho_2(\phi_1\partial\psi_1\cdot\partial\chi_2
+2\phi_2\partial\psi_2\cdot\partial\chi_1)\Big] \, ,
\eea
where we have assumed $\rho_1$ and $\rho_2$ to be constant, and where $|\phi_i|^2 \equiv \phi_i^2+\chi_i^2$.
We introduce the Fourier transformed fields via 
\be
\phi_i(X) = \frac{1}{\sqrt{TV}}\sum_K e^{-iK\cdot X}\phi_i(K) \, , \qquad \chi_i(X) = \frac{1}{\sqrt{TV}}\sum_K e^{-iK\cdot X}\chi_i(K) \, , 
\ee
with the spacetime four-vector $X=(-i\tau,\mathbf{x}) $ and four-momentum $K=(k_0,\mathbf{k})$, where $k_0=-i\omega_n$ with the bosonic Matsubara frequencies 
$\omega_n = 2\pi n T$, $n\in \mathbb{Z}$. Then, the second-order terms in the fluctuations can be written as  
\be
\int_X {\cal L}^{(2)}= -\frac{1}{2}\sum_K \Xi(-K)^T\frac{S^{-1}(K)}{T^2}\Xi(K) \, , 
\ee
where we have abbreviated 
\be
\Xi(K) = \left(\begin{array}{c} \phi_1(K) \\ \chi_1(K) \\ \phi_2(K) \\ \chi_2(K) \end{array}\right) \, .
\ee
In this basis, the inverse tree-level propagator in momentum space $S^{-1}(K)$ is a $4\times 4$ matrix, given in Eqs.\ (\ref{Sinv}), and the free energy (\ref{eq:OmegaZ}) becomes 
\be \label{Om1}
\Omega = U + \frac{1}{2}\frac{T}{V}\sum_K\ln\mathrm{ det}\frac{Q^{(0)}S^{-1}(K)}{T^2} \, , 
\ee
where the determinant is taken over $4\times 4$ space. The determinant of the inverse propagator is a polynomial in $k_0$ of degree 8, which we can write in terms of its zeros, $k_0 = \epsilon_{r,\mathbf{k}}$. 
As mentioned in the main text, these zeros can be grouped in 4 pairs, $\{\epsilon_{r,\mathbf{k}},-\epsilon_{r,-\mathbf{k}}\}$, and we can thus write 
\be
\mathrm{ det}\,S^{-1} = \left(1-\frac{G^2}{4}\rho_1^2\rho_2^2\right)\left(1-\frac{g^2}{4}\rho_1^2\rho_2^2\right)\prod_{r=1}^4(k_0-\epsilon_{r,\mathbf{k}})(k_0+\epsilon_{r,-\mathbf{k}}) \, .
\ee
The prefactor is exactly cancelled by $\mathrm{ det}\,Q^{(0)}$, see Eq.\ (\ref{Q0}). 
This is important since otherwise it would yield an unphysical, divergent contribution to the free energy. 
 
As an illustrative example for a thermodynamic quantity, let us write down the charge densities $n_i=-\partial \Omega/\partial\mu_i$ at nonzero temperatures. 
With the help of Eq.\ (\ref{Om1}) and $\ln\,\mathrm{ det} S^{-1} = \mathrm{Tr}\ln S^{-1}$ we find  
\bea \label{ni}
n_i&=& -\frac{\partial U}{\partial \mu_i}- \frac{1}{2}\frac{T}{V}\sum_K\mathrm{Tr}\left[S\frac{\partial S^{-1}}{\partial\mu_i}\right] \non[2ex] 
&=& -\frac{\partial U}{\partial \mu_i}- \frac{1}{2}\int\frac{d^3k}{(2\pi)^3}\;T\hspace{-0.2cm}\sum_{n=-\infty}^\infty \frac{F(k_0,\mathbf{k})}{\prod_{r=1}^4(k_0-\epsilon_{r,\mathbf{k}})
(k_0+\epsilon_{r,-\mathbf{k}})} \, ,
\eea
where, in the second step, we have rewritten the sum over four-momentum as a discrete sum over Matsubara frequencies $n$ and an integral over three-momentum $\mathbf{k}$, 
and $F(k_0,\mathbf{k})$ is a complicated function without poles in $k_0$ which obeys the symmetry $F(-k_0,-\mathbf{k})=F(k_0,\mathbf{k})$. 
The sum over Matsubara frequencies creates 8 terms, each for one of the poles. 
Due to the symmetries of $\epsilon_{r,\mathbf{k}}$ and $F(k_0,\mathbf{k})$ these terms give the same result pairwise under the $\mathbf{k}$-integral (this is easily seen with the new integration 
variable $\mathbf{k}\to -\mathbf{k}$ in one term of each pair). Therefore, we can restrict ourselves to 4 terms, 
\be
n_i =  -\frac{\partial U}{\partial \mu_i} + \frac{1}{2}\sum_{r=1}^4\int\frac{d^3k}{(2\pi)^3} \frac{F(\epsilon_{r,\mathbf{k}},\mathbf{k})}{(\epsilon_{r,\mathbf{k}}+\epsilon_{r,-\mathbf{k}})
\prod_s (\epsilon_{r,\mathbf{k}}-\epsilon_{s,\mathbf{k}})(\epsilon_{r,\mathbf{k}}+\epsilon_{s,-\mathbf{k}})}\coth\frac{\epsilon_{r,\mathbf{k}}}{2T} \, ,
\ee
where the product over $s$ runs over the three integers different from $r$. There are two
contributions due to $\coth\frac{\epsilon}{2T} = 1+2f(\epsilon)$ with the Bose distribution function $f(\epsilon)=1/(e^{\epsilon/T}-1)$. The first one is temperature independent 
(more precisely, there is no \textit{explicit} dependence on $T$, in general a temperature dependence enters through the condensates). This contribution is infinite and has to be 
regularized. The second contribution depends on temperature explicitly. It is the "usual" integral over the Bose distribution, here with a complicated, momentum-dependent prefactor.
This prefactor is trivial in the NOR phase, where there are no condensates, and particles and antiparticles carry positive and negative unit charges for each of the species 
separately. In the COE phase, where both fields condense, the 2 massive and 2 massless modes each contribute to both charge densities in a nontrivial way. Note that in the thermal 
contribution alone we do \textit{not} have 8 terms that are pairwise equal. Instead, now it is crucial to work with positive excitation energies, otherwise one obtains 
unphysical, negative occupation numbers. We do not further evaluate the charge densities or any other thermodynamic quantity since our main focus in this work is on 
the excitation energies $\epsilon_{r,\mathbf{k}}$ themselves. 

\section{Two Scalar Fields and one Gauge Field }
This derivation is slightly modified and extended from Ref.~\cite{Haber:2017kth}. We can now easily add the second, neutral, field. We can use the result for the single field (\ref{Lagsingle}) and add the respective terms for the neutral field, which can be obtained from the gauged field by setting $q=0$. For simplicity, we set $g=0$ as well. The coupling terms, especially the derivative coupling $G$ which also couples the scalar and the gauge fields, have to be computed separately, and we obtain
\bea
{\cal L}  &=&   {\cal L}_1(\varphi_1,A_\mu) +{\cal L}_2(\varphi_2) + {\cal L}_I(\varphi_1,\varphi_2,A_\mu) + {\cal L}_\mathrm{ YM}(A_\mu) \, ,
\eea
with 
\begin{subequations}
\bea
{\cal L}_1 &=& \frac{1}{2}\partial_\mu\phi_1\partial^\mu\phi_1 + \frac{1}{2}\partial_\mu\chi_1\partial^\mu\chi_1 + (qA_\mu-\delta_{0\mu}\mu_1)(\phi_1\partial^\mu\chi_1-\chi_1\partial^\mu\phi_1) 
 \non[2ex]
 &&+\frac{1}{2}(\phi_1^2+\chi_1^2)(\mu_1^2-m_1^2+q^2A_\mu A^\mu-2\mu_1 qA_0) -\frac{\lambda}{4}(\phi_1^2+\chi_1^2)^2 \, , \\[2ex]
  {\cal L}_2 &=&\frac{1}{2}\partial_\mu\phi_2\partial^\mu\phi_2 + \frac{1}{2}\partial_\mu\chi_2\partial^\mu\chi_2 -\mu_2(\phi_2\partial_0\phi_2-\chi_2\partial_0\phi_2) 
\\&&+\frac{1}{2}(\phi_2^2+\chi_2^2)(\mu_2^2-m_2^2) -\frac{\lambda}{4}(\phi_2^2+\chi_2^2)^2 \, , \hspace{0.5cm}\non[2ex]
  {\cal L}_I &=& \frac{h}{2}(\phi_1^2+\chi_1^2)(\phi_2^2+\chi_2^2) -\frac{G}{2}(\phi_1\partial_\mu\phi_1+\chi_1\partial_\mu\chi_1)(\phi_2\partial^\mu\phi_2+\chi_2\partial^\mu\chi_2) \non[2ex]
 {\cal L}_\mathrm{ YM} &=& -\frac{1}{16\pi}F_{\mu\nu}F^{\mu\nu} \, ,
\eea
\end{subequations}
We now allow for condensation of both fields by shifting $\phi_1\to \rho_1 + \phi_1$, $\phi_2\to \rho_2 + \phi_2$, where we once again have chosen the vacuum expectation value to be positive and real. The terms quadratic in the fluctuations then become
\bea
{\cal L}^{(2)}  &=&   {\cal L}_1^{(2)}(\varphi_1,A_\mu) +{\cal L}_2^{(2)}(\varphi_2) + {\cal L}_I^{(2)}(\varphi_1,\varphi_2,A_\mu) + {\cal L}_\mathrm{ YM}(A_\mu) \, ,
\eea
with 
\begin{subequations}
\bea
{\cal L}^{(2)}_1 &=& \frac{1}{2}\partial_\mu\phi_1\partial^\mu\phi_1 + \frac{1}{2}\partial_\mu\chi_1\partial^\mu\chi_1 +\rho_1 q A_\mu\partial^\mu\chi_1+\frac{\rho_1^2q^2}{2}A_\mu A^\mu\\ &&-\mu_1(\phi_1\partial_0\chi_1-\chi_1\partial_0\phi_1)-2\mu_1\rho_1 qA_0\phi_1+\frac{1}{2}(\phi_1^2+\chi_1^2)(\mu_1^2-m_1^2-\lambda_1\rho_1^2)-\lambda_1\rho_1^2\phi_1^2 \, , \non[2ex]
{\cal L}^{(2)}_2&=&\frac{1}{2}\partial_\mu\phi_2\partial^\mu\phi_2 + \frac{1}{2}\partial_\mu\chi_2\partial^\mu\chi_2 -\mu_2(\phi_2\partial_0\chi_2-\chi_2\partial_0\phi_2)\\ 
&&+\frac{1}{2}(\phi_2^2+\chi_2^2)(\mu_2^2-m_2^2-\lambda_2\rho_2^2)-\lambda_2\rho_2^2\phi_2^2\non[2ex]
{\cal L}^{(2)}_I&=& \frac{h}{2}(\rho_1^2|\varphi_2|^2+\rho_2^2|\varphi_1|^2+4\rho_1\rho_2\phi_1\phi_2)-G\frac{\rho_1\rho_2}{2}\partial_\mu\phi_1\partial^\mu\phi_2 \, .
\eea 
\end{subequations}
We now follow the same procedure as above and read off the following inverse tree-level propagator where the internal space is now spanned by the vector
\be
\Xi(K) = \left(\begin{array}{c} \phi_1(K) \\ \chi_1(K) \\ \phi_2(K) \\ \chi_2(K) \\ A_{\mu}(K) \end{array}\right) \, ,
\ee
and the propagator is given by
\bea
S^{-1} = \left(\begin{array}{cc} S_0^{-1} & I(K) \\ [2ex] I^T(-K) & D^{-1} \end{array}\right)\, ,
\eea
with $S^{-1}$ coupling the scalar fields;
\begin{align}
S_0^{-1}=\left(\begin{array}{cccc} -K^2+\eta_1(\rho_i)+2\lambda_1\rho_1^2 \hspace{-.2cm}&\hspace{-.2cm} 2ik_0\mu_1 \hspace{-.2cm}&\hspace{-.2cm} \frac{\rho_1\rho_2}{2}(GK^2-4h) \hspace{-.2cm}&\hspace{-.2cm} 0 \\[2ex] -2ik_0\mu_1 & -K^2+\eta_1(\rho_i) &0&0 \\[2ex]
\frac{\rho_1\rho_2}{2}(GK^2-4h) & 0 & -K^2+\eta_2(\rho_i)+2\lambda_2\rho_2^2 & 2ik_0\mu_2 \\[2ex] 0&0& -2ik_0\mu_2 & -K^2+\eta_2(\rho_i)\end{array}\right) \, , \non[.5ex]
\hfill
\end{align}
where $\eta_{1/2}(\rho_1,\rho_2)\equiv -(\mu_{1/2}^2-m_{1/2}^2)+\lambda_{1/2}\rho_{1/2}^2-h\rho_{2/1}^2$ and the off-diagonal blocks that couple the scalar fields to the gauge field,
\be
I(K) =\sqrt{4\pi}q\rho_1 \left(\begin{array}{cccc} 2i\mu_1  & 0&0&0 \\[2ex] -k_0 & ik_1 & ik_2 & ik_3 \\[2ex] 0&0&0&0 \\[2ex] 0&0&0&0 \end{array}\right) \, .
\ee

The inverse gauge-boson propagator $D^{-1}$ is identical by the one given in Eq.\ (\ref{gaugeprop}), where the replacement $\rho\to \rho_1$ has to be performed. 

We are interested in an effective potential for the condensates 
$\rho_1$ and $\rho_2$, and thus we need to keep these condensates general. Nevertheless, it is instructive
to first discuss the dispersions at the zero-temperature stationary point, i.e., we set  $\rho_1=\rho_{01}$ and $\rho_2=\rho_{02}$ with the condensates in the coexistence phase 
$\rho_{01}$ and $\rho_{02}$ from Eq.\ (\ref{COE}). 
Let us first set the cross-coupling between the scalar field to zero, $h=G=0$. The dispersion relations $k_0=\epsilon_k$ are given by the zeros of $\mathrm{det}\,S^{-1}$. 
Since this is a polynomial of degree 8 in $k_0^2$, we obtain 8 dispersions, 6 of which are physical. The two unphysical ones are of the form $\epsilon_k=k$. These are the usual
unphysical modes of the gauge field, whose contribution to the partition function is canceled by ghost fields. With the given gauge choice, ghosts do not couple to 
any of the fields and merely serve to cancel the unphysical modes. None of the modes depend on the gauge fixing parameter $\xi$, which only appears as a prefactor of the determinant $\mathrm{det}\,S^{-1}$ and thus does not have to be specified. The 6 physical dispersions are
\begin{subequations}
\begin{align}
\epsilon_k &=& &\sqrt{k^2+4\pi q^2\rho_{01}^2} \qquad \mbox{(2-fold)} \, ,  \label{eps1} \\[2ex]
\epsilon_k &=& &\sqrt{k^2+3\mu_1^2-m_1^2+2\pi q^2\rho_{01}^2\pm\sqrt{4\mu_1^2k^2+\left(3\mu_1^2-m_1^2-2\pi q^2\rho_{01}^2\right)^2}} \, , \label{eps2} \\[2ex]
\epsilon_k &=& &\sqrt{k^2+3\mu_2^2-m_2^2\pm\sqrt{4\mu_2^2k^2+(3\mu_2^2-m_2^2)^2}} \, . \label{eps3}
\end{align}
\end{subequations}
We have three gauge field modes with mass $\epsilon_{k=0}=\sqrt{4\pi}q\rho_{01}$ 
[the two modes of Eq.\ (\ref{eps1}) and the mode with the 
lower sign in Eq.\ (\ref{eps2})], two more massive modes from the scalar fields, and the Goldstone mode [the mode with the lower sign in Eq.\ (\ref{eps3})]. 

If we switch on the coupling constants, the dispersions, including the masses, i.e.\ the dispersions at $k=0$, become too complicated to write down, but can still be computed analytically. Note however that the first two gauge-boson dispersions remain unaffected by the coupling between the fluids. The same is true for the mass of the would-be Goldstone mode; It remains $q\rho_1$, even in the presence of $G$ and $h$. Consequently, there are always 3 modes with mass $\sqrt{4\pi}q\rho_1$, two of which have always dispersion $\epsilon_k^2 = k^2 + 4\pi q^2\rho_1^2$, while the dispersion of the third one is complicated and depends on the coupling between the two fields. 
The dispersions \textit{at large momenta} are independent of $\xi$ (the determinant of the inverse propagator turns out to be proportional to $\xi^{-1}$, but this is the only $\xi$ dependence. The zeros of the determinant are independent of $\xi$). This is no longer true if we go to smaller momenta (in particular it is not true for the masses). The determinant is a polynomial in $k_0$ of degree 16: in addition to the 6 physical dispersions shown above, the determinant also yields 2 unphysical modes, $\epsilon_k=k$, which cancel in a proper treatment including ghost fields. However, since our results are already gauge independent and we know which modes have to be considered unphysical (it is possible to use unitary gauge in which the massless modes vanish already on the level of the Lagrangian, showing that they have to be considered as pure gauge, see for instance Chap.\ 6 of \cite{superbook}), we do not perform this calculation. Additionally, all modes also appear with a negative sign. 

\begin{figure} [h]
\begin{center}
\includegraphics[width=0.5\textwidth]{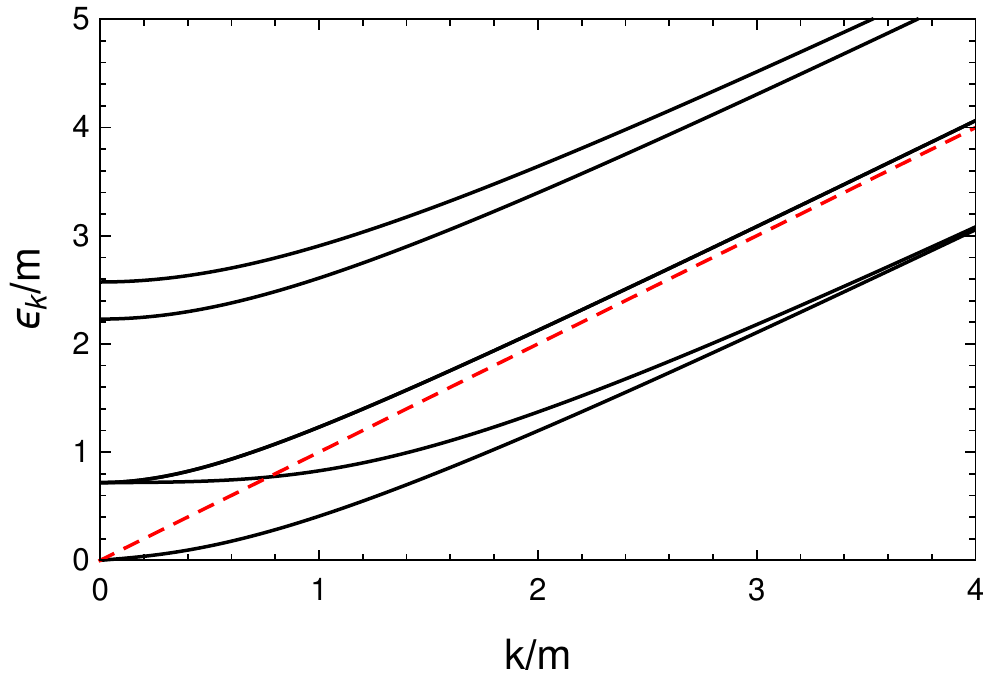}
\caption{Excitation energies for the COE phase, where both charged and neutral fields condense. The dashed (red) line is the diagonal $\epsilon_k=k$ to guide the eye. 
There are 6 modes in total, including one Goldstone mode and three massive gauge modes.  The excitation $\epsilon_k=\sqrt{k^2+4\pi q^2\rho_{01}^2}$, which approaches 
the diagonal from above, is 2-fold degenerate. All other dispersions have very complicated expressions due to the mixing of the gauge field with the scalar fields.
The parameters used for this plot are $m_1 =  m_2\equiv m$, $\mu_1 = 1.2m$, $\mu_2 = 1.1m$, $\lambda_1 = 0.3$, $\lambda_2 = 0.5$, $h = -0.1$, $G = 0$, $q=2e$. While these excitation energies are evaluated at the zero-temperature stationary point, the main purpose of this appendix is to derive an effective 
thermal potential, for which the dispersions for general values of the charged and neutral condensates are needed.}
\label{fig:expand1}
\end{center}
\end{figure}

For our nonzero-temperature calculation we want to derive an effective potential for the condensates, therefore we need to keep $\rho_1$ and $\rho_2$ general and compute the large-momentum behavior of the six dispersions. A large-momentum expansion of the dispersion relation can be written as 
\be\label{eq:epsc1c2}
\epsilon_{k}\approx c_1+k+\frac{c_2^2}{k}\, ,
\ee 
which leads for the two gauge field modes $\epsilon_k=\sqrt{k^2+4\pi q^2\rho_1^2}$ 
\be
c_1 = 0 \, , \qquad c_2 = 2\pi q^2\rho_1^2 \, .
\ee
For the other 4 modes we have 
\be
c_1 = e_1 \frac{\sqrt{\mu_1^2+\mu_2^2+e_2\sqrt{(\mu_1^2-\mu_2^2)^2+G^2\rho_1^2\rho_2^2\mu_1^2\mu_2^2}}}{\sqrt{2}\sqrt{1-\frac{G^2\rho_1^2\rho_2^2}{4}}} \, , \qquad e_1,e_2 = \pm \, , 
\ee
which goes to $\pm\mu_1$, $\pm\mu_2$ in the case without entrainment, $G=0$ ($h$ does not appear here). The coefficients $c_2$ are in general too complicated to write down. We show the numerical result of the expansion in Fig.\ \ref{fig:expand1}. The result for the nonzero-temperature potential however becomes comparably simple. Starting from Eq.~(\ref{eq:LOfiniteT}), we find 
\be \label{Tln}
T\int \frac{d^3k}{(2\pi)^3}\ln\left(1-e^{-\epsilon_k/T}\right) \simeq  -\frac{\pi^2T^4}{90} +\frac{c_1\zeta(3)T^3}{\pi^2} + \frac{(c_2^2-c_1^2)T^2}{12} \, .
\ee
In general, the dispersions now do depend on the gauge fixing parameter $\xi$. However, in the limit (\ref{eq:epsc1c2}) this dependence drops out, i.e., the coefficients 
$c_1$ and $c_2$ do not depend on $\xi$. Moreover, now the unphysical gauge modes no longer have the simple form $\epsilon_k=k$. Two of the physical gauge modes
keep their simple form (\ref{eps1}), while for the other 4 physical modes the coefficients $c_1$ and $c_2$ are (at least some of them) very lengthy. 
However, adding up the result for all 6 physical modes yields a relatively compact result,
\bea
&&\hspace*{-0.5cm}T\sum_{i=1}^6\int \frac{d^3k}{(2\pi)^3}\ln\left(1-e^{-\epsilon_{ki}/T}\right) \simeq \non[2ex]
&&\hspace*{-0.5cm}-\frac{\pi^2T^4}{15} -\frac{T^2}{12\left(1-\frac{G^2\rho_1^2\rho_2^2}{4}\right)}\Bigg\{2(\mu_1^2+\mu_2^2)-(m_1^2+m_2^2)-\left(2\lambda_1-h+6\pi q^2\right)\rho_1^2-(2\lambda_2-h)\rho_2^2 \non[2ex]
&&\hspace*{-0.5cm}+Gh\rho_1^2\rho_2^2-\frac{G^2\rho_1^2\rho_2^2}{8}\Big[\mu_1^2+\mu_2^2-(m_1^2+m_2^2)-(\lambda_1-h+12\pi q^2)\rho_1^2-(\lambda_2-h)\rho_2^2\Big]\Bigg\} \non[2ex]
&&\hspace*{-0.5cm}\simeq \frac{T^2}{12}\left[(2\lambda_1-h+6\pi q^2)\rho_1^2+(2\lambda_2-h)\rho_2^2-Gh\rho_1^2\rho_2^2\right] + \mathrm{const.} \, ,
\eea
where, in the second step, we have absorbed all terms that do not depend on $\rho_1$ or $\rho_2$ into "const.", and dropped all higher-order terms in the derivative coupling 
(i.e., we assume $G\mu^2\ll 1$, where $\mu$ stands for all energy scales $\mu_1$, $\mu_2$, $m_1$, $m_2$, $\rho_1$ $\rho_2$). Dropping the constant 
contribution, we add the $T^2$ terms to the potential (\ref{Ux}) and arrive at the potential (\ref{UxT}) in the main text.

Within our approximation of a small derivative coupling, we see that the temperature effect reduces to adding new quadratic terms in the condensates and a term similar to a density interaction. We therefore absorb this influence into a modified "thermal" mass\footnote{This should not be confused with the thermal mass known from thermal field theory in $\phi^4$-theory, which arises from loop corrections} and density coupling.

 \begin{subequations}
\bea 
m_{1,T}^2 &=& m_1^2+\frac{2\lambda_1-h+6\pi q^2}{6}T^2\, ,\\[2ex]
m_{2,T}^2 &=& m_2^2+\frac{2\lambda_2-h}{6}T^2\, , \\[2ex]
h_T&=& h\left(1+\frac{GT^2}{6}\right) \, .
 \eea
 \end{subequations}

This is in agreement for instance with Eq.\ (16) in Ref.\ \cite{Hebecker:1993rz} or Eq. (4.17) in reference \cite{Arnold:1992rz} 
(note the different convention of the self-coupling 
$\lambda$ in that paper and the use of Gaussian units here). On the one hand, our calculation is simpler than the one of these references because it does not include loops, but on the other hand it works with two scalar fields (in particular, a derivative coupling between them), and with chemical potentials (as opposed to negative squared masses). Our result shows that -- within our approximation --  the coupling between the fields does not have any non trivial effect in the gauged theory that it didn't already have in the ungauged case. In general, this is presumably not true, because the complicated terms proportional to the entrainment coupling $g$ are omitted. 
 
The condensates at finite temperature as well as the critical temperatures can now be computed as explained in the main part of this work. 
\chapter{Numerical Methods}
\label{App:num_methods}
In order to solve the various equations for the different profile
functions, several algorithms can be used. For the single flux tube,
the shooting method, where one has to iteratively determine the first
derivative at the origin, is quite popular. However, in the multicomponent
systems, guessing several derivatives for the various functions at
the same time is rather complicated. Consequently, we preferred a
version of Newton Simultaneous Over-Relaxation (SOR), described in
some detail in Ref.~\cite{num_recipies}. On top of that, a Gauss-Seidel
procedure is applied, which fastens the numerical calculation and
is described as well in Ref.~\cite{num_recipies}. The computation
of the profile functions itself is partially performed in FORTRAN
90, but the data post processing and most further calculations are
performed with Wolfram Mathematica. For some calculations in the color superconductivity
related chapters of this thesis, the solution of the system of equations
Eqs.~(\ref{eq:delta_y}) is of rank higher than three and is found
with help of the Linear Algebra Package LAPACK, see their users' guide
for more information \cite{lapack}. In this appendix, I explain the
general idea of the algorithm, the code itself can be found in the
corresponding subsections of this appendix. 

In the numerical procedure, we set up a one-dimensional lattice and
work with discrete functions. All derivatives in the equations of
motion are replaced with symmetric finite differences, i.e.~we transform
the continuous EOMs into FDEs (finite-differences equation). A first
and second derivative then read
\begin{align}
f^{'}(x) & =\frac{f(x_{i+1})-f(x_{i-1})}{2\Delta x}\,,\\
f^{''}(x) & =\frac{f(x_{i+1})-2f(x_{i})+f(x_{i-1})}{\Delta x^{2}}\,,
\end{align}
with the $i-th$ lattice point $x_{i}$ and the constant lattice spacing
$\Delta x$. This allows us to transform differential equations into
algebraic equations which can be solved on the computer. 

The Newton method works in general as follows: suppose we have a set
of coupled, non-linear, one-dimensional equations which we call 
\begin{equation}
F_{i}(y_{j})=0\,,
\end{equation}
where the vector $F_{i}$ denotes the $i-th$ equation and $y_{j}$
denotes the value of the $j-th$ function (the variables we are trying
to solve for). The idea is to compute the derivative of $F$ with
respect to a change of $y$. The change in the functions $y$ is the
next step of the iteration. In a $1^{st}$ order approximation, the
derivative reads
\begin{eqnarray}
\frac{F_{i}(y_{j}^{n+1})-F(y_{j}^{n})}{\Delta y_{j}^{n}} & \approx & \frac{\partial F_{i}}{\partial y_{j}^{n}}\,,\\
F_{i}(y_{j}^{n+1})-F(y_{j}^{n}) & \approx & \frac{\partial F_{i}}{\partial y_{j}^{n}}\Delta y_{j}^{n}\,.
\end{eqnarray}
In the end, the equations $F_{i}$ should be fulfilled after the (final)
iteration, i.e.~we want to achieve that $F_{i}(y_{j}^{n+1})=0$ for
all combinations of $i$ and $j$. Note that for a system of $n$
equations and $n$ unknowns, the derivative on the right hand side
is a matrix of dimension $n$. We will denote it by 
\begin{equation}
F_{ij}^{'}=\frac{\partial F_{i}}{\partial y_{j}^{n}}\,.
\end{equation}
Inverting this matrix allows us to compute the change in the functions
$y_{j}$:
\begin{equation}
\Delta y_{j}^{n}=-\left(F^{'}\right)_{ji}^{-1}F_{i}(y_{j}^{n})\,,\label{eq:delta_y}
\end{equation}
The Newton method allows us to write down the iteration procedure
for $y_{j}$,
\begin{eqnarray}
y_{j}^{n+1} & = & y_{j}^{n}+\Delta y_{j}^{n}\,.
\end{eqnarray}
The Newton method now has to be performed at each lattice point. Special
considerations have to be taken at the boundary of the lattice: it
is convenient to add two extra points at each end of the domain at
a fixed value in order to have a well defined derivative at both ends
of the calculation domain. Over relaxation means that we multiply
$\Delta y_{i}^{n}$ with a parameter $w$. For faster convergence,
it can be shown that $\omega$ has to be between $1<w<2$ with $w\in\mathbf{R}$.
\begin{equation}
y_{j}^{n+1}=y_{j}^{n}-w\left(F^{'}\right)_{ji}^{-1}F_{i}(y_{j}^{n})\,.
\end{equation}
Depending on the complexity of the problem, $\omega$ cannot be chosen
arbitrarily close to its maximum value. If it is chosen too large,
the iteration will diverge quickly.

In addition, we can use the Gauss-Seidel method for faster convergence.
This method uses the newest results of the current iteration for lattice
points that already have been computed. This means that for the $n-th$
iteration at the point $x_{i}$, the values of the same iteration
instead of the $(n-1)-th$ iteration for points at which the iteration
has already been performed are used. 

For the first iteration, we have to provide an initial guess for all
functions we want to compute, where a better initial guess usually
leads to a faster convergence. The initial guess has to obey the boundary
conditions of the final solution, for a superconducting flux tube
the use of the $\tanh(R)$ function has to be proven successful. For
a successive calculation of profile functions for various external
parameters, the full solution for the preceding parameter can be used
as an initial guess, at least if the step size in the external parameter
is small enough. 

The procedure for solving the profile functions can be summarized
as follows:
\begin{enumerate}
\item Define all external parameters and functions in Mathematica and write
them into a text file.
\item Start the computation of a given set of parameters by executing the
FORTRAN program, which can be started from within Mathematica with
the "RunProcess" command. The return code of the program is transferred
back to Mathematica automatically.
\item The FORTRAN SOR-program saves the data into a text file
\item Read in the results in Mathematica from the text file and perform
all data processing, including integration of curves and the calculation
of the free energy, critical fields and so on in Mathematica.
\end{enumerate}
\newpage
\section{FORTRAN Code for Neutron-Proton System}
This is the code used for solving the equations of motions Eqs.~(\ref{eomA}) of the two-component system. Note that this code uses the parameters from Ref.~\cite{Alford:2007np}, which can be easily translated by following the footnote below Eq.~(\ref{eomA}).
\begin{lstlisting}[language=Fortran]
program fluxtubesSOR

implicit none

integer:: nstart,nend,ninc,np,it,maxit,n,j,i
double precision:: eps, inf, h, nvev, pvev, bet
double precision:: kap, sig, w, ann, app, accgoal
character::tmp
double precision, DIMENSION(:), ALLOCATABLE :: f1,f2,g1,g2,a1,a2,r

!define input formats

200 format(A1)
400 format(I4)
300 format(F10.5)

!open parameter input file (input written into file from Mathematica)
open(2,file='para.dat',status='old', action='read')

!read physical parameters
read(2,*) nstart !start of winding number n scan
read(2,*) nend ! end of n
read(2,*) ninc !increment of scan in n
read(2,*) kap !kappa
read(2,*) bet !non entr coupling
read(2,*) sig !entraiment coupling
read(2,*) nvev ! vev of neutral field
read(2,*) pvev ! vev of charged field
read(2,*) app ! self coupling of charged field
read(2,*) ann ! self coupling of neutral field

read(2,200) tmp
!read numerical parameters
read(2,*) it !maximal number of iterations
read(2,*) np !number of gridpoints
read(2,*) w  !overshooting paramter
read(2,*) accgoal !accuracy goal
read(2,*) eps ! epsilon = "zero point" = start of radial coordinate
read(2,*) inf ! "infinity" end of radial coordinate

close(2)

!allocate tables for proton (f) neuton (g) and gauge field (a) profile functions
allocate(f1(1:np+1),g1(1:np+1),a1(1:np+1),f2(1:np+1),g2(1:np+1),a2(1:np+1))
allocate(r(1:np+1))

!first initial guess for n=1

h=(inf-eps)/np !grid stepsize

!initalize grid
do i=1,np+1,1
	r(i)=eps+(i-1)*h
end do

!initial guess for functions

do i=1,np+1,1
	g1(i)=1d0
	f1(i)=tanh(2d0*r(i)/3d0)
	a1(i)=f1(i)
end do

!BC's for first guess
f1(1:2)=0
a1(1:2)=0
f1(np:np+1)=1	
g1(np:np+1)=1
a1(np:np+1)=1


!call subroutine for relaxation
open(3, file='f.dat' , status='replace')
open(4, file='g.dat' , status='replace')
open(5, file='a.dat' , status='replace')

do n=nstart,nend,ninc

	call sor(n,kap,bet,sig,nvev,pvev,app,ann,it,np,w,accgoal,eps,inf,f1,f2,g1,g2,a1,a2,r)

	!safe results in files 	
	do j=1,np+1
		write(3,*) r(j),f2(j)
		write(4,*) r(j),g2(j)
		write(5,*) r(j),a2(j)
	end do
	write(3,*) !write empty lines to separate results for different winding numbers
	write(4,*) 
	write(5,*) 
	
end do

close(3)
close(4)
close(5)

end program

subroutine sor(n,kap,bet,sig,nvev,pvev,app,ann,it,np,w,accgoal,eps,inf,f1,f2,g1,g2,a1,a2,r)
implicit none

integer:: np,it,maxit,n,i,j
double precision:: eps, inf, h, nvev, pvev, bet 
double precision:: kap, sig, w, ann, app, accgoal, check
double precision, DIMENSION(1:np+1) :: f1,f2,g1,g2,a1,a2,r,fp,gp,ap,det,f,g,a,F11,F12,F21,F22,F23,F32,F33

maxit=0 !final iteration when accuaracy goal is reached

!relaxation, inital guess is provided by main program
h=(inf-eps)/np !grid stepsize

do i=1,it
	f1(1)=0
	a1(1)=0
	f1(np:np+1)=1	
	g1(np:np+1)=1
	a1(np:np+1)=1
	f2(1)=0
	a2(1)=0
	f2(np:np+1)=1	
	g2(np:np+1)=1
	a2(np:np+1)=1
	do j=np-1,2,-1

	a(j) =	(a1(j-1) - 2d0*a1(j) + a2(j+1) - (h*(-a1(j-1) + a2(j+1)))/(2d0*r(j)) + &
  	(h**2*(1 - a1(j))*f1(j)**2)/kap**2)


	f(j) = f1(j-1)-2d0*f1(j)+f2(1+j)+(h*(-f1(j-1)+f2(1+j)))/(2d0*r(j))+h**2* &
	(-((n**2*(1d0-a1(j))**2*f1(j))/r(j)**2)&
	-f1(j)*(-1d0+f1(j)**2) - (nvev**2/pvev**2*bet*f1(j)*(-1d0 + g1(j)**2)) - &
 	(nvev/pvev*sig*((f1(j)*(-g1(j-1) + g2(j+1))**2)/(4.*h**2) + &
 	f1(j)*g1(j)*((-g1(j-1) + g2(j+1))/(2d0*h*r(j)) + (g1(j-1) - 2d0*g1(j) + g2(j+1))/h**2))))

	g(j) = g1(j-1) - 2d0*g1(j) +h**2*(-(bet*(-1 + f1(j)**2)*g1(j)) - &
	(pvev*sig*(((-f1(j-1) + f2(j+1))**2*g1(j))/(4*h**2) + f1(j)*((-f1(j-1) + f2(j+1))/(2d0*h*r(j)) +& 
	(f1(j-1) - 2d0*f1(j) + f2(j+1))/h**2)*g1(j)))/nvev -(ann*nvev**2*g1(j)*(-1 + g1(j)**2))/&
	(app*pvev**2)) + g2(j+1) + (h*(-g1(j-1) + g2(j+1)))/(2d0*r(j))

	F11(j) = -2d0-h**2*f1(j)**2d0/kap**2
	F12(j) = h**2d0*2*f1(j)/kap**2d0*(1-a1(j))

	F21(j) = h**2d0*f1(j)*n**2d0/r(j)**2d0*2d0*(1-a1(j))	
	F22(j) = (-2-h**2*n**2*(1d0-a1(j))**2/r(j)**2-h**2*(3d0*f1(j)**2-1d0)-&
	h**2*sig*nvev/pvev*(g1(j)*(((g2(j+1)-2d0*g1(j)+g1(j-1))/ &
       	h**2)+(g2(j+1)-g1(j-1))/2d0/h/r(j)) + (g2(j+1)-g1(j-1))**2/(2d0*h)**2)-h**2*bet*nvev**2/pvev**2*&
		(g1(j)**2-1d0))
	F23(j) = -2*bet*(nvev/pvev)**2d0*f1(j)*g1(j)*h**2d0
	
	F32(j) = -h**2d0*2*bet*f1(j)**2d0*g1(j)**2d0
	F33(j) = -2 + h**2*(-(bet*(-1 + f1(j)**2)) - (pvev*sig*((-f1(j-1) + f2(j+1))**2/(4.*h**2) + &
	f1(j)*((-f1(j-1) + f2(j+1))/(2d0*h*r(j)) +  (f1(j-1) - 2d0*f1(j) + f2(j+1))/h**2)))/ &
	nvev-(ann*nvev**2*(-1d0+3d0*g1(j)**2))/(app*pvev**2))

	det(j)=-F11(j)*F23(j)*F32(j)-F12(j)*F21(j)*F33(j)+F11(j)*F22(j)*F33(j)

	a2(j) = a1(j) - w/det(j)*(F12(j)*F23(j)*g(j) - a(j)*F23(j)*F32(j)-F12(j)*f(j)*F33(j) + a(j)*F22(j)*F33(j))
	f2(j) = f1(j) - w/det(j)*(-F11(j)*F23(j)*g(j) + F11(j)*f(j)*F33(j) - a(j)*F21(j)*F33(j))
	g2(j) = g1(j) - w/det(j)*(-F12(j)*F21(j)*g(j) + F11(j)*F22(j)*g(j) - F11(j)*f(j)*F32(j) + a(j)*F21(j)*F32(j))
	end do

	g2(1)=g2(2) !enforce BC for g
	!check for convergence
	If(check(f1,f2,np).le.accgoal.and.check(g1,g2,np).le.accgoal.and.check(a1,a2,np).le.accgoal) then
		write(*,*) "Accuarcygoal of ",accgoal," reached at nit=",maxit
		write(*,*) "accuarcy is for proton ",check(f1,f2,np)," neutron ",check(g1,g2,np)," gauge field ",check(a1,a2,np)
		return
	end if 

	!"rewind" for next iteration step
	f1(1:np+1)=f2(1:np+1) 
	g1(1:np+1)=g2(1:np+1) 
	a1(1:np+1)=a2(1:np+1) 
	maxit=maxit+1 !counter for iteration
end do
write(*,*) maxit
end subroutine sor

double precision function check(f1,f2,np)
implicit none
double precision, DIMENSION(1:np+1) :: f1,f2
integer:: np

check = sqrt((sum(f2)-sum(f1))**2/(np+1))

end function


\end{lstlisting}

\chapter{Calculation of \texorpdfstring{$H_{c2}$}{Hc2} and Gibbs Free Energy just below \texorpdfstring{$H_{c2}$}{Hc2}}
\label{app:Hc2}

Here we derive Eqs.\ (\ref{HcHc2}) and (\ref{DeltaG}). 
To this end, we need the equations of motion for the scalar fields and the gauge field. We go back to the Lagrangian (\ref{L}), 
take the static limit and replace the parameters $m_i$ and $h$ by their $T$-dependent generalizations $m_{i,T}$ and $h_T$. 
This yields the potential 
\bea \label{freeener}
U&=&(\nabla-iq\mathbf{A})\vf_1\cdot(\nabla+iq\mathbf{A})\vf_1^*-(\mu_1^2-m_{1,T}^2)|\vf_1|^2+\lambda_1|\vf_1|^4+\nabla\vf_2\cdot \nabla\vf_2^*\non[2ex]
&&-(\mu_2^2-m_{2,T}^2)|\vf_2|^2+\lambda_2|\vf_2|^4-2h_T|\vf_1|^2|\vf_2|^2\non[2ex]&&
-\frac{G}{2}\left[\vf_1\vf_2(\nabla+iq\mathbf{A})\vf_1^*\nabla\vf_2^*+\vf_1\vf_2^*(\nabla+iq\mathbf{A})\vf_1^*\nabla\vf_2+c.c.\right] +\frac{B^2}{8\pi}\, ,
\eea
and the equations of motion for $\varphi_1^*$, $\varphi_2^*$, and $\mathbf{A}$ become
\begin{subequations}\label{eom} 
  \bea
 \hspace{-1cm} \left[(\nabla-iq\mathbf{A})^2 + \mu_1^2-m_{1,T}^2-2\lambda_1|\vf_1|^2+2h_T|\varphi_2|^2 \right]\vf_1-G \varphi_1\nabla\cdot\mathrm{Re}\,(\varphi_2\nabla\varphi_2^*)&=&0 \, , \label{eom1}\hspace{1cm}\\[2ex]
  \left(\Delta+\mu_2^2-m_{2,T}^2-2\lambda_2|\varphi_2|^2+2h_T|\varphi_1|^2\right)\varphi_2-G\varphi_2\nabla\cdot\mathrm{ Re}\,[\varphi_1(\nabla+iq\mathbf{A})\varphi_1^*]&=&0 \, , \label{eom2}\\[2ex]
 \nabla\times\mathbf{B}+8\pi q\,\mathrm{ Im}\,[\varphi_1(\nabla+iq\mathbf{A})\varphi_1^*] &=& 0 \, .\label{gauge}
  \eea
\end{subequations}
Since the transition from the flux tube phase to the 
normal-conducting phase is assumed to be of second order, the charged condensate becomes infinitesimally small just below $H_{c2}$, and we make the ansatz
$\varphi_1 = \bar{\varphi}_{1}+\delta\varphi_1$ with $\bar{\varphi}_{1} \propto (H_{c2}-H)^{1/2}$, and $\delta\varphi_1$ includes terms of order $(H_{c2}-H)^{3/2}$ and higher, i.e., 
is at least of order $\bar{\varphi}_1^3$.
We also introduce perturbations for the neutral condensate and the gauge field,
$\varphi_2=\bar{\varphi}_2+\delta\varphi_2$, $\mathbf{A} = (\bar{A}_{y}+\delta A_y) \mathbf{e}_y$, where $\delta A_y,\delta\varphi_2$ include terms of order $\propto H_{c2}-H$ and higher,
i.e., they are at least of order $\bar{\varphi}_{1}^2$.  As the magnetic field completely penetrates the superconductor at the phase transition, we can choose the 
unperturbed gauge field to be of the form 
$\bar{A}_{y} = x H_{c2}$, and we denote $\delta B = \partial_x\delta A_y$, such that $\mathbf{B} = (H_{c2}+\delta B)\mathbf{e}_z$. We assume all functions to be real and 
to depend on $x$ only, not on $y$ and $z$ (solutions with these properties are sufficient for our purpose, the derivation would also work without these
 restrictions but would be somewhat more tedious). We insert this ansatz into the equations 
of motion (\ref{eom}), and keep terms up to order $\bar{\varphi}_{1}^3$.
Then, the linear contributions from Eqs.\ (\ref{eom1}) and (\ref{eom2}) yield two equations for $\bar{\varphi}_1$ and $\bar{\varphi}_2$,
\begin{subequations} \label{lowest}
\bea
{\cal D}_1\bar{\varphi}_{1} &=& 0 \, ,\label{lowest1} \\[2ex]
{\cal D}_2\bar{\varphi}_2 &=& 0 \, ,\label{lowest2}
\eea
\end{subequations}
with
\begin{subequations} 
\bea
{\cal D}_1&\equiv& \partial_x^2-q^2\bar{A}_y^2+\mu_1^2-m_{1,T}^2+2h_T\bar{\varphi}_{2}^2-G\partial_x(\bar{\varphi}_2\partial_x\bar{\varphi}_2) \, , \\[2ex]
{\cal D}_2&\equiv& \partial_x^2+\mu_2^2-m_{2,T}^2-2\lambda_2\bar{\varphi}_{2}^2 \, ,
\eea
\end{subequations}
while the subleading contributions from Eqs.\ (\ref{eom1}) and (\ref{eom2}) and the leading contribution from Eq.\ (\ref{gauge}) yield the following equations for 
the perturbations $\delta\varphi_1$, $\delta\varphi_2$, and $\delta A_y$, 
\begin{subequations}\label{higher} 
\bea
{\cal D}_1\delta \varphi_1 &=& 
\Big[2(q^2\bar{A}_{y}\delta A_y+\lambda_1\bar{\varphi}_{1}^2
-2h_T\bar{\varphi}_2\delta\varphi_2)+G\partial_x^2(\bar{\varphi}_2\delta\varphi_2)\Big]\bar{\varphi}_1 \, , \label{phi31}\\[2ex]
{\cal D}_2\delta \varphi_2&=& \Big[2(2\lambda_2\bar{\varphi}_2\delta\varphi_2-h_T\bar{\varphi}_1^2)+G\partial_x(\bar{\varphi}_1\partial_x\bar{\varphi}_1)\Big]\bar{\varphi}_2 \, ,\label{rho31}\\[2ex]
\partial_x^2\delta A_y &=& -8\pi q^2\bar{A}_y\bar{\varphi}_1^2 \, .\label{gauge21}
\eea
\end{subequations}
Inserting our ansatz into the potential (\ref{freeener}), using partial integration and the equations of motion (\ref{lowest}) and (\ref{higher}), and keeping terms up to 
order $\bar{\varphi}_{1}^4$, we find after some algebra the free energy
\be \label{FGh}
F = \int d^3r \left\{\frac{B^2}{8\pi}-\lambda_1\bar{\varphi}_1^4-\lambda_2\bar{\varphi}_2^4 +\bar{\varphi}_2\delta\varphi_2[2h_T\bar{\varphi}_1^2-G\partial_x(\bar{\varphi}_1\partial_x\bar{\varphi}_1)]\right\} \, .
\ee
We will first compute $H_{c2}$ from Eqs.\ (\ref{lowest}) and afterwards compute the Gibbs free energy just below $H_{c2}$ from Eq.\ (\ref{FGh}). 

We assume the neutral condensate in the SF phase to be homogeneous, and thus Eq.\ (\ref{lowest2}) yields $2\bar{\varphi}_2^2=\rho_\mathrm{ SF}^2$, as expected. 
For the solution of Eq.\ (\ref{lowest1}) we can simply follow the textbook arguments because it has the 
same structure as for a single-component superconductor. It reads 
\be \label{z}
(-\partial_x^2 +q^2H_{c2}^2x^2)\bar{\varphi}_{1}=(\lambda_1\rho_\mathrm{ SC}^2+h_T\rho_\mathrm{ SF}^2)\bar{\varphi}_{1}   \, ,
\ee
and thus is equivalent to the Schr\"{o}dinger equation for the one-dimensional harmonic oscillator, $-\frac{\hbar^2}{2m}\psi''(x) +\frac{m}{2}\omega^2x^2\psi=E\psi$ 
with the identification $E/(\hbar\omega)=(\lambda_1\rho_\mathrm{ SC}^2+h_T\rho_\mathrm{ SF}^2)/(2q H_{c2} )$. Since the eigenvalues are $E_n =(n+\frac{1}{2})\hbar\omega$, 
the largest magnetic field for which the equation allows a physical solution is obtained by setting $n=0$,
\be
H_{c2} = \frac{\lambda_1\rho_\mathrm{ SC}^2}{q}\left(1+\frac{h_T\rho_\mathrm{ SF}^2}{\lambda_1\rho_\mathrm{ SC}^2}\right)
=\frac{1}{q\xi^2}\left(1-\frac{h_T^2}{\lambda_1\lambda_2}\right)\, ,
\ee
in agreement with Eq.\ (13) of Ref.\ \cite{Sinha:2015bva}.  In the second expression we have rewritten the condensates 
$\rho_\mathrm{ SC}$ and $\rho_\mathrm{ SF}$ in terms of the charged condensate in the coexistence phase $\rho_{01}$, see Eq.\ (\ref{COE}),
 and used the definition of the coherence length $\xi$ from Eq.\ (\ref{ellxi}). 
Since the relevant eigenvalue of Eq.\ (\ref{z}) is given by $n=0$, the corresponding eigenfunction is  a Gaussian,
\be 
\bar{\vf}_{1}(x) = C_0 e^{-x^2qH_{c2}/2} \, , \label{eq:SEsol}
\ee
where the exact value of the prefactor $C_0\propto (H_{c2}-H)^{1/2}$ is not relevant for the following. The result shows that, for $H$ just below $H_{c2}$, charged 
condensation with small magnitude of order $(H_{c2}-H)^{1/2}$ occurs in a slab confined in a direction perpendicular to the external magnetic field, here chosen to be the  
$x$-direction, with width $(qH_{c2})^{-1/2}$. Had we allowed for $y$ and $z$ dependencies of the condensate, we could have 
used this linearized approximation to discuss crystalline configurations and determine the preferred lattice structure. Here we continue by checking whether the solution (\ref{eq:SEsol}) is energetically preferred over the normal-conducting phase for $H$ below and close to $H_{c2}$.
To this end, we need to compute the Gibbs free energy, as defined in Eq.\ (\ref{Gibbsdef}), from the free energy (\ref{FGh}). We first solve Eq.\ (\ref{gauge21}) 
with the boundary condition $\delta B(\pm\infty)=H-H_{c2}$ (since $B=H$ in the normal-conducting phase) to find
\be 
\delta B(x) =-(H_{c2}-H)+4\pi q\bar{\varphi}_{1}^2(x) \, .
\ee
Inserting this result into Eq.\ (\ref{FGh}) and using Eq.\ (\ref{eq:SEsol}) yields the Gibbs free energy 
\be \label{gibbs_Hc2}
{\cal G}_\mathrm{ COE}= \mathcal{G}_\mathrm{ SF}+\int d^3r\left\{\left(\frac{1}{2\kappa^2}-1\right)\lambda_1\bar{\varphi}_1^4 +\bar{\varphi}_2\delta\varphi_2\Big[2h_T\bar{\varphi}_1^2-G\partial_x(\bar{\varphi}_1\partial_x\bar{\varphi}_1)\Big]\right\} \, , 
\ee
with $\mathcal{G}_\mathrm{ SF}$ from Eq.\ (\ref{GibbsSF}). 
It remains to compute $\delta\varphi_2$. We use Eq.\ (\ref{rho31}), which can be written as
\be
(\partial_t^2-p^2)\delta\varphi_2(t) = -\frac{h_T p^2 C_0^2}{2\sqrt{2}\lambda_2\rho_\mathrm{ SF}} e^{-t^2}(2+\gamma-2\gamma t^2) \, ,
\ee
with the dimensionless variable $t=\sqrt{q H_{c2}} \,x$ and the dimensionless quantities
\be
p^2 = \frac{2\lambda_2\rho_\mathrm{ SF}^2}{qH_{c2}} \, , \qquad \gamma = \frac{GqH_{c2}}{h_T} \, ,
\ee
where $p$ indicates the magnitude of the neutral condensate and $\gamma$ the magnitude of the gradient coupling $G$ relative to the density coupling $h_T$, both in units given by the critical magnetic field.  With the boundary conditions $\delta\varphi_2(\pm \infty)=0$, this equation has the solution
\be \label{dphi2t}
\delta\varphi_2(t) = \frac{1}{2}\frac{h_T p^2 C_0^2}{2\sqrt{2}\lambda_2\rho_\mathrm{ SF}}\left[\gamma e^{-t^2}+\frac{\sqrt{\pi}}{p}\left(1-\frac{p^2\gamma}{4}\right){\cal Z}(p,t)\right] \, , 
\ee
where we have abbreviated
\be
{\cal Z}(p,t)\equiv e^{p^2/4}\left\{
e^{pt}\left[1-\mathrm{ erf}\left(\frac{p}{2}+t\right)\right]+e^{-pt}\left[1-\mathrm{ erf}\left(\frac{p}{2}-t\right)\right]\right\} \, ,
\ee
with the error function erf. Inserting Eq.\ (\ref{dphi2t}) into Eq.\ (\ref{gibbs_Hc2}) yields 
\bea \label{DeltaGfull}
\frac{{\cal G}_\mathrm{ COE}}{V}\hspace*{-.2cm} &=& \hspace*{-.2cm}\lambda_1\langle\bar{\varphi}_1^4\rangle\left(\frac{1}{2\kappa^2}-1+\frac{h_T^2}{\lambda_1\lambda_2}
\left\{\frac{p^2\gamma}{4}\left(1+\frac{\gamma}{4}\right)+\left(1-\frac{p^2\gamma}{4}\right)\left[\left(1+\frac{\gamma}{2}\right){\cal I}_1(p)-\gamma{\cal I}_2(p)\right]\right\}\right) \non[2ex]
&&+\frac{{\cal G}_\mathrm{ SF}}{V}\, ,
\eea
where $\langle\ldots\rangle$ denotes spatial average, and 
\be
{\cal I}_1(p) \equiv \frac{p}{2\sqrt{2}}\int_{-\infty}^\infty dt\,e^{-t^2} {\cal Z}(p,t) \, , \qquad {\cal I}_2(p) \equiv \frac{p}{2\sqrt{2}}\int_{-\infty}^\infty dt\,t^2e^{-t^2} {\cal Z}(p,t) \, .
\ee
We discuss this result for the case without gradient coupling, $\gamma=0$, in the main text.

\chapter{Interaction between two Flux Tubes}
\label{app:fl_int}

In this appendix we derive the expression for the interaction energy Eq.\ (\ref{Fint}). We start from the definition (\ref{Fintdef}), i.e., we consider 
two parallel flux tubes $(a)$ and $(b)$ separated by the (dimensionless) distance $R_0$. We divide the total volume $V$ into
two half-spaces $V^{(a)}$ and $V^{(b)}$, which are the simplest versions of two Wigner-Seitz cells: we connect the two flux tubes by a line with length $R_0$, 
and the plane in the center of and 
perpendicular to that line divides $V$ into $V^{(a)}$ and $V^{(b)}$. The interaction free energy is then computed from 
\be \label{FintAB}
F_\mathrm{ int}^\circlearrowleft = 2\int_{V^{(a)}}d^3r \, \left[U_\circlearrowleft^{(a)+(b)} - U^{(a)}_\circlearrowleft - U^{(b)}_\circlearrowleft\right] \, , 
\ee
where, due to the symmetry of the configuration, we have restricted the integration to the half-space $V^{(a)}$, where $U^{(a)}_\circlearrowleft$, $U^{(b)}_\circlearrowleft$ are 
the free energy densities of the two flux tubes in the absence of the other flux tube, and where $U_\circlearrowleft^{(a)+(b)}$ is the total free energy of the flux tubes.
(Recall that by definition $U_\circlearrowleft$ denotes the pure flux tube energy density, with the free energy density of the homogeneous configuration already subtracted.)

We assume $R_0$ to be much larger than the widths of the flux tubes, such that the contribution of flux tube $(b)$ to the free energy is small in $V^{(a)}$. Therefore, we will now compute the free energy density of a "large" contribution that solves the full equations of motion plus a "small" contribution that solves the linearized equations of motion. We shall do so in a  general notation, not referring to the geometry of our two-flux tube setup. Only in Eq.\ (\ref{Fint3}), when we insert the results into the free energy (\ref{FintAB}), we shall come back to this setup and introduce a more explicit notation indicating the contributions of the two different flux tubes. 
Following Ref.\ \cite{Kramer:1971zza}, we define
\be
\mathbf{Q} \equiv \xi (q\mathbf{A}-\nabla\psi_1) = -\frac{n(1-a)}{R}\mathbf{e}_\theta \, ,
\ee
and write   
\begin{subequations}
\bea
\mathbf{Q} &=& \mathbf{Q}_0 + \delta\mathbf{Q} \, , \\[2ex]
f_1 &=& f_{10} + \delta f_1 \, , \\[2ex]
f_2 &=& f_{20} + \delta f_2 \, .
\eea
\end{subequations} 
The equations of motion for a single flux tube to leading order, $\delta\mathbf{Q} = \delta f_1 = \delta f_2 =0$, are (from now on, in this appendix, all gradients are taken with respect to the dimensionless coordinates)
\begin{subequations}
\bea
0&=&\nabla\times(\nabla\times \mathbf{Q}_0) + \frac{f_{10}^2}{\kappa^2}\mathbf{Q}_0  \, , \label{zeroQ}\\[2ex]
0&=&\Delta f_{10}+ f_{10}(1-f_{10}^2-Q_0^2) -\frac{h_T}{\lambda_1} x^2 f_{10}(1-f_{20}^2)- \frac{\Gamma x}{2}f_{10}\nabla\cdot(f_{20}\nabla f_{20}) \, , \label{zerof1}\\[2ex]
0&=&\Delta f_{20}+ \frac{\lambda_2}{\lambda_1}x^2 f_{20}(1-f_{20}^2) - \frac{h_T}{\lambda_1} f_{20}(1-f_{10}^2) -\frac{\Gamma}{2x}f_{20}\nabla\cdot(f_{10}\nabla f_{10})\, , \label{zerof2}
\eea
\end{subequations}
[equivalent to Eqs.\ (\ref{eomA}) in the main text], and the equations of motion of first order in the corrections $\delta\mathbf{Q}$, $\delta f_1$, $\delta f_2$ become
\begin{subequations}\label{oneQ12}
\bea
0&=&\nabla\times(\nabla\times \delta\mathbf{Q})+ \frac{f_{10}}{\kappa^2}(f_{10}\delta\mathbf{Q}+2\delta f_1\mathbf{Q}_0)    \, , \label{oneQ}\\[2ex]
0&=& -\mathbf{Q}_0\cdot(2f_{10}\delta\mathbf{Q}+\delta f_1\mathbf{Q}_0)+\Delta \delta f_{1}+ \delta f_1(1-3f_{10}^2)-\frac{h_T}{\lambda_1} x^2 [\delta f_1(1-f_{20}^2) -2f_{10}f_{20}\delta f_2]\non[2ex]
&&- \frac{{\Gamma }x}{2}[\delta f_1\nabla\cdot (f_{20}\nabla f_{20})+f_{10}\Delta(f_{20} \delta f_2)] \, ,\label{onef1} \\[2ex]
0&=&\Delta \delta f_{2}+ \frac{\lambda_2}{\lambda_1}x^2\delta f_2(1-3f_{20}^2)-\frac{h_T}{\lambda_1} [\delta f_2(1-f_{10}^2) -2f_{10}f_{20}\delta f_1]\non[2ex]
&&- \frac{{\Gamma }}{2x}[\delta f_2\nabla\cdot (f_{10}\nabla f_{10})+f_{20}\Delta(f_{10} \delta f_1)]\, . \label{onef2}
\eea
\end{subequations}
We denote the free energy density, up to second order and after using the equations of motions, by $U_0 + \delta U$, where 
\bea
U_0 \hspace{-.1cm}&=&\hspace{-.1cm} \frac{\rho_{01}^2}{2} \left\{\kappa^2(\nabla\times \mathbf{Q}_0)^2 +(\nabla f_{10})^2+f_{10}^2Q_0^2+\frac{(1-f_{10}^2)^2}{2} +x^2
\left[(\nabla f_{20})^2+ \frac{\lambda_2}{\lambda_1}x^2\frac{(1-f_{20}^2)^2}{2}\right]\right. \non[2ex]
&&\hspace{-.1cm} \left.-\frac{h_T}{\lambda_1} x^2 (1-f_{10}^2)(1-f_{20}^2)-{\Gamma}x f_{10}f_{20}\nabla f_{10}\cdot\nabla f_{20}\right\}  
\eea
is the free energy density of a single flux tube from Eq.\ (\ref{Efl}), and the first-order and second-order corrections can be written as a total derivative, 
\bea
\delta U &=&\rho_{01}^2 \nabla\cdot \Bigg\{\kappa^2 \delta\mathbf{Q}\times\left[\nabla\times\left(\mathbf{Q} _0+\frac{\delta\mathbf{Q}}{2}\right)\right] 
+\delta f_1\nabla\left(f_{10}+\frac{\delta f_1}{2}\right)
+x^2\delta f_2\nabla\left(f_{20}+\frac{\delta f_2}{2}\right) \non[2ex]
&&-\frac{{\Gamma}x}{2} \Bigg[\delta f_1\left(f_{10}+\frac{\delta f_1}{2}\right)f_{20}\nabla f_{20}+\delta f_2\left(f_{20}+\frac{\delta f_2}{2}\right)f_{10}\nabla f_{10}+\frac{1}{2}\nabla(f_{10}f_{20}\delta f_1 \delta f_2)\Bigg]
\Bigg\}   \, . \non[0.5ex]
\hfill
\eea
Notice that any explicit dependence on the density coupling $h_T$ has disappeared, while the derivative coupling $\Gamma$ does appear explicitly.

We can now go back to the interaction free energy (\ref{FintAB}) and identify the full free energy $U_\circlearrowleft^{(a)+(b)}$ in the half-space $V^{(a)}$with $U_0+\delta U$. 
In $V^{(a)}$, $U_\circlearrowleft^{(a)}$ is given by setting $\delta \mathbf{Q}=\delta f_1 = \delta f_2=0$ in $U_0 + \delta U$
(which simply leaves $U_0$), and
$U^{(b)}_\circlearrowleft$ is obtained by setting $\mathbf{Q}_0=0$, $f_{10} = f_{20}=1$ in $U_0 + \delta U$ (which leaves various terms from $\delta U$). Consequently, we find
\begin{align}\label{Fint3}
F_\mathrm{ int}^\circlearrowleft
\simeq&\ 2 \rho_{01}^2 \int_{\partial V^{(a)}} d\mathbf{S}\cdot\Bigg\{\kappa^2 \delta\mathbf{Q}^{(b)}\times\left(\nabla\times\mathbf{Q} _0^{(a)}\right) +\delta f_1^{(b)}\nabla f_{10}^{(a)}+x^2\delta f_2^{(b)}\nabla f_{20}^{(a)} \non[2ex]
&-\frac{\Gamma x}{2}\Bigg[\delta f_1^{(b)}\left(f_{10}^{(a)}
+\frac{\delta f_1^{(b)}}{2}\right)f_{20}^{(a)}\nabla f_{20}^{(a)}+\delta f_2^{(b)}\left(f_{20}^{(a)}+\frac{\delta f_2^{(b)}}{2}\right)f_{10}^{(a)}\nabla f_{10}^{(a)}\non[2ex]
&+\frac{1}{2}\nabla(f_{10}^{(a)}f_{20}^{(a)}\delta f_1^{(b)} \delta f_2^{(b)})-\frac{1}{2}\nabla(\delta f_1^{(b)}\delta f_2^{(b)}) \Bigg] \Bigg\} \, , 
\end{align}
where we have rewritten the volume integral as a surface integral and where we have made the contributions from the two flux tubes $(a)$ and $(b)$ explicit. 
Since the derivatives of all fields vanish at infinity, the integration surface is reduced to the 
plane that separates the two 
Wigner-Seitz cells. We now use the geometry of the setup to simplify this expression: we align the $z$-axis with flux tube $(a)$, such that this flux tube sits 
in the origin of the $x$-$y$ plane, with the $x$-axis connecting the two flux tubes. Therefore, $\mathbf{Q}^{(a)}$, $f_{10}^{(a)}$, $f_{20}^{(a)}$ are functions only of $R$, while 
$\delta \mathbf{Q}^{(b)}$, $\delta f_{1}^{(b)}$, $\delta f_{2}^{(b)}$ also depend on the azimuthal angle $\theta$. However, since we only need the functions and their gradients at the boundary between the two Wigner-Seitz cells and since this boundary is by assumption far away not only from flux tube $(b)$ but also from flux tube $(a)$, we can write ($i=1,2$)
\begin{subequations} \label{surface}
\bea
\mathbf{Q}_0^{(a)} \hspace*{-.3cm}&\simeq&\hspace*{-.3cm} \delta \mathbf{Q}^{(a)} \equiv - \delta Q \,\mathbf{e}_\theta = \delta Q (\sin\theta \, \mathbf{e}_x - \cos\theta \, \mathbf{e}_y) \, , \\[2ex]
\delta \mathbf{Q}^{(b)} \hspace*{-.3cm}&=&\hspace*{-.3cm} -\delta Q (\sin\theta \, \mathbf{e}_x + \cos\theta \, \mathbf{e}_y) \, ,  \
f_{i0}^{(a)} \simeq 1- \delta f_i^{(a)} \, , \quad \delta f_{i}^{(b)}  = \delta f_i^{(a)}  \equiv \delta f_{i} \, , \\[2ex]
\nabla f_{i0}^{(a)}\hspace*{-.3cm}&\simeq&\hspace*{-.3cm} -\nabla \delta f_i^{(a)} = -\delta f_i'\,\mathbf{e}_R = -\delta f_i'(\cos\theta \, \mathbf{e}_x + \sin\theta \, \mathbf{e}_y) \, , \ 
\nabla \delta f_{i}^{(b)} = \delta f_i' (-\cos\theta \, \mathbf{e}_x + \sin\theta \, \mathbf{e}_y) \, . \non[.5ex]
\hfill
\eea
\end{subequations}
Note in particular that, at the relevant surface, $\delta f_{i}^{(b)}  = \delta f_i^{(a)}$, but $d\mathbf{S}\cdot \nabla \delta f_i^{(a)} = -d\mathbf{S}\cdot \nabla \delta f_i^{(b)}$.
Now, $\delta Q$ and $\delta f_i$ are functions only of $R$. Inserting Eqs.\ (\ref{surface}) into Eq.\ (\ref{Fint3}) yields 
\bea
\frac{F_\mathrm{ int}^\circlearrowleft}{L} &=& 2 \rho_{01}^2R_0\int_{R_0/2}^\infty \frac{dR}{\sqrt{R^2-(R_0/2)^2}} \left\{-\kappa^2 \delta Q\left(\frac{\delta Q}{R}+\delta Q'\right) + \delta f_1 \delta f_1' + x^2\delta f_2\delta f_2' \right.\non[2ex]
&& \hspace{4.3cm}\left.-\frac{\Gamma x}{4}[2(1-\delta f_1-\delta f_2)+\delta f_1\delta f_2](\delta f_1\delta f_2)'\right\} \, .
\eea
We can employ this result by inserting the modified Bessel functions from Eq.\ (\ref{asympsol}),
\begin{subequations} 
\bea
\delta Q &\simeq&  -nC K_1(R/\kappa) \, , \\[2ex]
\delta f_1 &\simeq& -D_+ \gamma_+ K_0(\sqrt{\nu_+}R) -D_- \gamma_- K_0(\sqrt{\nu_-}R) \, , \\[2ex]
\delta f_2 &\simeq& -D_+  K_0(\sqrt{\nu_+}R) -D_-  K_0(\sqrt{\nu_-}R) \, . 
\eea
\end{subequations} 
We may also extrapolate this result down to 
smaller distances by reinstating the full numerical functions through $\delta Q \to Q = -n(1-a)/R$ and $\delta f_i \to 1-f_i$, which yields the result (\ref{Fint}) in the main text.

\section{Asymptotic Approximation of Flux Tube Interaction with Gradient Coupling}
\label{app:asymp}

In the main text, we discuss the large-distance behavior of the flux tube interaction without gradient coupling. In the presence of a gradient coupling,
the interaction is more complicated, but, as we show in this appendix,  an equally compact expression can be derived if we are only interested in the leading order contributions, i.e., 
the exponential behavior. 

We start by inserting the  asymptotic solutions (\ref{asympsol}) into the expression for the interaction free energy (\ref{Fint}). The result is an integral over a sum of many terms, 
each of which is a product of 2, 3, or 4 modified Bessel functions of the second kind. In each product, one factor is $K_1$ and the remaining ones are $K_0$. 
The integral over the terms with 2 Bessel functions that have the same argument can be expressed again as a Bessel function with the help of Eq.\ (\ref{Kint}).
For the integral over all other products we use the expansion,  
\be \label{Besselexp}
K_n(z) = \sqrt{\frac{\pi}{2z}} e^{-z}\left[1+\frac{4n^2-1}{8z}+{\cal O}\left(\frac{1}{z^2}\right)\right]  \, ,
\ee
and only keep terms with the smallest exponential suppression. These terms are found as follows. With Eq.\  (\ref{Besselexp}) we approximate
\be \label{K0K1}
e^{-\alpha R}\simeq \frac{\alpha R}{\pi}K_0(\alpha R/2)K_1(\alpha R/2) \, . 
\ee
Then, we approximate each product of Bessel functions $K_0K_1$, $K_0K_0K_1$, $K_0K_0K_0K_1$ by the leading order term, and re-express the exponential 
as a product $K_0K_1$ with the help of Eq.\ (\ref{K0K1}). If we have started with a product $K_0K_1$ with different arguments, we arrive
at an expression which we can integrate using Eq.\ (\ref{Kint}). If we have started with a product of 3 or 4 Bessel function, we do not exactly reproduce the 
integrand of Eq.\ (\ref{Kint}) because there is an additional factor $R^{-1/2}$ (for 3 Bessel functions) or $R^{-1}$ (for 4 Bessel functions). The resulting integral can be 
expressed in terms of the so-called Meijer G-function, which we expand again since we are anyway only interested in the asymptotic behavior. As a result, we 
obtain 
\begin{subequations}  
\bea
\int_{R_0/2}^\infty dR\,\frac{K_0(\alpha_1 R)K_1(\alpha_2  R)}{\sqrt{R^2-(R_0/2)^2}} &\sim& e^{-\frac{\alpha_1+\alpha_2}{2}R_0} \, ,\\[2ex]
\int_{R_0/2}^\infty dR\,\frac{K_0(\alpha_1 R)K_0(\alpha_2 R)K_1(\alpha_3  R)}{\sqrt{R^2-(R_0/2)^2}} &\sim& e^{-\frac{\alpha_1+\alpha_2+\alpha_3}{2}R_0} \, ,\\[2ex]
\int_{R_0/2}^\infty dR\,\frac{K_0(\alpha_1 R)K_0(\alpha_2 R)K_0(\alpha_3 R)K_1(\alpha_4  R)}{\sqrt{R^2-(R_0/2)^2}} &\sim& e^{-\frac{\alpha_1+\alpha_2+\alpha_3+\alpha_4}{2}R_0} \, .
\eea
\end{subequations}
For each of the terms in the interaction energy we need to replace $\alpha_i$ by either $\sqrt{\nu_+}$ or $\sqrt{\nu_-}$. From Eq.\ (\ref{nugam}) we see that
$\sqrt{\nu_+}>\sqrt{\nu_-}$. Therefore, the largest contribution we obtain is $\exp(-\sqrt{\nu_-}R_0)$, and this contribution is only created by the product of 2 Bessel functions
with the same argument $\sqrt{\nu_-}$ because 2 Bessel functions with different arguments give rise to $\exp[-(\sqrt{\nu_+}+\sqrt{\nu_-})R_0/2]$, which is suppressed  more strongly, 
3 Bessel functions give rise to suppressions of at least $\exp[-3\sqrt{\nu_-}R_0/2]$ etc. The largest contributions are thus given by the terms where we can apply the integral 
(\ref{Kint}), and we obtain 
\be 
\frac{F_\mathrm{ int}^{\circlearrowleft}(R_0)}{L} \simeq2\pi\rho_{01}^2[\kappa^2 n^2 C^2 K_0(R_0/\kappa) - D_+^2(\gamma_-^2+x^2-\Gamma x\gamma_-) K_0(R_0\sqrt{\nu_-})] \, .
\ee
Therefore, if $\gamma_-^2+x^2-\Gamma x\gamma_->0$, one can use the same arguments as in the main text for the discussion of the attractiveness of the flux tube 
interaction at large distances, only with a more complicated eigenvalue $\nu_-$, which now depends on the gradient coupling $\Gamma$.

\chapter{Gell-Mann Matrices}
\label{app:gell-mann}
The Gell-Mann matrices are a presentation of the generators for the special unitary group $SU(3)$. 
\begin{align*}
\lambda_{1} & =\left(\begin{array}{ccc}
0 & 1 & 0\\
1 & 0 & 0\\
0 & 0 & 0
\end{array}\right)\,,\qquad\lambda_{2}=\left(\begin{array}{ccc}
0 & -i & 0\\
i & 0 & 0\\
0 & 0 & 0
\end{array}\right)\,,\qquad\lambda_{3}=\left(\begin{array}{ccc}
1 & 0 & 0\\
0 & -1 & 0\\
0 & 0 & 0
\end{array}\right)\,,\\
\lambda_{4} & =\left(\begin{array}{ccc}
0 & 0 & 1\\
0 & 0 & 0\\
1 & 0 & 0
\end{array}\right)\,,\qquad\lambda_{5}=\left(\begin{array}{ccc}
0 & 0 & -i\\
0 & 0 & 0\\
i & 0 & 0
\end{array}\right)\,,\qquad\lambda_{6}=\left(\begin{array}{ccc}
0 & 0 & 0\\
0 & 0 & 1\\
0 & 1 & 0
\end{array}\right)\,,\\
\lambda_{7} & =\left(\begin{array}{ccc}
0 & 0 & 0\\
0 & 0 & -i\\
0 & i & 0
\end{array}\right)\,,\qquad\lambda_{8}=\frac{1}{\sqrt{3}}\left(\begin{array}{ccc}
1 & 0 & 0\\
0 & 1 & 0\\
0 & 0 & -2
\end{array}\right)\,.
\end{align*}

\chapter{Rotated Electromagnetism}
\label{app:rotated}

Following Ref.~\cite{Schmitt:2003aa}, we will derive the mixing angels and rotated gauge fields
of the CFL and 2SC phase. We work with the standard $SU(3)_{c}$ generators $T_{a}=\frac{1}{2}\lambda_{a}$
where $\lambda_{a}$ are the $8$ Gell-Mann matrices. Furthermore
we need the electrical charge generator $Q$ in flavor space, where we are going to use a different ordering in the two phases, and the eighth and the third gluon generator $T_{8}$ and $T_{3}$:
\begin{equation}
T_{8}=\frac{1}{2\sqrt{3}}\left(\begin{array}{ccc}
1\\
 & 1\\
 &  & -2
\end{array}\right)\,,\qquad T_{3}=\frac{1}{2}\left(\begin{array}{ccc}
1\\
 & -1\\
 &  & 0
\end{array}\right)\,.
\end{equation}
All non-diagonal entries are zero. We introduce the rotated gauge fields via 
\begin{equation}
\left(\begin{array}{c}
\tilde{A}_{\mu}\\
\tilde{A}_{\mu}^{8}
\end{array}\right)=\left(\begin{array}{cc}
\cos\t & \sin\t\\
-\sin\t & \cos\t
\end{array}\right)\left(\begin{array}{c}
A_{\mu}\\
A_{\mu}^{8}
\end{array}\right)\,,\Rightarrow\left(\begin{array}{c}
A_{\mu}\\
A_{\mu}^{8}
\end{array}\right)=\left(\begin{array}{cc}
\cos\t & -\sin\t\\
\sin\t & \cos\t
\end{array}\right)\left(\begin{array}{c}
\tilde{A}_{\mu}\\
\tilde{A}_{\mu}^{8}
\end{array}\right)\,.
\end{equation}
We now want to find new generators $\tilde{Q}$ and $\tilde{T}_{8}$
in such a way that $\tilde{Q}\left(\Phi\right)=0$ by forming linear
combinations of $T_{8}$ and $Q$ in the following way:
\begin{align}
\tilde{Q} & =Q\otimes\mathbf{1}+\eta\mathbf{1}\otimes T_{8}\,,\\
\tilde{T}_{8} & =\mathbf{1}\otimes T_{8}+\lambda Q\otimes\mathbf{1}\,,
\end{align}
which determines $\eta$ depending on the phase. Note that $Q$ is
acting in flavor and $T_{8}$ in color space. From now on we will
omit the direct product with the unit element for simplicity. By requiring
that 
\begin{equation}
gA_{\mu}^{8}T_{8}+eA_{\mu}Q=\tilde{g}\tilde{A}_{\mu}^{8}\tilde{T}_{8}+\tilde{e}\tilde{A}_{\mu}\tilde{Q}\,,
\end{equation}
and comparing the coefficients of the generators we can immediately
read off that 
\begin{equation}
\tilde{g}=g\cos\t,\qquad\tilde{e}=e\cos\t\,.
\end{equation}
Furthermore, we conclude $-g\sin\t=\tilde{e}\eta=e\cos\t\eta$ which
allows us to compute the mixing angle,
\begin{equation}
\tan\t=\frac{e}{g}\eta\,,\qquad\cos^{2}\t=\frac{g^{2}}{g^{2}+\eta^{2}e^{2}}\,,
\end{equation}
where we use basic trigonometric relations to compute the cosine from
the tan directly. The relations allow us to determine $\lambda$ as
well,
\begin{equation}
\lambda=-\frac{e}{g}\tan\theta=-\frac{e^{2}}{g^{2}}\eta\,.
\end{equation}
The parameter $\eta$ has to be determined for each phase separately.
\section{CFL}
In the CFL phase we assume a diagonal diquark order parameter 
\begin{equation}
\Phi=\frac{1}{2}\left(\begin{array}{ccc}
\phi_{1}\\
 & \phi_{2}\\
 &  & \phi_{3}
\end{array}\right)\,,
\end{equation}
and use the generator of electromagnetism in the following form, 
\begin{equation}
Q=\frac{1}{3}\left(\begin{array}{ccc}
-1\\
 & -1\\
 &  & 2
\end{array}\right)\,,
\end{equation}
which means that the flavor indices are ordered $\bar{d},\,\bar{s},\,\bar{u}$ such that $Q$ is proportional to $T_8$. This ordering allows us to find a rotated generator $\tilde{Q}$ to which all scalar fields are neutral, $\tilde{Q}(\Phi)=0$, without involving $T_3$ in the linear combination. 
The requirement
of $\tilde{Q}(\Phi)$ leads to 
\begin{equation}
\tilde{Q}(\Phi) = 0  \Rightarrow \eta=\frac{2}{\sqrt{3}}\,,
\end{equation}
in accordance with Tab.~(1) of \cite{Schmitt:2003aa}, which furthermore
yields 
\begin{equation}
\lambda=-\frac{2e^{2}}{\sqrt{3}g^{2}},\qquad\cos^{2}\theta_{CFL}=\frac{g^{2}}{g^{2}+\frac{4}{3}e^{2}}\,,
\end{equation}
and the new generators 
\begin{equation}
\tilde{Q}=\left(\begin{array}{ccc}
0\\
 & 0\\
 &  & 0
\end{array}\right)\,,\qquad\tilde{T}_{8}=-\frac{1}{2\sqrt{3}\cos^2\theta_{\mathrm{CFL}}}\left(\begin{array}{ccc}
1\\
 & \ 1\\
 &  & -2
\end{array}\right)\,.
\end{equation}

\section{2SC\texorpdfstring{$_{\mathrm{ud}}$}{ud}}

The "true" 2SC phase, which we will call 2SC$_{\mathrm{ud}}$ in the
following, is the phase where blue quarks of all flavors and strange
quarks of all colors are unpaired. The diquark order parameter and the charge generator read
\begin{equation}
\Phi=\frac{1}{2}\left(\begin{array}{ccc}
0\\
 & 0\\
 &  & \phi_{3}
\end{array}\right)\,, \qquad
Q=\frac{1}{3}\left(\begin{array}{ccc}
2\\
 & -1\\
 &  & -1
\end{array}\right)\,,
\end{equation}
which immediately shows that $T_{3}(\Phi)=0$. The flavor indices
are ordered $\bar{u},\,\bar{d}\,$and $\bar{s}$ in contrast to the derivation of the CFL mixing angle. The anti-strange quark
channel $\bar{s}$ can be interpreted as a pair one up and one down
quark (hence the abbreviation). The color indices are
labeled $\bar{r},\,\bar{g}$ and $\bar{b}$, which shows that no blue
(anti-blue does not contain blue but can be seen as a pair of a red
and a green quark) quark participates in the pairing. The requirement
of $\tilde{Q}(\Phi)$ leads to the equation
\begin{align}
\tilde{Q} & =Q+\eta T_{8}\,,\nonumber \\
\tilde{Q}(\Phi) & =\left(-\frac{1}{3}-\frac{\eta}{\sqrt{3}}\right)=0\Rightarrow\eta=-\frac{1}{\sqrt{3}}\,,
\end{align}
in accordance with Tab.~(1) of \cite{Schmitt:2003aa}, which furthermore
yields 
\begin{equation}
\lambda=\frac{1}{\sqrt{3}}\frac{e^{2}}{g^{2}},\qquad\cos^{2}\theta_{\mathrm{2SC}}=\frac{3g^{2}}{3g^{2}+e^{2}}\,,
\end{equation}
and the new generators 
\begin{equation}
\tilde{Q}=\frac{1}{2}\left(\begin{array}{ccc}
1\\
 & -1\\
 &  & 0
\end{array}\right)\,,\qquad\tilde{T}_{8}=\frac{1}{3\sqrt{3}g^{2}}\left(\begin{array}{ccc}
\frac{3}{2}g^{2}+2e^{2}\\
 & \frac{3}{2}g^{2}-e^{2}\\
 &  & -3g^{2}-e^{2}
\end{array}\right)\,.
\end{equation}

\section{2SC\texorpdfstring{$_{\mathrm{us}}$}{us}}

The order parameter 
\begin{equation}
\Phi=\frac{1}{2}\left(\begin{array}{ccc}
0\\
 & \phi_{2}\\
 &  & 0
\end{array}\right)\,
\end{equation}
is energetically identical to the one of 2SC$_{\mathrm{ud}}$ as long as we
neglect the strange quark mass $m_{s}$. In principle one can define
new mixing angles etc.~such that the requirement $\tilde{Q}\left(\Phi_{\mathrm{us}}\right)=0$
is fulfilled only in 2SC$_{\mathrm{us}}$. However, due to the similarity of
$\tilde{Q}$ of 2SC$_{\mathrm{ud}}$ defined in the previous section to $T_{3}$,
we can actually define a new set of generators $\bar{Q}$ and $\bar{T}_{3}$
such that $\bar{Q}\left(\Phi\right)=0$ is fulfilled in both phases,
which should facilitate any calculation which requires simultaneous
treatment of both phases. We trivially find that 
\begin{equation}
\bar{Q}=\tilde{Q}-T_{3}\,,
\end{equation}
which leads to $\bar{Q}=0$ for every general diagonal order parameter.
We now require 
\begin{equation}
gA_{\mu}^{8}T_{8}+eA_{\mu}Q+gA_{\mu}^{3}T_{3}=\tilde{g}\tilde{A}_{\mu}^{8}\tilde{T}_{8}+\tilde{e}\tilde{A}_{\mu}\tilde{Q}+gA_{\mu}^{3}T_{3}=\tilde{g}\tilde{A}_{\mu}^{8}\tilde{T}_{8}+\bar{e}\bar{A}_{\mu}\bar{Q}+\bar{g}\bar{A}_{\mu}^{3}\bar{T}_{3}\,,
\end{equation}
which means that we only rotate two fields in every step. In the main part of this thesis, we will perform both rotations simultaneously, rendering the introduction of $\bar{Q}$ unnecessary. Note that this leads to an inconsistency in notation. However, I decided to keep the standard notation in this appendix in order to facilitate comparison to the literature like Ref.~\cite{Schmitt:2003aa}.
Furthermore, we introduce the rotated generator $\bar{T}_{3}=T_{3}+\alpha\tilde{Q}$
and the further rotated gauge fields
\begin{equation}
\left(\begin{array}{c}
\bar{A}_{\mu}\\
\bar{A}_{\mu}^{3}
\end{array}\right)=\left(\begin{array}{cc}
\cos\bt & \sin\bt\\
-\sin\bt & \cos\bt
\end{array}\right)\left(\begin{array}{c}
\tilde{A}_{\mu}\\
A_{\mu}^{3}
\end{array}\right)\,,\Rightarrow\left(\begin{array}{c}
\tilde{A}_{\mu}\\
A_{\mu}^{3}
\end{array}\right)=\left(\begin{array}{cc}
\cos\bt & -\sin\bt\\
\sin\bt & \cos\bt
\end{array}\right)\left(\begin{array}{c}
\bar{A}_{\mu}\\
\bar{A}_{\mu}^{3}
\end{array}\right)\,.
\end{equation}
Inserting this into the latter equation yields
\begin{equation}
\tilde{e}\left(\cos\bt\bar{A}_{\mu}-\sin\bt\bar{A}_{\mu}^{3}\right)\tilde{Q}+g\left(\sin\bt\bar{A}_{\mu}+\cos\bt\bar{A}_{\mu}^{3}\right)T_{3}=\bar{e}\bar{A}_{\mu}\left(\tilde{Q}-T_{3}\right)+\bar{g}\bar{A}_{\mu}^{3}\left(T_{3}+\alpha\tilde{Q}\right)\,,
\end{equation}
from which we deduce that
\bea
&&\bar{g}=g\cos\bt\,,\qquad\bar{e}=\tilde{e}\cos\bt\,,\qquad\sin\bar{\t}=-\frac{\bar{e}}{g}\non[2ex]
&&\tan\bt=-\frac{\tilde{e}}{g}\,,\qquad\cos\bar{\theta}=\frac{\sqrt{3g^{2}+e^{2}}}{\sqrt{3g^{2}+4e^{2}}}\,,\qquad\alpha=\frac{\tilde{e}^{2}}{g^{2}}\,.
\eea
This leads to the generator
\begin{equation}
\bar{T}_{3}=T_{3}+\left(\frac{\tilde{e}}{g}\right)^{2}\tilde{Q}=\left(1+\frac{\tilde{e}^{2}}{g^{2}}\right)T_{3}=\frac{1}{\cos^{2}\bt}T_{3}\,.
\end{equation}

\chapter{Flux Tube Interaction in Color Superconductivity}
\label{app:inter_color}
The idea behind the derivation of the long-distance flux tube interaction energy (\ref{Fint2}) is to add a small correction 
to the gauge fields and the scalar fields, such that without that correction the resulting profiles are the ones for a single, 
isolated flux tube. Instead of the gauge fields themselves, one works with the following vectors, which go to zero 
as $R\to\infty$,
\begin{subequations}
\bea
\mathbf{Q}_3(R)&\equiv&  g\frac{a_3(\infty)-a_3(R)}{R}\mathbf{e}_\varphi \\[2ex]
\mathbf{Q}_8(R)&\equiv&  2\tilde{g}_8 \frac{\tilde{a}_8(\infty)-\tilde{a}_8(R)}{R}\mathbf{e}_\varphi\, .
\eea
\end{subequations}
The small perturbations are now introduced via $\mathbf{Q}_a=\mathbf{Q}_{a0}+\delta\mathbf{Q}_a$ ($a=3,8$) and $f_i=f_{i0}+\delta f_i$
($i=1,2,3$), and we can compute the equations of motion to zeroth and first order in the perturbations. 
Then, using these equations of motion, some tedious algebra yields the 
free energy density up to second order in the perturbations from Eq.\ (\ref{Uflux}). Writing $U_{\circlearrowleft}=U_{\circlearrowleft}^{(0)}+\delta U_{\circlearrowleft}$, we have the zeroth-order contribution
\bea
U_{\circlearrowleft}^{(0)} &=& \frac{\lambda\rho_\mathrm{ CFL}^4}{2}\left\{\frac{\kappa_3^2}{2}(\nabla\times\mathbf{Q}_{30})^2+\frac{3\tilde{\kappa}_8^2}{2}(\nabla\times\mathbf{Q}_{80})^2
+(\nabla f_{10})^2+f_{10}^2\frac{(\mathbf{Q}_{30}+\mathbf{Q}_{80})^2}{4}\right. \\[2ex]
&& \left.+\frac{(1-f_{10}^2)^2}{2} + (\nabla f_{20})^2+f_{20}^2\frac{(\mathbf{Q}_{30}-\mathbf{Q}_{80})^2}{4}+\frac{(1-f_{20}^2)^2}{2} + 
(\nabla f_{30})^2+f_{30}^2Q_{80}^2\right.\non[2ex]
&&\left.+\frac{(1-f_{30}^2)^2}{2}  -\frac{h}{\lambda}\left[(1-f_{10}^2)(1-f_{20}^2)+(1-f_{20}^2)(1-f_{30}^2)+(1-f_{10}^2)(1-f_{30}^2)\right]\right\} \, ,\nonumber
\eea
and the first- and second-order contributions, which can be written as a total derivative, 
\bea
\delta U_{\circlearrowleft} &=& \lambda\rho_\mathrm{ CFL}^4\nabla\cdot\left\{\frac{\kappa_3^2}{2}\delta\mathbf{Q}_3\times\left[\nabla\times\left(\mathbf{Q}_{30}+\frac{\delta\mathbf{Q}_3}{2}\right)\right]+
\frac{3\tilde{\kappa}_8^2}{2} \delta\mathbf{Q}_8\times\left[\nabla\times\left(\mathbf{Q}_{80}+\frac{\delta\mathbf{Q}_8}{2}\right)\right] \right.\non[2ex]
&&\left.+\delta f_1\nabla\left(f_{10}+\frac{\delta f_1}{2}\right)+
\delta f_2\nabla\left(f_{20}+\frac{\delta f_2}{2}\right)+\delta f_3\nabla\left(f_{30}+\frac{\delta f_3}{2}\right)\right\} \, .
\eea
We can now exactly follow the steps explained in Appendix C of Ref.\ \cite{Haber:2017kth} to find 
the interaction energy for two flux tubes in a distance $R_0$ from each other,
\bea 
\label{Fint1}
\frac{F_\mathrm{ int}^{\circlearrowleft}}{L} &=& \int_{R_0/2}^\infty\frac{2\rho_\mathrm{ CFL}^2 R_0dR}{\sqrt{R^2-(R_0/2)^2}}\Bigg[-\frac{\kappa_3^2}{2}\delta Q_3\left(\frac{\delta Q_3}{R}+\delta Q_3'\right)-\frac{3\tilde{\kappa}_8^2}{2}\delta Q_8\left(\frac{\delta Q_8}{R}+\delta Q_8'\right)\non[2ex]
&&\hspace{4cm} +\delta f_1\delta f_1'+\delta f_2\delta f_2'+\delta f_3\delta f_3'\Bigg]\non[2ex]
&=&\int_{R_0/2}^\infty\frac{2\rho_\mathrm{ CFL}^2 R_0dR}{\sqrt{R^2-(R_0/2)^2}}\Bigg[\frac{\kappa_3^2 g^2a_3'}{2}\frac{a_3(\infty)-a_3(R)}{R^2}+6\tilde{\kappa}_8^2\tilde{g}_8^2\tilde{a}_8'\frac{\tilde{a}_8(\infty)-a_8(R)}{R^2} \non
&&\hspace{4cm} -(1-f_1)f_1'-(1-f_2)f_2'-(1-f_3)f_3'\Bigg] \, .
\eea

where, in the second line, we have written the result in terms of the full (numerically determined) profile functions. This expression can 
be used to extrapolate the interaction energy down to smaller distances. Instead, we shall only work with the asymptotic 
result which is obtained by expressing the first line of Eq.\ (\ref{Fint1}) in terms of the asymptotic approximations to the profile functions. 
This is Eq.~(\ref{Fint}) in the main text. 

\newpage{}

\addcontentsline{toc}{chapter}{References}

\bibliographystyle{utphys}
\bibliography{refs}

\providecommand{\href}[2]{#2}\begingroup\raggedright\begin{thebibliography}{100}

\bibitem{Haber:2015exa}
A.~Haber, A.~Schmitt, and S.~Stetina, ``{Instabilities in relativistic
  two-component (super)fluids},''
  \href{http://dx.doi.org/10.1103/PhysRevD.93.025011}{{\em Phys. Rev.}
  {\bfseries D93} no.~2, (2016) 025011},
\href{http://arxiv.org/abs/1510.01982}{{\ttfamily arXiv:1510.01982 [hep-ph]}}.

\bibitem{Haber:2016ljn}
A.~Haber and A.~Schmitt, ``{Mixing of charged and neutral Bose condensates at
  nonzero temperature and magnetic field},''
  \href{http://dx.doi.org/10.1051/epjconf/201713709003}{{\em EPJ Web Conf.}
  {\bfseries 137} (2017) 09003},
\href{http://arxiv.org/abs/1612.01865}{{\ttfamily arXiv:1612.01865
  [astro-ph.HE]}}.

\bibitem{Haber:2017kth}
A.~Haber and A.~Schmitt, ``{Critical magnetic fields in a superconductor
  coupled to a superfluid},''
  \href{http://dx.doi.org/10.1103/PhysRevD.95.116016}{{\em Phys. Rev.}
  {\bfseries D95} no.~11, (2017) 116016},
\href{http://arxiv.org/abs/1704.01575}{{\ttfamily arXiv:1704.01575 [hep-th]}}.

\bibitem{Haber:2017oqb}
A.~Haber and A.~Schmitt, ``{New color-magnetic defects in dense quark
  matter},'' \href{http://dx.doi.org/10.1088/1361-6471/aabc1a}{{\em J. Phys.}
  {\bfseries G45} no.~6, (2018) 065001},
\href{http://arxiv.org/abs/1712.08587}{{\ttfamily arXiv:1712.08587 [hep-ph]}}.

\bibitem{chem_book}
P.~Flowers, K.~Theopold, R.~Langley, and W.~Robinson, {\em Chemistry}.
\newblock OpenStax, 2015.
\newblock \url{https://openstax.org/details/books/chemistry}.

\bibitem{pdg2017}
{\bfseries Particle Data Group} Collaboration, C.~Patrignani {\em et~al.},
  ``{Review of Particle Physics},''
\href{http://dx.doi.org/10.1088/1674-1137/40/10/100001}{{\em Chin. Phys.}
  {\bfseries C40} no.~10, (2016) 100001}.

\bibitem{quarks_pic}
{MissMJ/Wikipedia/CC BY 3.0}, ``Standard model of elementary particles,'' 2006.
\newblock
  \url{https://en.wikipedia.org/wiki/Quark#/media/File:Standard_Model_of_Elementary_Particles.svg}.
  [accessed 30-June-2018].

\bibitem{Brambilla:2014jmp}
N.~Brambilla {\em et~al.}, ``{QCD and Strongly Coupled Gauge Theories:
  Challenges and Perspectives},''
  \href{http://dx.doi.org/10.1140/epjc/s10052-014-2981-5}{{\em Eur. Phys. J.}
  {\bfseries C74} no.~10, (2014) 2981},
\href{http://arxiv.org/abs/1404.3723}{{\ttfamily arXiv:1404.3723 [hep-ph]}}.

\bibitem{Hands:2001ve}
S.~Hands, ``The phase diagram of qcd,'' {\em Contemp. Phys.} {\bfseries 42}
  (2001) 209--225,
\href{http://arxiv.org/abs/physics/0105022}{{\ttfamily physics/0105022}}.

\bibitem{NICA}
D.~Blaschke, J.~Aichelin, E.~Bratkovskaya, V.~Friese, M.~Gazdzicki, J.~Randrup,
  O.~Rogachevsky, O.~Teryaev, and V.~Toneev, ``Topical issue on exploring
  strongly interacting matter at high densities - nica white paper,''
  \href{http://dx.doi.org/10.1140/epja/i2016-16267-x}{{\em Eur. Phys. J. A}
  {\bfseries 52} no.~8, (2016) 267}.
  \url{https://doi.org/10.1140/epja/i2016-16267-x}.

\bibitem{Haber:2014ula}
A.~Haber, F.~Preis, and A.~Schmitt, ``{Magnetic catalysis in nuclear matter},''
  \href{http://dx.doi.org/10.1103/PhysRevD.90.125036}{{\em Phys. Rev.}
  {\bfseries D90} no.~12, (2014) 125036},
\href{http://arxiv.org/abs/1409.0425}{{\ttfamily arXiv:1409.0425 [nucl-th]}}.

\bibitem{Haber:2014zba}
A.~Haber, F.~Preis, and A.~Schmitt, ``{Baryon onset in a magnetic field},''
  \href{http://dx.doi.org/10.1063/1.4938699}{{\em AIP Conf. Proc.} {\bfseries
  1701} (2016) 080010},
\href{http://arxiv.org/abs/1412.6282}{{\ttfamily arXiv:1412.6282 [nucl-th]}}.

\bibitem{Gross:1973id}
D.~J. Gross and F.~Wilczek, ``Ultraviolet behavior of non-abelian gauge
  theories,''
{\em Phys. Rev. Lett.} {\bfseries 30} (1973) 1343--1346.

\bibitem{Alford:1997zt}
M.~G. Alford, K.~Rajagopal, and F.~Wilczek, ``Qcd at finite baryon density:
  Nucleon droplets and color superconductivity,'' {\em Phys. Lett.} {\bfseries
  B422} (1998) 247--256,
\href{http://arxiv.org/abs/hep-ph/9711395}{{\ttfamily hep-ph/9711395}}.

\bibitem{Alford:2007xm}
M.~G. Alford, A.~Schmitt, K.~Rajagopal, and T.~Sch{\"a}fer, ``{Color
  superconductivity in dense quark matter},''
  \href{http://dx.doi.org/10.1103/RevModPhys.80.1455}{{\em Rev.Mod.Phys.}
  {\bfseries 80} (2008) 1455--1515},
\href{http://arxiv.org/abs/0709.4635}{{\ttfamily arXiv:0709.4635 [hep-ph]}}.

\bibitem{NJL}
Y.~Nambu and G.~Jona-Lasinio, ``Dynamical model of elementary particles based
  on an analogy with superconductivity. i,''
  \href{http://dx.doi.org/10.1103/PhysRev.122.345}{{\em Phys. Rev.} {\bfseries
  122} (Apr, 1961) 345--358}.
  \url{https://link.aps.org/doi/10.1103/PhysRev.122.345}.

\bibitem{Glendenning:1997wn}
N.~K. Glendenning, {\em {Compact stars: Nuclear physics, particle physics, and
  general relativity}}.
\newblock
1997.
\newblock

\bibitem{Reddy:2004jg}
S.~Reddy, ``Neutron stars, supernova and phases of dense quark matter,''
{\em J. Phys.} {\bfseries G30} (2004) S879--S885.

\bibitem{Haensel:2007yy}
P.~Haensel, A.~Y. Potekhin, and D.~G. Yakovlev,
  \href{http://dx.doi.org/10.1007/978-0-387-47301-7}{{\em {Neutron stars 1:
  Equation of state and structure}}}, vol.~326.
\newblock Springer, New York, USA,
2007.
\newblock

\bibitem{Schmitt:2010pn}
A.~Schmitt, ``{Dense matter in compact stars - A pedagogical introduction},''
  \href{http://dx.doi.org/10.1007/978-3-642-12866-0}{{\em Lect. Notes Phys.}
  {\bfseries 811} (2010) 1--111},
\href{http://arxiv.org/abs/1001.3294}{{\ttfamily arXiv:1001.3294
  [astro-ph.SR]}}.

\bibitem{Watts:2014tja}
A.~Watts {\em et~al.}, ``{Probing the neutron star interior and the Equation of
  State of cold dense matter with the SKA},'' {\em PoS} {\bfseries AASKA14}
  (2015) 043,
\href{http://arxiv.org/abs/1501.00042}{{\ttfamily arXiv:1501.00042
  [astro-ph.SR]}}.

\bibitem{Demorest:2010bx}
P.~Demorest, T.~Pennucci, S.~Ransom, M.~Roberts, and J.~Hessels, ``{Shapiro
  Delay Measurement of A Two Solar Mass Neutron Star},''
  \href{http://dx.doi.org/10.1038/nature09466}{{\em Nature} {\bfseries 467}
  (2010) 1081--1083},
\href{http://arxiv.org/abs/1010.5788}{{\ttfamily arXiv:1010.5788
  [astro-ph.HE]}}.

\bibitem{Antoniadis:2013pzd}
J.~Antoniadis {\em et~al.}, ``{A Massive Pulsar in a Compact Relativistic
  Binary},'' \href{http://dx.doi.org/10.1126/science.1233232}{{\em Science}
  {\bfseries 340} (2013) 6131},
\href{http://arxiv.org/abs/1304.6875}{{\ttfamily arXiv:1304.6875
  [astro-ph.HE]}}.

\bibitem{Weber:2004kj}
F.~Weber, ``{Strange quark matter and compact stars},''
  \href{http://dx.doi.org/10.1016/j.ppnp.2004.07.001}{{\em Prog. Part. Nucl.
  Phys.} {\bfseries 54} (2005) 193--288},
\href{http://arxiv.org/abs/astro-ph/0407155}{{\ttfamily arXiv:astro-ph/0407155
  [astro-ph]}}.

\bibitem{NICER}
K.~C. {Gendreau}, Z.~{Arzoumanian}, and T.~{Okajima},
  \href{http://dx.doi.org/10.1117/12.926396}{``{The Neutron star Interior
  Composition ExploreR (NICER): an Explorer mission of opportunity for soft
  x-ray timing spectroscopy},''} in {\em Space Telescopes and Instrumentation
  2012: Ultraviolet to Gamma Ray}, vol.~8443 of {\em Proceedings of SPIE},
  p.~844313.
\newblock Sept., 2012.

\bibitem{Hessels1901}
J.~W.~T. Hessels, S.~M. Ransom, I.~H. Stairs, P.~C.~C. Freire, V.~M. Kaspi, and
  F.~Camilo, ``A radio pulsar spinning at 716 hz,''
  \href{http://dx.doi.org/10.1126/science.1123430}{{\em Science} {\bfseries
  311} no.~5769, (2006) 1901--1904},
  \href{http://arxiv.org/abs/http://science.sciencemag.org/content/311/5769/1901.full.pdf}{{\ttfamily
  http://science.sciencemag.org/content/311/5769/1901.full.pdf}}.
  \url{http://science.sciencemag.org/content/311/5769/1901}.

\bibitem{1999A&A...344..151H}
P.~{Haensel}, J.~P. {Lasota}, and J.~L. {Zdunik}, ``{On the minimum period of
  uniformly rotating neutron stars},'' {\em Astron. Astrophys.} {\bfseries 344}
  (Apr., 1999) 151--153.

\bibitem{Haensel:2009wa}
P.~Haensel, J.~L. Zdunik, M.~Bejger, and J.~M. Lattimer, ``{Keplerian frequency
  of uniformly rotating neutron stars and strange stars},''
  \href{http://dx.doi.org/10.1051/0004-6361/200811605}{{\em Astron. Astrophys.}
  {\bfseries 502} (2009) 605--610},
\href{http://arxiv.org/abs/0901.1268}{{\ttfamily arXiv:0901.1268
  [astro-ph.SR]}}.

\bibitem{Duncan:1992hi}
R.~C. Duncan and C.~Thompson, ``{Formation of very strongly magnetized neutron
  stars - implications for gamma-ray bursts},''
\href{http://dx.doi.org/10.1086/186413}{{\em Astrophys. J.} {\bfseries 392}
  (1992) L9}.

\bibitem{Rea2011}
N.~Rea and P.~Esposito, ``Magnetar outbursts: an observational review,'' in
  {\em High-Energy Emission from Pulsars and their Systems}, D.~F. Torres and
  N.~Rea, eds., pp.~247--273.
\newblock Springer Berlin Heidelberg, Berlin, Heidelberg, 2011.

\bibitem{Lai1991}
D.~{Lai} and S.~L. {Shapiro}, ``{Cold equation of state in a strong magnetic
  field - Effects of inverse beta-decay},''
  \href{http://dx.doi.org/10.1086/170831}{{\em Astrophys. J.} {\bfseries 383}
  (Dec., 1991) 745--751}.

\bibitem{Glampedakis:2017nqy}
K.~Glampedakis and L.~Gualtieri, ``{Gravitational waves from single neutron
  stars: an advanced detector era survey},''
\href{http://arxiv.org/abs/1709.07049}{{\ttfamily arXiv:1709.07049
  [astro-ph.HE]}}.

\bibitem{Potekhin:2015qsa}
A.~Y. Potekhin, J.~A. Pons, and D.~Page, ``{Neutron stars - cooling and
  transport},'' \href{http://dx.doi.org/10.1007/s11214-015-0180-9}{{\em Space
  Sci. Rev.} {\bfseries 191} no.~1-4, (2015) 239--291},
\href{http://arxiv.org/abs/1507.06186}{{\ttfamily arXiv:1507.06186
  [astro-ph.HE]}}.

\bibitem{PhysRevLett.119.161101}
{\bfseries LIGO Scientific Collaboration and Virgo Collaboration}
  Collaboration, B.~P. Abbott {\em et~al.}, ``Gw170817: Observation of
  gravitational waves from a binary neutron star inspiral,''
  \href{http://dx.doi.org/10.1103/PhysRevLett.119.161101}{{\em Phys.Rev.Lett.}
  {\bfseries 119} (Oct, 2017) 161101}.

\bibitem{Weisberg:2010zz}
J.~M. Weisberg, D.~J. Nice, and J.~H. Taylor, ``{Timing Measurements of the
  Relativistic Binary Pulsar PSR B1913+16},''
  \href{http://dx.doi.org/10.1088/0004-637X/722/2/1030}{{\em Astrophys. J.}
  {\bfseries 722} (2010) 1030--1034},
\href{http://arxiv.org/abs/1011.0718}{{\ttfamily arXiv:1011.0718
  [astro-ph.GA]}}.

\bibitem{Bauswein:2011tp}
A.~Bauswein and H.~T. Janka, ``{Measuring neutron-star properties via
  gravitational waves from binary mergers},''
  \href{http://dx.doi.org/10.1103/PhysRevLett.108.011101}{{\em Phys. Rev.
  Lett.} {\bfseries 108} (2012) 011101},
\href{http://arxiv.org/abs/1106.1616}{{\ttfamily arXiv:1106.1616
  [astro-ph.SR]}}.

\bibitem{Takami:2014zpa}
K.~Takami, L.~Rezzolla, and L.~Baiotti, ``{Constraining the Equation of State
  of Neutron Stars from Binary Mergers},''
  \href{http://dx.doi.org/10.1103/PhysRevLett.113.091104}{{\em Phys. Rev.
  Lett.} {\bfseries 113} no.~9, (2014) 091104},
\href{http://arxiv.org/abs/1403.5672}{{\ttfamily arXiv:1403.5672 [gr-qc]}}.

\bibitem{Agathos:2015uaa}
M.~Agathos, J.~Meidam, W.~Del~Pozzo, T.~G.~F. Li, M.~Tompitak, J.~Veitch,
  S.~Vitale, and C.~Van Den~Broeck, ``{Constraining the neutron star equation
  of state with gravitational wave signals from coalescing binary neutron
  stars},'' \href{http://dx.doi.org/10.1103/PhysRevD.92.023012}{{\em Phys.
  Rev.} {\bfseries D92} no.~2, (2015) 023012},
\href{http://arxiv.org/abs/1503.05405}{{\ttfamily arXiv:1503.05405 [gr-qc]}}.

\bibitem{Oppenheimer1939}
J.~R. Oppenheimer and G.~M. Volkoff, ``On massive neutron cores,''
  \href{http://dx.doi.org/10.1103/PhysRev.55.374}{{\em Phys. Rev.} {\bfseries
  55} (Feb, 1939) 374--381}.
  \url{https://link.aps.org/doi/10.1103/PhysRev.55.374}.

\bibitem{tolman_TOV}
R.~C. Tolman, ``Static solutions of einstein's field equations for spheres of
  fluid,'' \href{http://dx.doi.org/10.1103/PhysRev.55.364}{{\em Phys. Rev.}
  {\bfseries 55} (Feb, 1939) 364--373}.
  \url{https://link.aps.org/doi/10.1103/PhysRev.55.364}.

\bibitem{Chatterjee:2015pua}
D.~Chatterjee and I.~Vida{\~n}a, ``{Do hyperons exist in the interior of
  neutron stars?},'' \href{http://dx.doi.org/10.1140/epja/i2016-16029-x}{{\em
  Eur. Phys. J.} {\bfseries A52} no.~2, (2016) 29},
\href{http://arxiv.org/abs/1510.06306}{{\ttfamily arXiv:1510.06306 [nucl-th]}}.

\bibitem{wiki:xxx}
{Wikipedia}, ``Neutron star --- {W}ikipedia{,} the free encyclopedia,'' 2018.
\newblock \url{https://en.wikipedia.org/wiki/Neutron_star}. [accessed
  13-March-2018].

\bibitem{Baldo:2016jhp}
M.~Baldo and G.~F. Burgio, ``{The nuclear symmetry energy},''
  \href{http://dx.doi.org/10.1016/j.ppnp.2016.06.006}{{\em Prog. Part. Nucl.
  Phys.} {\bfseries 91} (2016) 203--258},
\href{http://arxiv.org/abs/1606.08838}{{\ttfamily arXiv:1606.08838 [nucl-th]}}.

\bibitem{Steiner:2010fz}
A.~W. Steiner, J.~M. Lattimer, and E.~F. Brown, ``{The Equation of State from
  Observed Masses and Radii of Neutron Stars},''
  \href{http://dx.doi.org/10.1088/0004-637X/722/1/33}{{\em Astrophys.J.}
  {\bfseries 722} (2010) 33--54},
  \href{http://arxiv.org/abs/arXiv:1005.0811}{{\ttfamily arXiv:arXiv:1005.0811
  [astro-ph.HE]}}.

\bibitem{Ozel:2015fia}
F.~Ozel, D.~Psaltis, T.~Guver, G.~Baym, C.~Heinke, and S.~Guillot, ``{The Dense
  Matter Equation of State from Neutron Star Radius and Mass Measurements},''
  \href{http://dx.doi.org/10.3847/0004-637X/820/1/28}{{\em Astrophys. J.}
  {\bfseries 820} no.~1, (2016) 28},
\href{http://arxiv.org/abs/1505.05155}{{\ttfamily arXiv:1505.05155
  [astro-ph.HE]}}.

\bibitem{RevModPhys.88.021001}
A.~L. Watts, N.~Andersson, D.~Chakrabarty, M.~Feroci, K.~Hebeler, G.~Israel,
  F.~K. Lamb, M.~C. Miller, S.~Morsink, F.~\"Ozel, A.~Patruno, J.~Poutanen,
  D.~Psaltis, A.~Schwenk, A.~W. Steiner, L.~Stella, L.~Tolos, and M.~van~der
  Klis, ``Colloquium: Measuring the neutron star equation of state using x-ray
  timing,'' \href{http://dx.doi.org/10.1103/RevModPhys.88.021001}{{\em Rev.
  Mod. Phys.} {\bfseries 88} (Apr, 2016) 021001}.
  \url{https://link.aps.org/doi/10.1103/RevModPhys.88.021001}.

\bibitem{1997A&A...326..924M}
D.~N. {Matsakis}, J.~H. {Taylor}, and T.~M. {Eubanks}, ``{A statistic for
  describing pulsar and clock stabilities.},'' {\em Astron. Astrophys.}
  {\bfseries 326} (Oct., 1997) 924--928.

\bibitem{Haskell:2015jra}
B.~Haskell and A.~Melatos, ``{Models of Pulsar Glitches},''
  \href{http://dx.doi.org/10.1142/S0218271815300086}{{\em Int. J. Mod. Phys.}
  {\bfseries D24} no.~03, (2015) 1530008},
\href{http://arxiv.org/abs/1502.07062}{{\ttfamily arXiv:1502.07062
  [astro-ph.SR]}}.

\bibitem{lyne_graham-smith_2012}
A.~Lyne and F.~Graham-Smith,
  \href{http://dx.doi.org/10.1017/CBO9780511844584}{{\em Pulsar Astronomy}}.
\newblock Cambridge Astrophysics. Cambridge University Press, 4~ed., 2012.

\bibitem{vela_glitch}
C.~Malmar, ``Neutronstjarnor,'' 1992.
\newblock \url{http://www.kosmologika.net/Stars/Neutronstjarnor.html}.
  [accessed 30-June-2018].

\bibitem{Anderson:1975}
P.~W. Anderson and N.~Itoh {\em Nature} {\bfseries 256} (1975) 25.

\bibitem{Wlazlowski:2016yoe}
G.~Wlaz{\l}owski, K.~Sekizawa, P.~Magierski, A.~Bulgac, and M.~M. Forbes,
  ``{Vortex pinning and dynamics in the neutron star crust},''
  \href{http://dx.doi.org/10.1103/PhysRevLett.117.232701}{{\em Phys. Rev.
  Lett.} {\bfseries 117} no.~23, (2016) 232701},
\href{http://arxiv.org/abs/1606.04847}{{\ttfamily arXiv:1606.04847
  [astro-ph.HE]}}.

\bibitem{Andersson:2002zd}
N.~Andersson, G.~L. Comer, and R.~Prix, ``{Are pulsar glitches triggered by a
  superfluid two-stream instability?},''
  \href{http://dx.doi.org/10.1103/PhysRevLett.90.091101}{{\em Phys. Rev. Lett.}
  {\bfseries 90} (2003) 091101},
\href{http://arxiv.org/abs/astro-ph/0210486}{{\ttfamily arXiv:astro-ph/0210486
  [astro-ph]}}.

\bibitem{Alpar:1984zz}
M.~A. Alpar, S.~A. Langer, and J.~A. Sauls, ``{Rapid postglitch spin-up of the
  superfluid core in pulsars},''
\href{http://dx.doi.org/10.1086/162232}{{\em Astrophys. J.} {\bfseries 282}
  (1984) 533--541}.

\bibitem{Ivanenko:1969gs}
D.~D. Ivanenko and D.~F. Kurdgelaidze, ``Remarks on quark stars,''
{\em Lett. Nuovo Cim.} {\bfseries IIS1} (1969) 13--16.

\bibitem{Ivanenko:1969bs}
D.~D. Ivanenko and D.~F. Kurdgelaidze, ``Quark stars,''
{\em Sov. Phys. J.} {\bfseries 13} (1970) 1015--1019.

\bibitem{Bodmer:1971we}
A.~R. Bodmer, ``Collapsed nuclei,''
{\em Phys. Rev.} {\bfseries D4} (1971) 1601--1606.

\bibitem{Witten:1984rs}
E.~Witten, ``Cosmic separation of phases,''
{\em Phys. Rev.} {\bfseries D30} (1984) 272--285.

\bibitem{Farhi:1984qu}
E.~Farhi and R.~L. Jaffe, ``Strange matter,''
{\em Phys. Rev.} {\bfseries D30} (1984) 2379.

\bibitem{Bauswein:2008gx}
A.~Bauswein, H.~T. Janka, R.~Oechslin, G.~Pagliara, I.~Sagert,
  J.~Schaffner-Bielich, M.~M. Hohle, and R.~Neuhauser, ``{Mass Ejection by
  Strange Star Mergers and Observational Implications},''
  \href{http://dx.doi.org/10.1103/PhysRevLett.103.011101}{{\em Phys. Rev.
  Lett.} {\bfseries 103} (2009) 011101},
\href{http://arxiv.org/abs/0812.4248}{{\ttfamily arXiv:0812.4248 [astro-ph]}}.

\bibitem{Alford:2008ge}
M.~G. Alford and D.~A. Eby, ``{Thickness of the strangelet-crystal crust of a
  strange star},'' \href{http://dx.doi.org/10.1103/PhysRevC.78.045802}{{\em
  Phys. Rev.} {\bfseries C78} (2008) 045802},
\href{http://arxiv.org/abs/0808.0671}{{\ttfamily arXiv:0808.0671 [nucl-th]}}.

\bibitem{Jaikumar:2005ne}
P.~Jaikumar, S.~Reddy, and A.~W. Steiner, ``{The Strange star surface: A Crust
  with nuggets},'' \href{http://dx.doi.org/10.1103/PhysRevLett.96.041101}{{\em
  Phys. Rev. Lett.} {\bfseries 96} (2006) 041101},
\href{http://arxiv.org/abs/nucl-th/0507055}{{\ttfamily arXiv:nucl-th/0507055
  [nucl-th]}}.

\bibitem{Chodos:1974pn}
A.~Chodos, R.~L. Jaffe, K.~Johnson, and C.~B. Thorn, ``{Baryon Structure in the
  Bag Theory},''
\href{http://dx.doi.org/10.1103/PhysRevD.10.2599}{{\em Phys. Rev.} {\bfseries
  D10} (1974) 2599}.

\bibitem{Chodos:1974je}
A.~Chodos, R.~L. Jaffe, K.~Johnson, C.~B. Thorn, and V.~F. Weisskopf, ``{A New
  Extended Model of Hadrons},''
\href{http://dx.doi.org/10.1103/PhysRevD.9.3471}{{\em Phys. Rev.} {\bfseries
  D9} (1974) 3471--3495}.

\bibitem{Alford:2001zr}
M.~G. Alford, K.~Rajagopal, S.~Reddy, and F.~Wilczek, ``The minimal cfl-nuclear
  interface,'' {\em Phys. Rev.} {\bfseries D64} (2001) 074017,
\href{http://arxiv.org/abs/hep-ph/0105009}{{\ttfamily hep-ph/0105009}}.

\bibitem{Palhares:2010be}
L.~F. Palhares and E.~S. Fraga, ``{Droplets in the cold and dense linear sigma
  model with quarks},''
  \href{http://dx.doi.org/10.1103/PhysRevD.82.125018}{{\em Phys. Rev.}
  {\bfseries D82} (2010) 125018},
\href{http://arxiv.org/abs/1006.2357}{{\ttfamily arXiv:1006.2357 [hep-ph]}}.

\bibitem{Pinto:2012aq}
M.~B. Pinto, V.~Koch, and J.~Randrup, ``{The Surface Tension of Quark Matter in
  a Geometrical Approach},''
  \href{http://dx.doi.org/10.1103/PhysRevC.86.025203}{{\em Phys. Rev.}
  {\bfseries C86} (2012) 025203},
\href{http://arxiv.org/abs/1207.5186}{{\ttfamily arXiv:1207.5186 [hep-ph]}}.

\bibitem{Alford:2017qgh}
M.~G. Alford and A.~Sedrakian, ``{Compact stars with sequential QCD phase
  transitions},'' \href{http://dx.doi.org/10.1103/PhysRevLett.119.161104}{{\em
  Phys. Rev. Lett.} {\bfseries 119} no.~16, (2017) 161104},
\href{http://arxiv.org/abs/1706.01592}{{\ttfamily arXiv:1706.01592
  [astro-ph.HE]}}.

\bibitem{Alford:2015gna}
M.~G. Alford and S.~Han, ``{Characteristics of hybrid compact stars with a
  sharp hadron-quark interface},''
  \href{http://dx.doi.org/10.1140/epja/i2016-16062-9}{{\em Eur. Phys. J.}
  {\bfseries A52} no.~3, (2016) 62},
\href{http://arxiv.org/abs/1508.01261}{{\ttfamily arXiv:1508.01261 [nucl-th]}}.

\bibitem{Alford:2015dpa}
M.~G. Alford, G.~F. Burgio, S.~Han, G.~Taranto, and D.~Zappal{\`a},
  ``{Constraining and applying a generic high-density equation of state},''
  \href{http://dx.doi.org/10.1103/PhysRevD.92.083002}{{\em Phys. Rev.}
  {\bfseries D92} no.~8, (2015) 083002},
\href{http://arxiv.org/abs/1501.07902}{{\ttfamily arXiv:1501.07902 [nucl-th]}}.

\bibitem{Andersson:1997xt}
N.~Andersson, ``A new class of unstable modes of rotating relativistic stars,''
  {\em Astrophys. J.} {\bfseries 502} (1998) 708--713,
\href{http://arxiv.org/abs/gr-qc/9706075}{{\ttfamily gr-qc/9706075}}.

\bibitem{Andersson:1998qs}
N.~Andersson, K.~D. Kokkotas, and N.~Stergioulas, ``On the relevance of the
  r-mode instability for accreting neutron stars and white dwarfs,'' {\em
  Astrophys. J.} {\bfseries 516} (1999) 307,
\href{http://arxiv.org/abs/astro-ph/9806089}{{\ttfamily astro-ph/9806089}}.

\bibitem{Andersson:2000mf}
N.~Andersson and K.~D. Kokkotas, ``{The R mode instability in rotating neutron
  stars},'' \href{http://dx.doi.org/10.1142/S0218271801001062}{{\em Int. J.
  Mod. Phys.} {\bfseries D10} (2001) 381--442},
\href{http://arxiv.org/abs/gr-qc/0010102}{{\ttfamily arXiv:gr-qc/0010102
  [gr-qc]}}.

\bibitem{Haskell:2015iia}
B.~Haskell, ``{R-modes in neutron stars: Theory and observations},''
  \href{http://dx.doi.org/10.1142/S0218301315410074}{{\em Int. J. Mod. Phys.}
  {\bfseries E24} no.~09, (2015) 1541007},
\href{http://arxiv.org/abs/1509.04370}{{\ttfamily arXiv:1509.04370
  [astro-ph.HE]}}.

\bibitem{Gusakov:2013jwa}
M.~E. Gusakov, A.~I. Chugunov, and E.~M. Kantor, ``{Instability windows and
  evolution of rapidly rotating neutron stars},''
  \href{http://dx.doi.org/10.1103/PhysRevLett.112.151101}{{\em Phys. Rev.
  Lett.} {\bfseries 112} no.~15, (2014) 151101},
\href{http://arxiv.org/abs/1310.8103}{{\ttfamily arXiv:1310.8103
  [astro-ph.HE]}}.

\bibitem{Alford:2013pma}
M.~G. Alford and K.~Schwenzer, ``{What the Timing of Millisecond Pulsars Can
  Teach us about Their Interior},''
  \href{http://dx.doi.org/10.1103/PhysRevLett.113.251102}{{\em Phys. Rev.
  Lett.} {\bfseries 113} no.~25, (2014) 251102},
\href{http://arxiv.org/abs/1310.3524}{{\ttfamily arXiv:1310.3524
  [astro-ph.HE]}}.

\bibitem{Alford:2014jha}
M.~G. Alford, S.~Han, and K.~Schwenzer, ``{Phase conversion dissipation in
  multicomponent compact stars},''
  \href{http://dx.doi.org/10.1103/PhysRevC.91.055804}{{\em Phys. Rev.}
  {\bfseries C91} no.~5, (2015) 055804},
\href{http://arxiv.org/abs/1404.5279}{{\ttfamily arXiv:1404.5279
  [astro-ph.SR]}}.

\bibitem{Glampedakis:2012qp}
K.~Glampedakis, D.~I. Jones, and L.~Samuelsson, ``{Gravitational waves from
  color-magnetic `mountains' in neutron stars},''
  \href{http://dx.doi.org/10.1103/PhysRevLett.109.081103}{{\em Phys. Rev.
  Lett.} {\bfseries 109} (2012) 081103},
\href{http://arxiv.org/abs/1204.3781}{{\ttfamily arXiv:1204.3781
  [astro-ph.SR]}}.

\bibitem{Meissner1933}
W.~Meissner and R.~Ochsenfeld, ``Ein neuer effekt bei eintritt der
  supraleitf{\"a}higkeit,'' \href{http://dx.doi.org/10.1007/BF01504252}{{\em
  Naturwissenschaften} {\bfseries 21} no.~44, (Nov, 1933) 787--788}.
  \url{https://doi.org/10.1007/BF01504252}.

\bibitem{meissner_pic}
P.~Jaworski, ``Meissner effect,'' 2005.
\newblock \url{https://commons.wikimedia.org/wiki/File:EfektMeisnera.svg}.
  [accessed 30-June-2018].

\bibitem{wolfke}
M.~Wolfke and W.~Keesom, ``On the change of the dielectric constant of liquid
  helium with the temperature,'' {\em Proc. Acad. Sci. Amsterdam} {\bfseries
  31} (1928) 81--89.

\bibitem{1938Natur.141...74K}
P.~{Kapitza}, ``{Viscosity of Liquid Helium below the {$\lambda$}-Point},''
  \href{http://dx.doi.org/10.1038/141074a0}{{\em Nature} {\bfseries 141} (Jan.,
  1938) 74}.

\bibitem{1938Natur.141...75A}
J.~F. {Allen} and A.~D. {Misener}, ``{Flow of Liquid Helium II},''
  \href{http://dx.doi.org/10.1038/141075a0}{{\em Nature} {\bfseries 141} (Jan.,
  1938) 75}.

\bibitem{BCS}
J.~Bardeen, L.~N. Cooper, and J.~R. Schrieffer, ``Theory of
  superconductivity,'' \href{http://dx.doi.org/10.1103/PhysRev.108.1175}{{\em
  Phys. Rev.} {\bfseries 108} (Dec, 1957) 1175--1204}.
  \url{https://link.aps.org/doi/10.1103/PhysRev.108.1175}.

\bibitem{GLtheory}
V.~Ginzburg and L.~Landau {\em Zh. Eksp. Teor. Fiz.} {\bfseries 20} no.~1064,
  (1950) .

\bibitem{tinkham2004introduction}
M.~Tinkham, {\em Introduction to Superconductivity}.
\newblock Dover Publications, New York, 2004.

\bibitem{superbook}
A.~Schmitt, \href{http://dx.doi.org/10.1007/978-3-319-07947-9}{{\em
  Introduction to Superfluidity}}, vol.~888 of {\em Lecture Notes in Physics}.
\newblock Springer International Publishing, 2015.
\newblock \url{http://arxiv.org/abs/1404.1284}.

\bibitem{Srednicki:2007qs}
M.~Srednicki, {\em {Quantum field theory}}.
\newblock Cambridge University Press, Cambridge, England,
2007.
\newblock

\bibitem{zee_qft-nut}
A.~Zee, {\em Quantum Field Theory in a Nutshell}.
\newblock Princeton University Press, 2~ed., 2010.

\bibitem{kapusta_tft}
J.~Kapusta and C.~Gale, {\em Finite-Tepmerature Field Theory - Principles and
  Applications}.
\newblock Cambridge University Press, 2~ed., 2006.

\bibitem{Alford:2013ota}
M.~G. Alford, S.~Kumar~Mallavarapu, A.~Schmitt, and S.~Stetina, ``{From field
  theory to superfluid hydrodynamics of dense quark matter},'' in {\em
  {Proceedings, Compact Stars in the QCD Phase Diagram III (CSQCD III):
  Guarujá, SP, Brazil, December 12-15, 2012}}.
\newblock 2013.
\newblock \href{http://arxiv.org/abs/1304.7102}{{\ttfamily arXiv:1304.7102
  [hep-ph]}}.
\newblock
\url{https://inspirehep.net/record/1230828/files/arXiv:1304.7102.pdf}.
\newblock

\bibitem{Alford:2013koa}
M.~G. Alford, S.~K. Mallavarapu, A.~Schmitt, and S.~Stetina, ``{Role reversal
  in first and second sound in a relativistic superfluid},''
  \href{http://dx.doi.org/10.1103/PhysRevD.89.085005}{{\em Phys. Rev.}
  {\bfseries D89} no.~8, (2014) 085005},
\href{http://arxiv.org/abs/1310.5953}{{\ttfamily arXiv:1310.5953 [hep-ph]}}.

\bibitem{Stetina:2015exa}
S.~Stetina, {\em {From Field Theory to the Hydrodynamics of Relativistic
  Superfluids}}.
\newblock PhD thesis, Technische Universit{\"a}t Wien, 2015.
\newblock \href{http://arxiv.org/abs/1502.00122}{{\ttfamily arXiv:1502.00122
  [hep-ph]}}.

\bibitem{BCSfromGL}
L.~Gorkov, ``Microscopic derivation of the ginzburg-landau equations in the
  theory of superconductivity,'' {\em Zh. Eksp. Teor. Fiz.} {\bfseries 36}
  no.~1918, (1959) .

\bibitem{Ravenhall:1983uh}
D.~G. Ravenhall, C.~J. Pethick, and J.~R. Wilson, ``Structure of matter below
  nuclear saturation density,''
  \href{http://dx.doi.org/10.1103/PhysRevLett.50.2066}{{\em Phys. Rev. Lett.}
  {\bfseries 50} no.~26, (Jun, 1983) 2066--2069}.

\bibitem{Caplan:2016uvu}
M.~E. Caplan and C.~J. Horowitz, ``{Colloquium : Astromaterial science and
  nuclear pasta},'' \href{http://dx.doi.org/10.1103/RevModPhys.89.041002}{{\em
  Rev. Mod. Phys.} {\bfseries 89} no.~4, (2017) 041002},
\href{http://arxiv.org/abs/1606.03646}{{\ttfamily arXiv:1606.03646
  [astro-ph.HE]}}.

\bibitem{landau_intermed}
L.~D. Landau, ``On the theory of superconductivity,'' {\em Phys. Z. Sowjet}
  {\bfseries 11} no.~129, (1937) .

\bibitem{London_book}
F.~London, {\em Superfluids}, vol.~I.
\newblock Wiley, New York, 1950.

\bibitem{abrikosov_magnetic}
A.~Abrikosov, ``On the magnetic properties of superconductors of the second
  group,'' {\em J.Exptl.Theoret.Phys. (U.S.S.R.)} {\bfseries 5} no.~6, (June,
  1957) 1442. \url{http://www.jetp.ac.ru/cgi-bin/dn/e_005_06_1174.pdf}.

\bibitem{Alford:2007np}
M.~G. Alford and G.~Good, ``{Flux tubes and the type-I/type-II transition in a
  superconductor coupled to a superfluid},''
  \href{http://dx.doi.org/10.1103/PhysRevB.78.024510}{{\em Phys. Rev.}
  {\bfseries B78} (2008) 024510},
\href{http://arxiv.org/abs/0712.1810}{{\ttfamily arXiv:0712.1810 [nucl-th]}}.

\bibitem{shubnikov}
J.~N. RJABININ and L.~W. SHUBNIKOW, ``Magnetic properties and critical currents
  of supra-conducting alloys,'' \href{http://dx.doi.org/10.1038/135581a0}{{\em
  Nature} {\bfseries 135} (04, 1935) 581}.

\bibitem{ft_lattice}
H.~F. Hess, R.~B. Robinson, R.~C. Dynes, J.~M. Valles, and J.~V. Waszczak,
  ``Scanning-tunneling-microscope observation of the abrikosov flux lattice and
  the density of states near and inside a fluxoid,''
  \href{http://dx.doi.org/10.1103/PhysRevLett.62.214}{{\em Phys. Rev. Lett.}
  {\bfseries 62} (Jan, 1989) 214--216}.
  \url{https://link.aps.org/doi/10.1103/PhysRevLett.62.214}.

\bibitem{PhysRev.133.A1226}
W.~H. Kleiner, L.~M. Roth, and S.~H. Autler, ``Bulk solution of ginzburg-landau
  equations for type ii superconductors: Upper critical field region,''
  \href{http://dx.doi.org/10.1103/PhysRev.133.A1226}{{\em Phys. Rev.}
  {\bfseries 133} (Mar, 1964) A1226--A1227}.

\bibitem{Higgs_orig}
P.~W. Higgs, ``Broken symmetries and the masses of gauge bosons,''
  \href{http://dx.doi.org/10.1103/PhysRevLett.13.508}{{\em Phys. Rev. Lett.}
  {\bfseries 13} (Oct, 1964) 508--509}.
  \url{https://link.aps.org/doi/10.1103/PhysRevLett.13.508}.

\bibitem{Elitzur}
S.~Elitzur, ``Impossibility of spontaneously breaking local symmetries,''
  \href{http://dx.doi.org/10.1103/PhysRevD.12.3978}{{\em Phys. Rev. D}
  {\bfseries 12} (Dec, 1975) 3978--3982}.
  \url{https://link.aps.org/doi/10.1103/PhysRevD.12.3978}.

\bibitem{Friederich:2011xs}
S.~Friederich, ``{Gauge Symmetry Breaking in Gauge Theories: In Search of
  Clarification},'' \href{http://dx.doi.org/10.1007/s13194-012-0061-y}{{\em
  Eur. J. Phil. Sci.} {\bfseries 3} (2013) 157--182},
\href{http://arxiv.org/abs/1107.4664}{{\ttfamily arXiv:1107.4664
  [physics.hist-ph]}}.

\bibitem{Bogolyubov1958}
N.~Bogoliubov {\em Doklady Akad. Nauk SSSR} {\bfseries 119} (1958) 52.

\bibitem{MIGDAL1959655}
A.~Migdal, ``Superfluidity and the moments of inertia of nuclei,'' {\em Nuclear
  Physics} {\bfseries 13} no.~5, (1959) 655 -- 674.

\bibitem{Sedrakian:2006xm}
A.~Sedrakian and J.~W. Clark, ``{Nuclear superconductivity in compact stars:
  BCS theory and beyond},''
  \href{http://dx.doi.org/10.1142/9789812773043_0006}{{\em Ser. Adv. Quant.
  Many Body Theor.} {\bfseries 8} (2006) 135--174},
\href{http://arxiv.org/abs/nucl-th/0607028}{{\ttfamily arXiv:nucl-th/0607028
  [nucl-th]}}.

\bibitem{Page2014}
D.~Page, J.~M. Lattimer, M.~Prakash, and A.~W. Steiner, ``{Stellar
  Superfluids},'' in {\em Novel Superfluids: Volume 2}, K.~H. Bennemann and
  J.~B. Ketterson, eds., p.~505.
\newblock Oxford University Press, New York, 2014.
\newblock \href{http://arxiv.org/abs/1302.6626}{{\ttfamily arXiv:1302.6626
  [astro-ph.HE]}}.

\bibitem{Gusakov:2009kc}
M.~E. Gusakov, E.~M. Kantor, and P.~Haensel, ``{The relativistic entrainment
  matrix of a superfluid nucleon-hyperon mixture at zero temperature},''
  \href{http://dx.doi.org/10.1103/PhysRevC.79.055806}{{\em Phys. Rev.}
  {\bfseries C79} (2009) 055806},
\href{http://arxiv.org/abs/0904.3467}{{\ttfamily arXiv:0904.3467
  [astro-ph.HE]}}.

\bibitem{Alford:1998mk}
M.~G. Alford, K.~Rajagopal, and F.~Wilczek, ``Color-flavor locking and chiral
  symmetry breaking in high density {QCD},'' {\em Nucl. Phys.} {\bfseries B537}
  (1999) 443--458,
\href{http://arxiv.org/abs/hep-ph/9804403}{{\ttfamily hep-ph/9804403}}.

\bibitem{Schafer:2000tw}
T.~Sch{\"a}fer, ``Quark hadron continuity in qcd with one flavor,'' {\em Phys.
  Rev.} {\bfseries D62} (2000) 094007,
\href{http://arxiv.org/abs/hep-ph/0006034}{{\ttfamily hep-ph/0006034}}.

\bibitem{Schmitt:2004et}
A.~Schmitt, ``The ground state in a spin-one color superconductor,'' {\em Phys.
  Rev.} {\bfseries D71} (2005) 054016,
\href{http://arxiv.org/abs/nucl-th/0412033}{{\ttfamily nucl-th/0412033}}.

\bibitem{1976JETP...42..164A}
A.~F. {Andreev} and E.~P. {Bashkin}, ``{Three-velocity hydrodynamics of
  superfluid solutions},'' {\em Zh. Eksp. Teor. Fiz.} {\bfseries 69} (1975)
  319.

\bibitem{2007JETP..105..135S}
S.~I. {Shevchenko} and D.~V. {Fil}, ``{The Andreev-Bashkin effect in a
  two-component Bose gas},''
  \href{http://dx.doi.org/10.1134/S106377610707028X}{{\em Soviet Journal of
  Experimental and Theoretical Physics} {\bfseries 105} (July, 2007) 135--137}.

\bibitem{Chamel:2006rc}
N.~Chamel and P.~Haensel, ``{Entrainment parameters in cold superfluid neutron
  star core},'' \href{http://dx.doi.org/10.1103/PhysRevC.73.045802}{{\em Phys.
  Rev.} {\bfseries C73} (2006) 045802},
\href{http://arxiv.org/abs/nucl-th/0603018}{{\ttfamily arXiv:nucl-th/0603018
  [nucl-th]}}.

\bibitem{2004MNRAS.354..101A}
N.~{Andersson}, G.~L. {Comer}, and R.~{Prix}, ``{The superfluid two-stream
  instability},''
  \href{http://dx.doi.org/10.1111/j.1365-2966.2004.08166.x}{{\em
  Mon.Not.Roy.Astron.Soc.} {\bfseries 354} (Oct., 2004) 101--110}.

\bibitem{Peralta:2006um}
C.~Peralta, A.~Melatos, M.~Giacobello, and A.~Ooi, ``{Transitions between
  turbulent and laminar superfluid vorticity states in the outer core of a
  neutron star},'' \href{http://dx.doi.org/10.1086/507576}{{\em Astrophys. J.}
  {\bfseries 651} (2006) 1079--1091},
\href{http://arxiv.org/abs/astro-ph/0607161}{{\ttfamily arXiv:astro-ph/0607161
  [astro-ph]}}.

\bibitem{PhysRevA.64.061603}
B.~Wu and Q.~Niu, ``Landau and dynamical instabilities of the superflow of
  bose-einstein condensates in optical lattices,''
  \href{http://dx.doi.org/10.1103/PhysRevA.64.061603}{{\em Phys. Rev. A}
  {\bfseries 64} (Nov, 2001) 061603}.

\bibitem{pethick2010superfluid}
C.~Pethick, N.~Chamel, and S.~Reddy, ``Superfluid dynamics in neutron star
  crusts,'' {\em Progress of Theoretical Physics Supplement} {\bfseries 186}
  (2010) 9--16.

\bibitem{Chamel:2012pk}
N.~Chamel, J.~M. Pearson, and S.~Goriely, ``{Superfluidity and entrainment in
  neutron-star crusts},'' {\em ASP Conf. Ser.} {\bfseries 466} (2012) 203,
\href{http://arxiv.org/abs/1206.6926}{{\ttfamily arXiv:1206.6926
  [astro-ph.HE]}}.

\bibitem{Andersson:2012iu}
N.~Andersson, K.~Glampedakis, W.~C.~G. Ho, and C.~M. Espinoza, ``{Pulsar
  glitches: The crust is not enough},''
  \href{http://dx.doi.org/10.1103/PhysRevLett.109.241103}{{\em Phys. Rev.
  Lett.} {\bfseries 109} (2012) 241103},
\href{http://arxiv.org/abs/1207.0633}{{\ttfamily arXiv:1207.0633
  [astro-ph.SR]}}.

\bibitem{tisza38}
L.~Tisza, ``{Transport Phenomena in Helium II},'' {\em Nature} {\bfseries 141}
  (1938) 913.

\bibitem{landau41}
L.~Landau, ``{Theory of the Superfluidity of Helium II},'' {\em Phys. Rev.}
  {\bfseries 60} (1941) 356.

\bibitem{2002JLTP..129..531T}
J.~{Tuoriniemi}, J.~{Martikainen}, E.~{Pentti}, A.~{Sebedash}, S.~{Boldarev},
  and G.~{Pickett} \href{http://dx.doi.org/10.1023/A:1021468614550}{{\em
  Journal of Low Temperature Physics} {\bfseries 129} (2002) 531--545}.

\bibitem{PhysRevB.85.134529}
J.~Rysti, J.~Tuoriniemi, and A.~Salmela, ``Effective ${}^{3}$he interactions in
  dilute ${}^{3}$he-${}^{4}$he mixtures,''
  \href{http://dx.doi.org/10.1103/PhysRevB.85.134529}{{\em Phys. Rev. B}
  {\bfseries 85} (Apr, 2012) 134529}.

\bibitem{2014Sci...345.1035F}
I.~{Ferrier-Barbut}, M.~{Delehaye}, S.~{Laurent}, A.~T. {Grier}, M.~{Pierce},
  B.~S. {Rem}, F.~{Chevy}, and C.~{Salomon}, ``{A mixture of Bose and Fermi
  superfluids},'' \href{http://dx.doi.org/10.1126/science.1255380}{{\em
  Science} {\bfseries 345} (Aug., 2014) 1035--1038},
  \href{http://arxiv.org/abs/1404.2548}{{\ttfamily arXiv:1404.2548
  [cond-mat.quant-gas]}}.

\bibitem{PhysRevLett.115.265303}
M.~Delehaye, S.~Laurent, I.~Ferrier-Barbut, S.~Jin, F.~Chevy, and C.~Salomon,
  ``Critical velocity and dissipation of an ultracold bose-fermi counterflow,''
  \href{http://dx.doi.org/10.1103/PhysRevLett.115.265303}{{\em Phys. Rev.
  Lett.} {\bfseries 115} (Dec, 2015) 265303}.
  \url{https://link.aps.org/doi/10.1103/PhysRevLett.115.265303}.

\bibitem{Schmitt:2013nva}
A.~Schmitt, ``{Superfluid two-stream instability in a microscopic model},''
  \href{http://dx.doi.org/10.1103/PhysRevD.89.065024}{{\em Phys. Rev.}
  {\bfseries D89} no.~6, (2014) 065024},
\href{http://arxiv.org/abs/1312.5993}{{\ttfamily arXiv:1312.5993 [hep-ph]}}.

\bibitem{PhysRevA.63.063612}
C.~K. Law, C.~M. Chan, P.~T. Leung, and M.-C. Chu, ``Critical velocity in a
  binary mixture of moving bose condensates,''
  \href{http://dx.doi.org/10.1103/PhysRevA.63.063612}{{\em Phys. Rev. A}
  {\bfseries 63} (May, 2001) 063612}.

\bibitem{2011PhRvA..83f3602I}
S.~{Ishino}, M.~{Tsubota}, and H.~{Takeuchi}, ``{Countersuperflow instability
  in miscible two-component Bose-Einstein condensates},''
  \href{http://dx.doi.org/10.1103/PhysRevA.83.063602}{{\em Phys. Rev. A}
  {\bfseries 83} no.~6, (June, 2011) 063602},
  \href{http://arxiv.org/abs/1106.0884}{{\ttfamily arXiv:1106.0884
  [cond-mat.quant-gas]}}.

\bibitem{2015EPJD...69..126A}
M.~{Abad}, A.~{Recati}, S.~{Stringari}, and F.~{Chevy}, ``{Counter-flow
  instability of a quantum mixture of two superfluids},''
  \href{http://dx.doi.org/10.1140/epjd/e2015-50851-y}{{\em European Physical
  Journal D} {\bfseries 69} (May, 2015) 126},
  \href{http://arxiv.org/abs/1411.7560}{{\ttfamily arXiv:1411.7560
  [cond-mat.quant-gas]}}.

\bibitem{2003JETPL..78..574A}
A.~F. {Andreev} and L.~A. {Melnikovsky}, ``{Thermodynamic Inequalities in a
  Superfluid},'' \href{http://dx.doi.org/10.1134/1.1641487}{{\em Soviet Journal
  of Experimental and Theoretical Physics Letters} {\bfseries 78} (Nov., 2003)
  574--577}, \href{http://arxiv.org/abs/cond-mat/0304019}{{\ttfamily
  cond-mat/0304019}}.

\bibitem{2006JETP..103..944A}
A.~F. {Andreev} and L.~A. {Melnikovsky}, ``{Two-fluid hydrodynamics: New
  aspects},'' \href{http://dx.doi.org/10.1134/S1063776106120120}{{\em Soviet
  Journal of Experimental and Theoretical Physics} {\bfseries 103} (Dec., 2006)
  944--961}.

\bibitem{2008JLTP..150..612K}
L.~Y. {Kravchenko} and D.~V. {Fil}, ``{Critical Velocities in Two-Component
  Superfluid Bose Gases},''
  \href{http://dx.doi.org/10.1007/s10909-007-9595-3}{{\em Journal of Low
  Temperature Physics} {\bfseries 150} (Feb., 2008) 612--617},
  \href{http://arxiv.org/abs/0807.0726}{{\ttfamily arXiv:0807.0726
  [cond-mat.other]}}.

\bibitem{Glampedakis:2010sk}
K.~Glampedakis, N.~Andersson, and L.~Samuelsson, ``{Magnetohydrodynamics of
  superfluid and superconducting neutron star cores},''
  \href{http://dx.doi.org/10.1111/j.1365-2966.2010.17484.x}{{\em Mon. Not. Roy.
  Astron. Soc.} {\bfseries 410} (2011) 805},
\href{http://arxiv.org/abs/1001.4046}{{\ttfamily arXiv:1001.4046
  [astro-ph.SR]}}.

\bibitem{Drischler:2016cpy}
C.~Drischler, T.~Krüger, K.~Hebeler, and A.~Schwenk, ``{Pairing in neutron
  matter: New uncertainty estimates and three-body forces},''
  \href{http://dx.doi.org/10.1103/PhysRevC.95.024302}{{\em Phys. Rev.}
  {\bfseries C95} no.~2, (2017) 024302},
\href{http://arxiv.org/abs/1610.05213}{{\ttfamily arXiv:1610.05213 [nucl-th]}}.

\bibitem{Graber:2016imq}
V.~Graber, N.~Andersson, and M.~Hogg, ``{Neutron Stars in the Laboratory},''
  \href{http://dx.doi.org/10.1142/S0218271817300154}{{\em Int. J. Mod. Phys.}
  {\bfseries D26} no.~08, (2017) 1730015},
\href{http://arxiv.org/abs/1610.06882}{{\ttfamily arXiv:1610.06882
  [astro-ph.HE]}}.

\bibitem{wambach1993quasiparticle}
J.~Wambach, T.~Ainsworth, and D.~Pines, ``Quasiparticle interactions in neutron
  matter for applications in neutron stars,'' {\em Nuclear Physics A}
  {\bfseries 555} no.~1, (1993) 128--150.

\bibitem{Iida:2004if}
K.~Iida, ``Magnetic vortex in color-flavor locked quark matter,'' {\em Phys.
  Rev.} {\bfseries D71} (2005) 054011,
\href{http://arxiv.org/abs/hep-ph/0412426}{{\ttfamily hep-ph/0412426}}.

\bibitem{Giannakis:2003am}
I.~Giannakis and H.-c. Ren, ``The ginzburg-landau theory and the surface energy
  of a colour superconductor,'' {\em Nucl. Phys.} {\bfseries B669} (2003)
  462--478,
\href{http://arxiv.org/abs/hep-ph/0305235}{{\ttfamily hep-ph/0305235}}.

\bibitem{Bedaque:2001je}
P.~F. Bedaque and T.~Sch{\"a}fer, ``High density quark matter under stress,''
  {\em Nucl. Phys.} {\bfseries A697} (2002) 802--822,
\href{http://arxiv.org/abs/hep-ph/0105150}{{\ttfamily hep-ph/0105150}}.

\bibitem{2009Natur.462..628L}
Y.-J. {Lin}, R.~L. {Compton}, K.~{Jim{\'e}nez-Garcia}, J.~V. {Porto}, and I.~B.
  {Spielman}, ``{Synthetic magnetic fields for ultracold neutral atoms},''
  \href{http://dx.doi.org/10.1038/nature08609}{{\em Nature} {\bfseries 462}
  (Dec., 2009) 628--632}, \href{http://arxiv.org/abs/1007.0294}{{\ttfamily
  arXiv:1007.0294 [cond-mat.quant-gas]}}.

\bibitem{2011RvMP...83.1523D}
J.~{Dalibard}, F.~{Gerbier}, G.~{Juzeli{\= u}nas}, and P.~{{\"O}hberg},
  ``{Colloquium: Artificial gauge potentials for neutral atoms},''
  \href{http://dx.doi.org/10.1103/RevModPhys.83.1523}{{\em Reviews of Modern
  Physics} {\bfseries 83} (Oct., 2011) 1523--1543},
  \href{http://arxiv.org/abs/1008.5378}{{\ttfamily arXiv:1008.5378
  [cond-mat.quant-gas]}}.

\bibitem{2014RPPh...77l6401G}
N.~{Goldman}, G.~{Juzeli{\= u}nas}, P.~{{\"O}hberg}, and I.~B. {Spielman},
  ``{Light-induced gauge fields for ultracold atoms},''
  \href{http://dx.doi.org/10.1088/0034-4885/77/12/126401}{{\em Reports on
  Progress in Physics} {\bfseries 77} no.~12, (Dec., 2014) 126401},
  \href{http://arxiv.org/abs/1308.6533}{{\ttfamily arXiv:1308.6533
  [cond-mat.quant-gas]}}.

\bibitem{Carlstrom:2010wn}
J.~Carlstrom, E.~Babaev, and M.~Speight, ``{Type-1.5 superconductivity in
  multiband systems: the effects of interband couplings},''
  \href{http://dx.doi.org/10.1103/PhysRevB.83.174509}{{\em Phys. Rev.}
  {\bfseries B83} (2011) 174509},
\href{http://arxiv.org/abs/1009.2196}{{\ttfamily arXiv:1009.2196
  [cond-mat.supr-con]}}.

\bibitem{Brandt2011}
E.~H. Brandt and M.~P. Das, ``Attractive vortex interaction and the
  intermediate-mixed state of??superconductors,'' {\em Journal of
  Superconductivity and Novel Magnetism} {\bfseries 24} no.~1, (2011) 57--67.

\bibitem{2012arXiv1206.6786B}
E.~Babaev and M.~Silaev, ``Type-1.5 superconductivity in multiband and other
  multicomponent systems,'' {\em Journal of Superconductivity and Novel
  Magnetism} {\bfseries 26} no.~5, (2013) 2045--2055,
  \href{http://arxiv.org/abs/1206.6786}{{\ttfamily 1206.6786}}.

\bibitem{Wu:2015sqk}
M.-S. Wu, S.-Y. Wu, and H.-Q. Zhang, ``{Vortex in holographic two-band
  superfluid/superconductor},''
  \href{http://dx.doi.org/10.1007/JHEP05(2016)011}{{\em JHEP} {\bfseries 05}
  (2016) 011},
\href{http://arxiv.org/abs/1511.01325}{{\ttfamily arXiv:1511.01325 [hep-th]}}.

\bibitem{babaev2004superconductor}
E.~Babaev, A.~Sudbo, and N.~Ashcroft, ``A superconductor to superfluid phase
  transition in liquid metallic hydrogen,'' {\em Nature} {\bfseries 431}
  no.~7009, (2004) 666--668.

\bibitem{PhysRevB.72.180502}
E.~Babaev and M.~Speight, ``Semi-meissner state and neither type-i nor type-ii
  superconductivity in multicomponent superconductors,''
  \href{http://dx.doi.org/10.1103/PhysRevB.72.180502}{{\em Phys. Rev. B}
  {\bfseries 72} (Nov, 2005) 180502}.

\bibitem{PhysRevLett.105.067003}
E.~Babaev, J.~Carlstr\"om, and M.~Speight, ``Type-1.5 superconducting state
  from an intrinsic proximity effect in two-band superconductors,''
  \href{http://dx.doi.org/10.1103/PhysRevLett.105.067003}{{\em Phys. Rev.
  Lett.} {\bfseries 105} (Aug, 2010) 067003}.

\bibitem{2013PhRvD..87f5001A}
M.~G. {Alford}, S.~K. {Mallavarapu}, A.~{Schmitt}, and S.~{Stetina}, ``{From a
  complex scalar field to the two-fluid picture of superfluidity},''
  \href{http://dx.doi.org/10.1103/PhysRevD.87.065001}{{\em Phys. Rev. D}
  {\bfseries 87} no.~6, (Mar., 2013) 065001},
  \href{http://arxiv.org/abs/1212.0670}{{\ttfamily arXiv:1212.0670 [hep-ph]}}.

\bibitem{Fejos:2016wza}
G.~Fejos and T.~Hatsuda, ``{Fixed point structure of the Abelian Higgs
  model},'' \href{http://dx.doi.org/10.1103/PhysRevD.93.121701}{{\em Phys.
  Rev.} {\bfseries D93} no.~12, (2016) 121701},
\href{http://arxiv.org/abs/1604.05849}{{\ttfamily arXiv:1604.05849
  [cond-mat.supr-con]}}.

\bibitem{Kapusta:1981aa}
J.~I. Kapusta, ``{Bose-Einstein Condensation, Spontaneous Symmetry Breaking,
  and Gauge Theories},''
\href{http://dx.doi.org/10.1103/PhysRevD.24.426}{{\em Phys. Rev.} {\bfseries
  D24} (1981) 426--439}.

\bibitem{Alford:2007qa}
M.~G. Alford, M.~Braby, and A.~Schmitt, ``{Critical temperature for kaon
  condensation in color-flavor locked quark matter},'' {\em J. Phys.}
  {\bfseries G35} (2008) 025002,
\href{http://arxiv.org/abs/arXiv:0707.2389 [nucl-th]}{{\ttfamily
  arXiv:0707.2389 [nucl-th]}}.

\bibitem{Schaefer:2014awa}
T.~Schaefer, ``{Fluid Dynamics and Viscosity in Strongly Correlated Fluids},''
  \href{http://dx.doi.org/10.1146/annurev-nucl-102313-025439}{{\em Ann. Rev.
  Nucl. Part. Sci.} {\bfseries 64} (2014) 125--148},
\href{http://arxiv.org/abs/1403.0653}{{\ttfamily arXiv:1403.0653 [hep-ph]}}.

\bibitem{Jeon:2015dfa}
S.~Jeon and U.~Heinz, ``{Introduction to Hydrodynamics},''
  \href{http://dx.doi.org/10.1142/S0218301315300106}{{\em Int. J. Mod. Phys.}
  {\bfseries E24} no.~10, (2015) 1530010},
\href{http://arxiv.org/abs/1503.03931}{{\ttfamily arXiv:1503.03931 [hep-ph]}}.

\bibitem{Kovtun:2012rj}
P.~Kovtun, ``{Lectures on hydrodynamic fluctuations in relativistic
  theories},'' \href{http://dx.doi.org/10.1088/1751-8113/45/47/473001}{{\em J.
  Phys.} {\bfseries A45} (2012) 473001},
\href{http://arxiv.org/abs/1205.5040}{{\ttfamily arXiv:1205.5040 [hep-th]}}.

\bibitem{Romatschke:2017ejr}
P.~Romatschke and U.~Romatschke, ``{Relativistic Fluid Dynamics In and Out of
  Equilibrium -- Ten Years of Progress in Theory and Numerical Simulations of
  Nuclear Collisions},''
\href{http://arxiv.org/abs/1712.05815}{{\ttfamily arXiv:1712.05815 [nucl-th]}}.

\bibitem{Alford:2012mv}
M.~G. Alford, S.~K. Mallavarapu, A.~Schmitt, and S.~Stetina, ``{Relativistic
  superfluid hydrodynamics from field theory},'' {\em PoS} {\bfseries
  ConfinementX} (2012) 256,
\href{http://arxiv.org/abs/1212.4410}{{\ttfamily arXiv:1212.4410 [hep-ph]}}.

\bibitem{Carter:1995if}
B.~Carter and D.~Langlois, ``{The Equation of state for cool relativistic two
  constituent superfluid dynamics},''
  \href{http://dx.doi.org/10.1103/PhysRevD.51.5855}{{\em Phys.Rev.} {\bfseries
  D51} (1995) 5855--5864},
\href{http://arxiv.org/abs/hep-th/9507058}{{\ttfamily arXiv:hep-th/9507058
  [hep-th]}}.

\bibitem{PhysRevB.77.144515}
I.~N. Adamenko, K.~E. Nemchenko, V.~A. Slipko, and A.~F.~G. Wyatt, ``Transverse
  sound in differentially moving superfluid helium,''
  \href{http://dx.doi.org/10.1103/PhysRevB.77.144515}{{\em Phys. Rev. B}
  {\bfseries 77} (Apr, 2008) 144515}.

\bibitem{Fukushima:2005gt}
K.~Fukushima and K.~Iida, ``Collective excitations in a superfluid of
  color-flavor locked quark matter,'' {\em Phys. Rev.} {\bfseries D71} (2005)
  074011,
\href{http://arxiv.org/abs/hep-ph/0501276}{{\ttfamily hep-ph/0501276}}.

\bibitem{2015JPSJ...84d4003Y}
H.~{Yamamura} and D.~{Yamamoto}, ``{Collective Excitation and Stability of
  Flowing Gapless Fermi Superfluids},''
  \href{http://dx.doi.org/10.7566/JPSJ.84.044003}{{\em Journal of the Physical
  Society of Japan} {\bfseries 84} no.~4, (Apr., 2015) 044003},
  \href{http://arxiv.org/abs/1403.5621}{{\ttfamily arXiv:1403.5621
  [cond-mat.quant-gas]}}.

\bibitem{PhysRevLett.24.611}
S.~Chandrasekhar, ``Solutions of two problems in the theory of gravitational
  radiation,'' \href{http://dx.doi.org/10.1103/PhysRevLett.24.611}{{\em Phys.
  Rev. Lett.} {\bfseries 24} (Mar, 1970) 611--615}.

\bibitem{1978ApJ...221..937F}
J.~L. {Friedman} and B.~F. {Schutz}, ``{Lagrangian perturbation theory of
  nonrelativistic fluids},'' \href{http://dx.doi.org/10.1086/156098}{{\em
  Astrophys. J.} {\bfseries 221} (May, 1978) 937--957}.

\bibitem{PhysRevA.87.063610}
T.~Ozawa, L.~P. Pitaevskii, and S.~Stringari, ``Supercurrent and dynamical
  instability of spin-orbit-coupled ultracold bose gases,''
  \href{http://dx.doi.org/10.1103/PhysRevA.87.063610}{{\em Phys. Rev. A}
  {\bfseries 87} (Jun, 2013) 063610}.

\bibitem{Buneman:1959zz}
O.~Buneman, ``{Dissipation of Currents in Ionized Media},''
\href{http://dx.doi.org/10.1103/PhysRev.115.503}{{\em Phys.Rev.} {\bfseries
  115} (1959) 503--517}.

\bibitem{1963PhRvL..10..279F}
D.~T. {Farley}, ``{Two-Stream Plasma Instability as a Source of Irregularities
  in the Ionosphere},''
  \href{http://dx.doi.org/10.1103/PhysRevLett.10.279}{{\em Physical Review
  Letters} {\bfseries 10} (Apr., 1963) 279--282}.

\bibitem{2001AmJPh..69.1262A}
D.~{Anderson}, R.~{Fedele}, and M.~{Lisak}, ``{A tutorial presentation of the
  two stream instability and Landau damping},''
  \href{http://dx.doi.org/10.1119/1.1407252}{{\em American Journal of Physics}
  {\bfseries 69} (Dec., 2001) 1262--1266}.

\bibitem{Samuelsson:2009up}
L.~Samuelsson, C.~Lopez-Monsalvo, N.~Andersson, and G.~Comer, ``{Relativistic
  Two-stream Instability},''
  \href{http://dx.doi.org/10.1007/s10714-009-0861-3}{{\em Gen.Rel.Grav.}
  {\bfseries 42} (2010) 413--433},
\href{http://arxiv.org/abs/0906.4002}{{\ttfamily arXiv:0906.4002 [gr-qc]}}.

\bibitem{Hawke:2013haa}
I.~Hawke, G.~Comer, and N.~Andersson, ``{The nonlinear development of the
  relativistic two-stream instability},''
  \href{http://dx.doi.org/10.1088/0264-9381/30/14/145007}{{\em
  Class.Quant.Grav.} {\bfseries 30} (2013) 145007},
\href{http://arxiv.org/abs/1303.4070}{{\ttfamily arXiv:1303.4070 [gr-qc]}}.

\bibitem{Gubankova:2006gj}
E.~Gubankova, A.~Schmitt, and F.~Wilczek, ``Stability conditions and fermi
  surface topologies in a superconductor,'' {\em Phys. Rev.} {\bfseries B74}
  (2006) 064505,
\href{http://arxiv.org/abs/cond-mat/0603603}{{\ttfamily cond-mat/0603603}}.

\bibitem{Deng:2006ed}
J.~Deng, A.~Schmitt, and Q.~Wang, ``{Relativistic BCS-BEC crossover in a
  boson-fermion model},''
  \href{http://dx.doi.org/10.1103/PhysRevD.76.034013}{{\em Phys. Rev.}
  {\bfseries D76} (2007) 034013},
\href{http://arxiv.org/abs/nucl-th/0611097}{{\ttfamily arXiv:nucl-th/0611097
  [nucl-th]}}.

\bibitem{Huang:2006kr}
X.~Huang, X.~Hao, and P.~Zhuang, ``{Phase diagram of asymmetric fermion
  superfluid including single-species pairing},''
  \href{http://dx.doi.org/10.1088/1367-2630/9/10/375}{{\em New J. Phys.}
  {\bfseries 9} (2007) 375},
\href{http://arxiv.org/abs/cond-mat/0610610}{{\ttfamily arXiv:cond-mat/0610610
  [cond-mat.supr-con]}}.

\bibitem{2004LaPhL...1...50Y}
V.~I. {Yukalov} and E.~P. {Yukalova}, ``{Stratification of moving
  multicomponent Bose-Einstein condensates},''
  \href{http://dx.doi.org/10.1002/lapl.200310012}{{\em Laser Physics Letters}
  {\bfseries 1} (Jan., 2004) 50--53},
  \href{http://arxiv.org/abs/cond-mat/0401234}{{\ttfamily cond-mat/0401234}}.

\bibitem{2009JPhCS.150c2057M}
L.~A. {Melnikovsky}, ``{Superfluid stability near Landau critical velocity},''
  \href{http://dx.doi.org/10.1088/1742-6596/150/3/032057}{{\em Journal of
  Physics Conference Series} {\bfseries 150} no.~3, (Feb., 2009) 032057}.

\bibitem{Landea:2014naa}
I.~S. Landea, ``{Inhomogeneous superfluids},''
\href{http://arxiv.org/abs/1410.7865}{{\ttfamily arXiv:1410.7865 [hep-th]}}.

\bibitem{1922}
E.~L. Rees, ``Graphical discussion of the roots of a quartic equation,'' {\em
  The American Mathematical Monthly} {\bfseries 29} no.~2, (1922) pp. 51--55.

\bibitem{2015arXiv150400570K}
D.~N. {Kobyakov} and C.~J. {Pethick}, ``{Two-component Superfluid Hydrodynamics
  of Neutron Star Cores},''
  \href{http://dx.doi.org/10.3847/1538-4357/836/2/203}{{\em Astrophys.\ J.}
  {\bfseries 836} (Feb., 2017) 203},
  \href{http://arxiv.org/abs/1504.00570}{{\ttfamily arXiv:1504.00570
  [cond-mat.quant-gas]}}.

\bibitem{Sinha:2015bva}
M.~Sinha and A.~Sedrakian, ``{Magnetar superconductivity versus magnetism:
  neutrino cooling processes},''
  \href{http://dx.doi.org/10.1103/PhysRevC.91.035805}{{\em Phys. Rev.}
  {\bfseries C91} no.~3, (2015) 035805},
\href{http://arxiv.org/abs/1502.02979}{{\ttfamily arXiv:1502.02979
  [astro-ph.HE]}}.

\bibitem{Kramer:1971zza}
L.~Kramer, ``{Thermodynamic Behavior of Type-II Superconductors with Small
  kappa near the Lower Critical Field},''
\href{http://dx.doi.org/10.1103/PhysRevB.3.3821}{{\em Phys. Rev.} {\bfseries
  B3} (1971) 3821--3825}.

\bibitem{Speight:1996px}
J.~M. Speight, ``{Static intervortex forces},''
  \href{http://dx.doi.org/10.1103/PhysRevD.55.3830}{{\em Phys. Rev.} {\bfseries
  D55} (1997) 3830--3835},
\href{http://arxiv.org/abs/hep-th/9603155}{{\ttfamily arXiv:hep-th/9603155
  [hep-th]}}.

\bibitem{Buckley:2003zf}
K.~B.~W. Buckley, M.~A. Metlitski, and A.~R. Zhitnitsky, ``{Neutron stars as
  type I superconductors},''
  \href{http://dx.doi.org/10.1103/PhysRevLett.92.151102}{{\em Phys. Rev. Lett.}
  {\bfseries 92} (2004) 151102},
\href{http://arxiv.org/abs/astro-ph/0308148}{{\ttfamily arXiv:astro-ph/0308148
  [astro-ph]}}.

\bibitem{Buckley:2004ca}
K.~B.~W. Buckley, M.~A. Metlitski, and A.~R. Zhitnitsky, ``{Vortices and type I
  superconductivity in neutron stars},''
  \href{http://dx.doi.org/10.1103/PhysRevC.69.055803}{{\em Phys. Rev.}
  {\bfseries C69} (2004) 055803},
\href{http://arxiv.org/abs/hep-ph/0403230}{{\ttfamily arXiv:hep-ph/0403230
  [hep-ph]}}.

\bibitem{Alford:2005ku}
M.~Alford, G.~Good, and S.~Reddy, ``{Isospin asymmetry and type-I
  superconductivity in neutron star matter},''
  \href{http://dx.doi.org/10.1103/PhysRevC.72.055801}{{\em Phys. Rev.}
  {\bfseries C72} (2005) 055801},
\href{http://arxiv.org/abs/nucl-th/0505025}{{\ttfamily arXiv:nucl-th/0505025
  [nucl-th]}}.

\bibitem{Bettencourt:1994kf}
L.~M.~A. Bettencourt and R.~J. Rivers, ``{Interactions between U(1) cosmic
  strings: An Analytical study},''
  \href{http://dx.doi.org/10.1103/PhysRevD.51.1842}{{\em Phys. Rev.} {\bfseries
  D51} (1995) 1842--1853},
\href{http://arxiv.org/abs/hep-ph/9405222}{{\ttfamily arXiv:hep-ph/9405222
  [hep-ph]}}.

\bibitem{MacKenzie:2003jp}
R.~MacKenzie, M.~A. Vachon, and U.~F. Wichoski, ``{Interaction between vortices
  in models with two order parameters},''
  \href{http://dx.doi.org/10.1103/PhysRevD.67.105024}{{\em Phys. Rev.}
  {\bfseries D67} (2003) 105024},
\href{http://arxiv.org/abs/hep-th/0301188}{{\ttfamily arXiv:hep-th/0301188
  [hep-th]}}.

\bibitem{Forgacs:2016ndn}
P.~Forgacs and {\'A}.~Luk{\'a}cs, ``{Vortices with scalar condensates in
  two-component Ginzburg-Landau systems},''
  \href{http://dx.doi.org/10.1016/j.physletb.2016.09.003}{{\em Phys. Lett.}
  {\bfseries B762} (2016) 271--275},
\href{http://arxiv.org/abs/1603.03291}{{\ttfamily arXiv:1603.03291 [hep-th]}}.

\bibitem{Forgacs:2016iva}
P.~Forgacs and {\'A}.~Luk{\'a}cs, ``{Vortices and magnetic bags in Abelian
  models with extended scalar sectors and some of their applications},''
  \href{http://dx.doi.org/10.1103/PhysRevD.94.125018}{{\em Phys. Rev.}
  {\bfseries D94} no.~12, (2016) 125018},
\href{http://arxiv.org/abs/1608.00021}{{\ttfamily arXiv:1608.00021 [hep-th]}}.

\bibitem{RevModPhys.89.041002}
M.~E. Caplan and C.~J. Horowitz, ``Colloquium: Astromaterial science and
  nuclear pasta,'' \href{http://dx.doi.org/10.1103/RevModPhys.89.041002}{{\em
  Rev. Mod. Phys.} {\bfseries 89} (Oct, 2017) 041002}.
  \url{https://link.aps.org/doi/10.1103/RevModPhys.89.041002}.

\bibitem{1995ApJ...447..305S}
A.~D. {Sedrakian} and D.~M. {Sedrakian}, ``{Superfluid Core Rotation in
  Pulsars. I. Vortex Cluster Dynamics},''
  \href{http://dx.doi.org/10.1086/175876}{{\em Astrophys.\ J.} {\bfseries 447}
  (July, 1995) 305}.

\bibitem{david_alex}
A.~Haber and D.~M{\"u}ller , in preparation.

\bibitem{Baym:1976yu}
G.~Baym and S.~A. Chin, ``Can a neutron star be a giant mit bag?,''
{\em Phys. Lett.} {\bfseries B62} (1976) 241--244.

\bibitem{Barrois:1977xd}
B.~C. Barrois, ``Superconducting quark matter,''
{\em Nucl. Phys.} {\bfseries B129} (1977) 390.

\bibitem{Barrois:1979pv}
B.~C. Barrois, {\em Non-perturbative effects in dense quark matter}.
\newblock PhD thesis, California Institute of Technology, Pasadena, California,
  1979.
\newblock {UMI} 79-04847.

\bibitem{Frautschi:1978rz}
S.~C. Frautschi, ``Asymptotic freedom and color superconductivity in dense
  quark matter,''. Presented at Workshop on Hadronic Matter at Extreme Energy
  Density, Erice, Italy, Oct 13-21, 1978.

\bibitem{Bailin:1979nh}
D.~Bailin and A.~Love, ``Superfluid quark matter,''
{\em J. Phys.} {\bfseries A12} (1979) L283.

\bibitem{Bailin:1983bm}
D.~Bailin and A.~Love, ``Superfluidity and superconductivity in relativistic
  fermion systems,''
{\em Phys. Rept.} {\bfseries 107} (1984) 325.

\bibitem{Schafer:1999jg}
T.~Sch{\"a}fer and F.~Wilczek, ``Superconductivity from perturbative one-gluon
  exchange in high density quark matter,'' {\em Phys. Rev.} {\bfseries D60}
  (1999) 114033,
\href{http://arxiv.org/abs/hep-ph/9906512}{{\ttfamily hep-ph/9906512}}.

\bibitem{Schmitt:2004hg}
A.~Schmitt, {\em Spin-one color superconductivity in cold and dense quark
  matter}.
\newblock PhD thesis, Johann-Wolfgang-Goethe-Universit{\"a}t, Frankfurt/Main,
  Germany, 2004.
\newblock
\href{http://arxiv.org/abs/nucl-th/0405076}{{\ttfamily nucl-th/0405076}}.
\newblock

\bibitem{Pisarski:1999av}
R.~D. Pisarski and D.~H. Rischke, ``{Superfluidity in a model of massless
  fermions coupled to scalar bosons},''
  \href{http://dx.doi.org/10.1103/PhysRevD.60.094013}{{\em Phys. Rev.}
  {\bfseries D60} (1999) 094013},
\href{http://arxiv.org/abs/nucl-th/9903023}{{\ttfamily arXiv:nucl-th/9903023
  [nucl-th]}}.

\bibitem{Eto:2013hoa}
M.~Eto, Y.~Hirono, M.~Nitta, and S.~Yasui, ``{Vortices and Other Topological
  Solitons in Dense Quark Matter},''
  \href{http://dx.doi.org/10.1093/ptep/ptt095}{{\em PTEP} {\bfseries 2014}
  no.~1, (2014) 012D01},
\href{http://arxiv.org/abs/1308.1535}{{\ttfamily arXiv:1308.1535 [hep-ph]}}.

\bibitem{1961PhRv..124..246N}
Y.~{Nambu} and G.~{Jona-Lasinio}, ``{Dynamical Model of Elementary Particles
  Based on an Analogy with Superconductivity. II},''
  \href{http://dx.doi.org/10.1103/PhysRev.124.246}{{\em Physical Review}
  {\bfseries 124} (Oct., 1961) 246--254}.

\bibitem{Buballa:2003qv}
M.~Buballa, ``Njl model analysis of quark matter at large density,'' {\em Phys.
  Rept.} {\bfseries 407} (2005) 205--376,
\href{http://arxiv.org/abs/hep-ph/0402234}{{\ttfamily hep-ph/0402234}}.

\bibitem{Warringa:2006dk}
H.~J. Warringa, ``The phase diagram of neutral quark matter with pseudoscalar
  condensates in the color-flavor locked phase,''
\href{http://arxiv.org/abs/hep-ph/0606063}{{\ttfamily hep-ph/0606063}}.

\bibitem{RevModPhys.82.109}
B.~Rosenstein and D.~Li, ``Ginzburg-landau theory of type ii superconductors in
  magnetic field,'' \href{http://dx.doi.org/10.1103/RevModPhys.82.109}{{\em
  Rev. Mod. Phys.} {\bfseries 82} (Jan, 2010) 109--168}.
  \url{https://link.aps.org/doi/10.1103/RevModPhys.82.109}.

\bibitem{Iida:2002ev}
K.~Iida and G.~Baym, ``Superfluid phases of quark matter. iii: Supercurrents
  and vortices,'' {\em Phys. Rev.} {\bfseries D66} (2002) 014015,
\href{http://arxiv.org/abs/hep-ph/0204124}{{\ttfamily hep-ph/0204124}}.

\bibitem{Iida:2000ha}
K.~Iida and G.~Baym, ``The superfluid phases of quark matter: Ginzburg-landau
  theory and color neutrality,'' {\em Phys. Rev.} {\bfseries D63} (2001)
  074018,
\href{http://arxiv.org/abs/hep-ph/0011229}{{\ttfamily hep-ph/0011229}}.

\bibitem{Iida:2001pg}
K.~Iida and G.~Baym, ``{Superfluid phases of quark matter. II: Phenomenology
  and sum rules},'' \href{http://dx.doi.org/10.1103/PhysRevD.65.014022}{{\em
  Phys. Rev.} {\bfseries D65} (2002) 014022},
\href{http://arxiv.org/abs/hep-ph/0108149}{{\ttfamily arXiv:hep-ph/0108149}}.

\bibitem{Alford:2010qf}
M.~G. Alford and A.~Sedrakian, ``{Color-magnetic flux tubes in quark matter
  cores of neutron stars},''
  \href{http://dx.doi.org/10.1088/0954-3899/37/7/075202}{{\em J. Phys.}
  {\bfseries G37} (2010) 075202},
\href{http://arxiv.org/abs/1001.3346}{{\ttfamily arXiv:1001.3346
  [astro-ph.SR]}}.

\bibitem{Ferrer:2005vd}
E.~J. Ferrer, V.~de~la Incera, and C.~Manuel, ``Magnetic color flavor locking
  phase in high density qcd,'' {\em Phys. Rev. Lett.} {\bfseries 95} (2005)
  152002,
\href{http://arxiv.org/abs/hep-ph/0503162}{{\ttfamily hep-ph/0503162}}.

\bibitem{Ferrer:2006vw}
E.~J. Ferrer, V.~de~la Incera, and C.~Manuel, ``Color-superconducting gap in
  the presence of a magnetic field,'' {\em Nucl. Phys.} {\bfseries B747} (2006)
  88--112,
\href{http://arxiv.org/abs/hep-ph/0603233}{{\ttfamily hep-ph/0603233}}.

\bibitem{Noronha:2007wg}
J.~L. Noronha and I.~A. Shovkovy, ``{Color-flavor locked superconductor in a
  magnetic field},'' \href{http://dx.doi.org/10.1103/PhysRevD.76.105030,
  10.1103/PhysRevD.86.049901}{{\em Phys. Rev.} {\bfseries D76} (2007) 105030},
  \href{http://arxiv.org/abs/0708.0307}{{\ttfamily arXiv:0708.0307 [hep-ph]}}.
[Erratum: Phys. Rev.D86,049901(2012)].

\bibitem{Fukushima:2007fc}
K.~Fukushima and H.~J. Warringa, ``{Color superconducting matter in a magnetic
  field},'' \href{http://dx.doi.org/10.1103/PhysRevLett.100.032007}{{\em Phys.
  Rev. Lett.} {\bfseries 100} (2008) 032007},
\href{http://arxiv.org/abs/0707.3785}{{\ttfamily arXiv:0707.3785 [hep-ph]}}.

\bibitem{Iida:2003cc}
K.~Iida, T.~Matsuura, M.~Tachibana, and T.~Hatsuda, ``Melting pattern of
  diquark condensates in quark matter,'' {\em Phys. Rev. Lett.} {\bfseries 93}
  (2004) 132001,
\href{http://arxiv.org/abs/hep-ph/0312363}{{\ttfamily hep-ph/0312363}}.

\bibitem{Iida:2004cj}
K.~Iida, T.~Matsuura, M.~Tachibana, and T.~Hatsuda, ``Thermal phase transitions
  and gapless quark spectra in quark matter at high density,'' {\em Phys. Rev.}
  {\bfseries D71} (2005) 054003,
\href{http://arxiv.org/abs/hep-ph/0411356}{{\ttfamily hep-ph/0411356}}.

\bibitem{Schmitt:2010pf}
A.~Schmitt, S.~Stetina, and M.~Tachibana, ``{Ginzburg-Landau phase diagram for
  dense matter with axial anomaly, strange quark mass, and meson
  condensation},'' \href{http://dx.doi.org/10.1103/PhysRevD.83.045008}{{\em
  Phys. Rev.} {\bfseries D83} (2011) 045008},
\href{http://arxiv.org/abs/1010.4243}{{\ttfamily arXiv:1010.4243 [hep-ph]}}.

\bibitem{Balachandran:2005ev}
A.~P. Balachandran, S.~Digal, and T.~Matsuura, ``Semi-superfluid strings in
  high density qcd,'' {\em Phys. Rev.} {\bfseries D73} (2006) 074009,
\href{http://arxiv.org/abs/hep-ph/0509276}{{\ttfamily hep-ph/0509276}}.

\bibitem{Eto:2009kg}
M.~Eto and M.~Nitta, ``{Color Magnetic Flux Tubes in Dense QCD},''
  \href{http://dx.doi.org/10.1103/PhysRevD.80.125007}{{\em Phys. Rev.}
  {\bfseries D80} (2009) 125007},
\href{http://arxiv.org/abs/0907.1278}{{\ttfamily arXiv:0907.1278 [hep-ph]}}.

\bibitem{Vinci:2012mc}
W.~Vinci, M.~Cipriani, and M.~Nitta, ``{Spontaneous Magnetization through
  Non-Abelian Vortex Formation in Rotating Dense Quark Matter},''
  \href{http://dx.doi.org/10.1103/PhysRevD.86.085018}{{\em Phys. Rev.}
  {\bfseries D86} (2012) 085018},
\href{http://arxiv.org/abs/1206.3535}{{\ttfamily arXiv:1206.3535 [hep-ph]}}.

\bibitem{Alford:2016dco}
M.~G. Alford, S.~K. Mallavarapu, T.~Vachaspati, and A.~Windisch, ``{Stability
  of superfluid vortices in dense quark matter},''
  \href{http://dx.doi.org/10.1103/PhysRevC.93.045801}{{\em Phys. Rev.}
  {\bfseries C93} no.~4, (2016) 045801},
\href{http://arxiv.org/abs/1601.04656}{{\ttfamily arXiv:1601.04656 [nucl-th]}}.

\bibitem{Vachaspati:1991dz}
T.~Vachaspati and A.~Achucarro, ``{Semilocal cosmic strings},''
\href{http://dx.doi.org/10.1103/PhysRevD.44.3067}{{\em Phys. Rev.} {\bfseries
  D44} (1991) 3067--3071}.

\bibitem{1991PhRvL..66.3071L}
F.~{Liu}, M.~{Mondello}, and N.~{Goldenfeld}, ``{Kinetics of the
  superconducting transition},''
  \href{http://dx.doi.org/10.1103/PhysRevLett.66.3071}{{\em Physical Review
  Letters} {\bfseries 66} (June, 1991) 3071--3074}.

\bibitem{Lin:2007rz}
L.-M. Lin, ``Constraining crystalline color superconducting quark matter with
  gravitational-wave data,''
\href{http://arxiv.org/abs/arXiv:0708.2965 [astro-ph]}{{\ttfamily
  arXiv:0708.2965 [astro-ph]}}.

\bibitem{Haskell:2007sh}
B.~Haskell, N.~Andersson, D.~I. Jones, and L.~Samuelsson, ``Is ligo already
  constraining the parameters of qcd?,''
\href{http://arxiv.org/abs/arXiv:0708.2984 [gr-qc]}{{\ttfamily arXiv:0708.2984
  [gr-qc]}}.

\bibitem{Knippel:2009st}
B.~Knippel and A.~Sedrakian, ``{Gravitational radiation from crystalline
  color-superconducting hybrid stars},''
  \href{http://dx.doi.org/10.1103/PhysRevD.79.083007}{{\em Phys. Rev.}
  {\bfseries D79} (2009) 083007},
\href{http://arxiv.org/abs/0901.4637}{{\ttfamily arXiv:0901.4637
  [astro-ph.SR]}}.

\bibitem{Anglani:2013gfu}
R.~Anglani, R.~Casalbuoni, M.~Ciminale, N.~Ippolito, R.~Gatto, M.~Mannarelli,
  and M.~Ruggieri, ``{Crystalline color superconductors},''
  \href{http://dx.doi.org/10.1103/RevModPhys.86.509}{{\em Rev. Mod. Phys.}
  {\bfseries 86} (2014) 509--561},
\href{http://arxiv.org/abs/1302.4264}{{\ttfamily arXiv:1302.4264 [hep-ph]}}.

\bibitem{Schmitt:2003aa}
A.~Schmitt, Q.~Wang, and D.~H. Rischke, ``Mixing and screening of photons and
  gluons in a color superconductor,'' {\em Phys. Rev.} {\bfseries D69} (2004)
  094017,
\href{http://arxiv.org/abs/nucl-th/0311006}{{\ttfamily nucl-th/0311006}}.

\bibitem{Schmitt:2002sc}
A.~Schmitt, Q.~Wang, and D.~H. Rischke, ``When the transition temperature in
  color superconductors is not like in bcs theory,'' {\em Phys. Rev.}
  {\bfseries D66} (2002) 114010,
\href{http://arxiv.org/abs/nucl-th/0209050}{{\ttfamily nucl-th/0209050}}.

\bibitem{Alford:2002kj}
M.~G. Alford and K.~Rajagopal, ``Absence of two-flavor color superconductivity
  in compact stars,'' {\em JHEP} {\bfseries 06} (2002) 031,
\href{http://arxiv.org/abs/hep-ph/0204001}{{\ttfamily hep-ph/0204001}}.

\bibitem{Forbes:2001gj}
M.~M. Forbes and A.~R. Zhitnitsky, ``Global strings in high density qcd,'' {\em
  Phys. Rev.} {\bfseries D65} (2002) 085009,
\href{http://arxiv.org/abs/hep-ph/0109173}{{\ttfamily hep-ph/0109173}}.

\bibitem{Ferrer:2010wz}
E.~J. Ferrer, V.~de~la Incera, J.~P. Keith, I.~Portillo, and P.~L. Springsteen,
  ``{Equation of State of a Dense and Magnetized Fermion System},''
  \href{http://dx.doi.org/10.1103/PhysRevC.82.065802}{{\em Phys. Rev.}
  {\bfseries C82} (2010) 065802},
\href{http://arxiv.org/abs/1009.3521}{{\ttfamily arXiv:1009.3521 [hep-ph]}}.

\bibitem{Son:2007ny}
D.~T. Son and M.~A. Stephanov, ``{Axial anomaly and magnetism of nuclear and
  quark matter},'' \href{http://dx.doi.org/10.1103/PhysRevD.77.014021}{{\em
  Phys. Rev.} {\bfseries D77} (2008) 014021},
\href{http://arxiv.org/abs/0710.1084}{{\ttfamily arXiv:0710.1084 [hep-ph]}}.

\bibitem{Chernodub:2010sg}
M.~N. Chernodub and A.~S. Nedelin, ``{Pipelike current-carrying vortices in
  two-component condensates},''
  \href{http://dx.doi.org/10.1103/PhysRevD.81.125022}{{\em Phys. Rev.}
  {\bfseries D81} (2010) 125022},
\href{http://arxiv.org/abs/1005.3167}{{\ttfamily arXiv:1005.3167 [hep-th]}}.

\bibitem{Cipriani:2012hr}
M.~Cipriani, W.~Vinci, and M.~Nitta, ``{Colorful boojums at the interface of a
  color superconductor},''
  \href{http://dx.doi.org/10.1103/PhysRevD.86.121704}{{\em Phys. Rev.}
  {\bfseries D86} (2012) 121704},
\href{http://arxiv.org/abs/1208.5704}{{\ttfamily arXiv:1208.5704 [hep-ph]}}.

\bibitem{Alford:2018mqj}
M.~G. Alford, G.~Baym, K.~Fukushima, T.~Hatsuda, and M.~Tachibana,
  ``{Continuity of vortices from the hadronic to the color-flavor locked phase
  in dense matter},''
\href{http://arxiv.org/abs/1803.05115}{{\ttfamily arXiv:1803.05115 [hep-ph]}}.

\bibitem{Chatterjee:2018nxe}
C.~Chatterjee, M.~Nitta, and S.~Yasui, ``{Quark-hadron continuity under
  rotation: vortex continuity or boojum?},''
\href{http://arxiv.org/abs/1806.09291}{{\ttfamily arXiv:1806.09291 [hep-ph]}}.

\bibitem{Penner:2011pd}
A.~J. Penner, N.~Andersson, L.~Samuelsson, I.~Hawke, and D.~I. Jones, ``{Tidal
  deformations of neutron stars: The role of stratification and elasticity},''
  \href{http://dx.doi.org/10.1103/PhysRevD.84.103006}{{\em Phys. Rev.}
  {\bfseries D84} (2011) 103006},
\href{http://arxiv.org/abs/1107.0669}{{\ttfamily arXiv:1107.0669
  [astro-ph.SR]}}.

\bibitem{Lau:2017qtz}
S.~Y. Lau, P.~T. Leung, and L.~M. Lin, ``{Tidal deformations of compact stars
  with crystalline quark matter},''
  \href{http://dx.doi.org/10.1103/PhysRevD.95.101302}{{\em Phys. Rev.}
  {\bfseries D95} no.~10, (2017) 101302},
\href{http://arxiv.org/abs/1705.01710}{{\ttfamily arXiv:1705.01710
  [astro-ph.HE]}}.

\bibitem{Ho:2017bia}
W.~C.~G. Ho, N.~Andersson, and V.~Graber, ``{Dynamical onset of
  superconductivity and retention of magnetic fields in cooling neutron
  stars},'' \href{http://dx.doi.org/10.1103/PhysRevC.96.065801}{{\em Phys.
  Rev.} {\bfseries C96} no.~6, (2017) 065801},
\href{http://arxiv.org/abs/1711.08480}{{\ttfamily arXiv:1711.08480
  [astro-ph.HE]}}.

\bibitem{baym_sfCS}
G.~Baym, C.~Pethick, and D.~Pines, ``Superfluidity in neutron stars,''
  \href{http://dx.doi.org/10.1038/224673a0}{{\em Nature} {\bfseries 224} (11,
  1969) 673--674}. \url{http://dx.doi.org/10.1038/224673a0}.

\bibitem{1964PhRvL..13..508H}
P.~W. {Higgs}, ``{Broken Symmetries and the Masses of Gauge Bosons},''
  \href{http://dx.doi.org/10.1103/PhysRevLett.13.508}{{\em Physical Review
  Letters} {\bfseries 13} (Oct., 1964) 508--509}.

\bibitem{1964PhRvL..13..321E}
F.~{Englert} and R.~{Brout}, ``{Broken Symmetry and the Mass of Gauge Vector
  Mesons},'' \href{http://dx.doi.org/10.1103/PhysRevLett.13.321}{{\em Physical
  Review Letters} {\bfseries 13} (Aug., 1964) 321--323}.

\bibitem{Hebecker:1993rz}
A.~Hebecker, ``{Finite temperature effective potential for the Abelian Higgs
  model to the order e**4, lambda**2},''
  \href{http://dx.doi.org/10.1007/BF01474623}{{\em Z. Phys.} {\bfseries C60}
  (1993) 271--276},
\href{http://arxiv.org/abs/hep-ph/9307268}{{\ttfamily arXiv:hep-ph/9307268
  [hep-ph]}}.

\bibitem{Arnold:1992rz}
P.~B. Arnold and O.~Espinosa, ``{The Effective potential and first order phase
  transitions: Beyond leading-order},''
  \href{http://dx.doi.org/10.1103/PhysRevD.50.6662,
  10.1103/PhysRevD.47.3546}{{\em Phys. Rev.} {\bfseries D47} (1993) 3546},
  \href{http://arxiv.org/abs/hep-ph/9212235}{{\ttfamily arXiv:hep-ph/9212235
  [hep-ph]}}.
[Erratum: Phys. Rev.D50,6662(1994)].

\bibitem{num_recipies}
W.~H. Press, S.~A. Teukolsky, W.~T. Vetterling, and B.~P. Flannery, {\em
  Numerical Recipes in FORTRAN; The Art of Scientific Computing}.
\newblock Cambridge University Press, New York, NY, USA, 2nd~ed., 1993.

\bibitem{lapack}
E.~Anderson, Z.~Bai, C.~Bischof, S.~Blackford, J.~Demmel, J.~Dongarra,
  J.~Du~Croz, A.~Greenbaum, S.~Hammarling, A.~McKenney, and D.~Sorensen, {\em
  {LAPACK} Users' Guide}.
\newblock Society for Industrial and Applied Mathematics, Philadelphia, PA,
  third~ed., 1999.

\end{thebibliography}\endgroup

\end{document}